%% file: thesis.tex
\documentclass[11pt,a4paper,twoside]{article}
\usepackage[bottom=20mm, inner=30mm, outer=30mm, top=20mm]{geometry} 
\usepackage{pdflscape}
\usepackage{parskip}    
\usepackage{graphicx}
\usepackage{subfigure}
\usepackage{atlasbiblatex}
\usepackage{mydefs}
\usepackage{atlasunit}
\usepackage{feynmp}
\usepackage{booktabs}
\usepackage{multirow}
\usepackage[colorlinks=true,linkcolor=thesisblue,citecolor=thesisred]{hyperref}
\usepackage{url}
\usepackage{slashed}
\usepackage[utf8]{inputenc}
\usepackage[T1]{fontenc}
\usepackage{tikz}
\usepackage{mdwlist}

\usepackage{libertine}
\usepackage[libertine]{newtxmath}
\usepackage{pdfpages}

\usepackage{lineno}

\title{Measurements of ZZ production with the ATLAS detector and simulation of loop-induced processes with the \HERWIG{} event generator}

\begin{document}
\begin{fmffile}{feynman_diagrams}



\newgeometry{bottom=20mm, inner=30mm, outer=30mm, top=20mm}

\thispagestyle{empty}
\phantom{.}
\vspace{-2.7cm}
\hspace{-4cm}\includegraphics[width=1.05\paperwidth]{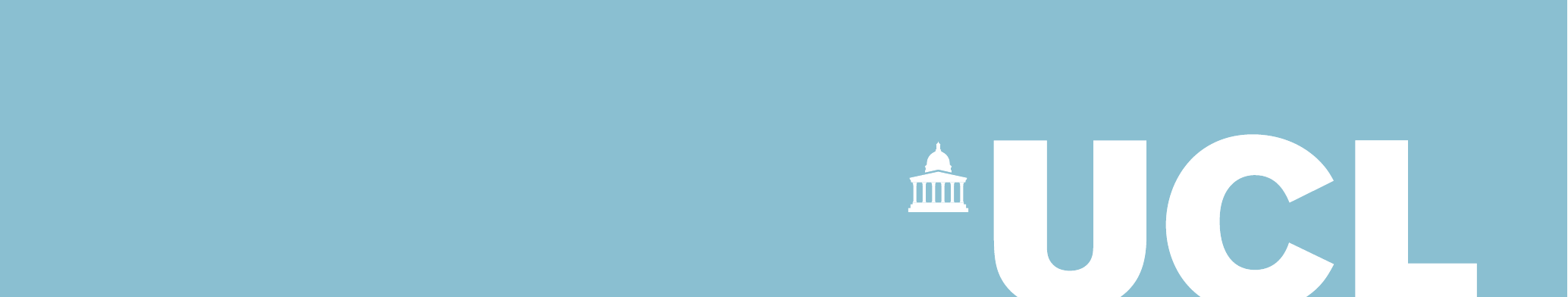}

\vspace{36mm}

\begin{center}
{\huge
\textsf{Measurements of ZZ production with the ATLAS\\ detector and simulation of loop-induced processes\\ with the \HERWIG{} event generator\\\phantom{.}}
}

{
\vspace{3cm}
\LARGE
\textsf{Stefan Richter}\\[2mm]
\textsf{December 7, 2017}
}

\vfill

{\large
Submitted to University College London in fulfilment of the requirements\\ for the award of the degree of Doctor of Philosophy.\\
}

\end{center}

\hfill

\clearpage
\restoregeometry



\phantom{.}
\vspace{4mm}

\section*{Declaration}
I, Stefan Richter, confirm that the work presented in this thesis is my own. Where information has been derived from other sources, I confirm that this has been indicated in the thesis.

\vspace{1cm}

\hspace{0.56\textwidth}Signed:

\vfill
\clearpage
\phantom{.}
\vspace{4mm}

\section*{Abstract}
This thesis presents results and method developments in both experimental and theoretical particle physics.
The main part shows measurements of \llll{} production (where $\ell$, $\ell'$ is either an electron or a muon) in proton-proton collisions at 13~\TeV{} centre-of-mass energy. 
The collisions were produced by the Large Hadron Collider in 2015 and 2016 and observed with the ATLAS detector.
In a phase space sensitive to \PZ{} boson pair production, the integrated cross section as well as differential cross sections with respect to twenty-one observables are measured.
Ten of these directly measure associated jet activity.
The measurements provide an important test of the Standard Model of particle physics.
A direct search for effects beyond the Standard Model affecting \ZZ{} production is performed in a generic effective field theory approach.
No significant deviations from the Standard Model predictions are observed.
Exclusion limits are set on the parameters describing new physics in the effective field theory.
In theoretical developments, the automated description of loop-induced processes with the \HERWIG{}~7 event generator is presented. These are processes that can only occur via a quantum loop of virtual particles.
Preliminary results in leading-order quantum chromodynamics are shown for the production of a Higgs boson, of a pair of Higgs bosons, and of four leptons.
The Higgs boson results show that the full loop-induced description can deviate significantly from the common approximation where the mass of the top quark is treated as infinitely large.
Thus, including loop effects is crucial to obtaining precise predictions to compare to measurements at the Large Hadron Collider.
Developments towards a next-to-leading-order description of arbitrary loop-induced processes are shown.

\vfill
\clearpage
\phantom{.}
\vspace{4mm}

\section*{Acknowledgements}
It has been quite a journey, and much has changed since it began. What you are reading here is its streamlined narrative: few of the failures, most of the successes, the work of weeks distilled into a paragraph. Of course, none of this would have been possible without the support of many people.

First, I would like to thank my primary supervisor, Emily Nurse. You were simply perfect for me. I'm grateful for your remarkable physical insight, unfailing support in all aspects of my work, and the independence you gave me to do what I wanted to. I only wish we could disagree more often, to make our discussions even more interesting.
I am also immensely grateful to my secondary supervisor, Keith Hamilton. I thoroughly enjoyed our work together, particularly the passion and brilliance with which you introduced me to QCD phenomenology. For space-time reasons, the project in which you supervised me (detailed studies of NLO parton shower matching in $\Ptop\APtop$ production) did not make it into this thesis, but its value is undiminished. I thank my collaborator Simon Pl\"{a}tzer for informally supervising my theoretical work with the \HERWIG{} event generator. Your encouragement and astonishingly versatile and deep knowledge was a great asset in that project.

I have benefitted greatly from the immense knowledge, versatility, and experience in the UCL high energy physics group and would like to extend my gratitude and best wishes to all its members. The same is true for the Standard Model and (therein) Electroweak working groups at ATLAS, whose input and discussions have contributed to my education as a physicist in a major way. In this context, I would like to particularly mention conveners Ulla Blumenschein, Jan Kretzschmar, Matthias Schott, \raisebox{-0.4pt}{\includegraphics{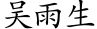}} (Yusheng Wu), Louis Helary, and Kristin Lohwasser for their continuous support of my work and much useful advice.
I thank all my analysis team partners, especially my long-time ATLAS collaborators Maurice Becker, Will Buttinger, Jonatan Rost\'{e}n, and \raisebox{-0.65pt}{\includegraphics{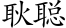}} (Cong Geng). You were a pleasure to work with and I am proud of what we have achieved together. I also thank Jon Butterworth and Andy Buckley for their guidance in many aspects of ATLAS physics, particularly QCD phenomenology and minimising model dependence in measurements, as well as Sarah Heim for her valuable mentorship and support.
My PhD studentship was enhanced immeasurably by being a member of MCnet. I greet and thank my many MCnet colleagues. Big thanks belong to Christian Reuschle and \raisebox{-3.1pt}{\includegraphics{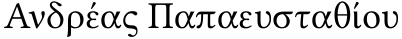}} (Andreas Papaefstathiou) for the close collaboration on \HERWIG{}. I would like to thank Marius Wiesemann and Stefan Kallweit for \matrixnnlo{} support and discussions and Jonas Lindert for \openloops{} support.

My PhD studies also included an internship in the company Blue Yonder based in the German city of Karlsruhe. I would like to express my gratitude to Manuel B\"{a}hr and Stefan Gieseke for providing me with this opportunity. It taught me more than I could ever have expected and had a large positive impact also on my academic research. My warmest regards go out to everyone at Blue Yonder, especially my friends in the former Tech Team (who collectively supervised me) and the company's founder, Michael Feindt, whom I thank for the enthusiastic discussions.


I thank the University of Vienna for hosting me for a very enjoyable week while working on \HERWIG{} and the University of Z\"{u}rich (in particular professors Florencia Canelli and Ben Kilminster) for accommodating me for several weeks while writing this thesis. The warm welcome I received from the CMS group there (technically our competitor) was astonishing. I am looking forward to repaying your kindness. I thank the administrative staff at all my institutions, especially Nadia Waller (UCL) as well as Nanie Perrin and Michelle Connor (CERN), for their excellent support.



I gratefully acknowledge generous funding from the European Union via the Marie Sk\l{}odowska-Curie Innovative Training Network \emph{MCnet}, which funded my academic work at UCL and CERN, as well as my industry internship in Karlsruhe, for a total of three years. UCL kindly supported me for one year with an \emph{Impact} studentship.
I thank many institutions for smaller grants for conference and workshop attendance, in particular the Wilhelm und Else Heraeus-Stiftung, IPPP, and STFC. Furthermore, I am indebted to David Grellscheid and Ivan Girotto for inviting me to teach scientific-software design at a summer school at Sharif University of Technology in Tehran and ICTP for organising the school and funding me.
Thanks to Mike Seymour and Gavin Hesketh for agreeing to examine my thesis.

To finish, I would like to thank the people I love. You have kept me going and shaped who I am. I have been privileged to have many wonderful friends, and apologise that I cannot mention them all by name here. I would particularly like to mention my physicist friends Thibaud Humair, Slavom\'{i}ra \v{S}tefkov\'{a}, Fr\'{e}d\'{e}ric Dreyer, Tevong You, \raisebox{-0.1pt}{\includegraphics{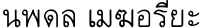}} (Noppadol Mekareeya), Rob Knoops, Hiu Fung Wong (\raisebox{-0.65pt}{\includegraphics{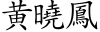}}), Kristian Gregersen, Alex Martyniuk, Maciej Pf\"{u}tzner, Stephen Jiggins, Suzanne Klaver, Giulio Dujany, Markus Hauru, and Carl-Johan Haster. In my northern home and beyond, I extend my warmest friendship to Stephan Schulz, Torsti Schulz, Joona and Oona Hulmi, Emma and Moritz Kortekangas, Soile Ylivuori, Pyry Kivisaari, Pia Niemi, Minna and the whole Lindfors family, as well as Hanna Mustaniemi. Lots of love goes out to my Europa-Kolleg friends Marta Garc\'{i}a Aliaga, Antje Rei\ss{}, Sara Garc\'{i}a Arteagoitia, Istv\'{a}n Dalicsek, D\'{a}niel Dalicsek, and the rest of you amazing people.
My wonderful parents Erika and Hans-Peter have supported me in every possible way my whole life. Whatever I may have achieved is also their achievement. My awesome siblings Annika and Martin have enriched my life beyond measure and are my heroes that I will always look up to. Big thanks also go out to my grandparents and extended family. I would like to end by thanking Annick, my best friend, my soul-mate, and my love. Every day with you is precious and makes me try to be the best I can.

\vspace{1cm}
\begin{flushright}
Stefan Richter\\
Helsinki, December 2017
\end{flushright}

\vfill
\clearpage

\phantom{.}
\vspace{7cm}
\begin{center}
{\Large
\emph{Omistettu Suomelle, joka juhlii 100-vuotiasta itsen\"{a}isyytt\"{a} t\"{a}n\"{a} vuonna}\\
\emph{Till\"{a}gnad Finland, som i \aa{}r firar 100 \aa{}r av sj\"{a}lvst\"{a}ndighet}\\
\emph{Dedicated to Finland, celebrating 100 years of independence this year}\\
}
\end{center}

\vfill
\clearpage

\tableofcontents
\vfill
\pagebreak
\listoffigures
\vfill
\pagebreak
\listoftables
\vfill

\clearpage\pagebreak
\part{Introduction}
\label{sec:overall_intro}

\section{Prologue: into the unknown}

Particle physics tries to answer the question, what the fundamental building blocks of Nature and their interactions are. Historically, it has been a matter of diving deeper into the structure of matter: many particles and interactions that were once believed elementary were later found to be emergent manifestations of more fundamental ingredients. In the early $19^{\text{th}}$ century, all chemical elements were discovered to be made of atoms, which were thought to be indivisible. However, around a hundred years later, the discovery of the electron (Thomson 1897) as well as the experimental demonstration that the positive charge inside an atom is located at its almost pointlike centre (Geiger, Marsden, Rutherford 1909--1911) led to the realisation that atoms were in fact composite objects, consisting of a central nucleus surrounded by electrons. The discovery of the neutron (Chadwick 1932) quickly led to a model of the atomic nucleus being made up of neutrons and protons (Ivanenko, Heisenberg 1932). Almost at the same time, the discovery of the positron (Anderson 1932) established the existence of antimatter. It is now known that all fundamental particles either have an antiparticle, or are their own antiparticle. The muon (Anderson, Neddermeyer, et al. 1936) and the pion (Lattes, Occhialini, Powell, et al. 1947) were the first discovered particle species that are not part of atoms. Both are unstable, with average lifetimes of the order of 1~\textmu{}s and 0.01~\textmu{}s, respectively. Starting in the years following the Second World War, many fresh particle discoveries led to an increasingly complex view of the structure of matter. By the 1960s, more and more powerful human-made particle accelerators replaced cosmic rays as the primary source of particle collisions to study. This enabled the performing of deep-inelastic-scattering experiments in the late 1960s that revealed many of the discovered particles to be composed of more fundamental ones, now called quarks. Particles composed of quarks are collectively referred to as \emph{hadrons}. Six types (`flavours') of quarks have been found to date, called up, down, charm, strange, top, and bottom. These are elementary particles \emph{to our current knowledge}. The other currently known particles that are thought to be elementary are three types of charged lepton (electron, muon, tau lepton) and their three corresponding neutrinos (electron neutrino, muon neutrino, tau neutrino), as well as four kinds of elementary boson. According to the Standard Model (SM) of particle physics, all known interactions except gravity can be described as the exchange of bosons between elementary particles. The electromagnetic interaction is described by the exchange of photons, the weak interaction by the exchange of charged \PW{} or neutral \PZ{} bosons, the strong interaction by gluon exchange, and the Higgs boson mediates Higgs interactions.\footnote{Technical note: the Higgs boson occupies a different role in the SM than the other bosons, because it does not correspond to a gauge group.} Gravity does not yet fit into this picture: there is currently no successful, tested quantum theory of gravity. However, it is so weak compared to the other forces that its effects are inconsequential in the study of particle collisions, as far as is known today. The SM is briefly summarised in \mypart~\ref{sec:theory_intro} of this thesis. The laws of Nature are quantum mechanical. While this is not immediately apparent in everyday situations due to the (typically) very small relative size of quantum effects and their `averaging out' in macroscopic systems made of countless particles, it determines the behaviour at the level of elementary particles. The SM can only predict \emph{probabilities} for events to occur, while events themselves are stochastic. This implies the need for data from many particle collisions and for statistical inference to draw conclusions about the underlying physical laws.

The research presented in this thesis is mainly experimental, although smaller contributions to the theoretical description of Nature are also presented. Even more than most scientific fields, experimental particle physics has organically grown to be driven by large international collaborations, with several dozen up to several thousand members. This is simply owing to the volume and complexity of the necessary instruments and the data they produce. While the relevance of pure particle physics to most people's everyday lives is limited --- although no more than that of, say, the opera or kickboxing --- particle physics plays a paramount role in revealing the fundamental secrets of the universe. It is the author's opinion that, being predominantly publicly funded, particle physicists are ultimately accountable to the general public, serving them by providing knowledge and a share in the adventure of exploring the unknown, pushing technologies beyond the state of the art, inspiring generations of young people to become scientists and engineers and educators, fostering international collaboration and global peace, challenging the status quo of human thought, promoting an egalitarian and rational worldview, and sometimes creating technological opportunities for the direct benefit of humankind as a side product (World Wide Web, proton cancer therapy, several medical diagnostics methods, data analysis tools, \ldots). The heart of the global particle physics community is undoubtedly CERN,\footnote{European Organisation for Nuclear Research, acronym derived from the historical name \emph{Conseil Europ\'een pour la Recherche Nucl\'eaire}.} at whose laboratory near the Swiss city of Geneva much of the author's research was conducted. The experimental results were obtained using data collected by the ATLAS detector at CERN's Large Hadron Collider. The experimental setup is described in \mypart~\ref{sec:experiment}.

\mypart~\ref{sec:analysis} gives a detailed account of a measurement of the production of \PZ{}-boson pairs in the highest-energy proton-proton collisions produced to date, with a centre-of-mass energy of thirteen teraelectronvolts. This is a rare process. The measured results are compared to the SM predictions. No statistically significant deviations are found. This allows excluding a priori possible values of parameters in a very generic parametrisation of non-SM physics effects. The measurement results and analysis are also preserved in human- and machine-readable format to allow comparisons to improved theoretical predictions or particular models for non-SM physics at any future time. \mypart~\ref{sec:analysis} also very briefly summarises other measurements that the author was involved in. All of the data analyses were done collaboratively in a small team, so any significant direct contributions that were not made by the author are clearly marked as such. Of course the work as a whole would not have been possible without the inputs of countless other people. 
The topic of \mypart~\ref{sec:loopinduced} is theoretical work towards improving the description of particle processes where the initial-state and final-state particles cannot interact with each other directly, but only via intermediate particles, in a quantum-mechanical effect called a \emph{loop}. More precise theory predictions are crucial for testing the validity of the SM, especially as increasing data and know-how will allow ever more precise and accurate measurements. The new predictions are implemented in the Herwig event generator software. Finally, \mypart~\ref{sec:appendix} is an appendix containing technical details and further results whose inclusion in earlier parts would harm the readability of the thesis.





\subsection{Notation, units, and definitions}

Three-vectors are denoted using boldface, while four-vectors are not: $p = (E, \text{\textbf{\emph{p}}})$. Loop-induced processes are indicated by an arrow with a loop (\looparrow{}) instead of an ordinary reaction arrow ($\to$). To be precise, the loop arrow means that the \emph{colourless} final-state particles must all be connected directly to a loop. For instance, the reaction $\Pproton\Pproton \,\looparrow{}\, \ZZ \Pgluon$ contains only those subprocesses where both \PZ{} bosons couple to a loop, whereas the final-state gluon may originate from anywhere in the hard process.

\paragraph{Natural units}\hfill\\
Natural units are frequently used in this thesis. While the relevant SI\footnote{\emph{Syst\`{e}me International (d'unit\'{e}s)}} base dimensions and units are length (m), time (s), and mass (kg), those of the natural-units system used in particle physics are speed (speed of light $c$), angular momentum (reduced Planck constant $\hbar$), and energy (gigaelectronvolts \GeV{}). The units $c$ and $\hbar$ are usually implied and not written for dimensionful quantities. In particular, energy (\GeV{}), momentum ($\GeV/c$), and mass ($\GeV/c^2$) are all expressed in $\GeV{}$, with implicit relevant powers of $c$. Charge is expressed in SI units, with the elementary charge $1e \approx 1.6 \times 10^{-19}$~C.

\paragraph{Detector frame of reference}\hfill\\
Coordinates in the rest frame of a detector at a circular collider are expressed using a right-handed coordinate system with its origin at the nominal interaction point in the centre of the detector and the $z$-axis along the beam pipe. The $x$-axis points to the centre of the collider ring, and the $y$-axis points upward. Cylindrical coordinates (\kern1pt$\rho$, $\phi$) are used in the transverse plane, $\phi$ being the azimuthal angle around the $z$-axis. Angles, notably the azimuthal angle $\phi$ in the detector frame, are given in radians unless otherwise noted.

\paragraph{Useful kinematical variables}\hfill\\
Rapidity is defined in terms of a particle's energy and longitudinal momentum as
\begin{equation*}
y = \frac{1}{2}\ln\frac{E + p_z}{E - p_z}.
\end{equation*}
If the particle has a small mass ($m \ll E$), the rapidity can be approximated by pseudorapidity
\begin{equation*}
\eta = \frac{1}{2}\ln\frac{|\text{\textbf{\textit{p}}}| + p_z}{|\text{\textbf{\textit{p}}}| - p_z} = \text{arctanh} \frac{p_z}{|\text{\textbf{\textit{p}}}|} = - \ln[\tan(\theta/2)],
\end{equation*}
which depends only on the polar angle $\theta$.
Transverse momentum $\pt$ is the projection of momentum onto the transverse plane,
\begin{equation*}
\pt = \sqrt{p_x^2 + p_y^2}.
\end{equation*}
The angular distance between two systems is given by their Pythagorean distance in the $\eta$-$\phi$ plane,
\begin{equation*}
\Delta R = \sqrt{(\Delta\eta)^2 + (\Delta\phi)^2}.
\end{equation*}
The invariant mass of a system is given by the norm of its four-momentum,
\begin{equation*}
m = |p| =  \sqrt{\bigg(\sum_{i}^{\text{constituents}} p_i \bigg)^2}.
\end{equation*}
The Mandelstam variables $s$, $t$, and $u$ in a $2 \to 2$ scattering processes with momenta with labels and directions as
\vspace{5mm}
\begin{center}
	\begin{fmfgraph*}(70,30)
		\fmfset{arrow_len}{3mm}
		\fmfstraight
		\fmfleft{i0,i1,i2}
		\fmfright{o0,o1,o2}
		\fmflabel{$p_1$}{o2}
		\fmflabel{$p_2$}{o0}
		\fmflabel{$p_a$}{i2}
		\fmflabel{$p_b$}{i0}
		\fmf{fermion}{i2,v0}
		\fmf{fermion}{i0,v0}
		\fmf{fermion}{v0,o0}
		\fmf{fermion}{v0,o2}
		\fmfblob{6.6mm}{v0}
	\end{fmfgraph*}
\end{center}
\vspace{5mm}
are defined as
\begin{equation*}
\begin{split}
s &= (\kern1ptp_a + p_b)^2 = (\kern1ptp_1 + p_2)^2,\\
t &= (\kern1ptp_a - p_1)^2 = (\kern1ptp_b - p_2)^2,\\
u &= (\kern1ptp_a - p_2)^2 = (\kern1ptp_b - p_1)^2.\\
\end{split}
\end{equation*}
The variants above without a caret ($\hat{\phantom{o}}$) refer to the proton momenta, while the variants with a caret ($\hat{s}$, $\hat{t}$, $\hat{u}$) refer to partonic momenta. Using the definition $s = (\kern1ptp_a + p_b)^2$, the Mandelstam $s$ ($\hat{s}$) is generalised to $2 \to N$ processes with final states of arbitrary multiplicity $N$ and is equal to the square of the hadronic (partonic) centre-of-mass energy.


\clearpage\pagebreak
\part{Theory}
\label{sec:theory_intro}

\section{Standard model of particle physics}\label{sec:theory_sm}


The SM is a quantum field theory describing the electromagnetic, weak, and strong interactions, as well as the interactions of the Higgs boson.
The electromagnetic interaction (affecting photons and all charged fermions), together with gravity, is the fundamental interaction governing most aspects of daily life. It determines such things as the structure of materials above subatomic size scales, chemical reactions, and the behaviour of light. The weak interaction affects all fermions, the weak bosons ($\PW$, $\PZ$), and the Higgs boson. It is responsible for radioactive $\beta$ decay. Despite its name, its coupling strength is larger than the electromagnetic one at short distances. However, its force carriers, the $\PW$ and $\PZ$ bosons are not massless like the photon, but have masses of $\mathcal{O}(100~\GeV{})$. This greatly suppresses the weak interaction at larger distance scales, so that its effects are practically unobservable in everyday life. The strong force affects all particles carrying \emph{colour charge}: quarks and gluons. It has massless force carriers, the gluons, but its effects are confined to very short distances for a different reason. The coupling strengths $g$ in quantum field theory in general satisfy the Callan-Symanzik equation \cite{Callan:1970yg,Symanzik:1970rt}
\begin{equation}\label{eq:callan_symanzik}
Q \frac{\partial g}{\partial Q} = \beta(g),
\end{equation}
where $Q$ is the energy scale of the coupling and $\beta$ is some function that can be expanded as a power series of $g$, whose leading terms can be computed. For the strong interaction, $\beta < 0$, while for the electromagnetic and weak interactions, $\beta > 0$. So contrarily to the electromagnetic and weak coupling strengths, the coupling strength of the strong interaction is small at high energies and large at low energies. In the limit $Q \to \infty$, the particles become decoupled. This is called asymptotic freedom. At low energies, the interaction strength is so high that strongly interacting particles can never be observed in isolation. This is called confinement. Trying to pull a strongly bound system apart, the energy between the particles becomes so large that new pairs are created between them, forming new bound states with the original particles. 
Despite confinement into colour-neutral objects, there is a \emph{residual} strong force called the \emph{nuclear force} acting at distances of around 1~fm ($10^{-15}$~m) that is responsible for binding protons and neutrons together to form atomic nuclei. The nuclear force is analogous to the van der Waals force exerted by electrically neutral molecules. Beyond a few femtometres, the nuclear force is negligible compared to the electromagnetic force, so it plays no role in e.g.~atoms binding to form molecules.

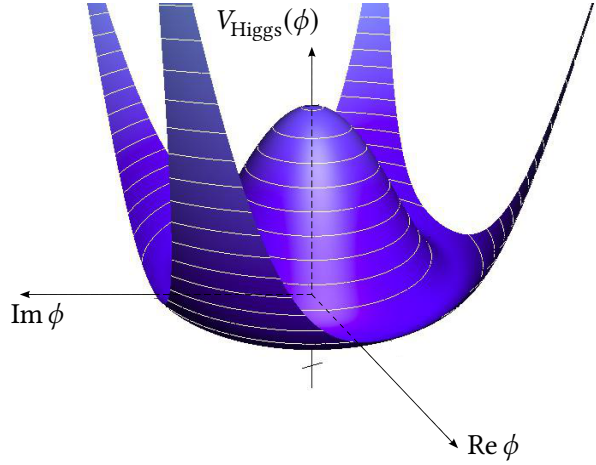
\begin{figure}[h!]
\centering
\begin{tikzpicture}
\node[anchor=south west,inner sep=0] (image) at (0,0) {\includegraphics[width=0.8\textwidth]{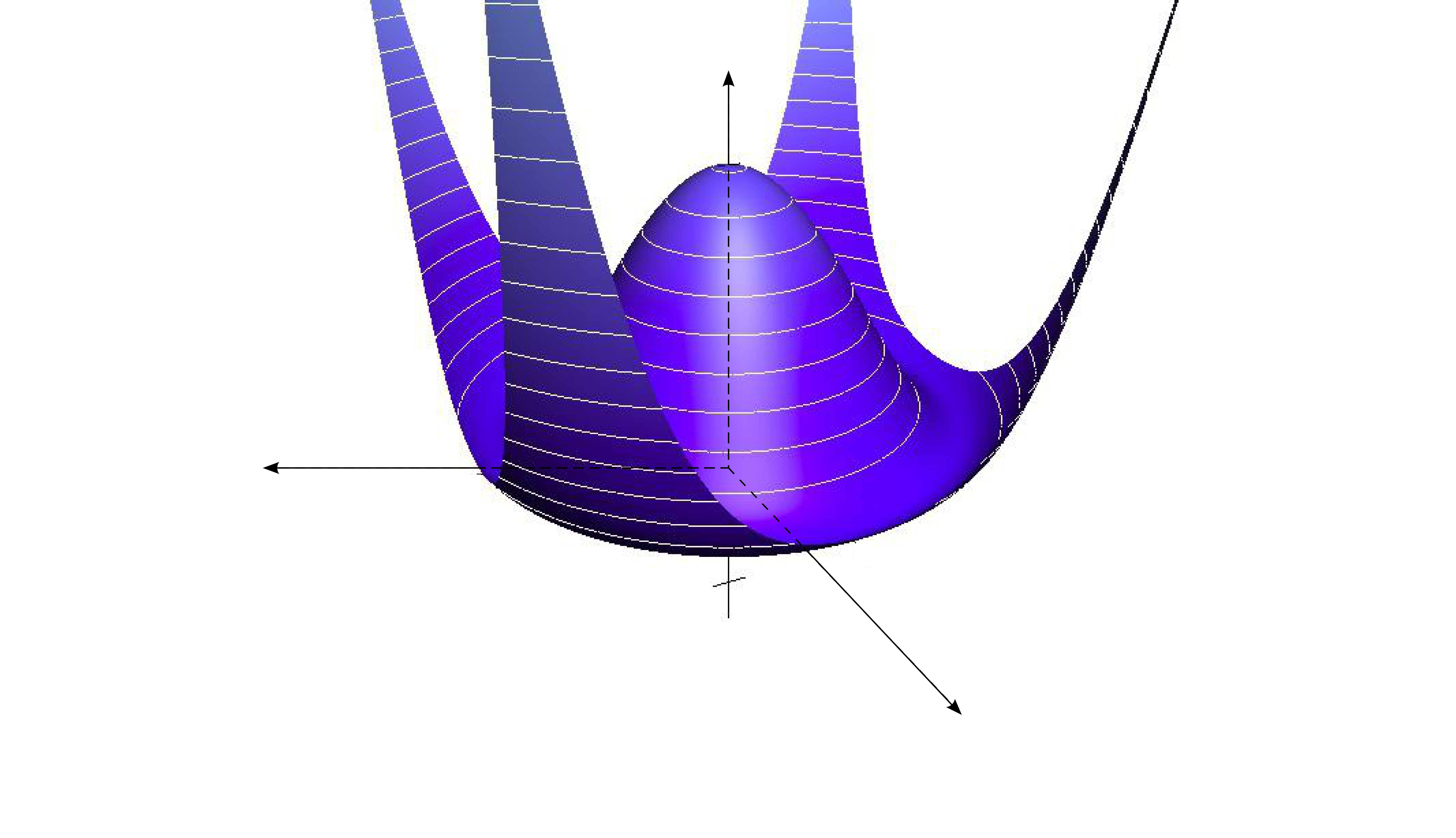}};
\begin{scope}[x={(image.south east)},y={(image.north west)}]
\node[anchor=west] at (0.66, 0.15) {$\text{Re}\, \phi$};
\node[anchor=west] at (0.16, 0.4) {$\text{Im}\, \phi$};
\node[anchor=east] at (0.52, 0.96) {$V_{\text{Higgs}}(\phi)$};
\end{scope}
\end{tikzpicture}
\vspace{-8mm}
\caption{Shape of the Higgs potential.}
\label{fig:higgs_potential}
\end{figure}

As a quantum field theory, the SM treats particles as excitations of quantum fields. Each quantum field, and therefore each type of particle, has well-defined quantum numbers and mass. Particles with zero mass are called massless.
Except for the interactions involving the Higgs boson, all interactions can be introduced by \emph{requiring} the Lagrangian to be invariant under a gauge transformation, i.e.~a local redefinition of the phases of the fields. Almost all of the SM can be encoded into its Lagrangian,\footnote{Actually, it is a Lagrangian \emph{density}, but in a field theory in which the fields span the entire universe, the Lagrangian $L(t) = \iiint_{-\infty}^{\infty} \mathcal{L}(t, x, y, z) \text{d}x\,\text{d}y\,\text{d}z$ is not interesting at least for studying local phenomena, so the density is simply termed `Lagrangian.'} from which the equations of motions of free fields as well as their interactions can be derived. The full SM Lagrangian with all fields and parameters written explicitly would span more than a page, so it is more useful to give the form of the Lagrangian here and explain its field content and symmetry groups separately. The Lagrangian has the form
\begin{equation}\label{eq:unbroken_lagrangian}
	\mathcal{L}_{\text{SM}} = -\frac{1}{4} (F^{a}_{\mu\nu})^2 + i\bar{\psi}\gamma^{\mu}D_{\mu}\psi + y\bar{\psi}\psi\phi + |D_{\mu}\phi|^2 - \mu^2 \phi^{\dagger}\phi - \lambda (\phi^{\dagger}\phi)^2,
\end{equation}
where $F_{\mu\nu}^a = \partial_{\mu}A_{\nu}^a - \partial_{\nu}A_{\mu}^a + g f^{abc} A_{\mu}^b A_{\nu}^c$ is the field strength tensor and $D_{\mu} = \partial_{\mu} - i g A_{\mu}^a t^a$ is the covariant derivative, with gauge fields $A$, fermion fields $\psi$ and the Lorentz-scalar Higgs field $\phi$. Furthermore, there are structure constants $f^{abc}$ and representation matrices $t^{a}$ corresponding to the gauge group of the considered interaction, and Dirac matrices $\gamma^{\mu}$ (satisfying a Clifford algebra) that relate to the Poincar\'{e} (space time) transformation properties of fermion fields. The electromagnetic and weak interactions are unified into the electroweak (EW) interaction and correspond to a $\text{SU(2)} \times \text{U(1)}$ gauge group \cite{Glashow:1961tr,Weinberg:1967tq,Salam:1968rm}, while the strong interaction is described by quantum chromodynamics (QCD) with a SU(3) gauge group \cite{qcd1,NEEMAN1961222,PhysRev.125.1067,GellMann:1964nj,Fritzsch:1973pi}. The gauge coupling strengths are denoted $g$, and the Yukawa coupling strengths between the Higgs and fermion fields are denoted $y$. The parameters $\mu$ and $\lambda$ govern the self-interactions of the Higgs field.
The terms in the Lagrangian have the following meanings:
\begin{itemize}
	\item $-\frac{1}{4} (F^{a}_{\mu\nu})^2 \equiv -\frac{1}{4} F^{a}_{\mu\nu}F_{a}^{\mu\nu}$ describes the free propagation of gauge fields, as well as the self-interactions of the gauge fields,
	\item $i\bar{\psi}\gamma^{\mu}D_{\mu}\psi$ describes the free propagation of fermion fields and their interactions with gauge bosons,
	\item $y\bar{\psi}\psi\phi$ describes the interactions of the Higgs field with the fermion fields,
	\item $|D_{\mu}\phi|^2 \equiv (D_{\mu}\phi)^{\dagger}(D^{\mu}\phi)$ describes the free propagation of the Higgs field and its interaction with gauge fields,
	\item $\mu^2 \phi^{\dagger}\phi$ and $\lambda (\phi^{\dagger}\phi)^2$ describe self-interactions of the Higgs field.
\end{itemize}
As was stated, only \emph{nearly} all of the SM is encoded into the above Lagrangian. The masses of fermions and weak bosons are missing. The only massive field appearing in \myeq~\ref{eq:unbroken_lagrangian} is the Higgs field. This is because fermion and gauge boson mass terms would break gauge invariance, leading to an unphysical theory. For instance, the gauge boson mass term $m^2A_{\mu}A^{\mu}$ is not invariant under a gauge transformation $A_{\mu} \to A_{\mu} - \partial_{\mu}\chi(x)$, so $m$ must be zero. 
On the other hand, it is known experimentally that some particles have masses, so they must be generated somehow. According to current understanding, this happens via the Higgs-Brout-Englert-Guralnik-Hagen-Kibble mechanism \cite{PhysRevLett.13.321,Higgs:1964ia,PhysRevLett.13.508,PhysRevLett.13.585,PhysRev.145.1156,PhysRev.155.1554} as follows. In the Higgs potential of the SM Lagrangian (\myeq~\ref{eq:unbroken_lagrangian}),
\begin{equation*} 
	V_{\mathrm{Higgs}}(\phi) = \mu^2\phi^{\dagger}\phi + \lambda (\phi^{\dagger}\phi)^2,
\end{equation*}
the complex constant $\mu$ and real constant $\lambda$ are chosen such that $\mu^2 < 0$ and $\lambda > 0$, so that the potential is bounded from below. With these choices, the Higgs potential has the shape shown in \myfig~\ref{fig:higgs_potential}.

The potential has an infinite number of minima satisfying
\begin{equation*}
	\phi^{\dagger}\phi = -\frac{\mu^2}{2\lambda} \equiv \frac{v^2}{2},
\end{equation*}
i.e.~lying on the circle $|\phi| = v \neq 0$.\footnote{The field $\phi$ is chosen to be an SU(2) doublet, so $\phi^{\dagger}\phi = |\phi|^2/2$.}
The minimum of the potential is by definition the \emph{vacuum}, the configuration with the lowest energy, so $v$ is called the vacuum expectation value. The crucial point is that the minimum of the potential is not at the origin ($|\phi| = 0$) and that the potential is independent of the phase $\text{mod}(\phi)$. This means that any phase can be chosen to find the vacuum, but once chosen, the term $|D_{\mu}\phi|^2$ in the SM Lagrangian coupling the Higgs and electroweak gauge fields causes the electroweak $\text{SU(2)} \times \text{U(1)}$ symmetry to be spontaneously broken. Writing out the field content and group structure explicitly, the electroweak covariant derivative reads
\begin{equation}\label{eq:covariant_derivative_ew}
	D_{\mu} = \partial_{\mu} - \frac{ig}{2}\tau_a W^a_{\mu} - \frac{ig'}{2}YB_{\mu},
\end{equation}
with the electroweak coupling strengths $g$ and $g'$, the $\mathrm{SU}(2)$ generator $\tau_a$ and gauge field $W_{\mu}^a$, and the $\mathrm{U}(1)$ generator $Y$ and gauge field $B_{\mu}$. The (arbitrary) conventional choice for the vacuum configuration of the field $\phi$ is
\begin{equation*}
	\phi_0 = \frac{1}{\sqrt{2}}\left(\begin{array}{c}0\\v\end{array}\right).
\end{equation*}
The field is expanded around the vacuum value to find the particle spectrum of the theory,
\begin{equation}\label{eq:phi_expanded_choice}
	\phi(x) = \frac{1}{\sqrt{2}}\left(\begin{array}{c}0\\v+h(x)\end{array}\right),
\end{equation}
so that $\phi^{\dagger} \phi = (v+h)^2/2$. Using \myeqs~\ref{eq:covariant_derivative_ew} and \ref{eq:phi_expanded_choice}, the last three terms of the SM Lagrangian after spontaneous symmetry breaking read
\begin{equation*}
	\begin{split}
		\mathcal{L}_{\mathrm{SM}} &\supset \frac{1}{2} (\partial_{\mu}h)(\partial^{\mu}h) + \mu^2 h^2 + \frac{g^2 v^2}{8} \left((W_1)_{\mu}(W_1)^{\mu} + (W_2)_{\mu}(W_2)^{\mu} \right)\\
			&\kern0.5pt+ \frac{v^2}{8}\left(g'B_{\mu} - g (W_3)_{\mu} \right)\left(g'B^{\mu} - g(W_3)^{\mu}\right) + \mbox{interaction terms}
	\end{split}
\end{equation*}
The interaction terms are of no interest in the present discussion. The terms containing squares of the gauge fields, $A_{\mu}A^{\mu}$, have the form of mass terms. The gauge fields $W_3$ and $B$ \emph{mix} in the last mass term. They do therefore not describe physical particles, which are mass eigenstates. The electrically neutral physical fields corresponding to the photon ($A_{\mu}$) and the \PZ boson ($Z_{\mu}$) are given by
\begin{equation}\label{eq:zgamma_superposition}
	\left(
	\begin{array}{c}
		A_{\mu}\\ Z_{\mu}
	\end{array}
	\right)
	= \left(
	\begin{array}{cc}
		\cos\theta_{\text{w}}		&	\sin\theta_{\text{w}}\\
		-\sin\theta_{\text{w}}	&	\cos\theta_{\text{w}}
	\end{array}
	\right)
	\left(
	\begin{array}{c}
		B_{\mu}\\ (W_3)_{\mu}
	\end{array}
	\right),
\end{equation}
where $\theta_{\text{w}}$ is the weak-mixing (or Weinberg) angle defined by
\begin{equation*}
	g\sin\theta_{\text{w}} = g'\cos\theta_{\text{w}} \equiv e,
\end{equation*}
($e$ is the elementary charge), while the physical charged boson fields are
\begin{equation}\label{eq:w_superposition}
	W^{\pm}_{\mu} = \frac{1}{\sqrt{2}}\left[(W_1)_{\mu} \mp i(W_2)_{\mu} \right].
\end{equation}
Using \myeqs~\ref{eq:zgamma_superposition} and \ref{eq:w_superposition}, the relevant terms in the Lagrangian become
\begin{equation*}
	\begin{split}
	\mathcal{L}_{\mathrm{SM}} &\supset \frac{1}{2} (\partial_{\mu}h)(\partial^{\mu}h) + \mu^2 h^2 + \left(\frac{gv}{2}\right)^2(W^+)_{\mu}(W^-)^{\mu}\\
			&\kern0.5pt+ \frac{1}{2} \left(\frac{gv}{2\cos\theta_{\text{w}}}\right)^2 Z_{\mu}Z^{\mu} + 0\times A_{\mu}A^{\mu} + \mbox{interaction terms},
	\end{split}
\end{equation*}
from which it can be seen that the \PWpm bosons have acquired the mass $m_{\PWpm} = gv/2$ and the \PZ boson the mass $m_{\PZ} = gv/(2\cos\theta_{\text{w}})$, whereas the photon remains massless.\footnote{The photon remains massless, because an unbroken gauge subgroup U(1) remains, whose generator $I_3 + Y/2 = Q$ is the electric charge. ($I_3$ is the third component of weak isospin and $Y$ is the hypercharge --- both are charges of the unbroken $\text{SU(2)} \times \text{U(1)}$ group.)} Thanks to experimental measurements of the weak-boson masses, the vacuum expectation value of the Higgs potential is known to be $v \approx 246~\GeV$ \cite{Olive:2016xmw}.
After spontaneous symmetry breaking, the Yukawa interactions in the SM Lagrangian become
\begin{equation*}
\mathcal{L}_{\mathrm{SM}} \supset y\bar{\psi}\psi\phi \to \frac{y v}{\sqrt{2}} \bar{\psi}\psi + \frac{y}{\sqrt{2}} \bar{\psi}\psi h,
\end{equation*}
where the first term is the mass term of a fermion with mass $y v / \sqrt{2}$ and the second term is a Yukawa interaction of fermions with the Higgs boson.
Of the $2 \times 2 = 4$ degrees of freedom of the complex doublet Higgs field, three are `absorbed' by the gauge fields as the \PWpm and \PZ bosons acquire mass, since a massive particle with spin has a longitudinal polarisation degree of freedom, while a massless particle does not. Therefore, the massive weak bosons have three polarisation degrees of freedom each, while the massless photon only has two. The remaining degree of freedom corresponds to a new physical scalar particle: the Higgs boson. It was discovered by the ATLAS and CMS collaborations in 2012 \cite{HIGG-2012-27,CMS-HIG-12-028} with properties in good agreement with the SM predictions \cite{HIGG-2015-07}, providing strong evidence that the mechanism described above indeed provides the origin of mass.\footnote{Of course, more fundamental questions about the origin of mass are now asked. Each discovery raises new, deeper questions.} The Higgs boson mass $m_{\PHiggs} = -2\mu^2$ has been determined experimentally to be approximately 125~\GeV{} \cite{ATLAS-CONF-2017-046}.



Historically, quantum field theory emerged from the need to reconcile quantum mechanics with Einsteinian relativity. Normally, adding symmetry (in this case Lorentz invariance) to a theory makes it simpler, but this was not the case for relativistic quantum mechanics. The main reason is that relativistic interactions, with particle energies greater than the masses of at least some of the massive elementary particles, imply that the possibility to create more particles must be taken into account. Another reason is that the wave functions of spin-$\frac{1}{2}$ particles, expressed as \emph{spinors}, turn out to have a complicated Lorentz structure. The fact that creation and annihilation of particles are possible means that quantum field systems have an infinite number of degrees of freedom. However, practical calculations can be made using \emph{perturbation theory}, in which the interaction terms of the Lagrangian are expanded in powers of the coupling strengths. If the coupling strengths are $\ll 1$, the perturbative series converges quickly, so that the leading-order (LO) or next-to-leading-order (NLO) approximation is adequate. At energy scales relevant at the LHC, the strong coupling strength \alphas{} is typically $\mathcal{O}(0.1)$ and the electromagnetic and weak coupling strengths are $\mathcal{O}(0.01)$, meaning that NLO EW corrections are typically much less important than QCD NLO corrections, with a size typically corresponding to that of next-to-next-to-leading order (NNLO) QCD corrections.
However, this is not always true. NLO EW corrections can be very sizeable, as will be demonstrated in \mysec~\ref{sec:zz_ew_corrections}. The energy dependence of \alphas{} given by \myeq~\ref{eq:callan_symanzik} is approximately 
\begin{equation*}
\alphas(Q) \propto \frac{1}{\ln(Q/\Lambda_{\text{QCD}})},
\end{equation*}
where $\Lambda_{\text{QCD}} \sim 200~\MeV$ is the QCD scale, so the coupling becomes too strong for perturbation theory to be applicable at energy scales below $\mathcal{O}(1~\GeV)$. Low-energy QCD effects require a non-perturbative treatment.
The most important observables in LHC physics are \emph{cross sections}, which are a measure of the probability for a process to occur.
Cross sections have dimensions of area and are often expressed in picobarn, $1~\text{pb} = 10^{-16}~\text{m}^2$. Since particle scattering processes obey quantum mechanics, cross sections are proportional to the absolute square of a quantum mechanical amplitude, interchangeably called \emph{matrix element} (of the scattering matrix), and exhibit interference effects if more than one amplitude can relate the same initial state to the same final state.

\subsection{Successes and failures} 

\begin{figure}[h!]
\centering
\includegraphics[width=\textwidth]{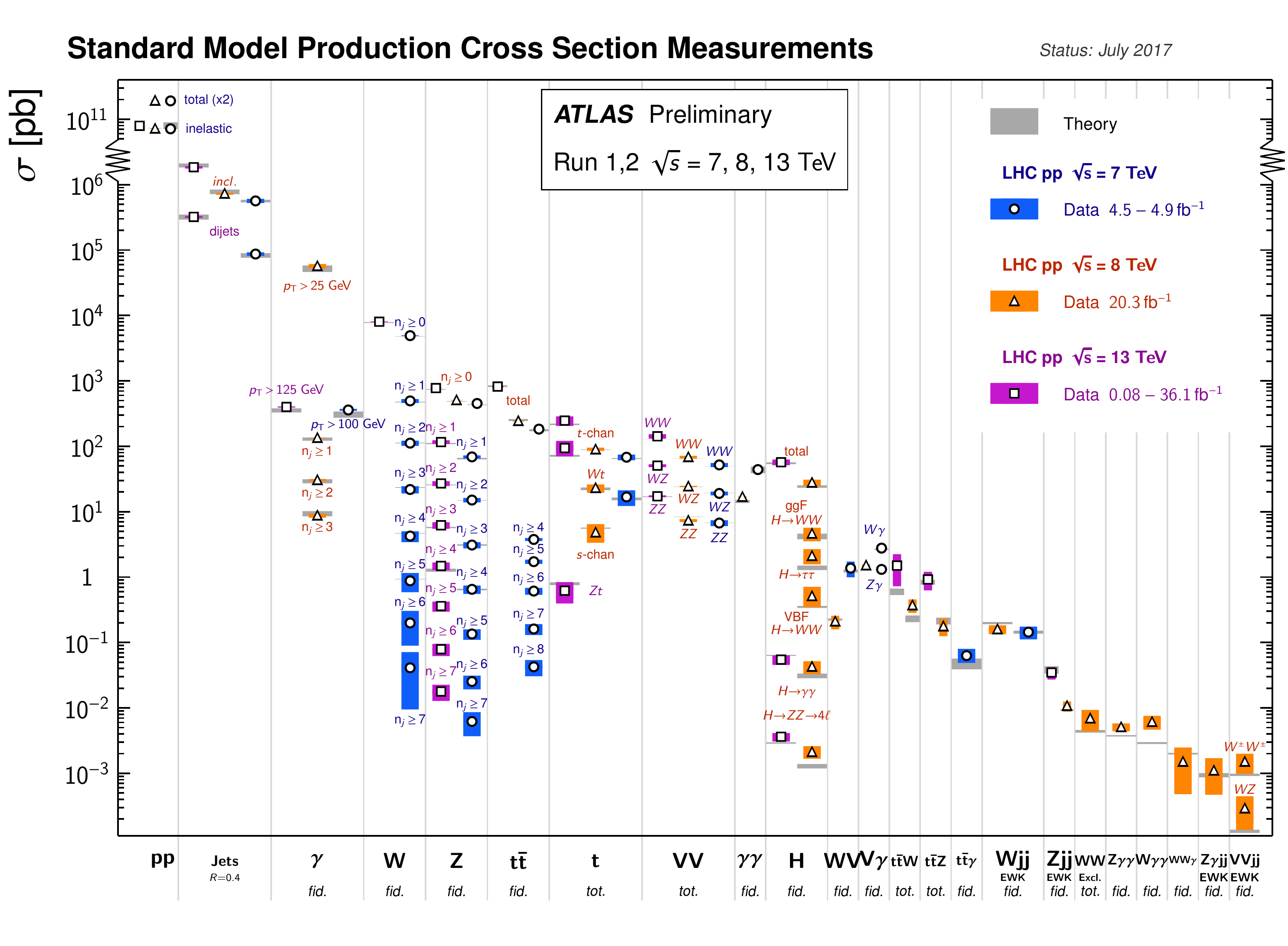}
\caption{Summary of fiducial (labelled \emph{fid.}) or total (\emph{tot.}) ATLAS measurements and SM predictions in $\Pproton\Pproton$ collisions at various centre-of-mass energies. The bands indicate the $1\sigma$ uncertainty range. Taken from \myref~\cite{atlas_sm_summaries}.}
\label{fig:sm_successes}
\end{figure}

The SM has been incredibly successful in predicting the results of previous and ongoing particle physics experiments. A summary of many LHC measurements compared to SM predictions is shown in \myfig~\ref{fig:sm_successes}, demonstrating the very good agreement of cross sections spanning at least nine orders of magnitude (even more if including the total $\Pproton\Pproton$ interaction cross sections). 

At the same time, it is known that the SM is not the final theory of Nature.
For one thing, it does not include a description of gravity.
Astronomical evidence also suggests that the universe contains substantial amounts of (predominantly cold) dark matter \cite{Zwicky:1933gu,1937ApJ....86..217Z,1970ApJ...159..379R,1980ApJ...238..471R,1974ApJ...193L...1O,Sofue:2000jx,Clowe:2006eq,Ade:2015xua} forming halos around the visible parts of galaxies. It does not seem to interact electromagnetically, but its gravitational effects can be measured. The dark matter is hypothesised to be composed of massive, still unknown particles. Cosmology favours dark matter that interacts via the weak interaction as well as gravity. The SM does not contain particles that are candidates for the observed dark matter (according to cosmological evidence, SM neutrinos can account for at most a small part of the dark matter), so it may need to be extended.
Furthermore, neutrinos are known experimentally to change their flavour with time (`oscillate') \cite{PhysRevLett.81.1562,PhysRevLett.89.011301,Eguchi:2002dm}. This means that their flavour and mass eigenstates are not aligned. However, this requires the three types of neutrino to have different masses, and therefore non-zero masses. Neutrinos may have mass in the SM (via the Higgs mechanism), but this requires them to be their own antiparticles. It is not known whether this is the case. If they are, some physicists argue that their masses are unnaturally small compared to the electroweak scale. So while neutrino oscillations do not necessarily contradict the SM, they require certain properties of the SM that are not yet experimentally confirmed.
There are further possible problems of the SM that are related to something called fine-tuning. Fine-tuning means that some parameter has exactly the right value to produce certain behaviour of the theory, without any apparent reason for the value, such as a symmetry of Nature requiring it. For instance, there is no known principle that prevents the strong interaction from breaking charge-parity (CP) symmetry, but this is not observed in nature, so the parameters controlling CP violation could be considered fine-tuned. Similarly, the observable Higgs boson mass is only $\mathcal{O}(100~\GeV)$, even though there is no known symmetry protecting its bare value from huge quantum corrections. \emph{Assuming} that the SM is a low-energy effective theory that is only valid up to some high energy scale (many, many orders of magnitude higher than the scale of LHC collisions), some physicists consider it strange that the Higgs boson mass happens to be of the same order of magnitude as the masses of the weak bosons --- rather than the size of the scale where the SM breaks down. This is called the hierarchy problem of the Higgs boson mass.

As of the writing of this thesis, the ATLAS and CMS collaborations have not observed any clear hints of physics beyond the SM. In the author's personal opinion, the clearest hints of new physics observed at the LHC come from the LHCb collaboration \cite{Alves:2008zz}, which has measured the decays of individual hadrons containing heavy (beauty or charm) quarks.
LHCb observes a $2.5\sigma$ disagreement between measurement and prediction in the decay lepton-flavour ratio $R_{\PKstar}$ \cite{Aaij:2017vbb}, defined as a ratio of branching fractions,
\begin{equation}\label{eq:rkstar}
R_{\PKstar} = \frac{\mathcal{B}(\PBzero \to \PKstar^0 \APmuon\Pmuon)}{\mathcal{B}(\PBzero \to \PKstar^0 \Ppositron\Pelectron)},
\end{equation}
and a $2.6\sigma$ disagreement in $R_{\PKplus}$ (replacing $\PBzero$ with $\PBplus$ and $\PKstar$ with $\PKplus$ in \myeq~\ref{eq:rkstar}) \cite{Aaij:2014ora}.
A more sensitive measurement of $R_{\PKplus}$ using Run 2 data is ongoing \cite{pichonet}. Since theoretical uncertainties largely cancel in the ratio, the SM predictions of $R_{\PKstar}$ and $R_{\PKplus}$ are very precise \cite{Hiller:2003js}. In addition, angular distributions of products of the decay $\PBzero \to \PKstar^0 \APmuon\Pmuon$ exhibit a $3.4\sigma$ disagreement from the SM prediction \cite{Aaij:2015oid}. Together with other similar measurements, these results have created a coherent picture hinting at new physics violating lepton flavour universality, which is analysed e.g.~in \myrefs~\cite{Capdevila:2017bsm,Allanach:2017bta}. Future measurements by the LHCb and Belle II \cite{Abe:2010gxa} collaborations will shed more light on these interesting deviations.
Furthermore, the gyromagnetic ratio of the muon (a measure of the precession frequency of the spin of a muon in a magnetic field) has been found experimentally \cite{Bennett:2006fi} to exhibit a $3.5\sigma$ deviation from the most precise SM prediction \cite{Olive:2016xmw}. In the SM, quantum corrections cause its value to differ from 2 by approximately 0.1\%. The very small deviation from the SM prediction could indicate the existence of new particles and interactions, which modify the quantum corrections. The new Muon g-2 experiment \cite{Grange:2015fou} will measure the value with unprecedented precision in the near future, in the hope of confirming or excluding the previously observed deviation.

\section{Phenomenology of proton collisions}\label{sec:theory_pheno}

This section explores the most relevant phenomenological aspects of the SM for the context of the original research presented in this thesis. 
The presented research is concerned with studying some of the smallest structures that are currently experimentally accessible. To probe small structures in particle collisions, high centre-of-mass energies are necessary, as given by the de Broglie relation
\begin{equation*}
\text{probe energy} \sim \frac{1}{\text{structure size}}.
\end{equation*}
In addition, new particles --- or other objects, such as quantum black holes --- having a large mass might exist. Producing these again requires a large centre-of-mass energy, at least equal to the mass of the particle(s) to be produced. In this thesis, proton-proton collisions produced by the LHC are studied. At LHC energies, the collisions resolve the internal constituents of the protons, collectively referred to as partons, so the collisions are really among \emph{partons}.
In addition to the three proton valence quarks (one down quark and two up quarks), quantum fluctuations inside the proton mean that the partons may be sea quarks or gluons, or (with very small probabilities) even non-coloured particles such as photons (\myapp~\ref{sec:photon_induced}).


The composite nature of the proton has far-reaching consequences for the nature of proton-proton collisions. Each parton carries a fraction $x \in [0, 1]$ of the proton momentum. The initial-state parton flavours and longitudinal momenta are not known in a given collision. This explains the need for transverse variables, such as the transverse momentum, that are invariant under longitudinal boost. Differences in (pseudo)rapidity also have this property. One distinguishes the \emph{hadronic} centre-of-mass energy $\sqrt{s}$, which is determined by the accelerator (e.g.~13~\TeV{} at the LHC), and the \emph{partonic} centre-of-mass energy $\sqrt{\hat{s}}$, which is the actual energy scale of the partonic interaction,
\begin{equation*}
\hat{s} = x_1 x_2 s,
\end{equation*}
where $x_1$ and $x_2$ are the momentum fractions of the incoming participating partons in the $+z$- and $-z$-direction, respectively. They are related to centre-of-mass kinematics by
\begin{equation}\label{eq:bjorkenx_rapidity}
x_{1,\,2} = \sqrt{\frac{\,\hat{s}\,}{s}} \, e^{\pm y},
\end{equation}
where $y$ is the rapidity of the centre of mass in the laboratory frame. The LHC effectively `scans' a wide range of hard-process scales. This makes it very suitable for discovering new particles of unknown mass $m_{\text{NP}}$, since high-cross-section events with $\hat{s} \sim m_{\text{NP}}$ will occur without changes in the experimental setup.

\subsection{Parton distribution functions}\label{sec:pdf} 

The cross section of the process $\Pproton\Pproton \to X$, where $X$ is an arbitrary final state, can be expressed using the partonic cross sections of producing $X$,
\begin{equation}\label{eq:pdf_factorisation}
\sigma_{\Pproton\Pproton \to X}(Q^2) = \sum_{i,\,j}^{\text{parton flavours}} \int_0^1 \int_0^1 \mathrm{d}x_1 \mathrm{d}x_2 f_i(x_1, Q^2) f_j(x_2, Q^2) \hat{\sigma}_{ij \to X}(Q^2),
\end{equation}
where the weights $f_i(x, Q^2)$, called parton distribution functions (PDFs), ensure the correct normalisation of the cross section. At LO, the PDFs give the probability of finding a parton of flavour $i$ carrying momentum fraction $x$ of the proton, given that the proton was probed at scale $Q$.\footnote{At higher orders, this interpretation does not hold. The reason is that hard-scattering matrix elements beyond LO QCD have a non-trivial low-energy behaviour themselves, so low-energy contributions can be shifted around between PDFs and matrix elements, depending on what matching scheme is chosen. PDFs beyond LO may even be negative.}
The form of \myeq~\ref{eq:pdf_factorisation} shows that the PDFs and hard-process cross section factorise. The transition from the PDF description of the physics to that by the hard-process cross section happens at the \emph{factorisation scale}, $Q^2 \equiv \muf^2$, which is chosen to represent the scale of the event. The dependence of $\sigma$ and $\hat{\sigma}$ on the final state and coupling strengths is implied.
The PDFs are governed by non-perturbative QCD and must be determined by fitting to experimental data, but their evolution from one scale $Q^2$ to another can be described perturbatively using the DGLAP evolution equations \cite{Gribov:1972ri,Dokshitzer:1977sg,Altarelli:1977zs} and the universal Altarelli-Parisi splitting functions \cite{Altarelli:1977zs}.

The shape of the PDFs means that the vast majority of LHC collisions take place at relatively small values of $x$ and therefore $\hat{s} \ll s$. Hard scattering processes, which are the ones that are primarily of interest, are rare occurrences.




\subsection{Parton radiation and hadronisation}
Partons emitted at large transverse momenta radiate further partons, causing a cascade. Additional parton emissions might be expected to be suppressed by powers of $\alphas \sim \mathcal{O}(0.1)$ for scales above $\sim$ 1~\GeV{}, but in fact they can be enhanced by large logarithms of the form $\ln(Q/q)$, where $Q$ is the scale of the hard process and $q$ is the scale of the parton splitting. The logarithms become large when $q \to 0$, which occurs when the two partons after splitting are nearly collinear (\emph{collinear limit}) or an emitted parton has very little energy (\emph{soft limit}). Collinear and soft radiation patterns have a universal structure that is independent of the details of the hard process and can be described using Altarelli-Parisi splitting functions \cite{Altarelli:1977zs}.
Radiation in the collinear limit factorises at the cross-section level, which allows describing arbitrary numbers of splittings by exponentiation. This yields a Sudakov form factor $\Delta_i (Q, q)$ \cite{sudakov_form_factor} giving the probability that parton $i$ produced at scale $Q$ does not split above a lower scale $q$,
\begin{equation*} 
\begin{split}
	\Delta_i (Q, q) &= \exp\left(-\int_{q^2}^{Q^2} \frac{\dee k^2}{k^2} \frac{\alphas(k^2)}{2\pi} \int_{q^2/k^2}^{1-q^2/k^2} \dee z\, P_{i\to jk}(z) \, F(z, k^2)\right)\\
	&\sim \exp\left(-\text{const.} \times \ln^2\frac{Q^2}{q^2}\right),
\end{split}
\end{equation*}
where $P_{i\to jk}(z)$ is the Altarelli-Parisi splitting function for the splitting $i \to jk$, such that parton $j$ carries fraction $z$ of the momentum of the original parton $i$. The form of the Sudakov form factor shows that there is a \emph{hierarchy} of logarithmically enhanced emissions. It is implied that all possible post-splitting flavours $j$, $k$ are summed over. For instance, a gluon can split as $\Pgluon \to \Pquark\APquark$ or $\Pgluon \to \Pgluon\Pgluon$.
The factor $F$ depends on whether the emission is from the initial state of the hard process or final-state radiation (FSR). Initial-state radiation (ISR) means that the parton with momentum fraction $x'$ taken from the proton PDF is no longer the same as the one entering the hard scattering with momentum fraction $x$. For this reason, the Sudakov form factor for ISR includes a ratio of momentum fractions and PDFs before and after the splitting. Using the ISR notations

\begin{equation*}
	\parbox{20mm}{
	\begin{fmfgraph*}(80,40)
		\fmfset{arrow_len}{3mm}
		\fmfstraight
		\fmfleft{i1}
		\fmfright{o0,o1,o2}
		\fmflabel{$i,~x/z$}{i1}
		\fmflabel{$k$}{o0}
		\fmf{plain}{i1,v1}
		\fmf{phantom}{v1,o1}
		\fmffreeze
		\fmf{plain, label=$j,,~x$, label.side=left}{v1,o2}
		\fmf{plain}{v1,o0}
		\fmfblob{6.6mm}{o2}
	\end{fmfgraph*}
	} \qquad \qquad ,
\end{equation*}

\vspace{2mm}
with the blob representing the rest of the evolution up to and including the hard process, gives
\begin{equation*} 
	F(z, k) = \left\{
	\begin{array}{cl}
		\frac{x/z}{x} \frac{f_i(x/z;\, k^2)}{f_j(x;\, k^2)} & \quad \mbox{if ISR,}\\
		1 & \quad \mbox{if FSR.}
	\end{array}
	\right.
\end{equation*}
Non-collinear radiation in the soft limit, however, factorises at the matrix-element level, which makes its treatment much more complicated. Constructing a Sudakov form factor for soft radiation requires exponentiating a (colour) matrix rather than a scalar quantity, making its correct description challenging.

To calculate the full parton shower, subsequent emissions with consecutively lower transverse momentum are then attached to the new final state iteratively. The transverse momentum integration in the Sudakov form factor is from that of the current splitting up to that of the previous one. There are many details and variations in how the scale of a splitting is defined, how colour charge and soft wide-angle emissions are treated, etc.~that are beyond the scale of this discussion. A good introduction can be found in \myref~\cite{Buckley:2011ms}. The parton shower effectively \emph{resums} important higher-order corrections enhanced by large logarithms. Resummation means making a perturbative expansion, calculating its important terms in some approximation (here: collinear approximation), and summing them up to all orders. 

The evolution is continued until scales where \alphas{} becomes large and perturbation theory fails, at approximately 1~\GeV{}. At this point, the final-state partons are mapped onto hadrons according to an empirical hadronisation model, such as the cluster model \cite{WEBBER1984492} or the Lund string model \cite{Andersson:1983ia}. 


\subsection{Underlying event}
\label{sec:ue}

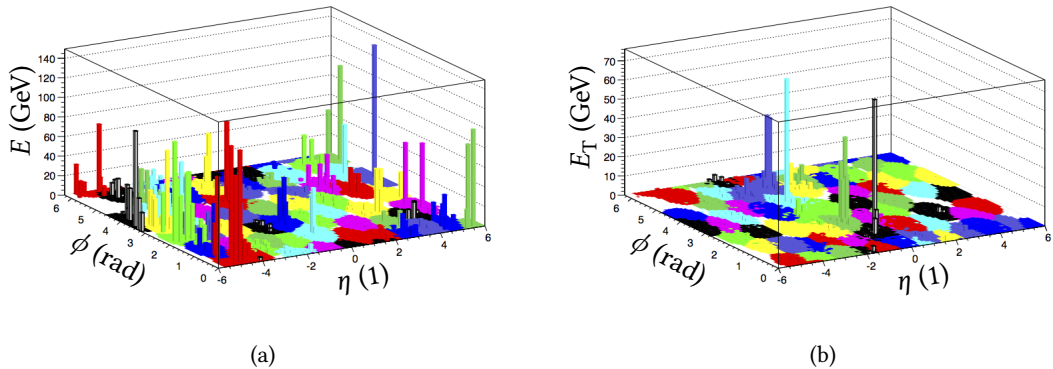
\begin{figure}[h!]
\centering
\subfigure[]{
\begin{tikzpicture}
\node[anchor=south west,inner sep=0] (image) at (0,0) {\includegraphics[width=0.46\textwidth]{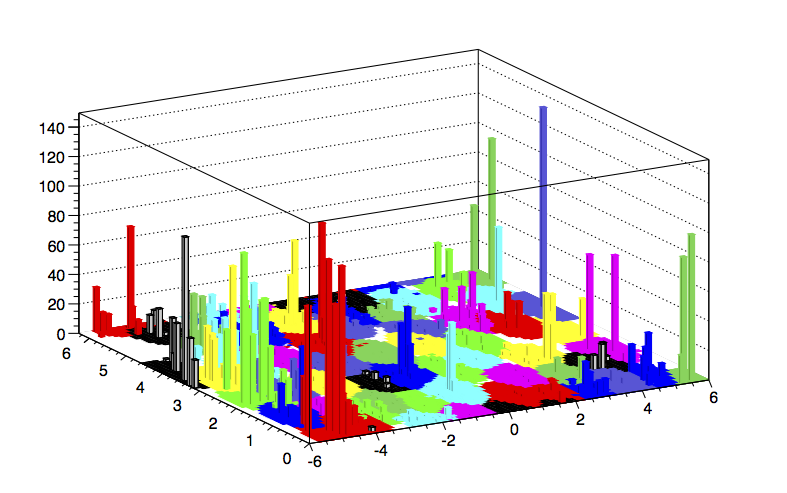}};
\begin{scope}[x={(image.south east)},y={(image.north west)}]
\node[anchor=west, rotate=9] at (0.6, 0.05) {$\eta$ (1)};
\node[anchor=west, rotate=-29] at (0.1, 0.2) {$\phi$ (rad)};
\node[anchor=east, rotate=90] at (0.02, 0.75) {$E$ (\GeV{})};
\end{scope}
\end{tikzpicture}
}
\subfigure[]{
\begin{tikzpicture}
\node[anchor=south west,inner sep=0] (image) at (0,0) {\includegraphics[width=0.46\textwidth]{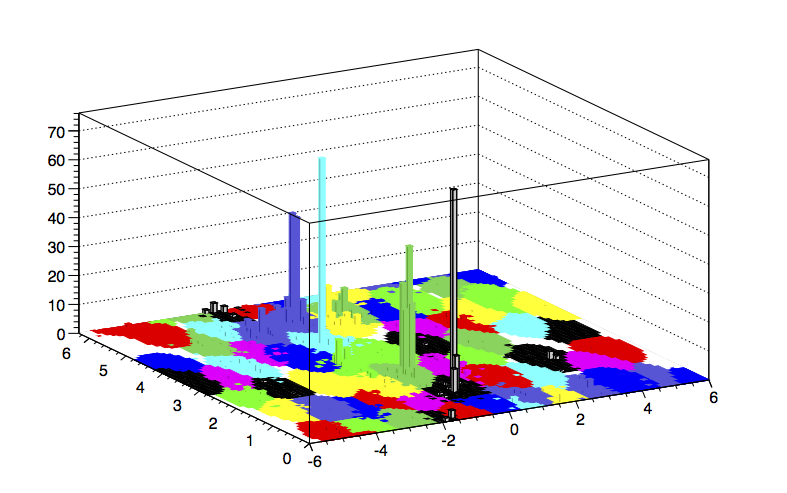}};
\begin{scope}[x={(image.south east)},y={(image.north west)}]
\node[anchor=west, rotate=9] at (0.6, 0.05) {$\eta$ (1)};
\node[anchor=west, rotate=-29] at (0.1, 0.2) {$\phi$ (rad)};
\node[anchor=east, rotate=90] at (0.02, 0.75) {$E_{\text{T}}$ (\GeV{})};
\end{scope}
\end{tikzpicture}
}
\caption{Distribution of (a) energy and (b) transverse energy in the $\eta$-$\phi$ plane in the same all-hadronic $\bar{\Ptop}\Ptop$ event generated with \PYTHIA{}~8 \cite{Sjostrand:2006za,Sjostrand:2007gs}. The colours have no physical meaning, they only served to distinguish energy deposits clustered together according to some algorithm. (The clustering is done using a new jet algorithm that the author invented and experimented with for fun, but this is irrelevant here. Jet algorithms are introduced in \mysec~\ref{sec:jets}.)}
\label{fig:energy_vs_transverse_energy}
\end{figure}

All particle production contributing to the event that is not part of the hard-process and its accompanying ISR and FSR is considered part of the \emph{underlying event}.\footnote{Except pileup, which will be introduced in \mysec~\ref{sec:experiment_pileup}.}
The two partons participating in the hard-scattering process are not alone, but accompanied by all the other partons of the colliding protons. As the hard scattering occurs, the protons break up into coloured remnants, which in turn radiate further partons and hadronise. The resulting hadrons typically have low transverse momentum, but large longitudinal momentum due to the protons' initial momentum. This is visualised in \myfig~\ref{fig:energy_vs_transverse_energy}, showing a comparison of the flow of energy and transverse energy in the same simulated all-hadronic top-antitop quark ($\bar{\Ptop}\Ptop \to \Pquark\APquark' \Pquark''\APquark''' \Pbottom\bar{\Pbottom}$) event. The forward region $|\eta| \gtrsim 4$ has significant energy flow, dwarfing the energy of the products of the hard-scattering process. However, the corresponding \emph{transverse} energy flow is very small compared to that from the hard scattering.
Additional interactions between further pairs of partons from the two protons may also take place, which is called multiple-parton scattering. Due to the shape of the PDFs, the scales of these interactions are typically soft, and again mainly lead to the production of soft hadrons. \mysec~\ref{sec:theory_double_parton_scattering} explores a rare scenario where this is not the case, but rather two roughly equally hard scattering processes take place in the same proton collision.

While itself an interesting probe of soft QCD effects, the underlying event may present an experimental challenge. It is a source of additional particles that overlay the process of interest and complicate its identification and measurement. The size of the impact depends very much on the analysis and ranges from an undisentanglable effect to a negligible nuisance.



\subsection{Jets}\label{sec:jets}
Since high-\pt{} partons radiate and hadronise, they are observed as a spray of hadrons, with those carrying the most energy typically quite collimated, and accompanied by mainly softer, less collimated hadrons. To reconstruct the momentum of the initiating parton, hadrons are clustered into \emph{jets} according to some algorithm. A careful jet definition yields a set of theoretically well-defined jets that is infrared-safe, i.e.~it remains unchanged if a parton splits into two collinear partons or emits an arbitrarily soft parton. The inputs of jet clustering algorithms are called proto-jets in this thesis. They may be partons (typically in fixed-order calculations), stable particles (Monte Carlo events), topological clusters (ATLAS), or e.g.~\emph{particle flow} objects (CMS \cite{CMS-PRF-14-001}, recently ATLAS \cite{PERF-2015-09}). This way, jets provide a common `language' to compare hard-scale physics at all these different levels. In Nature, jets consist mainly of pions, because these are the lightest hadrons, but also kaons, protons, neutrons, as well as the non-hadronic decay products of short-lived hadrons. For instance, the copiously produced neutral pions decay almost invariably to photon pairs.
Several jet definitions have been used. An overview can be found in \myref~\cite{Salam:2009jx}. At the LHC, the most commonly used is the (inclusive) anti-$k_t$ algorithm \cite{cacciari08}, which is also the one used in this thesis. 
It defines the distance between two proto-jets $i$ and $j$ as
\begin{equation*}
d_{ij} = \min\big(p_{\text{T},\,i}^{-2},\, p_{\text{T},\,j}^{-2}\big) \left(\frac{\Delta R_{ij}}{R}\right)^2
\end{equation*}
where $\Delta R_{ij}$ is the angular distance between $i$ and $j$ and $R$ is a dimensionless radius parameter chosen by the user. The distance between a proto-jet and the beam is
\begin{equation*}
d_{iB} = p_{\text{T},\,i}^{-2}.
\end{equation*}
At each clustering step, the proto-jets $i$ and $j$ with the smallest distance $d_{ij}$ are combined by adding their four-momenta to form a new proto-jet to replace $i$ and $j$. Alternatively, if $d_{iB} < d_{ij}$ for all $j$, the proto-jet $i$ is called a jet and removed from the list of inputs. This procedure is iterated until only jets are left.\footnote{The name `anti-$k_t$' refers to the fact that the transverse momentum (sometimes denoted $k_t$, though this thesis denotes it \pt{} throughout) enters with a power of \emph{negative} two in the jet measure, juxtaposed with the $k_t$ algorithm \cite{Catani:1992zp,Ellis:1993tq} where it enters with a power of positive two.}

\begin{figure}[h!]
\centering
\subfigure[]{\includegraphics[width=0.47\textwidth]{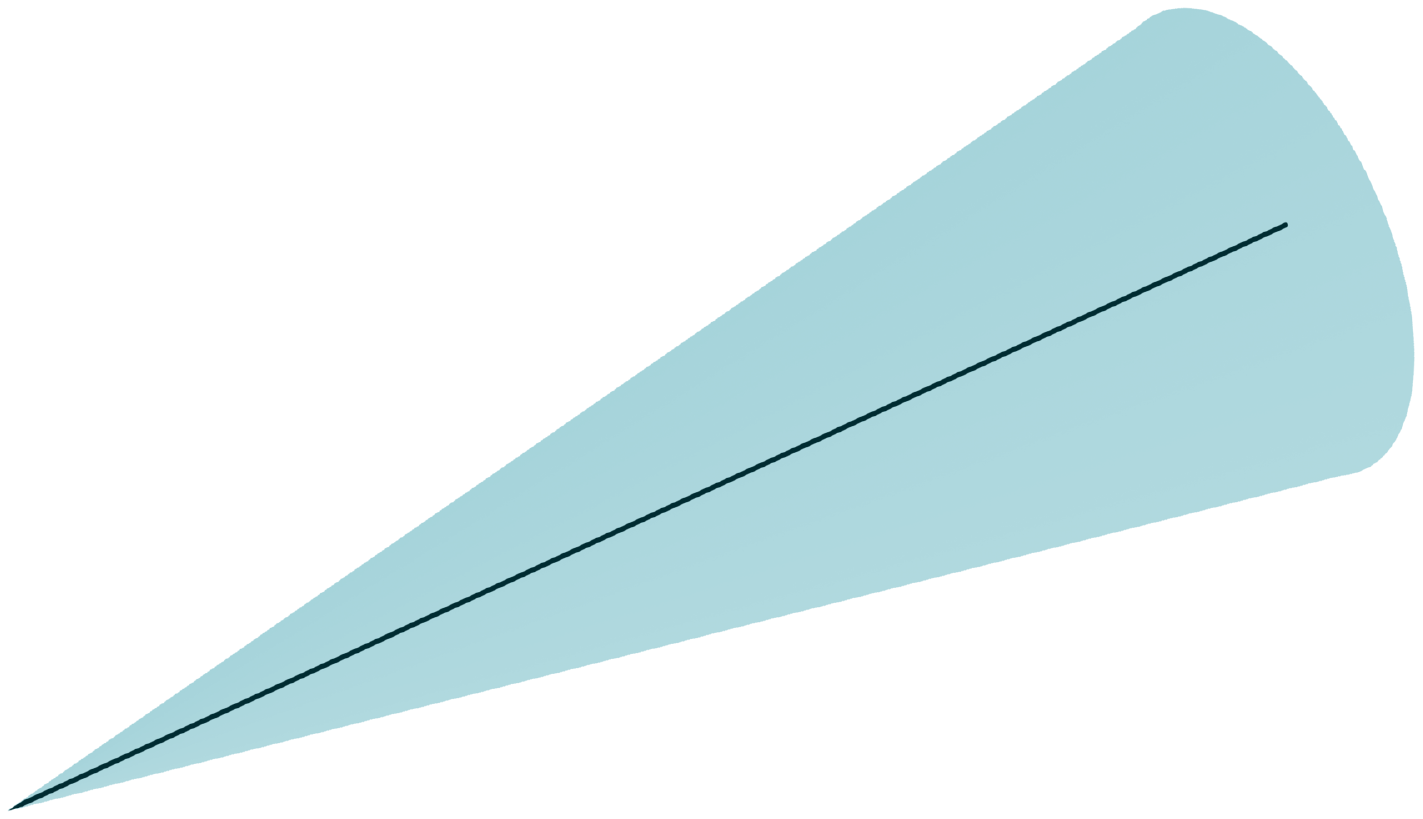}}
\hspace{1mm}
\subfigure[]{\includegraphics[width=0.47\textwidth]{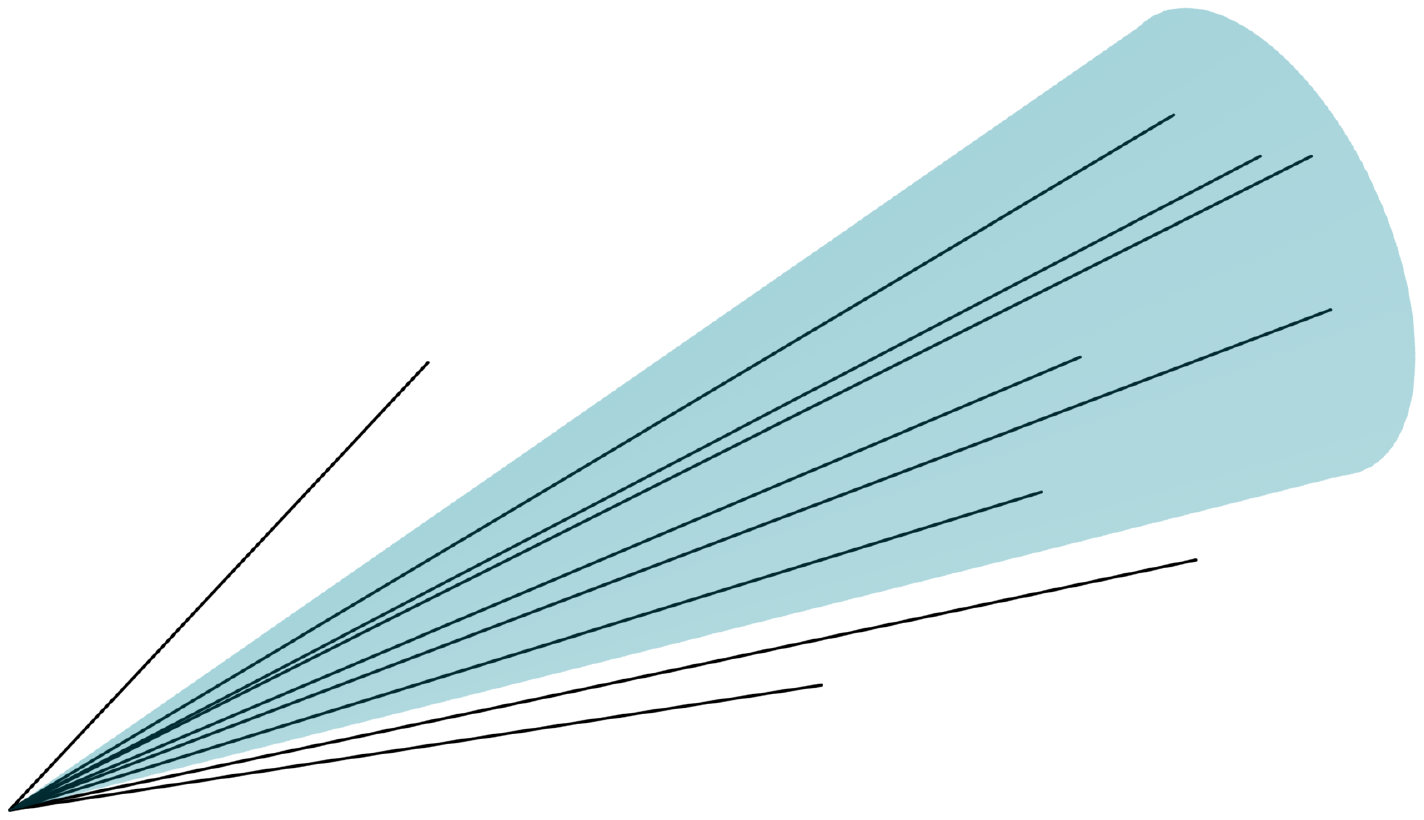}}
\caption{The out-of-cone radiation effect. (a) At the hard-parton level, there is perfect correspondence between a well-isolated parton and a jet. (b) After the parton shower and possibly hadronisation, part of the energy of the original parton may be radiated outside of the jet cone (or more generally: catchment area \cite{Cacciari:2008gn}).}
\label{fig:out_of_cone_radiation}
\end{figure}

Due to the fact that the algorithm combines the hardest proto-jets first, the direction of an anti-$k_t$ jet does not change much during the clustering. Therefore, the algorithm yields jets that are almost circular in shape in the $\eta$-$\phi$ plane, with an area of approximately $\pi R^2$. Both the shape and the area are insensitive to soft radiation. This makes the contributions to a jet due to the underlying event and pileup (defined in \mysec~\ref{sec:experiment_pileup}) more predictable and improves the experimental calibration of the jet energy.
The radius parameter $R$ can be tuned to balance between two effects. A larger $R$ captures more of the radiation associated with the parton initiating the jet by decreasing \textit{out-of-cone radiation}, as visualised in \myfig~\ref{fig:out_of_cone_radiation}. On the other hand, a smaller $R$, reduces the contributions from the underlying event and pileup. The transverse-momentum flow of these sources is relatively constant in (pseudo)rapidity and therefore across the $\eta$-$\phi$ plane, so that their unwanted contribution to the jet \pt{} is roughly proportional to the jet area. Throughout this thesis, the de-facto standard choice of $R = 0.4$ is used.

%

Jets with larger radii, typically $R \in \{1.0, 1.2, 1.5\}$, are used extensively to study their substructure. The hadronic decay products of a heavy resonance, such as $\PZ \to \Pquark\APquark$, $\PW \to \Pquark\APquark'$, or $\PHiggs \to \Pbottom\bar{\Pbottom}$ will be sufficiently collimated to be clustered into one large jet if the resonance has sufficient (transverse) momentum.
Jet substructure techniques were proposed 25 years ago for identifying hadronically decaying heavy resonances \cite{Seymour1994} (another early example is \myref{}~\cite{Butterworth:2002tt})
and have since grown into a very vibrant theoretical and experimental field with many applications, as laid out e.g.~in \myref~\cite{boost_review}.

\subsection{Scale variations}\label{sec:theory_scale_variations}

The calculation of a cross section in perturbative QCD involves (at least) two different scales: the renormalisation scale $\mur$ at which e.g.~\alphas{} is evaluated, and the factorisation scale $\muf$, at which the PDF evolution ends and the hard process takes over. Their value is chosen to represent the scale of the hard process. The scales are unphysical in the sense that an exact all-order calculation would eliminate the dependence on them. In a fixed-order calculation, recalculating observables after varying values of the scales by a factor $\mathcal{O}(1)$ may give a hint of the order of magnitude of missing higher orders. The deviations of the varied results from the nominal one are commonly taken as \emph{QCD scale uncertainties}. In this thesis, QCD scale uncertainties of predicted cross sections are evaluated by varying the factorisation scale $\muf$ and renormalisation scale 
up and down independently by a factor of two, ignoring however the extreme variations ($2 \muf$, $0.5 \mur$) and ($0.5 \muf$, $2 \mur$), and taking the largest deviations from the nominal value (i.e.~the envelope) as the systematic uncertainties.
Scale variations do not always represent an adequate measure of the size of missing orders. \mysec~\ref{sec:nnlo_predictions} will show an example where this is not the case.

%


\section{Monte Carlo event generation}

In practice, SM predictions are often obtained by computer-generating events using the Monte Carlo (MC) method. The MC method uses random numbers to sample the allowed kinematic configurations (the \emph{phase space}) of a process, evaluating the matrix element at each generated phase space point. In this way, it effectively numerically integrates the matrix element. The MC method is particularly suited for two main reasons. First, it converges faster than many other numerical integration methods if the phase space has more than a few dimensions. This is nearly always the case: in a given set of final-state particles, each contributes three phase-space dimensions, namely the three spatial components of its momentum. Second, the stochastic nature of particle collisions means that each sampled phase space point can be naturally interpreted as one generated \emph{event} with a matrix-element weight that is representative of the likelihood of that event. This is incredibly useful, because it means that the phase-space integration does not need to be repeated for each observable --- one simply generates a sample of events, stores their corresponding momentum configurations, and calculates arbitrary observables from the generated samples as needed. Many general-purpose and specialised \emph{event generator} softwares have been developed. An excellent overview of the theory and practice of MC event generation can be found in \myref~\cite{Buckley:2011ms}.
Since squared matrix elements are large in some regions of phase space and small in others, it is desirable to generate phase space points that sample important regions more densely than less important ones. The need for an efficient phase space integrator leads to a hen-and-egg problem: to efficiently sample the phase space, an importance sampling according to the squared matrix element is necessary, but evaluating the squared matrix element over the phase space of interest requires efficient phase space sampling. To get around this, heuristic techniques are used, such as making educated guesses about where the resonances and singularities of the matrix element lie (which can be inferred from the considered Feynman diagrams of the process) and sampling more densely in their vicinity. For instance, a photon propagator could suggest a sampling density proportional to $1/q^2$, the invariant mass of the virtual photon, while for a $\PW$ boson propagator it could be $\propto 1/(q^2 - m_{\PW}^2)$. Similarly, densities can be constructed for more complicated structures, such as loops. In practice, it is often simplest to sample each resonance and singularity structure separately, i.e.~in its own ``integration channel'', of course evaluating the full matrix element with all contributions at each point. This is called multi-channel integration. 

In addition to the matrix element for the hard-scattering process, many event generators can simulate the decays of heavy resonances (if not already included in the matrix element), parton showering, the underlying event, electromagnetic radiation off charged final-state particles, hadronisation, and decays of unstable hadrons and leptons.

In some contexts, it is desirable to predict and model the response of the experimental setup on the event. This can be achieved by MC-simulating the interactions of the final-state particles in a generated event with the materials they encounter as they fly away from the collision point. These material interactions will be described in \mysec~\ref{sec:particles_in_matter}. Material interactions and detector responses may be simulated with dedicated software, such as \GEANT{}~\cite{Agostinelli:2002hh} or \textsc{Delphes}~\cite{deFavereau:2013fsa}. 
The goal is usually that the MC samples after detector simulation can be treated on equal footing with the actual experimental data.

Experimental particle physics relies on MC events in nearly all aspects and stages of analysis --- from feasibility studies, over method development and analysis optimisation, to comparing SM predictions to the data. Even where data-driven methods can be used to measure e.g.~performance or background contributions, these are usually first developed with the help of MC simulation. The continued development of not just calculations, but also their practical implementation in reliable and user-friendly software is a key requirement for the success of the field.








\subsection{Next-to-leading-order matrix elements}

Beyond LO, combining matrix elements with parton showers is non-trivial, because parton emissions have already been generated at the matrix-element level and must not be double-counted by the parton shower. In addition, the approximate NLO virtual corrections included by the parton shower must be replaced by the full NLO result. The two commonly used methods for \emph{matching} parton showers and matrix elements are the MC@NLO method \cite{Frixione:2002ik} and the \POWHEG{} method \cite{Nason:2004rx}. Both have been algorithmically automated in event generators. 

Matching of parton showers with NNLO matrix elements is being developed actively. It has been achieved for individual processes, but is not yet routinely available in event generators. A brief summary of the status and some recent results are presented in \myref~\cite{Hoeche:2015vea}.

NLO corrections may be included in an \emph{approximate} way by multiplying the LO prediction with a $k$-factor, defined as
\begin{equation*}
k = \frac{\sigma^{\text{NLO}}}{\sigma^{\text{LO}}},
\end{equation*}
or, in general, with arbitrary orders in the numerator and denominator. The $k$-factor depends on the event kinematics and may be binned in some observable(s). In this thesis, the deviation of the $k$-factor from unity is denoted $\delta = k - 1$.

\subsection{Multijet merging}
Various techniques have been developed to combine the LO or NLO simulations of a hard process with varying numbers of additional partons generated at the matrix-element level into a single consistent sample. For instance, making a sample combining the processes $\Pproton\Pproton \to \PZ + 0~\text{partons}$ and $\Pproton\Pproton \to  \PZ + 1~\text{parton}$, both generated in NLO QCD. This is referred to as multijet merging, because the additional partons may give rise to associated jets. The methods laid out in \myrefs~\cite{Hoeche:2012yf,Gehrmann:2012yg,Hoeche:2010kg,Hoeche:2009rj} are used in this thesis.

\clearpage\pagebreak
\part{Experimental setup}
\label{sec:experiment}

To undertake the experimental research presented in this thesis, impressive machines built over a generation by many people are needed. The experimental setup consists of the Large Hadron Collider (LHC) \cite{Evans:2008zzb_modifiedauthors} for producing proton-proton collisions, and the ATLAS detector for observing the particles produced in these collisions. Electronics and software algorithms are used to reconstruct what happened in the event in terms of long-lived particles as well as objects describing the collective properties of multiple particles (jets, missing energy). These are the inputs to high-level analysis of the underlying physical processes, such as statistically inferring the production of short-lived particles from their decay products.

The measurements presented in this thesis all use data from LHC Run 2, which started in 2015 after upgrades to the LHC and the detectors.

\section{Large Hadron Collider}
The proton-proton collisions analysed in this thesis are produced by the LHC. The protons are produced by stripping hydrogen atoms of their electron. They are first accelerated by a linear accelerator to 50~\MeV{} energy (speed as fraction of the speed of light $\beta \approx 5\%$), before entering the Proton Synchrotron Booster, commissioned in 1972. This circular accelerator of 25~m radius accelerates the protons to 1.4~\GeV{} ($\beta \approx 83\%$). The booster allows injecting about 100 times more particles into the next accelerator, the Proton Synchrotron, than if they had come directly from the linear accelerator. The Proton Synchrotron was commissioned already in 1959 and was CERN's original flagship machine, measuring 628 metres in circumference. It accelerates the protons to 24~\GeV{} ($\beta \approx 99.9\%$). Next, the 7-km-long Super Proton Synchrotron accelerates them to 450~\GeV{} ($\beta \approx 99.9998\%$). This accelerator, commissioned in 1976, has delivered beams to many experiments and produced the proton-antiproton collisions that lead to the direct discovery of the \PW{} and \PZ{} bosons in 1983 by the UA1 and UA2 collaborations \cite{wboson_ua1,wboson_ua2,zboson_ua1,zboson_ua2}. Finally, the protons enter the Large Hadron Collider, which accelerates them to 6.5~\TeV{} ($\beta \approx 99.999999\%$) in about 20 minutes. It is a giant machine, 27~km in circumference and about a hundred metres underground. Its length is thus almost the same as that of the Circle Line of the London Underground. The acceleration is provided by electromagnetic fields inside superconducting radiofrequency cavities through which the beams pass. There are eight such cavities per beam along the LHC. A total of 1232 superconducting dipole magnets bend the paths of the protons to keep them in orbit. Magnets of higher multipolarity serve to focus, shape, and stabilise the beams. The limiting factor on the beam energy in the LHC (and other circular hadron colliders) is the field strength of the main bending dipoles. The field must be very strong and highly homogeneous to allow for a feasible accelerator radius and stable beams. Conversely, the limit on what field strengths are technically and economically feasible determines how large a collider needs to be to achieve a given collision energy. It is why the LHC is so large.

The protons in each beam are grouped into up to 2808 bunches. Each bunch comprises $\mathcal{O}(10^{11})$ protons. At four interaction points along the LHC, the proton bunches of the two beams are made to pass through each other, producing typically $\mathcal{O}(10)$ inelastic proton-proton interactions per bunch crossing (depending on the run conditions). The bunches are generally 7.5~m apart, so bunch crossings occur every 25~ns at each interaction point. The interaction points are instrumented with huge, sophisticated detectors, including the ATLAS detector, that observe the particles produced in the collisions.

Compared to fixed-target experiments, colliding two beams of equal energy $E_{\text{beam}}$ has the advantage that all the beam energy can be converted into mass for new particles. In fixed-target experiments, conservation of momentum means that much of the collision energy must go into kinetic energy of the produced particles, and is therefore lost to the production of interesting heavy particles. The LHC proton-proton centre-of-mass energy is $2 E_{\text{beam}} = 13$~\TeV{} (since 2015). If one of the beams were replaced with a fixed target of protons, the centre-of-mass energy would only be $\sqrt{2 m_{\text{proton}} E_{\text{beam}}} \approx 114$~\GeV{} --- not even sufficient to produce a SM Higgs boson!\footnote{Indeed, due to momentum sharing among partons (\mysec~\ref{sec:pdf}), in practice not even sufficient to produce $\PW/\PZ$ bosons at a reasonable rate.}

\subsection{Luminosity}
The rate at which the LHC delivers collisions is quantified by the instantaneous luminosity $\mathcal{L}$, measured in units of inverse cross section per time, such as $\text{cm}^{-2} \text{s}^{-1}$. The expected event rate of a process with cross section $\sigma$ is given by
\begin{equation*}
	\langle \text{rate} \rangle(t) = \sigma \mathcal{L}(t).
\end{equation*}
In ATLAS, the delivered instantaneous luminosity is measured by monitoring the total rate of inelastic collisions \cite{DAPR-2013-01}. This technique is suitable for monitoring relative variations, but not for fixing the absolute calibration scale of the instantaneous luminosity. To determine the calibration, van der Meer scans \cite{vdm_scans} are performed $\mathcal{O}(1)$ time per year using dedicated beam conditions. At the time of writing of this thesis, the LHC holds the record for the highest instantaneous luminosity (around $1.7 \times 10^{34}~\text{cm}^{-2}\text{s}^{-1}$, in 2017 \cite{luminosity_public_plots}) of any collider ever built.

\begin{figure}[h!]
\centering
\subfigure[]{\includegraphics[width=0.47\textwidth]{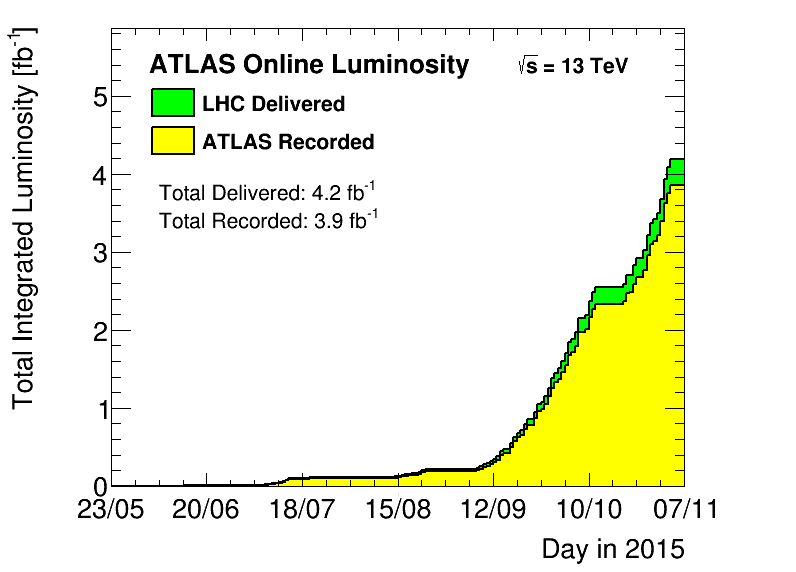}}
\subfigure[]{\includegraphics[width=0.47\textwidth]{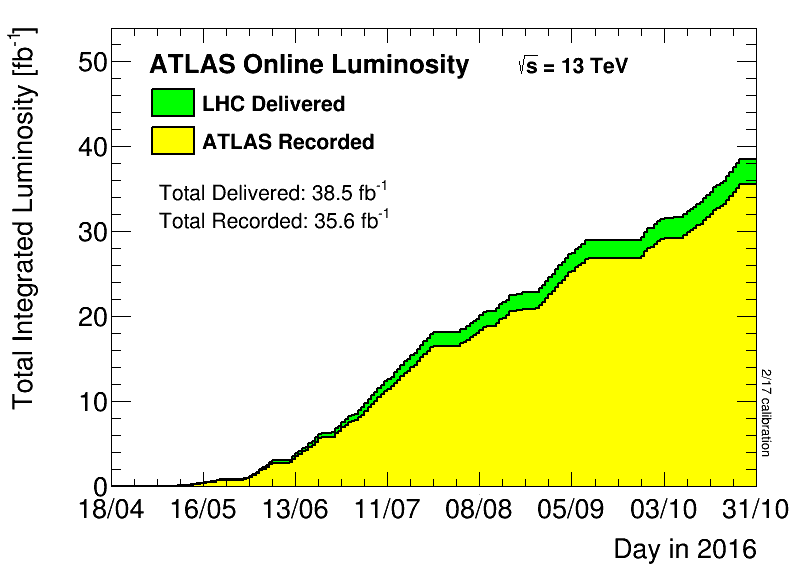}}
\caption{Evolution of the integrated luminosity in 2015 (a) and 2016 (b). Taken from \myref~\cite{luminosity_public_plots}.}
\label{fig:experiment_int_lumi_by_year} 
\end{figure}

The time integral of the instantaneous luminosity over the relevant data-taking period is called the integrated luminosity, $L = \int_{0}^{T} \kern-2pt\mathcal{L}(t)\,\text{d}t$. The recorded integrated luminosity is used as a measure of how much data is available for analysis. \myfig~\ref{fig:experiment_int_lumi_by_year} shows how the integrated luminosity was delivered by the LHC and recorded by ATLAS over the course of the 2015 and 2016 data takings. The difference between delivered and recorded integrated luminosities is due to detector inefficiency as well as the fact that the detector only records data during high-quality beam conditions. Precise knowledge of the integrated luminosity is crucial for analysis, because it is needed to correctly relate background, signal, and assumed BSM cross sections to the expected numbers of events.

\subsection{Pileup}
\label{sec:experiment_pileup}

\begin{figure}[h!]
\centering
\includegraphics[width=0.48\textwidth]{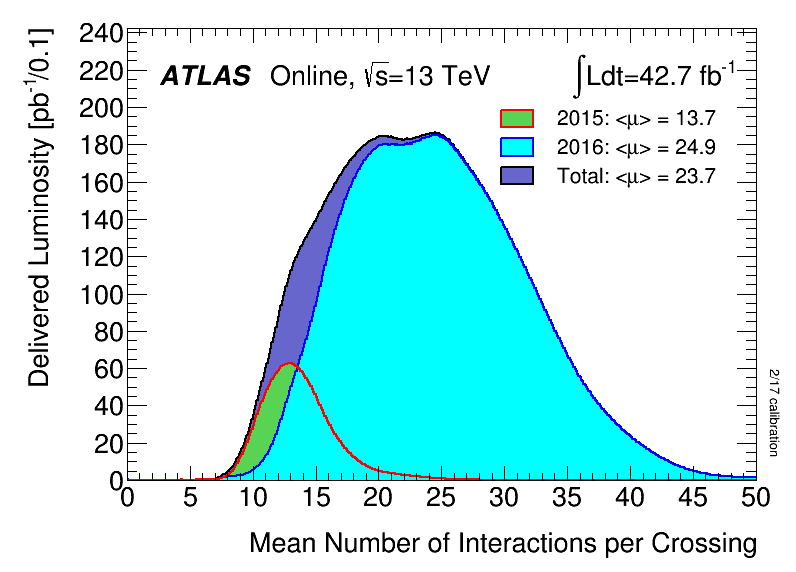}
\caption{Delivered integrated luminosity as a function of the mean number of $\Pproton\Pproton$ interactions per bunch crossing. The data for 2015, 2016, and the sum of both years are shown. Taken from \myref~\cite{luminosity_public_plots}.}
\label{fig:experiment_pileup_profile}
\end{figure}

Additional inelastic proton-proton interactions occurring in the same bunch crossing as the process of interest, or in nearby bunch crossings, are referred to as pileup. Particles from pileup overlay the process of interest in the detector, posing a challenge for event reconstruction. The vast majority of inelastic $\Pproton\Pproton$ collisions can be described by QCD scattering at low momentum transfer, so pileup predominantly produces soft hadrons.
The integrated luminosity delivered to ATLAS as a function of the mean number of inelastic $\Pproton\Pproton$ interactions per bunch crossing is shown in \myfig~\ref{fig:experiment_pileup_profile}. It is calculated for each bunch crossing assuming an inelastic $\Pproton\Pproton$ cross section of $\sigma_{\text{inel}} = 80$~mb at 13~\TeV{} as
\begin{equation*}
\langle \mu \rangle = \frac{\mathcal{L}_{\text{bunch}}\, \sigma_{\text{inel}}}{f_{\text{LHC}}},
\end{equation*}
where $\mathcal{L}_{\text{bunch}}$ is the instantaneous luminosity per bunch crossing and $f_{\text{LHC}}$ the revolution frequency of the LHC \cite{luminosity_public_plots}.
In the data analysed for this thesis, the average number of inelastic interactions per bunch crossing is $\langle \mu \rangle = 23.7$. 

\section{Interaction of particles with matter}\label{sec:particles_in_matter}
Particle detectors such as ATLAS exploit the interaction of particles with matter to identify them and measure their momenta, but also suffer from unwanted material interactions of the particles.

\begin{figure}[h!]
\centering
\includegraphics[width=0.5\textwidth]{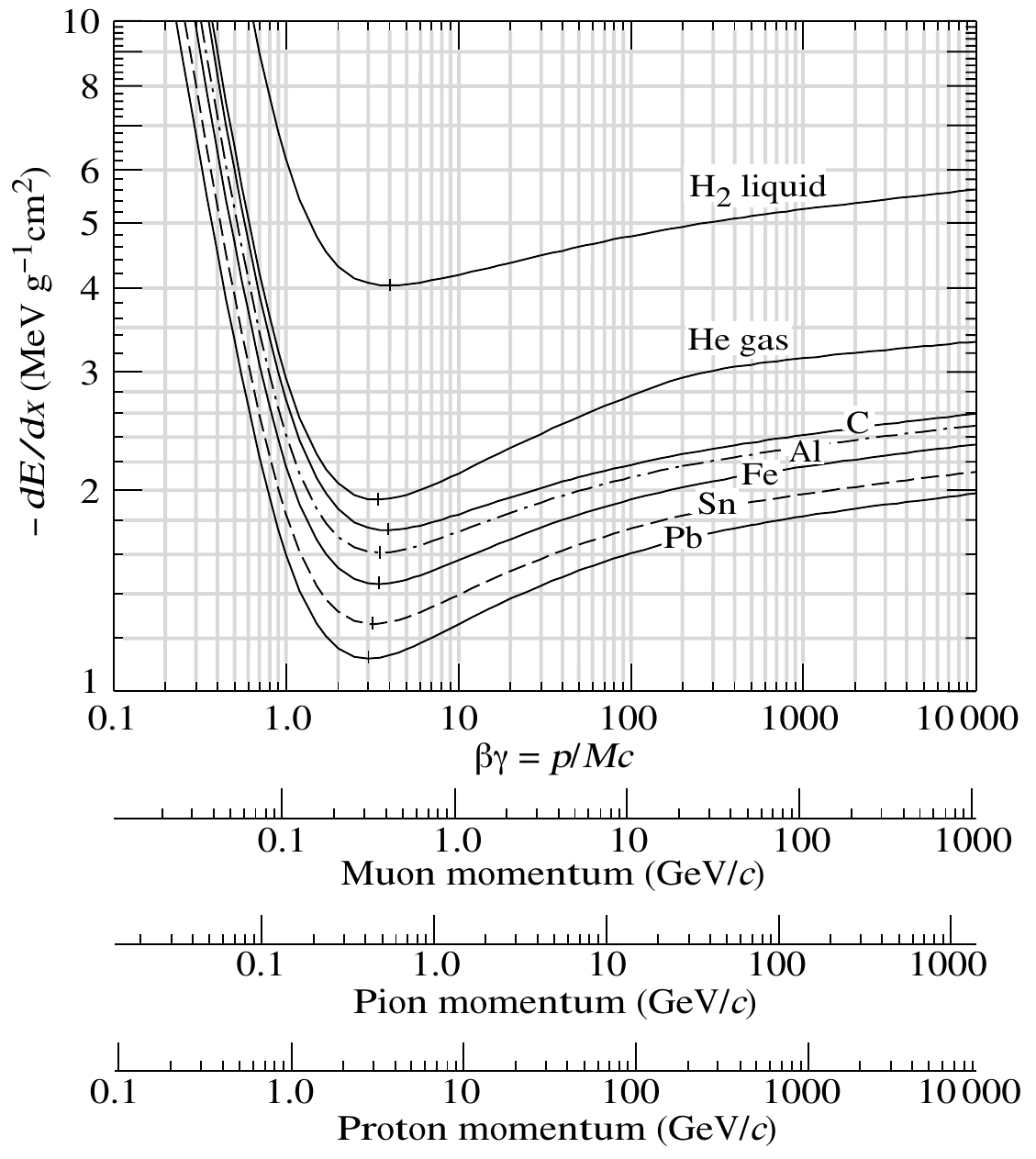}
\caption{Mean energy loss rate per distance travelled of charged particles passing through liquid hydrogen, gaseous helium, carbon, aluminium, iron, tin, and lead. Radiative effects, relevant for muons and pions, are not included. Taken from \myref~\cite{Olive:2016xmw}.}
\label{fig:mip}
\end{figure}

As charged particles travel through matter at relatively high velocities, they lose energy via ionisation of the material. The mean energy loss per distance travelled is approximately described by the Bethe formula and depends on the material (density, atomic number, etc.) and the speed $\beta$ of the particle. An interesting feature of the mean ionisation losses $\langle \text{d}E / \text{d}x \rangle$ is that they exhibit a minimum around $\beta\gamma \sim 3$ (depending slightly on the material) and that the mean ionisation loss experienced by a particle only rises logarithmically with its energy for several orders of magnitude, until radiative effects such as Bremsstrahlung become important. This behaviour is shown in \myfig~\ref{fig:mip}. It remains approximately constant between $\sim$$1m$ and $\sim$$1000m$, where $m$ is the mass of the particle (e.g. $m \approx 0.1$~\GeV{} for a muon and $m \approx 1$~\GeV{} for a proton). This means that at typical particle energies observed in ATLAS, most stable particles passing through the detector material can be approximated as \textit{minimum ionising particles}.



Unlike other charged particles at typical ATLAS energies of several to several hundred \GeV{}, electrons experience significant energy loss via Bremsstrahlung due to their low mass. Energetic electrons and photons hitting a bulk of high-density material initiate electromagnetic showers of secondary particles, in which electrons and positrons lose energy mainly via Brems\-strah\-lung ($\Pepm \Pphoton_{\text{m}} \to \Pepm \Pphoton$) and photons mainly via pair creation ($\Pphoton\Pphoton_{\text{m}} \to \Ppositron\Pelectron$), where the photons $\Pphoton_{\text{m}}$ are quanta of the electromagnetic fields of the atomic nuclei and (to a lesser degree \cite{photons_in_matter}) atomic electrons in the material. The electromagnetic radiation length $X_0$ defines the rate of energy loss via Bremsstrahlung of an electron in material, $\text{d}E/\text{d}x = E/X_0$. This means that after a distance $X_0$ in the material, the electron's energy is reduced to $1/e \approx 0.37$ of the original. The mean distance travelled by an energetic photon before undergoing pair production is related to the radiation length and around $30\%$ longer, $\frac{9}{7} X_0$ \cite{nuclei_and_particles}.

Similarly, energetic hadrons, both charged and neutral, will produce hadronic showers when hitting a bulk of high-density material. These are mainly initiated via inelastic hadron-nucleus scattering. They are very complex and involve a wide range of phenomena, such as elastic neutron scattering off nuclei, decays of secondary particles ($\Ppizero \to \Pphoton\Pphoton$, $\PJpsi \to \Pleptonplus\Pleptonminus$, etc.), as well as excitation of nuclei and subsequent relaxation via emission of photons ($\text{A}^* \to \text{A}\Pphoton$). The produced photons and electrons can in turn initiate electromagnetic subshowers. The number of particles produced in the shower increases approximately logarithmically with the energy of the initial hadron. An important quantity is the hadronic interaction length $\lambda_{\text{had}}$, after which a hadron beam has dropped to $1/e$ of the original intensity, satisfying $\text{d}E/\text{d}x = E/\lambda_{\text{had}}$. It depends on the material and is in general larger than the radiation length for a given material.

The length of an electromagnetic or hadronic shower is approximately proportional to the logarithm of the energy. This is fortunate, because it facilitates the design of a single, relatively compact calorimeter that is adequate for measuring energies over several orders of magnitude.

Neutrinos do not, for any practical purposes, interact with the detector at all and therefore escape it undetected. Their presence can at best be inferred from an analysis of the detected particles: because of conservation of momentum, the net momentum of all undetected particles must be equal to the negative net momentum of all detected particles (in the centre-of-mass system of the interacting initial-state particles). The presence of as of yet unknown exotic particles that are long-lived and barely interact with matter might also be inferred like this. Only the net momentum of all undetected particles can be inferred this way. Their number and nature cannot be determined: they may be any combination of neutrinos, possibly exotic particles that go undetected, and in principle detectable particles that simply fell outside the detector acceptance.

\section{ATLAS detector}
\begin{figure}[h!]
\centering
\includegraphics[width=0.5\textwidth]{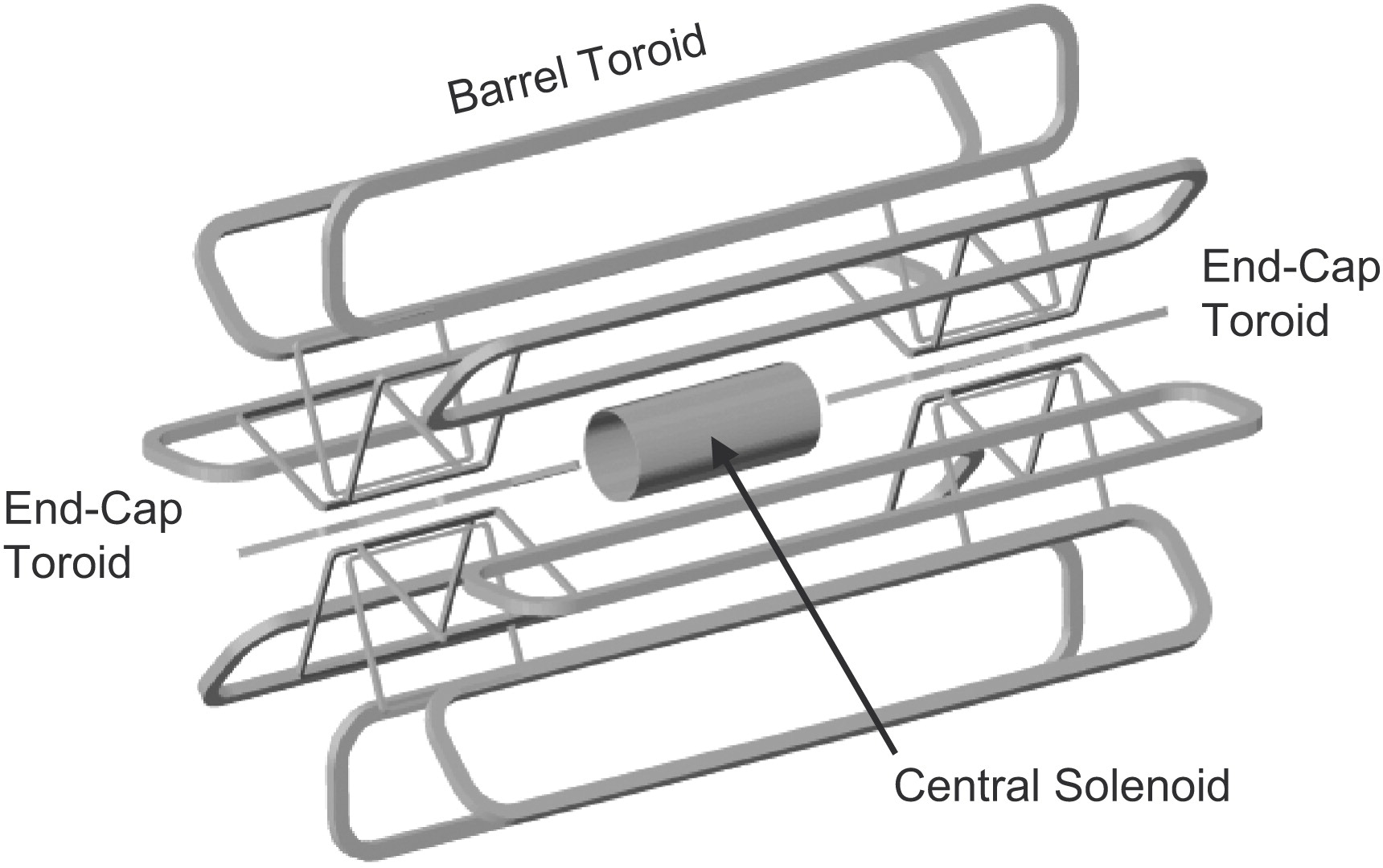}
\caption{Drawing of the magnet coils of the ATLAS detector. Taken from \myref~\cite{YAMAMOTO200853}.}
\label{fig:magnets_only_drawing}
\end{figure}

The ATLAS detector \cite{PERF-2007-01,ATLAS-TDR-19,ATLAS-TDR-19-ADD-1} is a multipurpose particle detector with a cylindrical geometry. It consists of layers of trackers in the inner detector (ID), electromagnetic calorimeters (ECAL), hadronic calorimeters (HCAL), and a muon spectrometer (MS). These are often divided into coaxial tubular `barrel' segments in the middle and circular disks called `endcaps' on each side. Additionally, the detector encompasses 25 superconducting magnet coils, shown in~\myfig~\ref{fig:magnets_only_drawing}. The detector is designed to measure a wide range of phenomena in proton and heavy-ion collisions that produce particles with high transverse momenta. It is operated, continuously developed, and its data exploited for physics by the international ATLAS collaboration. The sun literally never sets on the collaboration, whose institutes and members are based on all continents except Antarctica.

\begin{figure}[h!]
\centering
\begin{tikzpicture}
\node[anchor=south west,inner sep=0] (image) at (0,0) {\includegraphics[width=0.4\textwidth]{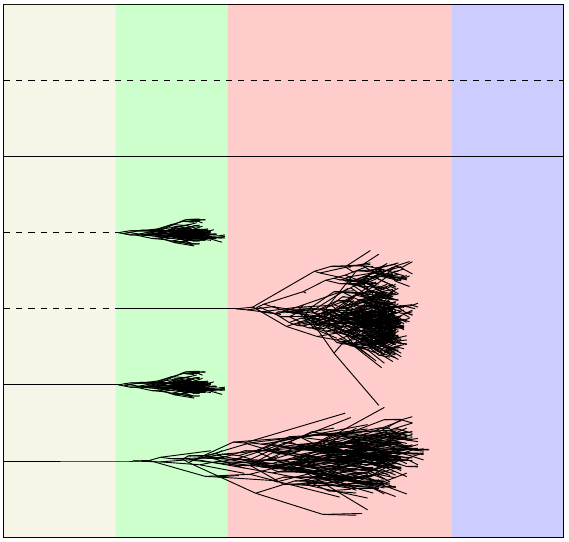}};
\begin{scope}[x={(image.south east)},y={(image.north west)}]
\node at (0.1, 1.05) {\footnotesize ID};
\node at (0.3, 1.05) {\footnotesize ECAL};
\node at (0.6, 1.05) {\footnotesize HCAL};
\node at (0.885, 1.05) {\footnotesize MS};
\node[anchor=west] at (-0.1, 0.85) {\small $\nu$};
\node[anchor=west] at (-0.1, 0.71) {\small $\Pmu^{\pm}$};
\node[anchor=west] at (-0.1, 0.57) {\small $\Pphoton$};
\node[anchor=west] at (-0.1, 0.43) {\small $\Pneutron$};
\node[anchor=west] at (-0.1, 0.29) {\small $\Pe^{\pm}$};
\node[anchor=west] at (-0.1, 0.15) {\small $\Ppi^{\pm}$};
\end{scope}
\end{tikzpicture}
\caption{Typical signatures of different particle species in the ATLAS detector: neutrinos $\Pnu$, muons $\Pmu^{\pm}$, photons $\gamma$, neutrons $\Pneutron$, electrons $\Pe^{\pm}$, and charged pions $\pi^{\pm}$. A dashed line indicates that the particle leaves no energy. A solid line indicates energy deposits. The detector proportions are not realistic. Drawing taken from \myref~\cite{Nachman:2016qyc}.}
\label{fig:experiment_signatures}
\end{figure}
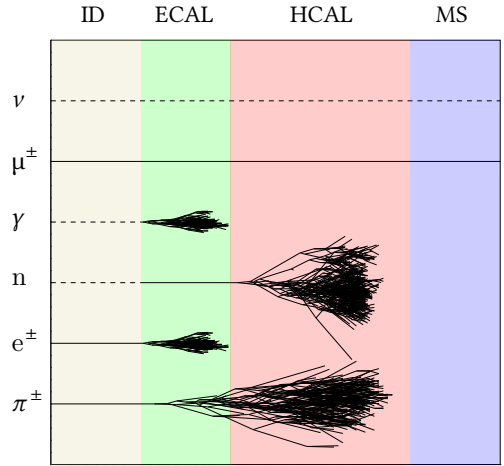

\myfig~\ref{fig:experiment_signatures} shows typical signatures of different particle species in the ATLAS detector that allow their reconstruction and identification. The different subdetectors are presented in the following sections, motivating some of the important design choices.

\subsection{Trigger}

A two-level \textit{trigger} system is used to select events of interest in real time \cite{TRIG-2016-01}. The Level-1 trigger is implemented in hardware and uses a subset of detector information to reduce the rate of potentially interesting events from 40~MHz to around 100~kHz. This is followed by a software-based high-level trigger system that reduces the event rate to about 1~kHz.

\subsection{Tracker}\label{sec:tracker}

\begin{figure}[p]
\centering
\includegraphics[width=0.7\textwidth]{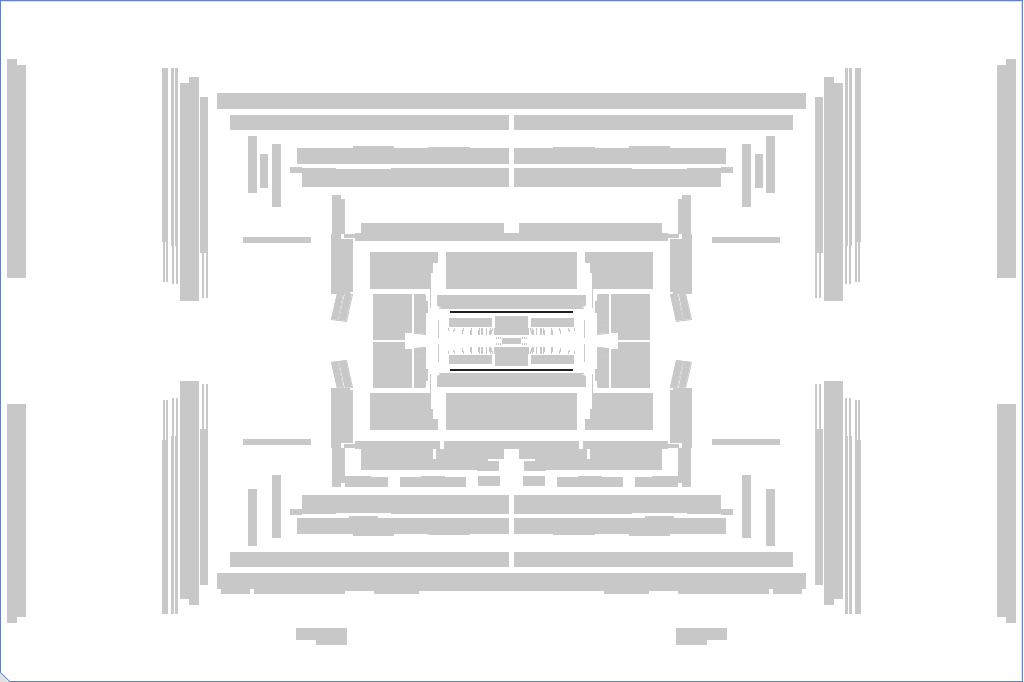}
\caption{The location of the solenoid magnet (highlighted black) in the ATLAS experiment, shown in the $\rho$-$z$ projection. Drawing made with \atlantis{} \cite{atlantis}.}
\label{fig:experiment_solenoid_location}
\end{figure}

The tracker measures points along the trajectory of a charged particle (``hits''), from which a track can be reconstructed. The signal is created by the ionisation the particle causes in the sensitive material. 
The tracker is embedded in a homogeneous magnetic field $B \approx 2$~T, parallel to the beam, provided by the central solenoid magnet shown in Figs.~\ref{fig:magnets_only_drawing} and \ref{fig:experiment_solenoid_location}. This forces the charged particles onto helical trajectories. The sign of the electric charge of the particle can be determined from the bending direction.\footnote{It is assumed that all trackable particles have electric charge $\pm e$. Known particles with electric charge $\pm 2e$, such as $\Sigma_{\text{c}}^{++}$, are too short-lived to travel a measurable distance in the tracker. Searches for exotic long-lived doubly-charged particles could perform a simple rescaling of the momentum of candidates for such particles.} Measuring the bending radius $R$ with the tracker allows calculating the transverse momentum of the particle,
\begin{equation*}
\pt = \sqrt{4\pi\alpha} B R \approx 0.3 \left(\frac{B}{\text{T}}\right) \left(\frac{R}{\text{m}}\right) \GeV
\end{equation*}
Tracks with momentum greater than about 1~\GeV{} are fairly straight lines in the tracker. In their case, the relative momentum resolution $\sigma_{\pt}/\pt$ is proportional to $\sigma_{s}\kern0.5pt\pt / (L^2 B)$, where $\sigma_{s}$ can be thought of as the position resolution of the tracker\footnote{It is actually the resolution of the track sagitta measurement.} and the lever arm $L$ is the distance over which measurements can be made. This demonstrates the need for a large tracker in a strong magnetic field. The momentum resolution is worse for high-\pt{} tracks, as intuitively expected from the decreasing track curvature.

\begin{figure}[p]
\centering
\includegraphics[width=0.6\textwidth]{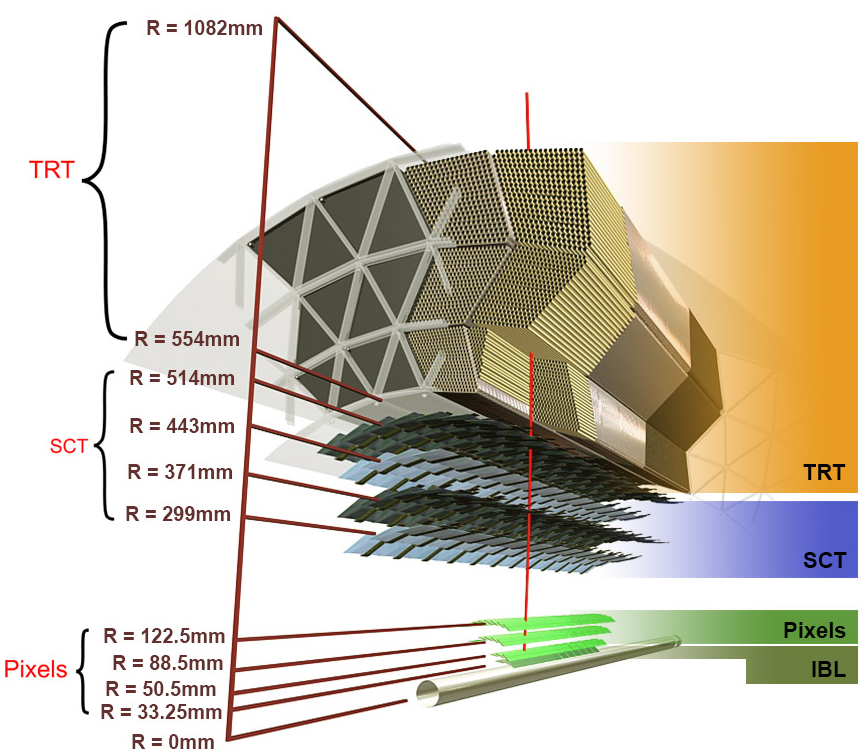}
\caption{Segment of the ATLAS inner detector, showing the tracker layers. The silicon strip detector is marked SCT, for \emph{semiconductor tracker}. For the pixel and strip layers, the radii correspond to the mean radius of the sensitive material in each layer. The innermost pixel layer, IBL, was added in 2014, during the first long shutdown of the LHC. Taken from \myref~\cite{Potamianos:2016ptf}.}
\label{fig:tracker_layers}
\end{figure}

\begin{figure}[h!]
\centering
\includegraphics[width=0.7\textwidth]{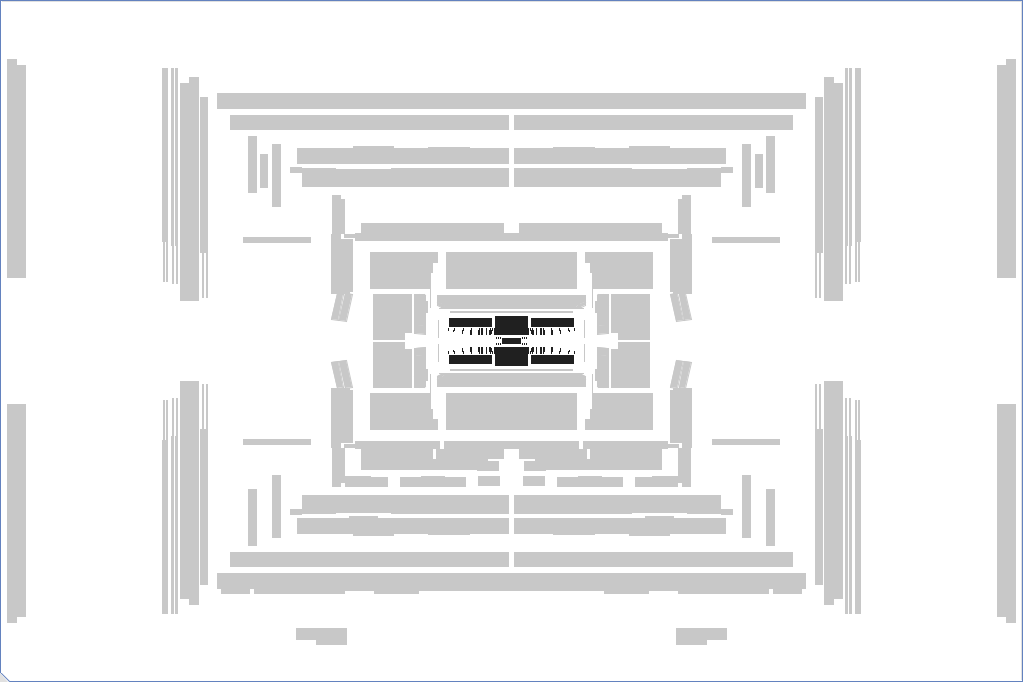}
\caption{The location of the inner detector (highlighted black) in the ATLAS experiment, shown in the $\rho$-$z$ projection. Drawing made with \atlantis{} \cite{atlantis}.}
\label{fig:experiment_id_location}
\end{figure}

The layout of the ID tracker is shown in \myfig~\ref{fig:tracker_layers}, and its location inside the ATLAS detector is indicated in \myfig~\ref{fig:experiment_id_location}. It consists of four inner layers of silicon pixels, four (nine) layers of silicon strips in the barrel (each endcap), and the transition radiation tracker (TRT). The entire inner detector is surrounded by the superconducting solenoid magnet coil.

Both the pixel and strip detector measurement is based on charged particles producing thousands of electron-hole pairs\footnote{A minimum ionising particle produces around 80~electron-hole pairs per micrometre of silicon traversed.} in a p-n junction of silicon. A voltage is applied across the silicon, collecting the electrons and holes on the faces of the pixel or strip. The electrical signal is read out and, if significant above the electronics noise, interpreted as a hit.

The pixel detector ($|\eta| < 2.5$) is needed for a high-precision measurement close to the beam. It has very high granularity to resolve the high particle trajectory density near the interaction point, consisting of $\mathcal{O}(100~\text{million})$ pixels. In the barrel, each pixel has a size of $50 \times 400~\text{\textmu{}m}^2$ in the $\phi \times z$ directions, achieving a resolution of 10~\textmu{}m ($\phi$) and 115~\textmu{}m ($z$). In the endcaps, the dimensions and resolutions are numerically identical, but with the $z$-direction replaced by the $\rho$-direction. Each charged particle typically produces four pixel hits. The pixel detector is crucial for reconstructing primary and secondary vertices.

The strip detector ($|\eta| < 2.5$) consists of four double layers of silicon strips in the barrel and nine double layers per side in the endcaps. Frontside-backside pairs of hits in each double layer are combined into a single space point measurement, with a resolution of 17~\textmu{}m in the $\phi$-direction and 580~\textmu{}m in the $z$-direction ($\rho$-direction) in the barrel (endcaps).

The TRT consists of many $4$-mm-thick polyimide straw tubes filled with a Xe-$\text{CO}_2$-$\text{O}_2$ gas mixture, with a gold-plated tungsten wire stretched through the middle of the tube. The tubes run parallel to the beam in the barrel and radially in the endcaps. The primary signal is generated by ionisation in the gas left by the primary particle, with acceleration towards the wire leading to an avalance of secondary ionisation and finally an electric signal in the wire. In addition, the space between the tubes is filled with polymer fibres or foils, creating many material boundaries. Highly relativistic particles passing through these emit $\mathcal{O}(10~\keV)$ photons as transition radiation \cite{tr_ginzburg_frank}.\footnote{An accessible discussion can be found in many textbooks, e.g.~Refs.~\cite{landau_lifshitz_ed,jackson_ed}.} These can be absorbed by the xenon atoms in the tube gas, amplifying the ionisation signal significantly. As the energy emitted via transition radiation is proportional to $\gamma = E/m$ of the particle, the TRT can help discriminate between electrons and hadrons, up to a particle energy of about 150~\GeV{}. The TRT only spans $|\eta| < 2.0$ and only measures the $\phi$ coordinate (and the sign of the $z$-coordinate) due to the extension of the tubes. Its measurement resolution is 130~\textmu{}m. This is better than would be expected from the tube radius, improved by considering the charge-carrier drift time. The poorer resolution compared to the silicon trackers is partly compensated by the high number of TRT hits per track, on average 36. The large radius of the TRT also extends the lever arm of the track measurement, improving the momentum resolution.

The tracker material is kept to a minimum to disturb the particles as little as possible before they hit the calorimeters. Nevertheless, in particular electrons can undergo significant Bremsstrahlung losses in the tracker, while photons can convert to $\Ppositron\Pelectron$ pairs.


In 2022--2023, the entire inner detector is scheduled to be replaced by the new inner tracker (ITk) using only silicon-based detector technologies \cite{phase2_letter_of_intent}. This is in preparation of the ATLAS detector for the High Luminosity LHC \cite{Rossi:1471000}.

\subsection{Calorimeters}\label{sec:experiment_calos}

Calorimeters measure the particle energy destructively, i.e.~stopping the particle. They are the only way ATLAS can measure neutral particles, and they help provide nearly hermetic coverage, important for measuring missing transverse energy. ATLAS uses sampling calorimeters, in which only a part of the incident particles' energy is measured directly, while the remaining part is absorbed. The original energy is calculated based on the measured value by using a calibration. The calorimeters contain layers of high-density metal to cause showering. This transfers the energy of a single particle to a cascade of many lower-energy particles. Showering is necessary to measure the energy of neutral particles, since they must first transfer it to charged particles. The metal also absorbs part of the energy, allowing for a more compact calorimeter design. In between the metal layers are layers of sensitive detector medium to read out the ionisation or light signal caused by the secondary particles. The downside of sampling calorimeters is that their energy resolution $\sigma_E / E$ is generally lower than that of homogeneous calorimeters (where the entire calorimeter volume is sensitive material), due to higher sensitivity to stochastic variations in the showering. The relative energy resolution can be parametrised as
\begin{equation}\label{eq:calores}
	\frac{\sigma_E}{E} = a\, \oplus \frac{b}{\sqrt{E}} \oplus \frac{c}{E},
\end{equation}
where $\oplus$ indicates summing in quadrature, e.g.~$x \oplus y \equiv \sqrt{x^2 + y^2}$. The coefficients $a$, $b$, and $c$ depend on the detector. The energy-independent term $a$ includes effects such as nonlinear response and imperfect calibration. The stochastic term $b/\sqrt{E}$ reflects statistical fluctuations in the shower development, whose relative importance is larger for less energetic showers. The term $c/E$ encompasses noise effects in the electronics/optics, but also due to low-energy particles e.g.~from pileup. Its importance decreases with energy, as the signal-over-noise ratio increases. As the form of \myeq{}~\ref{eq:calores} shows, the energy resolution \emph{improves} with the growing energy. This is the opposite case than for track momentum resolution.

Due to the different responses to electromagnetic and hadronic showers, it is beneficial to place an electromagnetic calorimeter producing only electromagnetic compact showers upstream of a hadronic calorimeter. In the hadronic calorimeter, the presence of secondary electromagnetic showers is unavoidable. The different response to these is corrected for using software algorithms.

\mytab~\ref{tab:radiation_and_hadint_lengths} shows the radiation lengths $X_0$ and hadronic interaction lengths $\lambda_{\text{had}}$ for the absorber materials used in ATLAS calorimeters. The ECAL uses absorber materials with a relatively high ratio $\lambda_{\text{had}} / X_0 \sim 30$ to avoid prompting hadronic showers.

\begin{table}[h!]
\centering
\begin{tabular}{lllll}
\toprule
Material & $X_0$ (cm) & $\lambda_{\text{had}}$ (cm) & $\lambda_{\text{had}} / X_0$ & Used in\\
\midrule
Lead & \phantom{0}0.56 & 18 & 31 & ECAL\\
Tungsten & \phantom{0}0.35 & \phantom{0}9.9 & 28 & ECAL\\
Copper & \phantom{0}1.4 & 15 & 11 & HCAL\\
Iron & \phantom{0}1.8 & 17 & 10 & HCAL (in steel)\\
\bottomrule
\end{tabular}
\caption{Radiation lengths $X_0$ and hadronic interaction lengths $\lambda_{\text{had}}$ for various materials used in ATLAS calorimeters. Also shown is their ratio, as it informs the choice of absorber materials for the electromagnetic calorimeter. All data taken from \myref~\cite{rad_and_hadint_lengths}.}
\label{tab:radiation_and_hadint_lengths}
\end{table}

\subsubsection*{Electromagnetic calorimeter}
\begin{figure}[h!]
\centering
\includegraphics[width=0.7\textwidth]{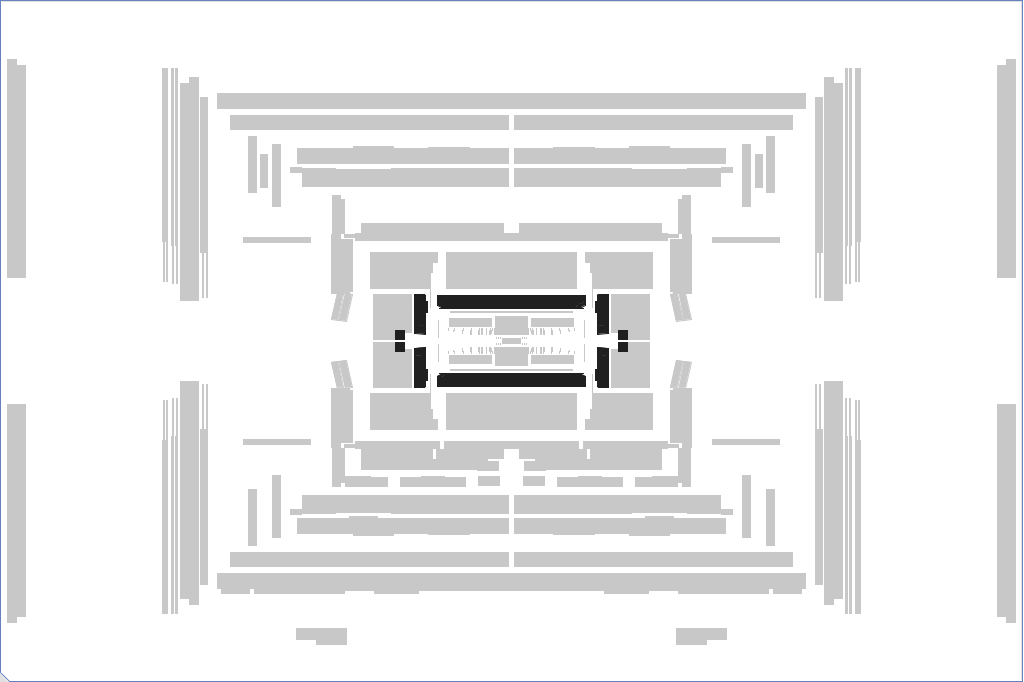}
\caption{The location of the electromagnetic calorimeter (highlighted black) in the ATLAS experiment, shown in the $\rho$-$z$ projection. Drawing made with \atlantis{} \cite{atlantis}.}
\label{fig:experiment_lar_location}
\end{figure}

The location of the ECAL is shown in \myfig~\ref{fig:experiment_lar_location}. It covers the region $|\eta| < 3.2$ and consists of layers of lead or tungsten absorber embedded in liquid argon. Liquid argon was chosen for the high intrinsic linearity and stability over time of its signal response, as well as its radiation hardness \cite{PERF-2007-01}. The ECAL has a total length of about $22 X_0$ to ensure good shower containment. Based on the values in \mytab~\ref{tab:radiation_and_hadint_lengths}, this corresponds to less than one hadronic interaction length, so no compact hadronic showers form in the ECAL, as desired.

\begin{figure}[h!]
\centering
\includegraphics[width=0.7\textwidth]{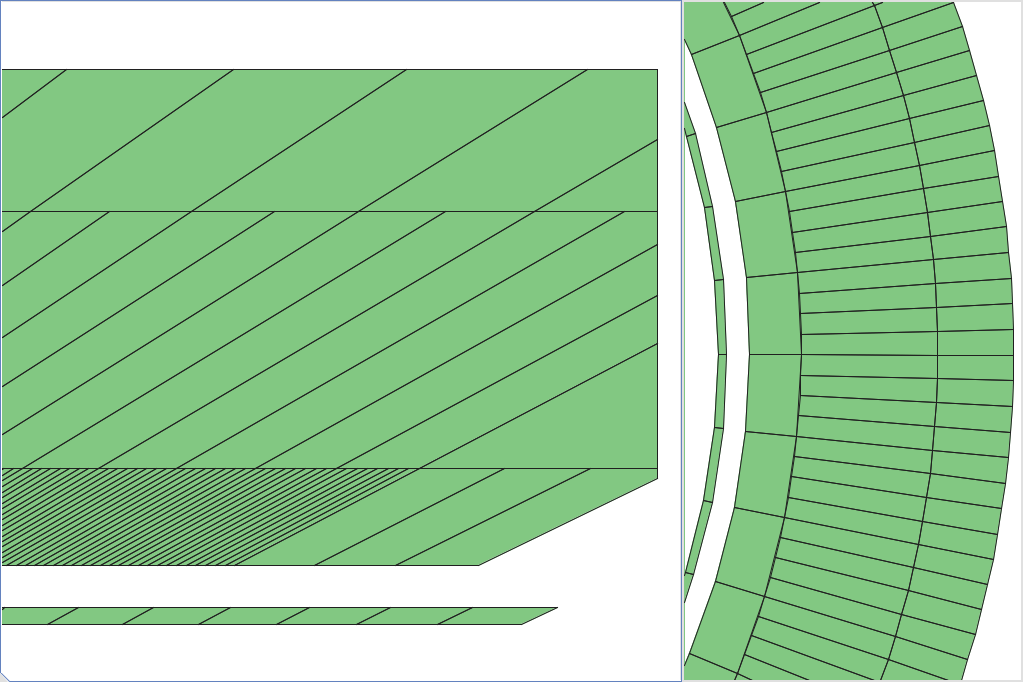}
\caption{Detail showing the cell geometry of part of the liquid-argon calorimeter barrel in the $\rho$-$z$ projection (left) and in the $x$-$y$ projection (right). The single presampler layer is distinctively visible. Drawing made with \atlantis{} \cite{atlantis}.}
\label{fig:experiment_lar_cells_detail}
\end{figure}

The ECAL consists of three layers. The first has a fine $\eta$-segmentation of $\Delta\eta = 0.0031$ to resolve photon pairs from $\Ppizero$ decays. The middle layer has a fine segmentation of $\Delta\eta \times \Delta\phi = 0.025 \times 0.025$ within $|\eta| < 2.47$ to enable electron and photon identification, together with inner-detector information. The outer layer is mainly needed to measure very energetic showers and has a segmentation of $\Delta\eta \times \Delta\phi = 0.050 \times 0.025$. The cell segmentation is shown in \myfig~\ref{fig:experiment_lar_cells_detail}. The ECAL achieves an energy resolution of about 2\% for electrons with $\pt \approx 50$~\GeV{}. It has reduced coverage where its barrel and endcap parts meet, leading to poor energy resolution in the region $1.37 < |\eta| < 1.52$.

A \textit{presampler} is installed immediately before the bulk calorimeter within $|\eta| < 1.8$. It is around 1~cm thick and consists of a single sensitive liquid-argon layer with no upstream absorber. The presampler improves the resolution of the energy measurement by measuring particles originating from showering in the material upstream of itself, such as Bremsstrahlung photons radiated by electrons. The energy losses can be estimated and corrected for.

\subsubsection*{Hadronic calorimeter}
\begin{figure}[h!]
\centering
\includegraphics[width=0.7\textwidth]{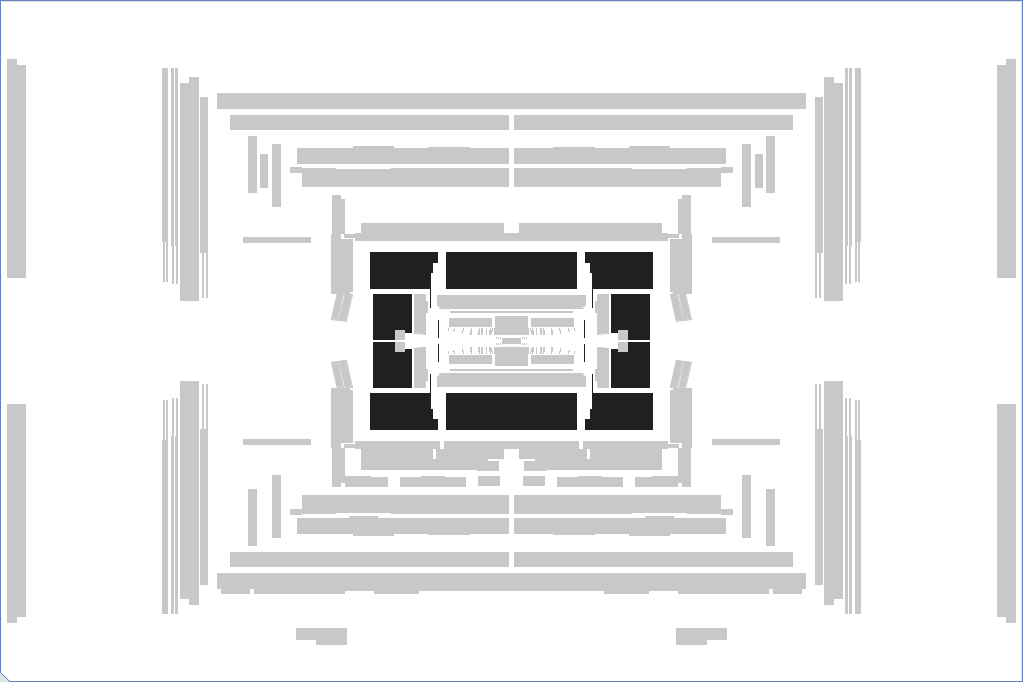}
\caption{The location of the hadronic calorimeter (highlighted black) in the ATLAS experiment, shown in the $\rho$-$z$ projection. Drawing made with \atlantis{} \cite{atlantis}.}
\label{fig:experiment_tile_location}
\end{figure}

The HCAL consists of a steel/scintillating-tile calorimeter, segmented into three barrel structures within $|\eta| < 1.7$, and of two copper/liquid-argon calorimeters within $1.7 < |\eta| < 3.2$. Its location and layout is shown in \myfig~\ref{fig:experiment_tile_location}. The measurement principle is the same as for the ECAL, except that scintillating tiles are used in the central barrel. These are made of plastic that emits light when ionising particles travel through it. Using optical fibres, the light is guided to photomultiplier tubes, in which the optical signal is converted to an electrical one and amplified. The HCAL has a thickness of about 10 hadronic interaction lengths (at $\eta \approx 0$) to ensure good shower containment and minimise punch-through of hadrons into the muon spectrometer. 

The HCAL has three cell layers in the barrel, with cell size $\Delta\eta \times \Delta\phi = 0.1 \times 0.1$ in the first two layers and $0.2 \times 0.1$ in the third. In the endcap, the number of layers increases to four, with a granularity of $0.1 \times 0.1$ in $1.5 < |\eta| < 2.5$ and $0.2 \times 0.2$ in $2.5 < |\eta| < 3.2$. The intrinsic energy resolution of the HCAL is about 15\% for a jet with 100~\GeV{} energy and around 3\% for a jet with 1~\TeV{} energy, depending on the pseudorapidity of the jet.




\subsubsection*{Forward calorimeter}
The calorimetry is extended to $|\eta| = 4.9$ by the forward calorimeter, which consists of one module of copper layers and two modules of tungsten layers, all in liquid argon. The first mainly measures electromagnetic showers while the two outer ones measure hadronic showers.\footnote{The choice of copper for electromagnetic and tungsten for hadronic showers seems unintuitive in light of the ratios $\lambda_{\text{had}} / X_0$ shown in \mytab~\ref{tab:radiation_and_hadint_lengths}, which could favour the opposite choice. However, they were chosen to optimise other factors, as explained in \myref~\cite{PERF-2007-01}.}

\subsection{Muon spectrometer}
\begin{figure}[h!]
\centering
\includegraphics[width=0.6\textwidth]{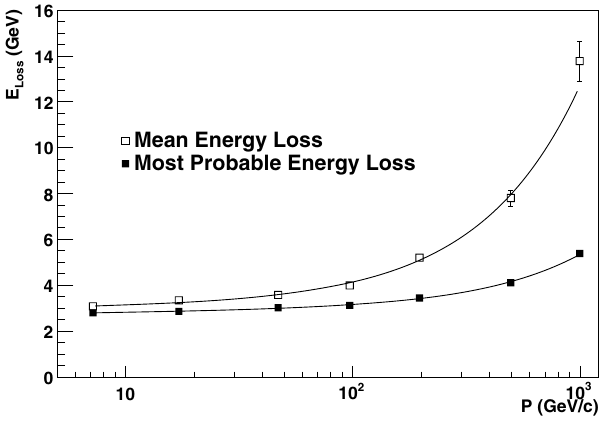}
\caption{Simulated mean and most probable energy loss of a muon traversing the ATLAS detector as a function of its momentum. The considered muons have $0.4 < |\eta| < 0.5$. Taken from \myref{}~\cite{muon_energy_loss_atlas}.}
\label{fig:muon_energy_loss}
\end{figure}

Muons pass through the detector material, including the calorimeters, without stopping. They behave roughly like minimum ionising particles, whose energy loss is via ionisation and depends only mildly on their momentum. \myfig{}~\ref{fig:muon_energy_loss} shows the simulated mean and most probable muon energy loss in the ATLAS detector as a function of the muon momentum. Statistically, the energy loss approximately follows a Landau distribution \cite{Landau:1944if}, and it can be seen that the long tail towards higher values means that the mean loss\footnote{The mean loss is calculated numerically from the simulated values. The analytical mean of a Landau distribution is undefined.} grows faster than the most probable loss.

\begin{figure}[h!]
\centering
\subfigure[Monitored drift tubes provide a precise position measurement in the bending direction.]{\includegraphics[width=0.47\textwidth]{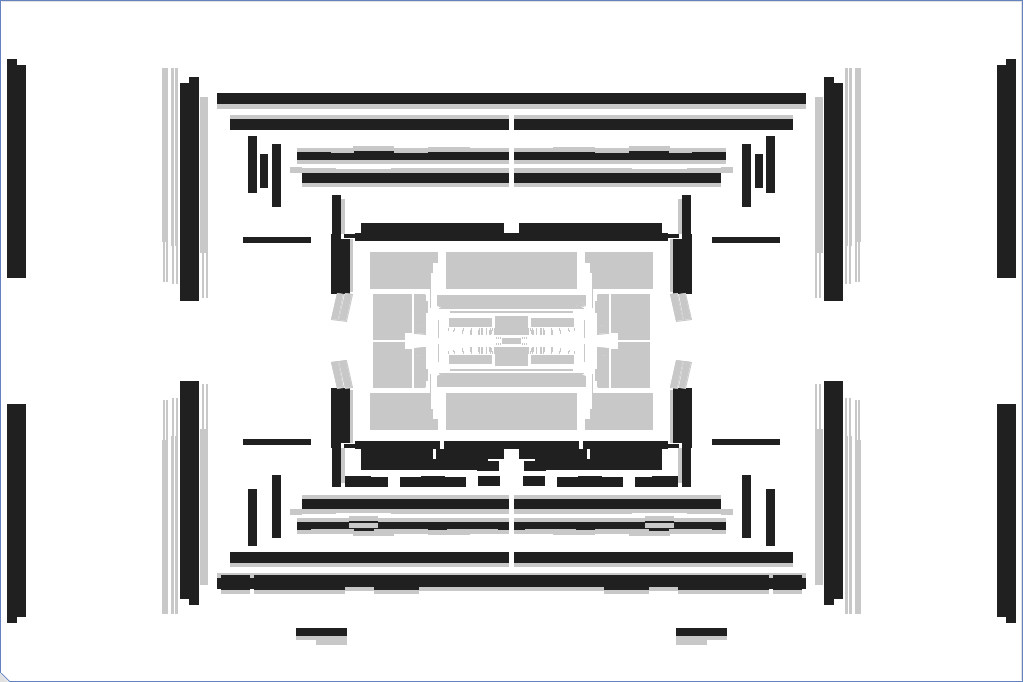}}
\subfigure[Cathode strip chambers provide a precise position measurement in the bending direction as well as a measurement in the non-bending direction.]{\includegraphics[width=0.47\textwidth]{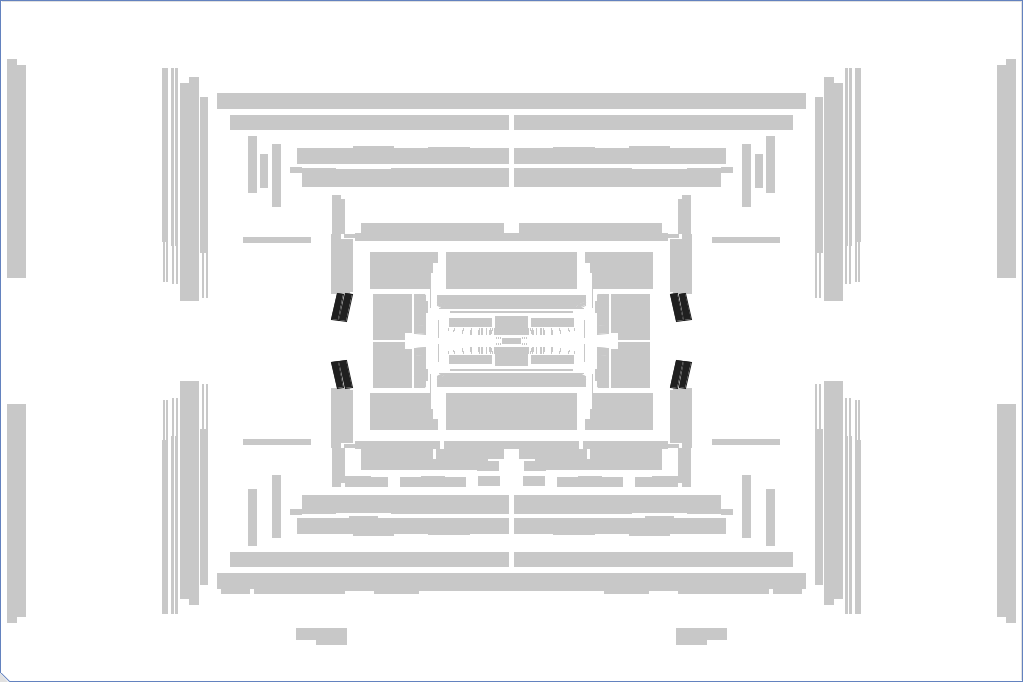}}
\hspace{5mm}
\subfigure[Resistive plate chambers provide a position measurement in the bending and non-bending direction, and fast enough response to be used for low-level triggering.]{\includegraphics[width=0.47\textwidth]{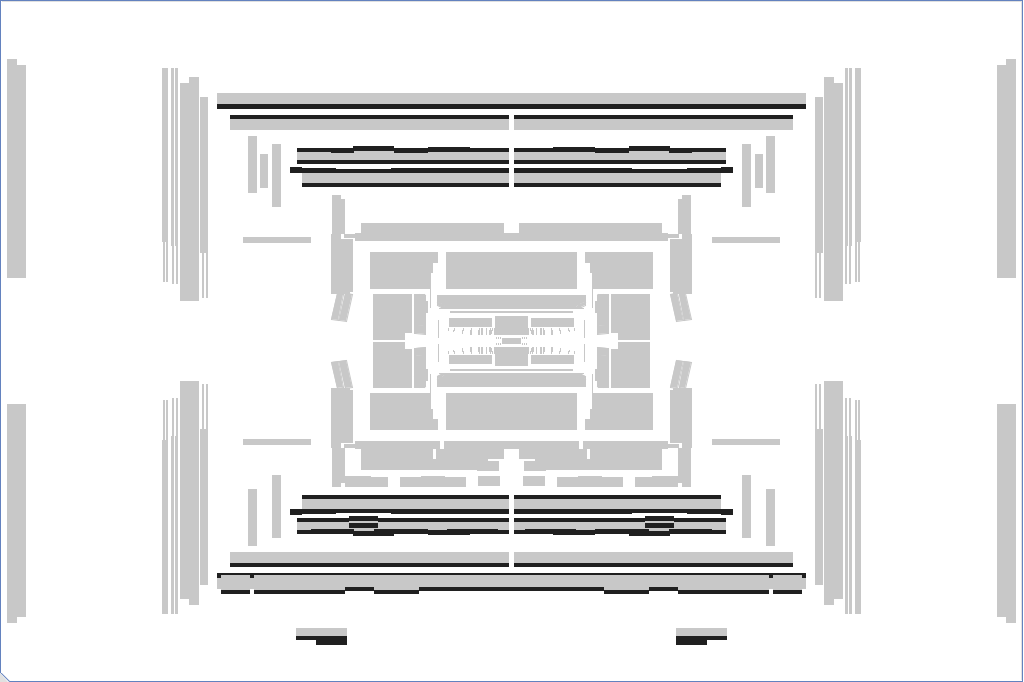}}
\subfigure[Thin gap chambers provide a position measurement in the bending and non-bending direction, and fast enough response to be used for low-level triggering.]{\includegraphics[width=0.47\textwidth]{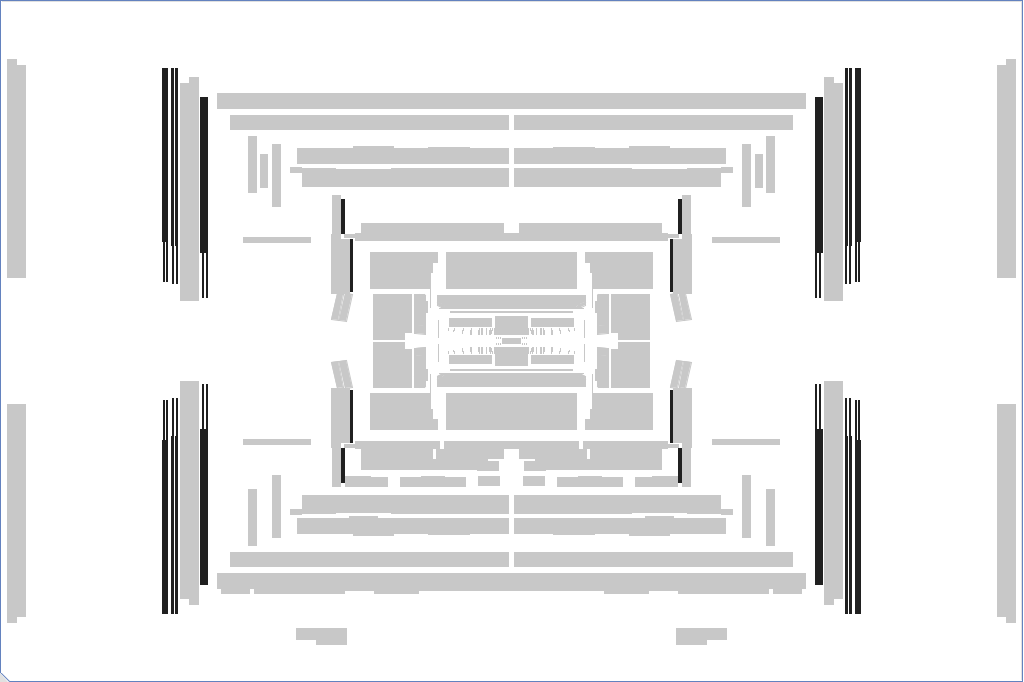}}
\caption{The location of the various muon detectors (highlighted black) in the ATLAS experiment, shown in the $\rho$-$z$ projection. Drawings made with \atlantis{} \cite{atlantis}.}
\label{fig:experiment_ms_location}
\end{figure}

Muons are unique in their even, ionisation-dominated energy loss: energetic hadrons interact strongly in the dense calorimeter material, electrons and photons experience significant radiative or pair-production losses, tau leptons decay before flying a long distance, and neutrinos don't interact with the detector at all in practice. Installing a tracking system beyond the calorimeters therefore gives muons a very distinctive detector signature, allowing their reconstruction with great purity and efficiency. This tracking system is called the muon spectrometer (MS). It is embedded in a non-homogeneous magnetic field generated by toroidal coils in the barrel and endcaps, so muon trajectories are bent in the $\rho$-$z$ plane. The magnetic flux density ranges between around 0.2~T and 2.5~T (3.5~T) in the barrel (endcaps), providing a bending power of approximately 1--7.5~Tm. The MS uses four different detector types, whose locations in the ATLAS detector are shown in \myfig{}~\ref{fig:experiment_ms_location}. In the barrel, the MS extends from a radius of around 5~m to around 10~m.

To measure the muon momentum, very precise tracking in the bending direction is required, so that even the very slight curvature of a high-\pt{} muon track can be measured. Precision tracking is provided by monitored drift tubes (MDT) over the full MS coverage of $|\eta| < 2.7$. These are aluminium tubes filled with 93\% argon and 7\% carbon dioxide with a wire stretched axially through them. They are mostly\footnote{Additional smaller MDTs were installed in 2013--2015 to improve coverage.} 3~cm in diameter and about one to six metres long. 
The MDTs measure hits with a spatial resolution of approximately $35$~\textmu{}m per chamber ($\sim$80~\textmu{}m per tube) in the bending direction, but cannot measure the hit position along the tube, in the non-bending $\phi$ direction. In the forward region $2.0 < \eta < 2.7$, the MDTs are supplemented by an innermost layer of cathode strip chambers (CSC), which are multiwire proportional chambers with cathodes segmented into strips. Their spatial resolution in the bending direction is comparable to that of the MDTs. The cathode segmentation furthermore allows them to measure the hit position in the non-bending direction with a resolution of 5~mm. As the particle flux in the innermost forward chambers is around 20 times higher than the average in the other MS regions \cite{PALESTINI2003337}, this second coordinate provided by the CSC is important for resolving track ambiguities.
The spatial coordinate in the non-bending and bending direction is measured by resistive plate chambers (RPCs) in the barrel and thin gap chambers (TGC) in the endcaps, up to $|\eta| = 2.4$. The measurement resolution is of the order of some millimetres. The RPCs are gas-filled plate capacitors. The TGCs are multiwire chambers operating in saturated mode, providing a finer granularity than the RPCs to accommodate for the higher track density and smaller trajectory bending for a given \pt{} in the forward region. 

The long drift time in the MDTs (up to 700~ns) and CSCs (typically $\sim$20~ns, but with a long tail due to the weak drift field in some regions) makes them unsuitable for low-level triggering, since collisions happen every 25~ns. However, the RPCs and TGCs have a sufficiently fast response to enable muon triggering within $|\eta| = 2.4$.

In the very central region ($|\eta| \approx 0$), there is a gap in the MS coverage to allow for services to the solenoid magnet, the calorimeters and the inner detector. Some of the used muon reconstruction algorithms are optimised to recover part of the lost efficiency in this region. These will be described below. 


%
%

\clearpage

\section{Event reconstruction}

This section briefly describes the reconstruction of objects used in this thesis, namely electrons, muons, and jets. It also describes the reconstruction of low-level objects on which the higher-level objects depend, namely tracks, vertices, and localised energy clusters in the calorimeters. Before doing so, it introduces the important concepts of \emph{efficiency} and \emph{scale factors}.



\subsection{Efficiencies and scale factors}
Efficiencies $\varepsilon$ express the probability of some genuine object (or event) to pass a given selection,
\begin{equation*}
\varepsilon = \frac{\text{number of genuine objects (or events) passing selection}}{\text{number of genuine objects (or events)}}
\end{equation*}
for objects in some sample. Calculating the efficiency is often easy in MC simulation, where the number of genuine objects is known. Here, only one complication arises occasionally, namely if the adequate matching of genuine objects and corresponding selected objects is non-trivial, because the reconstruction has heavily distorted the objects. On the other hand, efficiencies in simulation might differ from the true experimental efficiencies. If this is not taken into account, the MC simulation will model the data incorrectly, and the correction of measurements for detector effects will be wrong. So there is a strong interest to measure the efficiencies directly in data. One widespread way of doing so is the tag-and-probe method.

There are many variations of the tag-and-probe method, but the most relevant to this thesis requires a well-known process to serve as a standard candle, such as the resonance $\PJpsi \to \APmuon\Pmuon$ or $\PZ \to \Ppositron\Pelectron$. All but one of the final-state objects are required to pass a very tight selection, with minimal probability of misidentification. These objects are the \emph{tag}. For instance, when measuring muon identification efficiencies in $\PJpsi \to \APmuon\Pmuon$ events, the tag muon would be required to pass a tight selection. The remaining object is called the \emph{probe} and required to pass very loose selection criteria, whose efficiency is either approximated as one or measured in a complementary way (which could also be a tag-and-probe approach). It is also required to `complete' the standard candle (in the example: give a dimuon mass consistent with the \PJpsi{} resonance), to ensure that it is almost certainly a genuine object. The efficiency is given by the probability of the probe passing the (tighter) selection criteria whose efficiency is to be determined. For the method to be correct, this probability must, to a good approximation, be independent of the probability for the tag to pass its tight selection,
\begin{equation*}
P(\text{probe passes}\,|\,\text{tag passes}) \overset{!}{=} P(\text{probe passes}).
\end{equation*}

To correct the MC simulation for differences in the selection efficiencies with respect to data, \emph{scale factors} SF defined as
\begin{equation*}
\text{SF} = \frac{\varepsilon_{\text{data}}}{\varepsilon_{\text{simulation}}}
\end{equation*}
are used. These scale factors represent per-object weights and are usually measured as a function of the object kinematics, e.g.~its $\eta$ and \pt{}. If the selection of an event depends on multiple objects being selected, the event weight due to scale factors, $w_{\text{SF}}$, must be calculated such that the equation
\begin{equation*}
w_{\text{SF}} = \frac{P(\text{event passes}\,|\,\text{is data event})}{P(\text{event passes}\,|\,\text{is simulated event})}
\end{equation*}
is satisfied. This weight is multiplied onto the prior event weight.

\subsection{Tracks and vertices}\label{sec:tracking}

Tracks represent the trajectories of charged particles. Being able to reconstruct them is important for many reasons, such as
\begin{itemize}
	\item determining particle directions and momenta,
	\item finding primary as well as secondary vertices, such as those from relatively long-lived B-hadrons that decay after flying a macroscopic distance \cite{ATL-PHYS-PUB-2017-011},
	\item determining the vertex association of leptons and jets,
	\item particle identification, such as distinguishing electrons from photons (which have no track, or two tracks after undergoing $\gamma + \gamma_{\text{material}} \to \Ppositron\Pelectron$ conversion),
	\item providing a way of studying jet performance by allowing an alternative jet measurement that is independent of the calorimeters,
	\item event visualisation for physics analysis or pedagogical reasons (including outreach \cite{atlas_outreach}),
	\item and as direct input to analysis, e.g.~for minimum-bias studies or searches for hypothetical exotic particles.
\end{itemize}
Tracks are found by applying pattern recognition algorithms to the ensemble of hits in the ID and, for muon tracks, in the MS \cite{ATL-PHYS-PUB-2015-018}. In the ID, three-dimensional \emph{space points} are formed from clusters of silicon pixel hits and the intersections of frontside and backside strips in an SCT module. Track seeds are formed by looking for triplets of space points compatible with a track. Growing the tracks, compatible pixel and SCT hits are added using a Kalman filter \cite{Cornelissen:2007vba}, with a special alternative algorithm to find electron tracks with significant Bremsstrahlung energy losses \cite{Fruhwirth:1987fm}. Tracks are ranked by the number of associated hits and the absence of `holes', i.e.~missing hits where one would be expected. In the case of shared hits, they are associated with the higher ranked track. After this, tracks with less than seven hits are discarded. Tracks are extrapolated to the TRT and all TRT hits within 10~mm added to the track if adding them improves the track's goodness of fit. (A complementary strategy finding and growing tracks outside-in is used to recover late neutral-particle decays, $\gamma  + \gamma_{\text{material}} \to \Ppositron\Pelectron$ conversions, and tracks of particles that have undergone a very significant energy loss.) The final track parameters are obtained by a global precision fit to the associated hits. In the MS, seed segments are formed in each layer of chambers, which are then combined if a loose candidate matching and subsequent fit are successful.

The tracking efficiency in the ID is about 95\% for central pseudorapidities $|\eta| < 1.5$, dropping to about 80\% in the forward region due to the higher amount of material encountered there. However, for muons it is close to 100\% independently of their pseudorapidity, because they do not undergo hadronic interactions or significant Bremsstrahlung losses.

Primary vertices are formed from at least two associated tracks as described in \myrefs~\cite{PERF-2015-01,ATL-PHYS-PUB-2015-026}.


\subsection{Topological calorimeter clusters}

Topological clustering \cite{PERF-2014-07} is used to find energy deposits in the calorimeter that are likely from energetic deposits originating from the hard scattering process, while suppressing noise due to electronics and pileup particles. The signal-over-noise significance of the energy deposited in a cell is defined as
\begin{equation*}
\zeta = \frac{\text{energy deposited in cell}}{\text{average expected noise in cell}}.
\end{equation*}
Calorimeter cells whose measured energy exceeds $\zeta > 4$ are used as cluster seeds. Cells neighbouring a cluster are added to the cluster if their energy exceeds $\zeta > 2$. This is repeated until no more such neighbouring cells are found. The clustering is done in all three dimensions, so cells can be in the same calorimeter layer, different layers of the same calorimeter, or even different calorimeters. Finally, all cells with $\zeta > 0$ adjacent to a cluster are added to it (this is only done once at the end of the clustering). The procedure can lead to clusters merging. If significant local maxima exist within one cluster, the cluster is split according to an algorithm described in \myref~\cite{PERF-2014-07}. This is important for instance to preserve local structure information, for instance in the case of a $\PZ \to \qq$ decay whose resulting hadronic products end up in the same primary topological cluster. The advantage of the topological clustering (e.g.~compared to sliding-window clustering \cite{ATL-LARG-PUB-2008-002}) is that it lets clusters grow naturally by making no assumption about their shape or size. The corresponding disadvantage is that this complexity makes the calibration more challenging \cite{ATL-LARG-PUB-2008-002}.



\subsection{Muons}
\label{sec:experiment_muon_reconstruction}



\begin{figure}[h!]
\centering
\includegraphics[width=\textwidth]{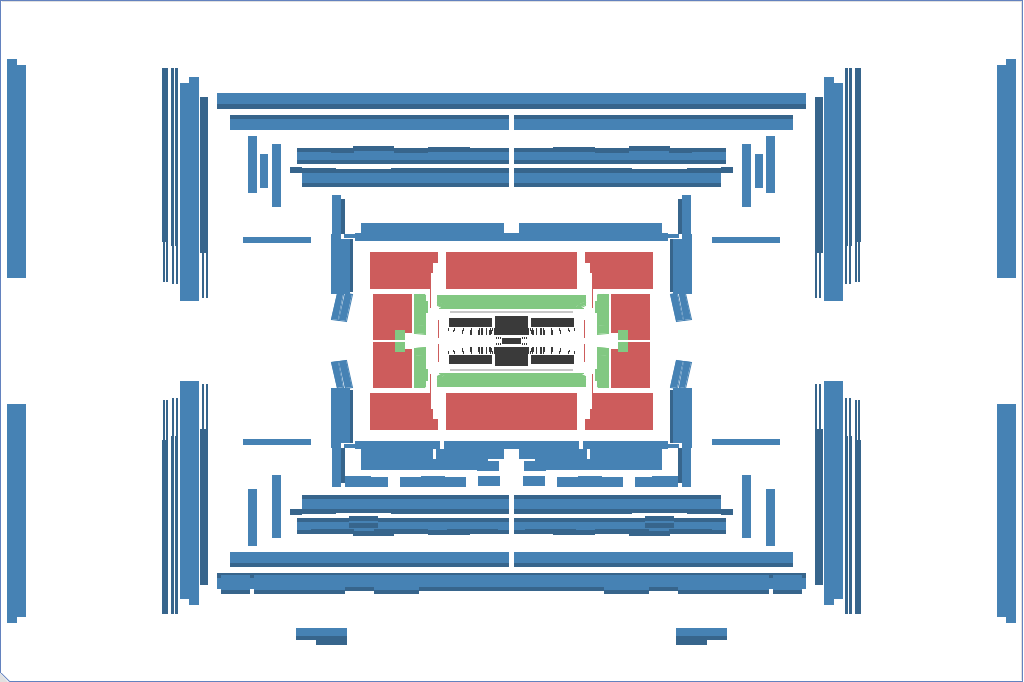}
\caption{Subdetectors participating in muon reconstruction. The inner detector is shown in black, the ECAL in green, the HCAL in red, and the MS in blue. Also shown is the central solenoid magnet in grey. Drawing made with \atlantis{} \cite{atlantis}.}
\label{fig:detector_muon_rapidity_ranges_display}
\end{figure}

Muon reconstruction and identification \cite{PERF-2015-10} uses primarily the inner detector and the muon spectrometer, supplemented by information from the calorimeters. The participating subdetectors are shown in \myfig~\ref{fig:detector_muon_rapidity_ranges_display}. Four different types of reconstructed muons can be considered:
\begin{enumerate*}
\item Combined muons,
\item Segment-tagged muons,
\item Calorimeter-tagged muons,
\item Standalone muons.
\end{enumerate*}

Most reconstructed muons are \textit{combined muons}, which are reconstructed by matching a track reconstructed in the MS to a track reconstructed in the ID. Their four-momenta are calculated by combining the information from the two systems, refitting the track using all the ID and MS hits that were previously assigned to the muon candidate. Energy deposited in the calorimeters is corrected for.

Some muons are expected to cross only one layer of MS chambers. This can be either in regions with reduced MS acceptance, or for low-\pt{} muons because of their strongly bent trajectory. In these cases, \textit{segment-tagged muons} are accepted. They are reconstructed by matching ID tracks to at least one local track segment in the MDT or CSC modules.

In the very central region $|\eta| < 0.1$, there is a gap in the coverage of the MS to allow electrical cables to pass as well as for service access to the inner detector. In this region, muon reconstruction efficiency is recovered by using \textit{calorimeter-tagged muons}. These are reconstructed by matching an ID track to energy deposits in the calorimeters that are consistent with a minimum ionising particle.

To extend the muon coverage to the forward region $2.5 < |\eta| < 2.7$, where there is MS acceptance but no ID acceptance, \textit{standalone muons} are reconstructed from MS tracks alone. The MS track is required to be compatible with originating from the interaction point. When computing the momentum, estimated energy losses in the calorimeters are taken into account.

The muon efficiencies in data are measured using the tag-and-probe method in $\PZ \to \APmuon \Pmuon$ and $\PJpsi \to \APmuon \Pmuon$ events. The \PZ{} events are used for muons with $\pt > 10$~\GeV{} and the \PJpsi{} events for muons with \pt{} between 5~\GeV{} and 20~\GeV{}. Based on the data and MC efficiencies, scale factors are determined in bins of $\eta$ and \pt{} of the muons for the muon reconstruction, isolation, and vertex association efficiencies.

The ATLAS detector simulation is not sufficiently accurate to allow the desired descripton of the muon momentum scale to the permille and momentum resolution to the percent level. However, this can be achieved by applying corrections to the simulated muons. To establish the size of the corrections, the muon momentum scale and resolution are studied by comparing the measured and predicted shape of the $\PZ \to \APmuon \Pmuon$ and $\PJpsi \to \APmuon \Pmuon$ mass peaks. Momentum scale corrections are parametrised as a function of the muon \pt{} in regions of $\phi$ and $\eta$ and applied to simulated muons. The momentum resolution is corrected to that observed in data by applying random Gaussian smearing to each simulated muon.
The muon momentum measurement relies on tracks and therefore has worse resolution for high-momentum muons, as described in \mysec~\ref{sec:tracker}.

\subsection{Electrons}
\label{sec:experiment_electron_reconstruction}

\begin{figure}[h!]
\centering
\includegraphics[width=\textwidth]{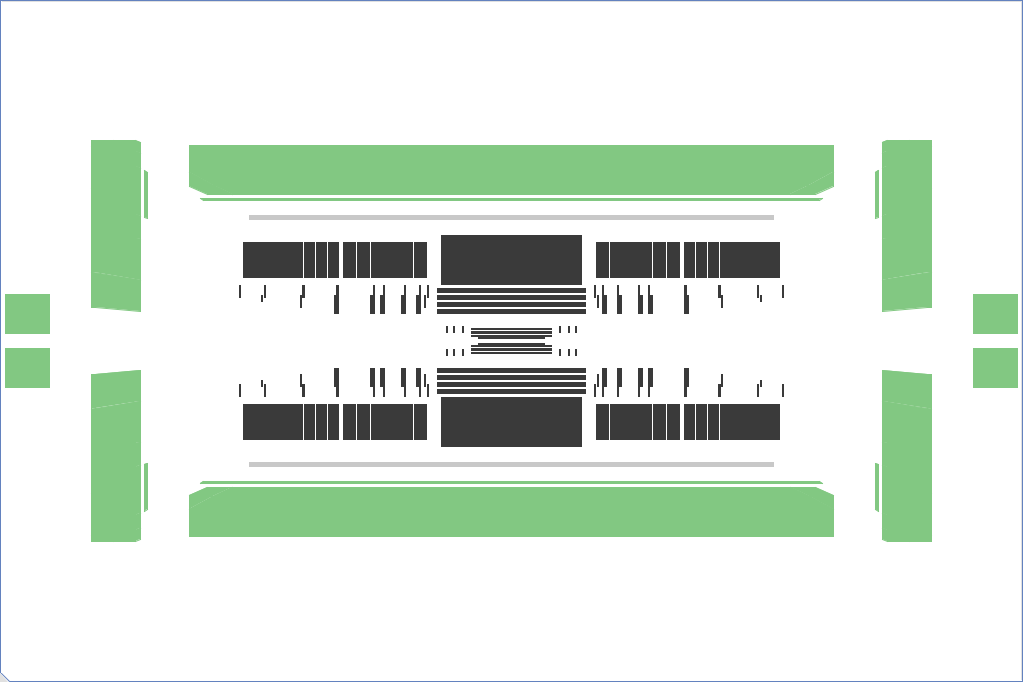}
\caption{Subdetectors participating in electron reconstruction. The inner detector is shown in black and the ECAL in green. Also shown is the central solenoid magnet in grey. Drawing made with \atlantis{} \cite{atlantis}.}
\label{fig:detector_electron_rapidity_ranges_display}
\end{figure}

An electron is reconstructed from a topological cluster in the electromagnetic calorimeter matched to a high-quality track in the ID. Its momentum is computed from the cluster energy and the direction of the track and calibrated \cite{ATL-PHYS-PUB-2016-015}. The subdetectors contributing to electron reconstruction are shown in \myfig{}~\ref{fig:detector_electron_rapidity_ranges_display}. Since the energy is measured in the calorimeter, the energy resolution is better for high-energy electrons, as discussed in \mysec~\ref{sec:experiment_calos}.
Electrons are distinguished from other particles using several identification criteria that rely on the shapes of electromagnetic showers as well as tracking and track-to-cluster matching quantities. Following the description in \myref{}~\cite{PERF-2016-01}, the output of a likelihood function taking these quantities as input is used to identify electrons.
The efficiencies of reconstructing and identifying electrons are measured using the tag-and-probe method in $\PZ \to \Ppositron \Pelectron$, $\PZ \to \Ppositron \Pelectron\Pphoton$, and $\PJpsi \to \Ppositron \Pelectron$ events \cite{PERF-2016-01}.

%


\subsection{Jets}
\label{sec:experiment_jet_reconstruction}
Jets \cite{ATL-PHYS-PUB-2015-036} are clustered from topological clusters in the calorimeters. At the clustering stage, the jets are assumed to originate from the origin of the detector. However, the luminous region in ATLAS extends $\mathcal{O}(10~\text{cm})$ in the $z$-direction, so the origin of the jet is corrected for at the calibration stage by assuming that the jet came from the hard-scattering vertex of the event, defined to be the primary vertex with the largest associated $\sum^{\text{tracks}}_i p_{\text{T},\,i}^2$, changing the $\eta$ of the jet axis.
The jet energy is calibrated as described in \myref{}~\cite{PERF-2016-04}. 
The jets in this thesis use no local calibration of topological clusters before jet clustering, only the final jets are calibrated. This means that energy deposits in the HCAL are underestimated by $\mathcal{O}(60\%)$ at the clustering stage, due to the sampling nature of the HCAL. The ECAL, while also a sampling calorimeter, has a response much closer to one. The reason why no local calibration is used is that it was not fully supported in 13~\TeV{} data when the analysis in \mypart~\ref{sec:analysis} was begun, and would not have provided much benefit anyway. The local calibration is mainly important for improving the energy calibration of energetic jets with $\pt > 100$~\GeV{} or so. The analysis in this thesis is mainly sensitive to low-\pt{} jets, where the performance with and without locally calibrated clusters is similar.




%

\clearpage\pagebreak
\part{Measurements of four-lepton production in 13~\TeV{} proton collisions with the ATLAS detector}
\label{sec:analysis}



\section{Introduction and motivation}\label{sec:zz_intro}
Three measurements of four-lepton production at the LHC are presented. All of them use data from proton-proton collisions with 13~\TeV{} centre-of-mass energy. The focus is on a measurement requiring two lepton pairs that are compatible with being the decay products of a $\PZ$ boson pair. It was published in \myref~\cite{STDM-2016-15}. The other two analyses, a quick measurement of the integrated cross section with the very first Run 2 data and an ongoing measurement of the four-lepton mass with only very loose selection requirements, are only discussed briefly in \mysecs{}~\ref{sec:early_zz_analysis} and \ref{sec:m4l_analysis}.

Studying the production of \PZ{} boson pairs at the LHC is an important test of the SM, probing electroweak and QCD predictions at the highest available collision energies. Any significant deviations from the SM predictions may point to new physics. 
In addition, $\PZ\PZ$ production is an important background in studies of the Higgs boson properties~\cite{HIGG-2014-11,HIGG-2014-10,CMS-HIG-14-036,CMS-HIG-14-028}. It is also a major background in searches for new physics processes producing pairs of \PZ{} bosons at high invariant mass~\cite{HIGG-2013-20,CMS-HIG-13-031,EXOT-2016-01,CMS-B2G-16-004}. Measuring $\PZ\PZ$ production cross sections can serve to constrain this background, or at least help understand how well it can be modelled with current predictions.

\begin{figure}[h!]
	\centering
	\subfigure[]{
		\centering
		\begin{fmfgraph*}(90,65)
			\fmfset{arrow_len}{3mm}
			\fmfstraight
			\fmfleft{idummy0,i0,idummy1,i1,idummy2,i2,idummy3}
			\fmfright{odummy0,o0,odummy1,o1,odummy2,o2,odummy3}
			\fmflabel{\Pquark}{i2}
			\fmflabel{\APquark}{i0}
			\fmflabel{\PZ}{o2}
			\fmflabel{\PZ}{o0}
			\fmf{fermion}{i2,v1}
			\fmf{fermion}{v0,i0}
			\fmf{photon}{v1,o2}
			\fmf{photon}{v0,o0}
			\fmffreeze
			\fmf{fermion}{v1,v0}
		\end{fmfgraph*}
        \label{fig:qqZZ}
	}
	\hspace{3cm}
	\subfigure[]{
		\centering
		\begin{fmfgraph*}(90,65)
			\fmfset{arrow_len}{3mm}
			\fmfstraight
			\fmfleft{idummy0,i0,idummy1,i1,idummy2,i2,idummy3}
			\fmfright{odummy0,o0,odummy1,o1,odummy2,o2,odummy3}
			\fmflabel{\Pgluon}{i2}
			\fmflabel{\Pgluon}{i0}
			\fmflabel{\PZ}{o2}
			\fmflabel{\PZ}{o0}
			\fmf{gluon}{i2,v21}
			\fmf{fermion}{v21,v22}
			\fmf{gluon}{i0,v01}
			\fmf{fermion}{v02,v01}
			\fmf{photon}{v22,o2}
			\fmf{photon}{v02,o0}
			\fmffreeze
			\fmf{fermion}{v01,v21}
			\fmf{fermion}{v22,v02}
		\end{fmfgraph*}
        \label{fig:ggZZ}  
	}\\[5mm]
	\subfigure[]{
		\centering
		\begin{fmfgraph*}(90,65)
			\fmfset{arrow_len}{3mm}
			\fmfstraight
			\fmfleft{i0,i1,i2,i3,i4}
			\fmfright{o0,o1,o2,o3,o4}
			\fmflabel{\PZ}{o3}
			\fmflabel{\PZ}{o1}
			\fmflabel{$\Pquark$}{i3}
			\fmflabel{$\Pquark'$}{i1}
			\fmflabel{$\Pquark''$}{o4}
			\fmflabel{$\Pquark'''$}{o0}
			\fmf{phantom}{i3,x3,v3,o3}
			\fmf{phantom}{i1,x1,v1,o1}
			\fmffreeze
			\fmf{fermion}{i3,x3,v3,o4}
			\fmf{fermion}{i1,x1,v1,o0}
			\fmffreeze
			\fmf{photon,label=\PZ,, \PWpm,label.side=right}{x1,x3}
			\fmf{photon}{v3,o3}
			\fmf{photon}{v1,o1}
		\end{fmfgraph*}
        \label{fig:nonvbsZZ}  
	}
	\hspace{3cm}
	\subfigure[]{
		\centering
		\begin{fmfgraph*}(90,65)
			\fmfset{arrow_len}{3mm}
			\fmfstraight
			\fmfleft{i0,i1,i2,i3,i4}
			\fmfright{o0,o1,o2,o3,o4}
			\fmflabel{\PZ}{o3}
			\fmflabel{\PZ}{o1}
			\fmflabel{$\Pquark$}{i3}
			\fmflabel{$\Pquark'$}{i1}
			\fmflabel{$\Pquark''$}{o4}
			\fmflabel{$\Pquark'''$}{o0}
			\fmf{phantom}{i3,v3,o3}
			\fmf{phantom}{i1,v1,o1}
			\fmffreeze
			\fmf{fermion}{i3,v3,o4}
			\fmf{fermion}{i1,v1,o0}
			\fmffreeze
			\fmf{photon,label.side=right}{v3,wk3}
			\fmf{photon,label.side=left}{v1,wk1}
			\fmf{photon,label=\PWpm,label.side=left}{wk1,wk3}
			\fmf{photon}{wk3,o3}
			\fmf{photon}{wk1,o1}
		\end{fmfgraph*}
        \label{fig:vbsZZ}  
	}
	\caption{Examples of leading-order SM Feynman diagrams for \ZZ{} (and $\ZZ jj$) production in proton--proton collisions: (a) $\Pquark\APquark$-initiated, (b) $\Pgluon\Pgluon$-initiated, (c) electroweak $\PZ\PZ jj$ production, (d) electroweak $\PZ\PZ jj$ production via weak-boson scattering. The decays to leptons are omitted for readability.}
  \label{fig:ZZ}
\end{figure}
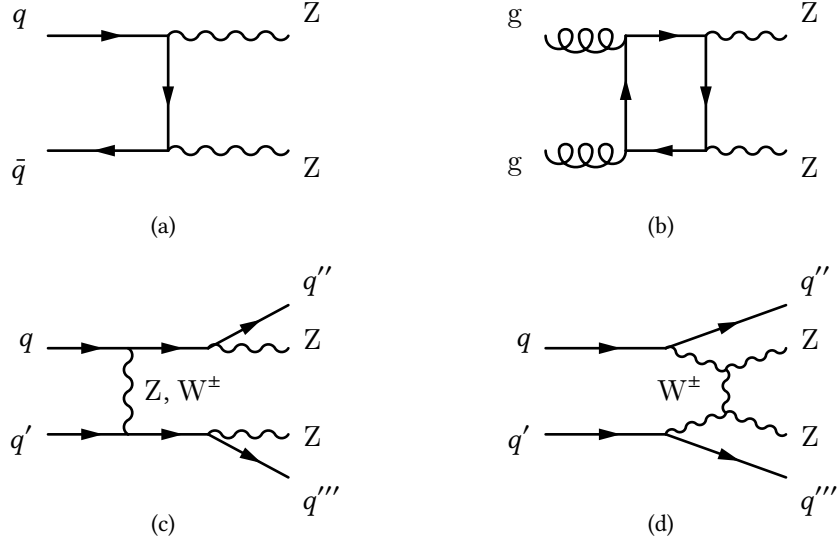

From the point of view of perturbative QFT, \ZZ{} production at the LHC is dominated by quark-antiquark ($\Pquark\APquark$) interactions, such as that shown in \myfig~\ref{fig:qqZZ}, with a smaller contribution of the order of 10\% from loop-induced gluon-gluon ($\Pgluon\Pgluon$) interactions, as in \myfig~\ref{fig:ggZZ}~\cite{Grazzini:2015hta,Caola:2015psa}. The author's ongoing work on describing the loop-induced contribution in next-to-leading-order QCD is the topic of~\mypart{}~\ref{sec:loopinduced}. The production of $\PZ\PZ$ in association with two electroweakly initiated jets, denoted EW-$\PZ\PZ jj$, includes the rare \ZZ{} weak-boson scattering process. Example Feynman diagrams are shown in \myfigs~\ref{fig:nonvbsZZ} and \ref{fig:vbsZZ}. Study of \ZZ{} production in association with jets is an important step in searching for \ZZ{} weak-boson scattering, which has so far not been experimentally observed by itself at the $3\sigma$ ($5\sigma$) significance level that is conventionally required to claim evidence (observation). However, a recent CMS measurement observed the process at the $2.7\sigma$ significance level and measured a cross section that is in agreement with the SM \cite{CMS-SMP-17-004}. \ZZ{} production can also proceed via a Higgs boson propagator, although this contribution is expected to be suppressed in the region where both \PZ{} bosons are produced nearly on-shell, as is the case in this analysis: the mass of the four-lepton system here is at least twice the required dilepton mass, $\mfourl > 132$~\GeV{}, which is greater than the Higgs boson mass of 125~\GeV{} \cite{ATLAS-CONF-2017-046}, so the Higgs boson resonance does not contribute.\footnote{The intrinsic width of the Higgs boson resonance is negligible for the purpose of this argument.
The SM Higgs boson has a predicted decay width of only $\mathcal{O}(1~\MeV)$ \cite{Heinemeyer:2013tqa}.
This also means that the reconstructed width is very much dominated by the experimental resolution.}

\begin{figure}[h!]
	\centering
	\begin{fmfgraph*}(90,65)
		\fmfset{arrow_len}{3mm}
		\fmfstraight
		\fmfleft{idummy0,i0,idummy1,i1,idummy2,i2,idummy3}
		\fmfright{odummy0,o0,odummy1,o1,odummy2,o2,odummy3}
		\fmflabel{\PZ}{o2}
		\fmflabel{\PZ}{o0}
		\fmflabel{\Pquark}{i2}
		\fmflabel{\APquark}{i0}
		\fmf{fermion}{i2,v0,i0}
		\fmf{photon,label=$\PZ$$/$$\gamma^{*}$,label.side=right}{v0,v1}
		\fmf{photon}{v1,o0}
		\fmf{photon}{v1,o2}
		\fmfv{decor.shape=circle,decor.filled=full,decor.size=2thick,fore=red}{v1}
	\end{fmfgraph*}
	\label{fig:atgcZZ}  
	\caption{Example Feynman diagram of $\PZ\PZ$ production containing an aTGC vertex, here indicated by a red dot, which is forbidden in the SM. The decays to leptons are omitted for readability.}
  \label{fig:atgc_feynman}
\end{figure}
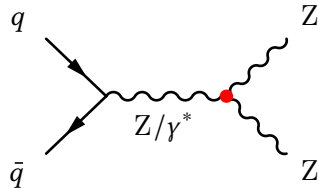

\ZZ{} production could be modified by anomalous triple gauge couplings (aTGCs) of neutral gauge bosons, which are not allowed in the SM~\cite{Baur:2000ae}. The SM does not have tree-level vertices coupling three neutral gauge bosons ($\PZ\PZ\PZ$, $\PZ\PZ\Pphoton$), because these would violate the underlying $\text{SU}(2)_{\text{L}} \times \text{U}(1)_{\text{Y}}$ symmetry. However, these couplings exist in some extensions of the SM, enhancing the $\PZ\PZ$ production cross section in regions where the energy scale of the interaction is high. An example Feynman diagram of $\PZ\PZ$ production via aTGC is shown in \myfig~\ref{fig:atgc_feynman}. 

\sloppypar
In the presented analysis, candidate events are reconstructed in the fully leptonic $\ZZllll$ decay channel, where $\ell$ and $\ell'$ can be an electron or a muon. Tau leptons are not included, because they decay before being measured by ATLAS and their momentum can never be fully reconstructed due to the presence of at least one neutrino in the decay. 
A \PZ{} boson decaying to charged leptons is not strictly distinguishable from a virtual photon. Therefore, throughout this analysis, the symbol \PZ{} denotes the combination of a \PZ{} boson and virtual photon, $\PZ$/$\gamma^*$. However, the invariant masses of the dileptons are required to lie between 66~GeV and 116~GeV, corresponding approximately to the \PZ{} boson pole mass plus/minus $25$~\GeV{}, meaning that the contribution of the \PZ{} boson dominates.

\subsection{Strategy and deliverables}

Both integrated and differential cross sections are measured, the latter with respect to twenty-one different observables. Ten of these directly measure associated jet activity in the events. A fiducial phase space is defined, where \emph{fiducial} means that non-trivial, or non-minimal, selection requirements are applied to the final-state particles, reflecting both the acceptance of the ATLAS detector and the selections imposed on the reconstructed leptons and jets in this analysis. The observed event yields are corrected to the fiducial phase space using simulated samples to model the detector effects. The integrated cross sections are inclusive with respect to associated jets. For easier comparison to other measurements, the combined integrated cross section is also extrapolated to a total phase space and to all \PZ{} boson decay modes. A search for aTGCs is performed in a generic effective-field-theory approach by looking for deviations of the data from the SM predictions at high values of the transverse momentum of the leading-\pt{} \PZ{} boson, which is one of the observables most sensitive to the energy scale of the interaction.

Differential fiducial cross sections are measured with respect to the following observables:
\begin{itemize}
\item Mass of the four-lepton system, $\mfourl$;
\item Transverse momentum of the four-lepton system, \ptfourl{};
\item Absolute rapidity of the four-lepton system, $|\yfourl{}|$;
\item Separation in azimuthal angle between the two \PZ{} boson candidates, \dphiZZ{}, defined such that it lies in the interval $[0, \pi]$;
\item Absolute difference in rapidity between the two \PZ{} boson candidates, $|\dyZZ{}|$;
\item Transverse momentum of the leading-\pt{} and the subleading-\pt{} \PZ{} boson candidates, \ptz{} and \ptzsub{};
\item Transverse momentum of each of the four leptons; 
\item Number of jets with $\pt > 30$~\GeV{} and $|\eta| < 4.5$, $N_{\text{jets}}$;
\item Number of jets with $\pt > 30$~\GeV{} and $|\eta| < 2.4$, referred to as central jets, $N_{\text{central jets}}$;
\item Number of jets with $\pt > 60$~\GeV{} and $|\eta| < 4.5$;
\item Scalar sum of the transverse momenta of all jets in the event with $\pt > 30$~\GeV{} and $|\eta| < 4.5$;
\item Absolute pseudorapidity of the leading-\pt{} and the subleading-\pt{} jets;
\item Transverse momentum of the leading-\pt{} and the subleading-\pt{} jets;
\item Absolute difference in rapidity between the two leading-\pt{} jets, $|\dyjj{}|$; 
\item Invariant mass of the two leading-\pt{} jets, \mjj{}. 
\end{itemize}

These measurements provide a detailed description of the kinematics in $\PZ\PZ{}$ events and allow comparisons and validations of current and future predictions. Regrettably, while included in this thesis, the four-lepton invariant mass is not published as a differential cross section in \myref~\cite{STDM-2016-15}, but only at the reconstruction level. It is one of the most interesting observables, as resonances due to new physics could show up as localised excesses --- or ``bumps'' --- in the mass spectrum. The reason for the omission is a decision by the ATLAS coordination to only publish this observable in a designated, but as of yet unpublished analysis (similar to the one in \myref~\cite{STDM-2014-15}), in which the author is also involved. Notwithstanding, the remaining above measurements provide a detailed description of the kinematics in $\PZ\PZ{}$ events and allow comparisons and validations of current and future predictions. Some of the differential measurements are particularly motivated: the transverse momentum of the four-lepton system directly measures the recoil against all other particles produced in the collision and therefore provides information about QCD and electroweak radiation across the entire range of scales. The rapidity of the four-lepton system is sensitive to the $z$-component of the total momentum of the initial-state partons involved in the \ZZ{} production. It may therefore be sensitive to the PDFs. The azimuthal-angle separation and rapidity difference between the \PZ{} boson candidates probe their angular correlations and may help extract the contribution of double-parton-scattering \ZZ{} production. The azimuthal-angle separation is also sensitive to radiation of partons and photons produced in association with the \ZZ{} pair. The scalar sum of the transverse momenta of all jets provides a measure of the overall jet activity that is independent of their azimuthal configuration. The measurements of $|\dyjj{}|$ and \mjj{} are particularly sensitive to the  EW-$\PZ\PZ jj$ process. They both tend to have larger values in weak-boson scattering than in other \ZZ{} production channels, providing an important step towards the study of \ZZ{} production via weak-boson scattering.

To maximise the impact of the analysis on the understanding of Nature, it is built on the following principles:
\begin{itemize}
\item \emph{Theoretical robustness}, e.g.~in the definitions of unstable particles and observables,
\item \emph{Model independence}, meaning that experimental methods are chosen such that they rely as little as possible on any particular model being a good description of reality,
\item \emph{Preservation} of results in a way that allows anyone in the world to reuse and reinterpret them later.
\end{itemize}

The four-lepton channel studied in the presented analysis only has a very small branching fraction of around $0.45\%$~\cite{Olive:2016xmw}. However, it has decisive advantages over other \ZZ{} decay channels, namely its low background, simple and complete reconstruction (whereas e.g.~the $\ell^+\ell^-\Pnu\APnu$ channel has missing momentum), and excellent experimental resolution of kinematic observables. It also provides a ``cleanroom'' for the study of jet production, in the sense that any associated QCD radiation can be unambiguously separated from the \ZZ{} decay products. This makes it an interesting process for studying associated jets. While single $\PZ \to \ell^+\ell^-$ production with associated jets provides a statistical uncertainty that is orders of magnitude smaller thanks to the much higher cross section \cite{STDM-2016-01}, the non-resonant nature of the \ZZ{} system means that different (and higher) hard-process scales are probed.

\subsection{Comparison to other analyses}

Integrated and differential \ZZ{} production cross sections have been previously measured at $\sqs=7$ and 8~\TeV{} by the ATLAS and CMS collaborations~\cite{STDM-2012-02,CMS-SMP-13-005,STDM-2014-16,CMS-SMP-12-007} and found to be consistent with SM predictions. The integrated $\Pproton\Pproton \to \ZZllll$ cross section at $\sqs=13$~\TeV{} was recently measured by the ATLAS \cite{STDM-2015-13} and CMS \cite{CMS-SMP-16-001} collaborations, each analysing data corresponding to an integrated luminosity of about 3~\ifb{}. The ATLAS measurement is summarised in~\mysec~\ref{sec:early_zz_analysis}. Searches for aTGCs have previously been performed at lower centre-of-mass energies by ATLAS \cite{STDM-2014-16}, CMS \cite{CMS-SMP-13-005,CMS-SMP-12-016}, D0 \cite{Abazov:2007ad}, and by the LEP experiments \cite{Alcaraz:2006mx}. Shortly after the publication of this analysis, CMS also published a measurement of integrated and differential cross sections as well as a search for aTGCs at 13~\TeV{} centre-of-mass energy~\cite{CMS-ZZ-13TEV}, again finding agreement with the SM predictions. A comparison of the differential cross sections considered in recent measurements is shown in \mytab{}~\ref{tab:zz_analysis_comparison}.

\begin{table}[p]
\centering
{\scriptsize
\begin{tabular}{llll}
\toprule
\textbf{Experiment} & \textbf{Data} & \textbf{Observable} & \textbf{Binning (in \GeV{} if energy-like)}\\
\midrule
ATLAS (this) & 13~\TeV{}, & 1.~lepton \pt{} & 20, 40, 50, 60, 70, 80, 90, 100, 110, 120, 130, 140, 150, 160, 180, 200, 230, 450\\
& 36.1~\ifb{} & \hl{2.~lepton \pt{}} & 15, 40, 50, 60, 70, 80, 90, 100, 120, 150, 300\\
& & \hl{3.~lepton \pt{}} & 10, 20, 25, 30, 35, 40, 45, 50, 55, 60, 65, 70, 80, 100, 200\\
& & \hl{4.~lepton \pt{}} & 5, 10, 15, 20, 25, 30, 35, 40, 45, 50, 60, 150\\
& & $\PZ_1$ \pt{} & 0, 10, 20, 30, 40, 50, 60, 70, 80, 90, 100, 120, 140, 160, 200, 250, 1500\\
& & \hl{$\PZ_2$ \pt{}} & 0, 10, 20, 30, 40, 50, 60, 70, 80, 90, 100, 120, 140, 160, 200, 250, 1500\\
& & $\mfourl$ (thesis only) & 140, 180, 200, 220, 240, 260, 280, 300, 325, 350, 400, 500, 600, 800, 1500\\
& & \ptfourl{} & 0, 5, 15, 25, 35, 45, 55, 65, 75, 85, 100, 125, 150, 200, 250, 1500\\
& & \hl{$\PZ\PZ$ $|y|$} & 0.0, 0.2, 0.4, 0.6, 0.8, 1, 1.2, 1.4, 1.6, 1.8, 2, 4.6\\
& & $\Delta\phi(\PZ_1, \PZ_2) / \pi$ & 0.0, 0.125, 0.25, 0.375, 0.5, 0.625, 0.6875, 0.75, 0.8125, 0.875, 0.9375, 1.0\\
& & $\Delta y(\PZ_1, \PZ_2)$ & 0.0, 0.2, 0.4, 0.6, 0.8, 1.0, 1.2, 1.5, 2.0, 2.5, 3.0, 10.0\\
& & $N_{\text{jets}}$ & 0, 1, 2, 3, $\geq 4$\\
& & \hl{$N_{\text{central jets}}$} & 0, 1, 2, 3, $\geq 4$\\
& & \hl{$N_{\text{jets}}$, $\pt > 60$} & 0, 1, 2, $\geq 3$\\
& & \hl{1.~jet \pt{}} & 30, 40, 50, 60, 80, 100, 120, 150, 200, 800\\
& & \hl{2.~jet \pt{}} & 30, 40, 60, 500\\
& & \hl{1.~jet $|\eta|$} & 0.0, 0.5, 1.0, 1.5, 2.0, 2.5, 4.5\\
& & \hl{2.~jet $|\eta|$} & 0.0, 0.5, 1.0, 1.5, 2.0, 2.5, 4.5\\
& & \hl{$\text{Dijet}_{12}$ mass} & 0, 50, 100, 200, 300, 1000\\
& & \hl{$\Delta y(j_1, j_2)$} & 0, 1, 2, 3, 9\\
& & \hl{Jets scalar \pt{} sum} & 30, 60, 90, 120, 150, 200, 400, 1000\\
\midrule
CMS \cite{CMS-ZZ-13TEV} & 13~\TeV{}, & \mfourl{} & 100, 200, 250, 300, 350, 400, 500, 600, 800\\
 & 35.9~\ifb{} & \ptfourl{} & 0, 25, 50, 75, 100, 150, 200, 300\\
 & & \hl{\PZ{} \pt{} (both)} & 0, 25, 50, 75, 100, 125, 150, 200, 300\\
 & & 1.~lepton \pt{} & 15, 30, 40, 50, 60, 75, 90, 105, 120, 135, 150, 165, 180, 195, 225\\
 & & $\Delta\phi(\PZ_1, \PZ_2)$ & 0.0, 1.5, 2.0, 2.25, 2.5, 2.75, 3.0, 3.25\\
 & & $\Delta R(\PZ_1, \PZ_2)$ & 0, 1, 2, 3, 4, 5, 6\\ 
\midrule
ATLAS \cite{Aaboud:2016urj} & 8~\TeV{}, & $\PZ_1$ \pt{} & 0, 30, 60, 100, 100, 200, 1500\\
& 20.3~\ifb{} & \hl{$\Delta\phi(\ell^+, \ell^-)$ in $\PZ_1$} & 0.0, 1.3, 1.9, 2.3, 2.7, 3.14\\
& & $\Delta y(\PZ_1, \PZ2)$ & 0.0, 0.4, 0.8, 1.2, 4\\
& & $N_{\text{jets}}$ & 0, 1, $\geq 2$\\
\midrule
CMS \cite{CMS:2014xja} & 8~\TeV{}, & \mfourl{} & 100, 200, 250, 300, 350, 400, 500, 600, 800\\
& 19.6~\ifb{} & $\PZ_1$ \pt{} & 0, 25, 50, 75, 100, 125, 150, 200, 250\\
& & \ptfourl{} & 0, 20, 40, 60, 80, 100, 150\\
& & 1.~lepton \pt{} & 20, 30, 40, 50, 60, 70, 80, 90, 100, 110, 120, 130, 140\\
& & $\Delta\phi(\PZ_1, \PZ_2)$ & 0.0, 1.9, 1.5, 2.0, 2.25, 2.5 2.75, 3.0, 3.25\\
& & $\Delta R(\PZ_1, \PZ_2)$ & 0, 1, 2, 3, 4, 5, 6\\
\bottomrule
\end{tabular}
}
\caption{Comparison of measured differential cross section in this analysis and selected others. All rankings such as `1.~lepton' are by \pt{}. Binnings include the upper edge of the highest bin. Only distributions in the four-lepton channel and for nearly ``on-shell'' \PZ{} bosons are listed. Those unique to one of the listed analyses are \hl{highlighted} (colour only).}
\label{tab:zz_analysis_comparison}
\end{table}

\subsection{Dataset}



The analysis uses a data sample of proton--proton collisions that was taken in 2015 and 2016 at a centre-of-mass energy of $\sqrt{s}$ = 13~\TeV. Only runs with 25~ns bunch spacing are used (i.e. ignoring the small 50~ns sample acquired at the beginning of Run 2 data-taking in 2015). Events are accepted for analysis based on data quality flags per luminosity block, using the \emph{good run lists} recommended for all analyses. The good run lists contain luminosity blocks during which all parts of the detector were functioning correctly. The usable integrated luminosity is 3.2~\ifb{} for 2015 and 32.9~\ifb{} for 2016, giving a total of 36.1~\ifb{}. The uncertainty of the total integrated luminosity is 3.2\%, corresponding to $\pm 1.1$~\ifb{}. The integrated luminosity and its uncertainty is derived, following a methodology similar to that detailed in \myref{}~\cite{Aaboud:2016hhf}, from a preliminary calibration of the luminosity scale using $x$-$y$ beam-separation scans performed in August 2015 and May 2016.



\clearpage
\section{Theoretical predictions}\label{sec:mc}
MC event samples are used to obtain corrections for detector effects and to estimate signal and background contributions. Fixed-order calculations are used as higher-order corrections, for additional comparisons to measurement results, and for extrapolation between phase spaces. Throughout this analysis, unless stated otherwise, orders of calculations refer to perturbative expansions in the strong coupling \alphas{} in QCD and all calculations use the \cten{} \cite{Lai:2010vv} PDFs with the evolution order in \alphas{} corresponding to the perturbative order in \alphas{} in the calculation. MC generator versions are only given the first time the generator is mentioned. Access to all PDFs is provided by the LHAPDF~6 interface \cite{Buckley:2014ana}. Electroweak parameters are set according to the $G_{\mu}$ scheme everywhere. In this scheme, the Fermi constant $G_{\mu}$ as well as the pole masses of the weak bosons are taken as independent input parameters \cite{Jegerlehner1990}. The electroweak coupling strength is then calculated using
\begin{equation*}
\alpha = \frac{\sqrt{2} G_{\mu} M_{\PW}^2 \sin^2 \theta_{\text{w}}}{\pi}.
\end{equation*}


\subsection{Event samples}

The nominal signal samples are generated with \SHERPA{}~2.2.1~\cite{Gleisberg:2008ta,Hoeche:2009rj,Gleisberg:2008fv,Schumann:2007mg,Schonherr:2008av,Cascioli:2011va,Hoeche:2012yf}, with the $\Pquark\APquark$-initiated process simulated at NLO for \ZZ{} plus zero or one additional parton and at LO for two or three additional partons generated at the matrix-element level. The different parton multiplicities are merged together into one consistent sample \cite{Hoeche:2012yf,Gehrmann:2012yg,Hoeche:2010kg,Hoeche:2009rj}.
A second \SHERPA{} sample is generated with the loop-induced \gluglu-initiated process simulated at LO using NLO PDFs, including subprocesses involving a Higgs boson propagator, with zero or one additional parton. 
The \gluglu-initiated process first enters at NNLO and is therefore not included in the NLO sample for the $\Pquark\APquark$-initiated process. (Due to different initial states, the \gluglu-initiated process does not interfere with the $\Pquark\APquark$-initiated process at NLO.) The loop-induced \gluglu-initiated process calculated at LO receives large corrections at NLO \cite{Caola:2015psa}. The cross section of the sample is therefore multiplied by an NLO/LO $k$-factor of $1.67 \pm 0.25$, which is based on the results presented in \myref~\cite{Caola:2015psa}.
The EW-$\PZ\PZ jj$ process is simulated using \SHERPA{} at its lowest contributing order in the electroweak coupling, $\alpha^6$ (including the decays of the $\PZ$ bosons). It includes the triboson subprocess $\ZZ V \to \llll jj$, where the third boson $V \in \{\PWpm, \PZ\}$ decays hadronically.
\SHERPA{} also simulates parton showering, electromagnetic radiation, underlying event, and hadronisation in the above samples.
Throughout this analysis, the prediction obtained by summing the above samples is referred to as the nominal \SHERPA{} setup.

An alternative prediction for the $\Pquark\APquark$-initiated process is obtained using the \POWHEG{} method and framework~\cite{Nason:2004rx,Frixione:2007vw} as implemented in \POWHEGBOX{} 2 \cite{Alioli:2010xd}, with a diboson event generator \cite{Melia:2011tj,Nason:2013ydw} used to simulate the $\PZ\PZ$ production process at NLO. The simulation of parton showering, electromagnetic radiation, underlying event, and hadronisation is performed with \PYTHIA{}~8.186 \cite{Sjostrand:2006za,Sjostrand:2007gs} using the \aznlo{} parameter tune~\cite{Aad:2014xaa}. This sample is used to estimate the systematic uncertainty due to modelling differences between the event generators. Another sample is created (without detector simulation) that is otherwise identical to the above \POWHEGpy{} sample, but using \PHOTOS{} \cite{Golonka:2005pn,Davidson:2010ew} to generate electromagnetic radiation. This sample is used to check the impact of differences in photon radiation modelling at the particle level.

Additional samples are generated to estimate the contribution from
background events.
Triboson events are simulated at LO with \SHERPA{}~2.1.1. Samples of $\Ptop\APtop\PZ$ events are simulated at LO with \MADGRAPH~2.2.2 \cite{Alwall:2014hca} +
\py{}~8.186 using the \nnpdf{} PDFs~\cite{Ball:2012cx} and the \afourteen{} tune~\cite{ATL-PHYS-PUB-2014-021}.

More information about the above diboson and triboson samples generated with \SHERPA{} and \POWHEGpy{} can be found in \myref{}~\cite{ATL-PHYS-PUB-2016-002}.

In all MC samples, pileup is simulated as inclusive inelastic $\Pproton\Pproton$ collisions with \py{} using
\mstw{} PDFs~\cite{Martin:2009iq} and the \atwo{} tune~\cite{ATL-PHYS-PUB-2012-003}. 
The samples are then passed through a simulation of the ATLAS detector~\cite{SOFT-2010-01}
based on \GEANT{} 4~\cite{Agostinelli:2002hh}. 
Weights are applied to the simulated events to correct for the
small differences from data in the reconstruction,
identification, isolation, and impact parameter efficiencies for electrons and
muons~\cite{PERF-2016-01,PERF-2015-10}. Furthermore, the lepton momentum or energy scales and resolutions are adjusted to match the data \cite{ATL-PHYS-PUB-2016-015,PERF-2015-10}.

Sometimes in the analysis, and only where explicitly stated, higher-order NNLO QCD and NLO weak corrections are applied to the predictions from the above samples. They are described in the following.

\subsection{NNLO QCD predictions}\label{sec:nnlo_predictions}
NNLO cross sections for $\Pproton\Pproton \to \ZZllll{}$ in the fiducial and total phase space are provided by \matrixnnlo{} \cite{Grazzini:2017mhc,Grazzini:2015hta}, also in bins of the jet-inclusive measured distributions. They include the \gluglu-initiated process at its lowest contributing order, which accounts for about 60\% of the cross section increase with respect to NLO \cite{Cascioli:2014yka}. The calculation is based on tree-level and one-loop amplitudes provided by \openloops{} \cite{Cascioli:2011va} and \collier{} \cite{Denner:2016kdg}, as well as two-loop calculations from \myref~\cite{Gehrmann:2015ora}. \matrixnnlo{} uses the $q_{\text{T}}$ NNLO subtraction method \cite{Catani:2007vq}. The calculation uses a dynamic QCD scale of $\mfourl / 2$ and the NNPDF 3.0 PDFs \cite{Ball:2014uwa} (with $\alphas = 0.118$ at the \PZ{} boson pole mass).  

\matrixnnlo{} LO, NLO, and NNLO results in the fiducial phase space are shown in \mytab~\ref{tab:matrix_results_atlas}. The relative statistical uncertainty of the calculations is set to be smaller than $10^{-3}$ on the integrated cross section. It is neglected throughout.\footnote{There is another small uncertainty in the calculation that is neglected, related to the NNLO subtraction implementation. \matrixnnlo{} evaluates the cross section for multiple different $q_{\text{T}}$ cutoffs, $q_{\text{cut}}$. These cross sections depend on $q_{\text{cut}}$. The dependence vanishes in the limit $q_{\text{cut}} \to 0$. This limit is taken by fitting the cross section as a function of $q_{\text{cut}}$ and extrapolating the fit to $q_{\text{cut}} = 0$. The extrapolation uncertainty is neglected here, as it was smaller than the statistical uncertainty.} The QCD scale uncertainty is evaluated as described in \mysec~\ref{sec:theory_scale_variations}. As of the writing of this thesis, \matrixnnlo{} has no internal PDF reweighting mechanism, making the evaluation of PDF uncertainties very computationally expensive, as $\mathcal{O}(100)$ complete recalculations of the cross sections would be necessary. Therefore, PDF uncertainties are not included for the NNLO predictions.

%
%
\begin{table}[h!]
\centering
\begin{tabular}{llll}
\toprule
\textbf{Order or subprocess} & \textbf{4e or 4\textmu{} (fb)} & \textbf{2e2\textmu{} (fb)} & \textbf{Scale uncertainty (\%)}\\
\midrule
LO & \phantom{0}6.066& 11.87 & $+5.7$, $-6.7$\\
NLO & \phantom{0}8.756 & 17.09 & $+2.5$, $-2.1$\\
NNLO & 10.45 & 20.38 &$+3.2$, $-2.7$\\
\midrule
Only $\gluglu\; \looparrow{}\; 4\ell$ & \phantom{0}0.9181 & \phantom{0}1.815 & $+23.5$, $-17.8$\\
\bottomrule
\end{tabular}
\caption{Integrated fiducial cross sections calculated with \matrixnnlo{}. LO and NLO PDFs are used for the LO and NLO cross sections, respectively. For the other cross sections, NNLO PDFs are used. The numerical accuracy is 0.1\%.} 
\label{tab:matrix_results_atlas}
\end{table}

The NNLO prediction does not lie within the NLO scale variation band. In addition, the NNLO scale uncertainty is larger than the NLO one. These observations are a testament to the fact that scale variations do not always provide a good estimate of the uncertainty. In this case, the contribution of the \gluglu-initiated loop-induced production mode entering at NNLO leads to a `jump' in the convergence of the perturbative series that the scale variations could not account for. No new flavour channels enter beyond NNLO, so the NNLO scale uncertainty can be considered relatively realistic.

The NNLO calculation is also used for extrapolation of the integrated cross section from the fiducial to a total phase space. The PDF uncertainty of the extrapolation factor was estimated by Eleni Skorda using an NLO (LO) calculation for the $\Pquark\APquark$-initiated ($\Pgluon\Pgluon$-initiated) process from \MCFM{}~6.8~\cite{Campbell:1999ah,Campbell:2011bn,Campbell:2015qma}, taking the mass of the four-lepton system, $\mfourl$, as the dynamic QCD scale. NLO PDFs are used for the \gluglu-initiated process and its contribution is multiplied by the NLO/LO $k$-factor of $1.67 \pm 0.25$.



\subsection{NLO weak corrections}\label{sec:zz_ew_corrections}
Weak corrections at next-to-leading order~\cite{Biedermann:2016lvg,Biedermann:2016yvs} are calculated in the fiducial phase space, also in bins of the jet-inclusive measured distributions. They fully include off-shell effects and non-resonant topologies. The weak corrections are a subset of the full electroweak NLO corrections, ignoring those corrections involving photon emission, loops involving photons, or photon-induced subprocesses (discussed in \myapp~\ref{sec:photon_induced}). The latter are phenomenologically negligible \cite{Biedermann:2016lvg}. Photonic corrections could technically have been included, but were deliberately excluded to allow more consistent reweighting of the fully simulated MC samples with the NLO weak $k$-factors in the search for aTGCs. These samples already contain approximate photonic corrections, so reweighting with the full EW $k$-factors could introduce a double counting. The EW corrections were finally also combined with fixed-order QCD calculations (as described in the next section), where a full EW NLO calculation including photonic corrections would have been more beneficial, but ended up not being feasible for time reasons. Future analyses can and should improve on this.

The NLO weak corrections are calculated with respect to the $\Pquark\APquark$-initiated process at LO in $\alphas$, meaning that they cannot be obtained differentially in observables that are trivial at LO in $\alphas$, e.g.~the transverse momentum of the four-lepton system. Where a differential calculation is not possible, the integrated value in the fiducial phase space is used. Excluding photonic corrections means that the NLO EW calculation contains no real-emission diagrams, since weak-boson emission is conventionally considered a separate process (triboson production), rather than a correction to \ZZ{} production. This means that the NLO EW cross section equally only populates the 0-jet bin, which in turn means that the $k$-factor for the `leading' and `subleading' dilepton is identical, as the two dileptons always recoil back-to-back and carry the same transverse momentum.

The NLO/LO weak $k$-factor integrated across the entire fiducial phase space is about 0.95. As is typical for electroweak corrections, the $k$-factor is further from unity in interactions at high energy scale, e.g.~at high four-lepton mass or dilepton transverse momentum. The differential $k$-factor for the dilepton transverse momentum is shown in \myfig~\ref{fig:differential_weak_kfactor}. At $\pt \sim 2~\TeV{}$, it is as small as around 0.2! Differential $k$-factors are available to this analysis as a function of the following observables: $|\yfourl|$, $|\dyZZ{}|$, \ptz{} and \ptzsub{} (which are identical at LO), and the \pt{} of each of the four leptons in the final selected quadruplet.
\begin{figure}[h!]
\centering
\includegraphics[width=0.68\textwidth]{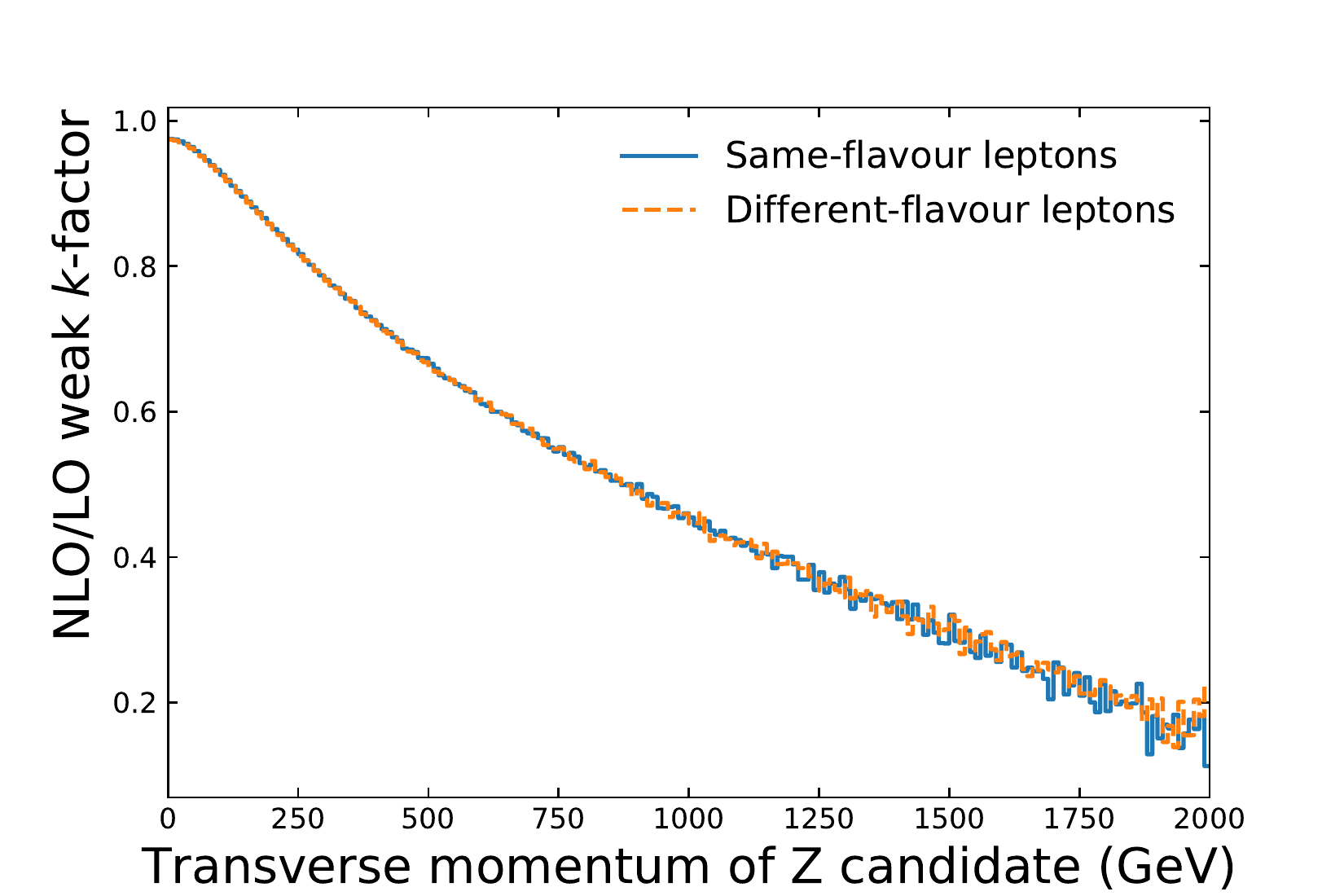}
\caption{NLO/LO weak $k$-factor as a function of the transverse momentum of the \PZ candidates for same-flavour final-state leptons (\eeee{} or \mmmm{} channel) as well as different-flavour leptons (\eemm{} channel). Data provided by Biedermann, Denner, Dittmaier, Hofer, and J\"ager~\cite{Biedermann:2016lvg,Biedermann:2016yvs}.}
\label{fig:differential_weak_kfactor}
\end{figure}

\subsection{Combining QCD and electroweak corrections}\label{sec:best_sm_prediction}

While both QCD and EW fixed-order corrections to \ZZllll{} have been calculated separately, no calculations of mixed QCD and EW corrections are available to date.
The principal difficulty is evaluating mixed two-loop diagrams \cite{Gieseke:2014gka}.
Mixed corrections were recently calculated for $\Pproton\Pproton \to \ell^+\ell^-\Pnu\APnu$ (via $\PWp\PWm /\, \PZ\PZ\, /\, \PZ\gamma^*$) using approximations \cite{Kallweit:2017khh}, which is perhaps a step towards their calculation for $\Pproton\Pproton\to\llll$.

In the presented analysis, QCD and EW corrections are combined multiplicatively, as e.g.~in \myref{}~\cite{Denner:2014bna},
\begin{equation}\label{eq:combining_qcd_ewk_corrections}
	\sigma^{\text{NNLO $\otimes$ NLO EW}} = \sigma^{LO} (1 + \delta_{\text{QCD}})(1 + \delta_{\text{EW}}) = \sigma^{\text{NNLO}}(1 + \delta_{\text{EW}}).
\end{equation}

Thus, the NNLO calculations serve as the basis of a SM prediction incorporating the formally most accurate available predictions. The contribution of the \gluglu-initiated process is multiplied by the NLO/LO $k$-factor of $1.67 \pm 0.25$. The NLO weak corrections are applied as multiplicative $k$-factors, differentially in the observable of interest if available, otherwise integrated over the fiducial phase space. In addition, the cross section of the EW-$\ZZ jj$ process calculated with \SHERPA{} is added.

\clearpage
\section{Signal definition}

\subsection{Fiducial phase space}
\label{sec:fiducial}
The fiducial phase space is defined using final-state particles, meaning particles whose average lifetime $\tau_0$ 
satisfies $c\tau_0 > 10~\text{mm}$ \cite{ATL-PHYS-PUB-2015-013}. (No parton-level properties are used. These are ill-defined, because coloured objects are not physically observable and their properties in simulation may depend on unphysical parameter choices.) A prompt lepton, photon, or neutrino refers to a final-state particle that does not originate from the decay of a hadron or $\tau$ lepton, or any material interaction (such as Bremsstrahlung or pair production) \cite{ATL-PHYS-PUB-2015-013}. Hadrons are never considered prompt in this analysis.\footnote{As in \rivet{} \cite{Buckley:2010ar}, promptness here is related to whether particles are directly connected to the hard process of the event, regardless of such factors as realistic reconstructibility of displaced vertices.}

The requirements used to define the fiducial phase space mirror the selections applied to the reconstructed leptons (described in \mysec~\ref{sec:selection}). This is done to ensure that the extrapolation from the observed data to the fiducial phase space is as model-independent as possible, ideally depending only on detector and reconstruction effects.


Events in the fiducial phase space contain at least four prompt electrons and/or prompt muons. The four-momenta of all prompt photons within $\Delta R = 0.1$ of a lepton are added to the four-momentum of the closest lepton. This \emph{dressing} of \emph{bare} leptons is done to reduce dependence to the modelling of photon radiation off charged leptons \cite{ATL-PHYS-PUB-2015-013}. 
Each dressed lepton is required to have transverse momentum $\pt > 5$~\GeV{} and absolute pseudorapidity $|\eta| < 2.7$. (This means a slight extrapolation for electrons, which in this analysis are only reconstructed within $\pt > 7$~\GeV{} and $|\eta| < 2.47$. The advantage is that the harmonised requirements make it easy to describe, compare, and combine the channels.)

All possible pairings of same-flavour opposite-charge dileptons are formed, referred to as quadruplets. In each quadruplet, the three highest-$\pt$ leptons must satisfy $\pt > 20$~\GeV{}, 15~\GeV{}, and 10~\GeV{}, respectively. If multiple selected quadruplets are present, the quadruplet minimizing $|m_{\ell\ell} - m_{\PZ}| + |m_{\ell'\ell'} - m_{\PZ}|$ is selected, where $m_{\ell^{(\prime)}\ell^{(\prime)}}$ is the mass of a given same-flavour opposite-charge dilepton and $m_{\PZ} = 91.1876$~\GeV{} is the \PZ{} boson pole mass \cite{Olive:2016xmw}. All remaining requirements are applied to the leptons in the final selected quadruplet. Any two different (same) flavour leptons $\ell_i$, $\ell'_j$ must be separated by $\Delta R (\ell_i,\ell'_j) > 0.2$~(0.1). This requirement emulates the reconstruction-level requirement that leptons be well-separated from each other and spatially isolated from other particles in the detector (to reduce the probability of misidentifying a lepton). All possible same-flavour opposite-charge dileptons must have an invariant mass greater than $5$~\GeV{}, to match the same requirement in the selection of reconstructed events, which is introduced to reduce the background from leptonically decaying hadrons, such as $\PJpsi \to \APmuon\Pmuon$. If all leptons are of the same flavour, the dilepton pairing that minimises $|m_{\ell\ell} - m_{\PZ}| + |m_{\ell'\ell'} - m_{\PZ}|$ is chosen. 
The selected dileptons are defined as the \PZ{} boson candidates. Each is required to 
have an invariant mass between 66~\GeV{} and 116~\GeV{}. Based on the leptons in the chosen quadruplet, events are classified into three signal channels: \eeee{}, \mmmm{}, and \eemm{}.

Jets are considered in several differential cross sections. They
are clustered from all final-state particles except prompt leptons, prompt neutrinos, and prompt photons using the anti-$k_t$ algorithm 
with radius parameter 0.4. Jets are required to have $\pt > 30$~\GeV{} and $|\eta| < 4.5$. Jets within $\Delta R = 0.4$ of any selected fiducial lepton (as defined above) are rejected.  

The fiducial selection is summarised in \mytab{}~\ref{tab:fiducial_selection}.

\begin{table}[!htbp]
\centering
{\small
\begin{tabular}{ll}
\toprule
\textbf{Type} & \textbf{Input or requirement}\\
\midrule
Leptons (\Pe{}, \Pmu{}) & Prompt\\
	& Dressed with prompt photons within $\Delta R = 0.1$\\
	& $\pt > 5~\GeV$\\
	& $|\eta|<2.7$\\
\midrule
Quadruplets & Two same-flavour opposite-charge lepton pairs\\
	& Three leading-\pt{} leptons satisfy $\pt > 20$~\GeV{}, 15~\GeV{}, 10~\GeV{}\\
\midrule
Events & Only quadruplet minimizing $|m_{\ell\ell} - m_{\PZ}| + |m_{\ell'\ell'} - m_{\PZ}|$ is considered\\
	& Any same-flavour opposite-charge dilepton has mass $m_{\ell\ell} > 5$~\GeV{}\\
	& $\Delta R > 0.1$ (0.2) between all same-flavour (different-flavour) leptons\\
	& Dileptons minimizing $|m_{\ell\ell} - m_{\PZ}| + |m_{\ell'\ell'} - m_{\PZ}|$ are taken as \PZ{} boson candidates\\
	& \PZ{} boson candidates have mass $66~\GeV{} < m_{\ell\ell} < 116$~\GeV{}\\ 
\midrule
Jets & Clustered from all non-prompt particles\\
	& Anti-$k_t$ algorithm with $R = 0.4$\\
	& $\pt > 30~\GeV$\\
	& $|\eta| < 4.5$\\
	& Rejected if within $\Delta R = 0.4$ of a fiducial lepton\\
\bottomrule
\end{tabular}
}
\caption{Summary of the selection criteria defining the fiducial phase space.}
\label{tab:fiducial_selection}
\end{table}


In the following, \myfigs~\ref{fig:dressing_pt}--\ref{fig:fiducial_jets} visualise the fiducial phase space and the performance of the fiducial selection criteria.

\myfig~\ref{fig:dressing_pt} shows the effect of dressing leptons with nearby photons as a function of the lepton \pt{}, in particle-level MC events with all other fiducial requirements applied. As expected, the dressing increases the hardness of the \pt{} spectra, by recovering radiation losses. Equally expectedly, the effect is larger for electrons than muons, which is explained by the smaller electron mass. Good agreement of the ratios dressed/bare is observed between the different MC generators and samples.

\begin{figure}[h!]
\centering
\subfigure[]{
\begin{tikzpicture}
\node[anchor=south west,inner sep=0] (image) at (0,0) {\includegraphics[width=0.7\textwidth]{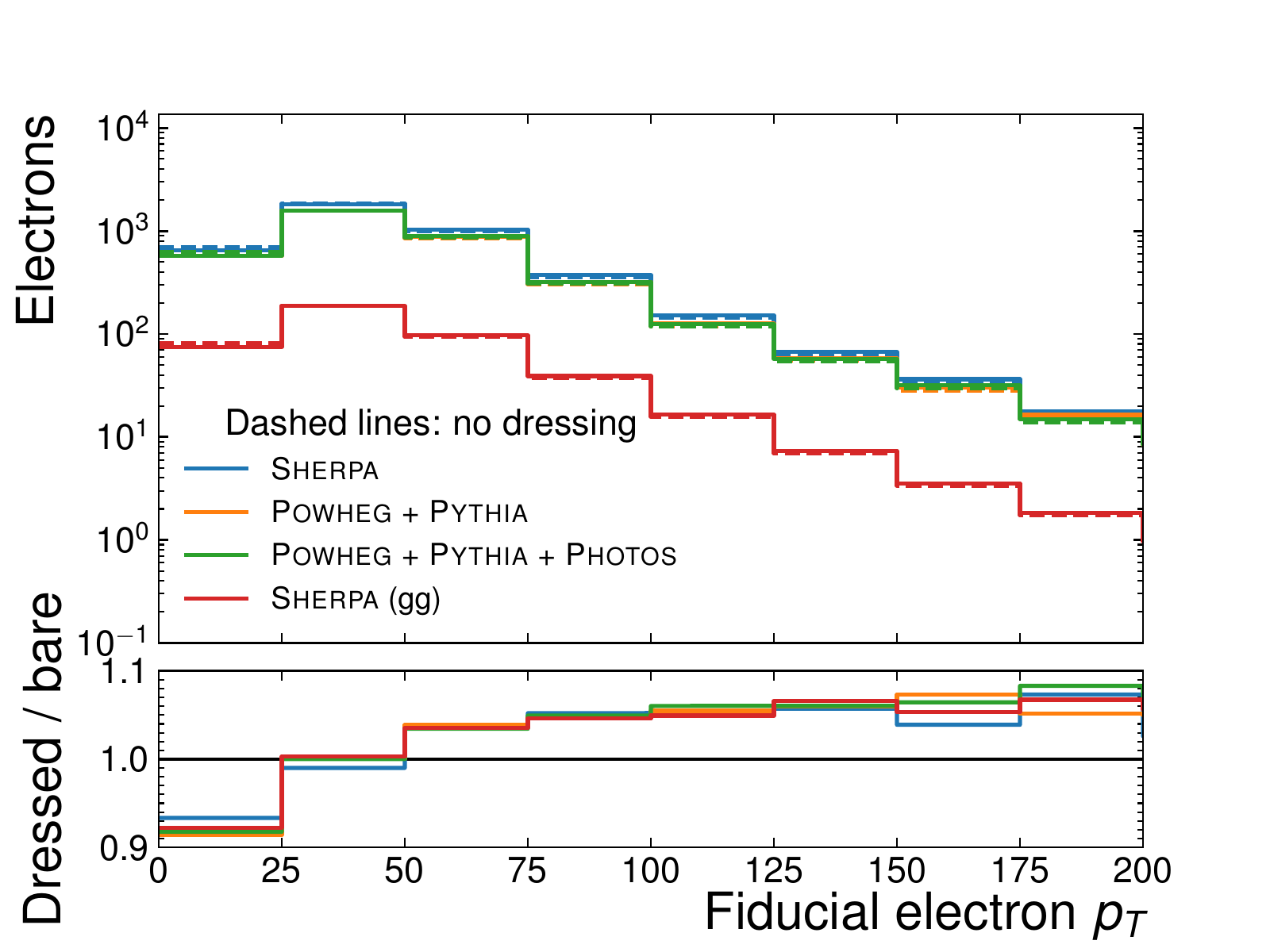}};
\begin{scope}[x={(image.south east)},y={(image.north west)}]
\draw[white, fill=white] (0.55,0.0) rectangle (0.7,0.07);
\end{scope}
\end{tikzpicture}
}
\subfigure[]{
\begin{tikzpicture}
\node[anchor=south west,inner sep=0] (image) at (0,0) {\includegraphics[width=0.7\textwidth]{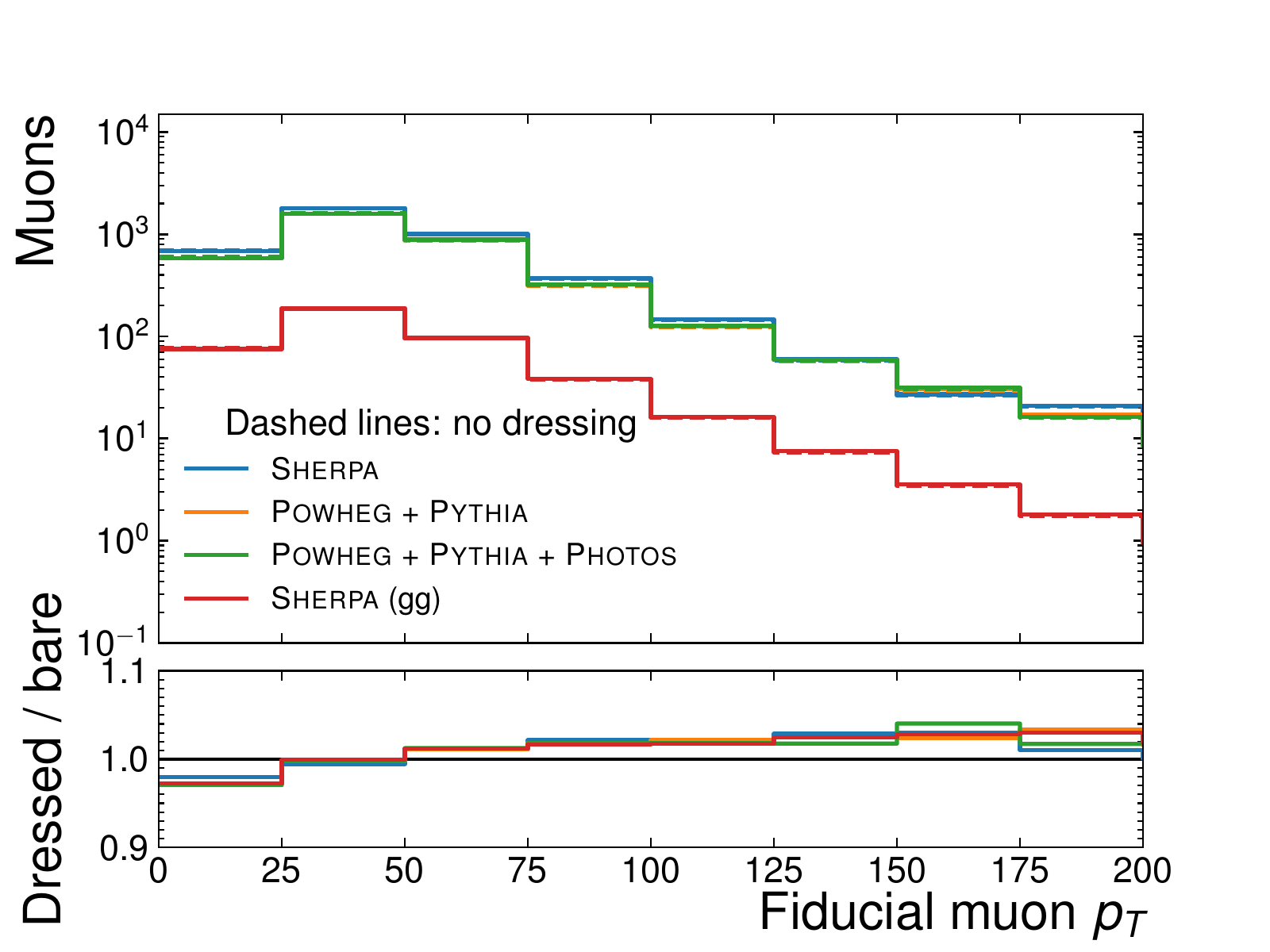}};
\begin{scope}[x={(image.south east)},y={(image.north west)}]
\draw[white, fill=white] (0.55,0.0) rectangle (0.74,0.07);
\end{scope}
\end{tikzpicture}

}
\caption{Effect of dressing as a function of the (a) electron and (b) muon transverse momentum. In all cases, all other fiducial criteria have been applied. In the upper panel, the dressed spectrum is shown as a solid line and the bare spectrum as a dashed line of the same colour. The lower panel shows the ratio of dressed to bare for each sample.}
\label{fig:dressing_pt}
\end{figure}

\myfig{}~\ref{fig:fiducial_m4l} shows the four-lepton mass before and after the dilepton mass requirements ($66~\GeV{} < m_{\ell\ell} < 116~\GeV{}$). In both cases, all other fiducial requirements are applied. This means that \myfig~\ref{fig:fiducial_m4l_fidPS} shows the final fiducial distribution. In \myfig~\ref{fig:fiducial_m4l_nomasscut}, the $\PZ{} \to \llll$ peak is clearly visible in the \qq-initiated processes, while the $\PH \to \llll$ peak is visible in the \gluglu-initiated process. \POWHEGpy{} with and without \PHOTOS{} agree very well, while \SHERPA{} predicts a significantly higher cross section at low mass. 

\begin{figure}[h!]
\centering
\subfigure[]{\includegraphics[width=0.7\textwidth]{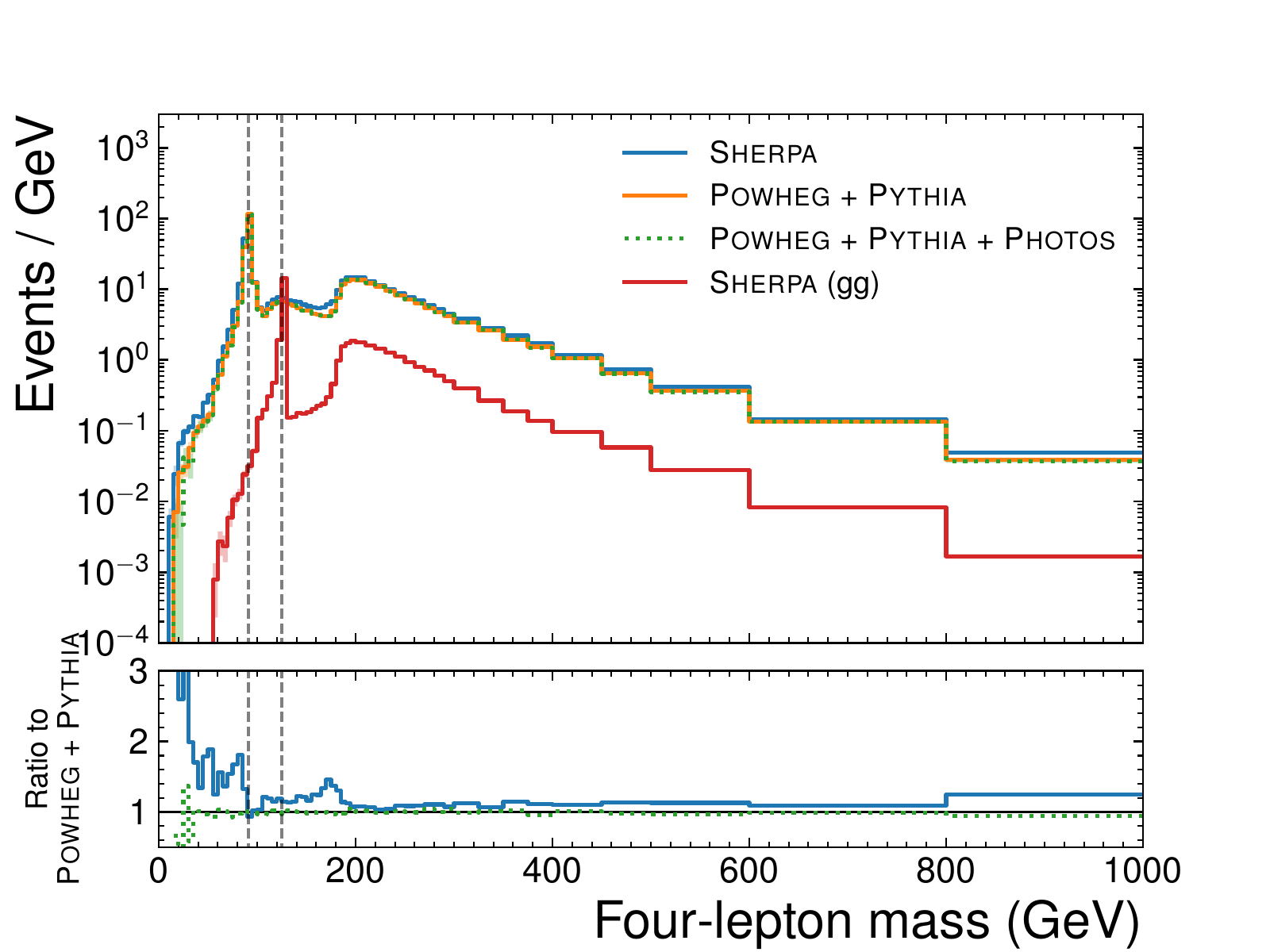}
\label{fig:fiducial_m4l_nomasscut}}
\hspace{1mm}
\subfigure[]{\includegraphics[width=0.7\textwidth]{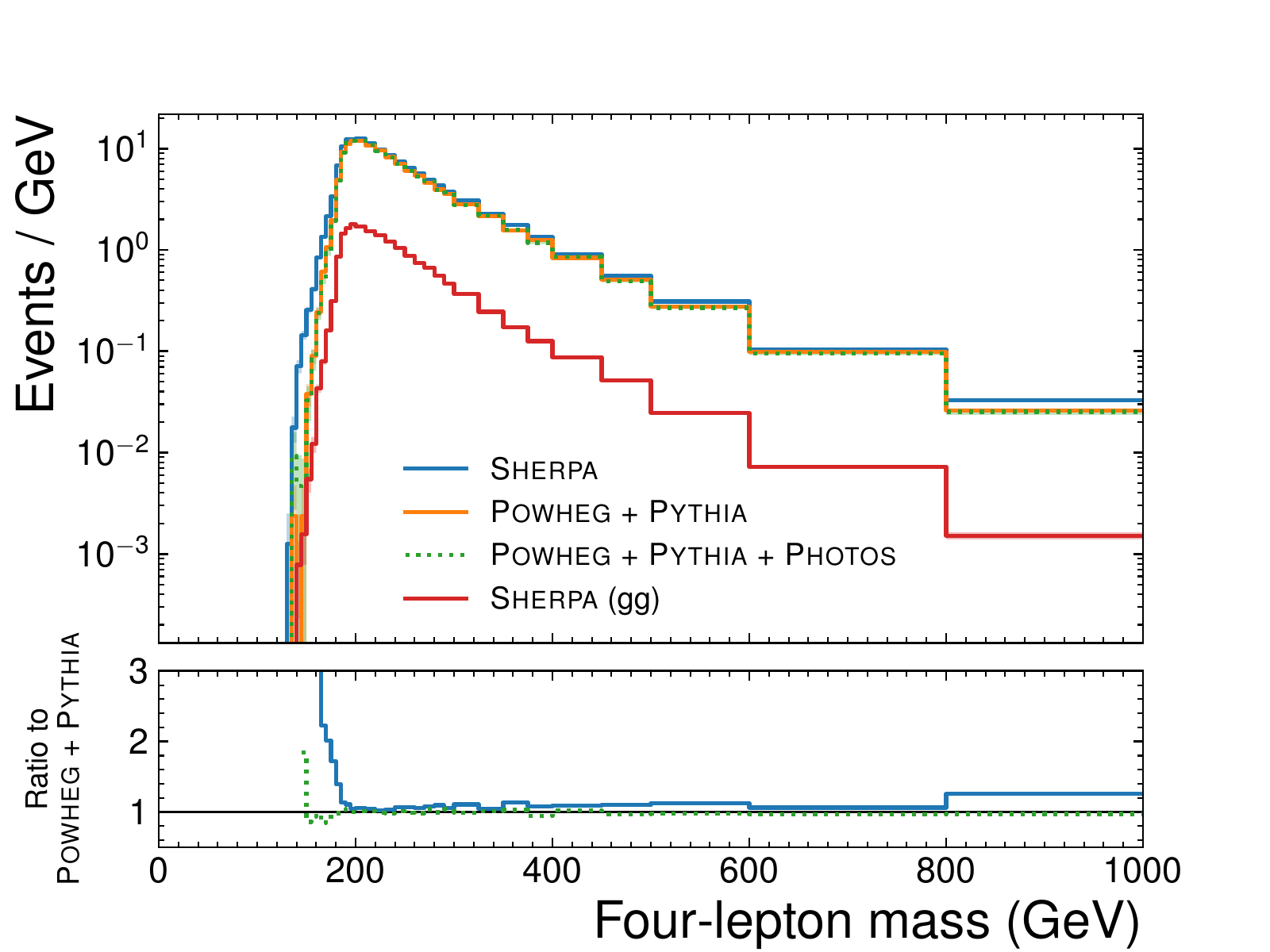}
\label{fig:fiducial_m4l_fidPS}}
\caption{Four-lepton mass (a) before and (b) after dilepton mass requirements. All other fiducial criteria have been applied. The dashed vertical lines at 91~\GeV{} and 125~\GeV{} mark the \PZ{} boson and Higgs boson peak, respectively. Shaded bands in the upper panels indicate the statistical uncertainties.}
\label{fig:fiducial_m4l}
\end{figure}

The predicted multiplicity and kinematics of fiducial jets in fiducial events are shown in \myfig~\ref{fig:fiducial_jets}. \SHERPA{} predicts more jets, which tend to have higher pseudorapidity and higher transverse momentum. This is (at least qualitatively) expected, because \SHERPA{} describes the three hardest jets at the matrix-element level, whereas \POWHEGpy{} only describes the first. Higher jet multiplicities are brought about by the parton shower in both cases.

\begin{figure}[h!]
\centering
\subfigure[]{\includegraphics[width=0.65\textwidth]{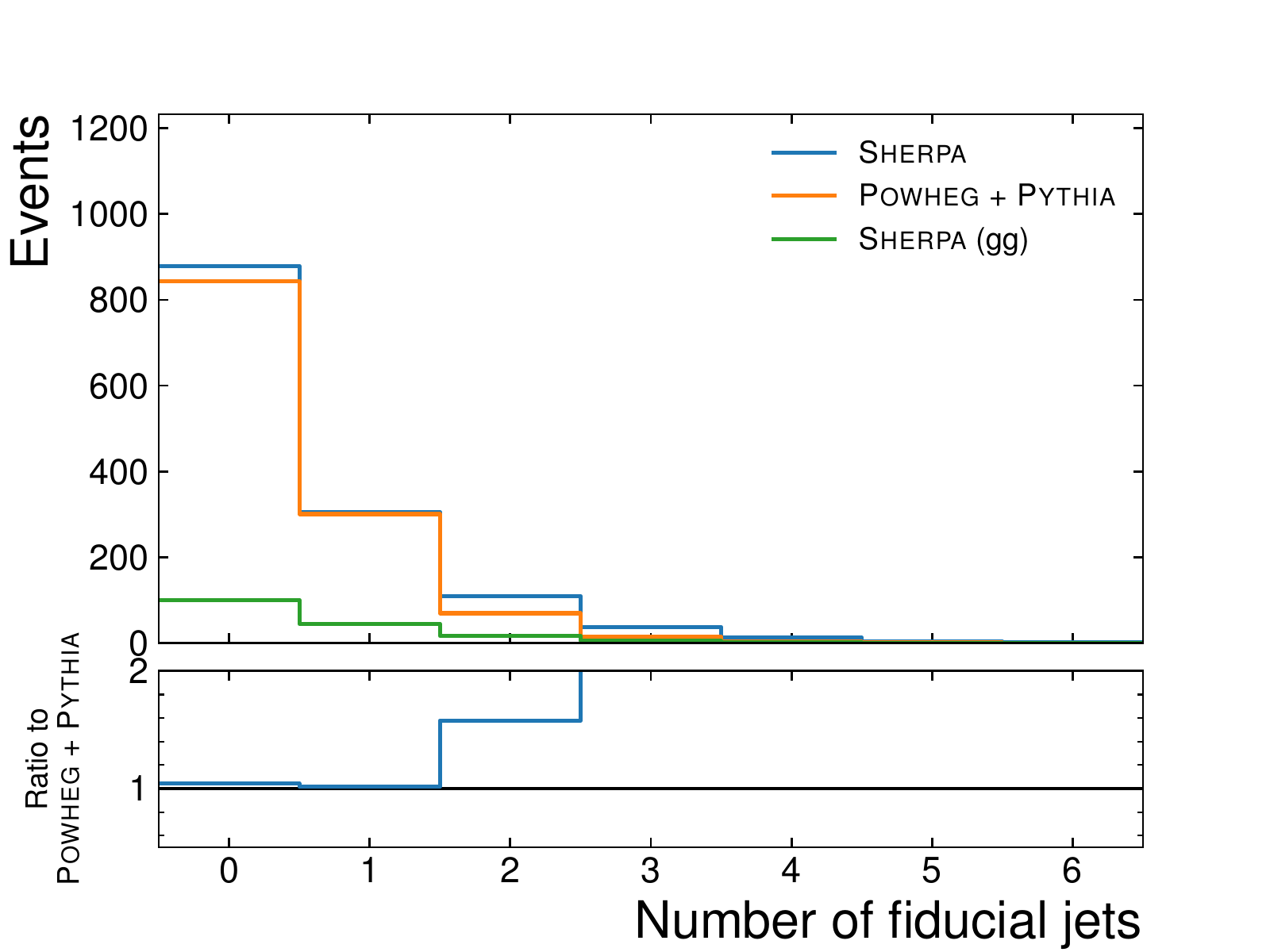}}
\vspace{-4mm}
\phantom{.}
\subfigure[]{\includegraphics[width=0.65\textwidth]{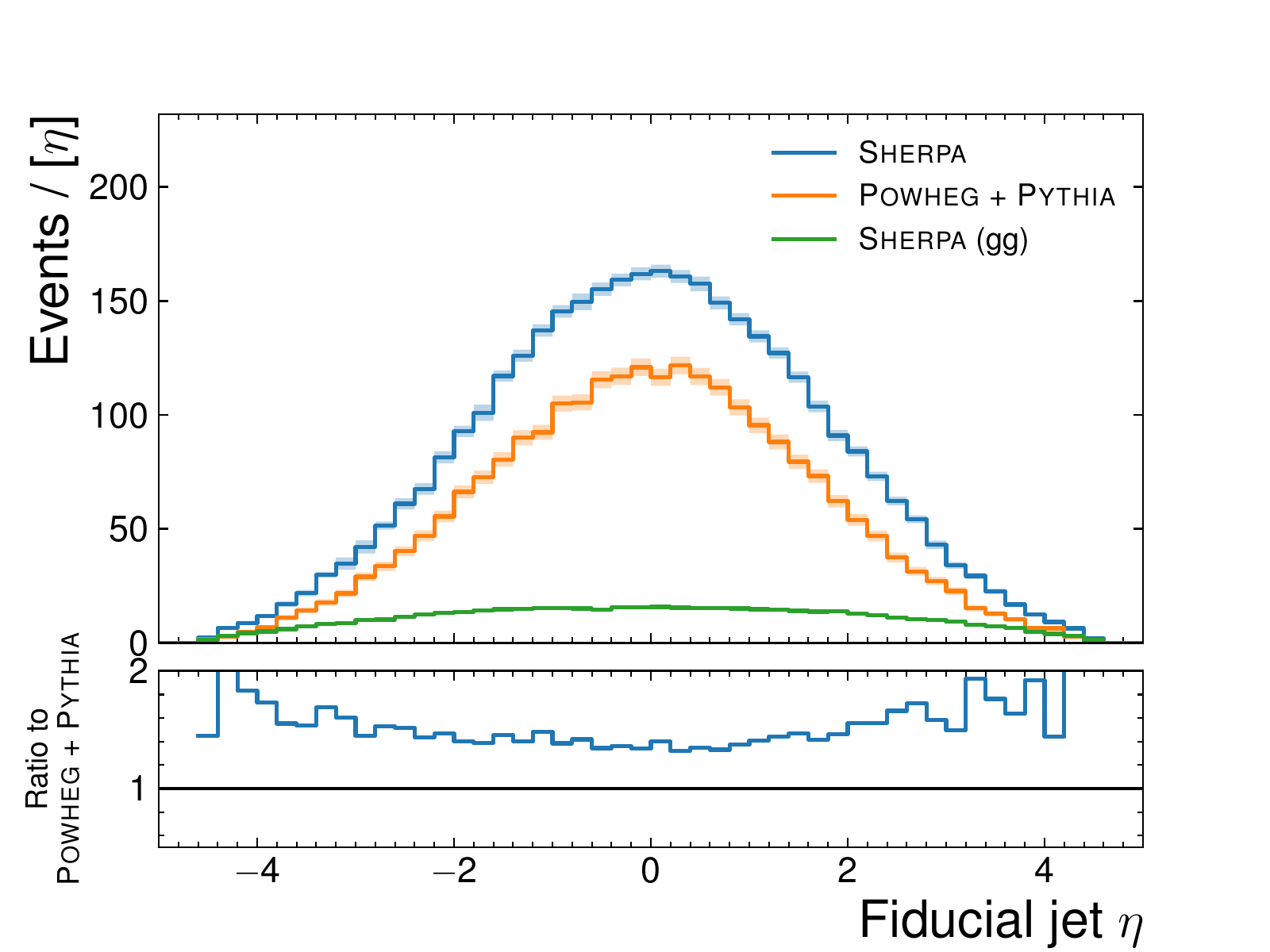}}
\vspace{-4mm}
\phantom{.}
\subfigure[]{\includegraphics[width=0.65\textwidth]{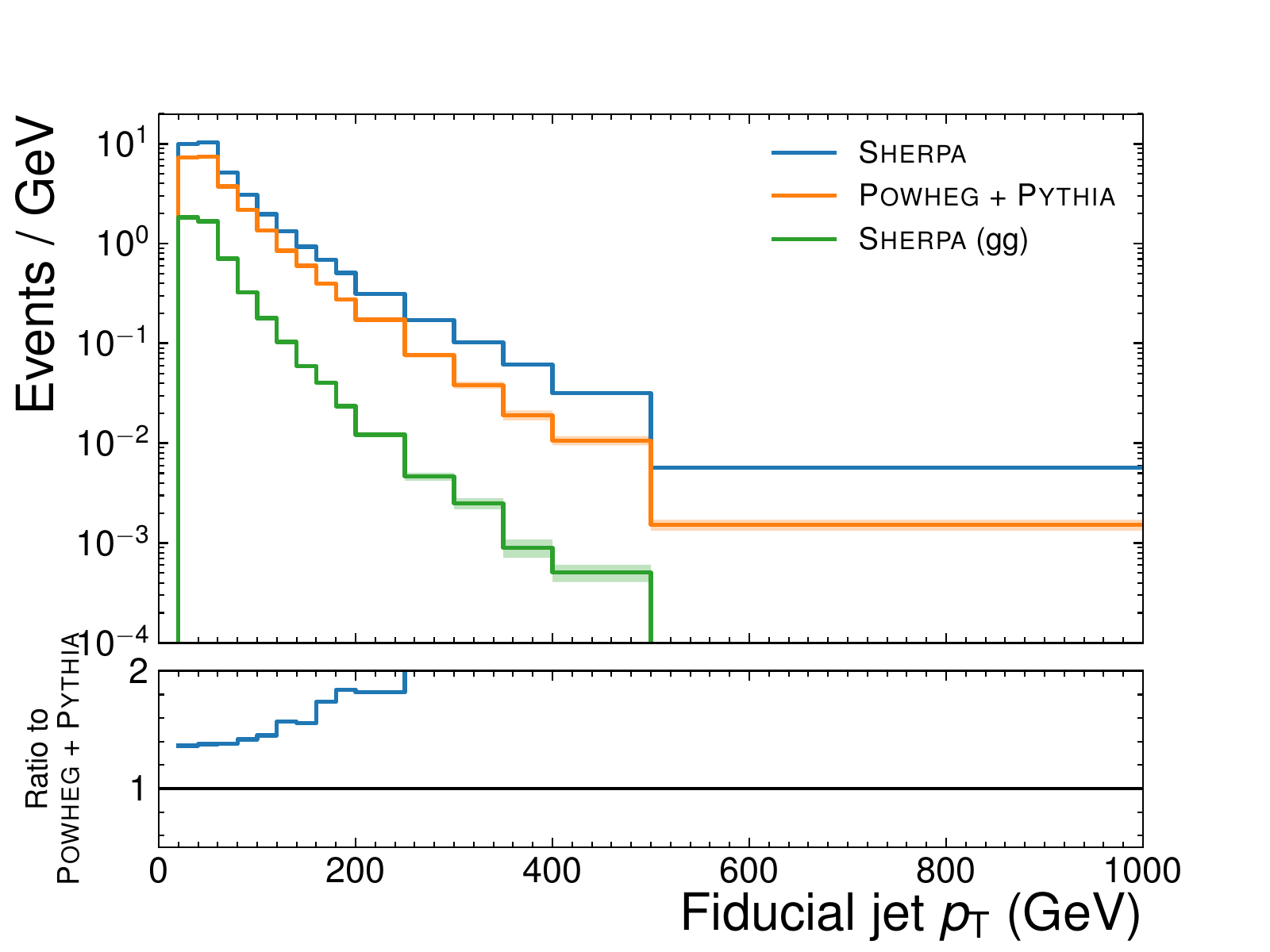}}
\caption{Predicted (a) multiplicity, (b) pseudorapidity, and (c) transverse momentum of fiducial jets in fiducial events falling in the fiducial phase space. Shaded bands in the upper panels indicate the statistical uncertainties.}
\label{fig:fiducial_jets}
\end{figure}

\clearpage

\subsection{On-shell phase space}
\label{sec:truthonshell}
Another particle-level phase space is defined, to which the integrated cross section is extrapolated (in addition to the fiducial measurement). This `on-shell' phase space is identical to the fiducial phase space \emph{except} that the lepton \pt{} and $\eta$ requirements and the lepton-lepton $\Delta R$ requirements are removed.\footnote{Technical detail: when leptons are treated as massless (such as in \matrixnnlo{}), a photon-pole divergence occurs as the splitting $\Pgamma^* \to \ell^+\ell^- $ ($\propto 1/q^2$) becomes collinear ($q^2 \to 0$). By keeping the requirement that \emph{any} same-flavour opposite-charge lepton pair has a mass of at least 5~\GeV{}, this divergence is still removed in the on-shell phase space, even after lifting the lepton-lepton $\Delta R$ requirements.}
The on-shell phase space is used to calculate an estimate for the production cross section of (nearly) on-shell $\PZ$ bosons, based on the fiducial measurement.

\myfig{}~\ref{fig:onshell_m4l} shows the four-lepton mass before and after the dilepton mass requirements after all other on-shell phase space requirement have been applied. As was the case in the fiducial distribution (possibly before dilepton mass requirements) shown in \myfig~\ref{fig:fiducial_m4l}, \POWHEGpy{} with and without \PHOTOS{} agrees very well, while \SHERPA{} predicts a higher cross section at low mass. In \myfig~\ref{fig:onshell_m4l_incl_offshell}, the \SHERPA{} cross section below $\sim$80~\GeV{} is lower than the \POWHEGpy{} prediction due to tighter lepton \pt{} requirements at the event-generation stage. These differing requirements are found to have no impact on the analysis. \myfigs~\ref{fig:prompt_leptons_eta} and \ref{fig:prompt_leptons_pt} show the $\eta$ and \pt{} distributions of prompt leptons in the on-shell events, with no fiducial $\pt$, $\eta$, or $\Delta R$ requirements applied. As expected, the distributions for electrons and muons agree very well.

\begin{figure}[h!]
\centering
\subfigure[]{\includegraphics[width=0.7\textwidth]{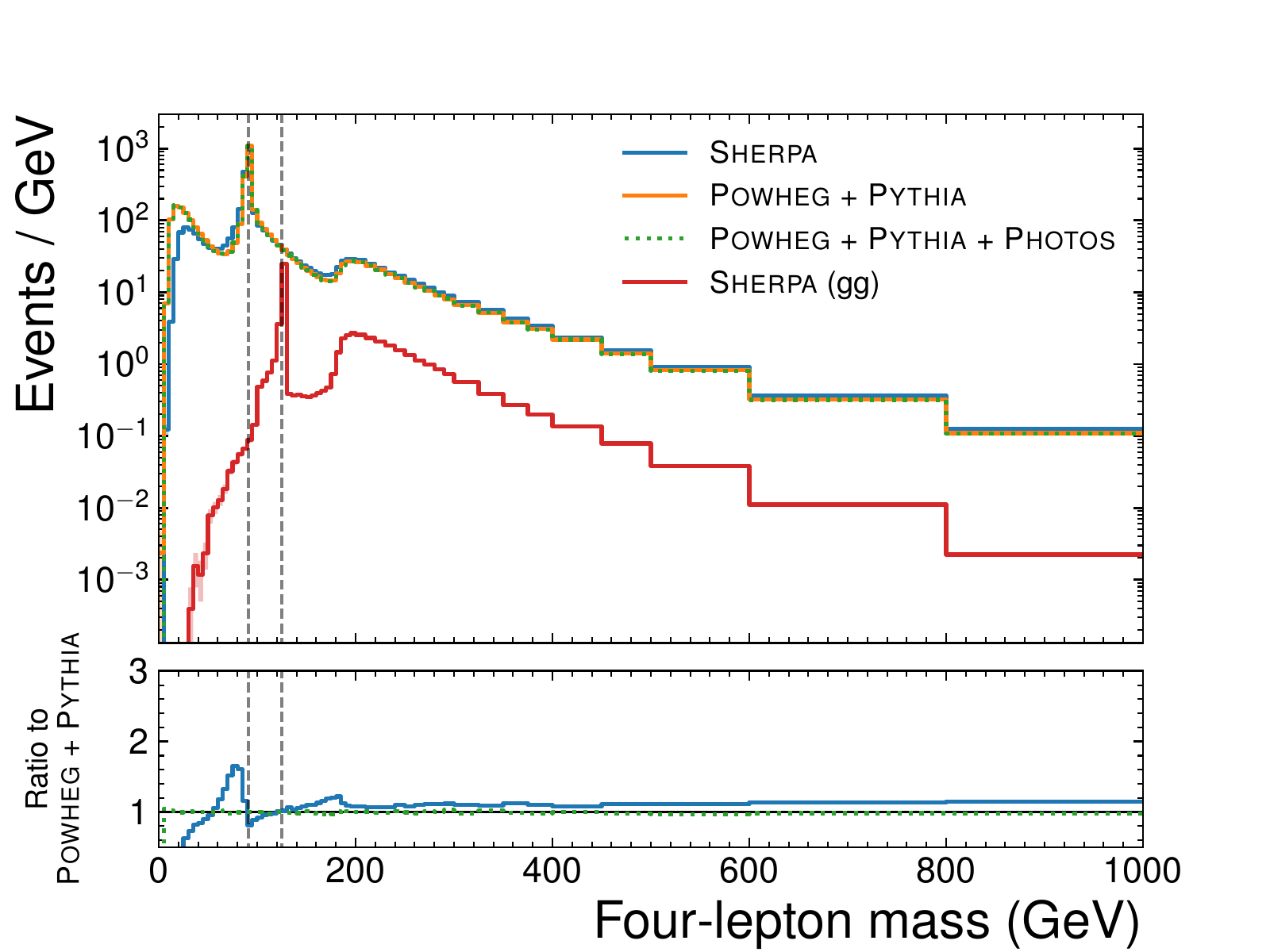}\label{fig:onshell_m4l_incl_offshell}}
\subfigure[]{\includegraphics[width=0.7\textwidth]{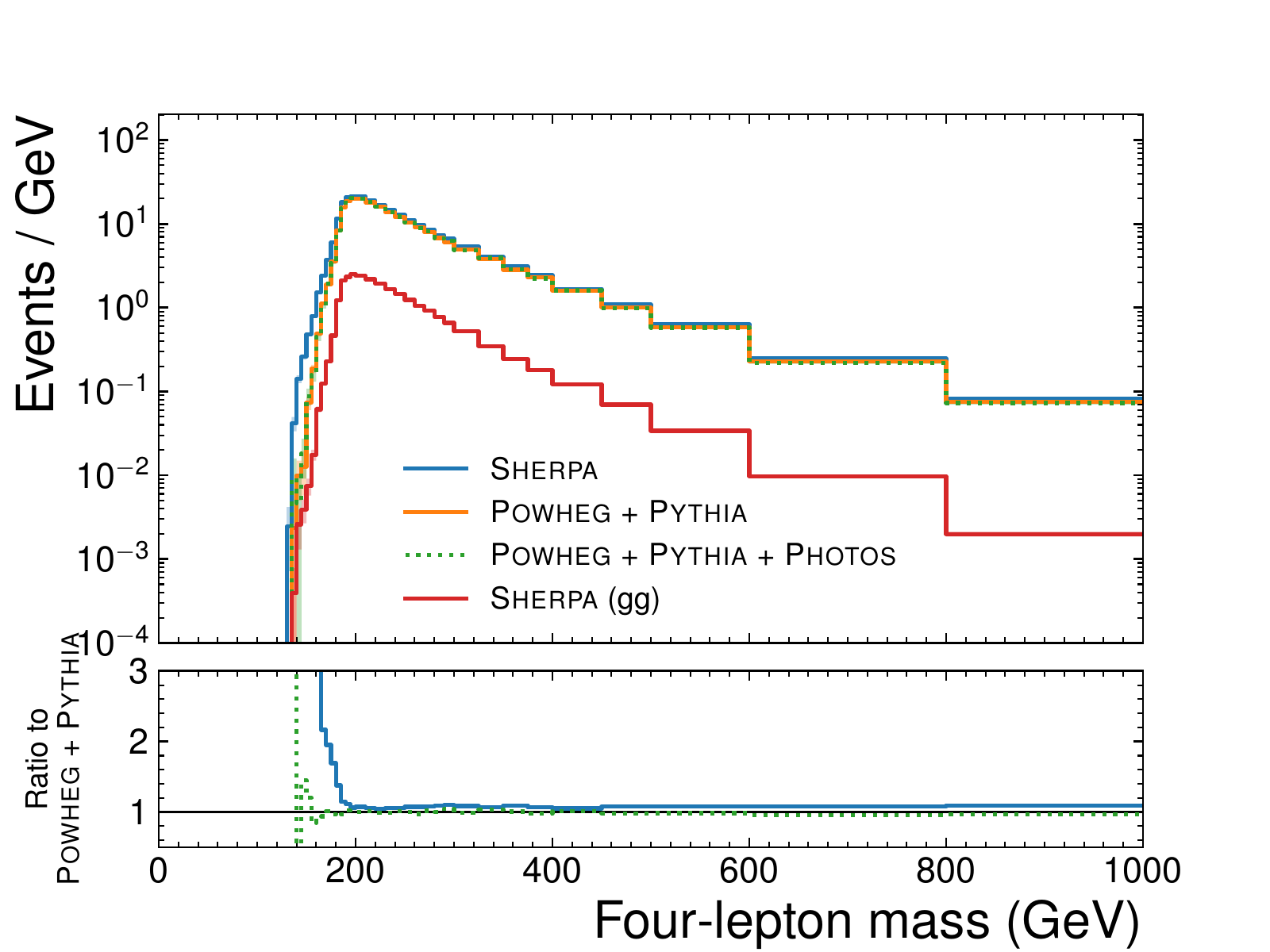}}
\caption{The particle-level four-lepton mass (a) before and (b) after on-shell requirement. All other on-shell phase space requirements have been applied. The dashed vertical lines at 91~\GeV{} and 125~\GeV{} mark the \PZ{} boson and Higgs boson peak, respectively. Shaded bands in the upper panels indicate the statistical uncertainties.}
\label{fig:onshell_m4l}
\end{figure}

\begin{figure}[h!]
\centering
\subfigure[]{\includegraphics[width=0.7\textwidth]{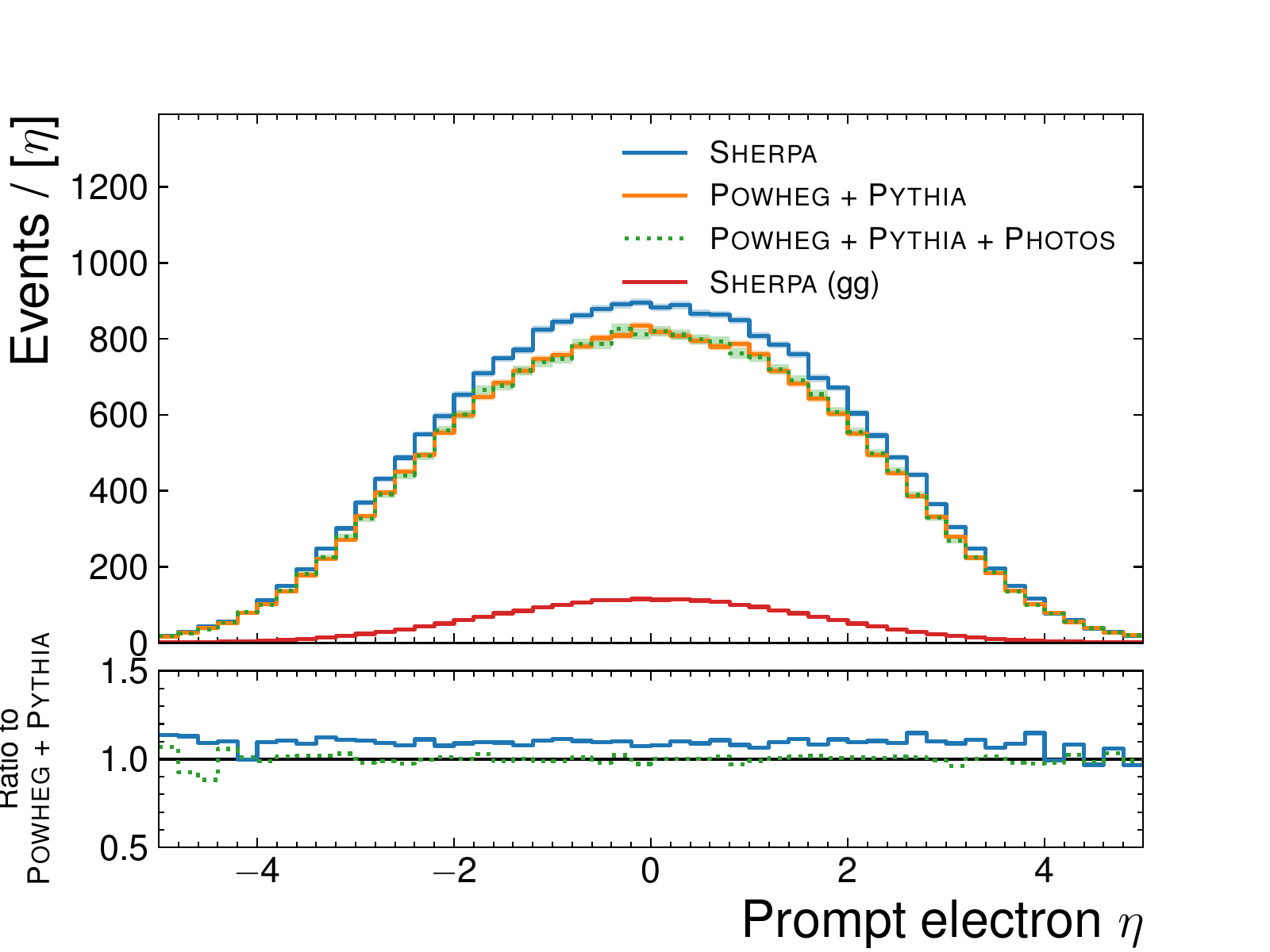}}
\hspace{1mm}
\subfigure[]{\includegraphics[width=0.7\textwidth]{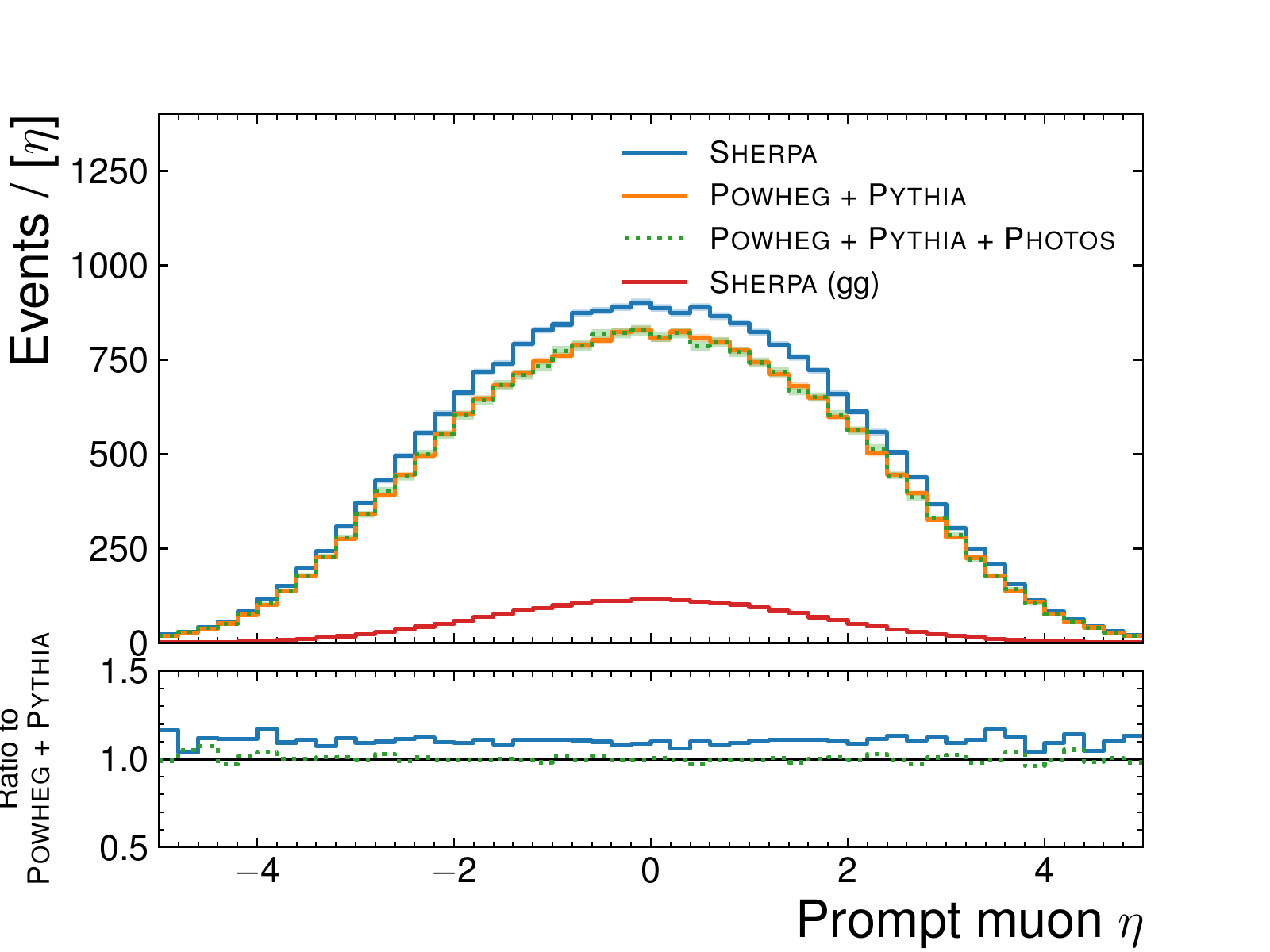}}
\caption{Pseudorapidity of prompt (a) electrons and (b) muons in on-shell events, before fiducial requirements. Shaded bands in the upper panels indicate the statistical uncertainties.}
\label{fig:prompt_leptons_eta}
\end{figure}

\begin{figure}[h!]
\centering
\subfigure[]{\includegraphics[width=0.7\textwidth]{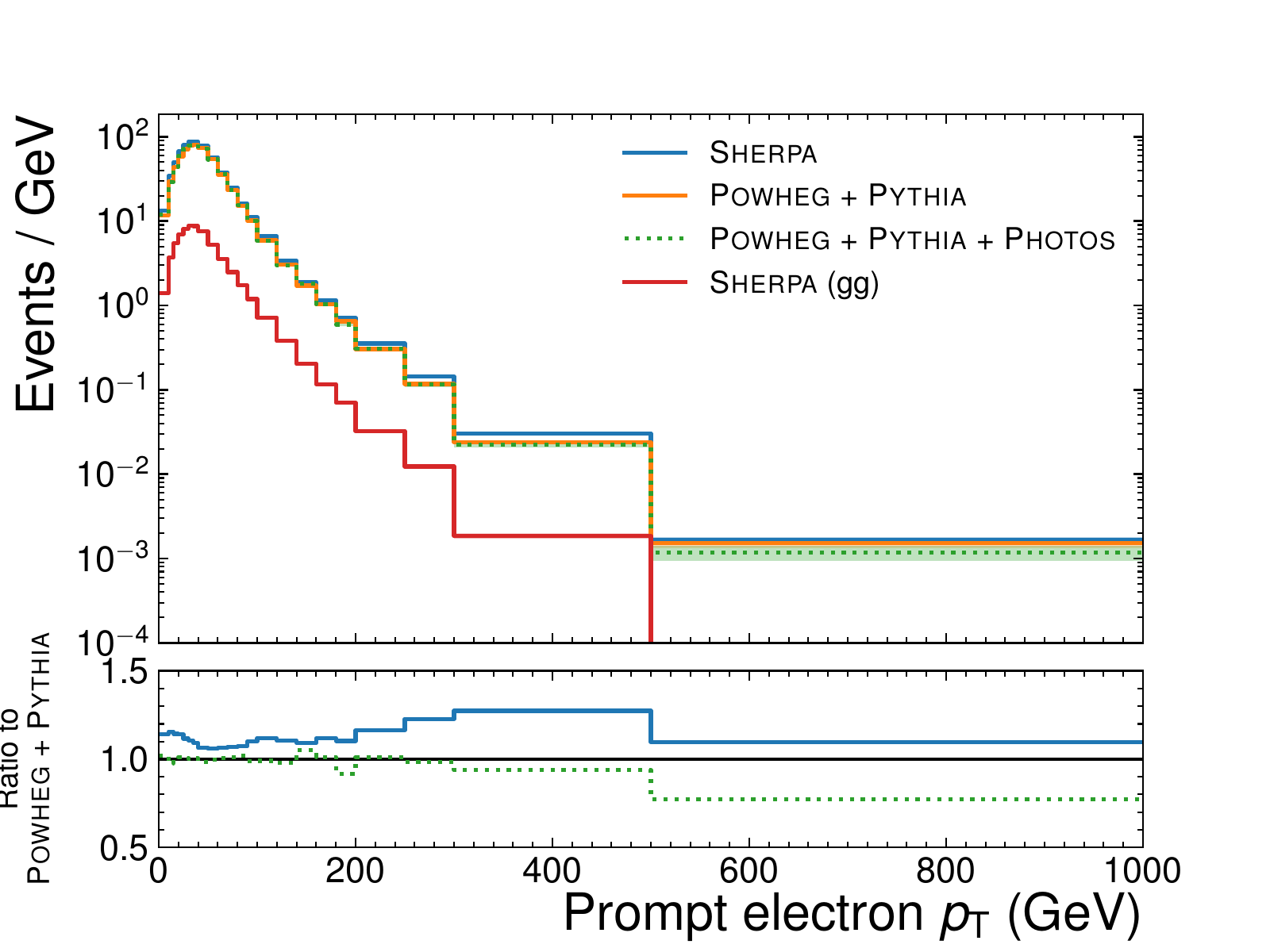}}
\hspace{1mm}
\subfigure[]{\includegraphics[width=0.7\textwidth]{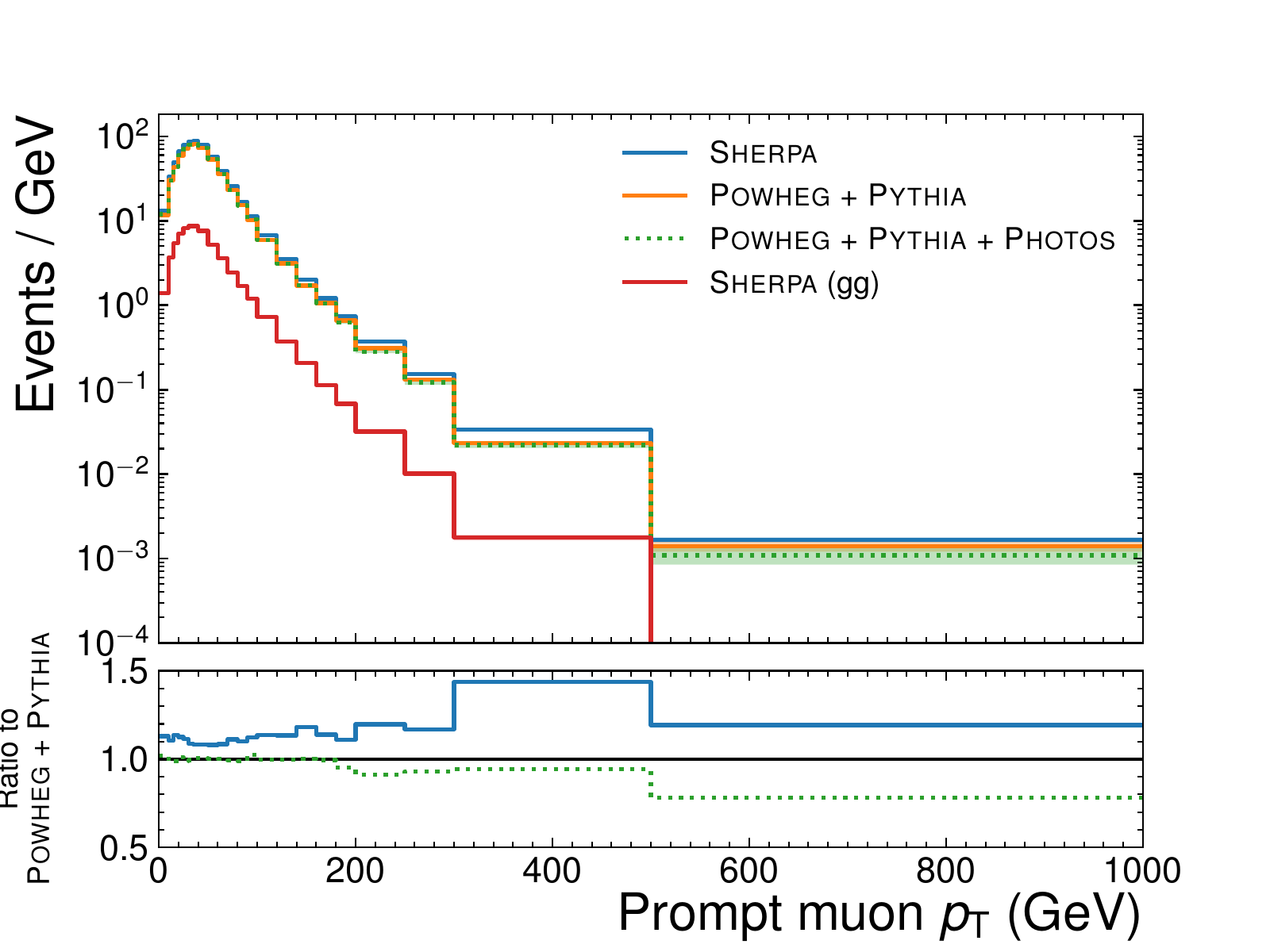}}
\caption{Transverse momentum of prompt (a) electrons and (b) muons in on-shell events, before fiducial requirements. Shaded bands in the upper panels indicate the statistical uncertainties.}
\label{fig:prompt_leptons_pt}
\end{figure}

\clearpage

\subsection{Signal-process definition}

Some SM processes can pass the fiducial selection but are still excluded from the signal. They are considered irreducible backgrounds and subtracted from the sample of selected candidate events. Any events containing four prompt leptons plus any additional leptons, neutrinos, or photons are considered \textit{irreducible backgrounds}. An example is the triboson process $\PZ\PZ\PWplus \to \ell^+\ell^-\ell^{\prime +}\ell^{\prime -}\ell^+\nu_{\ell}$. Formally, all such processes are subtracted as background. In practice, predictions only exist for a subset of the processes. The irreducible backgrounds that are actually subtracted in the analysis are discussed in \mysec{}~\ref{sec:bkg}. They are very small, approximately~1\% of the predicted signal. 

The fiducial phase space is inclusive with respect to jets, independently of their origin. Triboson (and higher boson-multiplicity) processes producing a $\PZ\PZ$ pair decaying leptonically with any additional electroweak bosons decaying hadronically are included in the signal, as are any other SM processes of the pattern $(\PZ\PZ \to \ell^+\ell^-\ell^{\prime +}\ell^{\prime -}) + (X \to \text{jets})$. In practice, only the process $\PZ\PZ V \to \llll jj$ (where $V = \PWpm,\,\PZ$) is included in the theoretical predictions, in the EW-$\ZZ jj$ sample generated with \SHERPA{}.

Production via double parton scattering (DPS) in the same $\Pproton\Pproton$ collision, explained in \mysec{}~\ref{sec:theory_double_parton_scattering}, is formally included in the signal. Its contribution is not included in the theoretical predictions, but is expected to be smaller than 1\% of the total signal yield, as calculated in \mysec{}~\ref{sec:theory_double_parton_scattering}.

The above signal-process definition deliberately leaves open the possibility of new physics contributing to the fiducial region, by making few assumptions of what processes occur in nature.

\subsection{Double parton scattering}
\label{sec:theory_double_parton_scattering}

A pair of \PZ{} bosons can be produced via two separate parton scatterings in a single proton-proton interaction, $\Pproton\Pproton \to \PZ \otimes \PZ \to \llll$. An introduction to the theory of DPS can be found e.g.~in \myref~\cite{Manohar:2012pe}. This production channel is formally included in the fiducial signal but is not modelled in the derivation of the results or in their interpretation.
As laid out in \myref~\cite{Sadeh:2013wka} and references therein, in particular \myref~\cite{Humpert:1983fy}, assuming incoherent DPS, the integrated contribution to the \ZZllll{} cross section from DPS can be estimated using
\begin{equation}
\label{eq:dps}
\sigma_{\PZ \otimes \PZ \to \llll} = \frac{\sigma_{\PZ \to \ell^+\ell^-}^2}{2 \sigma_{\text{eff}}},
\end{equation}
where $\sigma_{\PZ\to \ell^+\ell^-}$ is the cross section to produce one on-shell \Z{} boson and $\sigma_{\text{eff}}$ is an empirically determined effective cross section for DPS. The $\PZ{}\to \ell^+\ell^-$ production cross section was measured at 13~\TeV{} to be $1981 \pm 57$~pb ~\cite{STDM-2015-03}. 
The effective cross section was measured to be $\sigma_{\mathrm{eff}} = 15^{+6}_{-4}$~mb at 7~\TeV{} \cite{Aad:2013bjm}. Various other measurements of it were made~\cite{ALITTI1991145, Abe:1993rv, Abe:1997xk, Abazov:2009gc, Aaboud:2016dea, Aaij:2012dz,Chatrchyan:2013xxa,Abazov:2014fha} and suggest no significant dependence on the centre-of-mass energy nor the final state used to extract it.\footnote{Except for measurements by the D0 collaboration using charm- and bottom-quark final states (\kern1pt$\PJpsi \otimes \PUpsilon$), which suggest a smaller $\sigma_{\mathrm{eff}}$ \cite{Abazov:2015fbl} --- meaning a \emph{larger} DPS contribution. However, these measurements are outliers compared to all other experiments' results.} Using these values and \myeq~\ref{eq:dps}, an estimate of $\sigma_{\PZ \otimes \PZ \to \llll} = 0.13^{+0.06}_{-0.04}$~fb is obtained, which is $(0.5 \pm 0.2)$\% of the predicted integrated \ZZ{} cross section. So DPS is estimated to make a very small contribution to the signal.



\clearpage
\section{Event selection}
\label{sec:selection}

The event selection begins with trigger and data-quality requirements.
Candidate events are preselected by single-, di-, or trilepton triggers~\cite{TRIG-2016-01}, with a combined efficiency very close to 100\%. The exact list of triggers used is shown in \mytab{}~\ref{tab:zz_trigger_list}. 
Events must have at least one primary vertex with two or more associated tracks with $\pt > 400$~\MeV{} \cite{ATL-PHYS-PUB-2015-026}. Events must pass cleaning criteria \cite{ATLAS-CONF-2015-029} designed to reject events with excessive noise in the calorimeters. The data are subjected to quality requirements to reject events in which detector components were not operating correctly.

\begin{table}[!htbp]
	\begin{center}
		{\small
		\begin{tabular}{llll}
			\toprule
			\textbf{Type} & \textbf{Trigger name} & \textbf{Run range} & \textbf{Corresponding periods}\\
			\midrule
			\multirow{4}{*}{\Pe{}} & e24\_lhmedium\_L1EM20VH &   $\rightarrow$ 284484  & 2015\\
			& e60\_lhmedium &                         $\rightarrow$ 284484 & 2015\\
			& e26\_lhtight\_nod0\_ivarloose & 296939 $\rightarrow$ & 2016\\
			& e60\_lhmedium\_nod0 & 296939 $\rightarrow$ & 2016\\
			\midrule
			\multirow{5}{*}{\Pmu{}} & mu20\_iloose $^{\text{(see caption)}}$ &  $\rightarrow$ 300287 & 2015 + 2016 period A\\
			& mu24\_ivarmedium & 296939 $\rightarrow$ 302393 & 2016 periods A--C\\
			& mu26\_ivarmedium & 296939 $\rightarrow$ & 2016\\
			& mu40 &  $\rightarrow$ 300287 & 2015 + 2016 period A\\
			& mu50 & All runs & 2015 + 2016\\
			\midrule
			\multirow{2}{*}{$\Pe\Pe$} & 2e12\_lhloose\_L12EM10VH &  $\rightarrow$ 284484 & 2015\\
			& 2e17\_lhvloose\_nod0 & 296939 $\rightarrow$ & 2016\\
			\midrule
			\multirow{6}{*}{$\Pmu\Pmu$} & mu18\_mu8noL1 &  $\rightarrow$ 284484 & 2015\\
			& mu20\_mu8noL1 & $\rightarrow 302393$ & 2015 + 2016 periods A--C\\
			& 2mu10 &  $\rightarrow$ 300287 & 2015 + 2016 period A\\
			& 2mu14 & All runs & 2015 + 2016\\
			& mu22\_mu8noL1 & All runs & 2015 + 2016\\
			& mu20\_nomucomb\_mu6noL1\_nscan03 & 296939 $\rightarrow$ 302393 & 2016 periods A--C\\
			\midrule
			$\Pe\Pmu$ & e17\_lhloose\_mu14 &  $\rightarrow$ 284484 & 2015\\
			& e17\_lhloose\_nod0\_mu14 & 296939 $\rightarrow$ & 2016\\
			\midrule
			\multirow{2}{*}{$\Pe\Pe\Pe$}  & e17\_lhloose\_2e9\_lhloose &  $\rightarrow$ 284484 & 2015\\
			& e17\_lhloose\_nod0\_2e9\_lhloose\_nod0 & 296939 $\rightarrow$ & 2016 $^{\text{(see caption)}}$\\
			\midrule
			$\Pmu\Pmu\Pmu$ & 3mu6 & All runs & 2015 + 2016\\
			\bottomrule
		\end{tabular}
		}
		\caption{Triggers used in the analysis, and the range of runs over which they are used, corresponding to runs where the trigger was unprescaled. Trigger mu20\_iloose is not available in simulated samples, instead the mu20\_iloose\_L1MU15 trigger is used there. Trigger e17\_lhloose\_nod0\_2e9\_lhloose\_nod0 was prescaled for one or two luminosity blocks at the start of a few runs in 2016 period G (run 305291 $\rightarrow$).}
		\label{tab:zz_trigger_list}
	\end{center}
\end{table}

Following this preselection, muons, electrons and jets are selected in each event as described below. Based on these, the best lepton quadruplet is selected and required to pass further selection criteria.



\subsection{Selection of muons, electrons, and jets}
\label{sec:object_selection}
Muon quality requirements and the `loose' identification criteria are applied as described in \myref{}~\cite{Aad:2016jkr}. The `loose' identification uses all four types of reconstructed muons. All combined and standalone muons are considered, covering $|\eta| < 2.7$. Calorimeter- and segment-tagged muons are considered within $|\eta| < 0.1$. All muons are required to have $\pt > 5$~\GeV{}, calorimeter-tagged muons must have $\pt > 15$~\GeV{}. \myfig{}~\ref{fig:analysis_muon_types} shows the number of observed muons by type in events passing the trigger and vertex selections, as well as in events passing the final selection, as a function of their pseudorapidity. 

Electrons are identified following the `loose' criteria in \myref{}~\cite{PERF-2016-01}. They are required to have $|\eta| < 2.47$ and $\pt > 7$~\GeV{}. Electrons whose calorimeter cluster lies in the ECAL crack region ($1.37 < |\eta| < 1.52$) are included.

\begin{figure}[h!]
\centering
\subfigure[]{\includegraphics[width=0.48\textwidth]{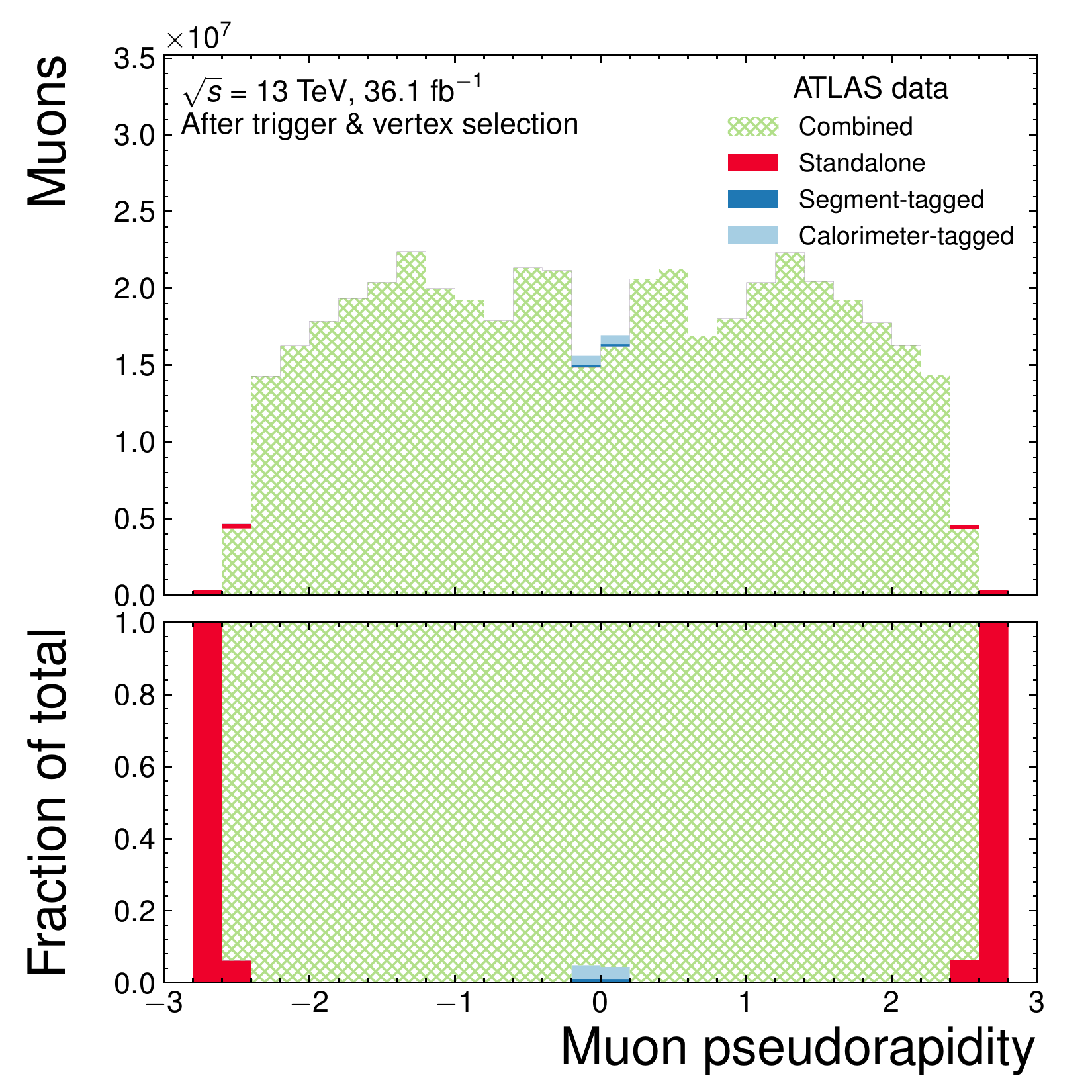}}
\hspace{1mm}
\subfigure[]{\includegraphics[width=0.48\textwidth]{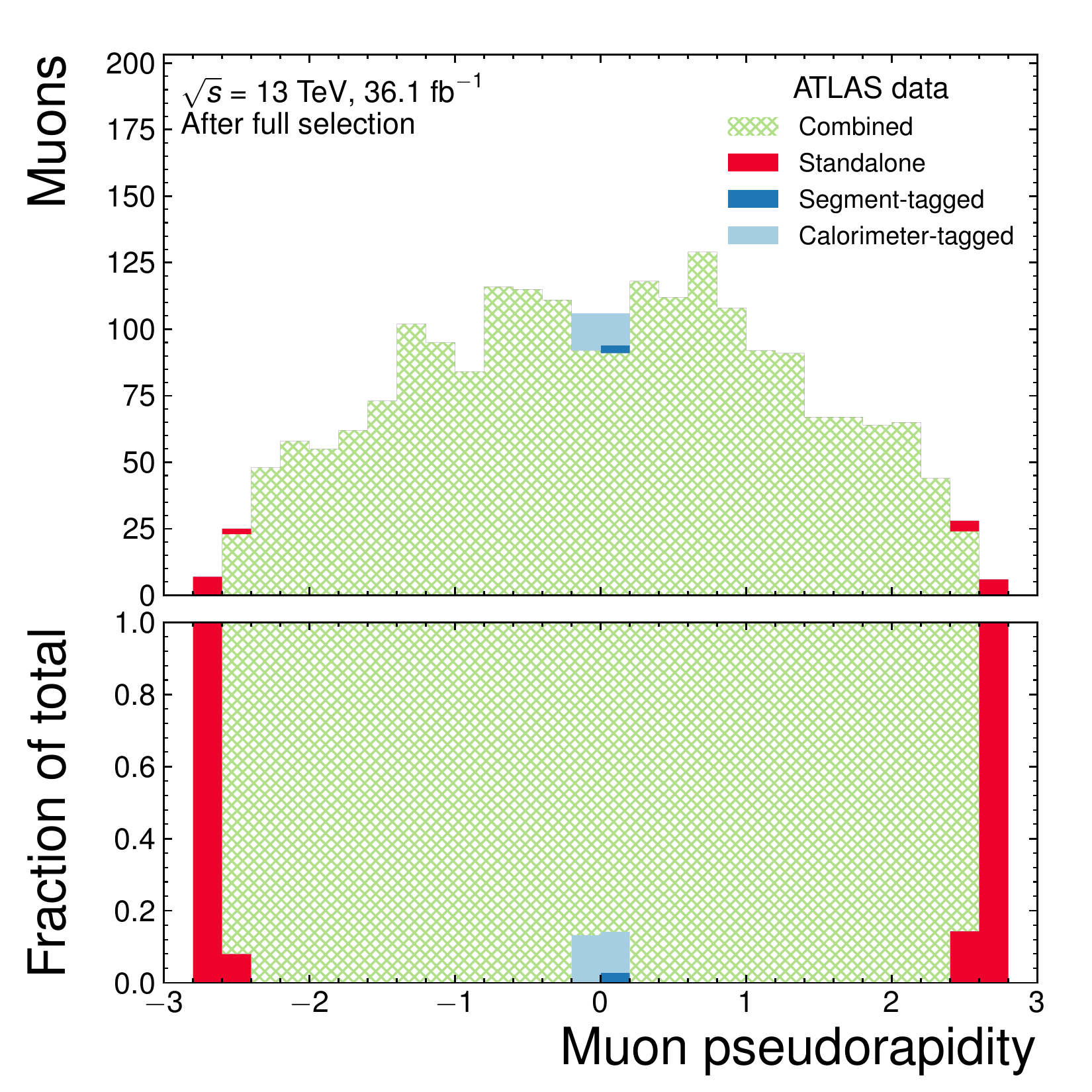}}
\caption{Identified muons in data by type (a) after trigger and vertex selection and (b) after the full event selection, as function of the muon pseudorapidity. The contributions of the different types are stacked. To better visualise the fractions in the bottom panel, the stacking order is inverted.}
\label{fig:analysis_muon_types}
\end{figure}





Leptons are required to originate from the hard-scattering vertex (HSV), defined as the primary vertex with the largest sum of the $\pt^{2}$ of the associated tracks. The longitudinal impact parameter of each lepton track, calculated with respect to the hard-scattering vertex and multiplied by $\sin\theta$ of the track, $|(z_0 - z_{\text{HSV}})\sin\theta|$ is required to be less than 0.5~mm. The multiplication by $\sin\theta$ can be understood as providing a measure of the significance of the difference $|z_0 - z_{\text{HSV}}|$. For very central tracks, a smaller difference is significant than for very forward tracks, since the distance travelled by the particle before encountering the first tracker layer is shorter for the former. This is illustrated in \myfig~\ref{fig:longitudinal_ip_analytical}.
Muons must have a transverse impact parameter calculated with respect to the beam line less than 1~mm in order to reject muons originating from cosmic rays. 
The significance of the transverse impact parameter, defined as the measured transverse impact parameter divided by its uncertainty, calculated with respect to the beam line is required to be less than three (five) for muons (electrons). The tighter requirements for muons reflect their better impact parameter resolution compared to electrons. Applying the same requirements to electrons would lead to too many genuine electrons being rejected.
Standalone muons are exempt from all three requirements, as they do not have an ID track. 

\begin{figure}[h!]
\centering
\includegraphics[width=0.7\textwidth]{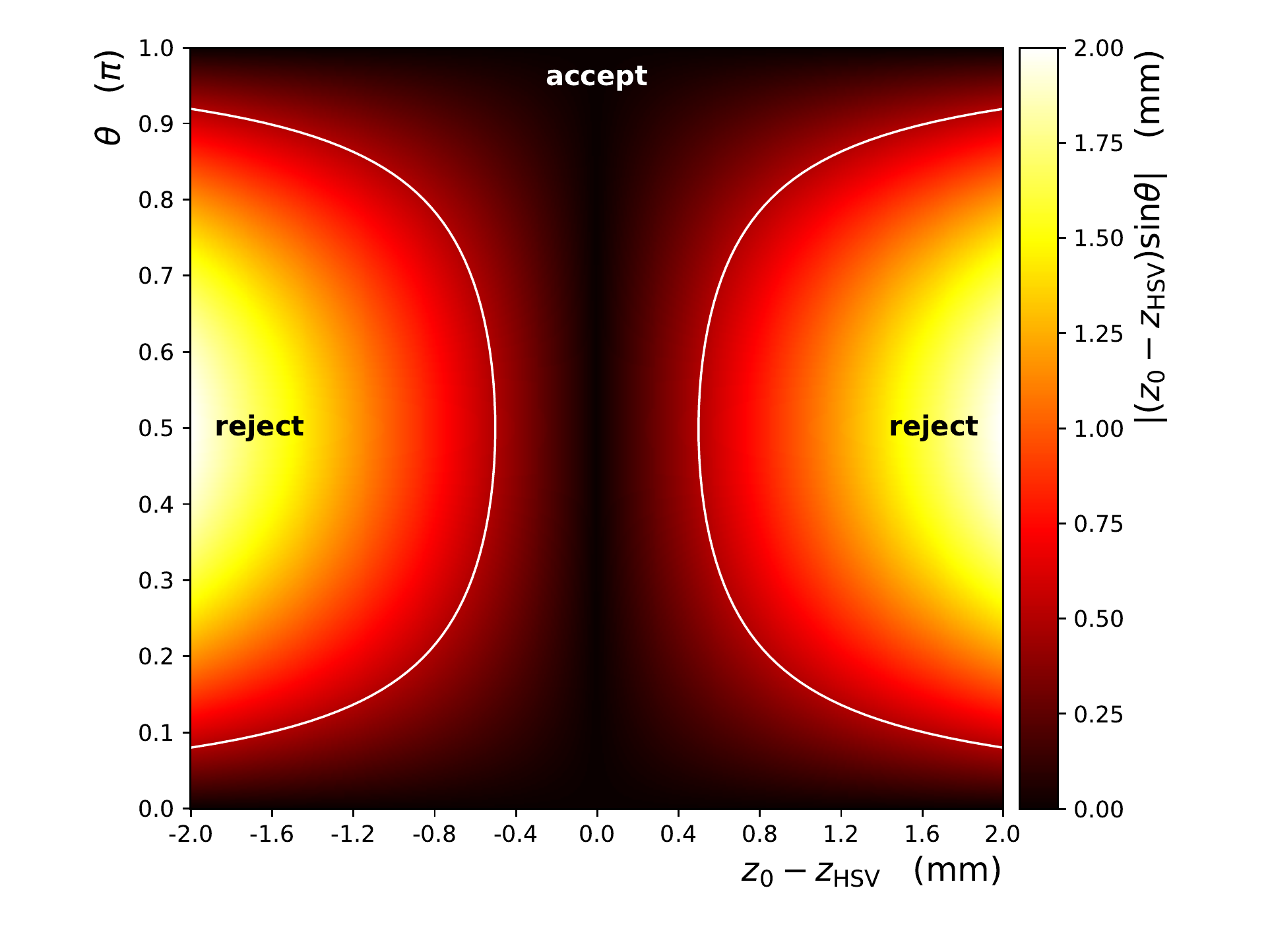}
\vspace{-7mm}
\caption{Calculated distribution of the longitudinal impact parameter as a function of the longitudinal distance between the track's point of closest approach of the $z$-axis from the hard-scattering vertex, $z_0 - z_{\text{HSV}}$, and the scattering angle $\theta$ of the track ($\theta = 0.5\pi$ corresponds to a perfectly central lepton). The white contour at $|(z_0 - z_{\text{HSV}})\sin\theta| = 0.5$~mm separates accepted and rejected leptons. In ATLAS, the limit of $\theta$ acceptance is approximately $0.05\pi$--$0.95\pi$ for electrons and $0.04\pi$--$0.96\pi$ for muons.}
\label{fig:longitudinal_ip_analytical}
\end{figure}

Leptons are required to be isolated from other particles using both ID-track and calorimeter-cluster information.
Muons (electrons) with transverse momentum $\pt$ are removed if the summed transverse momentum of other ID tracks within $\Delta R = \min[0.3, 10~\GeV / \pt]$ ($\min[0.2, 10~\GeV / \pt]$) of the lepton exceeds $0.15\, \pt$, or if the summed transverse energy of other topological clusters within $\Delta R = 0.2$ of the lepton exceeds $0.3\, \pt$ ($0.2\, \pt$).


Jets are clustered using the anti-$k_t$ algorithm \cite{cacciari08} with radius parameter 0.4, as implemented in \textsc{FastJet}~\cite{Cacciari:2011ma,Cacciari:2005hq}. They are required to have $|\eta| < 4.5$ and $\pt > 30$~\GeV{}, the same as for the fiducial definition. In order to reject jets originating from pileup interactions, they must either pass a jet vertex tagging selection \cite{ATLAS-CONF-2014-018,PERF-2016-06} or have $\pt > 60$~\GeV{}.
In the range $2.4 < |\eta| < 2.5$, neither the jet vertex tagging method for central nor for forward jets applies, so jets in this region are automatically required to have $\pt > 60~\GeV$.

In order to avoid the reconstruction of multiple objets from the same detector signals, all but one such overlapping objects are removed.
Electron candidates sharing an ID track with a selected muon are rejected, except if the muon is only calorimeter-tagged, in which case the muon is rejected instead. Electron candidates sharing their track or calorimeter cluster with a selected higher-\pt{} electron are rejected. Jets within $\Delta R = 0.4$ of a selected lepton are rejected.

\subsection{Quadruplet selection}
As for the fiducial definition, events must contain a quadruplet, formed of at least four leptons forming at least two pairs of same-flavour opposite-charge dileptons. All possible quadruplets in a given event are considered for further selection.
At most one muon in each quadruplet may be a calorimeter-tagged or standalone muon. The three highest-$\pt$ leptons in each quadruplet must satisfy $\pt > 20$~\GeV{}, 15~\GeV{}, 10~\GeV{}, respectively.
If multiple selected quadruplets are present, the best quadruplet is chosen as in the fiducial phase-space selection (\mysec{}~\ref{sec:fiducial}). 
Only the best quadruplet is considered further and the following requirements are applied on the leptons in that quadruplet.
Any two different (same) flavour leptons $\ell_i$, $\ell'_j$ must be separated by $\Delta R (\ell_i,\ell'_j) > 0.2$~(0.1). All possible same-flavour opposite-charge dileptons must have an invariant mass greater than $5$~\GeV{}, to reduce background from leptonic hadron decays. 
The two \PZ{} boson candidates, formed as in the fiducial definition, are required to have an invariant mass between 66~\GeV{} and 116~\GeV{}. \myfig{}~\ref{fig:2d_mass_plots} shows the distribution of invariant masses of the $\PZ$ boson candidates in selected data events. Based on the leptons in the chosen quadruplet, events are classified into the \eeee{}, \mmmm{}, and \eemm{} signal channels.





\begin{figure}[h!]
\centering
\subfigure[]{\includegraphics[width=0.48\textwidth]{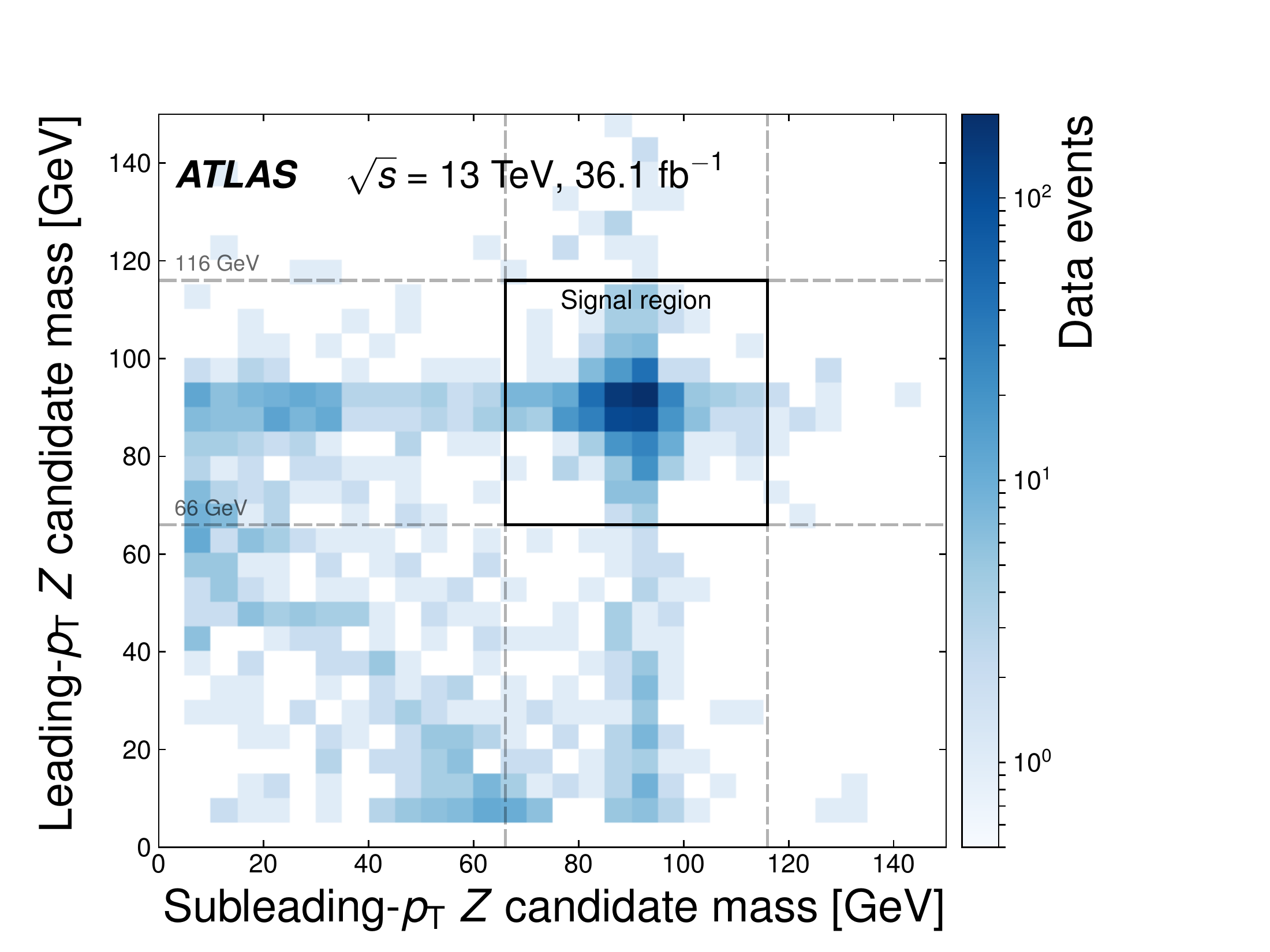}}
\hspace{1mm}
\subfigure[]{\includegraphics[width=0.48\textwidth]{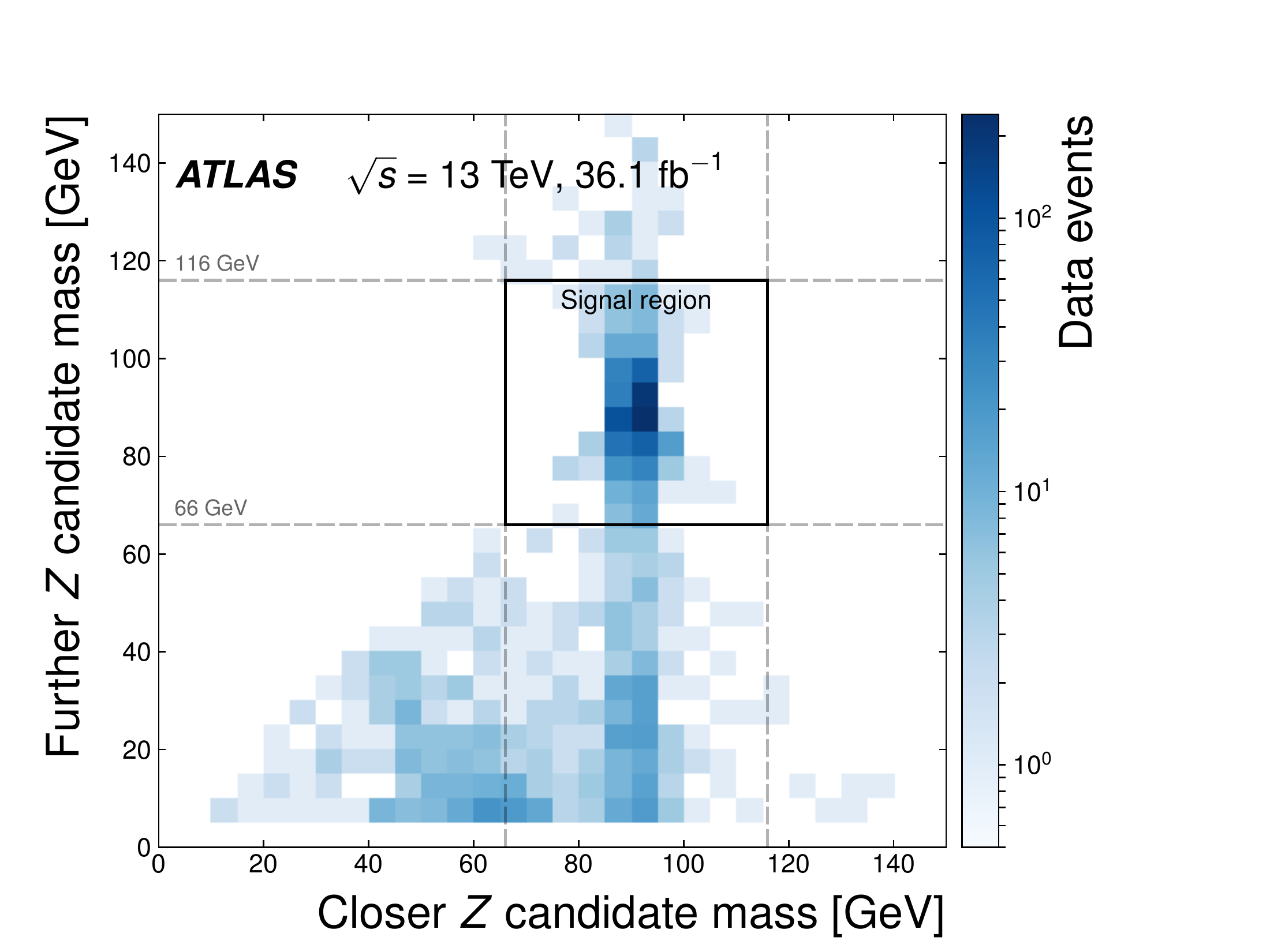}}
\caption{Invariant mass of one selected \PZ{} boson candidate dilepton vs.~the other, in the selected data events before the \PZ{} boson candidate mass requirement. All other selections have been applied. (a) shows the \PZ{} boson candidates arranged by transverse momentum. 
(b) shows the \PZ{} boson candidates arranged by proximity of their mass to the \PZ{} boson pole mass. The solid rectangle shows the signal region. Dashed gray lines mark the positions of the \PZ{} boson candidate mass cuts for each pair, 66~\GeV{} to 116~\GeV{}. Only data are shown. Published in \myref~\cite{STDM-2016-15}.}
\label{fig:2d_mass_plots}
\end{figure}

\subsection{Remarks on the event selection}

The analysis is statistically limited and almost background-free, so the event selection is optimised towards maximising signal acceptance. At the same time, it joins an ATLAS effort to harmonise event selections among similar analysis, so that many selection requirements are identical to those used in recent and ongoing ATLAS Higgs-boson measurements in the four-lepton channel.

According to MC predictions, the hierarchical lepton \pt{} requirements in each quadruplet candidate do not lower the acceptance. However, a lower \pt{} requirement for the softest lepton would have been beneficial. Currently ongoing analyses may benefit from a lower minimum electron \pt{} requirement of 4.5~\GeV{}, whereas 7~\GeV{} was the lowest supported value at the time of this analysis.

The use of so-called \emph{forward electrons} was considered. These are reconstructed using only calorimeter information and have $2.5 < |\eta| < 4.9$. Including forward electrons could have increased the signal acceptance in the \eeee{} channel by as much as $\sim$10--20\% (depending on assumptions). However, at the time of the analysis, their use was not yet fully supported and e.g.~the energy calibration had not yet been derived. Due to limited time and person power, the analysis team ultimately decided not to use them.

As the mass of the $\PZ\PZ$ system (and correspondingly the \pt{} of each \PZ{} boson) increases, the leptons in each \PZ{} boson candidate tend to become more collimated in the laboratory frame. This could lead to decreased acceptance of signal events in this region of phase space, as such events are more likely to fail the lepton $\Delta R$ or isolation requirements, which would be problematic for the aTGC search, whose sensitivity depends on the acceptance of high-scale events. However, using MC predictions, the requirements in place were found to be sufficiently loose for this not to be a problem in practice.

As is the case for recent ATLAS Higgs-boson analyses, no trigger efficiency scale factors are applied to the predictions. Due to the very high trigger efficiency, well above 99\%, they are found to be negligible. Trigger matching, i.e.~requiring the objects that triggered the recording of the event to have corresponding objects after the final selection, is not performed. This has no appreciable effect on the event selection.

This is the first published ATLAS analysis using jet vertex tagging for jets with $|\eta| > 2.5$, thanks to the new methods presented in \myref~\cite{PERF-2016-06}. 

\clearpage
\section{Background estimation}
\label{sec:bkg}

The expected total background is very small, approximately 2\% of the total predicted yield in each channel. 

\subsection{Simulation of irreducible background}

Irreducible backgrounds from processes with at least four prompt leptons in the final state are estimated with the simulated samples described in \mysec{}~\ref{sec:mc}, including uncertainties from the cross section predictions, luminosity measurement, and experimental effects. 
Non-hadronic triboson processes (15\% of the total background estimate) and fully leptonic $\Ptop\APtop\PZ$ processes (19\%) are considered:
\begin{equation*}
\begin{split}
\PZ\PWplus\PWminus &\to \llll \Pnu_{\ell'}\APnu_{\ell'}\\
\PZ\PZ\PWpm &\to \llll \ell^{\pm} \Pnu_{\ell}\\
\PZ\PZ\PZ &\to \llll \ell^+\ell-\\
\PZ\PZ\PZ &\to \llll \Pnu_{\ell}\APnu_{\ell}\\
\Ptop\APtop\PZ &\to \llll \Pnu_{\ell}\APnu_{\ell} \Pbottom\APbottom.
\end{split}
\end{equation*}

Simulated samples are also used to estimate the background from \ZZ{} processes where at least one \Z{} boson decays to $\tau$ leptons (8\% of the total background estimate),
\begin{equation*}
\begin{split}
\ZZ &\to \tau^+\tau^-\ell^+\ell^- \to \llll \Pnut\APnut\Pnu_{\ell}\APnu_{\ell}\\
\ZZ &\to \tau^+\tau^-\tau^+\tau^- \to \llll \Pnut\APnut\Pnut\APnut\Pnu_{\ell}\APnu_{\ell}\Pnu_{\ell'}\APnu_{\ell'}.
\end{split}
\end{equation*}

CMS \cite{CMS-ZZ-13TEV} (and earlier ATLAS \cite{STDM-2014-16}) includes leptons from $\tau$ lepton decay at the detector level, but not in the fiducial definition. This means that they are first treated as signal, leading to small distortions in the kinematic distributions due to incomplete reconstruction, then corrected for. In this analysis, the approach was changed to treat them as background instead. This was considered a more consistent classification, given that the final states differ, affecting their kinematics. The argument for including \Ptau{} lepton background is that, assuming lepton universality, it scales as the signal. Perhaps the best option would be to treat it as background, but scale the background contribution according to the measurement. Such a scaling was not done in this analysis for simplicity, as the \Ptau lepton background is extremely small.

\subsection{Data-driven estimation of reducible background}
\label{sec:dd_bkg}

Events from processes with two or three prompt leptons, e.g. \PZ{}, \WW{}, \WZ{}, $\Ptop\APtop$, and \ZZ{} events where one \PZ{} boson decays hadronically, can pass the event selection if associated jets, non-prompt leptons, or photons are misidentified as prompt leptons. This background is termed reducible, because an ideal detector for electrons and muons would eliminate it.
Maurice Becker used the data-driven technique briefly summarised in this section (\mysec~\ref{sec:dd_bkg}) to estimate the reducible background with misidentified leptons. The details of the method can be found in \myref~\cite{thesis_maurice}.

A lepton selection that is orthogonal to the nominal selection in \mysec{}~\ref{sec:object_selection} is defined by reversing some of its requirements. Muons must fail the transverse impact parameter requirement or the isolation requirement, or both. Electrons must fail either the isolation requirement or the likelihood-based identification, but not both. Electrons failing the likelihood-based identification must still pass quality criteria applied to their track (which are a subset of the likelihood-based identification).
A high-purity data sample of events containing a $\PZ$ boson candidate decaying to a pair of electrons or muons is selected. Any additional reconstructed leptons in this sample are \emph{assumed} to be misidentified, after the approximately~4\% contamination from genuine third leptons from $\PW\PZ$ and $\PZ\PZ$ production has been subtracted using MC simulation. Using the observed rates of third leptons passing the nominal or the reversed selection, $n_l$ and $n_r$, transfer factors $f$ are defined as

\vspace{-1\baselineskip}
\begin{equation}\label{eq:transfer_factor}
	f = \frac{n_l}{n_r}
\end{equation}
and measured in bins of $\pt$ and $\eta$ of the third leptons.
A background control sample of data events is then selected, satisfying all the $\PZ\PZ$ selection criteria described in \mysec{}~\ref{sec:selection}, \emph{except} that one or two leptons in the final selected quadruplet are required to only pass the reversed criteria and not the nominal criteria. The number of observed events with one lepton (two leptons) passing only the reversed criteria is denoted $N_{lllr}$ ($N_{llrr}$). The events originate predominantly from processes with two or three prompt leptons. Using MC simulation, the contamination of genuine $\PZ\PZ$ events is estimated to be approximately~36\% of $N_{lllr}$ and approximately~1\% of $N_{llrr}$. The number of background events with one or two misidentified leptons can be calculated as

\vspace{-1\baselineskip}
\begin{equation}\label{eq:ddbkg}
	N_{\text{misid.}} = \sum_{i}^{N_{lllr}} f_i - \sum_{i}^{N_{lllr}^{\PZ\PZ}} w_i f_i - \sum_{i}^{N_{llrr}} f_i f^{\prime}_i + \sum_{i}^{N_{llrr}^{\PZ\PZ}} w_i f_i f^{\prime}_i,
\end{equation}
where the superscript $\PZ\PZ$ indicates the MC-simulated contributing events from $\PZ\PZ$ production, $w_i$ indicates the simulated weight of the $i$th event, and $f_i$ and $f^{\prime}_i$ are the transfer factors depending on $\pt$ and $\eta$ of the leptons passing the reversed selection. 
In differential distributions, the yields in \myeq{}~\ref{eq:ddbkg} are considered separately in each bin.
Systematic uncertainties are applied to account for statistical fluctuations of the measured transfer factors, and for the simplification that the origins, rates and selection efficiencies of misidentified leptons are assumed equal in the sample where the transfer factors are determined and the background control sample. The latter uncertainties are derived using transfer factors obtained from simulation for the different background processes and taking the difference between the result and the nominal method as uncertainty. An additional uncertainty due to the modelling of the $\PZ\PZ$ contamination in the background control sample is estimated by varying $N_{lllr}^{\PZ\PZ}$ and $N_{llrr}^{\PZ\PZ}$ up and down by 50\%. The final total uncertainty is $100\%$ ($71\%$, $95\%$) in the \eeee{} (\eemm{}, \mmmm{}) channel. The misidentified-lepton background is $2.1\pm 2.1$, ($4.9\pm 3.9$, $5.3\pm 5.2$) in the \eeee{} (\eemm{}, \mmmm{}) channel and $12.3\pm 8.3$ in the combination of all three channels (also shown in \mytab~\ref{tab:yields}). The uncertainties of the different channels are partially correlated, so that the combined uncertainty is not simply the sum (linear or in quadrature) of the uncertainties in the individual channels.
It amounts to 58\% of the total background estimate. Background with three or more misidentified leptons is considered negligible and ignored.

Despite reversing selection criteria to enhance the size of the background control sample, the small number of control events pose a problem when determining the differential backgrounds as a function of some observable. To remedy this, the following approximation is made. The background shape is that of $N_{llrr} f_i f^{\prime}_i$ measured in data (i.e.~ignoring $N_{lllr}$ events). The same relative integrated uncertainty is applied in each bin as systematic uncertainty, while the statistical uncertainty in each bin $j$ is taken to be,
\begin{equation*}
\sqrt{\big[N_{llrr} f_i f^{\prime}_i\big]_{\text{bin}\,j}}\;.
\end{equation*}



\subsection{Independent cross-check of data-driven estimation}
\label{sec:zz_samesign_background}



To increase the confidence in the measured data-driven background, it is also measured using an independent method (by the author). This other method is based on selecting events in which one dilepton is formed of a same-flavour \emph{same}-charge pair, rather than a same-flavour opposite-charge pair. 
Assuming that there are as many misidentified leptons with negative as with positive charge and that their rates are independent, the number of events with 1--2 misidentified leptons passing the event selection (i.e.~the background) is equal to the number of events passing a modified selection, which is identical to the signal selection except that one opposite-sign dilepton is replaced by a same-sign dilepton. Using a notation where
\begin{itemize}
	\item $N$ denotes a number of events passing the signal selection,
	\item $C$ denotes a number of events passing the same-sign selection,
	\item $\text{g}$ denotes a genuine lepton,
	\item $\text{m}$ denotes a misidentified or non-prompt lepton,
	\item $\text{c}$ denotes a genuine lepton whose charge was reconstructed with the wrong sign ($\pm e$ as $\mp e$),
\end{itemize}
the background estimate can be written as
\begin{equation*}
\begin{split}
N_{\text{misid.}} &= N_{\text{g}^+\text{g}^-\text{g}^{\pm}\text{m}^{\mp}} + N_{\text{g}^+\text{g}^-\text{m}^{\pm}\text{m}^{\mp}}\\
&= C_{\text{g}^+\text{g}^-\text{g}^{\pm}\text{m}^{\pm}} + C_{\text{g}^+\text{g}^-\text{m}^{\pm}\text{m}^{\pm}}\\
&= C_{\ell^{+}\ell^{-}\ell^{\pm}\ell^{\pm}} - C^{\ZZ}_{\text{g}^+\text{g}^-\text{g}^{\pm}\text{c}^{\pm}},
\end{split}
\end{equation*}
where the terms after the last equality are both observable quantities: $C_{\ell^{+}\ell^{-}\ell^{\pm}\ell^{\pm}}$ is the observed number of same-sign events in data, and $C^{\ZZ}_{\text{g}^+\text{g}^-\text{g}^{\pm}\text{c}^{\pm}}$ is the contribution of genuine $\ZZ$ signal events that pass the same-sign selection due to charge mis-measurement, which can be estimated using MC simulation. With respect to the transfer-factor method described in \mysec~\ref{sec:dd_bkg}, the same-sign method has the advantages that it is simpler and requires no extrapolation between control regions. Its big disadvantage in the presented form is that it offers no decrease of the statistical uncertainty of the background estimate by using a relaxed control selection. Hence the statistical uncertainty of $n$ background events will be approximately an uncertainty of $\sqrt{n}$, which is a very large relative uncertainty in the case of a small background, such as in this analysis. The numbers of events passing the control selection as well as the misidentified-lepton background predicted by the same-sign method are shown in \mytab~\ref{tab:samesign_results}. The uncertainty of $C^{\ZZ}_{\text{g}^+\text{g}^-\text{g}^{\pm}\text{c}^{\pm}}$ is neglected here, since only a simple cross-check of the transfer-factor method is intended. The results are highly compatible with those predicted by the transfer-factor method, shown above as well as in \mytab~\ref{tab:yields}.

\begin{table}[h!]
\centering
\begin{tabular}{llll}
\toprule
\multirow{2}{*}{\textbf{Channel}} & \textbf{Data events} & \textbf{Signal leakage} & \textbf{Background prediction}\\
 & $C_{\ell^{+}\ell^{-}\ell^{\pm}\ell^{\pm}}$ & $C^{\ZZ}_{\text{g}^+\text{g}^-\text{g}^{\pm}\text{c}^{\pm}}$ & \textbf{(events)}\\
\midrule
\eeee{} & 7 & 8.2 & $0.0 \pm 2.6$ \\
\eemm{} & 14 & 11.1 & $2.9 \pm 3.7$\\
\mmmm{} & 7 & 2.4 & $4.6 \pm 2.6$\\
\midrule
Combined & 28 & 21.7 & $6.3 \pm 5.3$\\
\bottomrule
\end{tabular}
\caption{Measured yields in the same-sign control region and background prediction given by the same-sign method based on those numbers. The signal leakage is obtained using the nominal \SHERPA{} setup. The uncertainty estimate of the background prediction is the statistical uncertainty of the data.}
\label{tab:samesign_results}
\end{table}

\subsection{Single-\PZ{} pileup background}

In a high-pileup environment, background could arise from two (or more) single \PZ bosons being produced in independent proton--proton collisions in the same bunch crossing (``pileup-\ZZ{}''). 
The lepton impact parameter requirements will reject such background if the two primary vertices are well-separated. However, there is a non-zero probability that the vertices lie so nearby that they effectively overlap. In this case, pileup-\ZZ{} events could pass the event selection. In this section, an estimate of such background is obtained in two steps:
\begin{enumerate}
	\item Calculating the effective production cross section of pileup-$\PZ\PZ$ in the four-lepton channel,
	\item Estimating what fraction of pileup-$\PZ\PZ$ events pass the lepton vertex association criteria.
\end{enumerate}

To calculate the effective pileup-$\PZ\PZ$ cross section, the $\PZ{}\to \ell^+\ell^-$ production cross section is taken to be $1981 \pm 57$~pb ~\cite{STDM-2015-03}, and the total fiducial inelastic $\Pproton\Pproton$ collision cross section to be $\sigma_{\text{inel}} = 78.1 \pm 2.9$~mb \cite{STDM-2015-05}. Furthermore, the average number of inelastic interactions per bunch crossing during the considered data-taking period, rounded up to the nearest integer value, is taken to be $\langle \mu \rangle \approx 24$. The effective pileup-\ZZ{} production cross section $\sigma_{\text{eff.~pileup-}\PZ\PZ}$ is given by the cross section for single \PZ{} production times the probability of a second \PZ{} production occurring in the same bunch crossing.
For a single $\Pproton\Pproton$ collision, the probability of a $\PZ{}\to \ell^+\ell^-$ event occurring given that an inelastic collision occurred is
\begin{equation*}
P_{\PZ} \equiv P(\PZ \to \ell^+\ell^-\, |\, \text{inel}) = \frac{\sigma_{\PZ \to \ell^+\ell^-}}{\sigma_{\text{inel}}} \approx 2.5 \times 10^{-8}.
\end{equation*}
The probability of one or more \PZ{} events occurring in any one of the $\langle \mu \rangle - 1$ inelastic pileup collisions accompanying an event on average is
\begin{equation*}
1 - (1-P_{\PZ})^{\langle\mu\rangle - 1} \approx 5.8 \times 10^{-7},
\end{equation*}
so the effective pileup-\ZZ{} cross section is
\begin{equation*}
	\sigma_{\text{eff.~pileup-}\PZ\PZ} = \sigma_{\PZ \to \ell^+\ell^-} \left(1 - (1-P_{\PZ})^{\langle\mu\rangle - 1}\right) \approx 1.2~\text{fb},
\end{equation*}
or about 3\% of the predicted signal cross section. The probability of observing two or more \PZ{} events in pileup collisions (for a total of three or more) is
\begin{equation*}
\sum_{n = 2}^{\langle\mu\rangle - 1} \binom{\langle\mu\rangle - 1}{n} P_{\PZ}^n (1 - P_{\PZ})^{\langle\mu\rangle - 1 - n} \approx 1.6 \times 10^{-13},
\end{equation*}
where $\binom{\cdot}{\cdot}$ is a Binomial coefficient. This probability is negligibly small, so this contribution is ignored.
To estimate the fraction of pileup-\ZZ{} events surviving the lepton impact parameter requirements, the $z$-distribution of reconstructed primary vertices in data events is considered and the probability for two primary vertices to lie within a $z$-distance of 0.5~mm is calculated. The distance corresponds to the longitudinal impact parameter requirement for signal leptons (excluding standalone muons, but only events with at most one of such muon are accepted).
The $z$ distribution of primary vertices, shown in \myfig~\ref{fig:pv_z}, is approximated by a Gaussian with a fitted standard deviation of 37~mm.\footnote{The goodness of fit is very poor, $\chi^2/\text{\#(degrees of freedom)} \sim 10^3$, because the distribution corresponds to a sum of Gaussians with different means and standard deviations, corresponding to the conditions during different runs. However, while statistically significant, the deviations from a Gaussian are very small compared to the impact parameter requirement, and can therefore be safely ignored.} Random numbers distributed according to this Gaussian are generated in (independent) pairs and the vertex-overlap probability is calculated as the fraction of trials in which the vertices in the pair lie within 0.5~mm of each other.
Generating ten million random-number pairs yields a vertex overlap probability of approximately 0.7\%.
Assuming that the detector acceptance and reconstruction efficiency for pileup-$\PZ\PZ$ events with overlapping vertices is equal to that of signal events, the expected relative background contribution due to pileup-$\PZ\PZ$ events is
\begin{equation}
	\frac{1.2~\text{fb} \times 0.7\%}{42.6~\text{fb}} \approx 0.02\%,
\end{equation}
where 42.6~fb is the predicted $\ZZllll$ production cross section (details can be found in \mytab~\ref{tab:integrated_cross_sections}). It can be concluded that the pileup-\ZZ{} background is negligible. Conservatively considering vertices as overlapping if they are within $\Delta z = 1$~mm yields a relative background estimate of $0.04\%$, which is equally negligible.

\begin{figure}[h!]
\centering
\begin{tikzpicture}
\node[anchor=south west,inner sep=0] (image) at (0,0) {\includegraphics[width=0.8\textwidth]{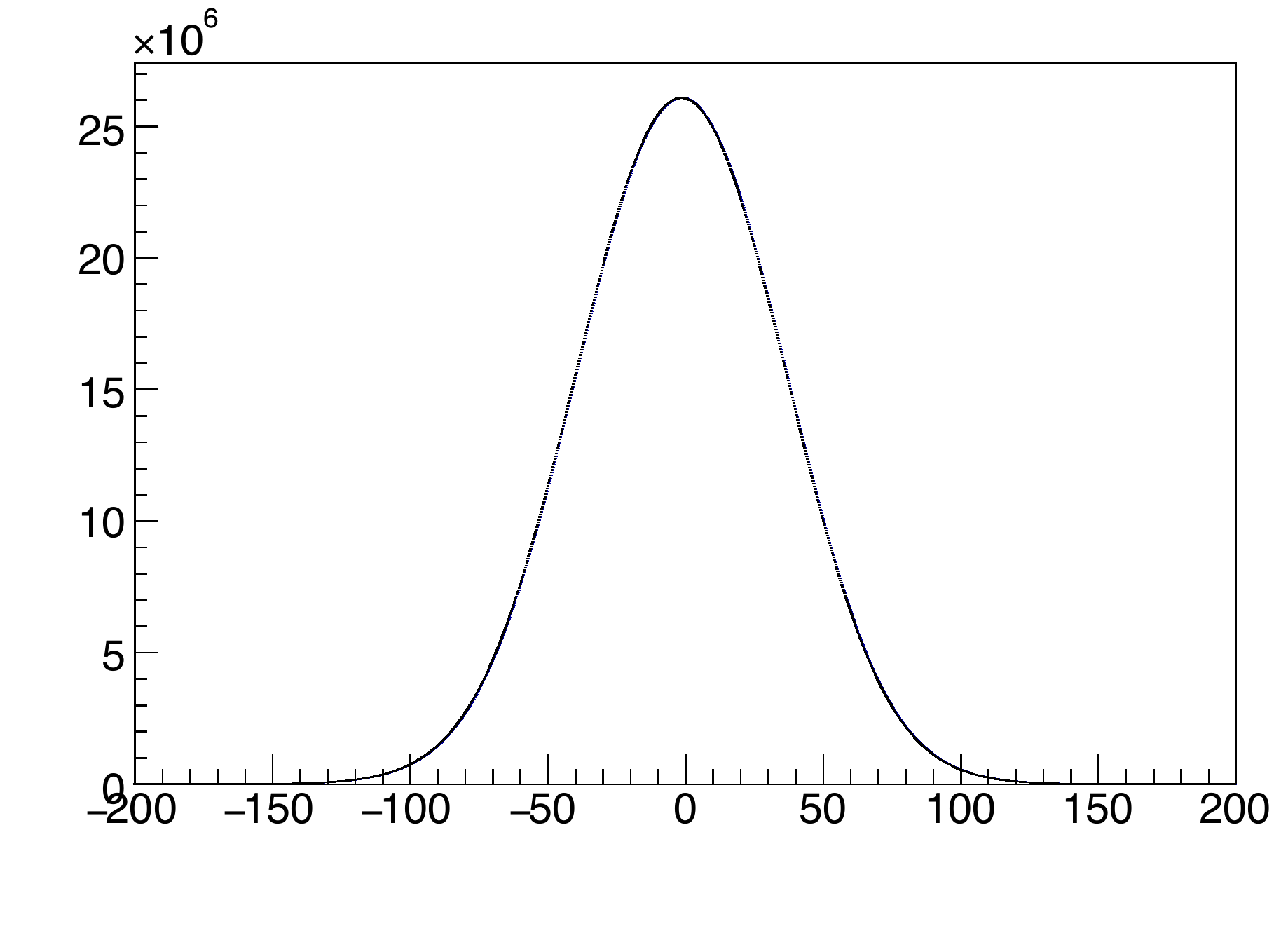}};
\begin{scope}[x={(image.south east)},y={(image.north west)}]
\node[anchor=east] at (0.97, 0.05) {Primary vertex $z$ \quad (mm)};
\node[anchor=east, rotate=90] at (0.02, 0.97) {Number of vertices};
\end{scope}
\end{tikzpicture}
\caption{Position of primary vertices (including hard-scattering and pileup vertices) along the beam axis in the 2015 and 2016 ATLAS data. Vertices are shown for events passing the trigger requirements of this analysis and a loose pre-selection of at least three reconstructed leptons. However, these selections are not expected to bias the distribution significantly. As only the relative distribution is of interest, the $y$-axis units can be considered arbitrary.}
\label{fig:pv_z}
\end{figure}


\clearpage
\section{Observed and predicted yields}

The observed and predicted event yields for signal and background are shown in \mytab{}~\ref{tab:yields}. The prediction uncertainties are discussed in \mysec{}~\ref{sec:uncertainties}. \myfig{}~\ref{fig:data_mc_plots} shows the distributions of data and predictions for the mass and transverse momentum of the four-lepton system, the transverse momentum of the leading \PZ{} boson candidate, and the jet multiplicity. Similar distributions for further observables can be found in \myapp~\ref{sec:zz_aux_datamc}. The agreement between data and the nominal \SHERPA{} prediction is good. The prediction using \POWHEGpy{} to simulate the $\Pquark\APquark$-initiated process tends to underpredict the normalisation slightly, which can be understood from its lack of (partial) higher-order corrections that \SHERPA{} implements. \POWHEGpy{} also provides a worse description of high jet multiplicities, as it only describes one parton emission at matrix-element level.

\begin{table}[h!]
\centering
{\small
\begin{tabular}{lllll}
\toprule
\textbf{Contribution} & \textbf{4e} & \textbf{2e2\Pmu{}} & \textbf{4\Pmu{}} & \textbf{Combined}\\
\midrule
Data & \input{data.tex}
\midrule
Total prediction (\SHERPA{}) & \input{sherpaprediction}
\midrule
Signal ($\Pquark\APquark$-initiated) & \input{sherpaqqbar}
Signal ($\Pgluon\Pgluon$-initiated) & \input{gg}
Signal (EW-$\ZZ jj$) & \input{zzjj}
$\PZ\PZ\to \tau^{+}\tau^{-}[\ell^{+}\ell^{-}, \tau^{+}\tau^{-}]$ & \input{taubkg}
Triboson & \input{tribosonbkg}
$\Ptop\APtop\PZ$ & \input{ttzbkg}
Misid.~lepton background & \input{fakebkg}
\midrule
Total prediction (\POWHEG{} + & \input{powhegprediction_withKfac}
\PYTHIA{} with higher-order & \\
corrections, \SHERPA{}) & \\
\bottomrule
\end{tabular}
}
\caption{Observed and predicted yields, using the nominal \SHERPA{} setup for the signal predictions. All statistical and systematic uncertainties are included in the prediction uncertainties. An alternative total prediction using \POWHEGpy{} with NNLO QCD and NLO weak corrections applied to simulate the $\Pquark\APquark$-initiated process is shown at the bottom.}
\label{tab:yields}
\end{table}

\begin{figure}[p]
\centering
\subfigure[]{\includegraphics[width=0.48\textwidth]{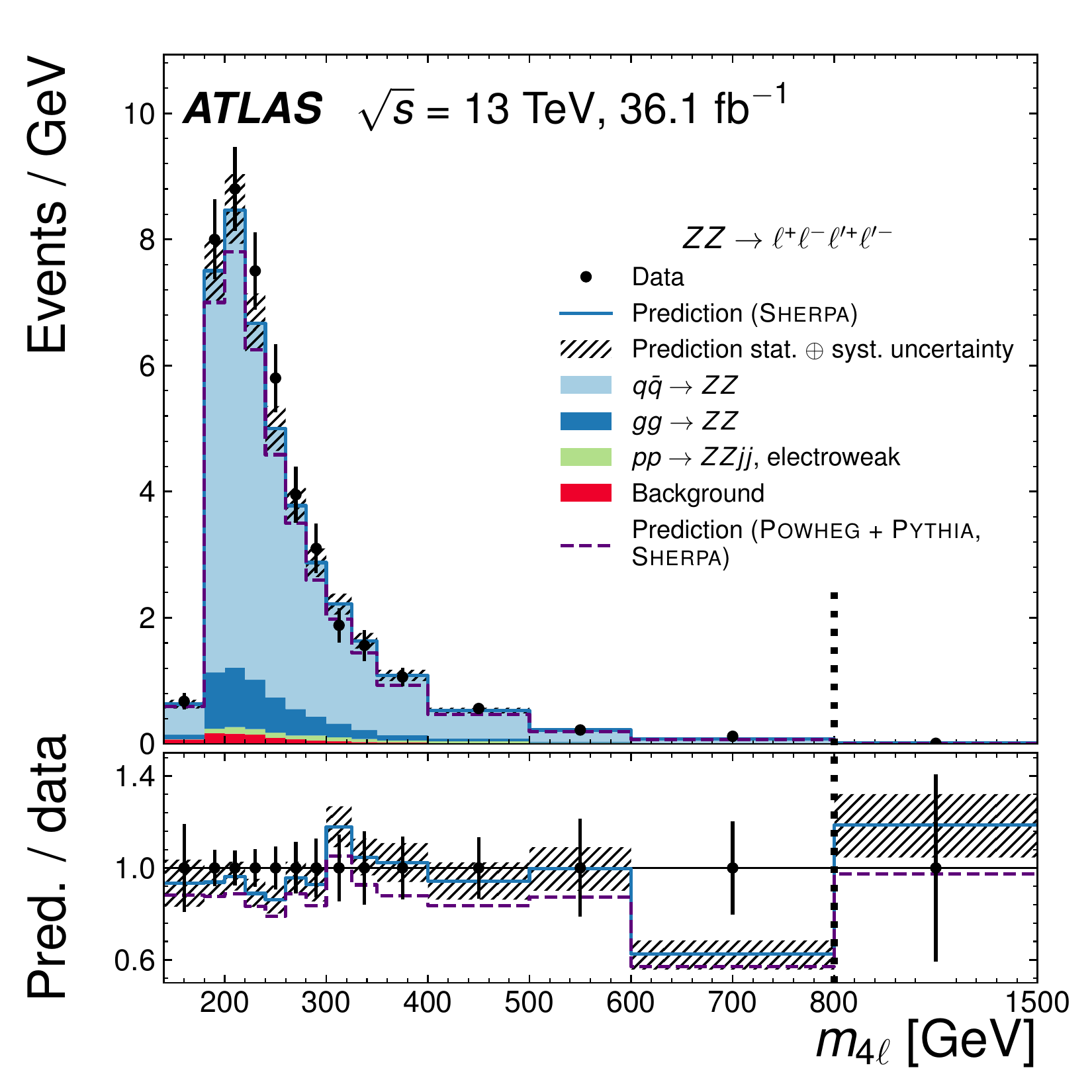}}
\hspace{1mm}
\subfigure[]{\includegraphics[width=0.48\textwidth]{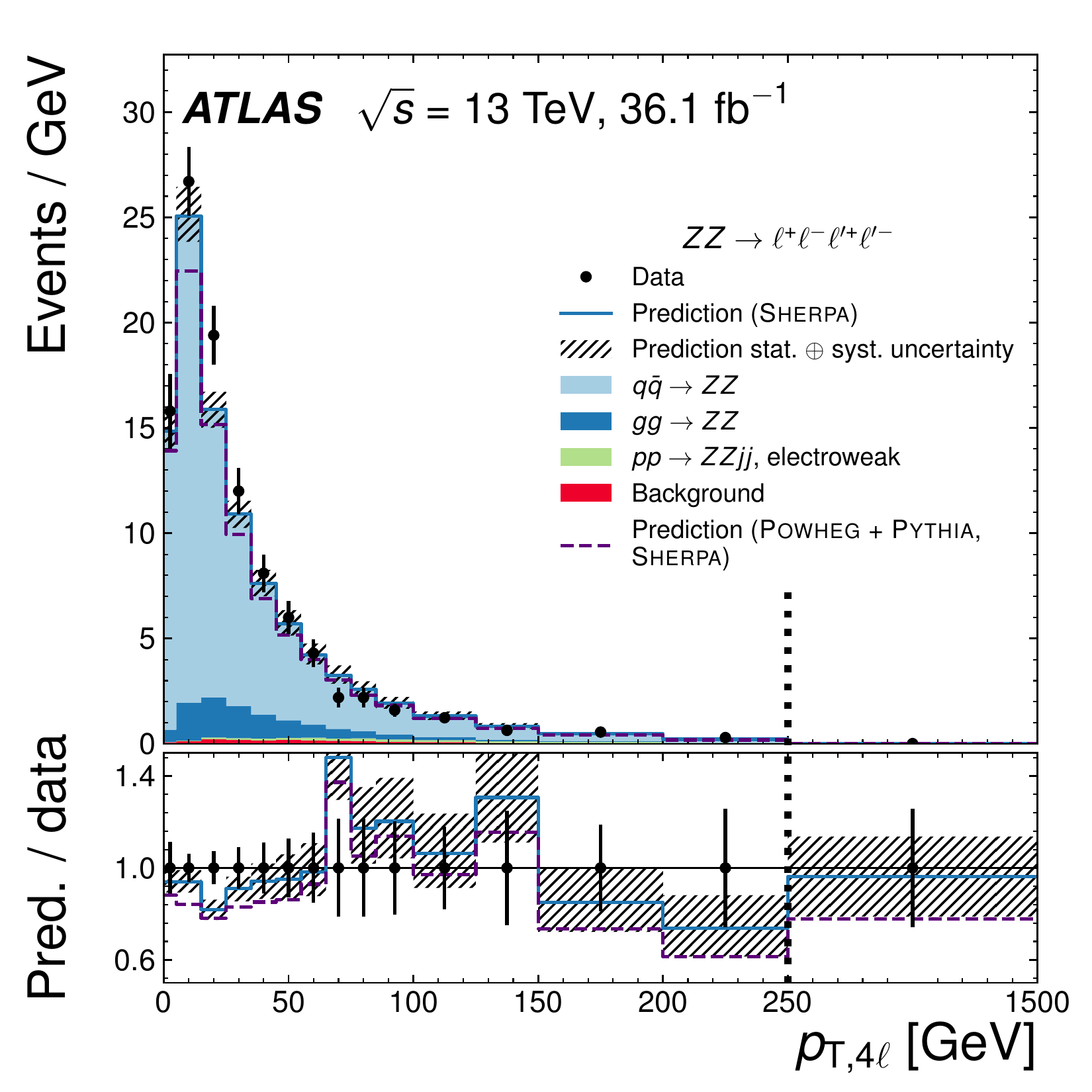}}
\subfigure[]{\includegraphics[width=0.48\textwidth]{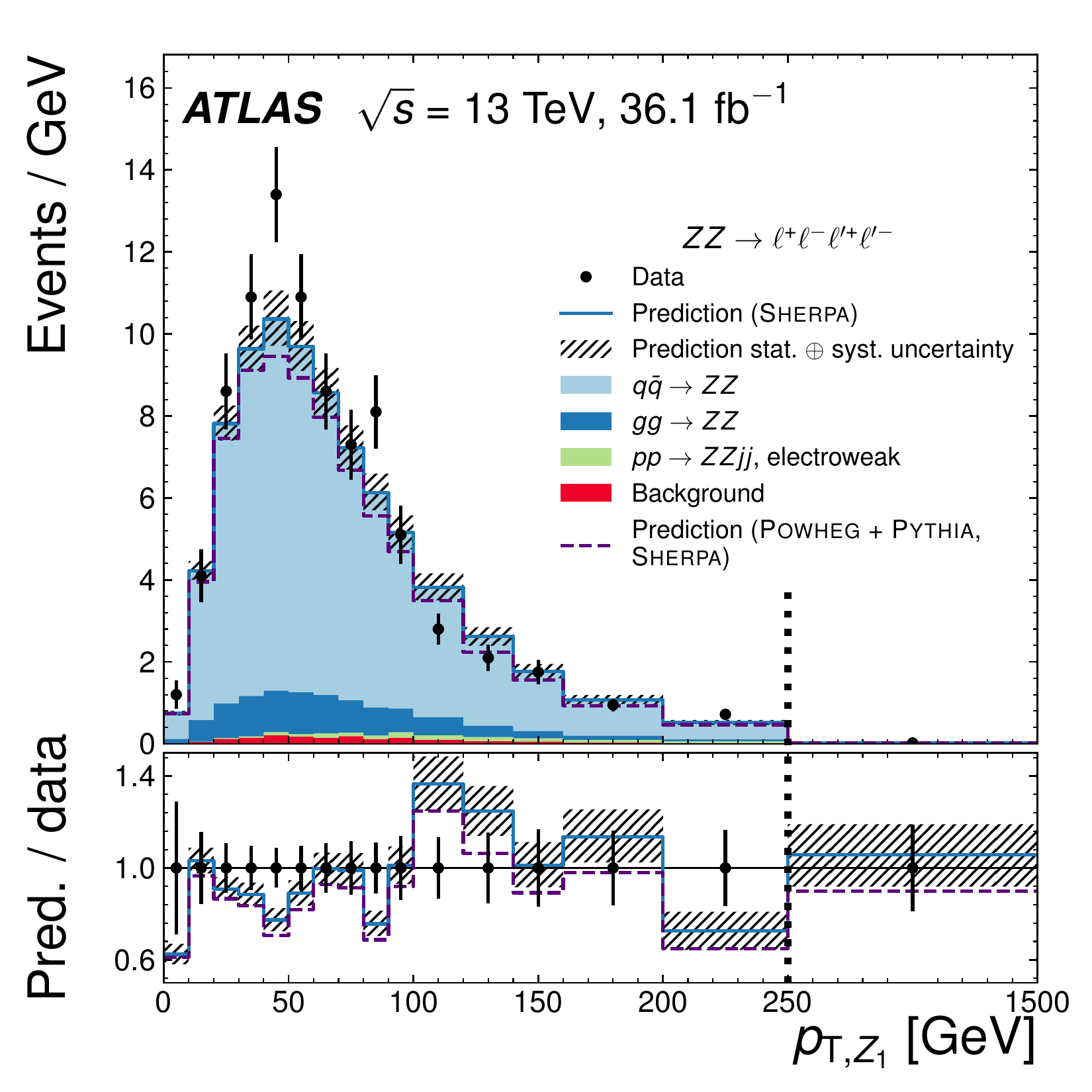}}
\hspace{1mm}
\subfigure[]{\includegraphics[width=0.48\textwidth]{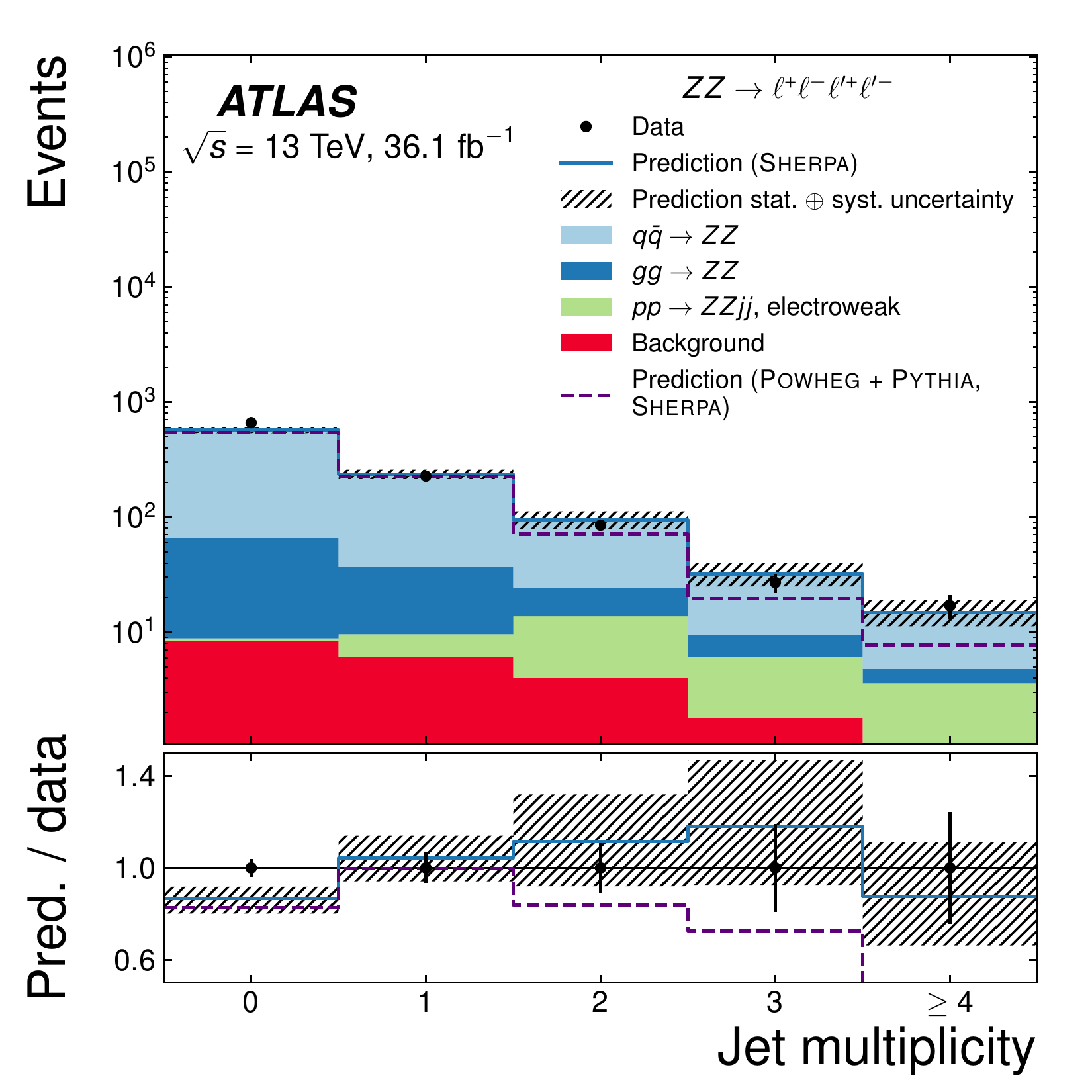}}
\caption{Measured distributions of the selected data events along with predictions in bins of (a) the four-lepton mass, (b) the four-lepton transverse momentum, (c) the transverse momentum of the leading \PZ{} boson candidate, and (d) the jet multiplicity. The main prediction uses the nominal \SHERPA{} setup. The prediction uncertainty includes the statistical and systematic components, all summed in quadrature. Different signal contributions and the background are shown, as is an alternative prediction that uses \POWHEGpy{} to generate the $\Pquark\APquark$-initiated subprocess. In (a), (b), and (c), the last bin is shown using a different $x$-axis scale for better visualisation. The scale change is indicated by the dashed vertical line. Published in \myref~\cite{STDM-2016-15}.}
\label{fig:data_mc_plots}
\end{figure}

A slight excess of events is observed in the \eeee{} channel. Its statistical significance in the integrated yield is $2.3\sigma$ with respect to the \SHERPA{} and $2.9\sigma$ with respect to the \POWHEGpy{} + \SHERPA{} prediction. It is not clearly localised in the four-lepton mass or any other control observable, although it has large contributions in the approximate range 10--20~\GeV{} of the transverse momentum of the four-lepton system. \myfig~\ref{fig:pt4l_channelwise} demonstrates this by showing the $\ptfourl$ distribution in each channel. Many studies were performed to validate the electron reconstruction, identification, and selection, as well as the event selection in the \eeee{} channel. No unexpected behaviour or hints at a problem were found that could have explained the slight excess. The conclusion is therefore that it is simply caused by a statistical fluctuation.

\begin{figure}[h!]
\centering
\subfigure[]{
\begin{tikzpicture}
\node[anchor=south west,inner sep=0] (image) at (0,0) {\includegraphics[width=0.48\textwidth]{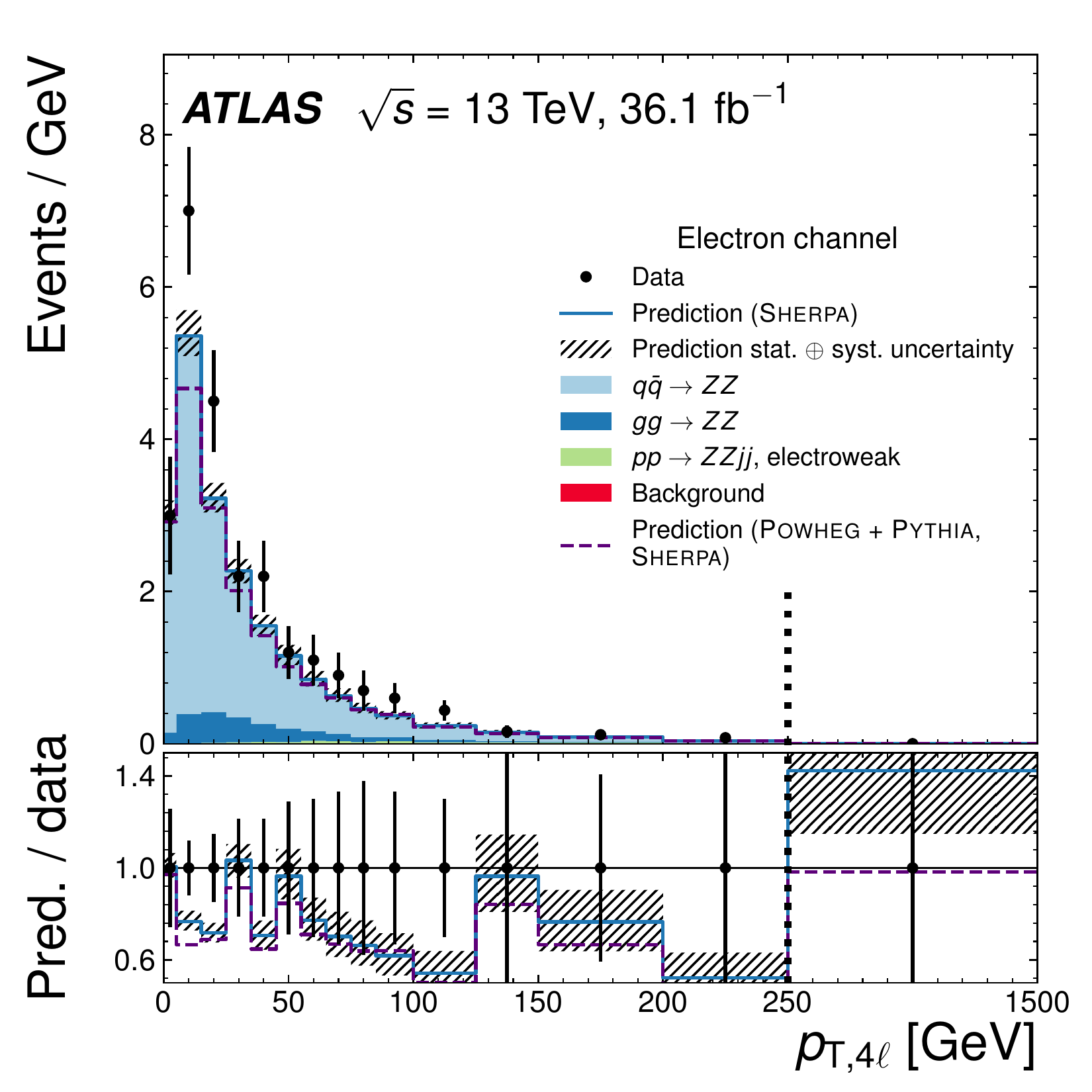}};
\begin{scope}[x={(image.south east)},y={(image.north west)}]
\draw[white, fill=white] (0.16,0.87) rectangle (0.3,0.93);
\draw[white, fill=white] (0.55,0.76) rectangle (0.85,0.82);
\node[anchor=west] at (0.59, 0.79) {\footnotesize \eeee{} channel};
\draw[white, fill=white] (0.5,0.53) rectangle (0.85,0.557);
\end{scope}
\end{tikzpicture}
}
\subfigure[]{
\begin{tikzpicture}
\node[anchor=south west,inner sep=0] (image) at (0,0) {\includegraphics[width=0.48\textwidth]{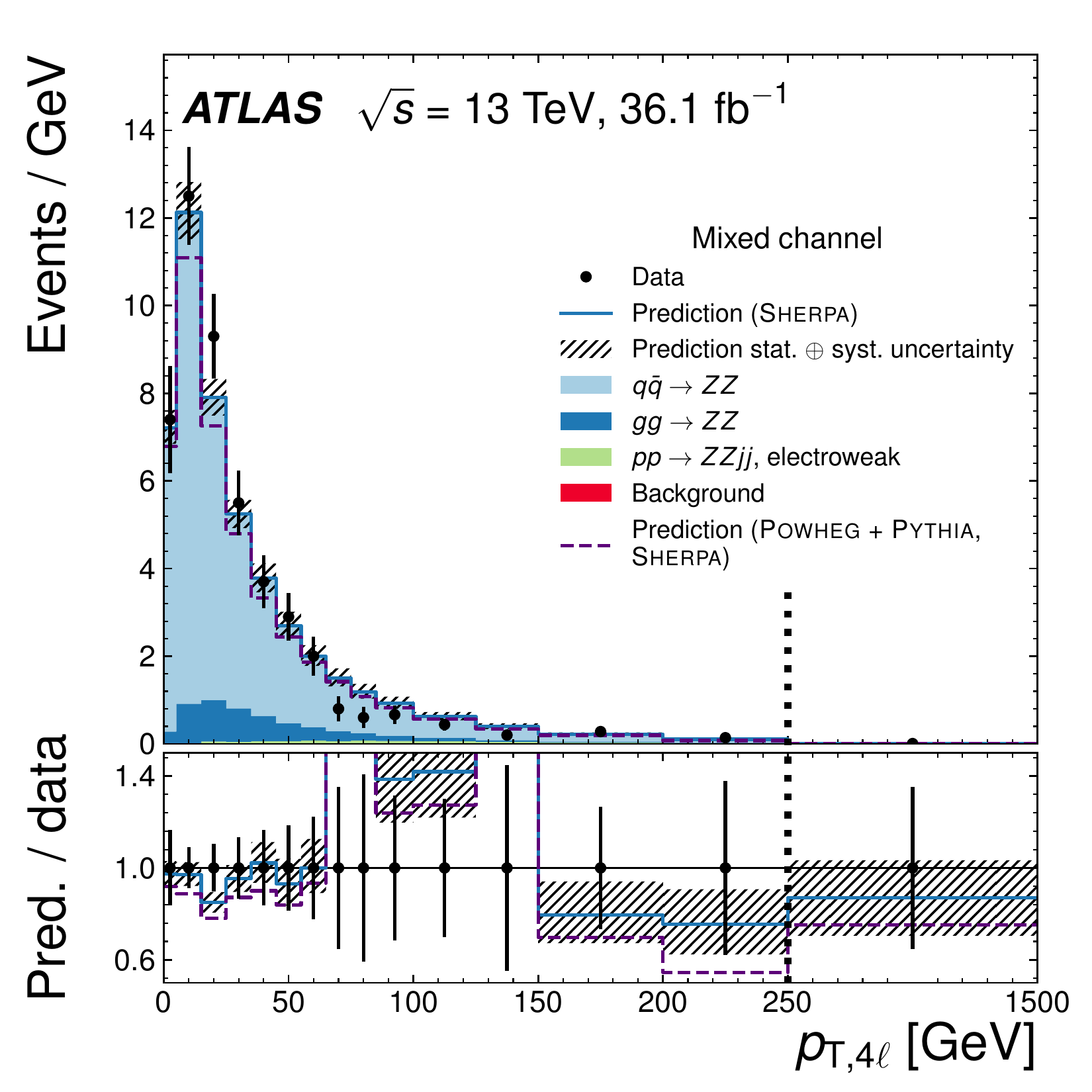}};
\begin{scope}[x={(image.south east)},y={(image.north west)}]
\draw[white, fill=white] (0.16,0.87) rectangle (0.3,0.93);
\draw[white, fill=white] (0.55,0.76) rectangle (0.85,0.82);
\node[anchor=west] at (0.59, 0.79) {\footnotesize \eemm{} channel};
\draw[white, fill=white] (0.5,0.53) rectangle (0.85,0.557);
\end{scope}
\end{tikzpicture}
}
\subfigure[]{
\begin{tikzpicture}
\node[anchor=south west,inner sep=0] (image) at (0,0) {\includegraphics[width=0.48\textwidth]{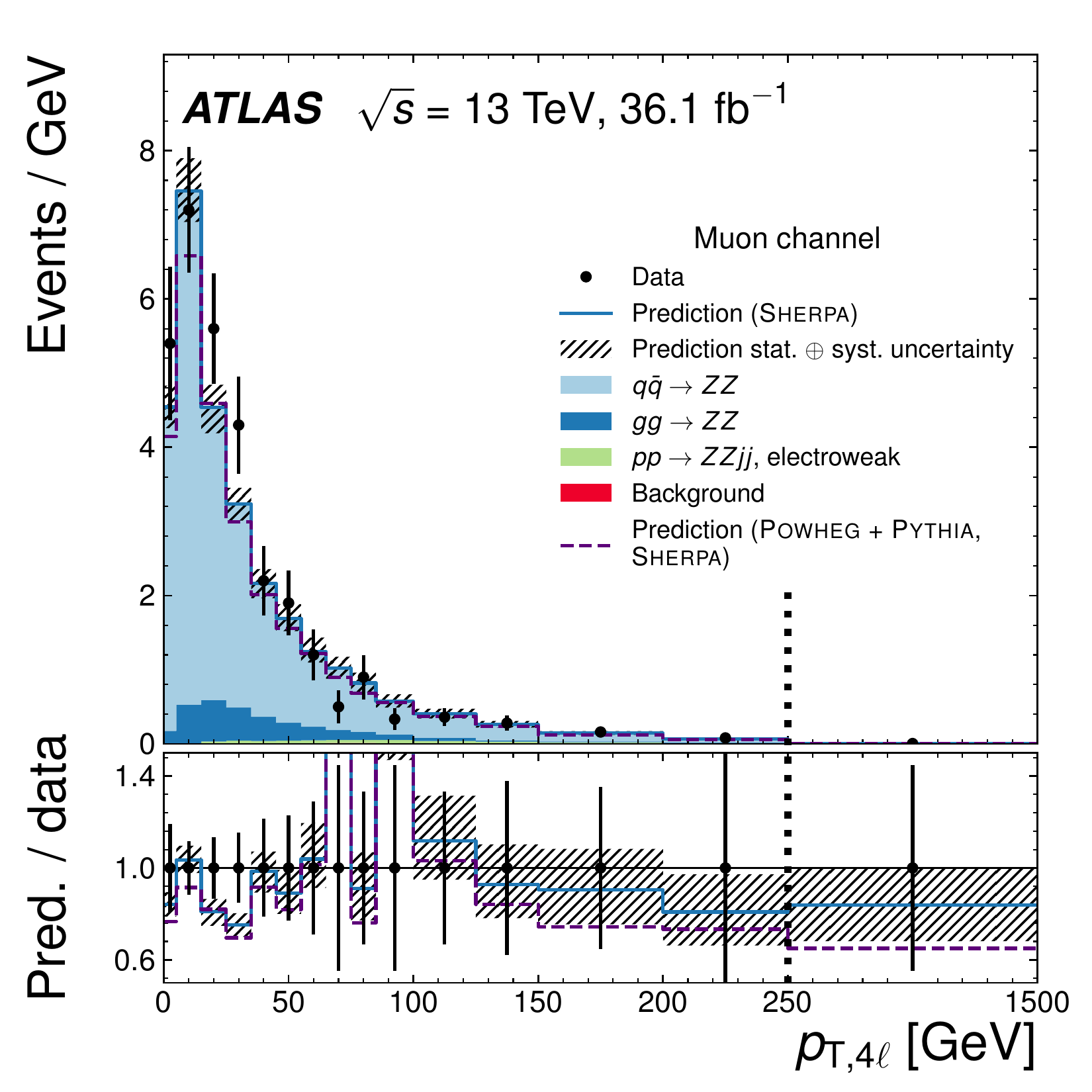}};
\begin{scope}[x={(image.south east)},y={(image.north west)}]
\draw[white, fill=white] (0.16,0.87) rectangle (0.3,0.93);
\draw[white, fill=white] (0.55,0.76) rectangle (0.85,0.82);
\node[anchor=west] at (0.59, 0.79) {\footnotesize \mmmm{} channel};
\draw[white, fill=white] (0.5,0.53) rectangle (0.85,0.557);
\end{scope}
\end{tikzpicture}
}
\caption{Measured distributions of the selected data events along with predictions in bins of the four-lepton transverse momentum, shown separately for the (a) \eeee{}, (b) \eemm{}, and (c) \mmmm{} channel. The background is not shown, because the reducible background was not measured differentially in separate channels, but is essentially negligible (as shown e.g.~in \mytab~\ref{tab:yields} and \myfig~\ref{fig:data_mc_plots}). The last bin is shown using a different $x$-axis scale for better visualisation. The scale change is indicated by the dashed vertical line.}
\label{fig:pt4l_channelwise}
\end{figure}
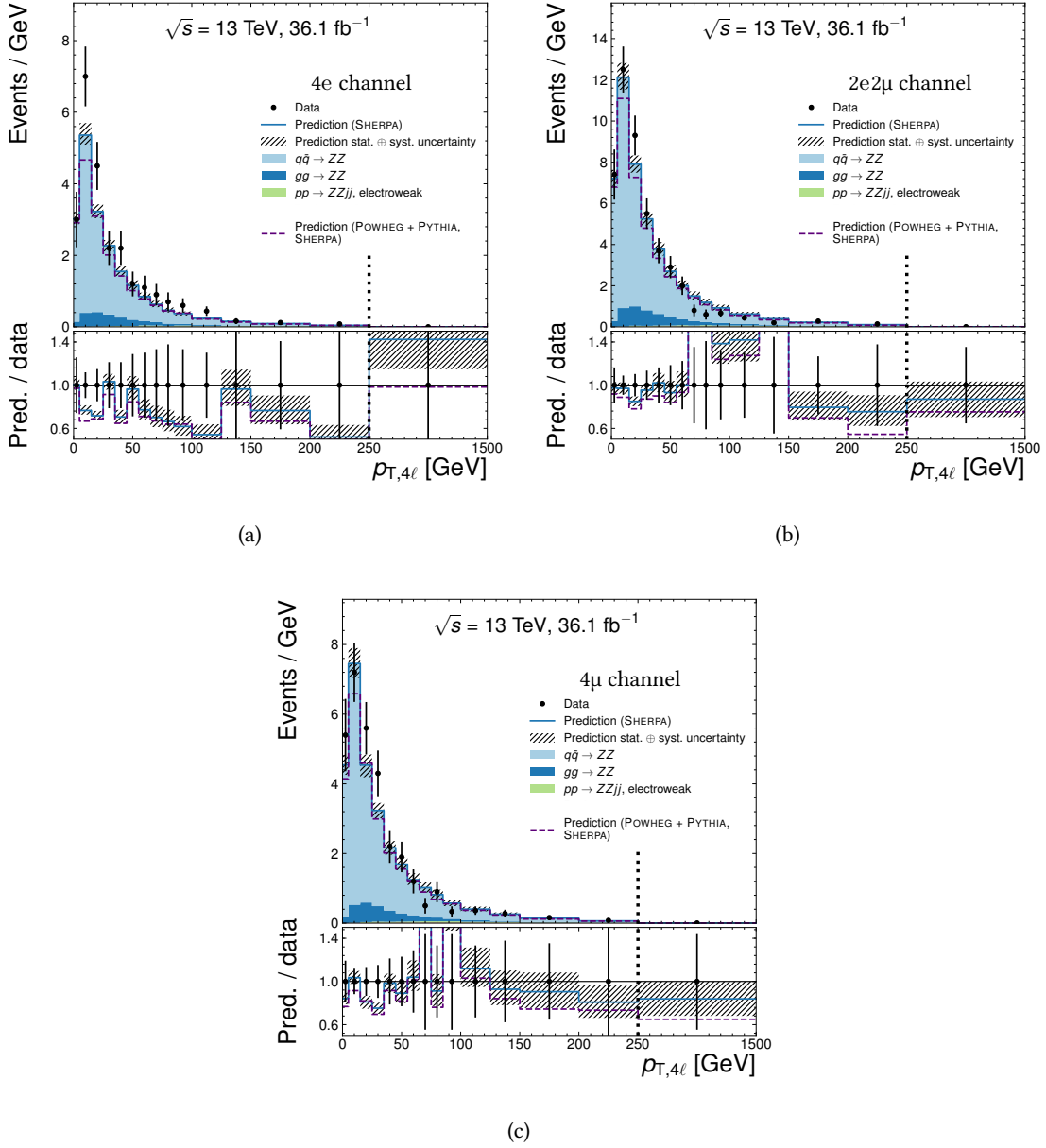

%
%
%
%

\clearpage
\section{Systematic uncertainties}
\label{sec:uncertainties}
The sources of systematic uncertainty are introduced below. Their effects on the predicted integrated signal yields after event selection are shown in \mytab{}~\ref{tab:yield_uncertainties}. 

For leptons and jets, uncertainties of the momentum or energy scale and resolution are considered. Uncertainties of the lepton reconstruction and identification efficiencies as well as the efficiency of the jet vertex tagging requirements in the simulation are taken into account. All of the above depend on the kinematics of the lepton or jet. The electron efficiency uncertainties contain contributions associated with the basic reconstruction, the identification, and the isolation. Each is split into $\mathcal{O}(10)$ components that are uncorrelated between individual electrons. For muons, the efficiency uncertainties associated with individual muons are treated as fully correlated, leading to a larger uncertainty compared to electrons. The uncertainties associated with the efficiencies of the muon reconstruction and the track-to-vertex association both amount to approximately 1\% per muon, and those associated with the isolation efficiency to approximately 0.2\% per muon. As the selection is fully jet-inclusive, jet uncertainties do not affect the integrated yields and are therefore not shown in \mytab{}~\ref{tab:yield_uncertainties}, but will affect those differential cross sections that depend explicitly on jets.

The pileup modelling uncertainty is assessed by performing variations in the number of simulated pileup interactions designed to cover the uncertainty of the ratio between the predicted and measured cross section of non-diffractive inelastic events producing a hadronic system of mass $m_X > 13$~\GeV{} \cite{Aaboud:2016mmw}. 

The uncertainty of the integrated luminosity is $3.2\%$. It is derived from a preliminary calibration of the luminosity scale using a pair of $x$--$y$ beam-separation scans performed in August 2015 and May 2016, following a methodology similar to that detailed in \myref{}~\cite{Aad:2013ucp}.

PDF uncertainties of predicted cross sections are evaluated considering the uncertainty of the used set, as well as by comparing to two other reference sets. This is similar to the method proposed in \myref~\cite{Botje:2011sn}, with the exception that the set-internal variations are only performed for the nominal set, and only the nominal PDFs of the two reference sets are taken into account. The reference sets are MMHT 2014 \cite{Harland-Lang:2014zoa} and NNPDF 3.0 (CT10), if CT10 (NNPDF 3.0) is the nominal set. The envelope of the nominal set's uncertainty band and the deviation of the reference sets from the nominal set is used as the uncertainty estimate.
The theoretical uncertainties due to PDFs and QCD scales along with the luminosity uncertainty dominate the total uncertainty of the integrated yields, as shown in \mytab{}~\ref{tab:yield_uncertainties}. However, they only cause a small uncertainty of the actual measurement, since the detector corrections essentially depend on ratios of MC yields before and after detector simulation. The PDF and QCD scale uncertainties are highly correlated among the numerator and denominator, so their effect cancels mostly. This will be shown in \mysec~\ref{sec:zz_integrated_xs}.


A predicted theoretical modelling uncertainty is applied in some contexts by using \POWHEGpy{} instead of \SHERPA{} to generate the $\Pquark\APquark$-initiated subprocess, and taking the absolute deviation of the result obtained with this setup from the one obtained with the nominal \SHERPA{} setup as an uncertainty, symmetrising it with respect to the nominal value. This contribution is not shown in \mytab{}~\ref{tab:yield_uncertainties}. In the longer term, ``event generator'' uncertainties constructed like this should be phased out. A more careful approach would begin by studying \emph{why} different generators disagree. For instance, the amount of additional QCD radiation generated might affect the predicted efficiency of isolation requirements. It can then be studied whether the differences are already covered by other uncertainties, e.g.~the isolation efficiency scale factor uncertainty.
In addition or alternatively, one could use control distributions of the data to find which generator provides the best description and whether generator differences cover any disagreement. This could inform the construction of an uncertainty band.

A further source of uncertainty are statistical fluctuations in the used MC samples.


The uncertainty of the misidentified-lepton background is described in \mysec{}~\ref{sec:bkg}. A 30\% normalisation uncertainty is applied for triboson and $\Ptop\APtop\PZ$ backgrounds with four genuine leptons to account for the cross section uncertainty. The value of 30\% is an order of magnitude estimate for the size of missing QCD and EW higher-order corrections (the processes are predicted at LO).

The propagation of uncertainties in the unfolding as well as the estimation of unfolding-specific uncertainties is described in \mysec{}~\ref{sec:unfolding}.

\begin{table}[h!]
\centering
\begin{tabular}{lr}
\toprule
\textbf{Source} &   \textbf{Effect on total predicted yield (\%)}\\
\midrule
MC statistical uncertainty    & $0.4$\\ 
Electron efficiency            & $0.9$\\
Electron energy scale \& resolution   & $0.0$\\
Muon efficiency                & $1.7$\\
Muon momentum scale \& resolution       & $0.0$\\
Pileup modelling           & $1.2$\\
Luminosity				& $3.2$\\
QCD scales           & $^{+5.2}_{-4.7}$\\
PDFs                      & $^{+2.7}_{-1.7}$\\
Background prediction & $0.9$\\
\midrule
Total                          & $^{+7.4}_{-6.6}$\\
\bottomrule
\end{tabular}
\caption{Relative uncertainties in percent of the predicted integrated signal yields after event selection, derived using the nominal \SHERPA{} setup. All uncertainties are rounded to one decimal place.} 
\label{tab:yield_uncertainties}
\end{table}

\clearpage
\section{Integrated cross sections}\label{sec:zz_integrated_xs}
 

The integrated fiducial cross section \sigmafid{} is determined by a maximum-likelihood fit in each channel separately as well as for all channels combined. The fit itself was performed by Jonatan Rost\'{e}n. The full methodology is explained in \myref~\cite{thesis_jonatan}. The expected yield in each channel $i$ is given by

\vspace{-1\baselineskip}
\begin{equation*}
	N_{\text{exp}}^i = \intL \czz^i \sigmafid^i + N_{\text{bkg}}^i
\end{equation*}
where $\intL$ is the integrated luminosity, and $N_{\text{bkg}}$ is the expected background yield. The factor \czz{} is applied to correct for detector inefficiencies and resolution effects. It relates the background-subtracted number of selected events to the number in the fiducial phase space. \czz{} is defined as the ratio of generated signal events passing the selection criteria using reconstructed objects to the
number passing the fiducial criteria using the particle-level objects defined in \mysec{}~\ref{sec:fiducial}. It is determined with the nominal \sherpa{} setup. 
The \czz{} value and its total uncertainty is determined to be $0.494 \pm 0.015$ ($0.604 \pm 0.017$, $0.710 \pm 0.027$) in the \eeee{} (\eemm{}, \mmmm{}) channel. 
The higher \czz{} values in the \mmmm{} and \eemm{} channels reflect the higher reconstruction efficiency of muons with respect to electrons. Muons have a cleaner detector signature and the entire muon spectrometer dedicated only to their reconstruction, whereas electrons need to be reconstructed and distinguished from a large background of hadronic jets based on the shape of their calorimeter shower and properties of their track. Electrons can also undergo significant Bremsstrahlung losses in the inner detector, which may further complicate their reconstruction.
The dominant \czz{} uncertainties come from the uncertainties of the lepton reconstruction and identification efficiencies in the simulation, the choice of MC event generator, QCD scales and PDFs, and the modelling of pileup effects. Other smaller uncertainties come from the scale and resolution of the lepton momenta as well as statistical fluctuations in the MC sample. Table~\ref{tab:czz_uncert_overview} gives a breakdown of the systematic uncertainties. The statistical uncertainty of \czz{} needs to be calculated taking the correlations between numerator and denominator into account. The contributing events are classified into three disjoint categories: events passing only the reconstruction-level selection (labelled $r$), events passing only the fiducial selection (\kern0.5pt$f$), and events passing both of the above ($r\kern-1ptf$). After this, standard Gaussian error propagation can be used to compute the statistical uncertainty,
\begin{equation*}
\delta \czz = \frac{\sqrt{ {\delta w^2_{{r}}} \left(  w_{{f}} + w_{{rf}} \right)^2 +
          {\delta w^2_{{f}}} \left(  w_{{r}} + w_{{rf}} \right)^2
      +  {\delta w^2_{{rf}}} \left(  w_{{f}} - w_{{r}} \right)^2}}{{\left(w_{{f}} + w_{{rf}} \right)}^2},
\end{equation*}
where the quantities $w_i$ are the sums of weights of events in category $i$ and the quantities $\delta w$ are their corresponding statistical uncertainties. 
The value of \czz{} for different MC samples is shown in \mytab~\ref{tab:czz_by_prodchannel}. Differences between different production channels (\qq{}-initiated, \gluglu-initiated, EW $\ZZ jj$) as well as \SHERPA{} versus \POWHEGpy{} are observed. These are mainly due to the different lepton kinematics, leading to a different reconstruction efficiency, in particular for leptons with $2.5 < |\eta| < 2.7$, where the efficiency is reduced for muons and zero for electrons. The lepton isolation efficiencies are also slightly different between samples and generators due to the different predictions of associated QCD radiation.
%

\begin{table}[h!]
\centering
\begin{tabular}{lrrr}
\toprule
\textbf{Source} &   \textbf{4e} & \textbf{2e2\textmu{}} & \textbf{4\textmu{}} \\
\midrule
MC statistical uncertainty    & $0.4$ & 0.2 & 0.1 \\
Electron efficiency            & $2.0$ & 1.0 & $0.0$ \\
Electron energy scale \& resolution   & $0.1$ & $0.0$ & $0.0$ \\
Muon efficiency                & $0.0$ & 1.6 & 3.2 \\
Muon momentum scale \& resolution       & $0.0$ & $0.0$ & $0.1$ \\
Pileup modelling           & $1.3$ & 0.8 & 2.0 \\
QCD scales \& PDFs           & $^{+0.4}_{-0.8}$ & $^{+0.3}_{-0.4}$ & $^{+0.3}_{-0.6}$ \\
Event generator                      & $1.8$ & 1.8 & $0.2$ \\
\midrule
Total                          & $3.1$ & $2.8$ & $3.8$ \\
\bottomrule
\end{tabular}
\caption{Relative uncertainties of the correction factor \czz{} by channel, given in percent. All uncertainties are rounded to one decimal place.}
\label{tab:czz_uncert_overview}
\end{table}

\begin{table}[h!]
	\centering
	{\footnotesize
	\begin{tabular}{lllll}
		\toprule
		\textbf{Sample} & \textbf{4e} & \textbf{2e2\textmu} & \textbf{4\textmu} & \textbf{Combined}\\
		\midrule
		\POWHEGpy{} $\Pquark\APquark$-initiated & \input{CZZ_powhegqqbar}
		\SHERPA{} $\Pquark\APquark$-initiated & \input{CZZ_sherpaqqbar}
		\SHERPA{} $\Pgluon\Pgluon$-initiated & \input{CZZ_sherpagg}
		\SHERPA{} EW $\PZ\PZ jj$ production & \input{CZZ_sherpazzjj}
		\midrule
		Nominal \SHERPA{} setup & \input{CZZ_line}
		\bottomrule
	\end{tabular}
	}
	\caption{\czz{} with its statistical uncertainty by sample and channel, and the final values used in the analysis on the last row. The reason why only the statistical uncertainty is shown here is that it allows for comparisons between the different production modes without having correlated uncertainties between them.}
	\label{tab:czz_by_prodchannel}
\end{table}

The likelihood function to be minimised in the cross section fit is defined as

\vspace{-1\baselineskip}
\begin{equation}\label{eq:likelihood}
	\mathcal{L} = \mathcal{L}_{\text{stat}}\; \mathcal{L}_{\text{corr}}\; \mathcal{L}_{\text{uncorr}},
\end{equation}
where

\vspace{-1\baselineskip}
\begin{equation*}
	\mathcal{L}_{\text{stat}} = \text{Poisson}(N_{\text{obs}} | N_{\text{exp}})
\end{equation*}
is the probability of observing $N_{\text{obs}}$ events given that the yield follows a Poisson distribution with mean $N_{\text{exp}}$, and $\mathcal{L}_{\text{corr}}$ and $\mathcal{L}_{\text{uncorr}}$ are products of Gaussian nuisance parameters corresponding to the uncertainties of $\intL$, \czz{}, and $N_{\text{bkg}}$. $\mathcal{L}_{\text{corr}}$ contains the nuisance parameters that are fully correlated between channels, i.e.~all except the statistical uncertainties, while $\mathcal{L}_{\text{uncorr}}$ contains those that are uncorrelated, i.e.~the statistical uncertainties of \czz{} and $N_{\text{bkg}}$ in each channel. Nuisance parameters corresponding to different sources of systematic uncertainty are considered uncorrelated.
In the combined cross section fit, the product over channels is taken in the likelihood function shown in \myeq{}~\ref{eq:likelihood}, fixing the relative contributions of the signal channels to their theoretically predicted values.

\subsection{Results}
\label{sec:integrated_xs_results}

\mytab{}~\ref{tab:integrated_cross_sections} shows the integrated fiducial cross sections for each channel as well as all channels combined, along with a theoretical prediction. Measurements and predictions agree within one standard deviation, except for the \eeee{} channel, where the agreement is within approximately 1.4 standard deviations. The sum of the \eeee{} and \mmmm{} cross sections is not equal to the \eemm{} cross section. This is because of interference in the \eeee{} and \mmmm{} channels (visible in \mytab~\ref{tab:sherpa_channel_reweighting} in \myapp~\ref{sec:sherpa_fixes}) and the bias caused by the pairing prescription in the fiducial definition. \myfig{}~\ref{fig:fiducial_cross_sections} shows the ratio of measured over predicted cross sections. The goodness of the combined cross section fit is assessed, taking as hypothesis that the relative contributions of the channels are as predicted. This assumes lepton universality in the decay $\PZ \to \ell^+ \ell^-$, which is experimentally confirmed to high precision \cite{ALEPH:2005ab,Aaboud:2016btc}. Using the maximum likelihood for the observed yields, $\mathcal{L}_{\text{obs}}$, and for the expected yields, $\mathcal{L}_{\text{exp}}$, the ratio $-2 \ln(\mathcal{L}_{\text{obs}} / \mathcal{L}_{\text{exp}})$ is found to be 8.7. The $p$-value is calculated as the fraction of $10^5$ MC pseudoexperiments giving a larger ratio than the fit to data, and found to be $2.3\%$. This relatively low $p$-value is driven by the compatibility of the $\eeee$ channel with the other two channels.

\begin{table}[h!]
\centering
\begin{tabular}{lll}
\toprule
\textbf{Channel}			& \textbf{Measurement (fb)} & \textbf{Prediction (fb)}\\
\midrule
$\eeee$	& $13.8^{+1.1}_{-1.0}$ $[\pm 0.9$ (stat.) $\pm 0.3$ (syst.) $^{+0.5}_{-0.4}$ (lumi.)$]$			& $10.8^{+0.5}_{-0.4}$\\
$\eemm$	& $21.1^{+1.3}_{-1.2}$ $[\pm 1.0$ (stat.) $^{+0.5}_{-0.4}$ (syst.) $^{+0.7}_{-0.6}$ (lumi.)$]$	& $21.0^{+0.9}_{-0.8}$\\
$\mmmm$	& $11.5^{+0.9}_{-0.8}$ $[\pm 0.7$ (stat.) $\pm 0.4$ (syst.) $^{+0.4}_{-0.3}$ (lumi.)$]$			& $10.8^{+0.5}_{-0.4}$\\
\midrule
Combined & $46.4^{+2.4}_{-2.2}$ $[\pm 1.5$ (stat.) $\pm 1.0$ (syst.) $^{+1.5}_{-1.4}$ (lumi.)$]$		& $42.6^{+1.8}_{-1.5}$\\
\bottomrule
\end{tabular} 
\caption{Measured and predicted integrated fiducial cross sections. The prediction is based on an NNLO calculation from \matrixnnlo{} with the $\Pgluon\Pgluon$-initiated contribution multiplied by a global NLO correction factor of $1.67$. A global NLO weak correction factor of $0.95$ is applied, and the contribution of around 2.5\% from EW-$\ZZ jj$ generated with \sherpa{} is added. For the prediction, the QCD scale uncertainty is shown.
}
\label{tab:integrated_cross_sections}
\end{table}

\begin{figure}[h!]
\centering
\includegraphics[width=0.68\textwidth]{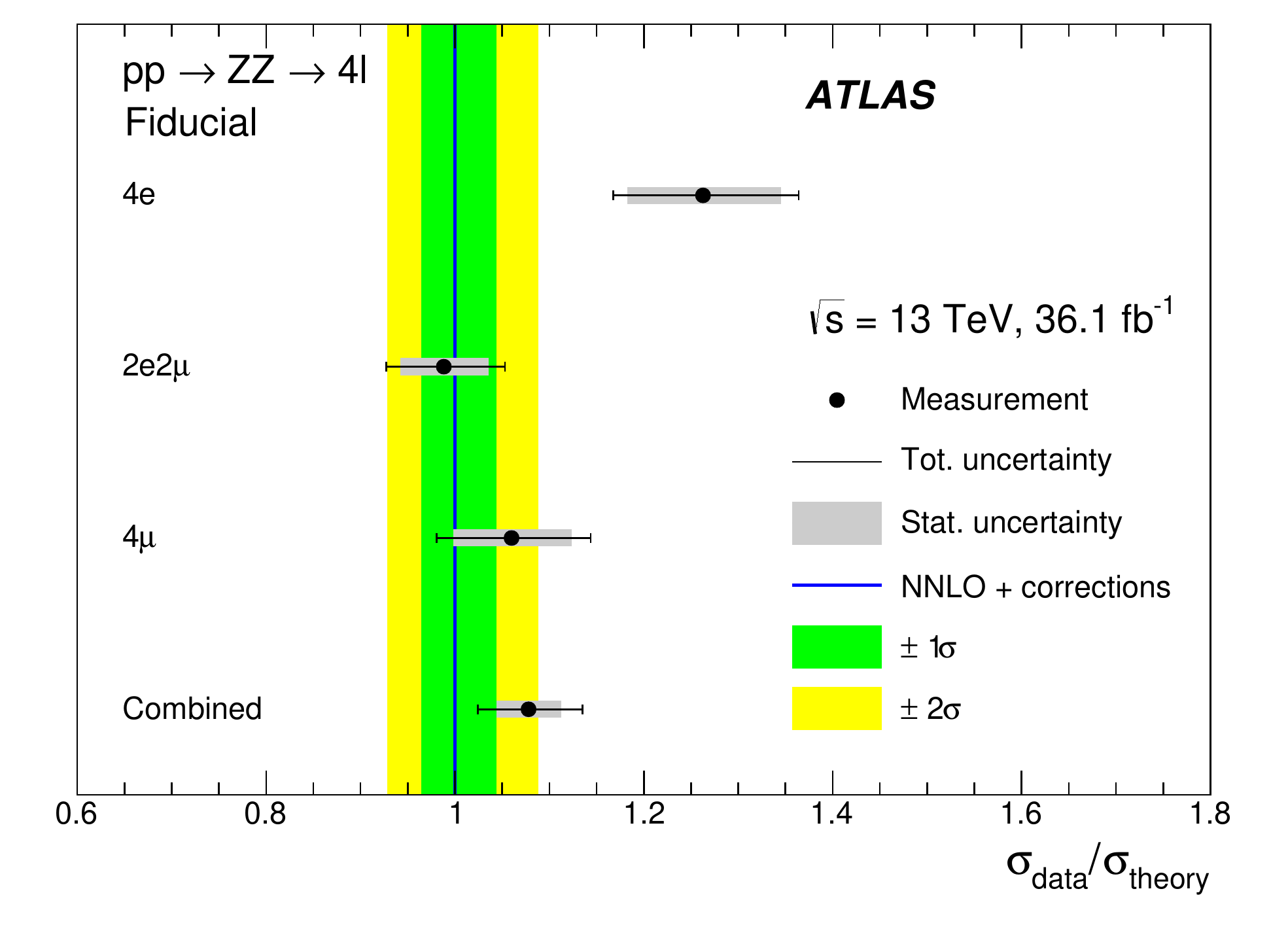}
\caption{Comparison of measured integrated fiducial cross sections to a SM prediction based on calculation from \matrixnnlo{} with the $\Pgluon\Pgluon$-initiated contribution multiplied by a global NLO correction factor of $1.67$. A global NLO weak correction factor of $0.95$ is applied, and the contribution of around 2.5\% from EW-$\ZZ jj$ generated with \sherpa{} is added. For the prediction, the QCD scale uncertainty is shown as a one- and two-standard-deviation band. Figure produced by Jonatan Rost\'{e}n and published in \myref~\cite{STDM-2016-15}.}
\label{fig:fiducial_cross_sections}
\end{figure}

\subsection{Extrapolation to total phase space and all $\PZ$ boson decay modes}

To allow easy comparison to other measurements, extrapolation of the cross section to the on-shell phase space (\mysec~\ref{sec:truthonshell}) and any SM \PZ{} boson decay is performed. The total phase space is the same as the fiducial phase space (\mysec{}~\ref{sec:fiducial}), except that no $\pt$ and $\eta$ requirements are applied to the leptons.
The ratio of the fiducial to on-shell cross section is determined using the \matrixnnlo{} setup described in Section~\ref{sec:nnlo_predictions} and found to be $\azz = 0.58\pm 0.01$, 
where the uncertainty includes the following contributions. A similar value is found when the calculation is repeated with the nominal \sherpa{} setup, and the difference between these (1.0\% of the nominal value) is included in the uncertainty of \azz{}. Other included uncertainties are derived from PDF variations (0.4\%, calculated with \MCFM{}) and QCD scale variations (0.8\%).

To calculate the extrapolated cross section, the combined fiducial cross section is divided by \azz{} and by the leptonic branching fraction $4 \times (3.3658\%)^{2}$ \cite{Olive:2016xmw}, where the factor of four accounts for the different flavour combinations of the decays. In the \eeee{} and \eemm{} channel, the pairing prescription as well as quantum-mechanical interference lead to a net increase of the cross sections by around $2.6\%$ with respect to the \eemm{} channel. This difference is corrected for by applying an additional factor in the same-flavour channels. The final cross section is obtained using the same maximum-likelihood method as for the combined fiducial cross section, but now including the uncertainties of \azz{} as additional nuisance parameters. The used leptonic branching fraction value excludes virtual-photon contributions. Based on a calculation with \PYTHIA{}, including these would lead to a branching fraction for $\ZZllll$ that is around 1.01--1.02 times larger. This difference could have been corrected for and/or a corresponding systematic uncertainty applied to \azz{}, but this was overlooked at the time of the analysis. Given that the extrapolation has other intrinsic problems, which will be explained in the next section, the author would not consider this a significant problem.

The extrapolated cross section is found to be \extrapolatedxsectshort{}~$[$\extrapolatedxsectstatuncert{}~(stat.)\;\extrapolatedxsectsystuncert{}~(syst.)\;\extrapolatedxsectlumiuncert{}~(lumi.)$]$~pb. The NNLO prediction from \matrixnnlo{}, with the $\Pgluon\Pgluon$-initiated process multiplied by a global NLO correction factor of $1.67$ is $16.9^{+0.6}_{-0.5}$~pb, where the uncertainty is estimated by performing QCD scale variations. A comparison of the extrapolated cross section to the NNLO prediction as well as to previous measurements is shown in \myfig{}~\ref{fig:extrapolated_cross_sections}. Since the publication of \myfig~\ref{fig:extrapolated_cross_sections}, CMS has measured the cross section at $\sqrt{s} = 13$~\TeV{} with more data and hence smaller uncertainty \cite{CMS-ZZ-13TEV}, comparable to that of the ATLAS measurement. In the next section, comparisons to the new CMS measurement will be shown.

\begin{figure}[h!]
\centering
\includegraphics[width=0.68\textwidth]{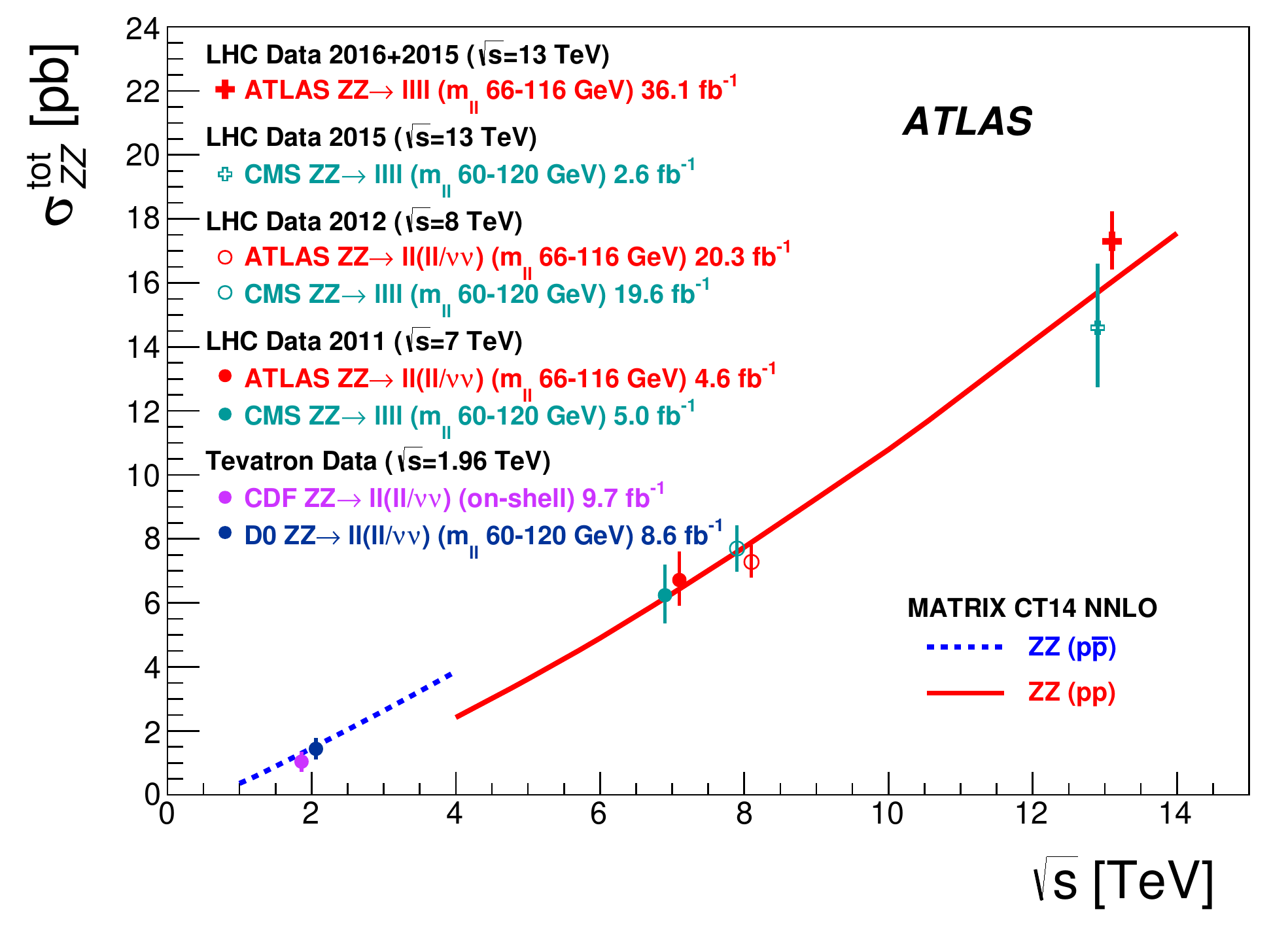}
\caption{Extrapolated cross section compared to other measurements at various centre-of-mass energies by ATLAS, CMS, CDF, and D0~\cite{Aad:2012awa,CMS:2014xja,Chatrchyan:2012sga,Aaltonen:2014yfa,Abazov:2012cj,Khachatryan:2016txa}, and to NNLO predictions from \matrixnnlo{} (excluding NLO corrections for the $\Pgluon\Pgluon$-initiated process, because these were not available to the analysis for all centre-of-mass energies and collision types). The total uncertainties of the measurements are shown as bars. Some data points are shifted horizontally to improve readability. In the context of the CDF measurement, ``on-shell'' means that extrapolation to zero-width \PZ{} bosons is performed, ignoring the contribution of virtual photons and $\PZ/\Pphoton^*$ interference. Figure produced by Jonatan Rost\'{e}n and published in \myref~\cite{STDM-2016-15}.}
\label{fig:extrapolated_cross_sections}
\end{figure}

\subsection{Comparison to fiducial CMS results and combination}
\mysec~\ref{sec:integrated_xs_results} showed a simple comparison of results from various experiments. However, the extrapolation introduces model-dependence and often is not done in the same way in different analyses: different predictions and definitions for the on-shell phase space (\mysec~\ref{sec:truthonshell}) are used. Perhaps worse than these \emph{known} unknowns are the \emph{unknown} unknowns: the modelling of hard electroweak processes in the forward region might be poor due to unknown new physical effects (SM or non-SM) setting in. To circumvent these problems, this section shows comparison of the integrated \emph{fiducial} cross section to the latest $\ZZllll$ results from CMS \cite{CMS-ZZ-13TEV}. A simple combination of the ATLAS and CMS results is also carried out in the intersection of the respective fiducial phase spaces to avoid extrapolation outside of the experimental coverage. Intersection here means the phase space whose events fall in both the ATLAS and CMS fiducial phase space.

\paragraph{Joint phase space}\hfill\\[1.5mm]
The ATLAS and CMS results are combined by defining a joint phase space that is approximately the intersection of their respective phase spaces and labeled $\text{ATLAS} \cap \text{CMS}$. \mytab{}~\ref{tab:fiducial_selection_ATLASandCMS} shows its definition. Using the intersection of both phase spaces has the advantage that neither experiment extrapolates outside of its fiducial region, so the model-dependence of the combination is minimised. However, for two reasons, it is not strictly speaking the intersection. Firstly, the CMS pairing algorithm is used, which does not select a subset of events of those selected by the ATLAS pairing algorithm. Secondly, there are two requirements that are slightly looser than in the ATLAS phase space:
\begin{itemize}
\item the third-highest-\pt{} lepton is required to have $\pt > 5$~\GeV{} (ATLAS: $> 10$~\GeV{}),
\item $\Delta R > 0.1$ is required between different-flavour leptons (ATLAS: $> 0.2$).
\end{itemize}
Based on studies with the simulated \SHERPA{} samples used in the analysis, both simplifications are expected to have a small impact, but it was not quantified exactly. The simplifications are due to a purely technical reason: the author calculated the joint phase space with a later version of \matrixnnlo{} than was used in the analysis and did not reimplement custom selections in the new source code. The remaining requirements are supported by \matrixnnlo{} out of the box.

\begin{table}[!htbp]
\centering
\begin{tabular}{ll}
\toprule
\textbf{Type} & \textbf{Input or requirement}\\
\midrule
Leptons (\Pe{}, \Pmu{}) & Prompt\\
	& Dressed with prompt photons within $\Delta R = 0.1$\\
	& $\pt > 5~\GeV$\\
	& $|\eta|<2.5$\\
\midrule
Events & Exactly four leptons\\
	& Two leading-\pt{} leptons satisfy $\pt > 20$~\GeV{}, 15~\GeV{}\\
	& Any same-flavour opposite-charge dilepton has mass $m_{\ell\ell} > 5$~\GeV{}\\
	& $\Delta R > 0.1$ between all leptons\\
	& Dileptons giving minimum $|m_{\ell\ell} - m_{\PZ}|$ are taken as \PZ{} boson candidates\\
	& \PZ{} boson candidates have mass $66~\GeV{} < m_{\ell\ell} < 116$~\GeV{}\\ 
\bottomrule
\end{tabular}
\caption{Summary of the selection criteria defining the joint $\text{ATLAS} \cap \text{CMS}$ fiducial phase space.}
\label{tab:fiducial_selection_ATLASandCMS}
\end{table}

\matrixnnlo{} results in the CMS and joint $\text{ATLAS} \cap \text{CMS}$ fiducial phase spaces are shown in \mytabs~\ref{tab:matrix_results_cms} and \ref{tab:matrix_results_combination}. More information about the calculation and the corresponding results for ATLAS can be found above in \mytab~\ref{tab:matrix_results_atlas}. The total integrated cross section calculated at NNLO with \matrixnnlo{} in the CMS (joint) phase space is 37.5~fb (36.0~fb). To extrapolate the ATLAS measurements to the joint phase space, they are multiplied by extrapolation factors calculated at NNLO of 0.856 for the \eeee{} and \mmmm{} channels and 0.904 for the \eemm{} channel. (Assuming the contributions of the channels to be as predicted by the SM, this would correspond to an extrapolation factor of 0.871 for the combination of the three channels.) CMS does not publish the cross sections measured in the individual channels, so an overall extrapolation factor of 0.960 is used.
The extrapolation factors have negligible uncertainties.

%
%
\begin{table}[h!]
\centering
\begin{tabular}{llll}
\toprule
\textbf{Order or subprocess} & \textbf{4e or 4\textmu{} (fb)} & \textbf{2e2\textmu{} (fb)} & \textbf{Scale uncertainty (\%)}\\
\midrule
LO & 5.484 & 10.68 & $+5.8$, $-6.8$\\
NLO & 7.947 & 15.41 & $+2.6$, $-2.1$\\
NNLO & 9.519 & 18.42 &$+3.2$, $-2.7$\\
\midrule
Only $\gluglu\; \looparrow{}\; 4\ell$ & 0.8491 & \phantom{0}1.676 & $+23.5$, $-17.7$\\
\bottomrule
\end{tabular}
\caption{Integrated fiducial cross sections in the CMS phase space, calculated with \matrixnnlo{}.}
\label{tab:matrix_results_cms}
\end{table}

\begin{table}[h!]
\centering
\begin{tabular}{llll}
\toprule
\textbf{Order or subprocess} & \textbf{4e or 4\textmu{} (fb)} & \textbf{2e2\textmu{} (fb)} & \textbf{Scale uncertainty (\%)}\\
\midrule
LO & 5.168 & 10.45 & $+5.8$, $-6.8$\\
NLO & 7.470 & 15.11 & $+2.6$, $-2.1$\\
NNLO & 8.949 & 18.06 &$+3.3$, $-2.7$\\
\midrule
Only $\gluglu\; \looparrow{}\; 4\ell$ & 0.8165 & \phantom{0}1.653 & $+23.5$, $-17.8$\\
\bottomrule
\end{tabular}
\caption{Integrated fiducial cross sections in the joint $\text{ATLAS} \cap \text{CMS}$ phase space, calculated with \matrixnnlo{}.}
\label{tab:matrix_results_combination}
\end{table}


\paragraph{Combination and uncertainties}\hfill\\[1.5mm]
The phase space extrapolation factors are applied to the nominal values of the cross sections. The \emph{relative} uncertainties of the cross sections are left unchanged. 
The extrapolated $\ZZllll$ cross section is $40.5^{+2.3}_{-2.2}$~fb for both ATLAS and CMS.
The nominal value of the combined cross section is found by simply taking the average of the extrapolated ATLAS and CMS result,
\begin{equation*}
\sigma^{\text{combined}} = \frac{0.856 \left(\sigma^{\text{ATLAS}}_{\eeee} + \sigma^{\text{ATLAS}}_{\mmmm}\right) + 0.904\, \sigma^{\text{ATLAS}}_{\eemm} + 0.960\, \sigma^{\text{CMS}}}{2} \approx 40.5~\text{fb}. 
\end{equation*}
This is justified, because the two experiments report very similar relative uncertainties. Otherwise, a weighted average taking into account the different uncertainties might have been more adequate.

When combining the uncertainties, the luminosity uncertainty is treated as fully correlated between ATLAS and CMS, while the statistical and all other systematic uncertainties are treated as uncorrelated. The combined uncertainty is found by summing the absolute per-experiment uncertainties linearly (luminosity) or in quadrature (all others) and dividing the result by the sum of the nominal cross sections to obtain the total relative uncertainty. The final result is $\sigma^{\text{combined}} = 40.5^{+1.9}_{-1.8}$~fb.

\paragraph{Results}\hfill\\[1.5mm]
The fiducial comparison and combination of the latest ATLAS and CMS results is shown in \myfig{}~\ref{fig:atlas_cms_fiducial_xs_comparison}. The measured values are compared to NNLO predictions, with partial NNNLO corrections, namely the $\alphas^3$ corrections to the loop-induced \gluglu-initiated contribution, as well as NLO weak corrections applied, as explained in \mysec~\ref{sec:best_sm_prediction}. While the NNLO predictions are calculated in the respective phase space using \matrixnnlo{}, all other corrections are computed in the ATLAS phase space. The ATLAS and CMS results agree well with the prediction, except for the $\sim$$2.5\sigma$ excess in the \eeee{} channel. They are also very compatible with each other, in the sense that their deviation from the prediction, approximately $+1\sigma$, is very similar.

\begin{figure}[h!]
\centering
\includegraphics[width=0.65\textwidth]{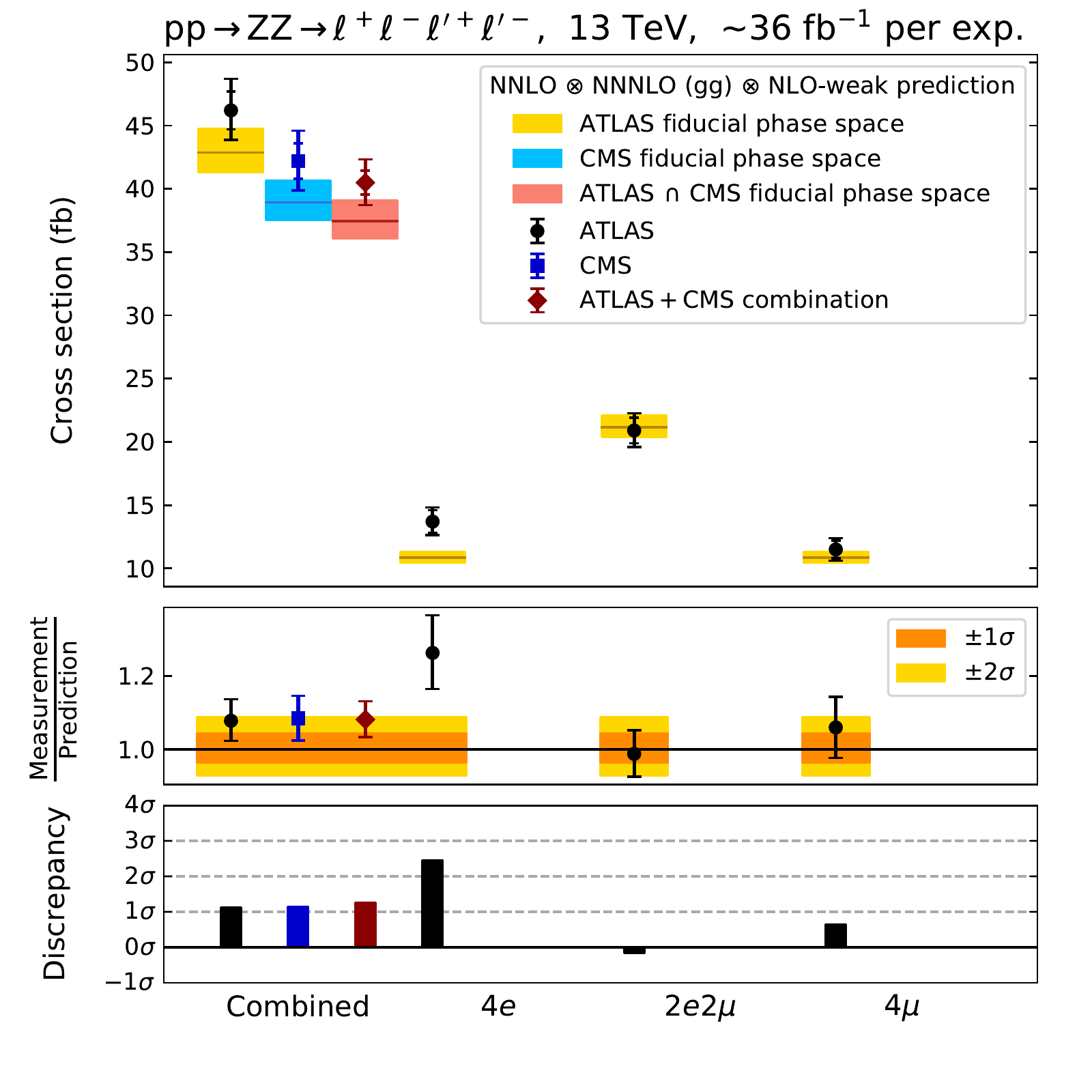}
\caption{Comparison of ATLAS, CMS, and combined results to the most formally accurate predictions available. The three different measurement sources correspond to different fiducial phase spaces, so their absolute values are not directly comparable to each other. Only ATLAS has published per-channel cross sections. The shown discrepancy takes into account the uncertainty of both measurement and prediction.}
\label{fig:atlas_cms_fiducial_xs_comparison}
\end{figure}

\clearpage
\section{Differential cross sections}
\label{sec:unfolding}

If the response of the detector and the reconstruction algorithms can be simulated, e.g.~with \GEANT{} or \textsc{Delphes} \cite{deFavereau:2013fsa}, particle-level theoretical predictions can be compared directly to the reconstructed differential distributions of events, such as those shown in \myfig{}~\ref{fig:data_mc_plots}. However, this approach is cumbersome and prone to inaccuracy at best, and not applicable to parton-level predictions, such as fixed-order calculations. To allow comparisons to theory, differential cross sections are determined. This is done by counting candidate events in each bin of the studied observable, subtracting the expected background, and using \emph{unfolding} to correct the measured distributions for the following experimental effects:

\begin{description}
\item[Acceptance and efficiency] Particles might escape measurement by falling into an uninstrumented region of the detector. Even particles hitting sensitive regions of the detector might not be measured because they fail quality or identification criteria, such as the number of tracker hits required for identification.
\item[Resolution] Every detector has a finite resolution, smearing reconstructed quantities with respect to their true values.
\item[Calibration] (Also called `scale.') The detector calibration is imperfect, meaning that even the calibrated measurement is not an unbiased estimate of the true corresponding quantity.
\item[Combinatorics] Objects passing some reconstruction-level criteria might not correspond to those that would pass the same criteria in truth. For instance, the reconstructed highest-\pt{} lepton might not be the true highest-\pt{} lepton. (Combinatorics effects are actually a consequence of the three other classes of effects above.)
\end{description}



All of the above steps can be encoded into a single response matrix $R$ relating the measured histogram $m$ to the true\footnote{In the context of unfolding, ``true'' and ``particle-level'' are used interchangeably.} histogram $t$ and the reconstruction-level background histogram $b$,
\begin{equation}\label{eq:response_matrix_definition}
m_i = R_{ij} t_j + b_i,
\end{equation}
where $i$ and $j$ are bin indices and summation over repeated indices is implied. The response matrix is determined using MC simulation. Its element $R_{ij}$ gives the probability of finding an event in measured bin $i$ given that it was in true bin $j$ \emph{or} in no true bin at all. It can be decomposed as
\begin{equation*}
R_{ij} = M_{ij}\,\varepsilon_{j}\,\phi_{i},
\end{equation*}
where the matrix $M_{ij}$ describes the \emph{bin migrations},
\begin{equation}\label{eq:migration_matrix}
M_{ij} = \frac{P(\text{reconstructed in bin $i$}~\cap~\text{in true bin $j$})}{\sum_{i'} P(\text{reconstructed in bin $i'$}~\cap~\text{in true bin $j$})},
\end{equation}
$\varepsilon_j$ corresponds to the \emph{reconstruction efficiency},
\begin{equation*}
\begin{split}
\varepsilon_j &= \frac{\sum_{i'} P(\text{reconstructed in bin $i'$}~\cap~\text{in true bin $j$})}{P(\text{in true bin $j$})} \\
&= P(\text{reconstructed in any bin}~|~\text{in true bin $j$}) \leq 1,
\end{split}
\end{equation*}
and $\phi_i$ is the correction for $\ZZllll + X$ signal events failing the fiducial selection but passing the reconstructed selection (\kern1.2pt\emph{fake correction}),
\begin{equation*}
\phi_i = \frac{P(\text{reconstructed in bin $i$})}{\sum_{j}P(\text{reconstructed in bin $i$}~\cap~\text{in true bin $j$})} \geq 1.
\end{equation*}
`Fake' contributions happen for instance due to resolution effects leading to particles passing reconstruction-level selections that they fail at particle level. Alternatively, the fake events could be subtracted as another background contribution $f$, in which case \myeq~\ref{eq:response_matrix_definition} is replaced by
\begin{equation*}
m_i = \tilde{R}_{ij} t_j + b_i + f_i,
\end{equation*}
where the modified response matrix element $\tilde{R}_{ij}$ represents the probability of finding an event in measured bin $i$ given that it was in true bin $j$,
\begin{equation}\label{eq:modified_response_matrix_definition}
\begin{split}
\tilde{R}_{ij} &= M_{ij}\,\varepsilon_i\\
&= \frac{P(\text{reconstructed in bin $i$}~\cap~\text{in true bin $j$})}{P(\text{in true bin $j$})} \\
&= P(\text{reconstructed in bin $i$}~|~\text{in true bin $j$}).
\end{split}
\end{equation}
This (\myeq~\ref{eq:modified_response_matrix_definition}) is perhaps the most common definition for the response matrix and used e.g.~in Cowan's textbook \cite{cowan}. The reason why the fake correction is absorbed into the response matrix in this analysis is that its contribution to each reconstructed bin should be scaled to the actual observed number of events in order to minimise model dependence, $f_i \propto (m_i - b_i)$. Performing the correction multiplicatively ($\phi_i$) rather than additively (\kern0.5pt$f_i$) automates this scaling in a handy way.\footnote{Strictly speaking, including the fake contribution in the response matrix means that the response matrix elements no longer represent a probability, since they could be greater than 1, but rather a `generalised efficiency', like \czz{}.} 
Unfolding could also be used to `correct' for physical effects such as hadronisation. However, this reduces the general validity of the measurement and makes it much more dependent on the used model (e.g.~the hadronisation model), and is therefore not used here. Unfolding is done to the particle level and the fiducial phase space, meaning that the correlation between the measured and corrected distribution is stronger and there is very little extrapolation to the outside of the detector acceptance. This makes the unfolding more robust and model-independent. In the following, `true' and `particle-level' are synonyms and will be used interchangeably.

Throughout this analysis, the same binning is used for the reconstruction- and particle-level histograms, meaning that the response matrix is square. (Some unfolding methods can deal with a higher number of reconstruction-level than particle-level bins and can in some situations even benefit from this overdetermination \cite{Blobel:2002pu, Cousins:2016ksu}.) Some of the used response matrices are shown in \myfig{}~\ref{fig:response_matrices}, the rest is shown in \myapp~\ref{sec:zz_aux_response}. It can be observed that despite wider bins, bin migrations are more important for some jet-exclusive observables. This is expected and due to the poorer jet energy resolution and calibration compared to leptons, as well as jets originating from or contaminated by pileup activity. As will be discussed below, the size of bin migrations determines what unfolding method is adequate. One useful figure in quantifying it is the purity $\pi$, defined as the probability for an event to fall in the same particle-level and reconstruction-level bin, given that it passes both the particle-level and reconstruction-level selection. The purity of bin $i$ corresponds to the diagonal element $M_{ii}$ of the migration matrix defined in defined in \myeq~\ref{eq:migration_matrix}. \myfig~\ref{fig:pef} shows the purity, efficiency, and fake correction for several of the unfolded observables. Their uncertainties are dominated by the difference between the nominal \SHERPA{} setup and the \POWHEGpy{} + \SHERPA{} setup. The fact that the generator uncertainty is one-sided also causes the total uncertainties to be very asymmetric in many bins. After unfolding, the generator uncertainty propagated to the final result is symmetrised as described in \mysec~\ref{sec:uncertainties}.

Efficiencies are typically around 60\%. They decrease noticeably in regions of phase space where leptons tend to be very forward, such as high values of $\yfourl$ or $\Delta y(\PZ_1, \PZ_2)$. This is driven by the diminished forward lepton acceptance and efficiency: fiducial leptons extend to $|\eta| = 2.7$, whereas reconstructed electrons and \emph{combined} muons only extend to $|\eta| \approx 2.5$. The purity is usually greater than 70\%, except in a few bins. For \pt{}- and mass-like observables, it often drops towards higher values if the bin width is constant, reflecting the poorer \emph{absolute} resolution (though the \emph{relative} resolution may increase). Increases in the bin width counter this effect, often leading to upward `jumps' in purity. The purity is generally very high for angular observables i.e.~those depending on $\eta$ and $\phi$, because these quantities can be measured with finer resolution than energies. The fake correction typically differs from unity by less than 5\% for jet-inclusive and jet multiplicity observables. For jet kinematics observables, it increases up to $\sim$2. 

\begin{figure}[p]
\centering
\subfigure{\includegraphics[width=0.49\textwidth]{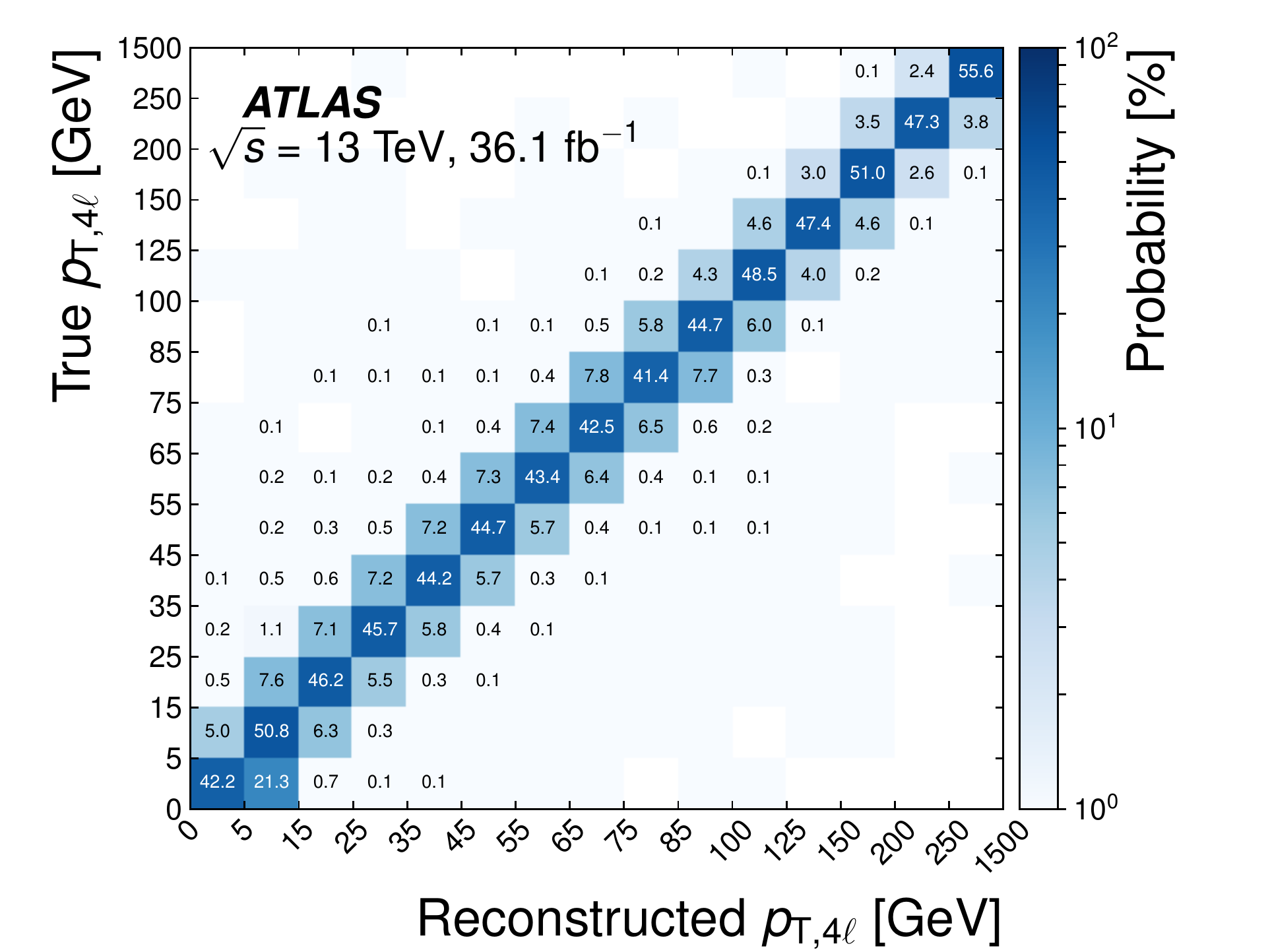}} 
\subfigure{\includegraphics[width=0.49\textwidth]{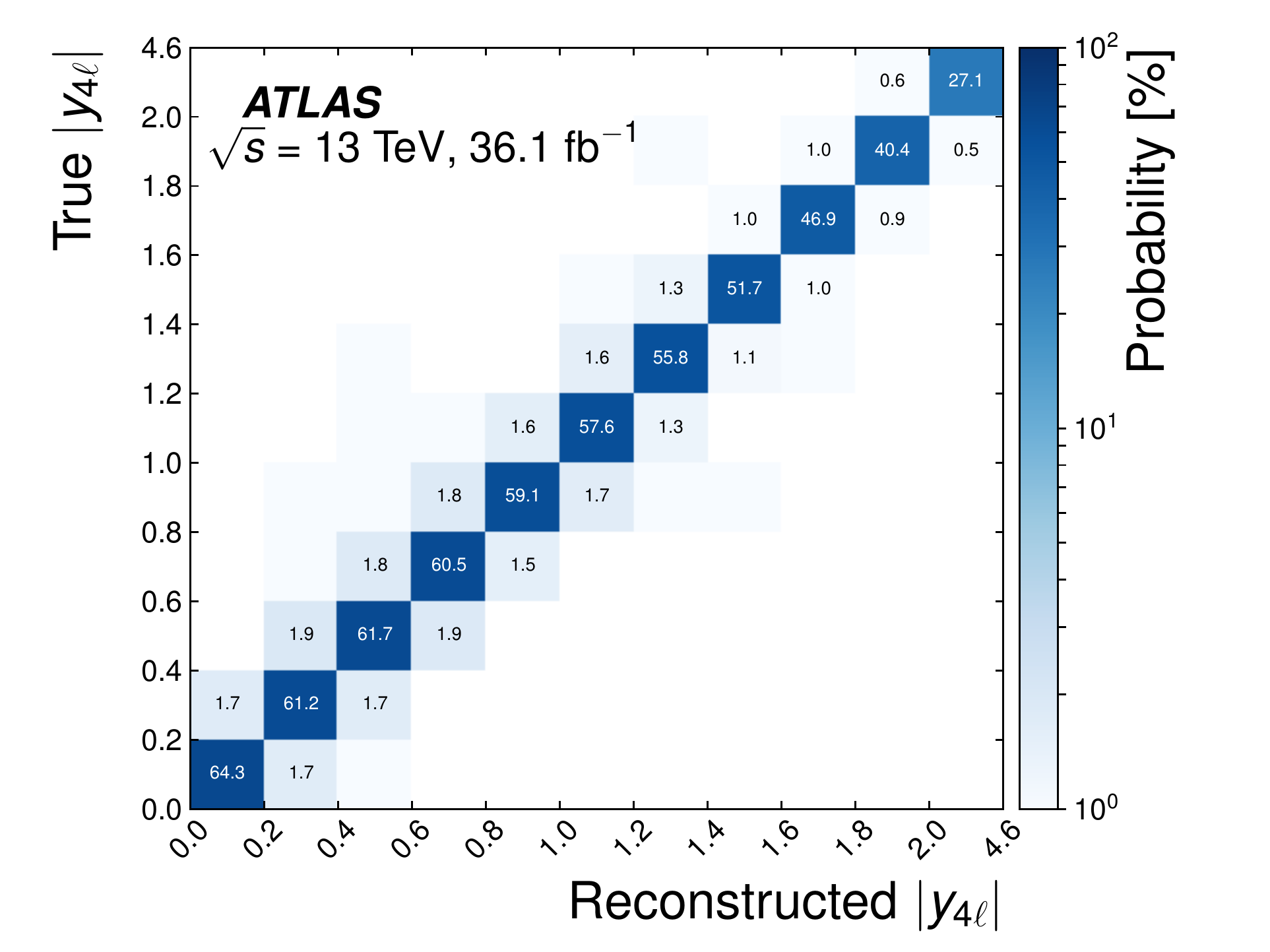}} 
\subfigure{\includegraphics[width=0.49\textwidth]{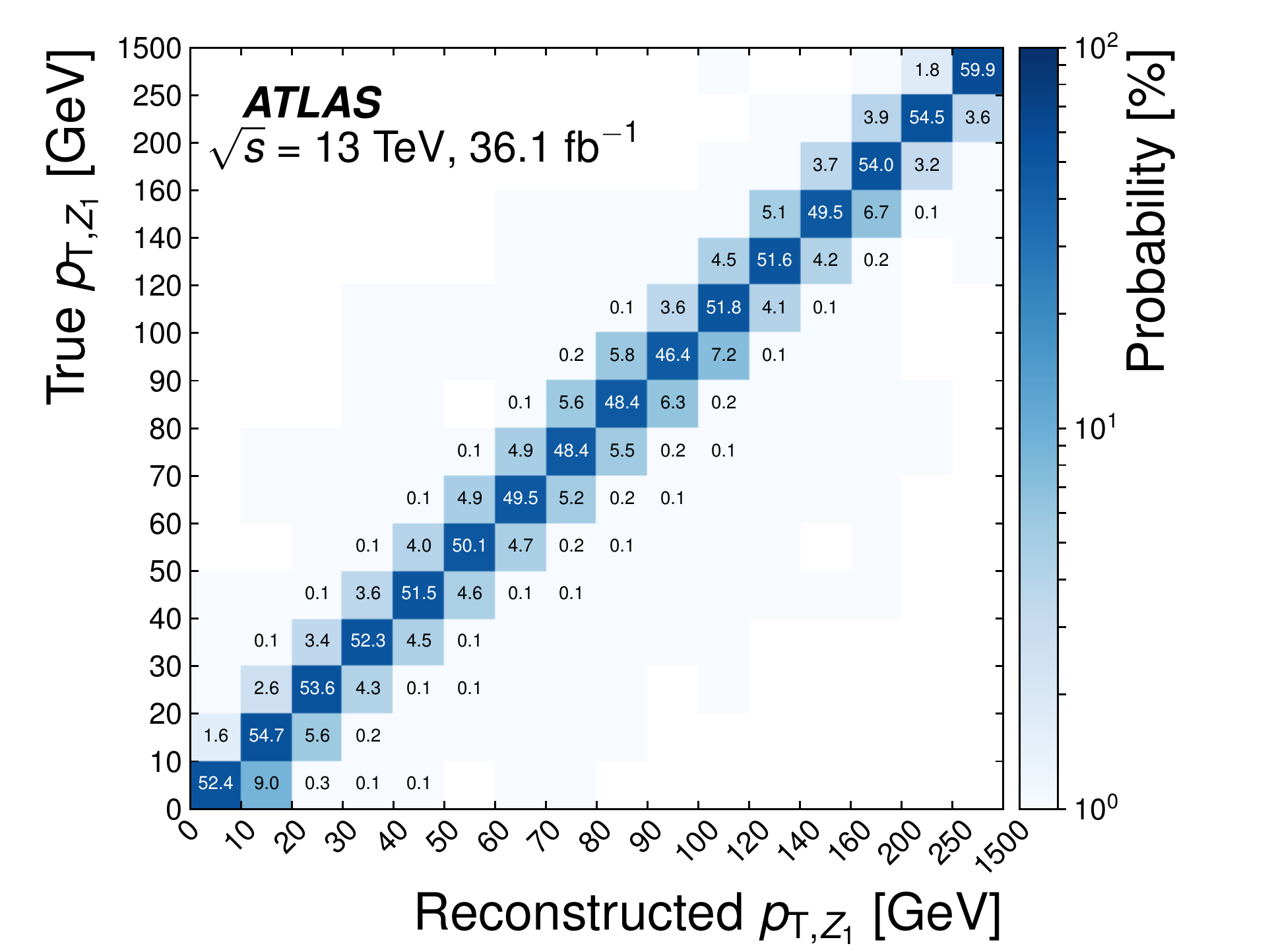}} 
\subfigure{\includegraphics[width=0.49\textwidth]{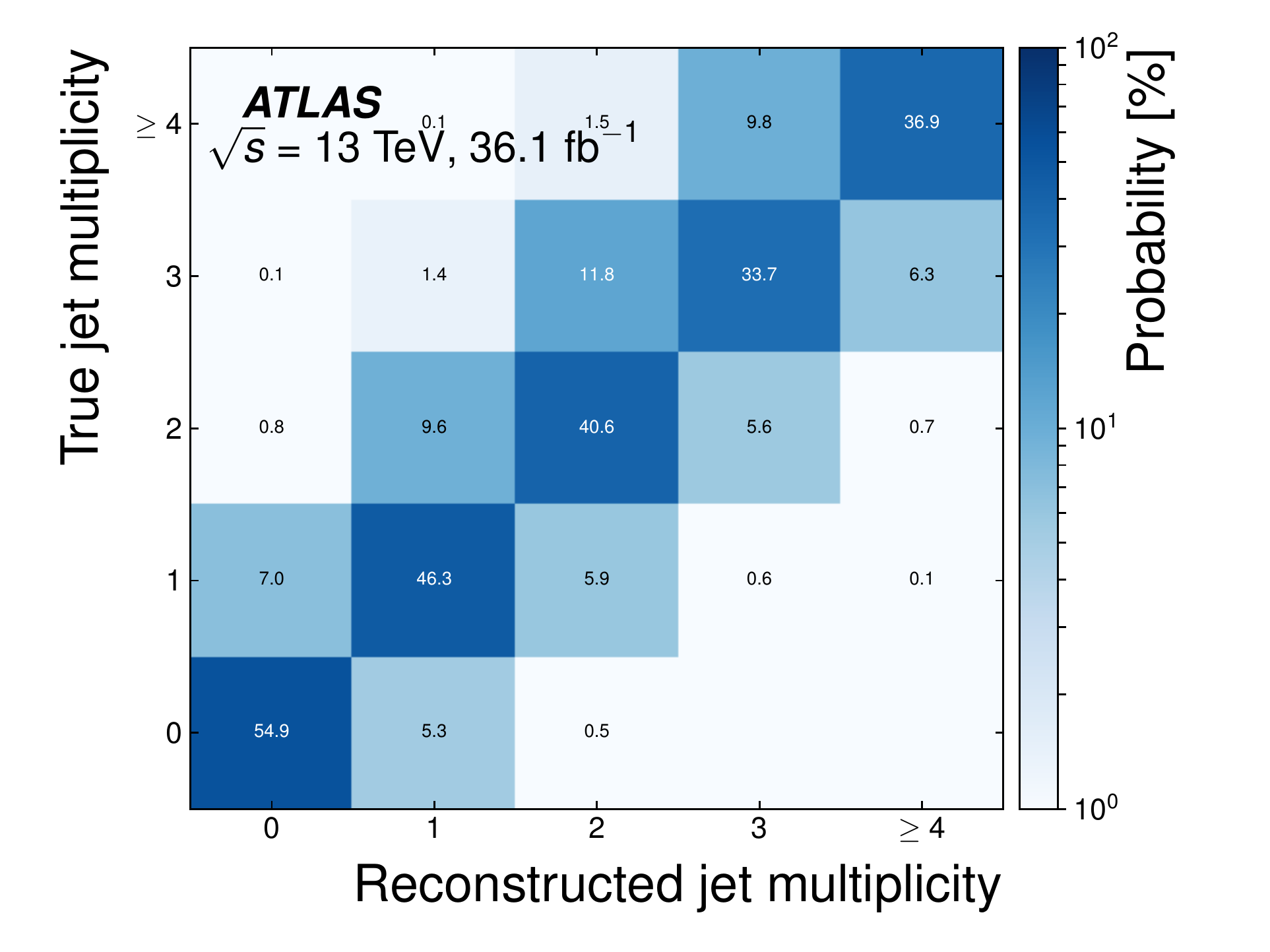}} 
\subfigure{\includegraphics[width=0.49\textwidth]{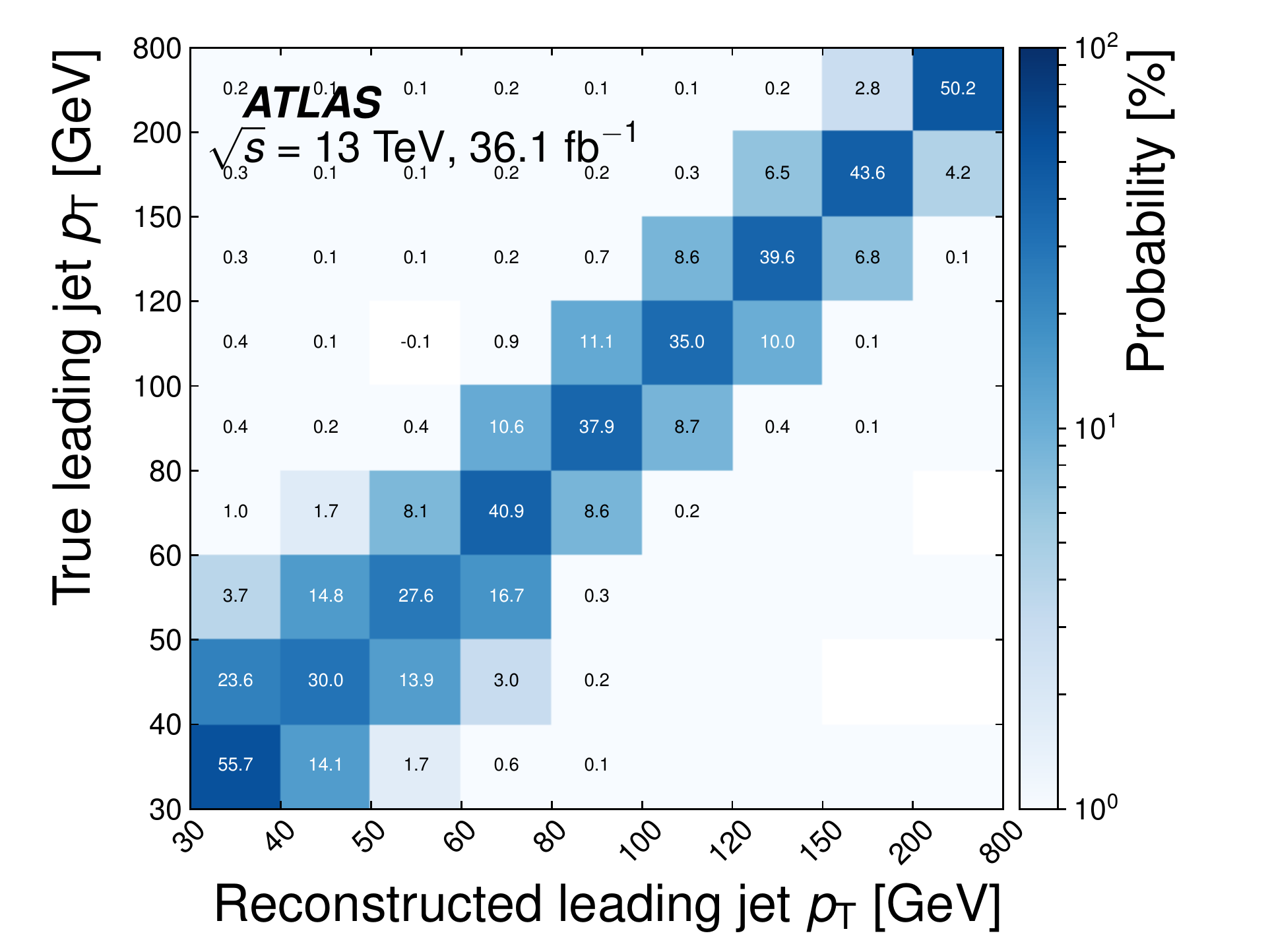}} 
\subfigure{\includegraphics[width=0.49\textwidth]{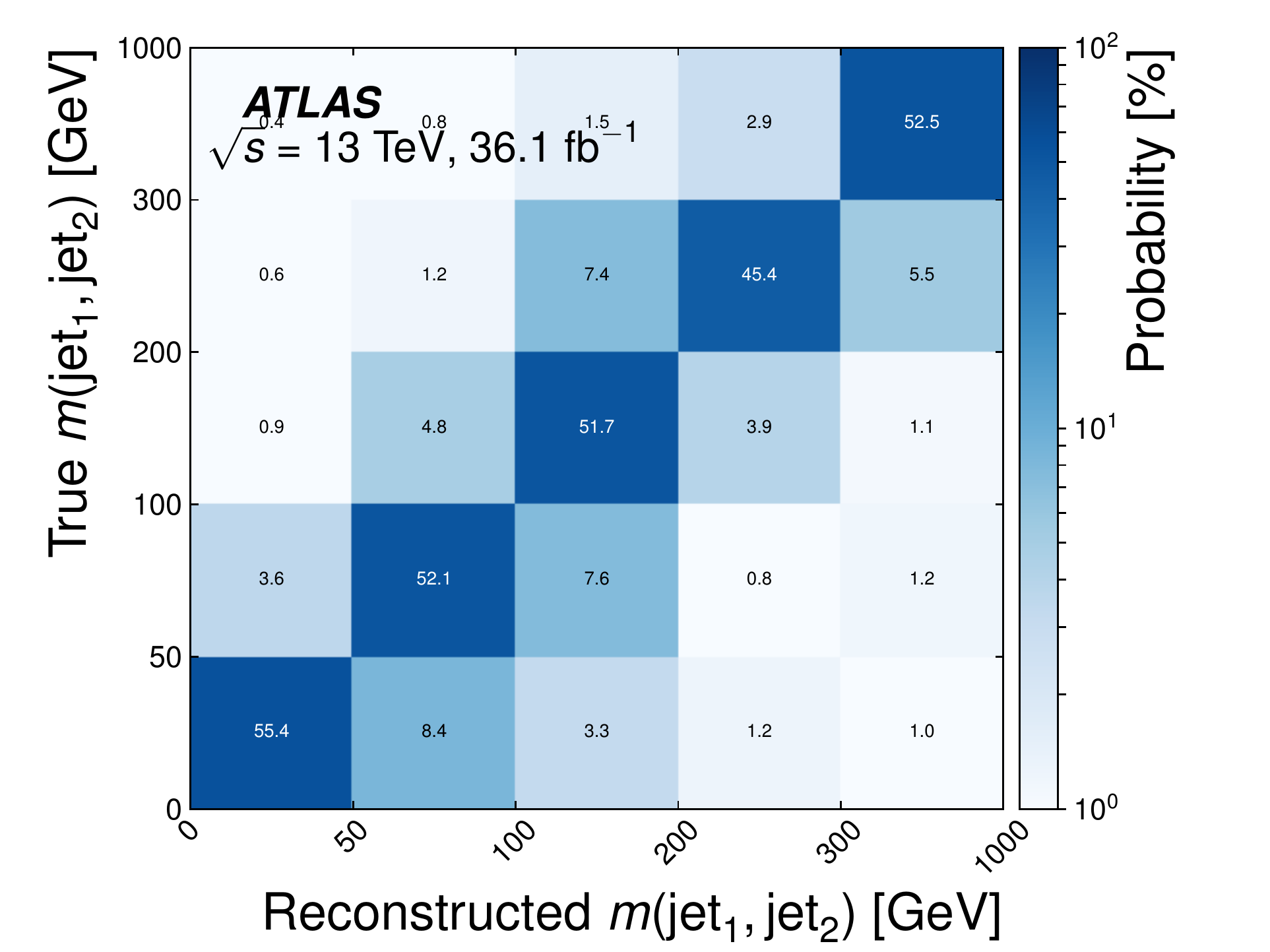}} 
\caption{Example response matrices used in the unfolding for several different observables, obtained using the nominal \SHERPA{} setup. Published in \myref~\cite{STDM-2016-15} without the bin contents written out as numbers.}
\label{fig:response_matrices}
\end{figure}

\begin{figure}[p]
\centering
\subfigure{\includegraphics[width=0.4\textwidth]{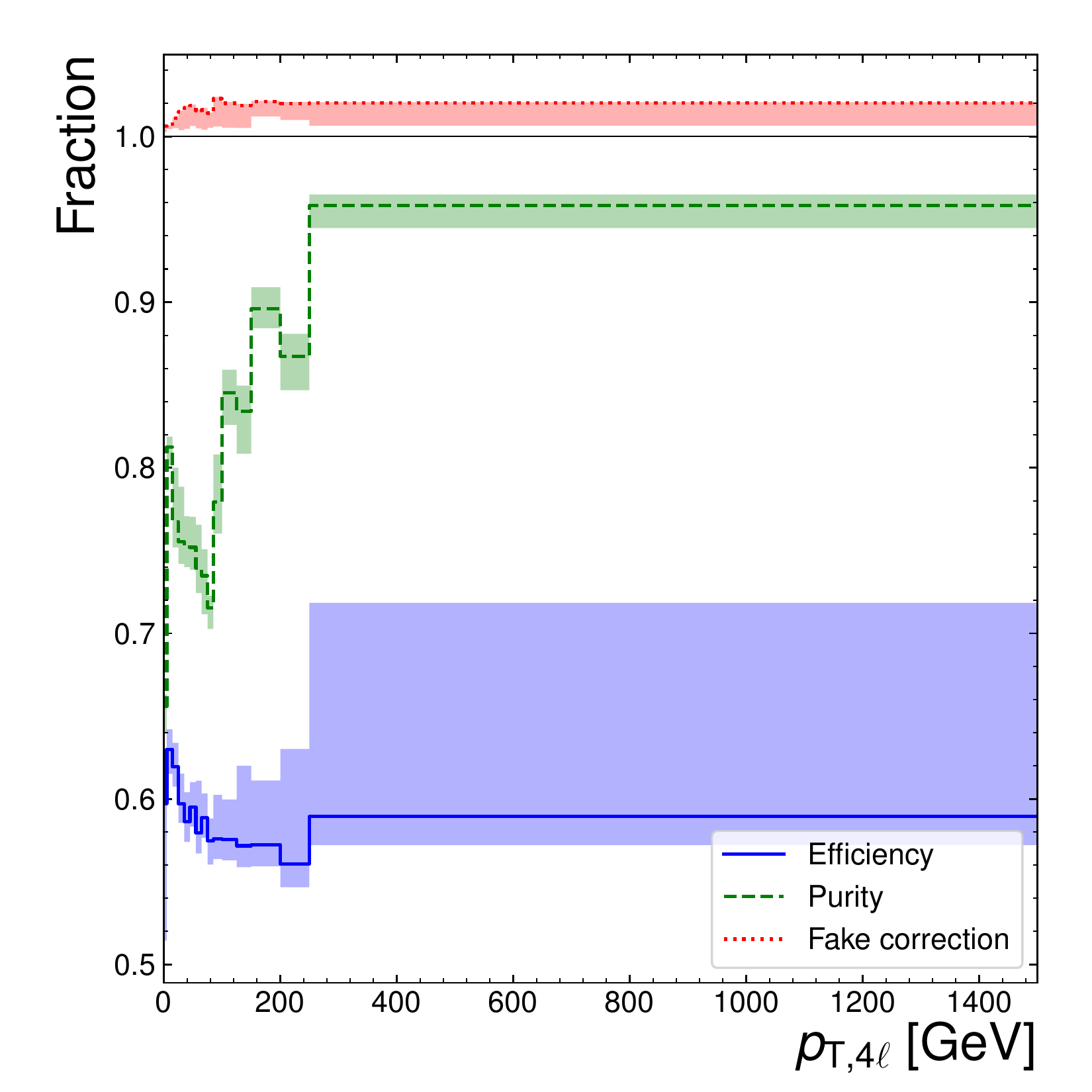}}
\subfigure{\includegraphics[width=0.4\textwidth]{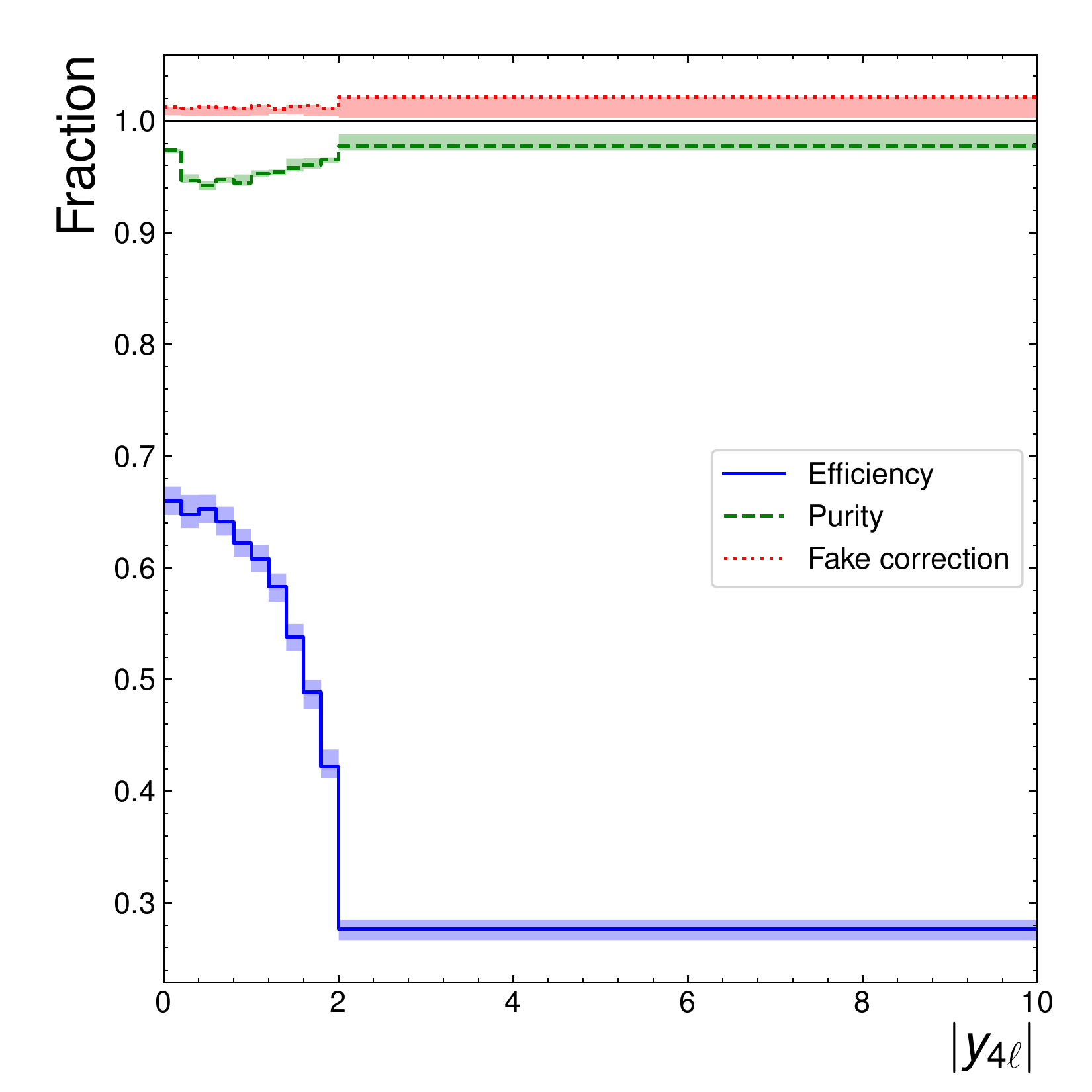}}
\subfigure{\includegraphics[width=0.4\textwidth]{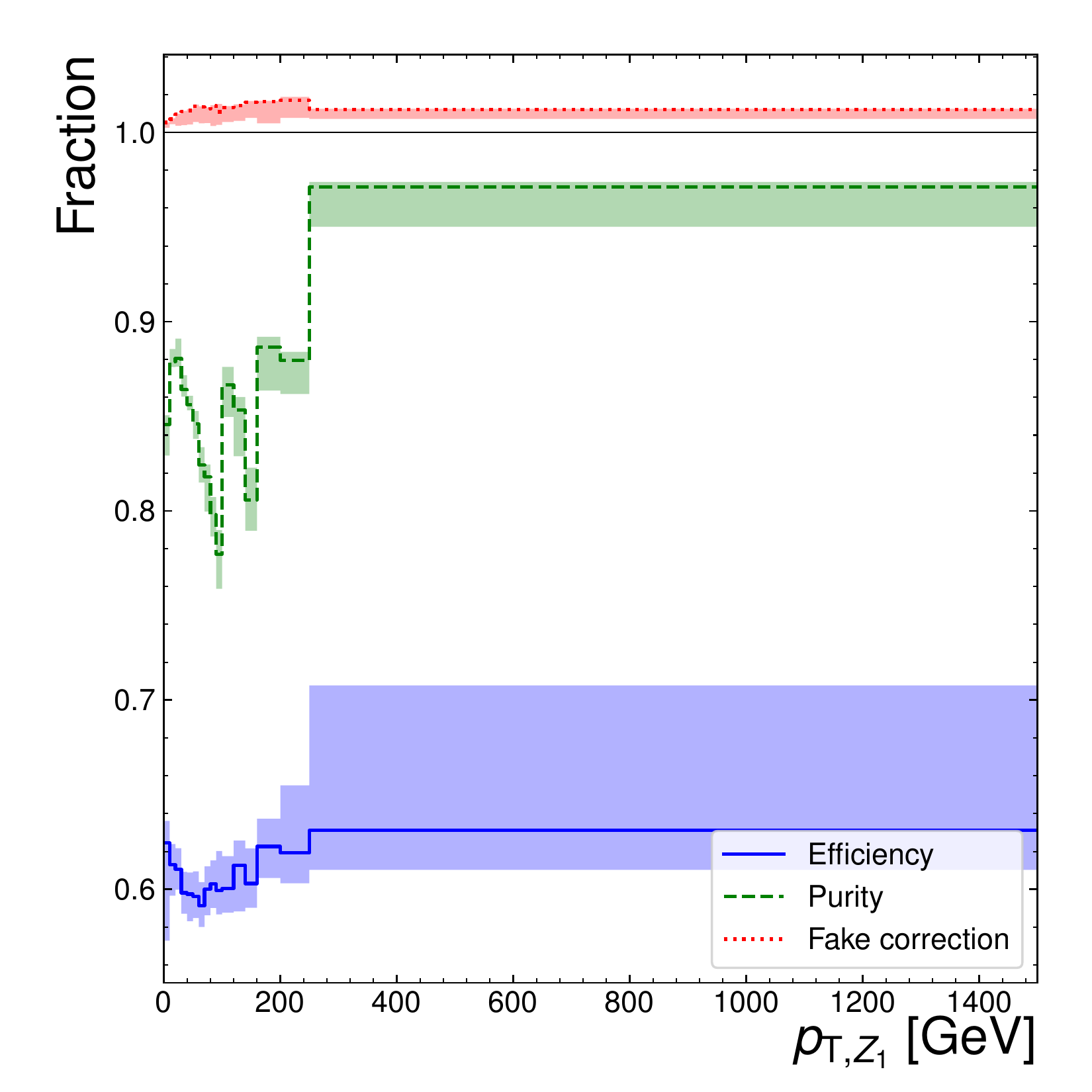}}
\subfigure{\includegraphics[width=0.4\textwidth]{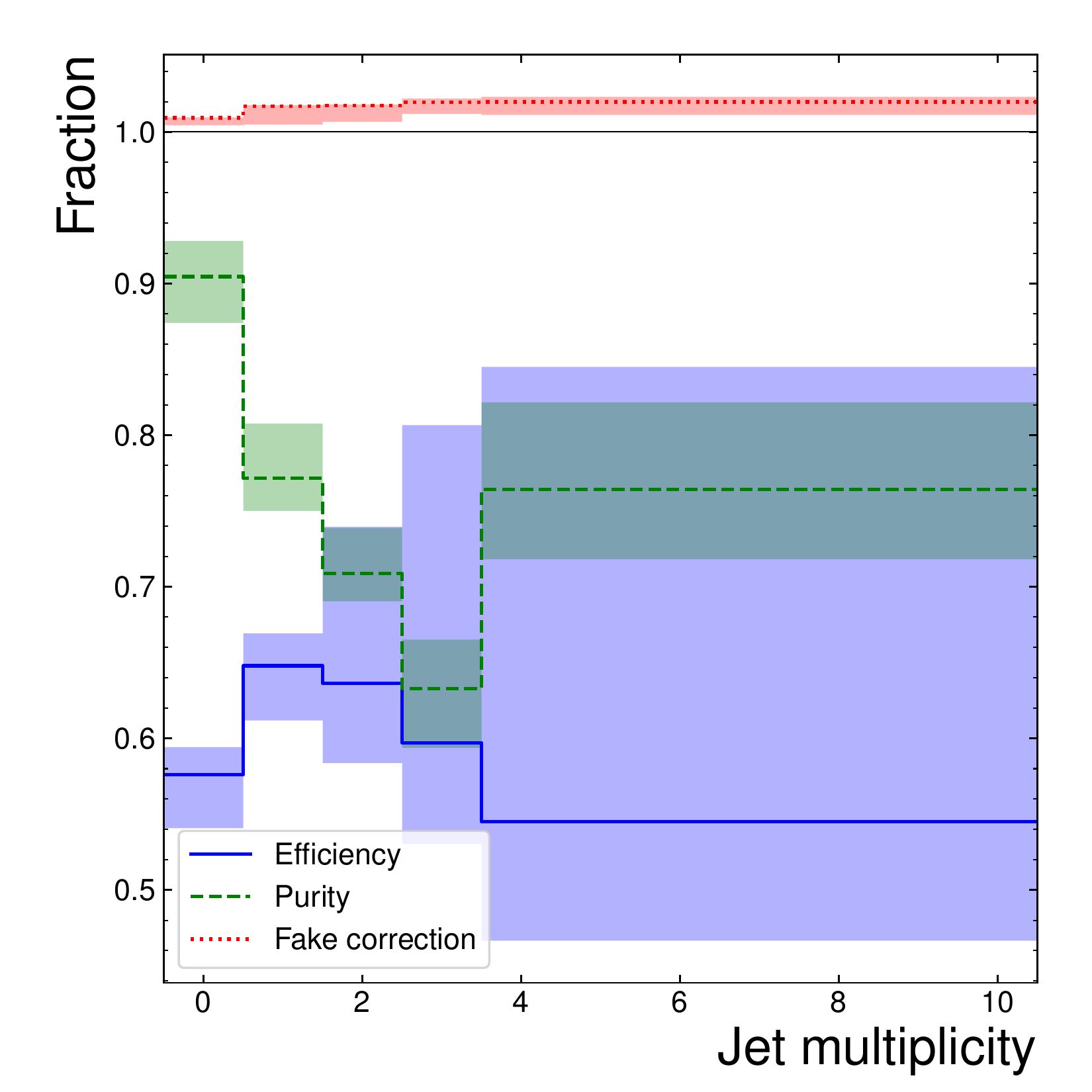}}
\subfigure{\includegraphics[width=0.4\textwidth]{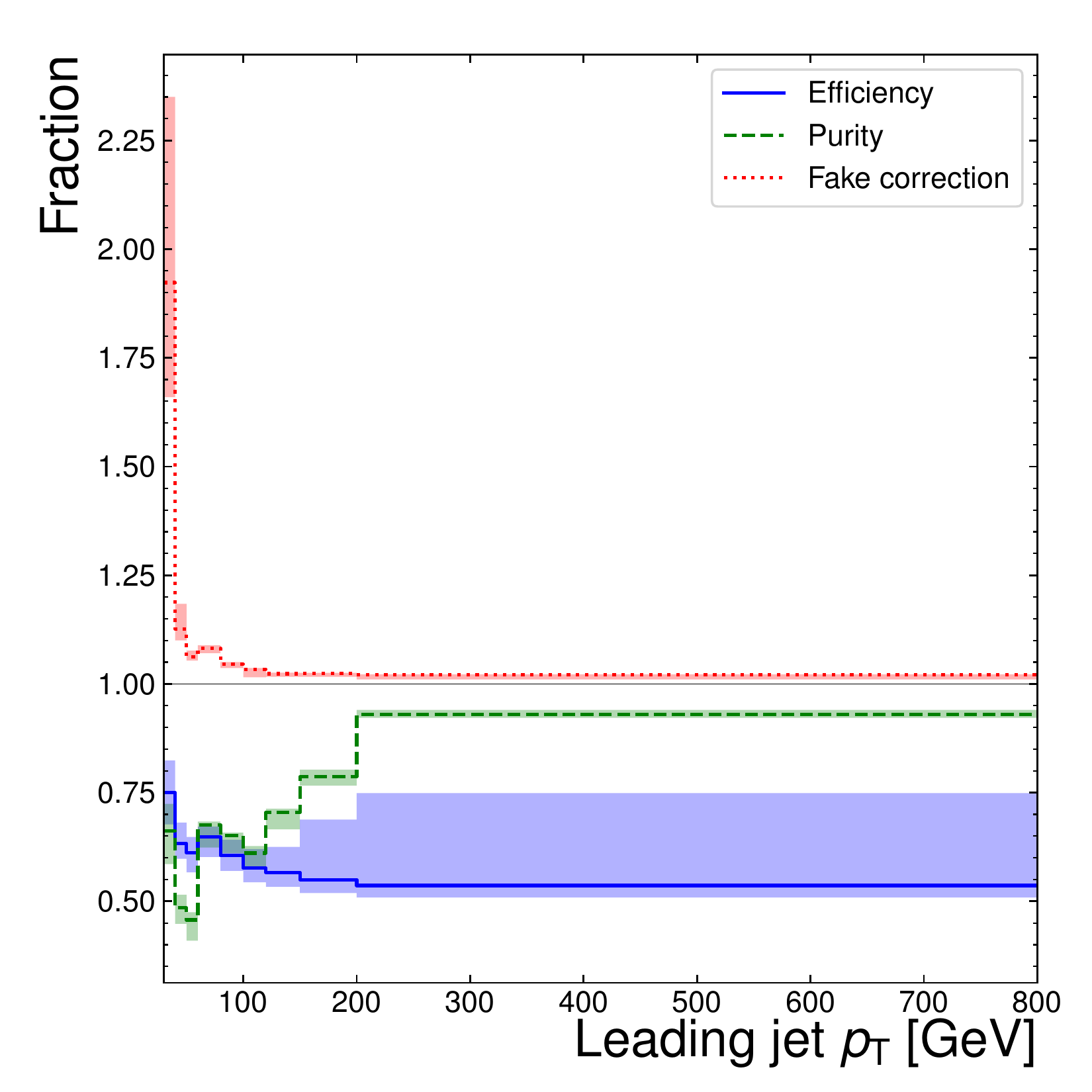}}
\subfigure{\includegraphics[width=0.4\textwidth]{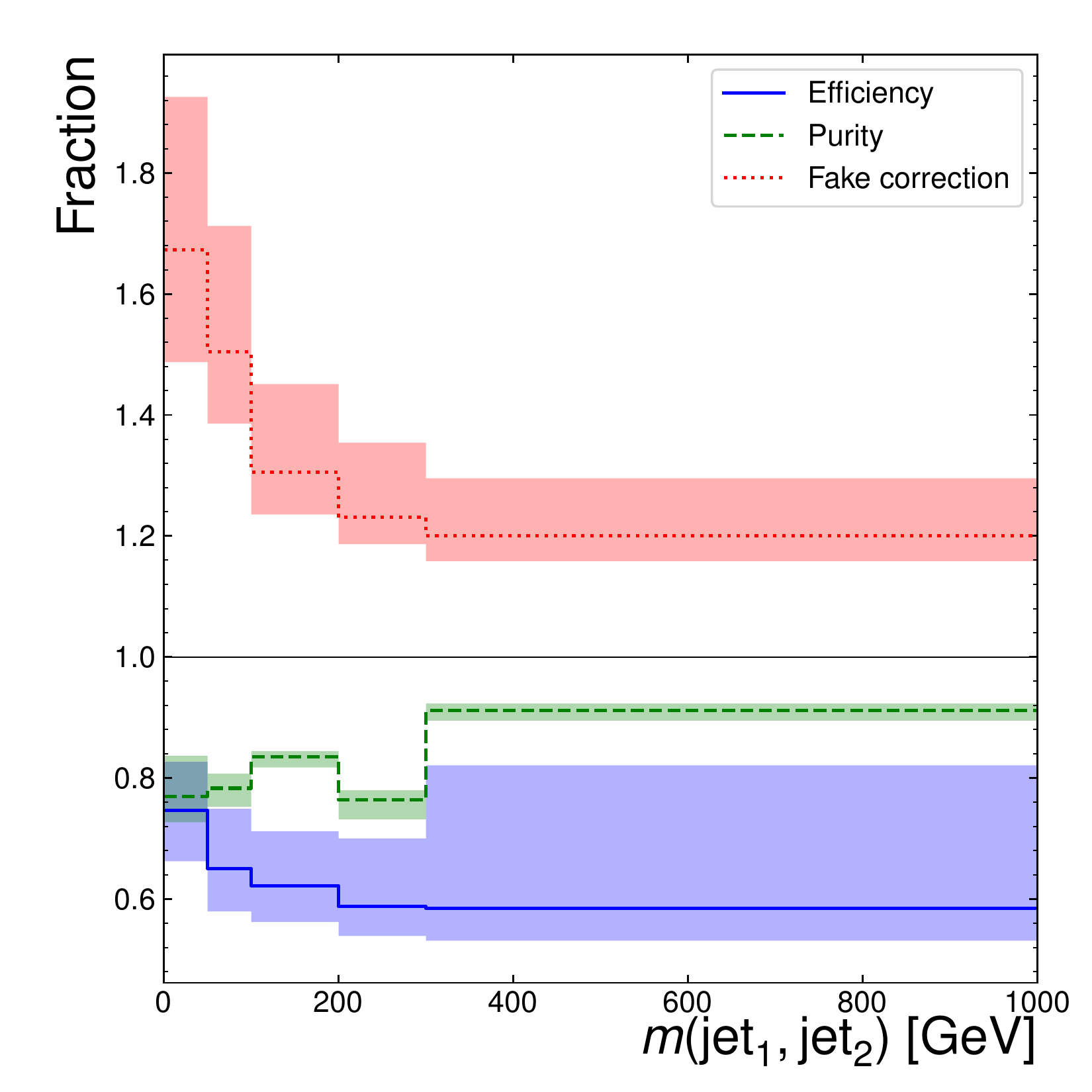}}
\caption{Purity, efficiency, and fake correction as a function of several observables. The shaded bands represent the total uncertainty.}
\label{fig:pef}
\end{figure}

\subsection{Unfolding method}

Once the response matrix is determined, the problem of unfolding consists of estimating the true histogram $t_j$ indexing the bins, given the background-subtracted measured histogram $(m_i - b_i)$. As the form of \myeq~\ref{eq:response_matrix_definition} shows, this can be achieved by inverting the response matrix,
\begin{equation}\label{eq:matrix_inversion}
\hat{t}_j = R^{-1}_{ji} (m_i - b_i),
\end{equation}
where the caret ($\hat{\phantom{o}}$) denotes an estimator, as long as $R$ is invertible, so that the corresponding system of linear equations is determinate.
While \myeq~\ref{eq:matrix_inversion} provides an unbiased estimate of $t$, it is not numerically stable, even if the response matrix is not singular. The following example, adapted from \myref~\cite{Hocker:1995kb}, highlights this. The simple $2\times 2$ response matrix
\begin{equation*}
R =
\begin{pmatrix} 
1-\varepsilon & \varepsilon\\ 
\varepsilon & 1-\varepsilon
\end{pmatrix}
\end{equation*}
is invertible as long as $0 \leq \varepsilon < 0.5$, since then $\det(R) = 1 - 2\varepsilon > 0$. However, estimating $\hat{t}$ by inverting $R$ as shown in \myeq~\ref{eq:matrix_inversion} can lead to wildly fluctuating solutions when the off-diagonal elements $\varepsilon$ are large, with the variance of $\hat{t}$ being proportional to $1 / \det(R)$, which diverges as $\varepsilon \to 0.5$. Of course \myeq~\ref{eq:matrix_inversion} is a mathematical equality, but the instability comes from the fact that the measured values are random variables $\mu_i$, following, in the case of yields, Poisson distributions:
\begin{equation*}
\mu_i \sim \text{Poisson}(m_i).
\end{equation*}
Here, the tilde ($\sim$) denotes that the random variable follows the given distribution.
So in the given example, the unfolding by matrix inversion is very sensitive to fluctuations of the observed histogram $\mu$. This problem is discussed with great clarity in \myref~\cite{cowan}. The result is very sensitive to statistical fluctuations, so that the statistical uncertainty of a given bin may be greatly increased by the unfolding. In general, the larger the off-diagonal elements are, the less stable is the unfolding by inverting the response matrix directly. In the limit where the response is exactly diagonal, unfolding becomes trivial as \myeq~\ref{eq:response_matrix_definition} reduces to a system of linear uncoupled equations and $\hat{t}_i = (m_i - b_i) / (\varepsilon_i \phi_i)$ is trivially the optimal (i.e.~unbiased and efficient) estimator.

The solution to finding a well-behaved solution when dealing with a non-diagonal response matrix is to add a regularisation procedure that disfavours strongly fluctuating solutions. Several approaches have been proposed \cite{Hocker:1995kb,Blobel:2002pu,DAgostini:1994zf,Malaescu:2009dm}. In this analysis, the \emph{Bayesian iterative unfolding} method \cite{DAgostini:1994zf} is used due to its simplicity and robustness. The method inverts the response matrix in \myeq~\ref{eq:response_matrix_definition} by using Bayes' theorem:
\begin{equation}\label{eq:bayesian_unfolding_derivation}
\begin{split}
	\hat{t}_i &= \frac{1}{\phi_j}\, \sum_j P(\text{in true bin $i$}~|~\text{reconstructed in bin $j$})\, (m_j - b_j)\\
	&= \frac{1}{\phi_j}\, \sum_j \frac{P(\text{reconstructed in bin $j$}~|~\text{in true bin $i$})\; P(\text{in true bin $i$})}{P(\text{reconstructed in bin $j$})}\, (m_j - b_j)
\end{split}
\end{equation}
Introducing a \emph{prior} $t^{(0)}$ for the true distribution, which is taken to be the true distribution predicted by MC, the estimator of \myeq~\ref{eq:bayesian_unfolding_derivation} can be rewritten as
\begin{equation}\label{eq:bayesian_iterative_unfolding}
	\hat{t}^{(1)}_i= \frac{1}{\phi_j}\, \sum_j \frac{\tilde{R}_{ji} t^{(0)}_i}{\sum_k \tilde{R}_{jk} t^{(0)}_k}\, (m_j - b_j).
\end{equation}
The normalisation of the prior is irrelevant, as it cancels in \myeq~\ref{eq:bayesian_iterative_unfolding}.
The dependence of the result on the prior is reduced by iterating the procedure, making the substitution $t^{(0)} \leftarrow \hat{t}^{(1)}$ to compute $\hat{t}^{(2)}$, and so on. In this unfolding method, the number of iterations is the regularisation parameter. Fewer iterations $n$ mean smaller statistical fluctuations of the result $\hat{t}^{(n)}$, but at the cost of a larger dependence on the prior, i.e.~bias. The number of iterations is optimised for each observable according to criteria defined below and the residual bias is quantified. The software implementation of the Bayesian iterative unfolding used in this analysis is based on the \textsc{RooUnfold} libraries \cite{roounfold,Adye:2011gm}. \myfig~\ref{fig:prior_convergence} illustrates the convergence of the result with the number of iterations, using the example of the central-jet multiplicity distribution. It shows the background-subtracted data divided by the fake correction, $(m - b) / \phi$, as well as the result after $n = 1$, 2, 3, 5, 10 iterations multiplied by the response matrix (excluding the fake correction) to transform it back to the reconstruction-level, $\tilde{R}\, t^{(n)}$. It can be seen that the unfolded result transformed back to the reconstruction-level agrees better with the background-subtracted data with more iterations. In machine-learning language terms, one could say that the unfolding procedure \emph{learns} more and more \emph{features} of the data with each iteration. The cost, as discussed above, is less stability of the result in the presence of statistical fluctuations.

\begin{figure}[h!]
\centering
\includegraphics[width=0.6\textwidth]{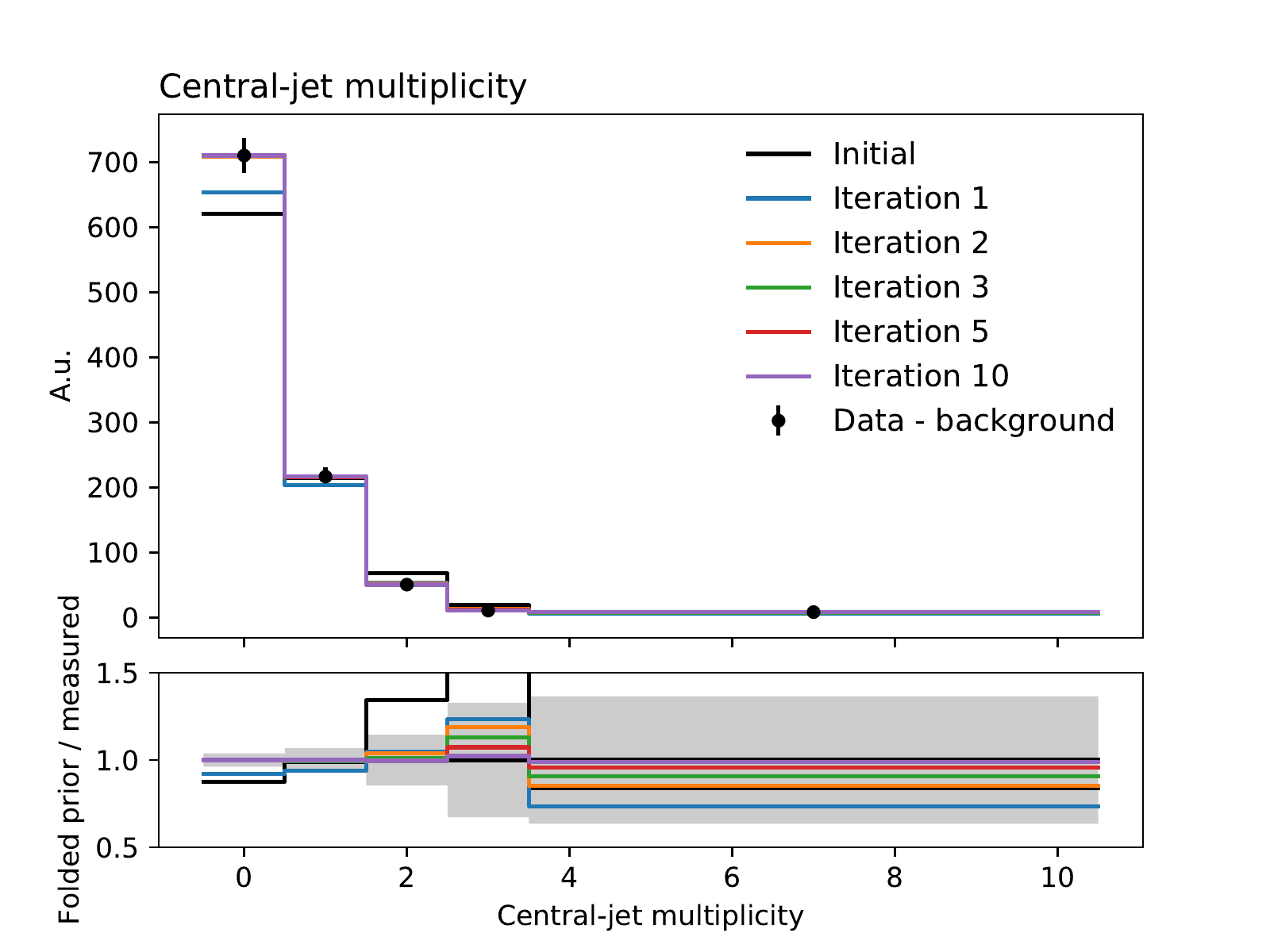}
\caption{Convergence of the unfolded result multiplied by the response matrix (excluding the fake correction) $\tilde{R}$ with the number of iterations. The bottom panel shows the ratio to the background-subtracted data after each considered number of iterations. The error bars (top panel) and grey band (bottom panel) indicate the statistical uncertainty of the data.}
\label{fig:prior_convergence}
\end{figure}


\subsection{Binning optimisation}
\label{sec:binning_optimisation}

The binning in itself constitutes a regularisation: wider bins will decrease bin migrations and relative statistical uncertainties. Finer bins have the advantage of giving a more detailed view of the distribution.
They also decrease the effect that the observable of interest might be distributed differently inside a given bin in data and prediction, which could lead to differences between the predicted and actual response, if the response depends on the observable of interest.
The binning of each unfolded observable is optimised under the following rough criteria, in order of precedence:
\begin{enumerate}
	\item The bin range covers all observed data events, except in jet observables, where bins containing less jets than required for a non-trivial value are excluded (e.g.~the scalar \pt{} sum of all jets distribution begins at 30~\GeV{} due to the jet \pt{} threshold --- all events below 30~\GeV{} lie exactly at 0~\GeV{}),
	\item Each bin has at least 10 expected events, corresponding to a predicted probability of the bin ending up not containing any data events of $< 10^{-4}$ as well as an expected statistical uncertainty of $\lesssim 30\%$,
	\item All expected major features (rises, drops, peaks) of the distribution are resolved,
	\item The purity in each bin is at least $\sim$70\%, except in individual bins that are motivated by the previous criterion.
\end{enumerate}
Within the above approximate constraints, the binning is chosen as fine as reasonable.

A remark: the unfolding performed in this analysis is one-dimensional, i.e.~of only one observable at a time. The detector response might depend on observables other than the observable of interest, so if the underlying kinematic configurations differ between data and prediction, there might be a mismodelling in the response. This is true \emph{even if} the observable of interest is modelled well by the prediction. The effect may be worse for wider bins, since they integrate over more underlying phase space and are therefore more dependent on the predicted distributions themselves being accurate, whereas finer bins are more determined by the detector response alone. 

\subsection{Number of unfolding iterations}
\label{sec:iteropt}

Once the binning is fixed (\mysec~\ref{sec:binning_optimisation}), the number of iterations in the Bayesian iterative unfolding is optimised to balance two competing effects.
\begin{enumerate}
	\item Fewer iterations mean a higher degree of regularisation, so the unfolded distribution tends to be smoother and less prone to fluctuation. The price is a larger regularisation bias of the unfolded result.
	\item More iterations mean a less biased estimator for the true contents of each bin, but an unfolded distribution that is more prone to large bin-by-bin fluctuations and larger statistical uncertainty of the unfolded result.
\end{enumerate}

The regularisation bias is estimated using a data-driven method \cite{Malaescu:2009dm}. The initial priors are reweighted by a smooth polynomial function such that the agreement between the prior folded with the response matrix and the observed data is very good. The folded reweighted prior is unfolded using the nominal response matrix. The deviations of the obtained unfolded distribution from the reweighted prior are used as the unfolding bias uncertainty in each bin.
The idea of the smoothing by fitting a polynomial is to make the reweighting correct for physical differences between the predicted and observed distribution, rather than reweighting random statistical fluctuations in the data. It also serves to reduce the binning bias by allowing the true bin centre to shift.
First- to fifth-order polynomials are fitted to the ratio of data over the reconstruction-level prediction, with the order chosen for each observable to model the ratio reasonably. This relies on the analyser's subjective judgment of what is a physical feature and what is a statistical fluctuation. Examples of fitted polynomials used for the reweighting are shown in \myfig~\ref{fig:ddclosure_fits}. The corresponding figures for the other unfolded observables are shown in \myapp~\ref{sec:zz_aux_reweighting}.

\begin{figure}[h!]
\centering
\subfigure{\includegraphics[width=0.49\textwidth]{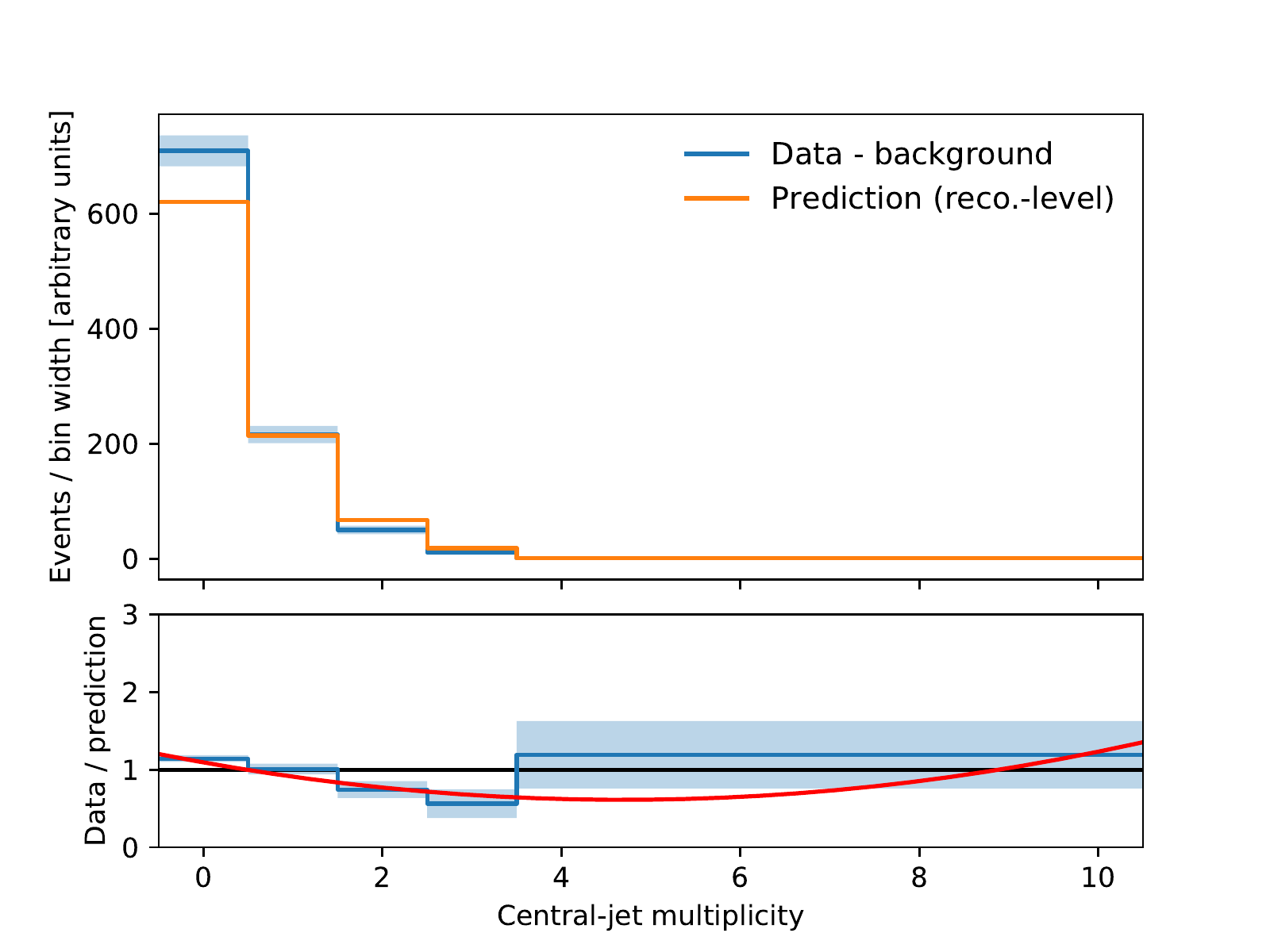}}
\subfigure{\includegraphics[width=0.49\textwidth]{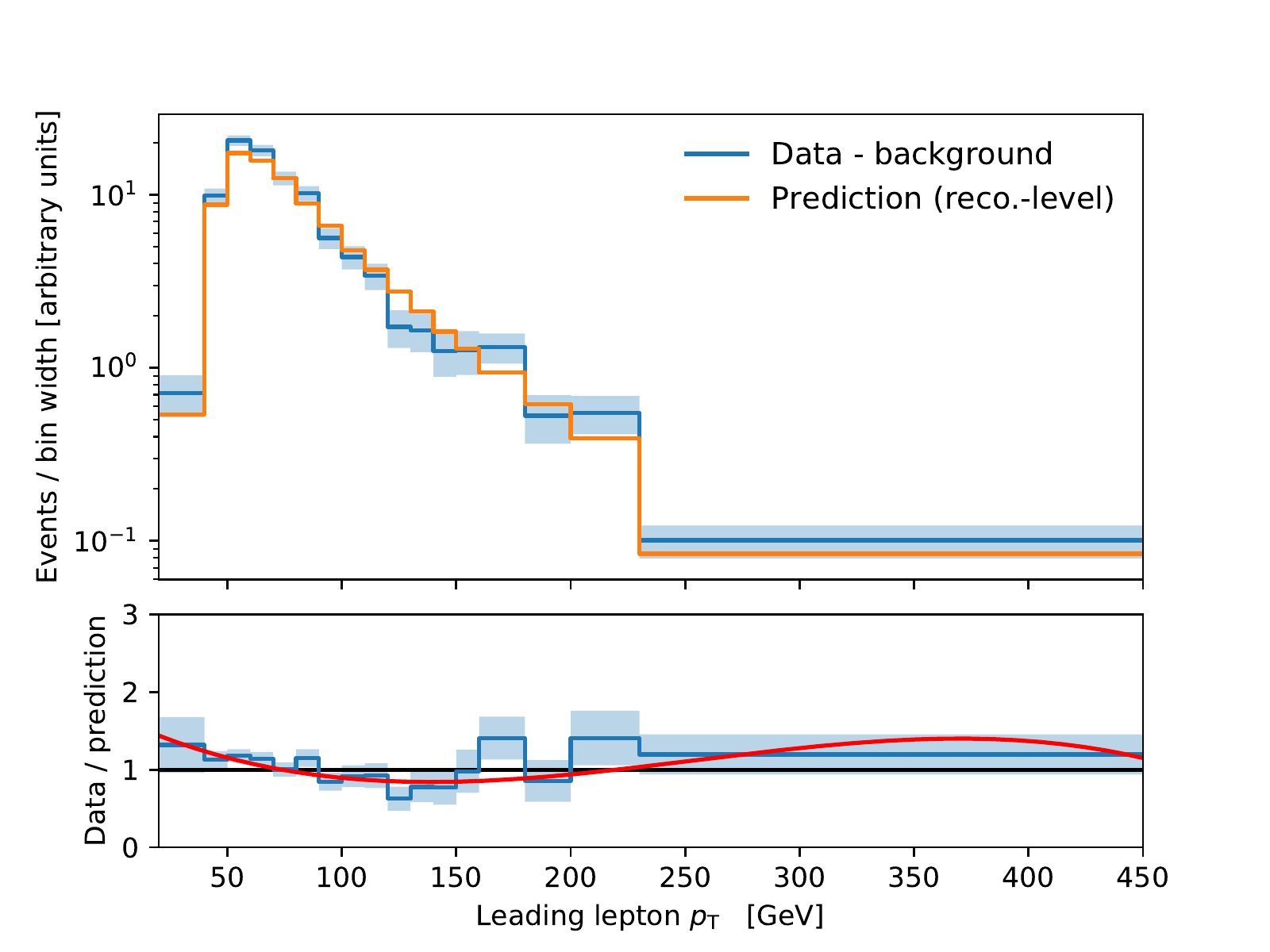}}
\caption{Background-subtracted data compared to the reconstruction-level prediction for two different observables. The ratio is fitted with a polynomial, visualised as a red curve.}
\label{fig:ddclosure_fits}
\end{figure}

The statistical and unfolding-method uncertainties for various numbers of iterations are shown in \myfig~\ref{fig:stat_bias} for a number of observables. The number of iterations for each observable is chosen to ensure a very small bias, around 1--2\% wherever possible. Furthermore, for the chosen number of iterations, the unfolding result re-folded by the response matrix must have converged to the background-subtracted data to within the statistical uncertainty of the data. An example of this convergence is shown in \myfig~\ref{fig:prior_convergence} above. The corresponding figures for the other unfolded observables are shown in \myapp~\ref{sec:zz_aux_statbias}.

\begin{figure}[h!]
\centering
\includegraphics[width=0.48\textwidth]{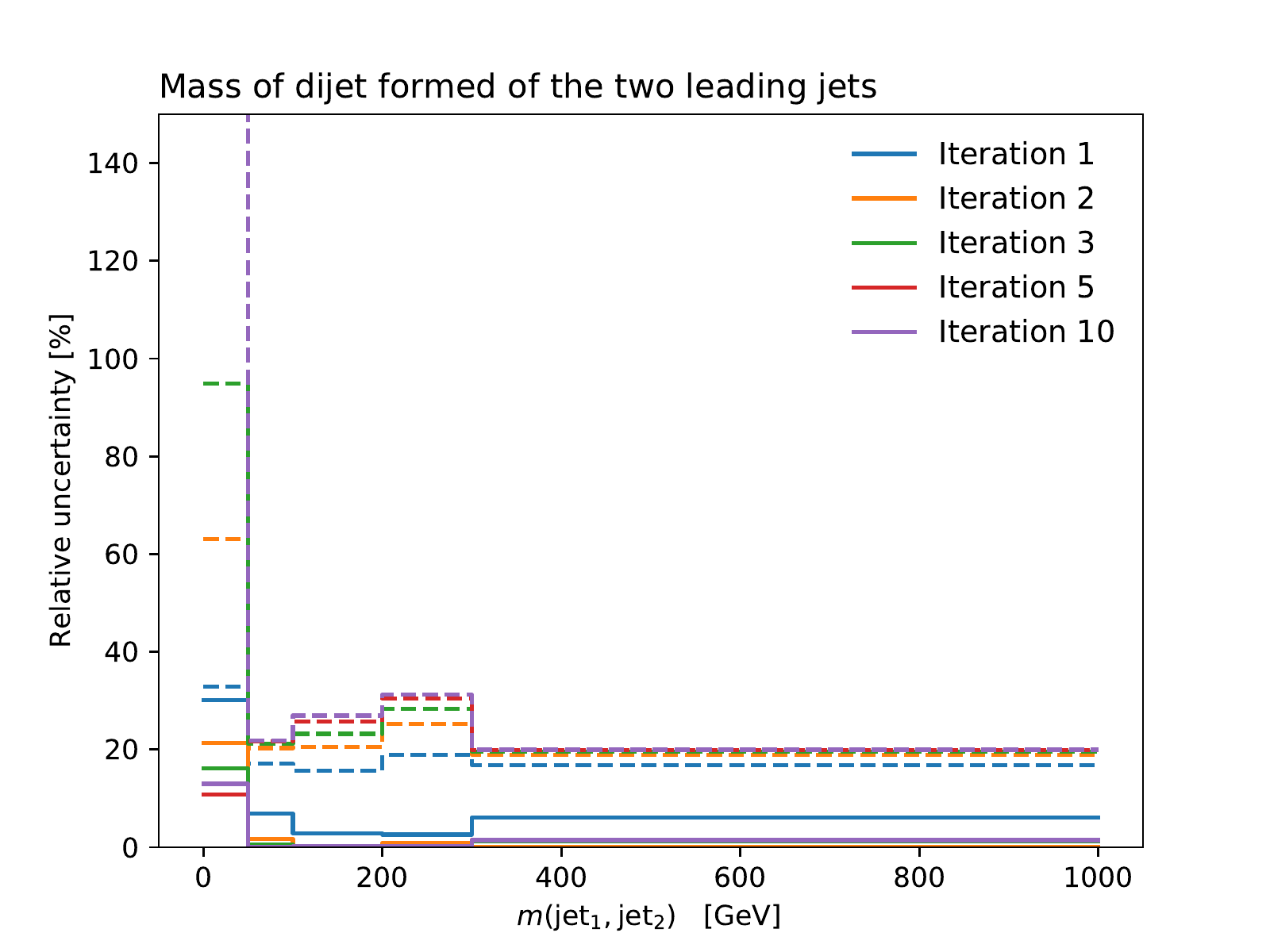}
\includegraphics[width=0.48\textwidth]{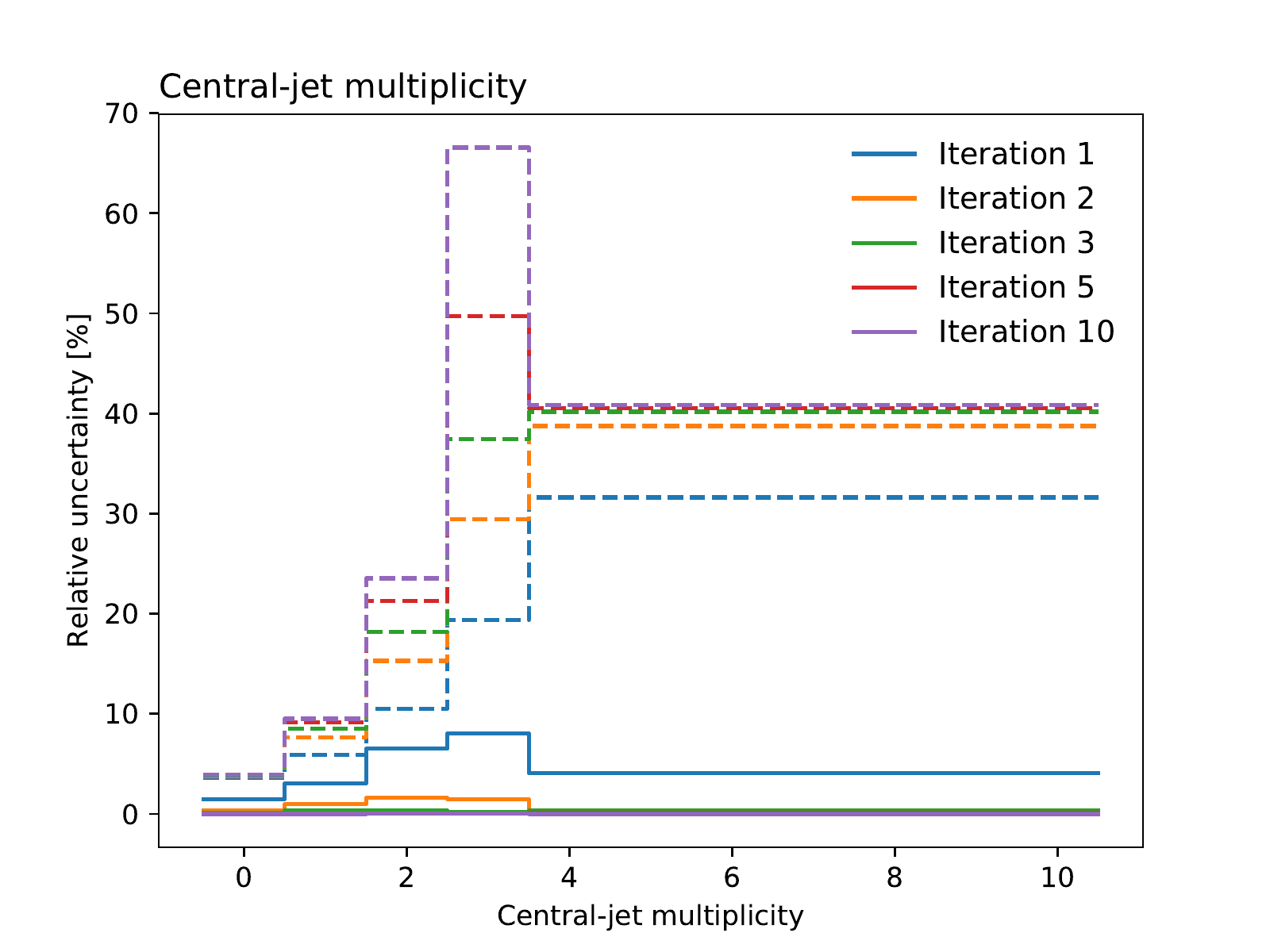}
\caption{Statistical uncertainty (dashed) and unfolding method uncertainty (solid) for various numbers of iterations for two different observables.} 
\label{fig:stat_bias}
\end{figure}

To make sure that the statistical uncertainty is not overly constrained by the unfolding procedure, it is verified that it is not significantly smaller than it would be when using a simple bin-by-bin unfolding with a MC-derived correction factor in each bin, ignoring bin migrations.\footnote{There is ongoing discussion among the interested members of ATLAS about whether this criterion is overly conservative or valid. In the author's view, it is overly conservative. Per-bin statistical uncertainties could be expected to be reduced when taking into account bin migrations, as each bin value helps constrain its neighbouring bins. This happens at the `cost' of statistical uncertainties that are (quantifiably) correlated between nearby bins. In bin-by-bin unfolding, useful information is lost by ignoring bin correlations.} The bin-by-bin unfolding corresponds to inverting the approximated, diagonal (and therefore in a sense maximally regularised) response matrix
\begin{equation*}
\tilde{R}^{\text{\kern1ptbin-by-bin}}_{ij} = \frac{w^{\text{reco.}}_i}{w^{\text{true}}_j} \delta_{ij},
\end{equation*}
where $w^{\text{reco.}}_i$ ($w^{\text{true}}_j$) designates the sum of weights of MC events passing the reconstruction-level (particle-level) selection and falling in bin $i$ (\kern1pt$j$), and $\delta_{ij}$ is the Kronecker delta. \myfig~\ref{fig:unfolding_statistical_uncertainty_vs_binbybin} compares the statistical uncertainty obtained with one or two iterations to that from bin-by-bin unfolding, as a function of the purity of the bin.\footnote{The definition of purity used in this figure is slightly different from that used in the rest of the thesis, giving it a tendency towards lower values, but that does not affect the conclusions.} It shows that at least two iterations should be performed according to the above criterion. It also reveals a clear dependency on the purity. As the purity tends towards unity, so must the ratio. For one iteration, the ratio decreases with decreasing purity, for which a possible explanation is that taking into account bin migration constrains the content of one bin via its correlation with the contents of nearby bins. This effect would be more pronounced for low-purity than high-purity bins, since the former exhibit a stronger correlation between neighbouring bins.

\begin{figure}[h!]
\centering
\includegraphics[width=0.6\textwidth]{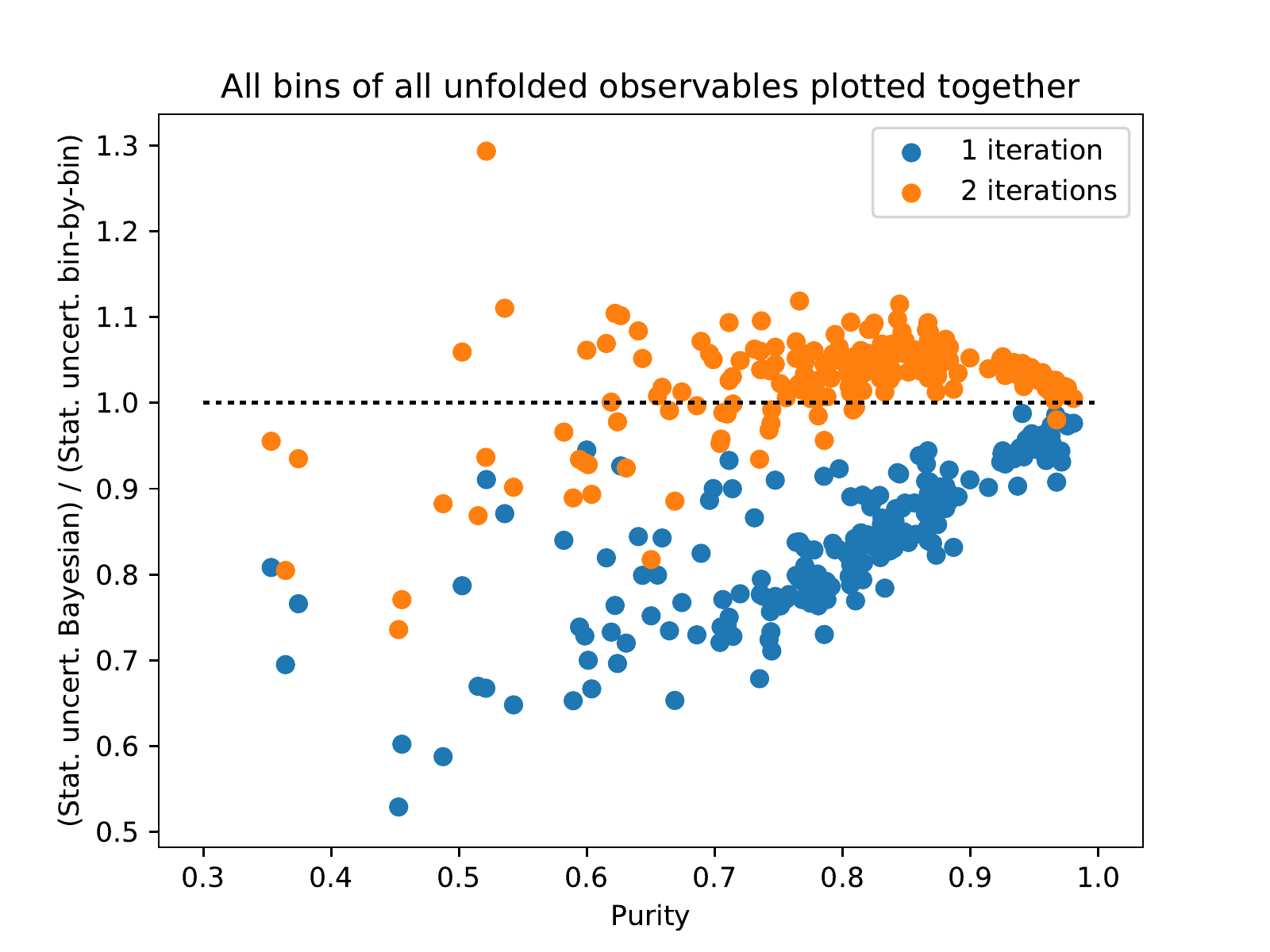}
\caption{Comparison of statistical uncertainties obtained with Bayesian iterative unfolding (one or two iterations) to those from bin-by-bin unfolding, as a function of the purity in that bin. Each dot corresponds to one bin of one unfolded observable, all of which are shown together.}
\label{fig:unfolding_statistical_uncertainty_vs_binbybin}
\end{figure}

The final number of iterations chosen for each distribution following the above guidelines is shown in \mytab~\ref{tab:unfolding_iterations}.

\begin{table}[h!]
\centering
\begin{tabular}{ll}
\toprule
\textbf{Distribution} & \textbf{Iterations}\\
\midrule
1.~lepton \pt{} & 2\\
2.~lepton \pt{} & 2\\
3.~lepton \pt{} & 2\\
4.~lepton \pt{} & 2\\
$\ptz$ & 2\\
$\ptzsub$ & 2\\
\ptfourl{} & 2\\
$|\yfourl{}|$ & 2\\
$\dphiZZ / \pi$ & 2\\
$|\dyZZ{}|$ & 2\\
$N_{\text{jets}}$ & 2\\
$N_{\text{central jets}}$ & 3\\
$N_{\text{jets}}$, $\pt > 60$ & 3\\
1.~jet \pt{} & 2\\
2.~jet \pt{} & 3\\
1.~jet $|\eta|$ & 2\\
2.~jet $|\eta|$ & 3\\
\mjj{} & 2\\
$|\dyjj{}|$ & 2\\
Jets scalar \pt{} sum & 3\\
\bottomrule
\end{tabular}
\caption{Number of unfolding iterations for each observable.}
\label{tab:unfolding_iterations}
\end{table}


\subsection{Combining channels}

The differential cross sections are only measured in the combination of all three signal channels, due to the large statistical uncertainties in the individual channels. The channels are combined by summing the reconstruction-level data and predictions and using a single response matrix. In the future, one response matrix per channel should be used and the results combined by summing only after the unfolding. This way, the overall response of the detector is modelled more accurately, since the actually observed contribution of each channel is taken into account. The reason why this was not done in this analysis is the low statistical precision in the individual channels, which makes the unfolding less stable. A combination before unfolding alleviates this effect.

\subsection{Propagation of uncertainties}
\label{sec:unfolding_uncertainties}

The statistical uncertainty due to fluctuations in the data is estimated by generating 2000 sets of random pseudodata following a Poisson distribution in each bin whose expectation value is the number of observed data events in that bin. The unfolding is repeated with the pseudodata sets, taking the root mean square of the deviation of the resulting unfolded spectrum from the actual unfolded data as the statistical uncertainty in each bin. The statistical uncertainty of the data is in the range 5--41\%, except in the bin from 2.0 to 2.5 of the absolute pseudorapidity of the subleading jet, where it is 85\%. It dominates the total uncertainty in most bins. The uncertainty due to statistical fluctuations in the MC simulations used to obtain the response matrix is obtained the same way, repeating the unfolding using randomly fluctuated copies of the response matrix.
Experimental and theoretical systematic uncertainties are estimated by repeating the unfolding with the varied response matrix and taking the deviation from the nominal of the resulting unfolded distribution as the uncertainty. In jet-inclusive observables, the largest systematic uncertainty comes from the theoretical modelling of the response matrix, composed of the PDF and QCD scale variations as well as the difference between using \POWHEGpy{} and \SHERPA{} to model the \qq{}-initiated prodution, added in quadrature (up to approximately~25\%). In jet-exclusive observables, the jet energy scale uncertainty is an additional large contribution (3--23\%).
Background uncertainties are estimated by subtracting the varied background predictions from the data before unfolding and are small.
The uncertainty due to the unfolding method is determined as described in \mysec~\ref{sec:iteropt}. It is smaller than 1\% in almost all bins, but reaches up to 22\% in individual bins (such as the first bin of the mass of the two leading-\pt{} jets, where the modelling of the data is poor).
\myfig{}~\ref{fig:unfolding_uncertainties} shows a bin-by-bin breakdown of uncertainties for selected observables. The corresponding distributions for the remaining unfolded observables are shown in \myapp~\ref{sec:zz_aux_uncerts}. In the figures, the theory uncertainty contains the contributions of the generator, PDF, and QCD scale variations, summed in quadrature. The jet, electron, and muon reconstruction uncertainties contain the efficiency and calibration uncertainties of the corresponding objects. Uncertainties from different systematic sources, as well as the statistical uncertainty, are added in quadrature in each bin in the final results.
Bin-by-bin statistical- and total-uncertainty correlation matrices for all observables are included in \myapp~\ref{sec:zz_aux_correlations}. 
The statistical uncertainties of the data are typically only weakly correlated between (mainly adjacent) bins, which is expected, because bin migrations are mostly small. The systematic uncertainties, on the other hand, may be strongly correlated between bins, as expected. As a result, correlations of $\mathcal{O}(50\%)$ and more between adjacent bins occur where the systematic uncertainties dominate.

\begin{figure}[h!]
\centering
\subfigure[Transverse momentum of the four-lepton system.]{\includegraphics[width=0.48\textwidth]{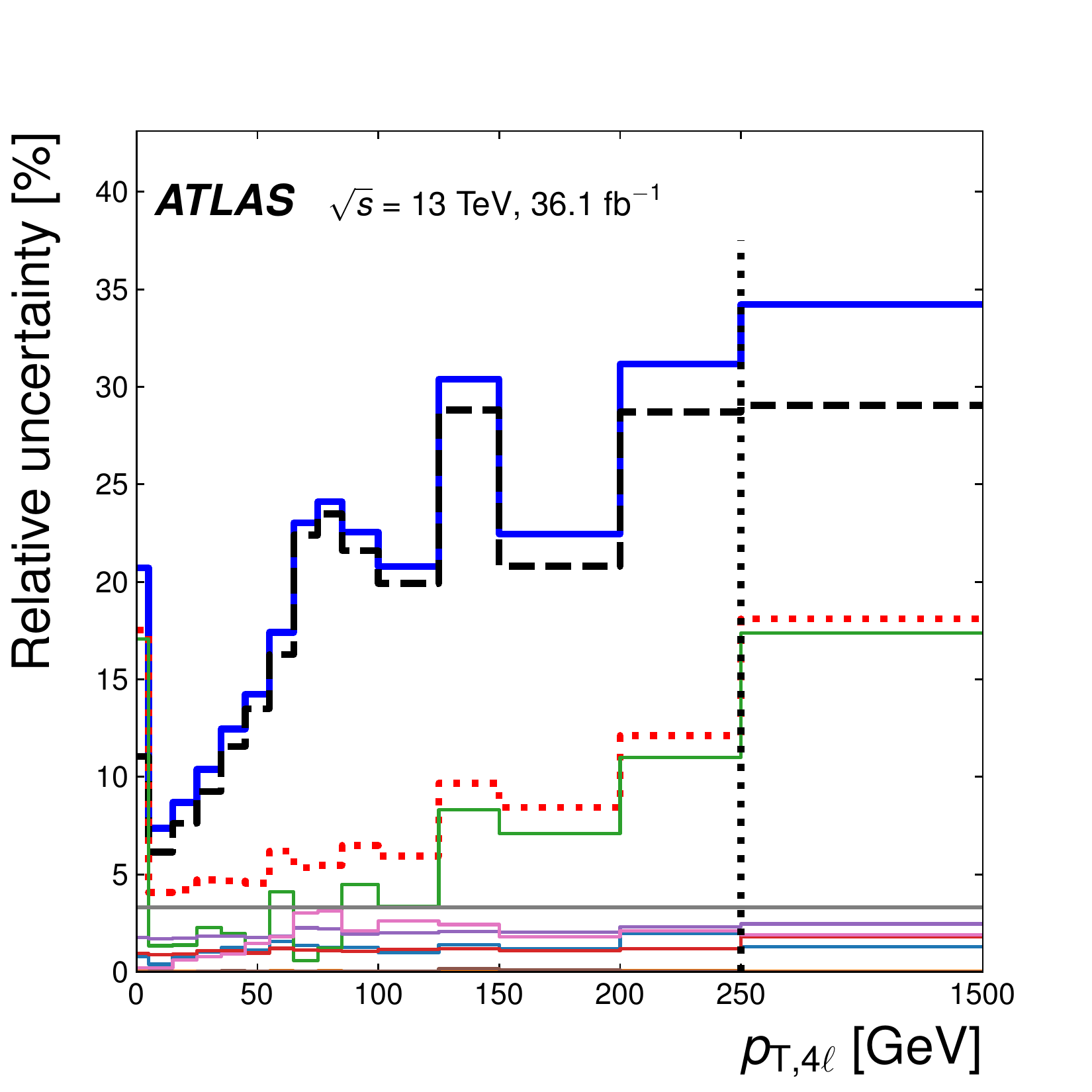}}
\subfigure[Jet multiplicity, considering all selected jets.]{\includegraphics[width=0.48\textwidth]{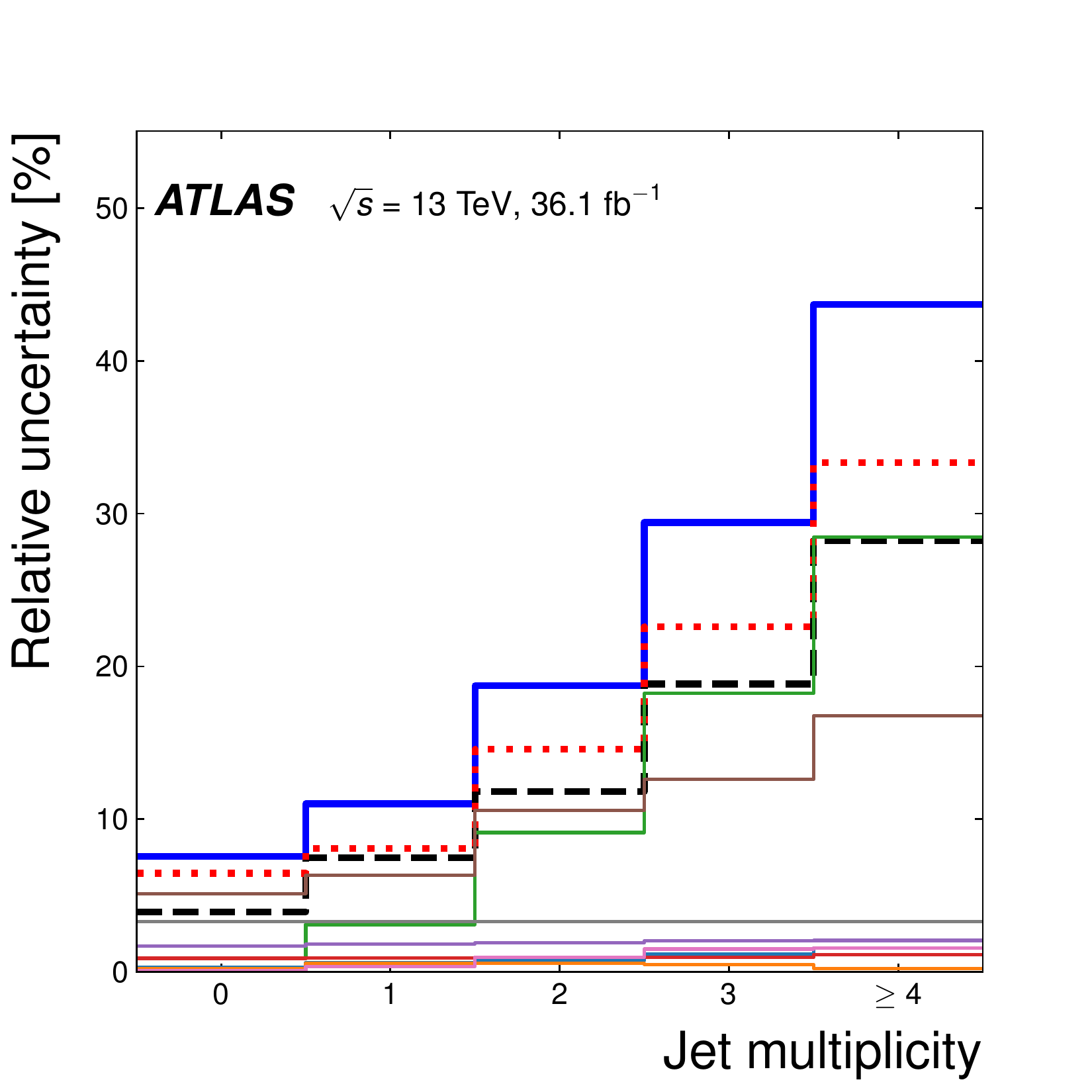}}
\subfigure[Invariant mass of the two leading-\pt{} jets.]{\includegraphics[width=0.48\textwidth]{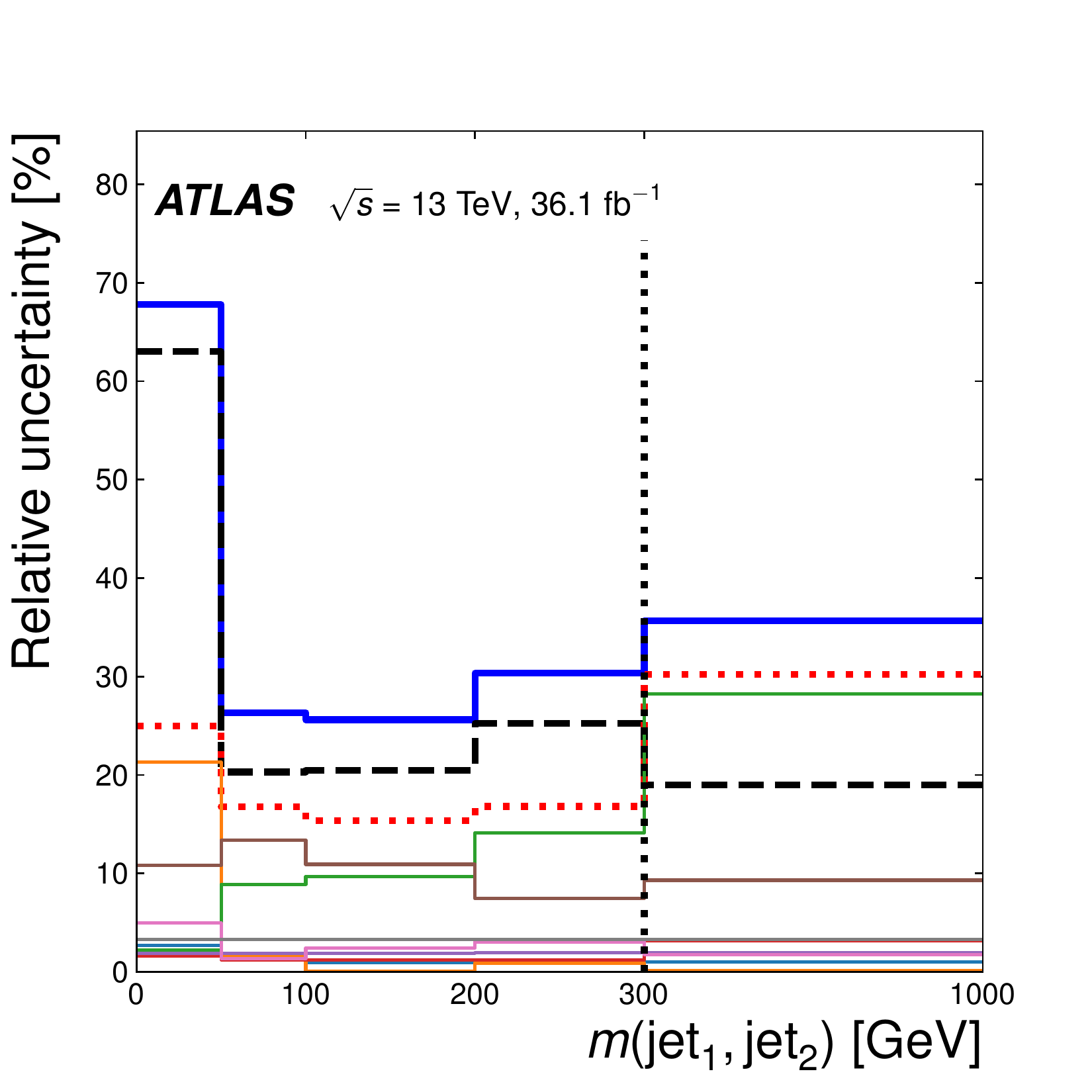}}
\subfigure{\includegraphics[width=0.48\textwidth]{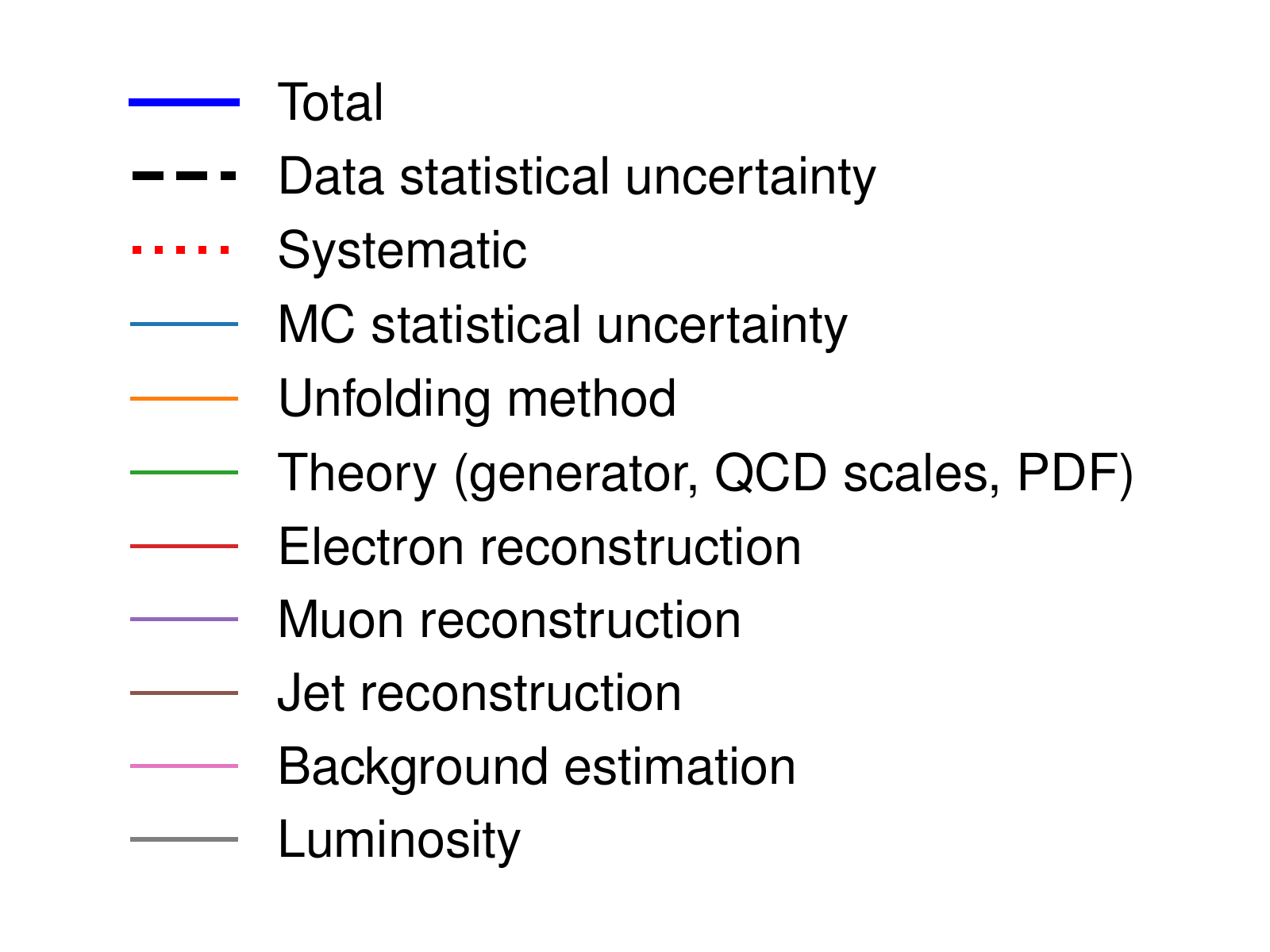}}
\caption{Uncertainty contributions after unfolding in each bin of three representative observables. The total systematic uncertainty contains all uncertainties except the statistical uncertainty of the data, summed in quadrature. For better visualisation, the last bin is shown using a different $x$-axis scale where indicated by the dashed vertical line. Published in \myref~\cite{STDM-2016-15}.}
\label{fig:unfolding_uncertainties}
\end{figure}



\pagebreak
\subsection{Results}

This section shows those results published in \myref~\cite{STDM-2016-15}. Following sections show additional results, notably including the unfolded four-lepton mass spectrum.

Figures~\ref{fig:zz_kinematics}--\ref{fig:dijets} present the unfolded cross sections, along with comparisons to various fixed-order and parton-showered theoretical predictions. Reasonable agreement of the various predictions with the data is observed. The fixed-order predictions are only shown for jet-inclusive observables. While the NNLO predictions can include up to two jets (from the real-emission matrix elements), jet-exclusive distributions were not supported out-of-the-box by \matrixnnlo{} at time of the analysis, and were not implemented by the author due to time constraints.

For the bins in which the greatest discrepancy is observed, the significance is estimated approximately. To profit from the simplification that Poisson statistics bring, the reconstruction-level yields in the corresponding bins are used, rather than the differential cross section. While bin migrations mean that this is not always the case in general, the largest discrepancies observed in the differential cross sections here have corresponding discrepancies in the reconstruction-level distributions. (In the future, differential cross sections may become more commonly used for statistical hypothesis testing. However, this is more complicated than using measured yields: yields are uncorrelated between different bins, they obey Poisson statistics, and all systematic uncertainties can be treated as only affecting the predictions, not the measurement. None of the above is true for differential cross sections obtained by unfolding.)
The reconstruction-level distributions for the observables in question can be found in \myfig~\ref{fig:data_mc_plots} or \myapp~\ref{sec:zz_aux_datamc}. The $p$-value is estimated as \cite{Cranmer:PhyStat2005,Linnemann:PhyStat2003}
\begin{equation}\label{eq:significance_estimation}
p = B\left(\frac{y\delta^2}{y\delta^2 + 1},\; x,\; \frac{\delta^{2}+1}{\delta^2}\right),
\end{equation}
where $B(\cdot, \cdot, \cdot)$ is the incomplete Beta function, $x$ ($y$) is the observed (predicted) number of events in the bin of interest, and $\delta$ is the corresponding relative (systematic and statistical) uncertainty of the prediction uncertainty. The observed yield $x$ is a random variable following a Poisson distribution and does not have any associated systematic uncertainties.


\newcommand{\datauncertcomment}{The statistical uncertainty of the measurement is shown as error bars, and shaded bands indicate the systematic uncertainty and the total uncertainty obtained by summing the statistical and systematic components in quadrature.}

\clearpage

\myfig~\ref{fig:zz_kinematics}(a) shows the transverse momentum of the four-lepton system, $\ptfourl$. The cross section has a peak around 10~\GeV{} and drops rapidly towards both lower and higher values. At low \ptfourl{}, the resummation of low-\pt{} parton emissions is important and fixed-order descriptions are inadequate. For this reason, the fixed-order predictions are not shown in the first two bins, 0--5~\GeV{} and 5--15~\GeV{}. The region below $\ptfourl = 60~\GeV{}$ is modeled slightly better by predictions that include a parton shower, again suggesting the importance of resummation. Above 60~\GeV{}, the fixed-order NNLO predictions describe the data slightly better. \myfig~\ref{fig:zz_kinematics}(b) shows the absolute rapidity of the four-lepton system, which drops gradually towards high values. The highest kinematically possible value can be found using \myeq~\ref{eq:bjorkenx_rapidity}. It is maximised when one of the incoming partons carries all of its proton's momentum, $x = 1$, and the partonic centre-of-mass energy is minimal, $\sqrt{\hat{s}} = 2 \times 66$~\GeV, which is the kinematic threshold for producing two dileptons of mass $m_{\ell^+\ell^-} > 66$~\GeV{}. This yields $y_{\text{max}} = \log\frac{2 \times 66~\GeV{}}{13\,000~\GeV{}} \approx 4.6$. The \yfourl{} distribution is potentially sensitive to a different choice of PDF, describing the momentum distribution of the incoming partons. Fixed-order calculations and predictions including a parton shower model this observable reasonably well, within the statistical and systematic uncertainties. The predictions tend to slightly underestimate the cross sections for small values of $|\yfourl{}|$.

\begin{figure}[h!]
\centering
\subfigure[]{\includegraphics[width=0.48\textwidth]{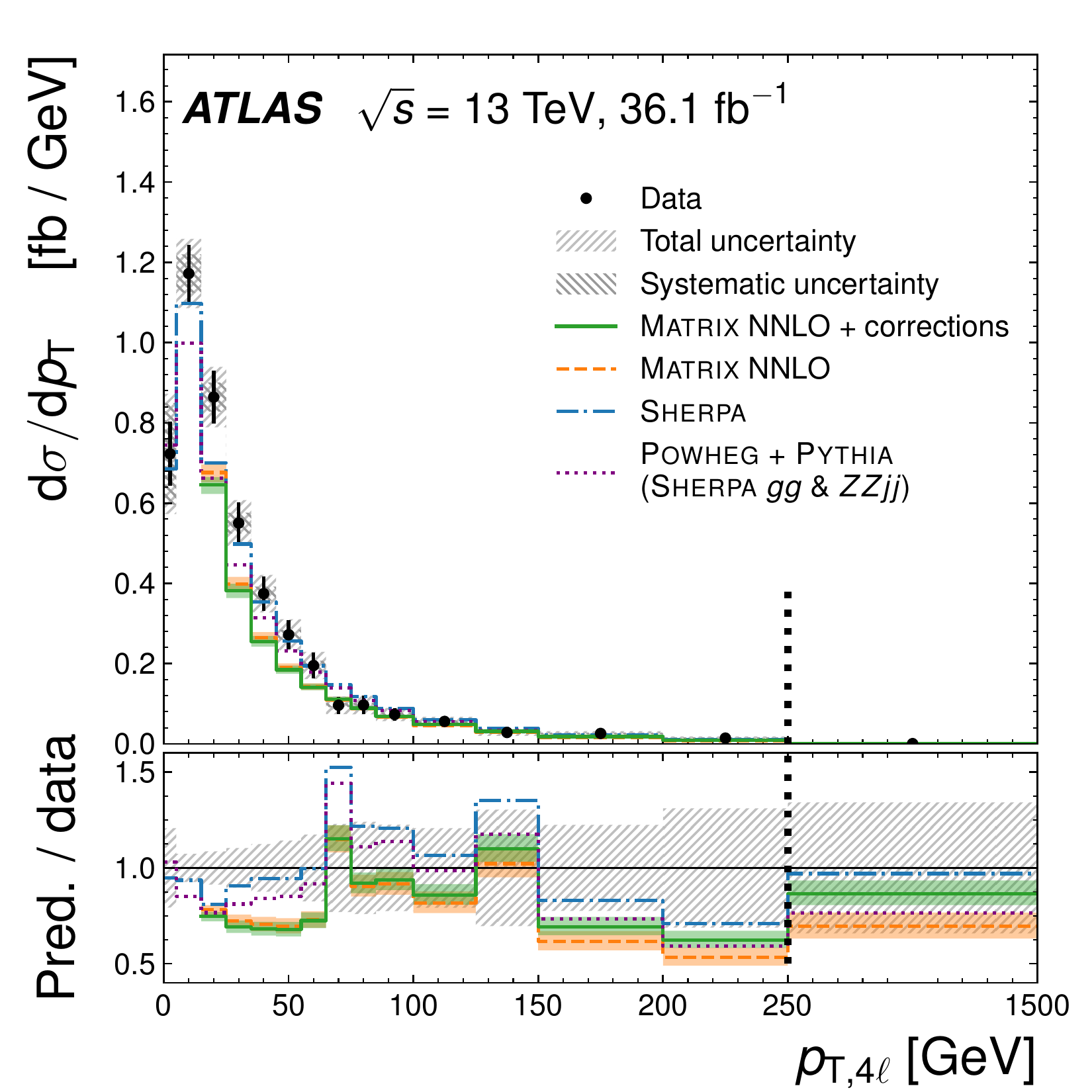}} 
\subfigure[]{\includegraphics[width=0.48\textwidth]{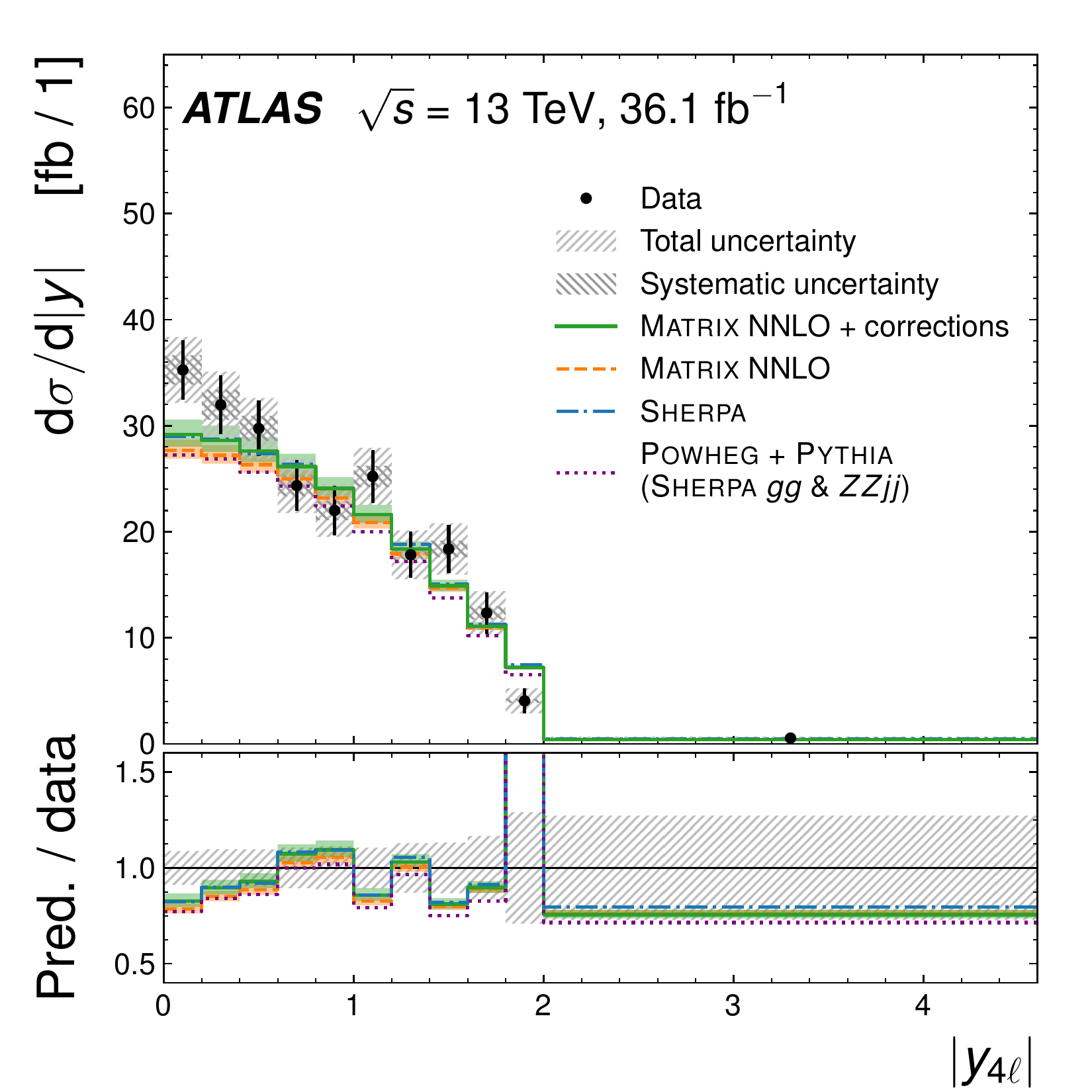}} 
\caption{Measured and predicted differential cross sections for (a) the transverse momentum and (b) the absolute rapidity of the four-lepton system. \datauncertcomment{} A pure NNLO calculation from \matrixnnlo{} is shown with no additional corrections applied. Another prediction is shown based on this NNLO calculation, with the $\Pgluon\Pgluon$-initiated contribution multiplied by a global NLO correction factor of $1.67$. For the $\ptfourl$ distribution in (a), the NLO weak correction is applied as a global factor of $0.95$ as a differential calculation is not available. For the $|\yfourl|$ distribution in (b), the EW correction factor is applied in each bin. The contribution from EW-$\ZZ jj$ generated with \sherpa{} is added. For the fixed-order predictions, the QCD scale uncertainty is shown as a shaded band. Parton-showered \POWHEGpy{} and \SHERPA{} predictions are also shown.  \brokenaxiscomment{} Published in \myref~\cite{STDM-2016-15}.}
\label{fig:zz_kinematics}
\end{figure}

\clearpage
\myfig{}~\ref{fig:zz_angles}(a) presents the azimuthal angle separation between the two \PZ{} boson candidates. The fixed-order predictions only describe the shape of the $\gluglu$-initiated process at LO and therefore predict a distribution that is more peaked at $\pi$ than those from \SHERPA{} and \POWHEGpy{}, where the parton shower shifts some events towards lower values. In this distribution, the way the $k$-factors for missing higher orders are applied actually gives a misleading result: the $k$-factor is calculated entirely in the bin containing $\dphiZZ = \pi$, which is the only possible value at LO. At NLO, this bin \emph{decreases}, as events migrate to values below $\pi$, even though the total cross section increases. The constant $k$-factor applied here does not capture this effect (or any other shape change of the distribution), so it wrongly predicts an \emph{increase} of the highest bin when adding higher-order corrections. This illustrates a fundamental shortcoming of the constant-$k$-factor approach.
\myfig{}~\ref{fig:zz_angles}(b) shows the absolute rapidity difference of the two \PZ{} boson candidates, which drops towards high values and is modeled by all calculations within the uncertainties.

\begin{figure}[h!]
\centering
\subfigure[]{\includegraphics[width=0.48\textwidth]{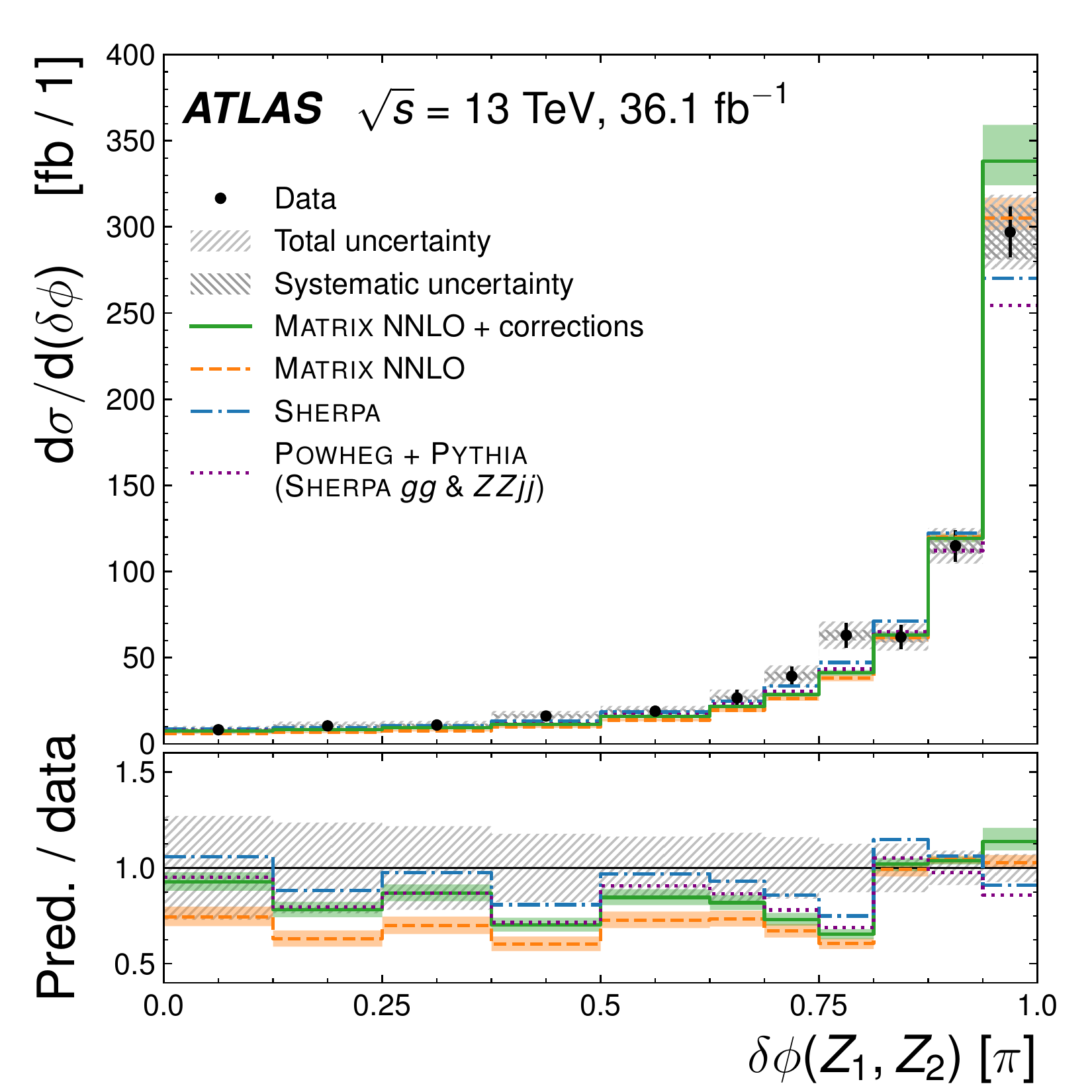}} 
\subfigure[]{\includegraphics[width=0.48\textwidth]{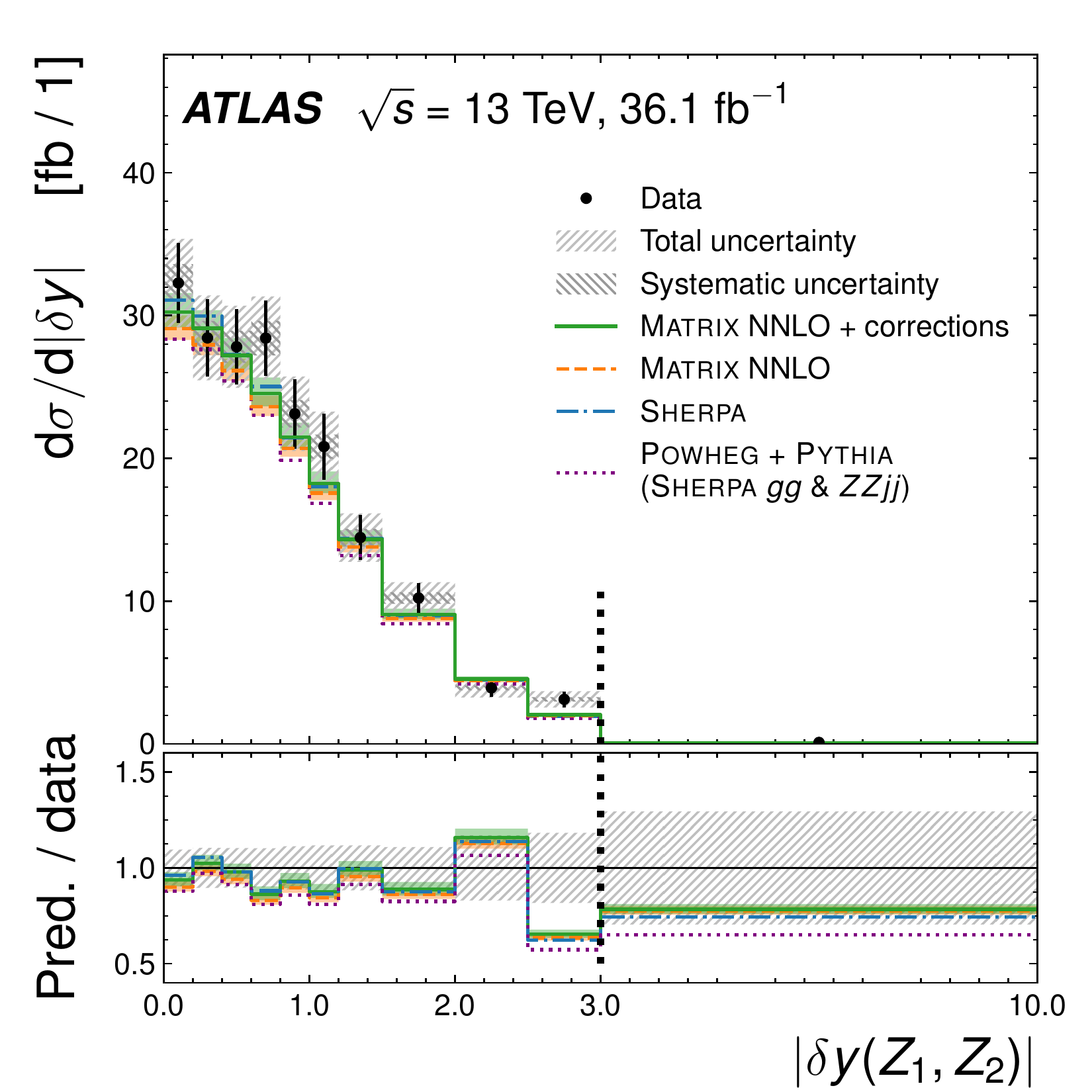}} 
\caption{Measured and predicted differential cross sections for (a) the azimuthal angle separation and (b) the absolute rapidity difference between the two $\PZ$ boson candidates. \datauncertcomment{} A pure NNLO calculation from \matrixnnlo{} is shown with no additional corrections applied. Another prediction is shown based on this NNLO calculation, with the $\Pgluon\Pgluon$-initiated contribution multiplied by a global NLO correction factor of $1.67$. For the $\dphiZZ$ distribution in (a), the NLO weak correction is applied as a global factor of $0.95$ as a differential calculation is not available. For the $|\dyZZ|$ distribution in (b), the EW correction factor is applied in each bin. The contribution from EW-$\ZZ jj$ generated with \sherpa{} is added. For the fixed-order predictions, the QCD scale uncertainty is shown as a shaded band. Parton-showered \POWHEGpy{} and \SHERPA{} predictions are also shown. \brokenaxiscomment{} Published in \myref~\cite{STDM-2016-15}.}
\label{fig:zz_angles}
\end{figure}

\clearpage
\myfig~\ref{fig:z_pts} shows the transverse momentum of the leading-\pt{} and subleading-\pt{} \PZ{} boson candidates, exhibiting a wide peak around 50~\GeV{} and 30~\GeV{}, respectively. Anomalous triple gauge couplings (as discussed in Section~\ref{sec:anomalous}) would manifest as an excess in the cross section at large values of the transverse momentum of the \PZ{} bosons, which is not observed in these differential cross section distributions (the last bin in each distribution is consistent with the SM predictions). The discrepancies at \pt{} of about 50~\GeV{}, 90~\GeV{} in the leading \PZ{} boson candidate are related to the excesses seen in \myfig~\ref{fig:data_mc_plots}(c). The local significance of these excesses with respect to the \SHERPA{} prediction is estimated to be 2.3 and 2.0 standard deviations respectively, calculated using \myeq~\ref{eq:significance_estimation}. 


\begin{figure}[h!]
\centering
\subfigure[]{\includegraphics[width=0.48\textwidth]{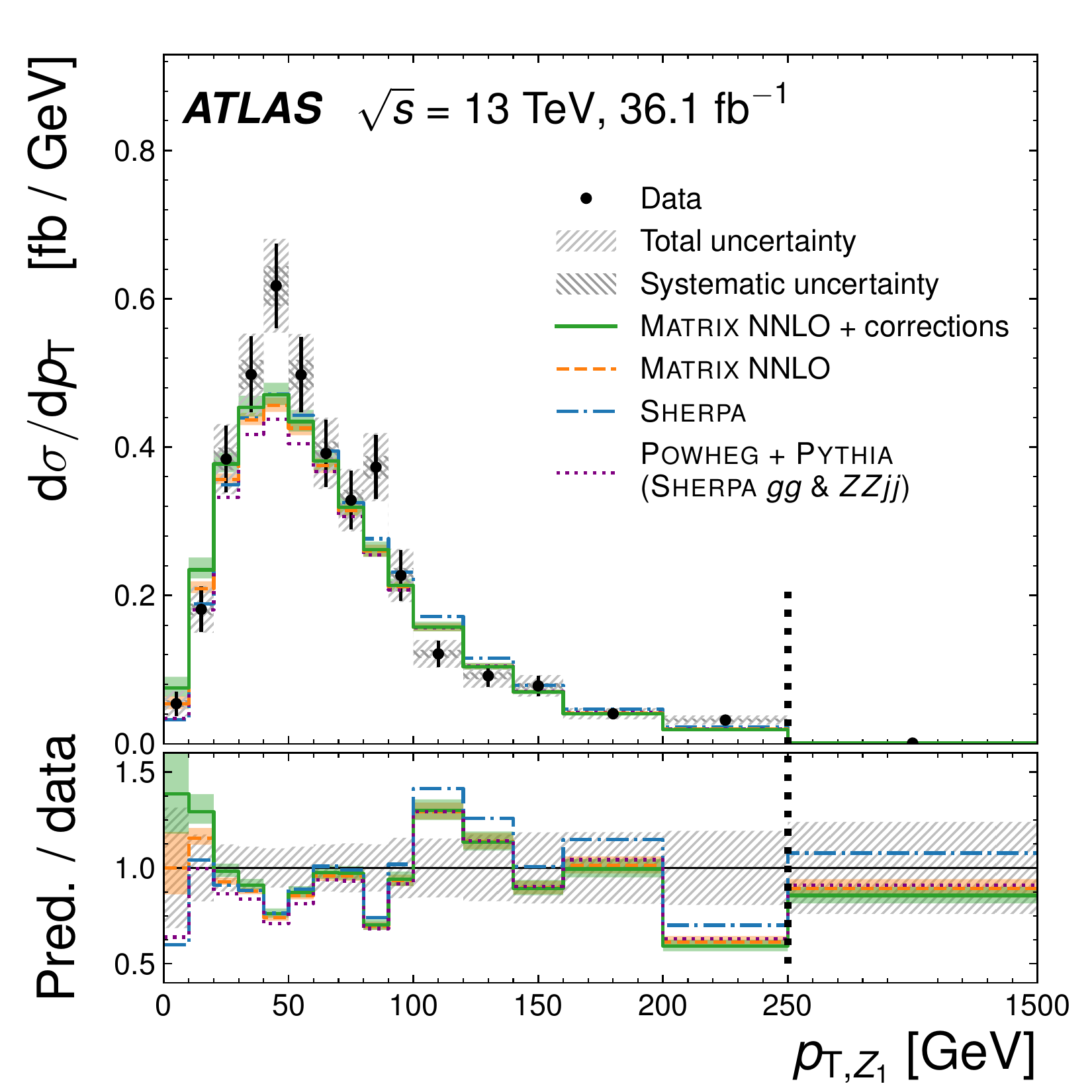}}
\subfigure[]{\includegraphics[width=0.48\textwidth]{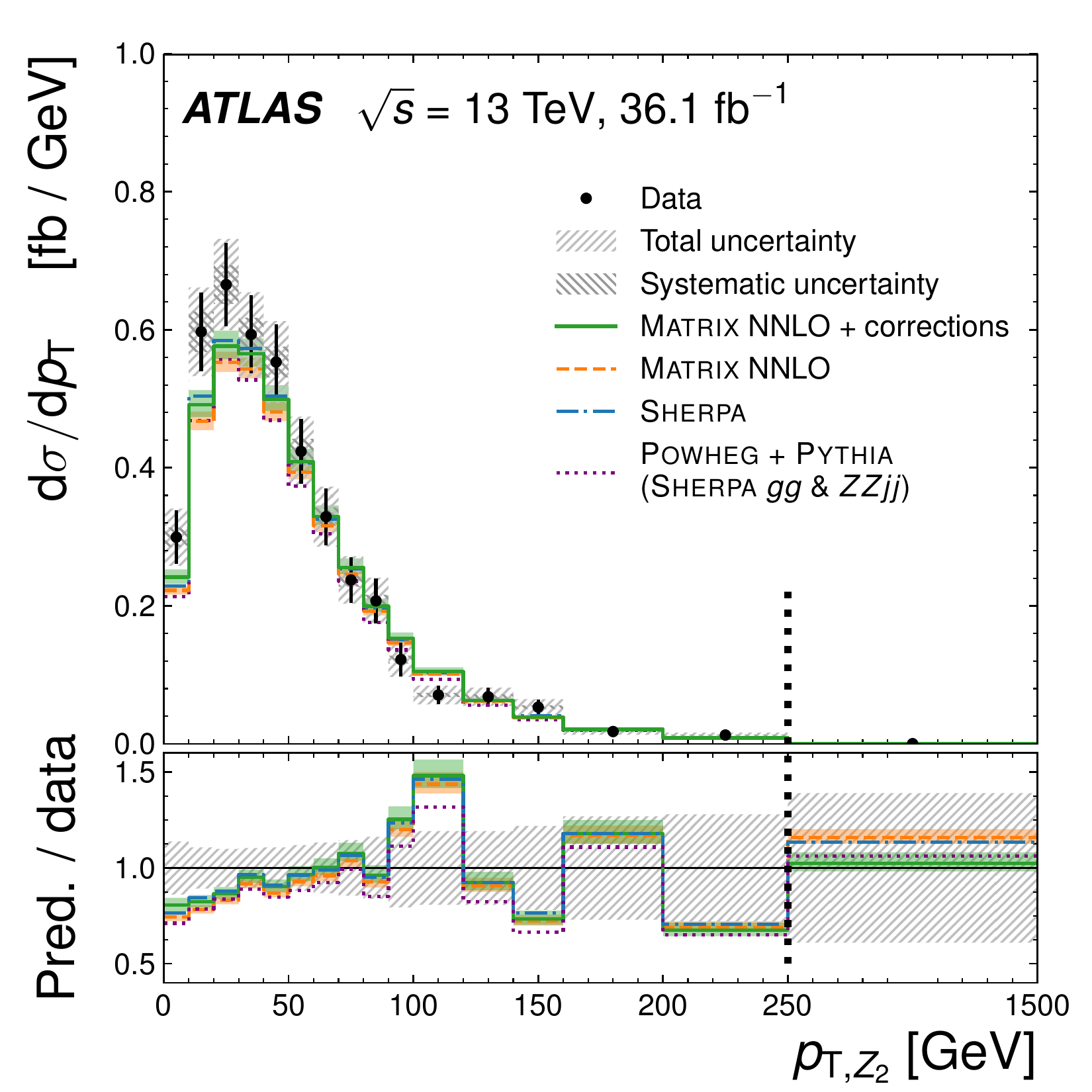}} 
\caption{Measured and predicted differential cross sections for the transverse momentum of (a) the leading-\pt{} and (b) the subleading-\pt{} \PZ{} boson candidate. \datauncertcomment{} A pure NNLO calculation from \matrixnnlo{} is shown with no additional corrections applied. Another prediction is shown based on this NNLO calculation, with the $\Pgluon\Pgluon$-initiated contribution multiplied by a global NLO correction factor of $1.67$. The EW correction factor is applied in each bin. The contribution from EW-$\ZZ jj$ generated with \sherpa{} is added. For the fixed-order predictions, the QCD scale uncertainty is shown as a shaded band. Parton-showered \POWHEGpy{} and \SHERPA{} predictions are also shown. \brokenaxiscomment{} Published in \myref~\cite{STDM-2016-15}.}
\label{fig:z_pts}
\end{figure}

\clearpage
\myfig{}~\ref{fig:lepton_pts} presents the transverse momenta of the leptons in the final selected quadruplet. From the highest-\pt{} to the lowest-\pt{} lepton, the distribution becomes less peaked and more symmetric about the peak, while the position of the peak shifts from $\sim$60~\GeV{} to $\sim$50~\GeV{}, then $\sim$35~\GeV{}, and finally $\sim$25~\GeV{}. All lepton \pt{} distributions agree well with the predictions. 
\vspace{-5mm}

\begin{figure}[h!]
\centering
\subfigure[]{\includegraphics[width=0.48\textwidth]{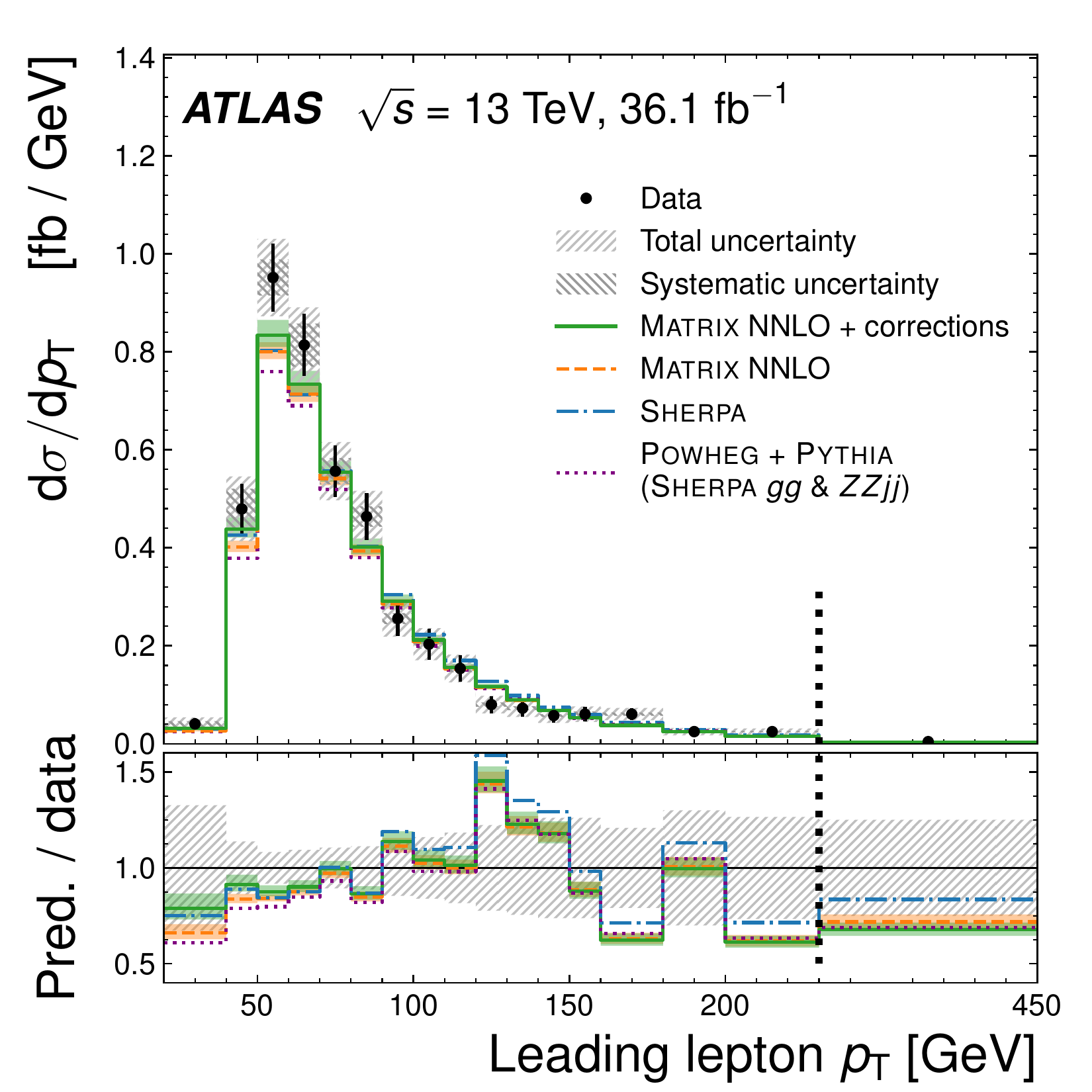}}
\subfigure[]{\includegraphics[width=0.48\textwidth]{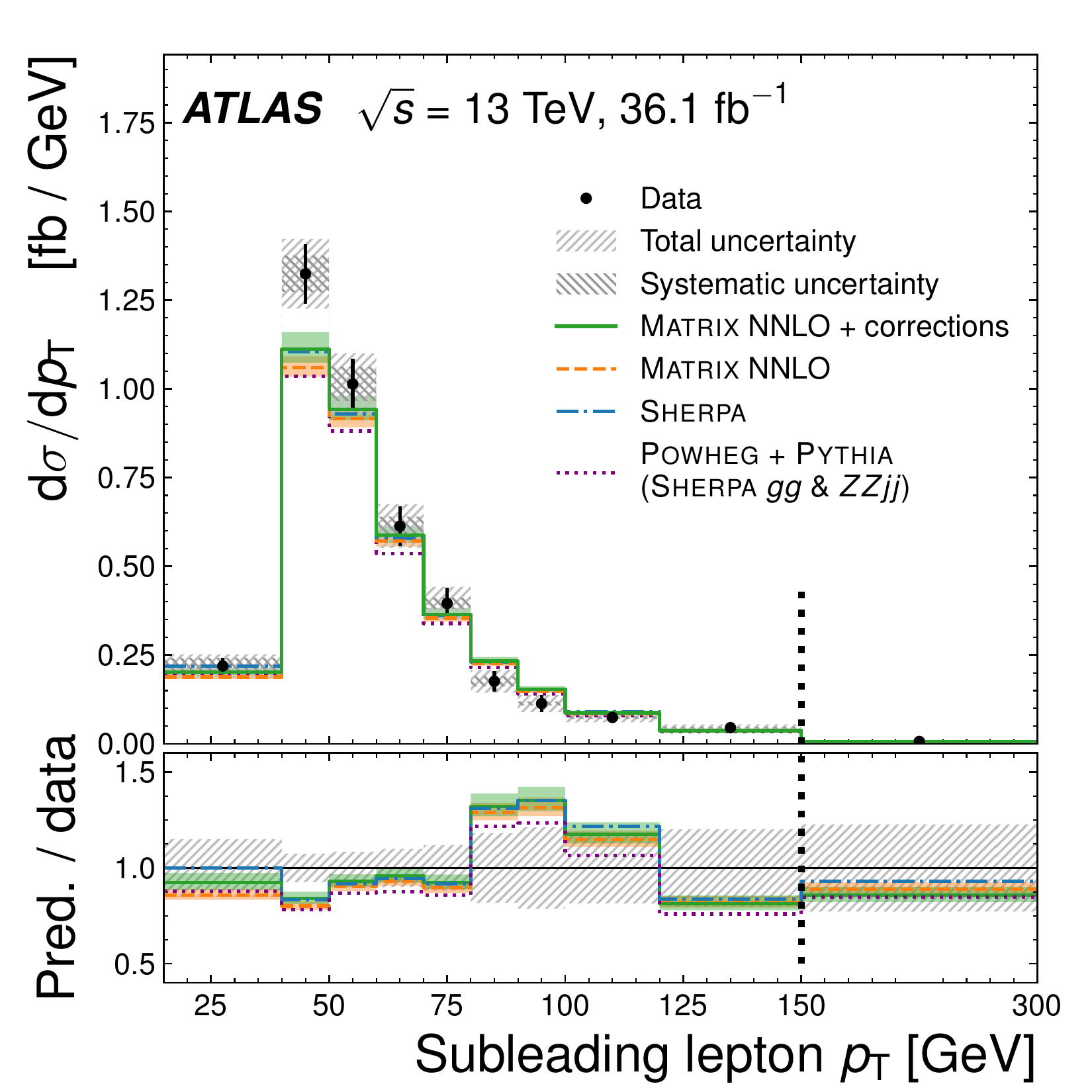}}
\vspace{-5mm}
\subfigure[]{\includegraphics[width=0.48\textwidth]{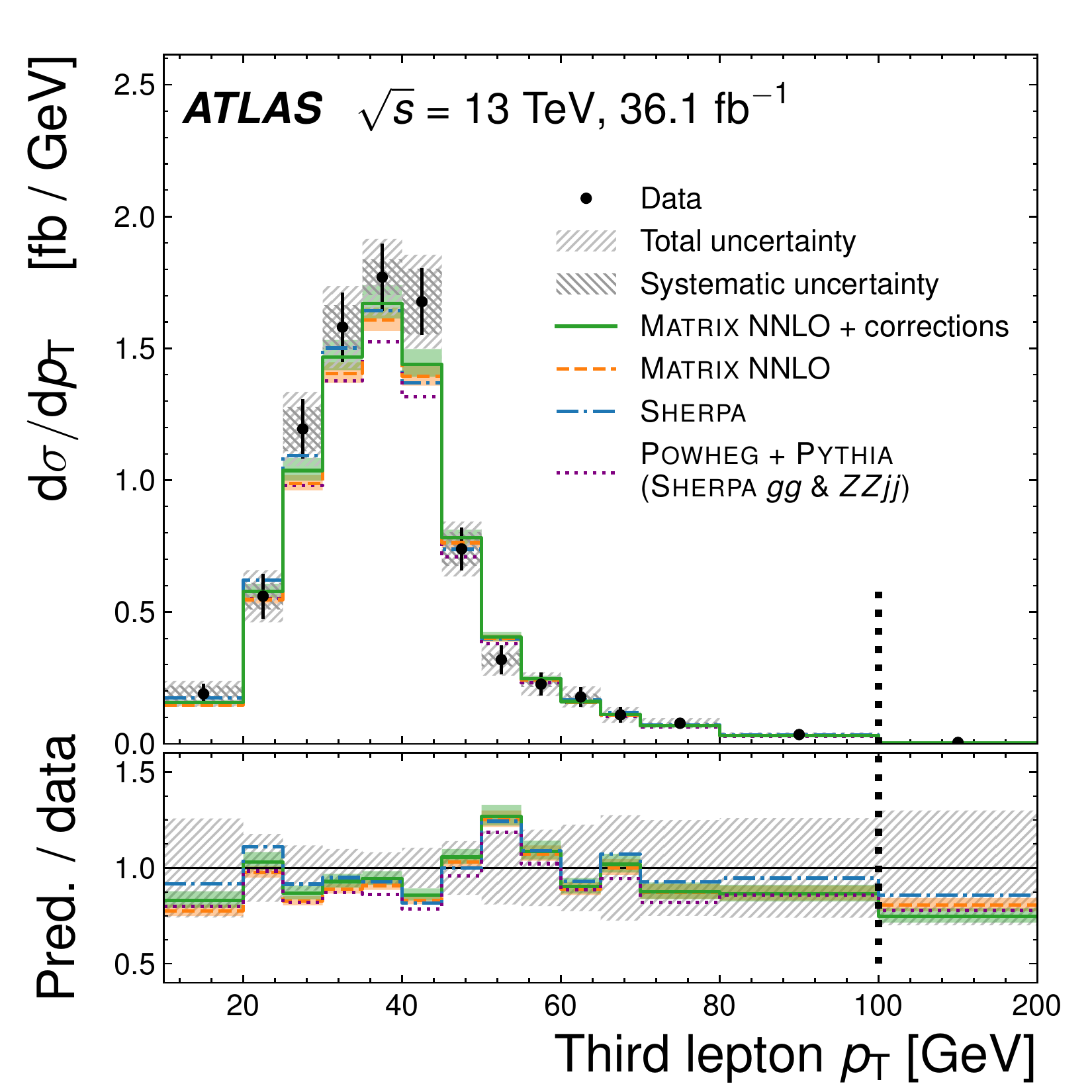}}
\subfigure[]{\includegraphics[width=0.48\textwidth]{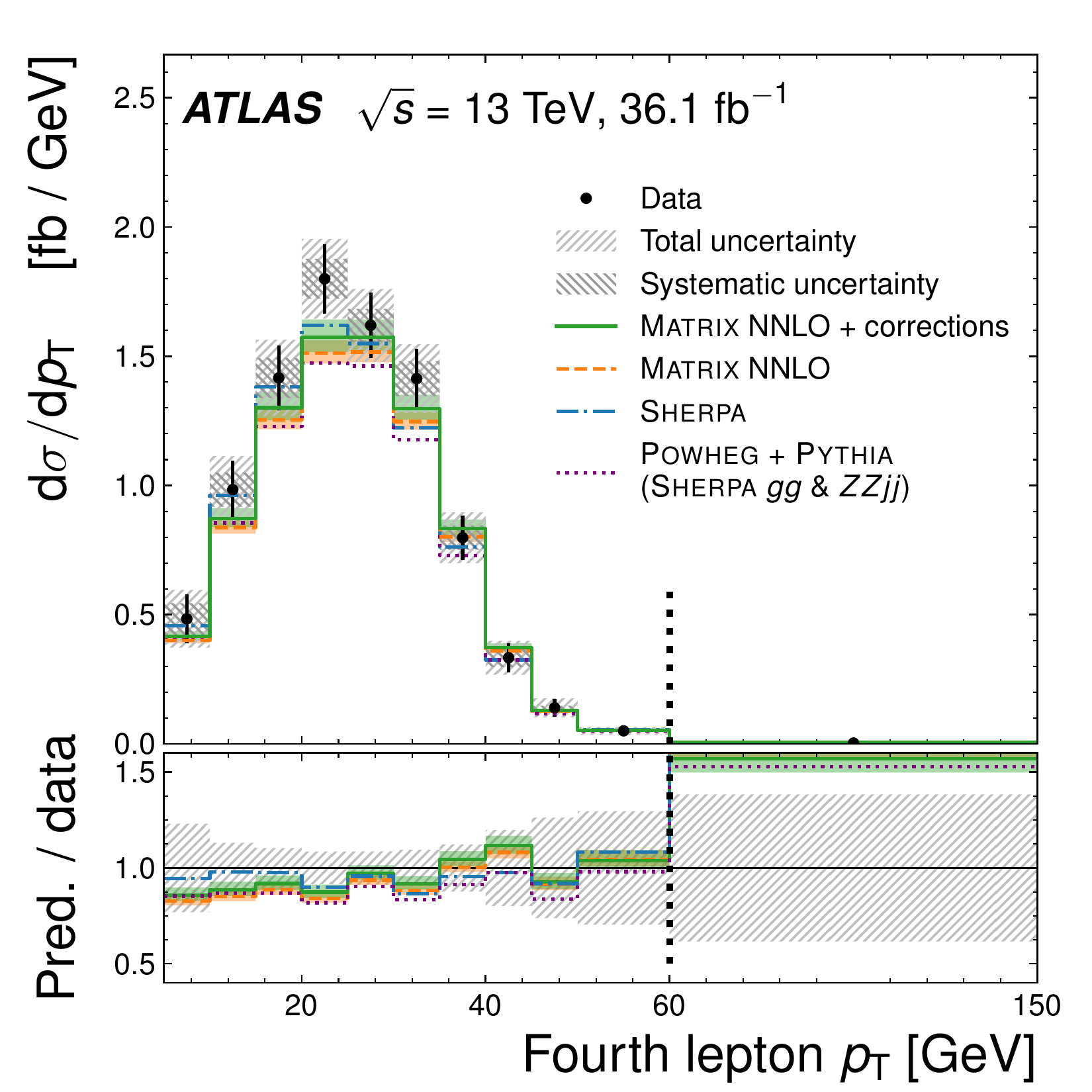}}
\caption{Measured and predicted differential cross sections with respect to the transverse momenta of the leptons in the final selected quadruplet, in descending order of transverse momentum. A pure NNLO calculation from \matrixnnlo{} is shown with no additional corrections applied. Another prediction is shown based on this NNLO calculation, with the $\Pgluon\Pgluon$-initiated contribution multiplied by a global NLO correction factor of $1.67$. The EW correction factor is applied in each bin. The contribution from EW-$\ZZ jj$ generated with \sherpa{} is added. For the fixed-order predictions, the QCD scale uncertainty is shown as a shaded band. Parton-showered \POWHEGpy{} and \SHERPA{} predictions are also shown. \brokenaxiscomment{} Published in \myref~\cite{STDM-2016-15}.}
\label{fig:lepton_pts}
\end{figure}

\clearpage
\myfig~\ref{fig:multijets} shows the jet multiplicity distributions as well as the scalar sum of the transverse momenta of all selected jets. \POWHEGpy{} shows a clear trend towards underestimating the cross section at jet multiplicities greater than one and large jet scalar~\pt{} sum, which is expected, because in \POWHEGpy{} only the hardest parton emission is included at the matrix-element level. \SHERPA{}, however, includes up to three parton emissions at the matrix-element level, and exhibits good agreement with the measurements for these higher jet multiplicities. The central-jet multiplicity in \myfig~\ref{fig:multijets}(b) is an exception, as \POWHEGpy{} describes it slightly better than \SHERPA{}. It seems that \SHERPA{} predicts somewhat too central jets beyond the leading jet, which can also be seen in the $|\eta|$ distribution of the subleading jet in \myfig~\ref{fig:jet_kinematics}(d) below.
The most significant observed disagreement is the deficit in the bin $60~\GeV{} < \sum\pt < 90~\GeV{}$ of the jet scalar~\pt{} sum. It has a local significance of 2.3 standard deviations with respect to the \SHERPA{} prediction, estimated from the corresponding bins in the measured distribution before unfolding.


\begin{figure}[h!]
\centering
\subfigure[]{\includegraphics[width=0.48\textwidth]{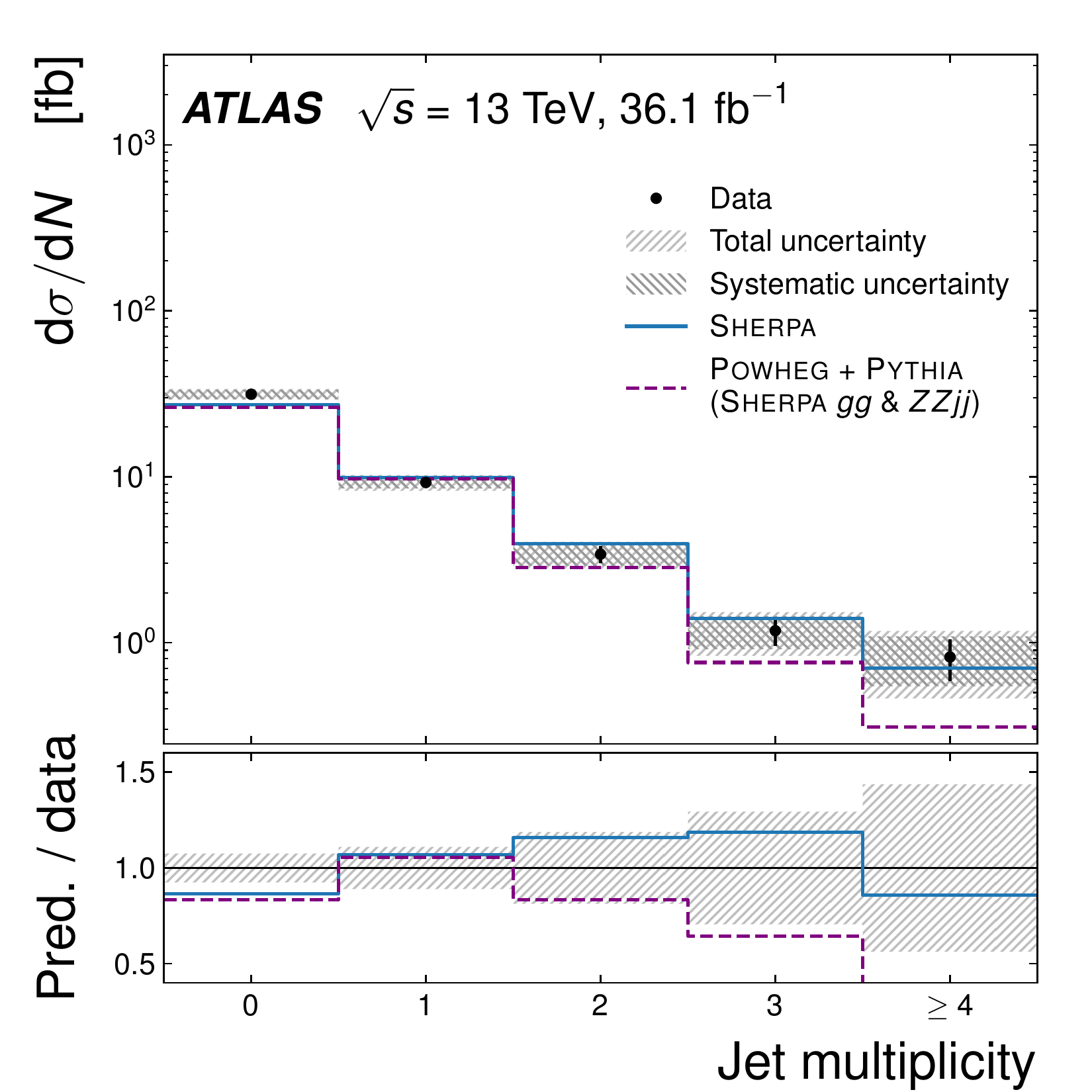}}
\subfigure[]{\includegraphics[width=0.48\textwidth]{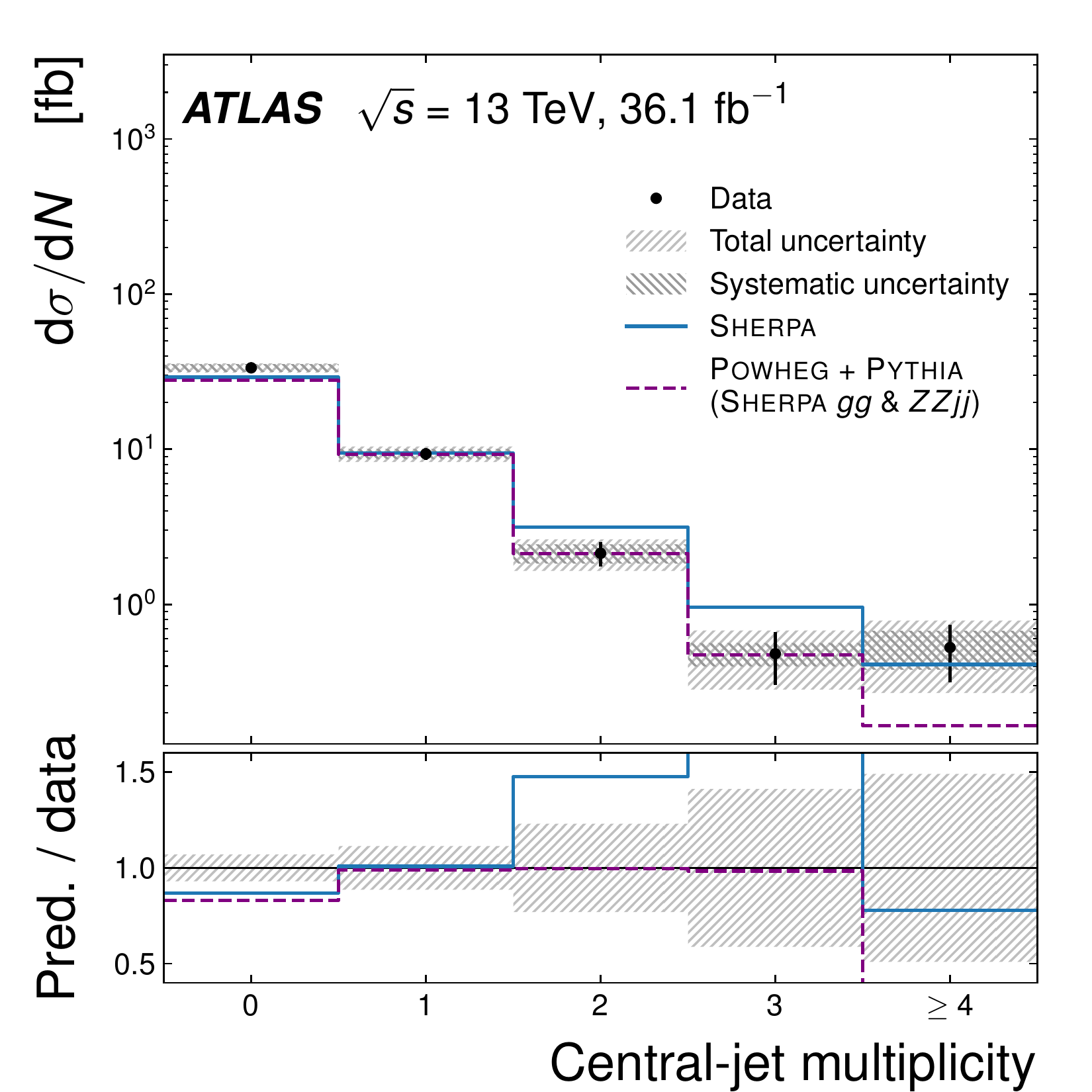}}
\vspace{-5mm}
\subfigure[]{\includegraphics[width=0.48\textwidth]{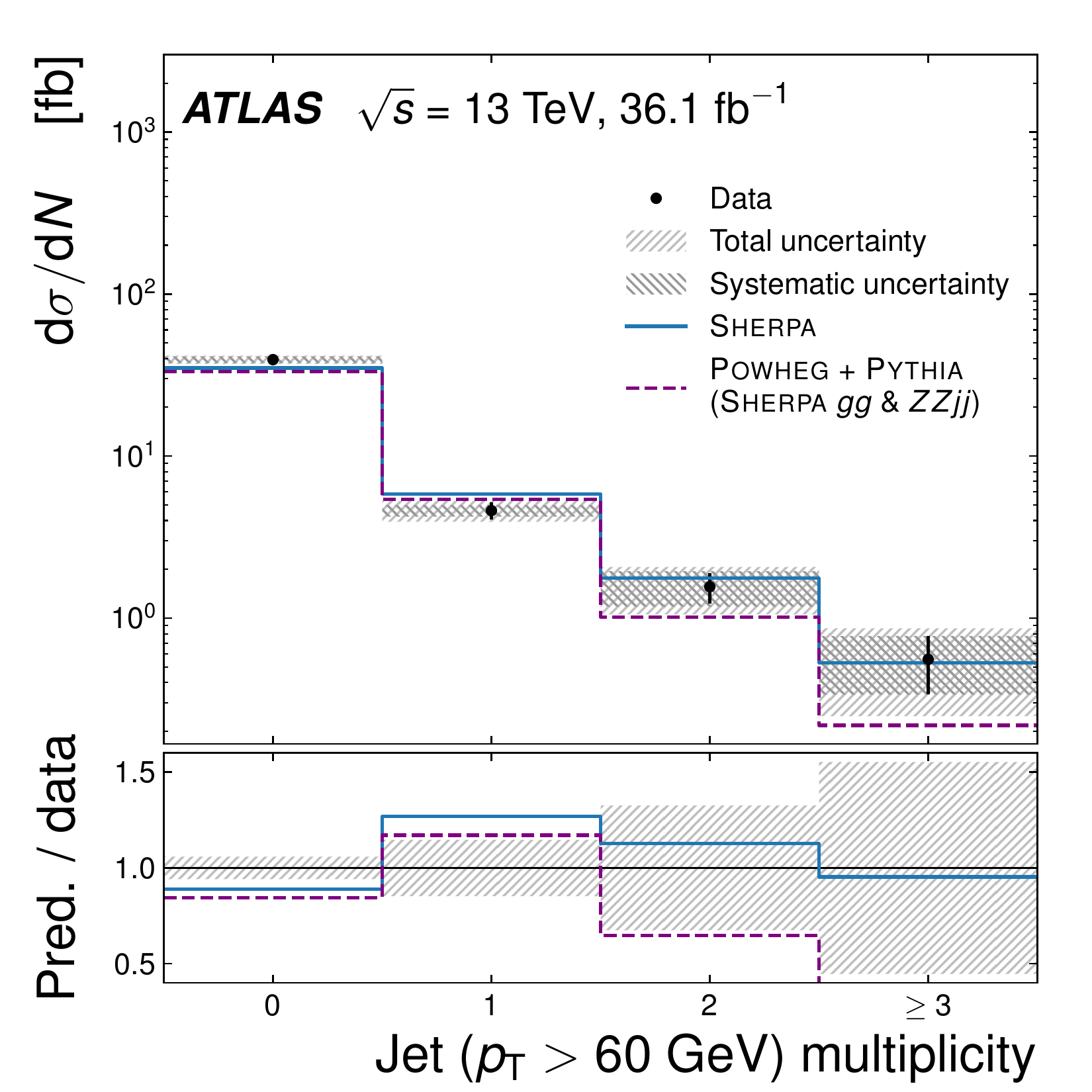}}
\subfigure[]{\includegraphics[width=0.48\textwidth]{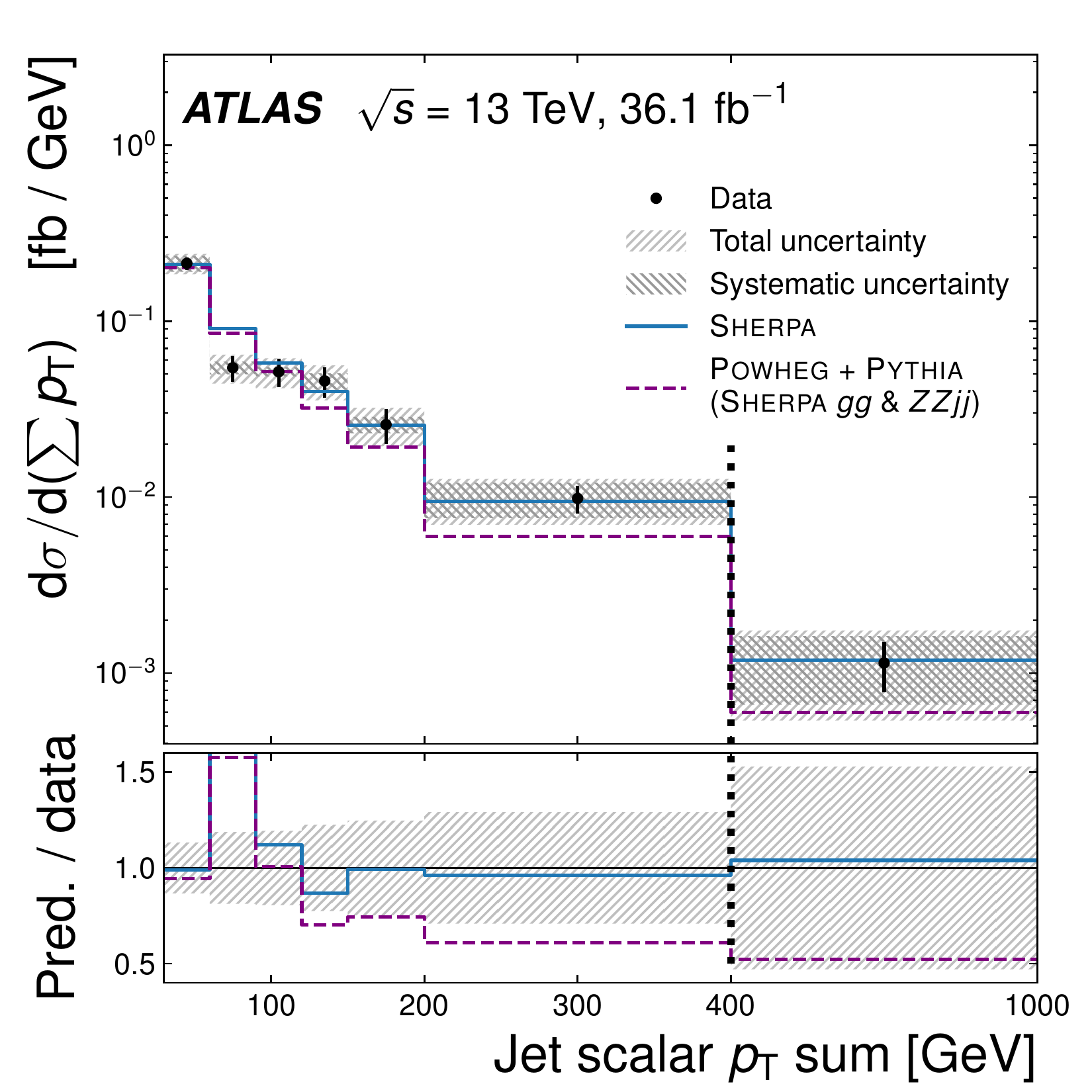}}
\caption{Measured and predicted differential cross sections for (a) the jet multiplicity considering all selected jets, (b) the central-jet multiplicity considering jets with $|\eta| < 2.4$, (c) the jet multiplicity considering jets with $\pt > 60$~\GeV{}, and (d) the scalar sum of the transverse momenta of all selected jets. \datauncertcomment{} For better visualisation, the last bin is shown using a different $x$-axis scale where indicated by the dashed vertical line. Published in \myref~\cite{STDM-2016-15}.}
\label{fig:multijets}
\end{figure}

\clearpage
\myfig~\ref{fig:jet_kinematics} shows the transverse momentum and absolute pseudorapidity of the leading-\pt{} and subleading-\pt{} jets. Within the relatively large uncertainties, \SHERPA{} provides a good description of the kinematics. \POWHEGpy{} also describes the shapes of the $|\eta|$ distributions well, while its normalisation is too low for the subleading-\pt{} jet. \POWHEGpy{} does not describe the \pt{} distribution of the subleading-\pt{} jet very well, predicting too few jets at high \pt{}. This is not surprising, given that subleading jets in \POWHEGpy{} are usually generated in the parton shower approximation, which is not adequate for describing high-\pt{} emissions. A deficit of events is observed in the bin $2.0 < |\eta| < 2.5$ of the subleading-\pt{} jet. Its local significance with respect to the \SHERPA{} prediction is estimated to be 3.2 standard deviations, based on the corresponding bins in the measured distribution before unfolding.

\begin{figure}[h!]
\centering
\subfigure[]{\includegraphics[width=0.48\textwidth]{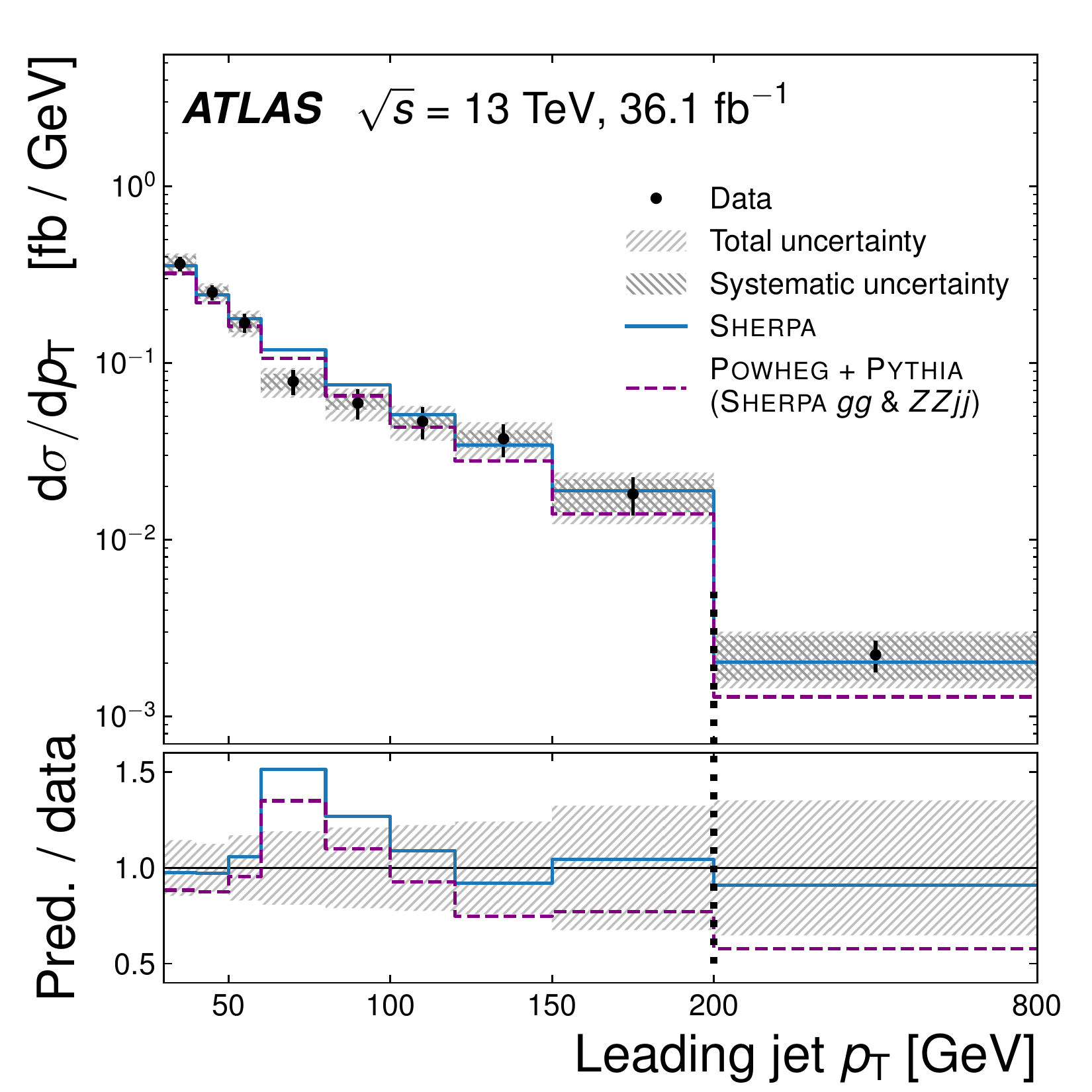}}
\subfigure[]{\includegraphics[width=0.48\textwidth]{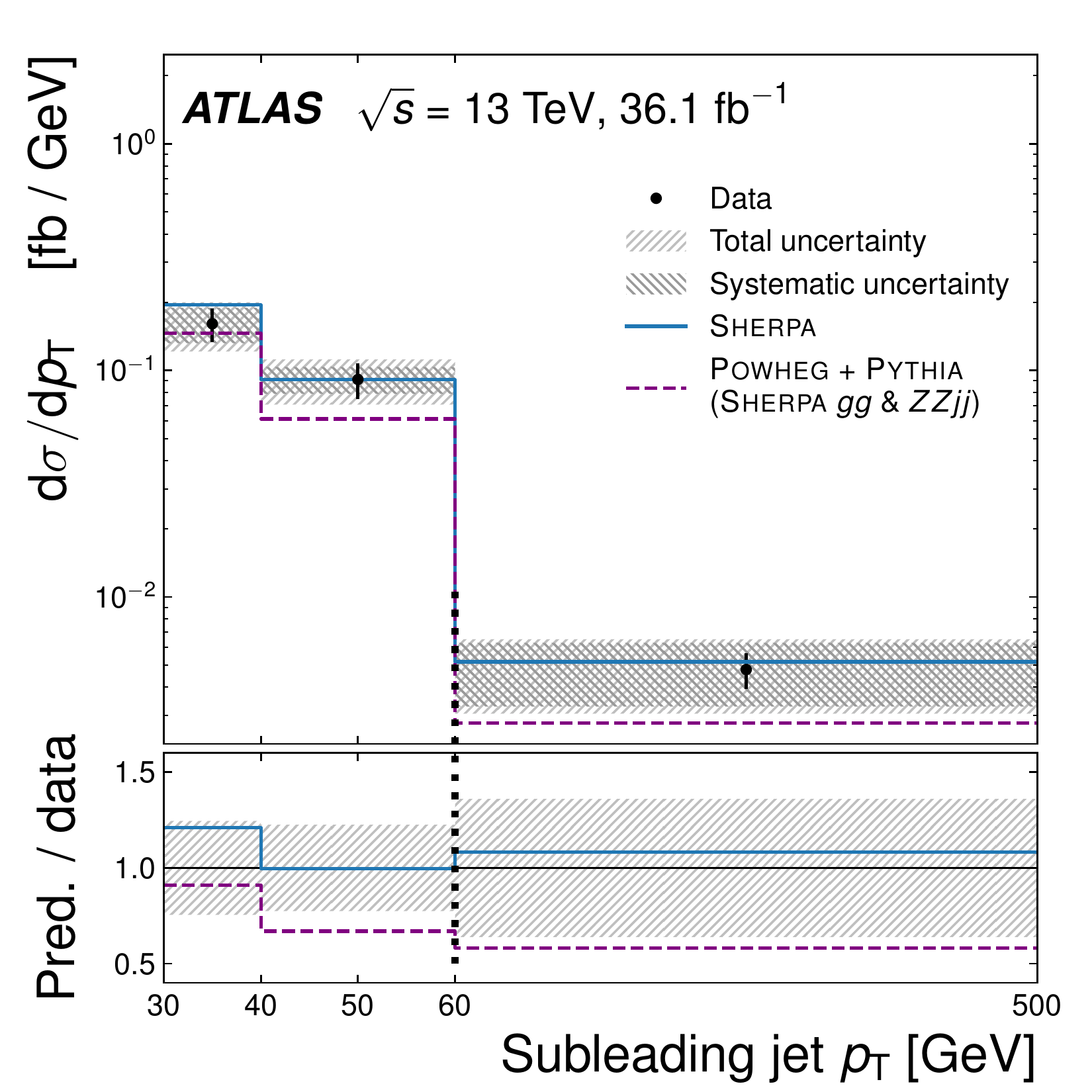}}
\vspace{-5mm}
\subfigure[]{\includegraphics[width=0.48\textwidth]{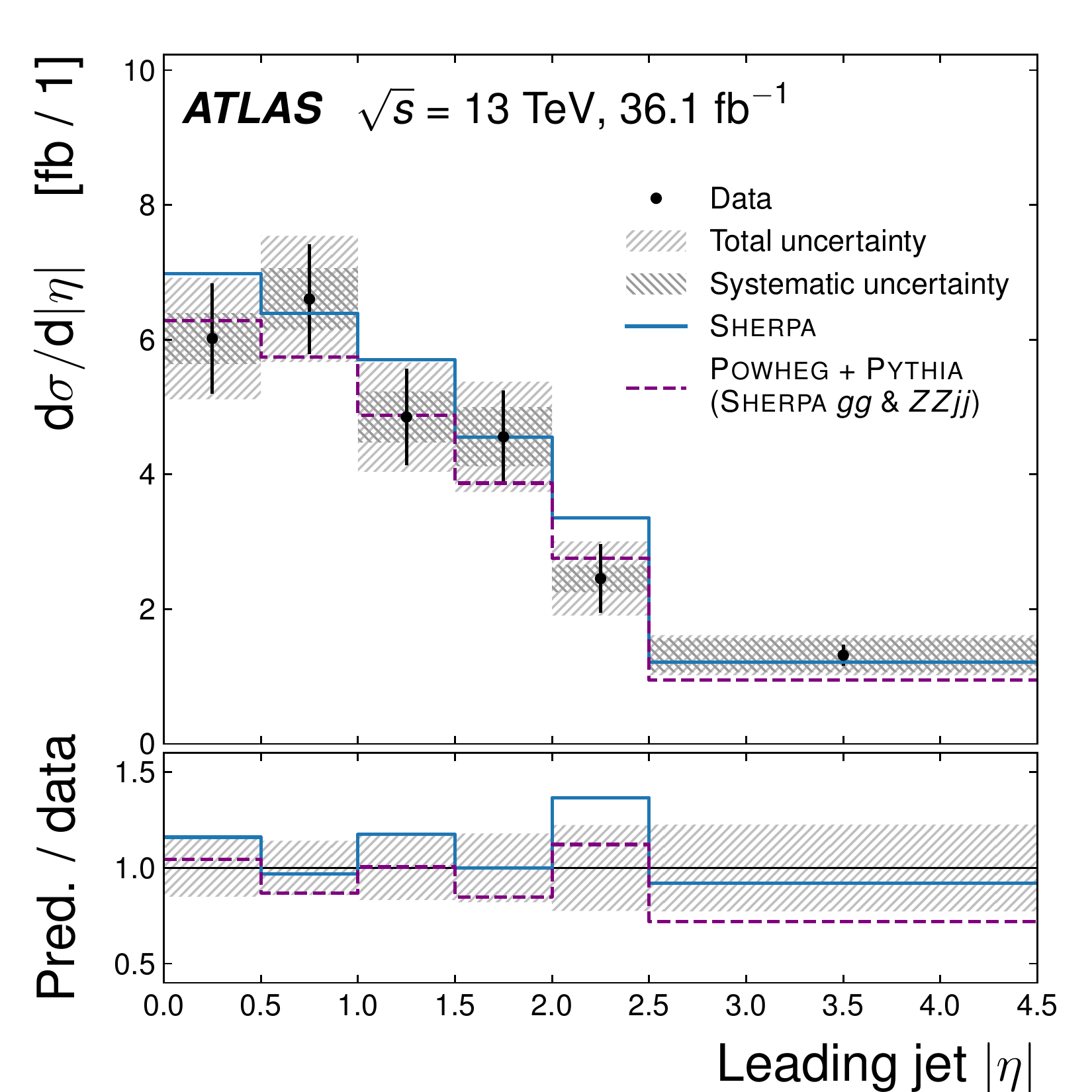}}
\subfigure[]{\includegraphics[width=0.48\textwidth]{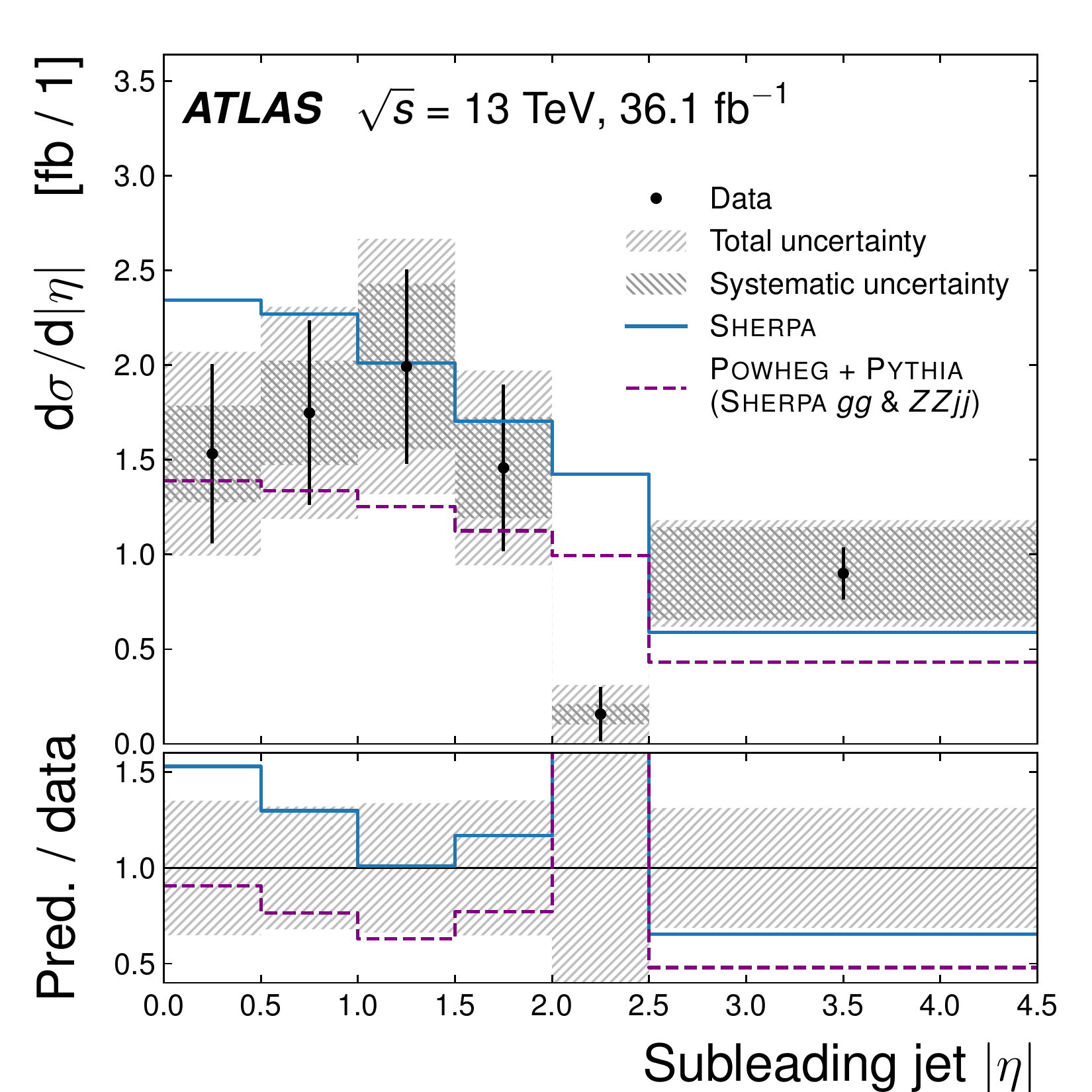}}
\caption{Measured and predicted differential cross sections for the transverse momentum of the (a) leading-\pt{} and (b) subleading-\pt{} jet, as well as the absolute pseudorapidity of the (c) leading-\pt{} and (d) subleading-\pt{} jet. \datauncertcomment{} \brokenaxiscomment{} Published in \myref~\cite{STDM-2016-15}.}
\label{fig:jet_kinematics}
\end{figure}

\clearpage
\myfig~\ref{fig:dijets} shows the rapidity difference and invariant mass of the two leading-\pt{} jets. The EW-$\ZZ jj$ production process predicted by \SHERPA{} is shown separately, in addition to the process-inclusive predictions from \SHERPA{} and \POWHEGpy{}. This contribution falls much less steeply towards higher values of the presented observables. The contribution from this process in the last bins in each distribution improves the agreement between prediction and measurement, demonstrating the importance of this process at these ends of the phase space. 


\begin{figure}[h!]
\centering
\subfigure[]{\includegraphics[width=0.48\textwidth]{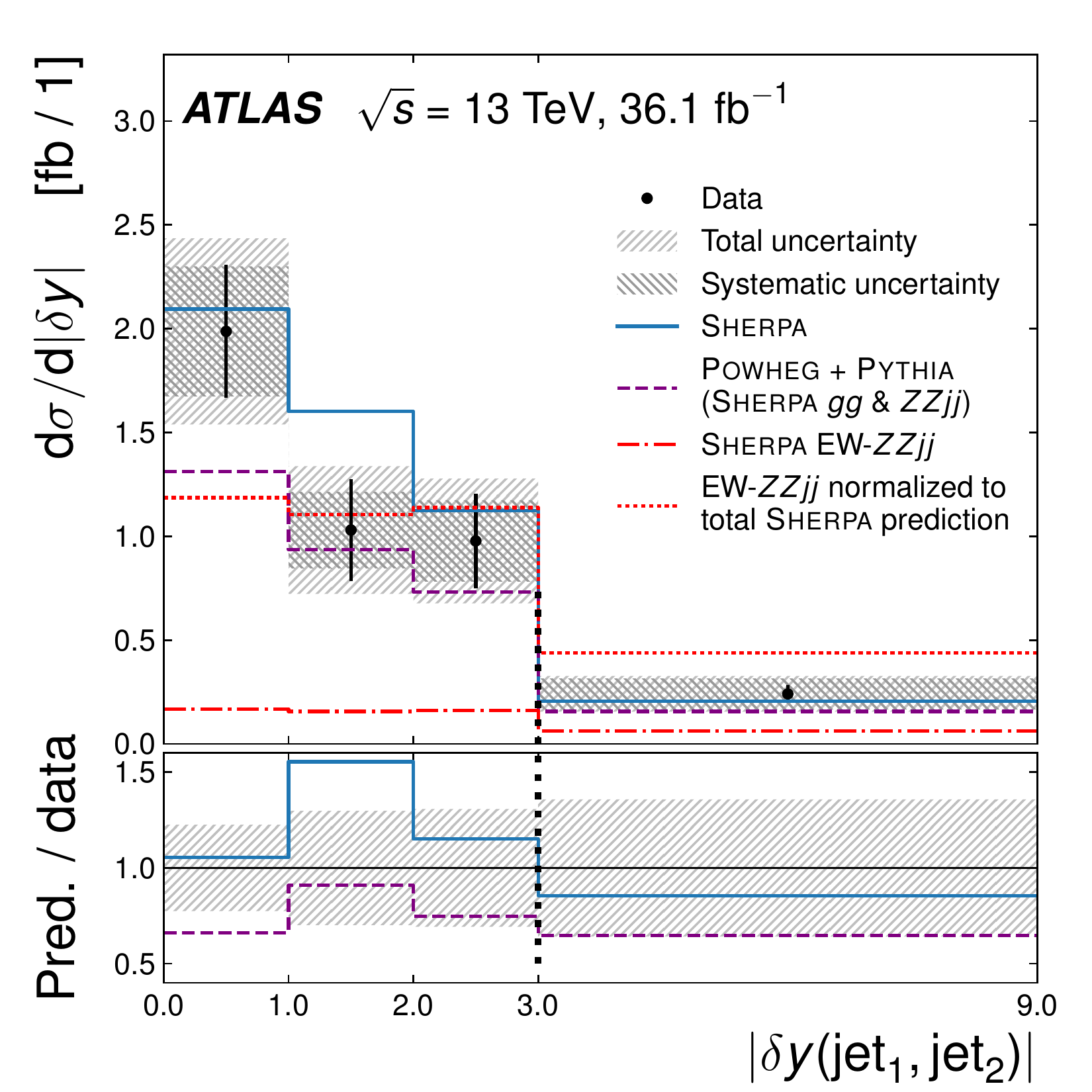}}
\subfigure[]{\includegraphics[width=0.48\textwidth]{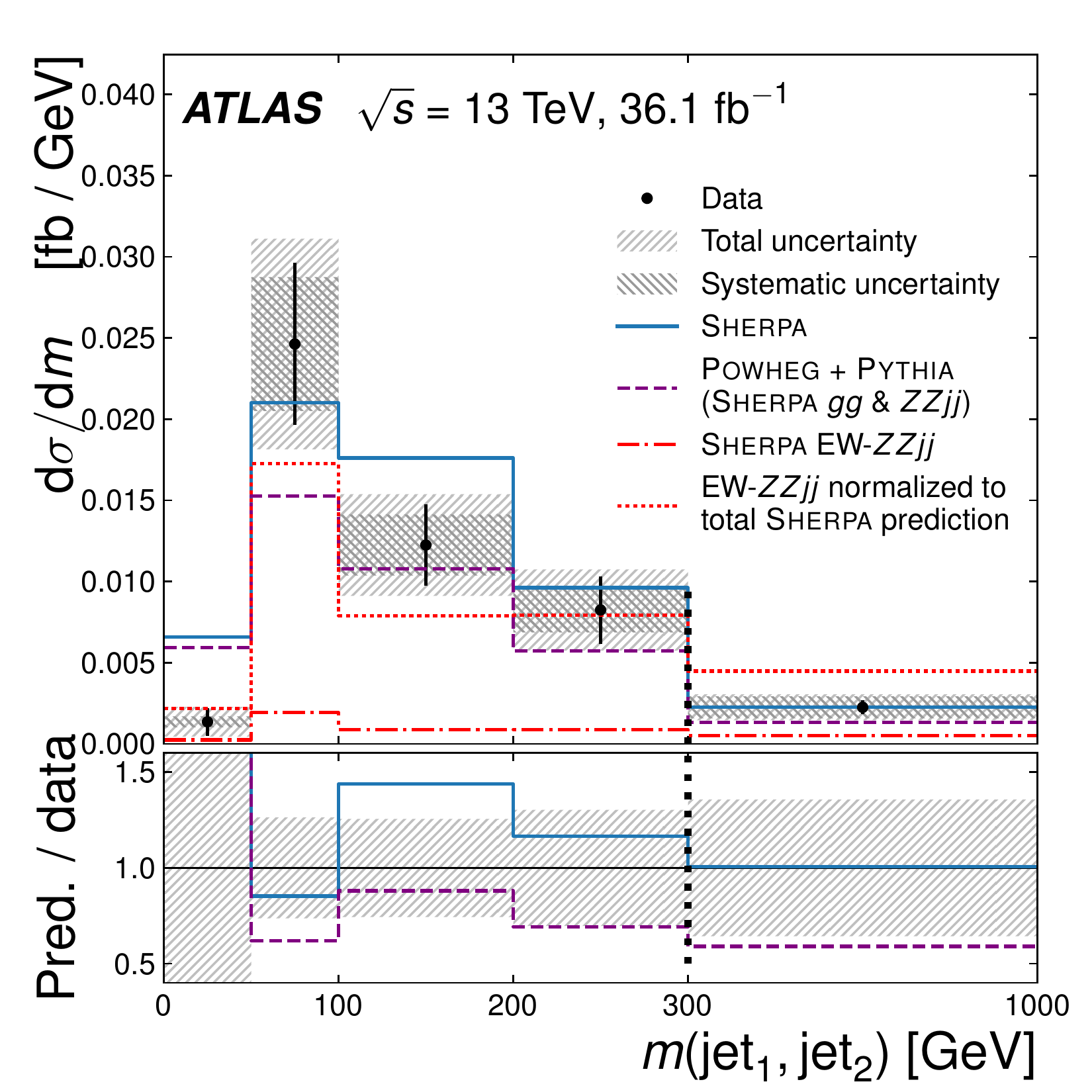}}
\caption{Measured and predicted differential cross sections for (a) the absolute difference in rapidity between the two leading-\pt{} jets and (b) the invariant mass of the two leading-\pt{} jets. \datauncertcomment{} In addition to the process-inclusive predictions from \SHERPA{} and \POWHEGpy{}, the EW-$\ZZ jj$ production process predicted by \SHERPA{} is shown separately. \brokenaxiscomment{} Published in \myref~\cite{STDM-2016-15}.}
\label{fig:dijets}
\end{figure}


\subsection{PDF and \alphas{} variations}

In order not to clutter the figures too much, the PDF uncertainty of the predictions is not indicated in any of the figures above showing the measured differential cross sections. Here, the PDF uncertainties are studied. \myfig~\ref{fig:pdfdep} shows the measured differential cross sections as a function of two observables that are expected to be relatively sensitive to the PDFs: the absolute rapidity of the four-lepton system, $|\yfourl{}|$, and the transverse momentum of the leading-\pt{} jet. They are compared to predictions from the nominal \SHERPA{} setup with different choices of NNLO PDF set and strong coupling strength \alphas{}. The shown theoretical uncertainties are much smaller than the current experimental uncertainties, so the sensitivity to them is very limited. However, a few trends among the theoretical predictions can be observed. Interestingly, for $|\yfourl{}|$, varying \alphas{} (evaluated at the \PZ{} pole mass) by $\pm 0.001$ (i.e.~0.117--0.119) leads to a spread in the predictions that is almost identical to the NNPDF 3.0 uncertainty band. For the leading-jet \pt{}, this is only true at the lower end of the spectrum, but above $\pt \sim 60$~\GeV{}, the \alphas{} variations give a larger spread than the PDF uncertainty. This is not surprising, as the value of \alphas{} governs the hardness of parton emissions (or, phrased differently, the rate of parton emissions of a given hardness). The spread between the nominal MMHT 2014 and CT14 PDF increases with growing $|\yfourl|$, but shows almost no dependence on the leading-jet \pt{}.


\begin{figure}[h!]
\centering
\subfigure[]{\includegraphics[width=0.49\textwidth]{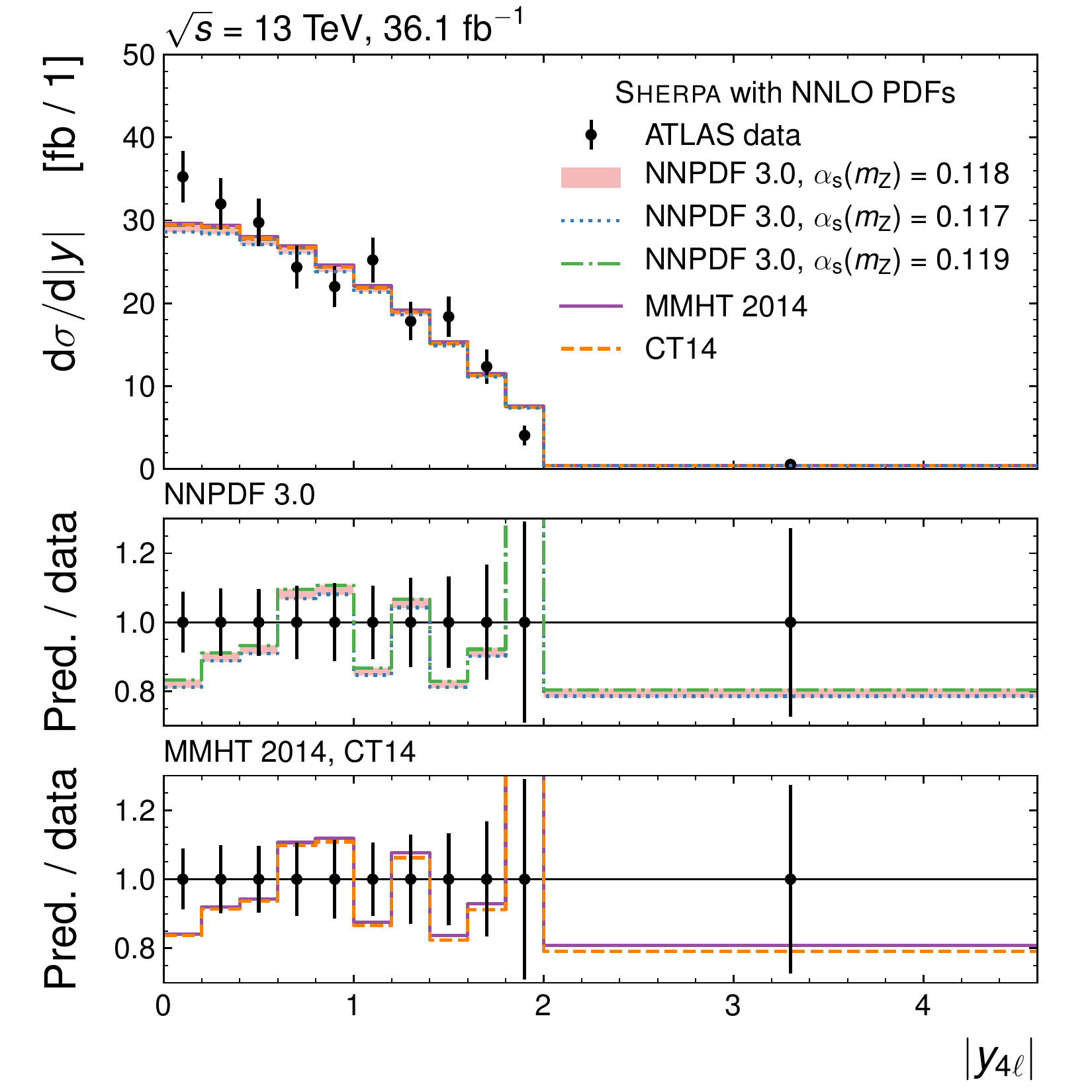}}
\subfigure[]{\includegraphics[width=0.49\textwidth]{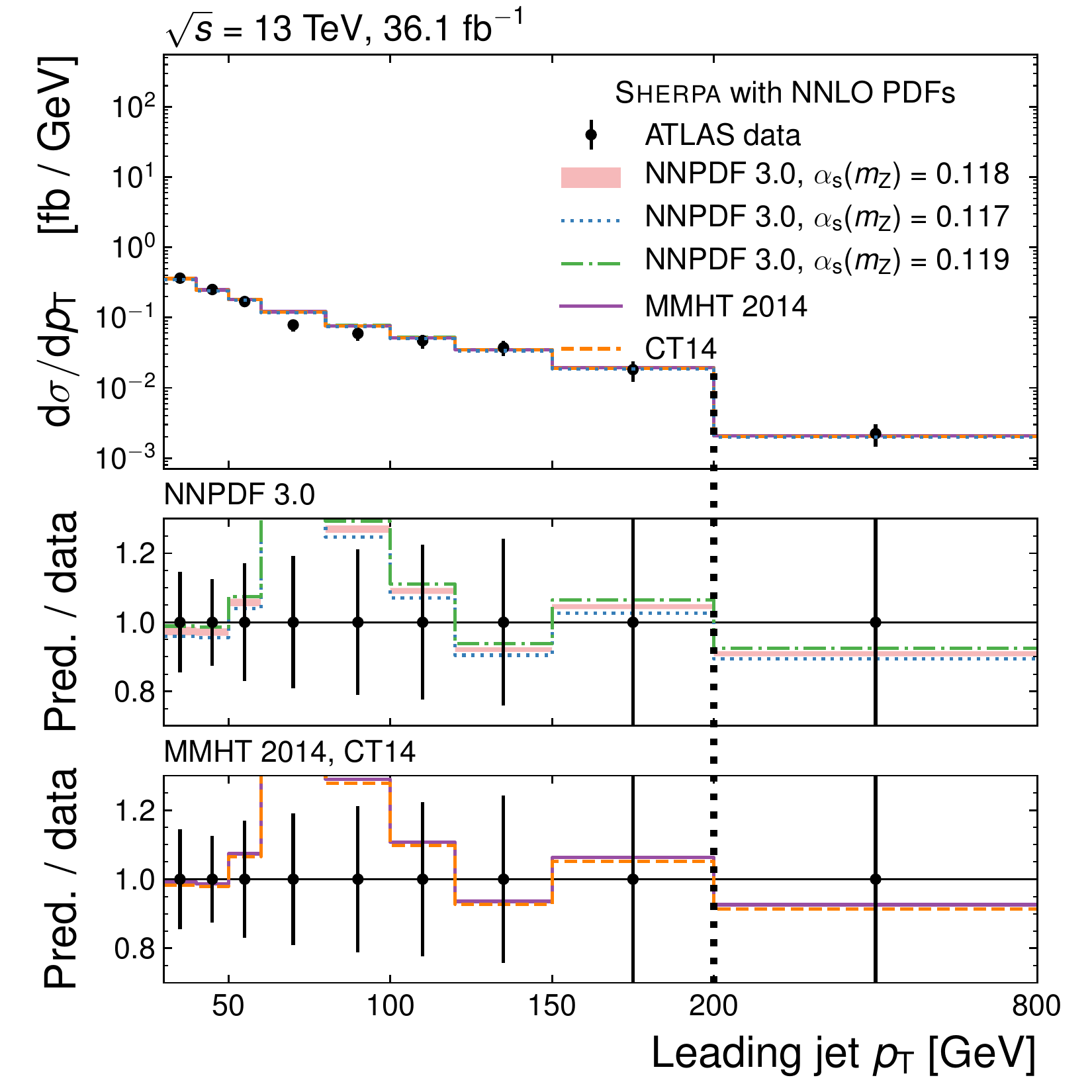}}
\caption{Measured and predicted differential cross sections for (a) the absolute rapidity of the four-lepton system and (b) the transverse momentum of the leading-\pt{} jet. The error bars indicate the total uncertainty of the measurement. Predictions from the nominal \SHERPA{} setup with various NNLO PDF set and \alphas{} choices are shown. For NNPDF 3.0, the uncertainty band is shown, calculated from the nominal and 100 variation PDFs. For all other predictions, the lines correspond to the nominal PDF. For better visualisation, the last bin in (b) is shown using a different $x$-axis scale where indicated by the dashed vertical line. The measured data are published in \myref~\cite{STDM-2016-15}.}
\label{fig:pdfdep}
\end{figure}

\subsection{Unfolded four-lepton mass}

The four-lepton mass is unfolded by the same techniques as the differential cross sections discussed above, using three unfolding iterations. The corresponding response matrix and bin-by-bin efficiencies, purities, and fake corrections are shown in \myfig~\ref{fig:m4l_inputs}. The differential cross section as a function of the four-lepton mass is shown in \myfig~\ref{fig:m4l} along with fixed-order and particle-level predictions. Measurement and prediction agree well.

\begin{figure}[h!]
\centering
\subfigure{
\begin{tikzpicture}
\node[anchor=south west,inner sep=0] (image) at (0,0) {\includegraphics[width=0.49\textwidth]{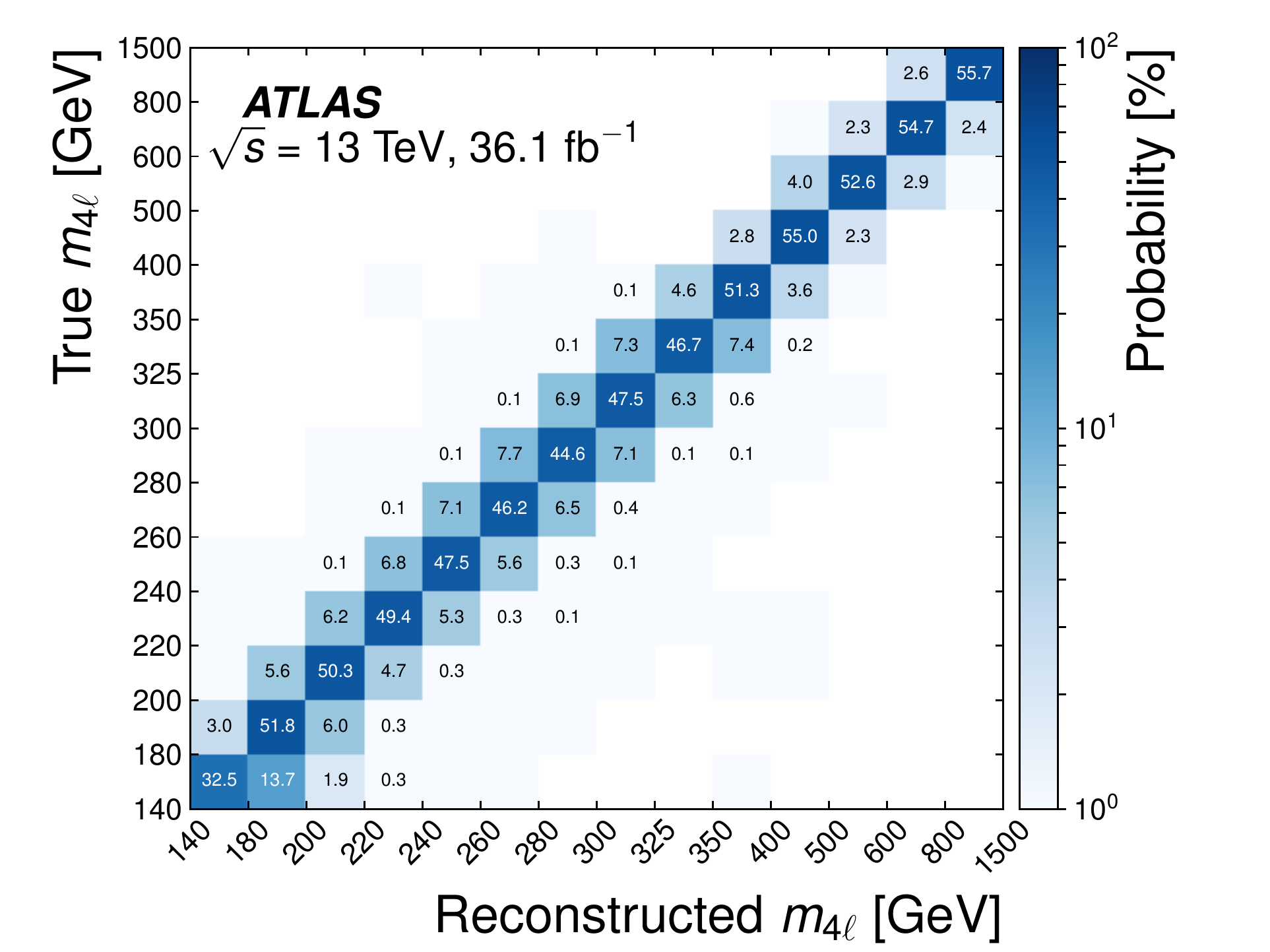}};
\begin{scope}[x={(image.south east)},y={(image.north west)}]
\draw[white, fill=white] (0.16,0.87) rectangle (0.3,0.93);
\end{scope}
\end{tikzpicture}
}
\subfigure{\includegraphics[width=0.4\textwidth]{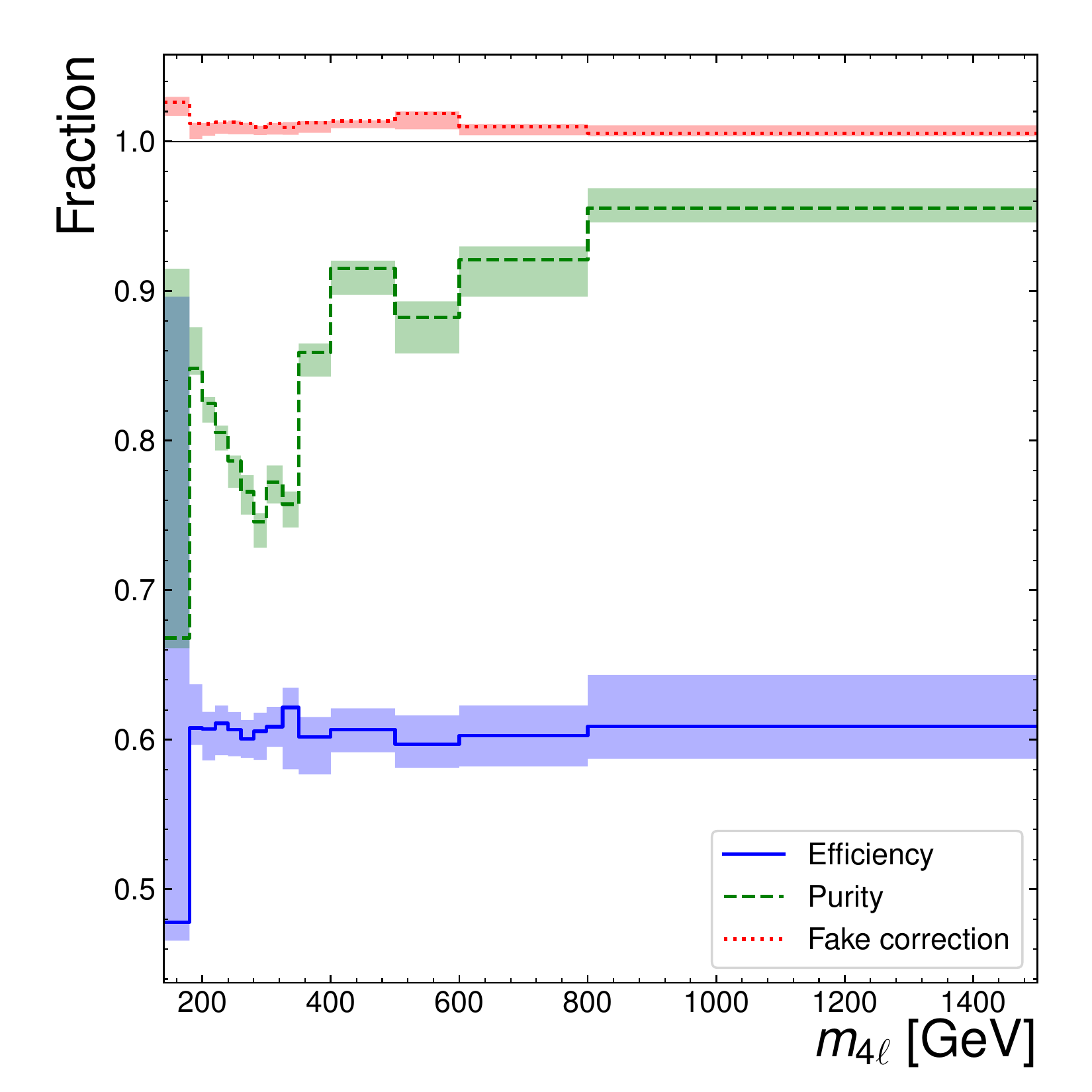}}
\caption{Response matrix, unfolding corrections, and bin-by-bin purity for the four-lepton mass.}
\label{fig:m4l_inputs}
\end{figure}

\begin{figure}[h!]
\centering
\begin{tikzpicture}
\node[anchor=south west,inner sep=0] (image) at (0,0) {\includegraphics[width=0.65\textwidth]{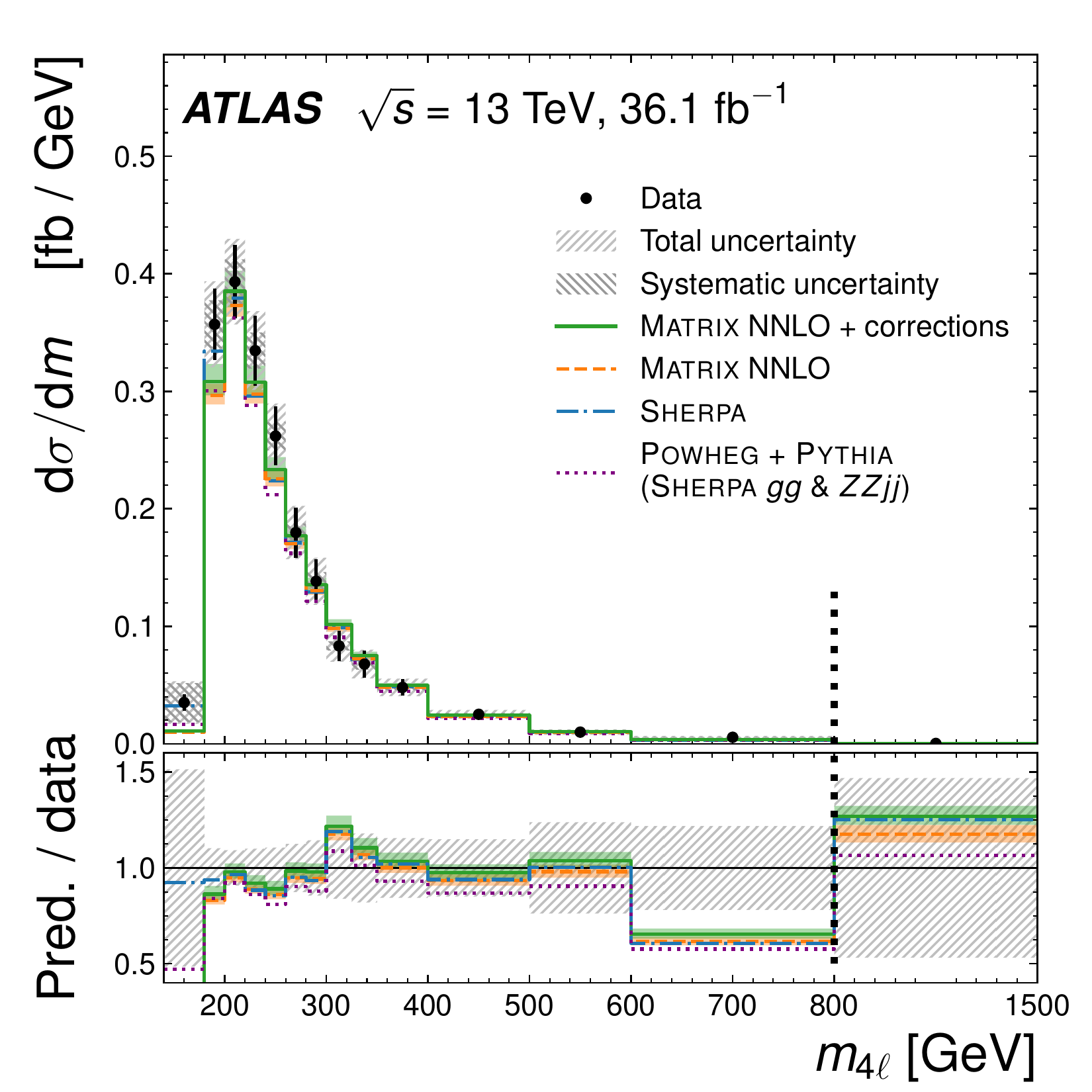}};
\begin{scope}[x={(image.south east)},y={(image.north west)}]
\draw[white, fill=white] (0.16,0.85) rectangle (0.3,0.93);
\end{scope}
\end{tikzpicture}
\caption{Measured and predicted differential cross sections for the mass of the four-lepton system. \datauncertcomment{} A pure NNLO calculation from \matrixnnlo{} is shown with no additional corrections applied. Another prediction is shown based on this NNLO calculation, with the $\Pgluon\Pgluon$-initiated contribution multiplied by a global NLO correction factor of $1.67$. The NLO weak correction is applied as a global factor of $0.95$ as a differential calculation was not requested from the theorists. The contribution from EW-$\ZZ jj$ generated with \sherpa{} is added. For the fixed-order predictions, the QCD scale uncertainty is shown as a shaded band. Parton-showered \POWHEGpy{} and \SHERPA{} predictions are also shown. \brokenaxiscomment{}}
\label{fig:m4l}
\end{figure}


\clearpage
\section{Search for anomalous triple gauge couplings}
\label{sec:anomalous}


A search for aTGCs was performed by Maurice Becker~\cite{thesis_maurice}, using systematic uncertainties and some simulated predictions provided by the author. An introduction to the topic can be found in \myref~\cite{thesis_kristian}. The search uses the reconstructed transverse momentum of the leading-$\pt$ \PZ{} boson candidate (\ptz{}) to look for deviations between the data and the SM, as this variable is found to provide the highest sensitivity to their predicted effects.
The four-lepton mass provides similar sensitivity (almost identical according to studies), but is not used, because no dedicated calculation of NLO weak corrections for $\Pproton\Pproton \to \ZZllll$ production binned in the four-lepton mass was requested in time for use in the analysis.
The considered aTGC signal model uses an effective vertex function approach~\cite{Baur:2000cx}. It includes two coupling strengths that violate charge-parity (CP) symmetry, \fyfour{} and \fzfour{}, as well as two CP-conserving ones, \fyfive{} and \fzfive{}. No unitarising form factor is used, as the sensitivity of the measurement is well within the unitarity bounds. In addition, the form factor may make the interpretation of the results in terms of an ultraviolet-complete (i.e.~non-effective) theory of new physics more difficult, as it has no physical correspondence there.


The expected aTGC signal yield $N$ is parameterised in terms of the coupling strengths, on which it depends both linearly and quadratically,

\vspace{-1\baselineskip}
\begin{equation}\label{eq:atgc_yield_parametrisation}
\begin{split}
N\left(\fyfour, \fzfour, \fyfive, \fzfive\right) &= N_{\text{SM}} + \fyfour N_{01} + \fzfour N_{02} + \fyfive N_{03} + \fzfive N_{04}\\
&+ \left(\fyfour\right)^2 N_{11} + \fyfour \fzfour N_{12} + \fyfour \fyfive N_{13} + \fyfour \fzfive N_{14}\\
&+ \left(\fzfour\right)^2 N_{22} + \fzfour\fyfive N_{23} + \fzfour\fzfive N_{24}\\
&+ \left(\fyfive\right)^2 N_{33} + \fyfive \fzfive N_{34}\\
&+ \left(\fzfive\right)^2 N_{44},
\end{split}
\end{equation}
where $N_{\text{SM}}$ is the SM expectation and the $N_{ij}$ are yield coefficients that depend on the final-state particle momenta.
To determine the coefficients $N_{ij}$, $2\times 10^5$ events with aTGC are generated at LO with one fixed reference set of coupling strengths using \sherpa{} and the CT10 PDF set. Based on the kinematic properties of each event, the coefficients $N_{ij}$ are extracted using a framework \cite{Bella:2008wc} based on the BHO program \cite{Baur:1997kz}. The yield for all other values of the coupling strengths can then be calculated using \myeq{}~\ref{eq:atgc_yield_parametrisation}.

The SM prediction $N_{\text{SM}}$ is constructed separately using the highest-order calculations available, so that the aTGC search results are as realistic as possible. The nominal \SHERPA{} setup is used, except that the $\Pquark\APquark$-initiated process is generated with \POWHEGpy{} and each event reweighted by NNLO and NLO weak corrections binned in \ptz{}. The SM \ZZ{} predictions, estimated backgrounds, as well as observed yields are shown in \mytab{}~\ref{tab:atgc_yields} as a function of \ptz{}. These contributions are also shown in \myfig{}~\ref{fig:atgc_input} together with two different aTGC predictions. The considered systematic uncertainties of the predictions are the same as in the integrated cross section measurement. An additional uncertainty due to the factorisation approximation of NNLO QCD and NLO weak corrections for the SM \ZZllll{} process is assigned as follows. Employing a criterion motivated in \myref{}~\cite{Gieseke:2014gka}, events are classified as having high QCD activity if $\left|\sum_{i} \vec{p}_{\mathrm{T},\,i}\right| > 0.3 \sum_{i} |\vec{p}_{\mathrm{T},\,i}|$, where the sums are over fiducial leptons. In events with high QCD activity, the NLO weak $k$-factors are in turn not applied and applied with doubled deviation from unity, as $1 + 2 (\text{$k$-factor} - 1)$. The deviations from the nominal result are taken as uncertainties, ranging from $\sim$1\% in the lowest to $\sim$10\% in the highest \ptz{} bin. The \ptz{} binning is optimised using the predictions to maximise the expected sensitivity.

\begin{table}[h!]
\centering
{\small
\begin{tabular}{llllll}
\toprule
\textbf{$\text{\emph{\textbf{p}}}_{\mathbf{T},\,\PZ_1}$ range (\GeV{})} &  \textbf{0--295} & \textbf{295--415} &  \textbf{415--555} & \textbf{555--3000}\\
\midrule
Data & 998 & 16 & 3 & 0\\
\midrule
Total SM prediction & $950\pm 40$& $10.6\pm 0.9$& $2.50\pm 0.33$& $1.18\pm 0.21$\\
\midrule
SM \ZZllll{} & $930 \pm 40$& $10.0\pm 0.9$& $2.34\pm 0.33$& $1.10\pm 0.21$\\ 
Triboson, $\Ptop\APtop\PZ$, $\PZ\PZ\to \tau^{+}\tau^{-}[\ell^{+}\ell^{-}, \tau^{+}\tau^{-}]$ & $9.2 \pm 2.8$ & $0.43\pm 0.13$ & $0.15\pm 0.05$ & $0.078 \pm 0.028$ \\
Misidentified leptons & $12\pm 8$ & $0.17\pm 0.11$ & $< 0.1$ & $< 0.1$\\
\bottomrule 
\end{tabular}
}
\caption{Observed and predicted yields in bins of the transverse momentum of the leading-\pt{} \PZ{} boson candidate. All statistical and systematic uncertainties are included in the prediction uncertainties, including the uncertainty associated with the combination of NNLO and NLO weak corrections for the SM \ZZllll{} process.}
\label{tab:atgc_yields}
\end{table}

\begin{figure}[h!]
\centering
\includegraphics[width=0.5\columnwidth]{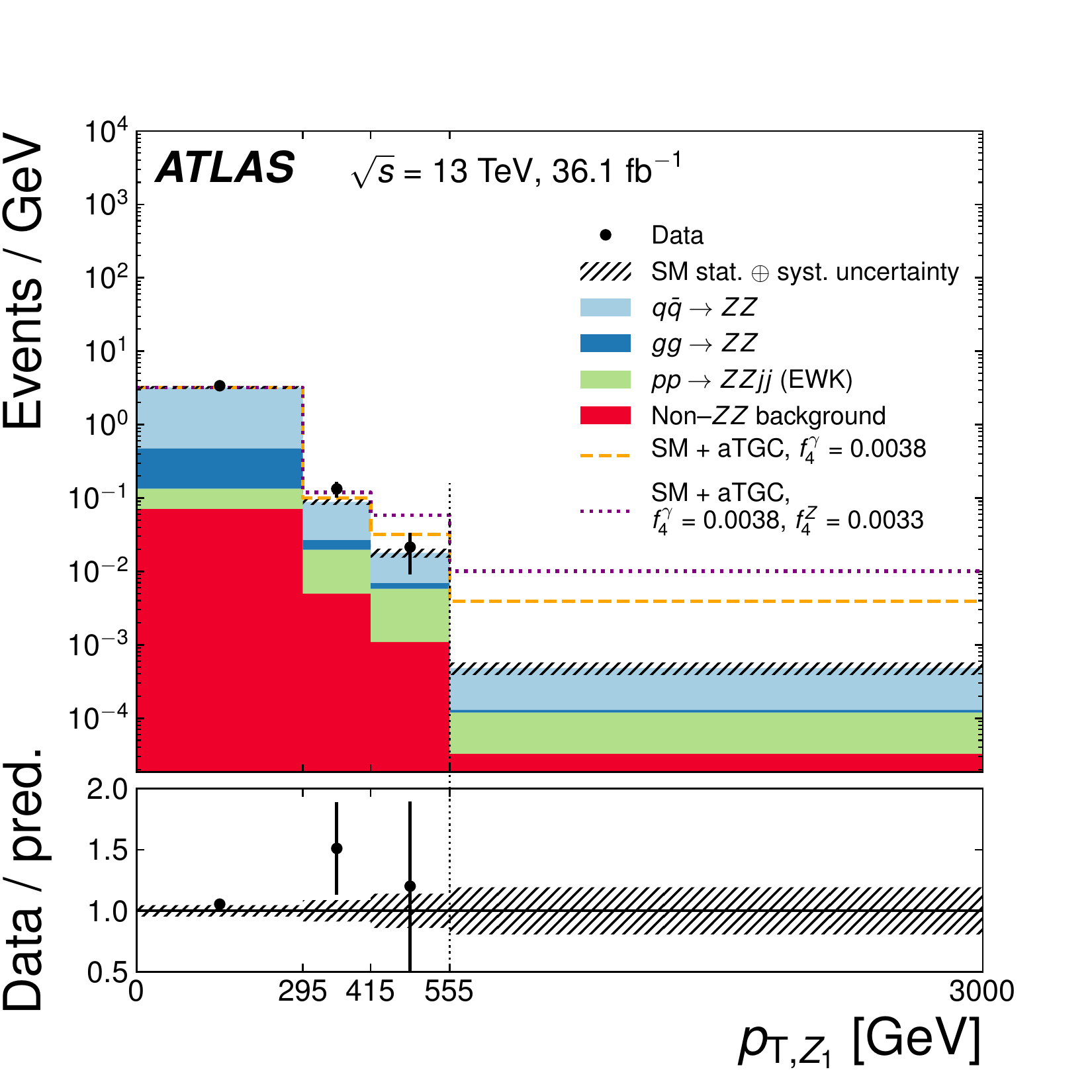}
\caption{Data and SM predictions as function of the transverse momentum of the leading-$\pt$ $\PZ$ boson candidate. Also shown is the SM plus aTGC signal prediction with $f_4^{\gamma} = 3.8 \times 10^{-4}$ as well as with $f_4^{\gamma} = 3.8 \times 10^{-4}$ and $f_4^{\PZ} = 3.3 \times 10^{-4}$. In both cases all other aTGC coupling strengths are set to zero. The shaded band shows the total SM prediction uncertainty including the statistical and all systematic uncertainties. \brokenaxiscomment{} Published in \myref~\cite{STDM-2016-15}.}
\label{fig:atgc_input}
\end{figure}

The data are found to be consistent with the SM predictions, and no indication of aTGCs is observed. Confidence intervals of aTGC parameters are determined using the expected and observed yields in bins of \ptz{} as reconstructed by the detector.
A frequentist method~\cite{Feldman:1997qc} is used to find the 95\% confidence level (CL) intervals for the aTGC parameters. The predicted and observed yields are assumed to follow Poissonian probability density functions, while the systematic uncertainties are treated as nuisance parameters constrained by Gaussian functions. The expected confidence intervals and their one- and two-standard-deviation confidence bands are established using many independent sets of randomly generated pseudodata following a Poisson distribution whose expectation value is the SM prediction in each bin. 
Confidence intervals are set for each coupling strength individually, setting all others to zero. The expected and observed 95\% CL intervals are listed in \mytab{}~\ref{tab:oneD_results}.
The one-dimensional confidence intervals are more stringent than those derived in previous measurements at lower $\sqrt{s}$ by the ATLAS and CMS collaborations \cite{Aaboud:2016urj,CMS:2014xja,Khachatryan:2015pba} and at the Tevatron and LEP colliders  \cite{Abazov:2007ad,Alcaraz:2006mx} and 
comparable to recent results from the CMS collaboration at $\sqrt{s}=13$~TeV~\cite{CMS-ZZ-13TEV}.
In addition, two-dimensional 95\% CL intervals are obtained by allowing pairs of aTGC parameters to vary simultaneously, while setting the others to zero. They are shown in \myfig{}~\ref{fig:2D_results}. No significant deviations from the SM are observed.

\begin{table}[h!]
\centering
\begin{tabular}{lll} 
\toprule
\textbf{Coupling strength} & \textbf{Expected 95\% CL} $\mathbf{(\times 10^{-3})}$ & \textbf{Observed 95\% CL} $\mathbf{(\times 10^{-3})}$\\
\midrule
\fyfour{} & $-2.4$, 2.4 & $-1.8$, 1.8 \\ 
\fzfour{} & $-2.1$, 2.1 & $-1.5$, 1.5 \\ 
\fyfive{} & $-2.4$, 2.4 & $-1.8$, 1.8 \\ 
\fzfive{} & $-2.0$, 2.0 & $-1.5$, 1.5\\ 
\bottomrule
\end{tabular}
\caption{One-dimensional expected and observed 95\% CL intervals on the aTGC coupling strengths. Each limit is obtained setting all other aTGC coupling strengths to zero.}
\label{tab:oneD_results}
\end{table}

\begin{figure}[p]
 \centering
\subfigure[]{\includegraphics[width=0.48\textwidth]{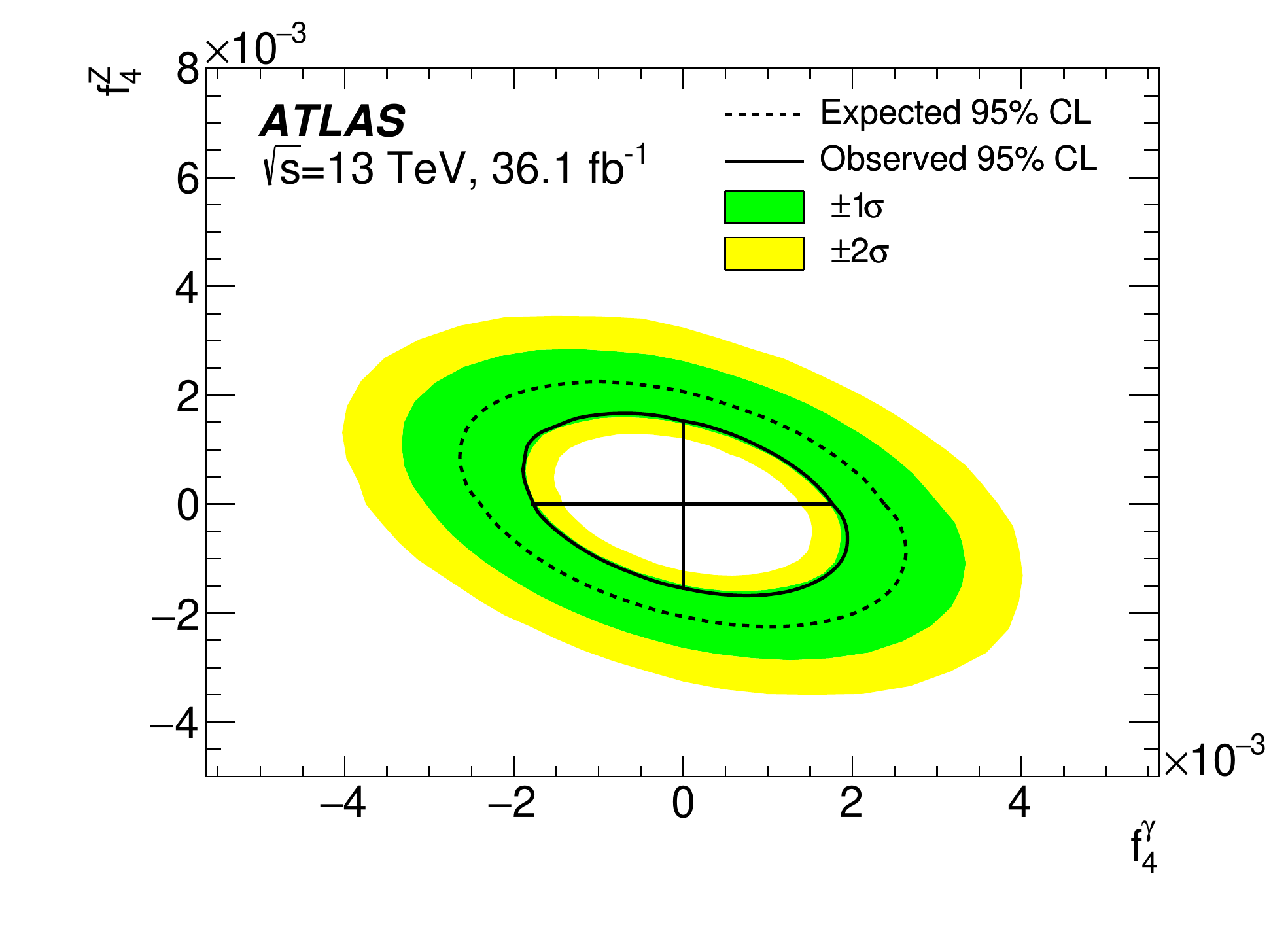}}
\subfigure[]{\includegraphics[width=0.48\textwidth]{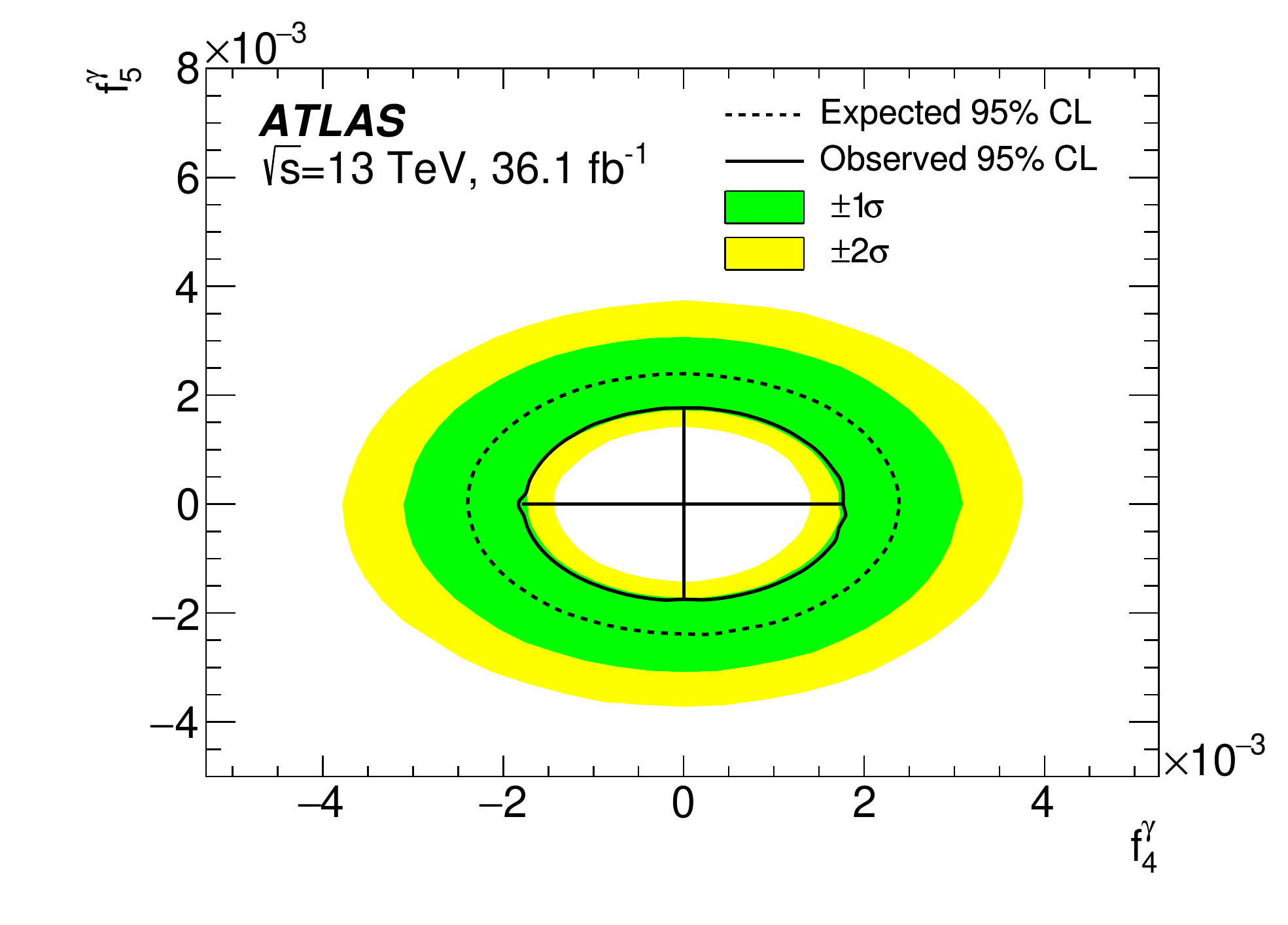}}
\subfigure[]{\includegraphics[width=0.48\textwidth]{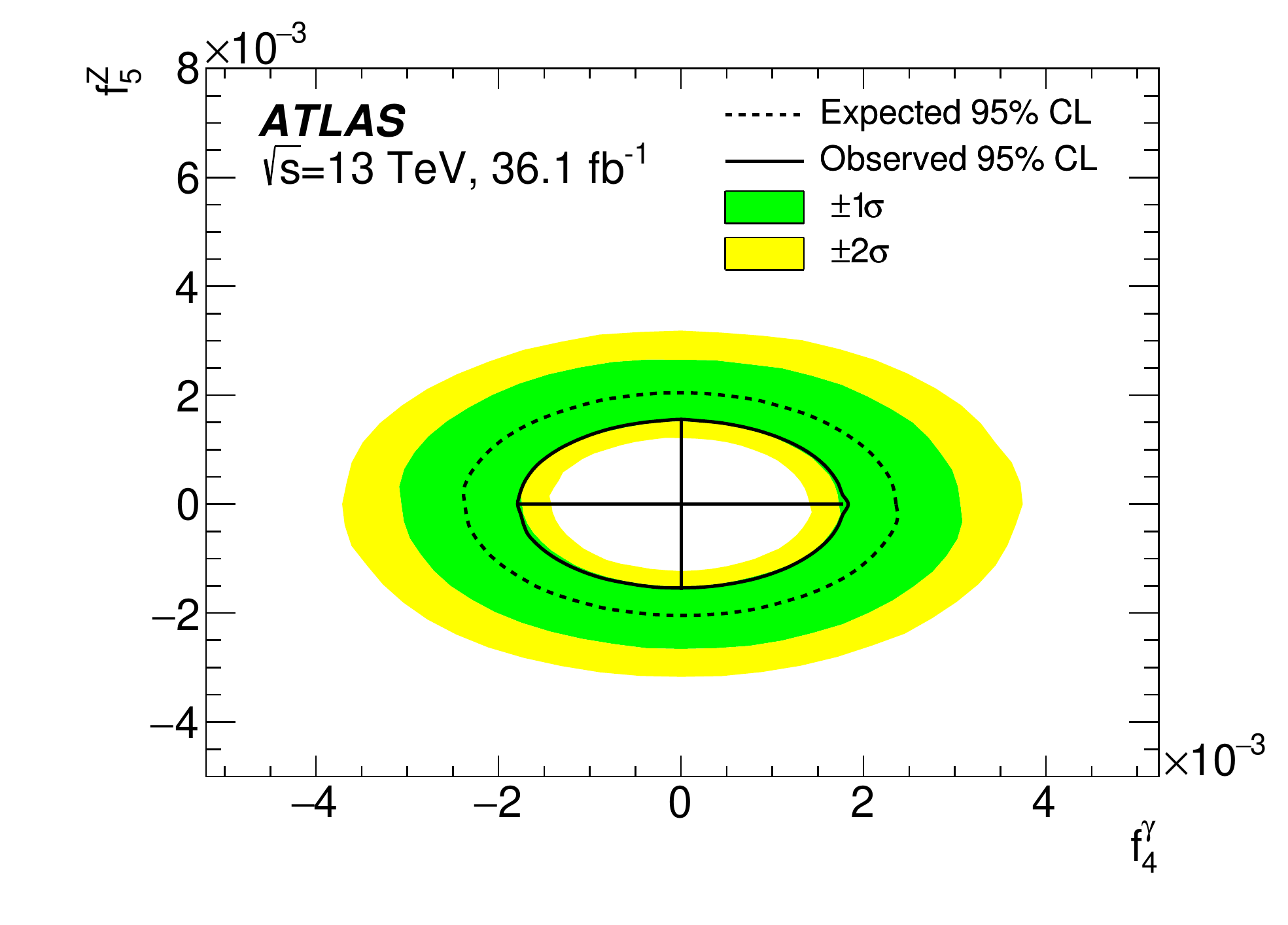}}
\subfigure[]{\includegraphics[width=0.48\textwidth]{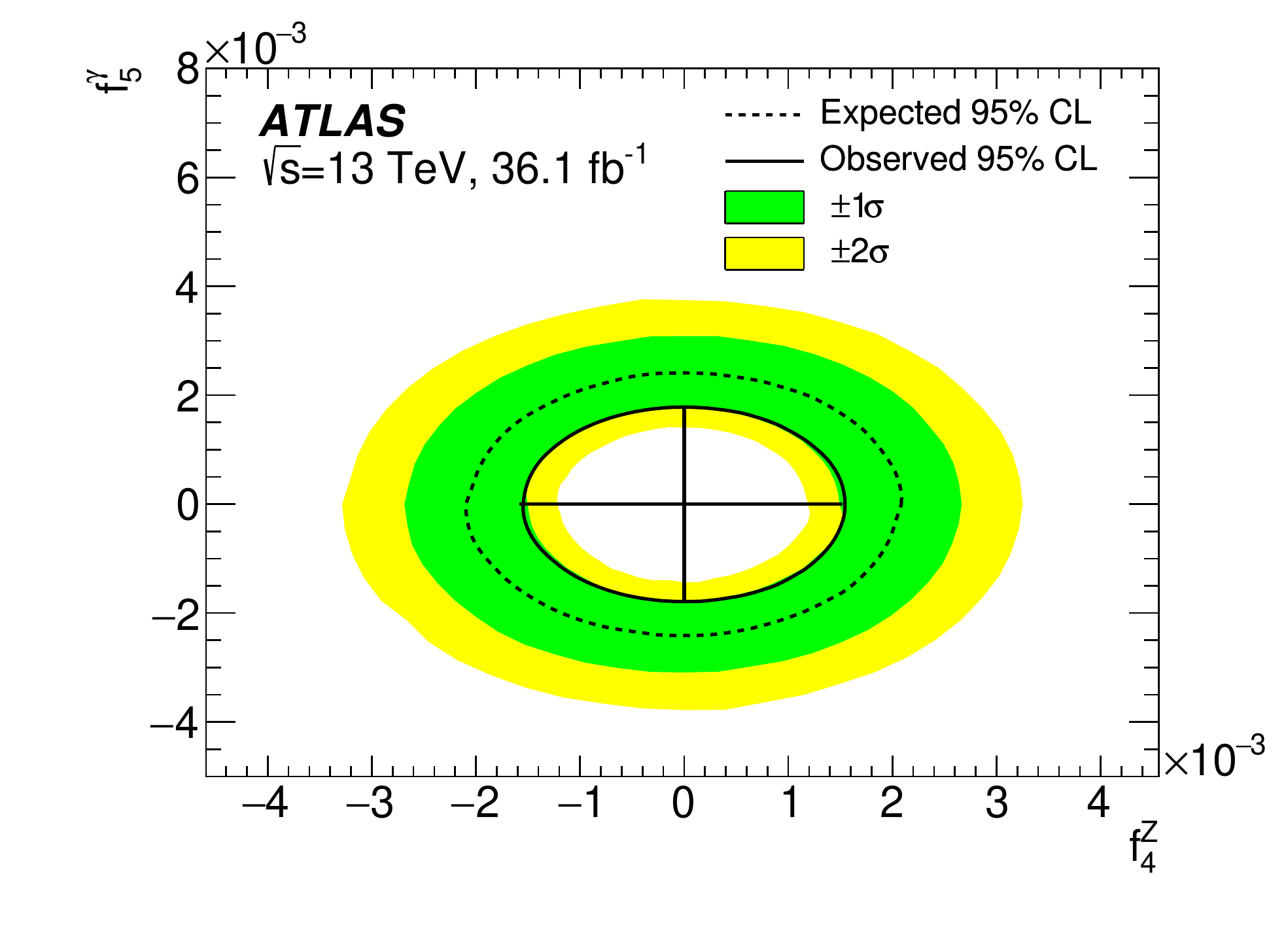}}
\subfigure[]{\includegraphics[width=0.48\textwidth]{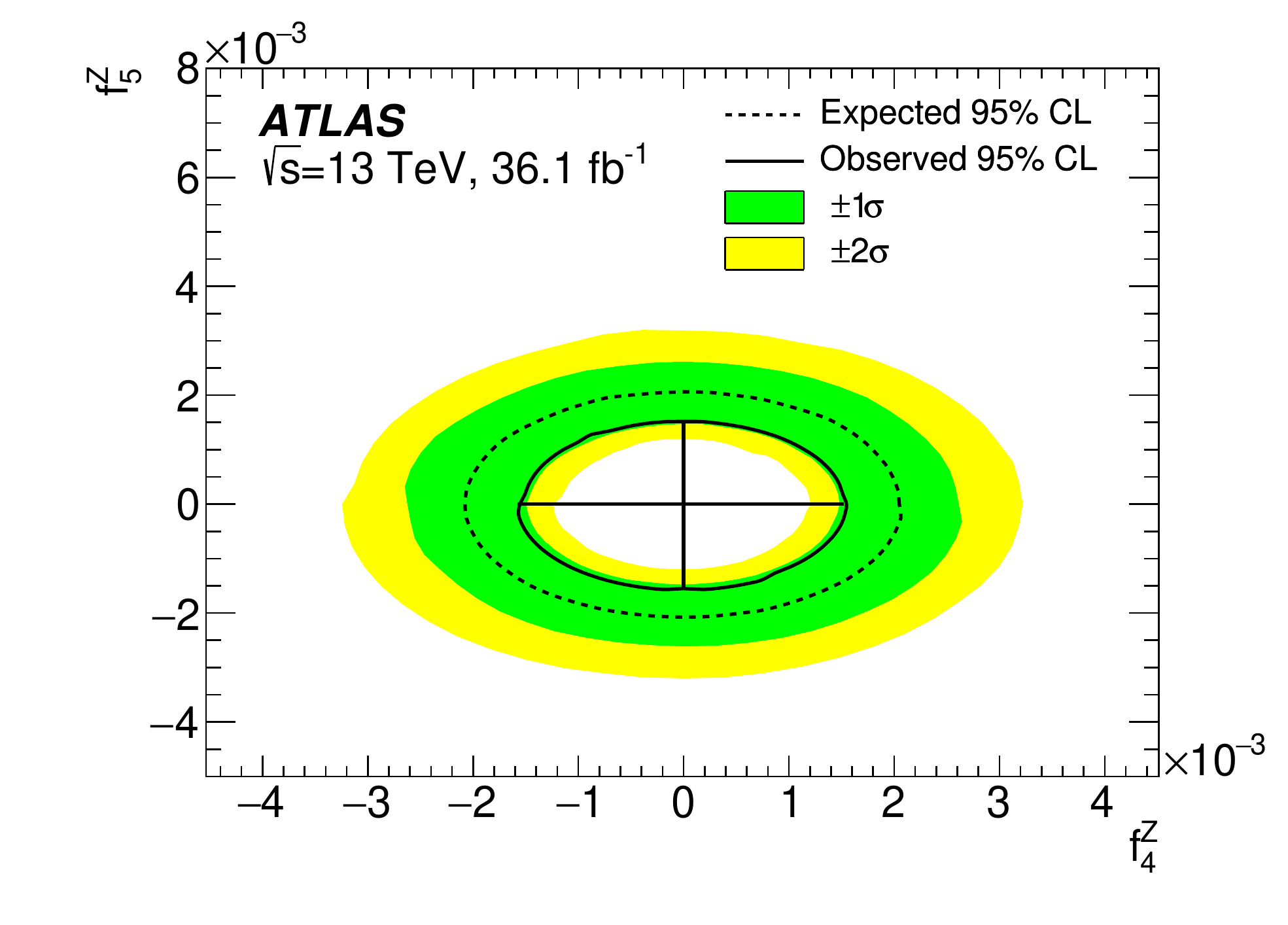}}
\subfigure[]{\includegraphics[width=0.48\textwidth]{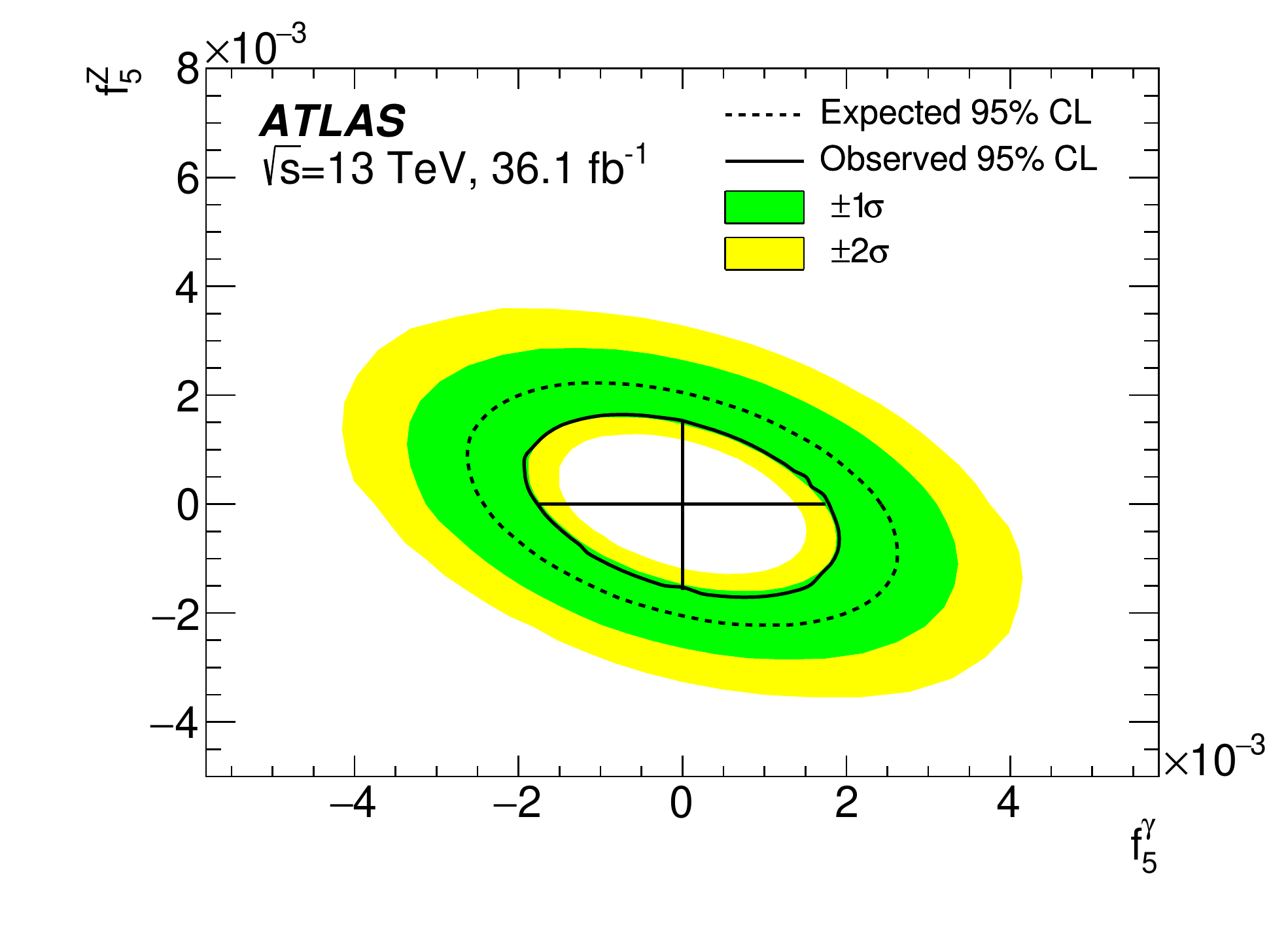}}
 \caption{Observed and expected two-dimensional 95\% CL intervals in planes of different pairs of aTGC coupling strengths. The aTGC coupling strengths other than those shown are set to zero. The black straight lines indicate the observed one-dimensional confidence intervals at 95\% CL. Figures produced by Maurice Becker and published in \myref~\cite{STDM-2016-15}.}
 \label{fig:2D_results}
\end{figure}

Confidence intervals are also provided for parameters of the effective field theory (EFT) of \myref{}~\cite{Degrande:2012wf}, which includes four dimension-8 operators describing aTGC interactions of neutral gauge bosons. The coefficients of the operators are denoted $C_{\tilde{B}W}/\Lambda^{4}$, $C_{BW}/\Lambda^{4}$, $C_{WW}/\Lambda^{4}$, and $C_{BB}/\Lambda^{4}$, where $\Lambda$ is the energy scale of the new physics described by the EFT. They can be linearly related to the parameters \fyfour{}, \fzfour{}, \fyfive{}, and \fzfive{} as described in \myref{}~\cite{Degrande:2013kka}. Thus \myeq{}~\ref{eq:atgc_yield_parametrisation} can be reformulated in terms of the EFT coefficients and confidence intervals set in the same way as for the coupling strengths. The resulting one-dimensional EFT confidence intervals can be found in \mytab{}~\ref{tab:oneD_EFTLimits}.
Two-dimensional EFT confidence intervals are shown in \myfig{}~\ref{fig:2D_results_EFT}.

\begin{table}[h!]
\centering
\begin{tabular}{lll} 
\toprule
\textbf{EFT parameter} & \textbf{Expected 95\% CL ($\mathbf{\TeV^{-4}}$)} & \textbf{Observed 95\% CL ($\mathbf{\TeV^{-4}}$)}\\
\midrule
$C_{\tilde{B}W}/\Lambda^{4}$ & $-8.1$, 8.1 & $-5.9$ ,  5.9 \\ 
$C_{WW}/\Lambda^{4}$ & $-4.0$, 4.0 & $-3.0$ ,  3.0 \\ 
$C_{BW}/\Lambda^{4}$ & $-4.4$, 4.4 & $-3.3$ ,  3.3 \\ 
$C_{BB}/\Lambda^{4}$ & $-3.7$, 3.7 & $-2.7$ ,  2.8 \\ 
\bottomrule
\end{tabular}
\caption{One-dimensional expected and observed 95\% CL intervals on EFT parameters using the transformation from \myref{}~\cite{Degrande:2013kka}. Each limit is obtained setting all other EFT parameters to zero.}
\label{tab:oneD_EFTLimits}
\end{table}

\begin{figure}[p]
 \centering
\subfigure[]{\includegraphics[width=0.48\textwidth]{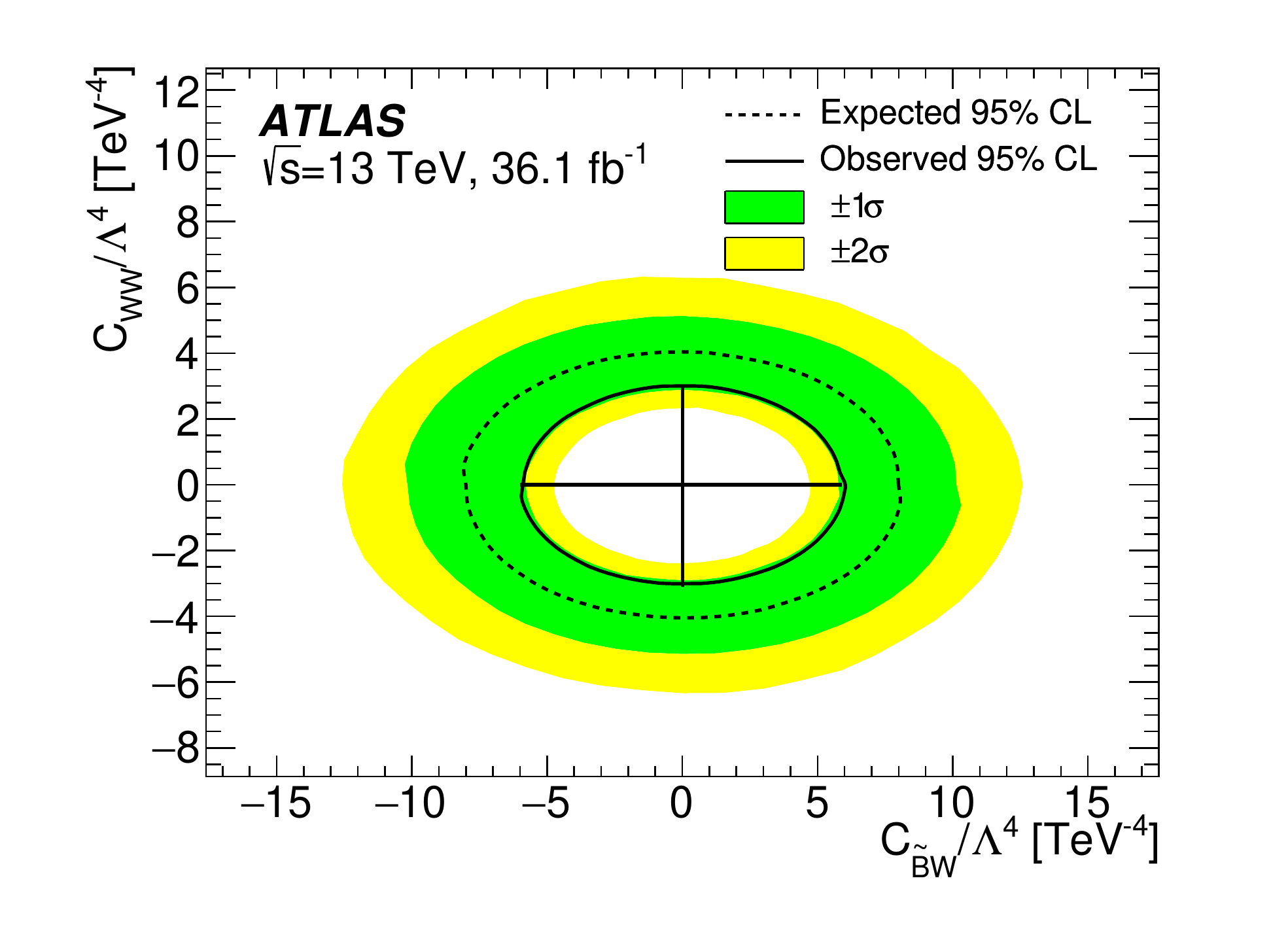}}
\subfigure[]{\includegraphics[width=0.48\textwidth]{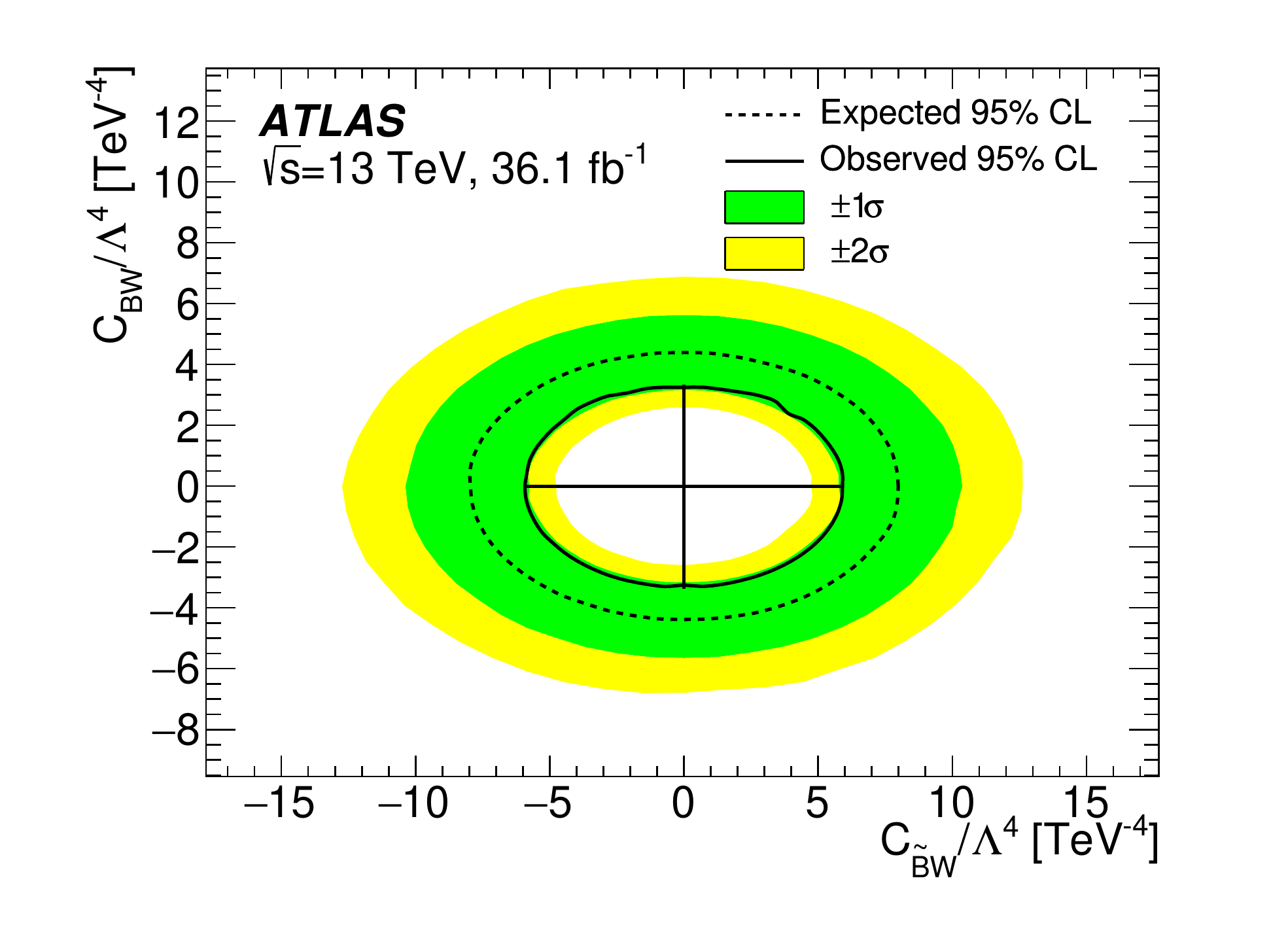}}
\subfigure[]{\includegraphics[width=0.48\textwidth]{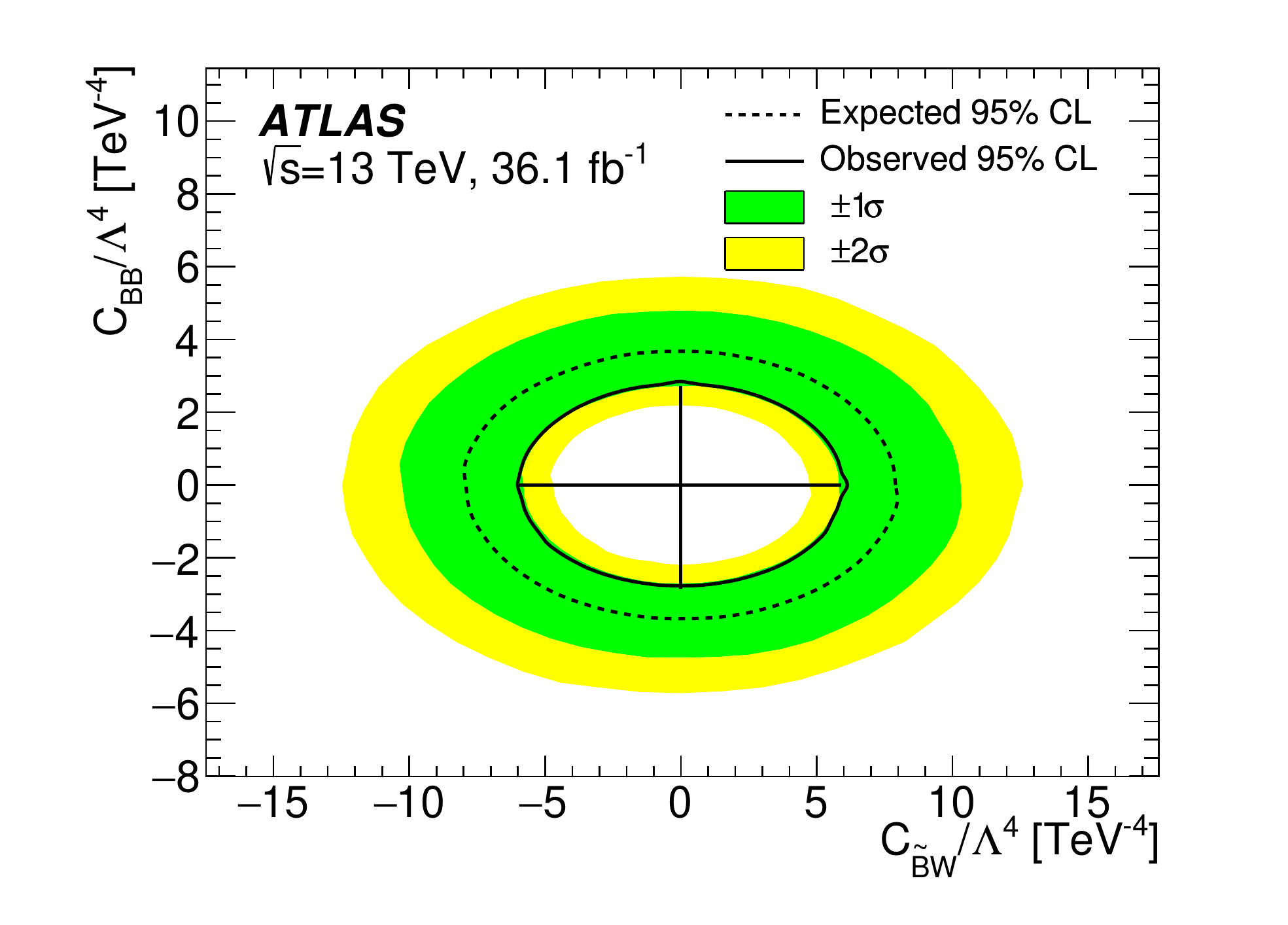}}
\subfigure[]{\includegraphics[width=0.48\textwidth]{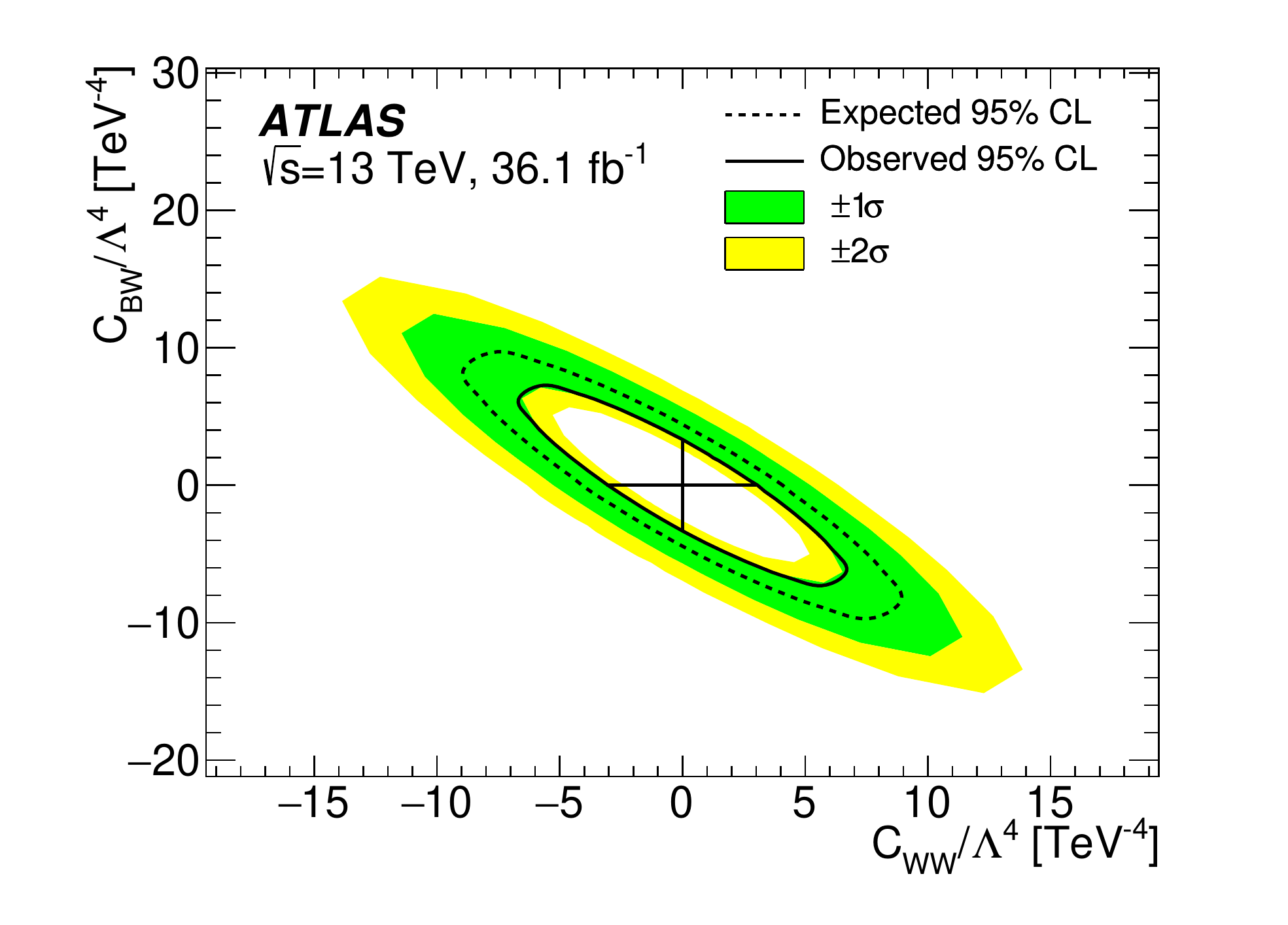}}
\subfigure[]{\includegraphics[width=0.48\textwidth]{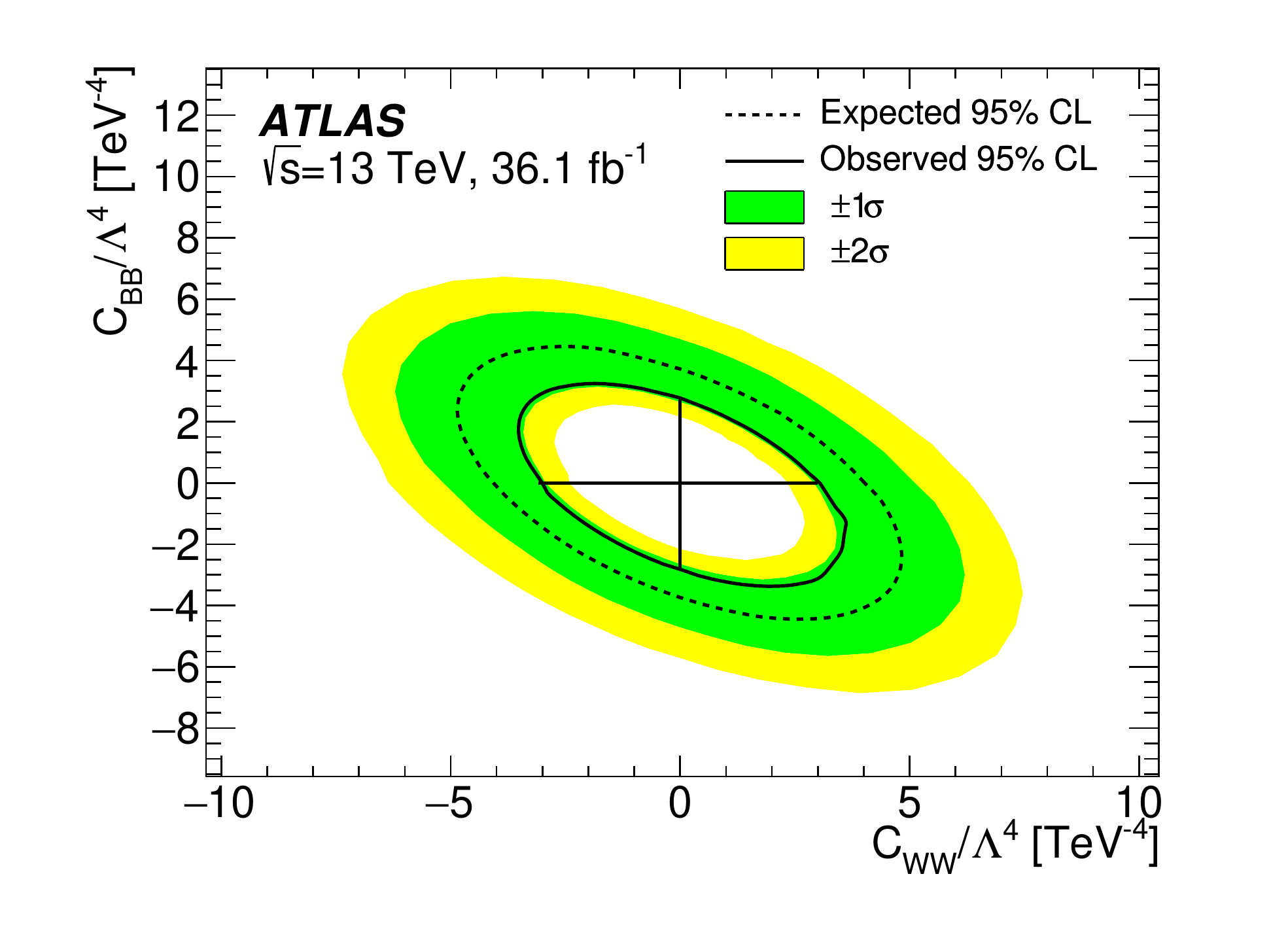}}
\subfigure[]{\includegraphics[width=0.48\textwidth]{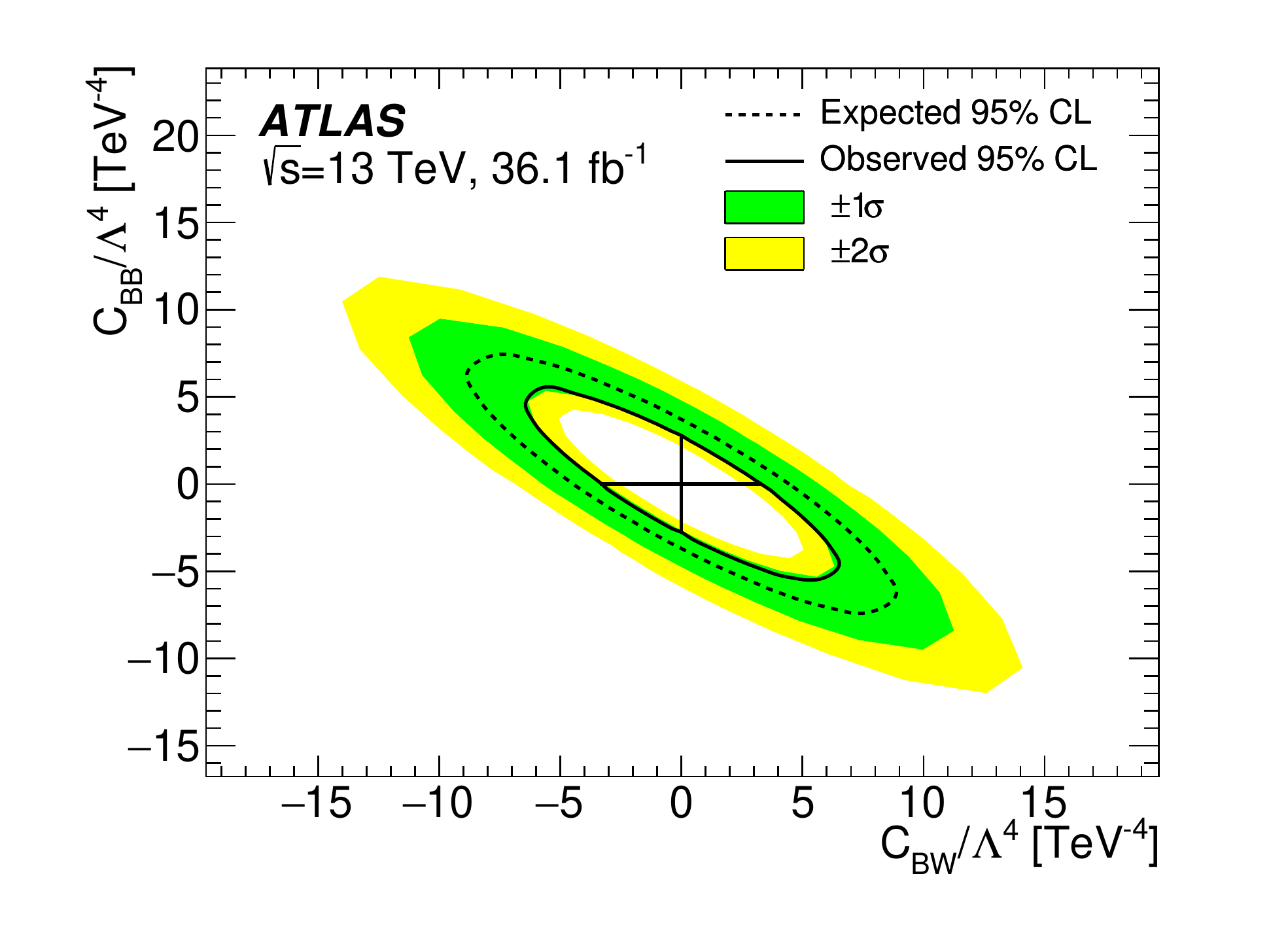}}
 \caption{Observed and expected two-dimensional 95\% CL intervals in planes of different pairs of EFT parameters using the transformation from \myref{}~\cite{Degrande:2013kka}. The EFT parameters other than those shown are set to zero. The black straight lines indicate the observed one-dimensional confidence intervals at 95\% CL. Figures produced by Maurice Becker and published in \myref~\cite{STDM-2016-15}.}
 \label{fig:2D_results_EFT}
\end{figure}

\clearpage

\clearpage
\section{Discussion and outlook}
The production of pairs of $\PZ$ bosons were studied in the \ZZllll{} channel in 13~\TeV{} proton--proton collisions produced at the LHC and recorded with the ATLAS detector, using data corresponding to an integrated luminosity of $(36.1 \pm 1.1)$~\ifb{}. Integrated fiducial cross sections were measured separately in the three decay channels \eeee{}, \eemm, and \mmmm{} as well as in their combination. They were found to agree well with NNLO SM predictions with NLO-QCD corrections for the \gluglu-initiated production process as well as NLO weak corrections applied. A slight excess of events in the \eeee{} channel led to poor compatibility of the channels, reflected by the low $p$-value of \pVal{} of the hypothesis that the relative contributions of the channels are as predicted by the SM.
The combined cross section was extrapolated to a total phase space and all SM \PZ{} boson decays. Differential cross sections were measured for twenty-one observables. They were compared to (possibly multiplicity-merged) NLO predictions with parton shower, to fixed-order NNLO predictions, and to fixed-order predictions combining predictions at the highest known orders for the different subprocesses (NNLO $\Pproton\Pproton \to \ZZ$, NLO $\gluglu \to \ZZ$, NLO weak corrections, electroweak $\Pproton\Pproton \to \ZZ jj$). In general, the predictions describe the observables reasonably well, within one standard deviation of the measurement in most bins. At the same time, the measurement precision is starting to be sufficiently high to reveal hints at small deviations from the predictions. Future measurements using the full Run 2 dataset, projected to be $\mathcal{O}(100~\text{fb})$, will shed more light on these trends.

A comparison and simple combination of integrated fiducial cross sections measured by ATLAS and CMS results was shown. The measurements are found to be very compatible with each other, and the combination is in good agreement with the SM. Such comparisons and combinations could in the future also be done for differential cross sections. One approach would be to construct a ``response'' matrix between each experiment's and the combined fiducial phase space using MC predictions, keeping track of different phase space acceptances as well as bin migrations (e.g.~due to the different pairing algorithm). Multiplication of each experiment's differential measurement by the ``response'' matrix can then be used to extrapolate it to the combined phase space. Thereupon, the extrapolated ATLAS and CMS results can be combined, with a more or less careful treatment of correlated systematic uncertainties. As the measurement is still quite statistically limited, even a conservative treatment of correlations should yield satisfactory results.

Using the transverse momentum of the leading-\pt{} \PZ{} boson candidate, confidence intervals were obtained for parameters of aTGCs forbidden at tree-level in the SM, both parametrised as aTGC coupling strengths and in an effective field theory approach. No significant deviations from the SM are observed. 
Again, future combinations of ATLAS and CMS results are possible, analogous to those done with the 7~\TeV{} searches \cite{ATLAS:2016hao}.

The results of the measurement along with a vast amount of metadata are preserved in human- and machine-readable format on \textsc{HepData} (\url{https://hepdata.net}) \cite{Maguire:2017ypu}. Preservation of the fiducial selection and data in \textsc{Rivet} \cite{Buckley:2010ar} is in preparation.




%



\clearpage
\section{Further analyses}

In addition to the analysis presented in this part so far, the author was directly involved in the following two analyses.

\subsection{Very early Run 2 measurement of $\PZ\PZ$ production}
\label{sec:early_zz_analysis}

The author led a team performing an early measurement of the integrated \ZZllll{} production cross section in 13~\TeV{} $\Pproton\Pproton$ collisions, published in \myref~\cite{STDM-2015-13} (and discussed further in conference proceedings in \myref~\cite{Richter:2016thg}). The methodology is essentially the same as that described in \mysecs~\ref{sec:zz_intro}--\ref{sec:zz_integrated_xs}, though less refined in many places. Using 3.1~\ifb{} of data, 63 candidate events are observed, allowing a measurement of the integrated cross sections with a statistical precision of around 28\% (\eeee{} channel) to 13\% (combination of all three channels). A comparison of the measured fiducial cross sections to NNLO predictions from \matrixnnlo{} is shown in \myfig~\ref{fig:early_fiducial_xs}. The results are in excellent agreement with the SM, deviating by less than one standard deviation from the NNLO prediction.
The early measurement has a narrower scope and several simplifications than the analysis presented in the sections above. Being one of the first analyses in Run 2 of the LHC, many essential reconstruction-performance studies were still ongoing in parallel, so the event selection was limited to using lepton selection requirements whose performance could already be estimated and relied upon. Each lepton is required to have $\pt > 20~\GeV{}$, which lowers the acceptance with respect to the later analysis by around 30\%.
\myref~\cite{STDM-2015-13} is the first published SM measurement of 13~\TeV{} $\Pproton\Pproton$ collisions from either the ATLAS or CMS collaboration, published around the same time as a few first direct searches for new physics by the two collaborations.

\begin{figure}[h!]
 \centering
\includegraphics[width=0.6\textwidth]{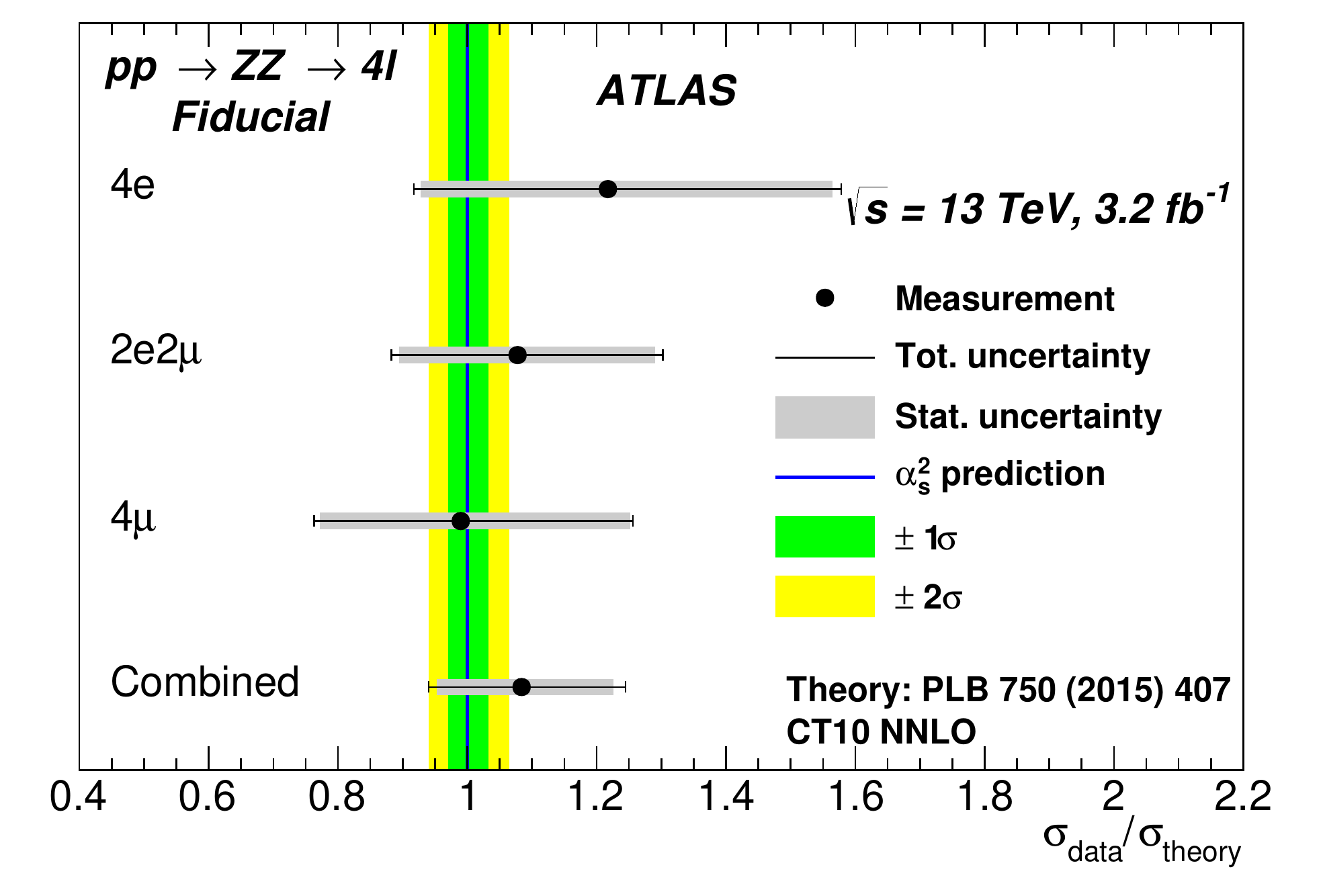}
\caption{Comparison of measured integrated fiducial cross sections to the NNLO SM prediction from \matrixnnlo{}. For the prediction, the QCD scale uncertainty is shown as a one- and two-standard-deviation band. Figure produced by Jonatan Rost\'{e}n and published in \myref~\cite{STDM-2015-13}.}
\label{fig:early_fiducial_xs}
\end{figure}

\subsection{Inclusive measurement of the four-lepton mass}
\label{sec:m4l_analysis}

The author also played a direct role in an as of yet unpublished measurement of the four-lepton mass in the interval $[75~\GeV{},\,\infty)$ with very loose dilepton-mass requirements, similar to \myref~\cite{STDM-2014-15}. He initiated the analysis and contributed to its first exploratory studies and preliminary optimisations, but only took on an advisory role in later stages because of time constraints. The four-lepton mass is corrected for experimental effects by unfolding and compared to SM predictions. The signal strength (relative to the SM prediction) of off-shell Higgs boson production is measured using events with $\mfourl > 180$~\GeV{}. Direct searches for modifications to Higgs-boson production by physics beyond the SM are performed in a generic effective-field-theory approach, similar to the search for aTGCs in the electroweak sector presented in \mysec~\ref{sec:anomalous}.
Both the off-shell signal strength measurement and the searches are performed using the detector-corrected measurement. This is currently not the norm, as almost all ATLAS and CMS searches use the reconstructed information directly. If done carefully, searches using unfolded distributions have the advantage that the impact of imperfect detector modelling can be reduced. However, careful studies of the unfolding uncertainties are required to ensure that the unfolding does not bias the conclusions.
At lower masses, the measurement is sensitive to the on-shell contributions from $\PZ \to \llll$ and $\PHiggs \to \llll$. Example Feynman diagrams for these processes are shown in \myfigs~\ref{fig:feynman_z4l} and \ref{fig:feynman_h4l}, respectively. These contributions give rise to a peak in the cross section at around $\mfourl \approx 91$~\GeV{} and 125~\GeV{}, as shown the predictions in \myfig~\ref{fig:fiducial_m4l_nomasscut}. The relative contribution of the \gluglu-initiated loop-induced process can be extracted in a model-dependent fashion thanks to its different relative contribution in different $\mfourl$ bins (also visible in \myfig~\ref{fig:fiducial_m4l_nomasscut}), by fitting the normalisation of templates constructed from the MC predictions. This allows an assessment of how compatible the predicted relative contribution is with the measurement.

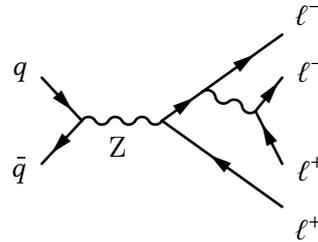
\begin{figure}[h!]
\centering
\begin{fmfgraph*}(90,65)
\fmfset{arrow_len}{3mm}
\fmfstraight
\fmfleft{i0,i1,i2,i3,i4}
\fmfright{o0,o1,o2,o3,o4}
\fmflabel{$\ell^-$}{o4}
\fmflabel{$\ell^-$}{o3}
\fmflabel{$\ell^+$}{o1}
\fmflabel{$\ell^+$}{o0}
\fmflabel{$\Pquark$}{i3}
\fmflabel{$\APquark$}{i1}
\fmf{phantom}{i2,z2,dummy2,d2,dummy3,dummy4,o2}
\fmffreeze
\fmf{fermion}{i3,z2,i1}
\fmf{photon,label=\PZ,label.side=right}{z2,d2}
\fmf{phantom}{o0,d2,p3,foo3,o4}
\fmffreeze
\fmf{fermion}{o0,d2,p3,o4}
\fmffreeze
\fmf{photon}{p3,s2}
\fmf{fermion}{o1,s2,o3}
\end{fmfgraph*}
\caption{\PZ-boson production with subsequent decay to four leptons via $\PZ^*/\gamma^*$ radiation off a lepton.}
\label{fig:feynman_z4l}
\end{figure}

\begin{figure}[h!]
\centering
\begin{fmfgraph*}(110,65)
\fmfset{arrow_len}{3mm}
\fmfleft{i0,i1,i2,i3,i4}
\fmfright{o0,o1,o2,o3,o4}
\fmflabel{\PZ}{o3}
\fmflabel{\PZ}{o1}
\fmflabel{$\Pgluon$}{i3}
\fmflabel{$\Pgluon$}{i1}
\fmflabel{$\ell^-$}{o4}
\fmflabel{$\ell^+$}{o3}
\fmflabel{$\ell^-$}{o1}
\fmflabel{$\ell^+$}{o0}
\fmf{phantom}{i3,t3,a3,b3,z3,o3}
\fmf{phantom}{i1,t1,a1,b1,z1,o1}
\fmffreeze
\fmf{gluon}{i3,t3}
\fmf{gluon}{i1,t1}
\fmf{fermion}{t3,t2,t1,t3}
\fmf{dashes,label=\PHiggs{},label.side=right}{t2,h2}
\fmf{photon}{h2,z3}
\fmf{photon}{h2,z1}
\fmf{fermion}{o3,z3,o4}
\fmf{fermion}{o0,z1,o1}
\end{fmfgraph*}
\caption{Higgs-boson production via a heavy-quark loop with subsequent decay to four leptons via $\PZ^{(*)}$ bosons.}
\label{fig:feynman_h4l}
\end{figure}
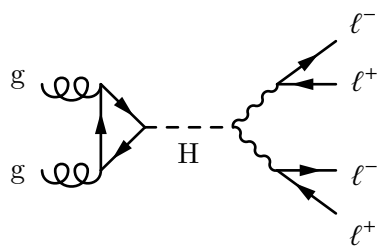


\clearpage\pagebreak
\part{Simulation of loop-induced processes with the \HERWIG{} event generator}
\label{sec:loopinduced}

\section{Introduction}
Loop-induced (LI) particle scattering processes are those that can only occur via a quantum loop even at their lowest order in perturbation theory, rather than at tree level.
Especially as the LHC now enters its precision-measurement era, many LI processes are phenomenologically important. The main reason for this is that they couple initial-state gluons to final-state colourless particles, such as Higgs or electroweak bosons. For processes that can occur at tree-level with a $\Pquark\APquark$ initial state, \gluglu-initiated LI production represents a gauge-invariant subset of the NNLO corrections. In the case of the $\PZ\PZ$ measurement presented in \mypart~\ref{sec:analysis}, it was shown that this accounts for approximately 15\% of the predicted signal, despite being formally an NNLO effect. This can be understood by considering the high gluon luminosity at the LHC due to the high hadronic centre-of-mass energy $\sqrt{s}$. The gluon luminosity at a given scale (e.g.~the Higgs boson mass) increases with $\sqrt{s}$, since the corresponding momentum fractions $x_1$, $x_2$ of the incoming partons decrease, and the gluon PDF is larger for low $x$ values. 
Indeed, due to the Higgs boson's small coupling to light quarks and large coupling to top quarks, the \gluglu-initiated LI process is the \emph{dominant} production mode for a Higgs boson or a pair of Higgs bosons at the LHC, $\gluglu \,\looparrow\, \PHiggs(\PHiggs)$ . The same might be true for further Higgs bosons beyond the SM, if they exist.

This part describes ongoing research into enabling and improving theoretical predictions for the LI production of colourless particles in proton collisions with the \HERWIG{}~7 event generator \cite{Bahr:2008pv,Bellm:2015jjp,Bellm:2017bvx}. \mysec~\ref{sec:lilo} discusses the LO description and \mysec~\ref{sec:linlo} the ongoing work towards an NLO description, in both cases matched with a parton shower if desired by the user. The matrix elements for the hard process are provided by external software to which \HERWIG{} is interfaced. \HERWIG{}, among other things, provides the user interface and control over all relevant parameters, performs phase space integration, combines the different partonic subprocesses, ensures the cancellation of infrared divergences among NLO matrix elements, performs the matching to a parton shower, and generates the parton shower and hadronisation as well as the underlying event.
Some preliminary results are shown for three processes that are particularly important to the current LHC physics programme: Higgs boson production, production of pairs of Higgs bosons, and four-lepton production. However, almost all of the new developments are applicable to arbitrary LI processes, so other processes can be generated as soon as matrix elements become available.


\section{Implementation}

The innovations presented here rely on the \matchbox{} framework \cite{Reuschle:2016ndi,Platzer:2011bc}, which is part of \HERWIG{} 7. \matchbox{} manages the matrix elements for the hard process and provides interfaces to several external matrix-element providers, currently: \openloops{} \cite{Cascioli:2011va}, \MADGRAPH{} \cite{Alwall:2011uj}, \gosam{} \cite{Cullen:2011ac}, VBFNLO \cite{Arnold:2008rz,Arnold:2011wj}, and \textsc{NJet} \cite{Badger:2012pg,Badger:2010nx}. In the present work, all matrix elements are provided by \openloops{} unless explicitly mentioned otherwise. \openloops{} in turn relies on \textsc{Collier} \cite{Denner:2016kdg} for the fast and numerically stable evaluation of one-loop integrals \cite{Denner:2002ii,Denner:2005nn,Denner:2010tr}. In the future, the functionality will also be implemented using \gosam{} and \MADGRAPH{}, to allow cross-checks. Preliminary tests with \gosam{} have revealed it to be very slow compared to \openloops{}, both in the compilation of the matrix element code generated by \gosam{} (\openloops{} keeps the process libraries in a central location) and in the evaluation of the matrix elements. The \HERWIG{} interface to \openloops{} was partially rewritten by the author. As before, the updated Binoth Les Houches Accord (BLHA2) interface \cite{Alioli:2013nda} (based on the original BLHA \cite{Binoth:2010xt}) is used to request matrix elements from \openloops{}, but now the native \openloops{} C++ interface is used to evaluate them. This allows retrieving information from \openloops{} that is needed for NLO generation of LI processes. 

The NLO functionality of \matchbox{} will be discussed in \mysec~\ref{sec:linlo}. To make use of the full functionality of \matchbox{}, the author changed the bookkeeping of LI matrix elements in the \HERWIG{} software, to bring it onto an equal footing as that for tree-level processes.
Whether or not a process is LI can be inferred by counting its vertices and \emph{legs} (i.e.~incoming and outgoing particles at the matrix-element level) at the lowest contributing order,
\begin{equation*}
c = \#\text{(legs)} - \#\text{(vertices)}.
\end{equation*}
The classifier $c$ is 0 for a LI process and $\geq 2$ for a tree process (it is 2 unless quartic vertices such as $\gluglu\gluglu$ or $\PWplus\PWminus\ZZ$ are involved, in which case it can be higher). 
The user interface for LI processes was also unified, so that the only difference with respect to tree processes is that an additional configuration file (provided as part of \HERWIG{}) must be read in the \HERWIG{} configuration file:
\begin{verbatim}
read Matchbox/LoopInduced.in
\end{verbatim}
This sets up the generic building blocks from which loop-induced topologies are constructed internally. For instance, generating the production of a pair of Higgs bosons in gluon fusion, $\gluglu\, \looparrow \, \PHiggs\PHiggs$ at LO gives rise to the following four topologies:

%
%
\begin{quote}
{\footnotesize
\begin{verbatim}
 (0)                             (0)
  |                               |
 [g]                             [g]
  |                               |
  |        |--[h0]--(2)           |         |--[h0]--(2)
  |--[h0]--|                      |--[nLP]--|
  |        |--[h0]--(3)           |         |--[h0]--(3)
  |                               |
 [g]                             [g]
  |                               |
 (1)                             (1)

 (0)                             (0)
  |                               |
 [g]                             [g]
  |                               |
  |--[h0]--(3)                    |--[h0]--(2)
  |                               |
[cLP]                           [cLP]
  |                               |
  |--[h0]--(2)                    |--[h0]--(3)
  |                               |
 [g]                             [g]
  |                               |
 (1)                             (1)
\end{verbatim}
}
\end{quote}
where the numbers in parentheses label the external momenta, \texttt{h0} denotes a Higgs boson, and \texttt{nLP} (\texttt{cLP}) denotes an electrically neutral (charged) ``loop particle'', which is simply a bookkeeping device for the types of interactions that exist and the kinematic invariants involved.

\subsection{Event generation setup}

All event generation results shown in this part are preliminary and require further validation. Proton-proton collision events at 13~\TeV{} centre-of-mass energy are generated.
The renormalisation and factorisation scale is set to $m_{X}/2$, where $m_X$ is the invariant mass of the colourless system generated in the hard process. Doing this in \HERWIG{} makes use of a new generic scale choice implemented by the author, the invariant mass of all \emph{colourless} particles:
\begin{verbatim}
cd /Herwig/MatrixElements/Matchbox
set Factory:ScaleChoice Scales/ColourlessSHatScale
\end{verbatim}
The alternative choice $\hat{s}$ (\texttt{SHatScale}) that includes coloured particles must not be used for NLO generation, since it is not collinear safe: a collinear initial-state emission changes the scale $\hat{s}$, but not the observable final state. The PDF4LHC15\_nlo\_100 \cite{Butterworth:2015oua} NLO PDF set is used, accessed via the LHAPDF 6 interface \cite{Buckley:2014ana}. QCD scale and PDF uncertainties are not yet assessed. \HERWIG{} has two independent parton shower algorithms, a dipole \cite{Platzer:2009jq} and an angular-ordered shower \cite{Gieseke:2003rz}. In this work, the angular-ordered shower is used unless mentioned otherwise. 
Where associated jets are included, they are anti-$k_t$ jets with radius parameter $R = 0.4$ and required to have $\pt{} > 20$~\GeV{} and $|y| < 5.0$.
The generated events are analysed with \rivet{} \cite{Buckley:2010ar} to fill histograms.

%
%
%

\section{Leading-order results}\label{sec:lilo}

The first automated LO event generation of LI processes matched with a parton shower was performed in the \MGMCatNLO{} generator \cite{Hirschi:2015iia}. Such processes are also supported in \SHERPA{} \cite{Buschmann:2014sia}. The following sections show preliminary example results obtained with \HERWIG{}. 
Some include one additional jet generated at the matrix-element level. \matchbox{} is capable of merging matrix elements with various jet multiplicities (generated at LO or NLO) into one sample, but this functionality was not used in the results presented here.


\subsection{Higgs boson production}
To further highlight the need for a description of LI processes, it is instructive to consider the process $\gluglu \, \looparrow\, \PHiggs$ in some detail. The LO Feynman diagrams are shown in \myfig~\ref{fig:ggF_higgs_feynman_diagrams}.
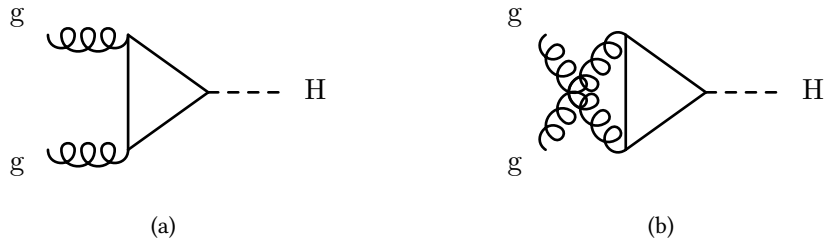
\begin{figure}[h!]
	\centering
	\subfigure[]{
		\centering
		\begin{fmfgraph*}(90,65)
			\fmfset{arrow_len}{3mm}
			\fmfstraight
			\fmfleft{idummy0,i0,idummy1,i1,idummy2,i2,idummy3}
			\fmfright{odummy0,o0,odummy1,o1,odummy2,o2,odummy3}
			\fmflabel{\Pgluon}{i2}
			\fmflabel{\Pgluon}{i0}
			\fmflabel{\PHiggs}{o1}
			\fmf{phantom}{i2,v2,dummy2,o2}
			\fmf{phantom}{i0,v0,dummy0,o0}
			\fmffreeze
			\fmf{gluon}{i2,v2}
			\fmf{plain}{v2,vh}
			\fmf{gluon}{i0,v0}
			\fmf{plain}{v0,vh}
			\fmf{dashes,tension=2}{vh,o1}
			\fmffreeze
			\fmf{plain}{v0,v2}
		\end{fmfgraph*}
	}
	\hspace{3cm}
	\subfigure[]{
		\centering
		\begin{fmfgraph*}(90,65)
			\fmfset{arrow_len}{3mm}
			\fmfstraight
			\fmfleft{idummy0,i0,idummy1,i1,idummy2,i2,idummy3}
			\fmfright{odummy0,o0,odummy1,o1,odummy2,o2,odummy3}
			\fmflabel{\Pgluon}{i2}
			\fmflabel{\Pgluon}{i0}
			\fmflabel{\PHiggs}{o1}
			\fmf{phantom}{i2,v2,dummy2,o2}
			\fmf{phantom}{i0,v0,dummy0,o0}
			\fmffreeze
			\fmf{gluon}{i2,v0}
			\fmf{plain}{v2,vh}
			\fmf{gluon}{v2,i0}
			\fmf{plain}{v0,vh}
			\fmf{dashes,tension=2}{vh,o1}
			\fmffreeze
			\fmf{plain}{v0,v2}
		\end{fmfgraph*}
	}
\caption{LO Feynman diagrams for Higgs boson production via gluon fusion.}
\label{fig:ggF_higgs_feynman_diagrams}
\end{figure}

Assuming that only one quark flavour (with mass $m_{\Pquark}$) contributes to the fermion loop, the partonic cross section is given by\footnote{Calculated by the author and in agreement with \myref~\cite{Dawson:1990zj}.}
\begin{equation}\label{eq:higgs_lo_xsect}
\hat{\sigma}(\Pg\Pg\to\PHiggs) = \frac{\alphas^2m_{\PHiggs}^2}{256\pi v^2} \delta(\hat{s}-m_{\PHiggs}^2) \left|\tau\left[1+(1-\tau) f(\tau)\right]\right|^2,
\end{equation}
where $\tau = 4m_{\Pquark}^2/m_H^2$ and
\begin{equation*} 
	f(\tau) = \left\{
	\begin{array}{cl}
		\arcsin^2\sqrt{1/\tau} & \quad \mbox{if $\tau \geq 1$,}\\
		-\frac{1}{4}\left[\ln \frac{1 + \sqrt{1- \tau}}{1 - \sqrt{1-\tau}} - i\pi \right]^2 & \quad \mbox{if $\tau < 1$.}
	\end{array}
	\right.
\end{equation*}
The cross section is visualised as a function of $m_{\Pquark}$ in \myfig~\ref{fig:higgs_lo_xsect_fig}. It can be seen that the contribution of the top quark is much larger than that of the bottom quark (by a factor of approximately 150), ignoring interference between the two flavours. Taking the top quark mass to be infinite, the cross section becomes
\begin{equation}
\lim_{\tau \to \infty} \hat{\sigma}(\Pg\Pg\to\PHiggs) = \frac{\alphas^2m_{\PHiggs}^2}{576\pi v^2} \delta(\hat{s}-m_{\PHiggs}^2)
\end{equation}
(which is a constant up to the renormalisation scale dependence of \alphas{}), where the expansion
\begin{equation}
	\arcsin \sqrt{1/\tau} = \frac{1}{\sqrt{\tau}} + \frac{1}{3! \left(\sqrt{\tau}\right)^3} +
	\mathcal{O}\left(\frac{1}{\left(\sqrt{\tau}\right)^5}\right)
\end{equation}
valid for large $\tau$ and hence
\begin{equation*}
\lim_{\tau \to \infty} \left|\tau\left[1+(1-\tau) f(\tau)\right]\right|^2 = \frac{4}{9}
\end{equation*}
has been used. Despite the fact that $\tau$ is only approximately $(2 \times 175~\GeV)^2 / (125~\GeV)^2 \approx 8$, the $m_{\Ptop} \to \infty$ approximation yields satisfactory results in many situations and has been used extensively. Its appeal lies in the fact that the $m_{\Ptop} \to \infty$ approximation means that the structure of the loop is not resolved, effectively shrinking it into a pointlike $\gluglu\PHiggs$ vertex. In the approximation, the two LO Feynman diagrams of \myfig~\ref{fig:ggF_higgs_feynman_diagrams} are replaced by that in \myfig~\ref{fig:ggF_heft_feynman_diagram}. This allows a much easier calculation of higher-order corrections, since one loop has been eliminated from the calculation. The advantage comes at the cost of less good modelling, particularly in the presence of high-\pt{} associated jets: hard radiation resolves the finite-$m_{\Ptop}$ loop, so the $m_{\Ptop}\to \infty$ approximation is poor for instance at high \pt{} of the Higgs boson (which recoils against high-\pt{} radiation). This is shown quantitatively below.

\begin{figure}[h!]
	\centering
	\begin{tikzpicture}
	    \node[anchor=south west,inner sep=0] (image) at (0,0) {\includegraphics[width=0.6\textwidth]{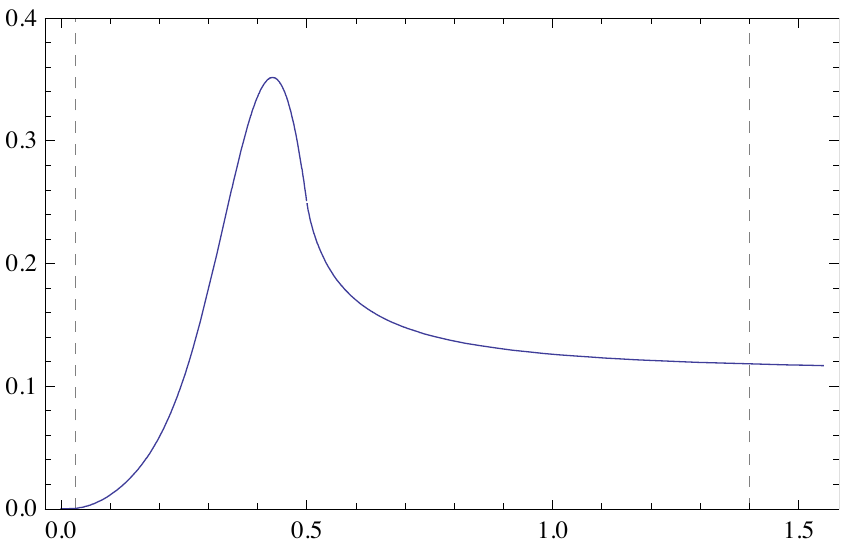}};
	    \begin{scope}[x={(image.south east)},y={(image.north west)}]
			\node [align=left, rotate=90] at (-0.04,0.56) {$\hat{\sigma}(\Pg\Pg\to\PHiggs)$ (arbitrary units)};
			\node [align=left] at (0.7,-0.01) {$\text{Fermion mass}\, /\, (m_{\PHiggs} = 125~\GeV)$};
	        \node [align=left, text width=10em, rotate=90] at (0.91,0.71) {Top quark mass};
	        \node [align=left, text width=10em, rotate=90] at (0.12,0.71) {Bottom quark mass};
	    \end{scope}
	\end{tikzpicture}
	\caption{LO partonic $\Pg\Pg\, \to\, \PHiggs$ cross section as a function of the loop fermion mass. The dashed vertical lines indicate the mass of the bottom and top quark, $m_{\Pbeauty} = 4.5$~\GeV{} and $m_{\Ptop} = 175$~\GeV{}.}
	\label{fig:higgs_lo_xsect_fig}
\end{figure}
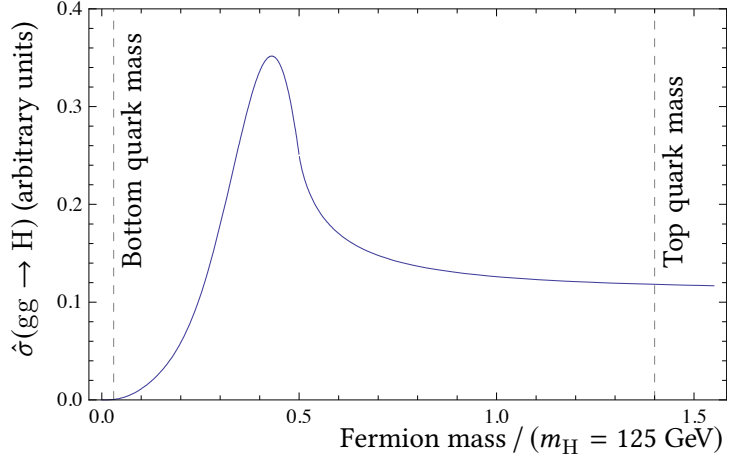

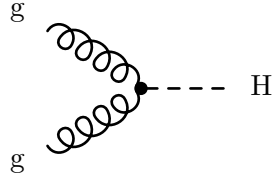
\begin{figure}[h!]
	\centering
		\begin{fmfgraph*}(70,65)
			\fmfset{arrow_len}{3mm}
			\fmfstraight
			\fmfleft{idummy0,i0,idummy1,i1,idummy2,i2,idummy3}
			\fmfright{odummy0,o0,odummy1,o1,odummy2,o2,odummy3}
			\fmflabel{\Pgluon}{i2}
			\fmflabel{\Pgluon}{i0}
			\fmflabel{\PHiggs}{o1}
			\fmf{gluon}{h1,i2}
			\fmf{gluon}{i0,h1}
			\fmf{dashes,tension=2}{h1,o1}
			\fmfv{decor.shape=circle,decor.filled=full,decor.size=2thick}{h1}
		\end{fmfgraph*}
\caption{LO Feynman diagram for Higgs boson production via gluon fusion in the $m_{\Ptop} \to \infty$ approximation. The large dot indicates the effective $\gluglu\PHiggs$ vertex.}
\label{fig:ggF_heft_feynman_diagram}
\end{figure}

\myfig~\ref{fig:lilo_h_hj} shows differential LO $\Pproton\Pproton \to \PHiggs$ cross sections calculated with \HERWIG{}, comparing the full LI cross sections and those in the $m_{\Ptop} \to \infty$ approximation. The matrix elements in the latter case are provided by \MADGRAPH{}. The Higgs boson is treated as a stable particle and its mass is taken to be 125~\GeV{}. For a more realistic description it is of course desirable to simulate the Higgs boson decay (at the matrix-element level), but the presented studies focus on the \emph{production} mechanism, so the decay is neglected.
\myfig~\ref{fig:lilo_h_hj__HpT} shows the transverse momentum of the Higgs boson. By conservation of transverse momentum, this observable can only differ from zero if there is something for the Higgs boson to recoil against, in this case QCD radiation. (The primordial \pt{} of the partons inside the protons is negligible at the scales of interest here.) The cross section is shown with all QCD radiation provided by the parton shower as well as with one additional jet generated at the matrix element level in addition to the parton shower contributions. The parton shower alone, by design only capable of describing quasi-collinear radiation adequately, predicts a steeply falling \pt{} spectrum, essentially vanishing around 200~\GeV{}. Including a jet at the matrix-element level produces a more gently falling distribution. As expected, the full LI prediction and the $m_{\Ptop} \to \infty$ approximation agree relatively well at low \pt{}, within approximately 5\%. However, they start to diverge enormously from around 200~\GeV{} upwards. This is due to aforementioned the effect of hard radiation resolving the loop structure. The $m_{\Ptop} \to \infty$ approximation leads to a too hard \pt{} spectrum. At around 1~\TeV{}, the disagreement has reached as much as a factor of ten. These findings are in agreement with a similar study in \myref~\cite{Buschmann:2014sia}.
\myfig~\ref{fig:lilo_h_hj__j1pT} shows the transverse momentum of the leading-\pt{} jet. Above the jet \pt{} threshold of 20~\GeV{} it resembles the \pt{} distribution of the Higgs boson closely, demonstrating that much of the recoil of the Higgs boson in the generated samples is due to the leading jet.

\begin{figure}[h!]
\centering
\subfigure[]{\includegraphics[width=0.6\textwidth]{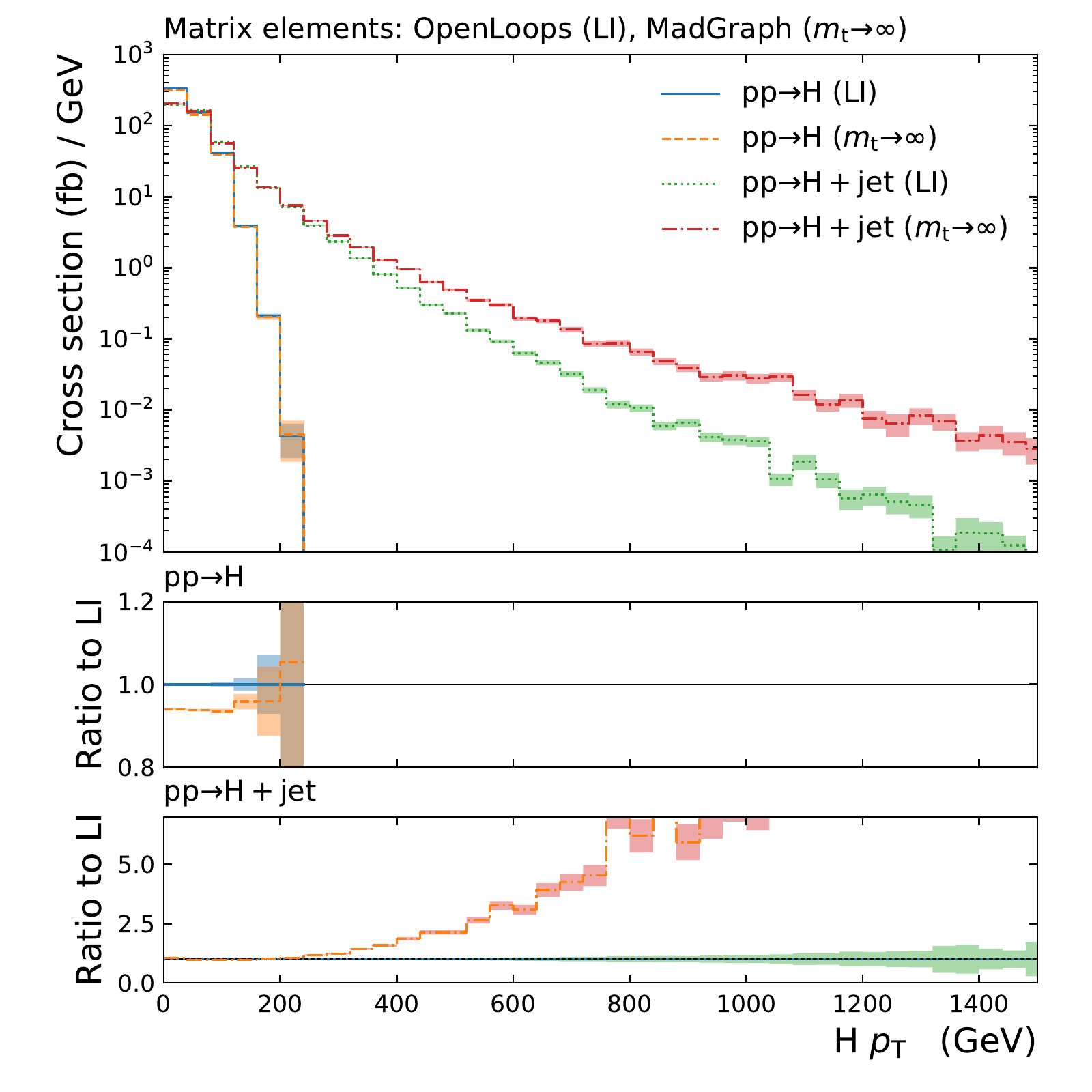}\label{fig:lilo_h_hj__HpT}}
\vspace{-3mm}
\hspace{1mm}
\subfigure[]{\includegraphics[width=0.6\textwidth]{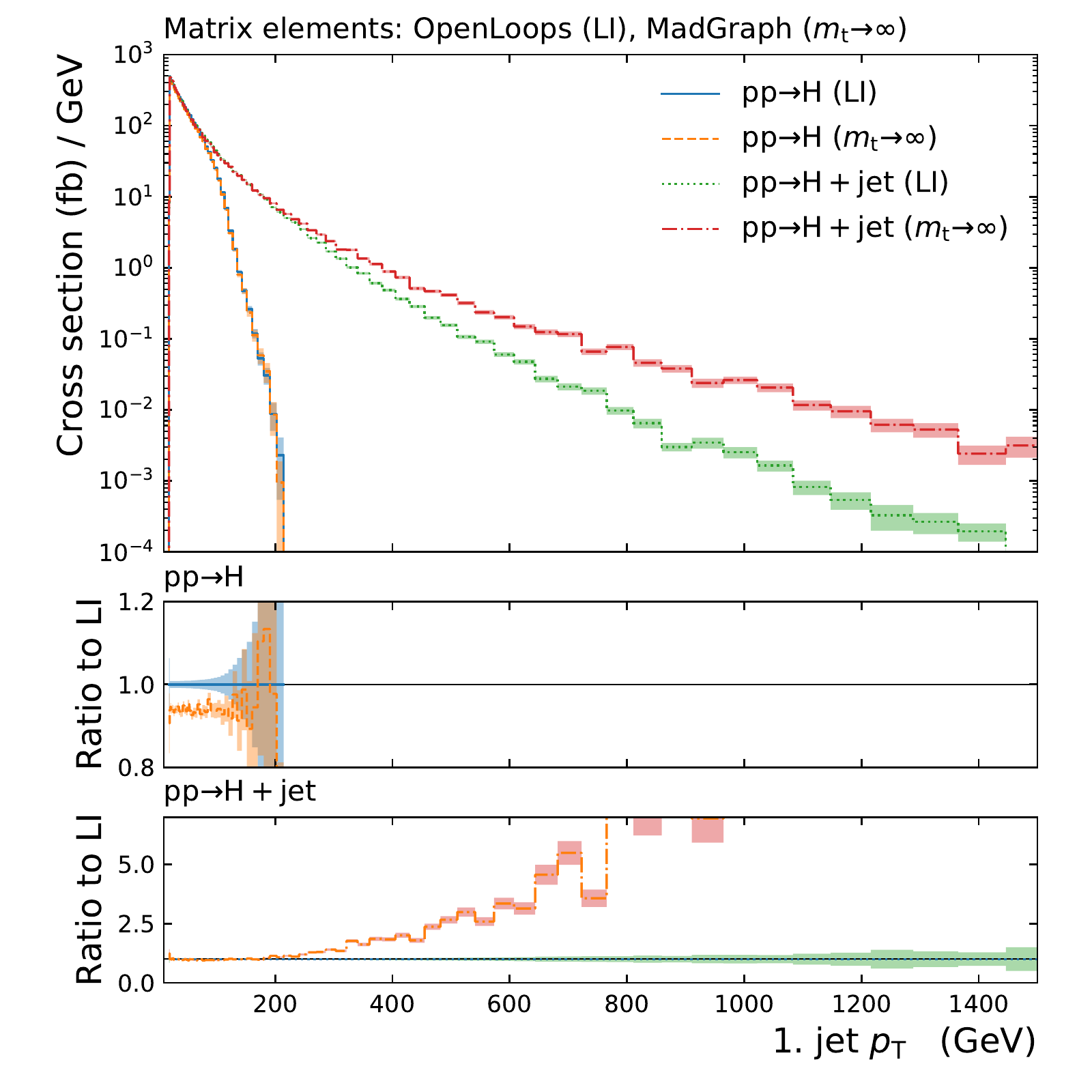}\label{fig:lilo_h_hj__j1pT}}
\vspace{-2mm}
\caption{Higgs boson production cross sections with zero or one additional jet generated at the matrix-element level, as a function of the transverse momentum of the (a)  Higgs boson and (b) leading-\pt{} jet. Both the LI production with full quark mass dependence and production in the $m_{\Ptop} \to \infty$ limit are shown. The shaded bands indicate the statistical uncertainty of the MC integration.} 
\label{fig:lilo_h_hj}
\end{figure}

It might seem surprising that the cross section for $\PHiggs j$ production is not much lower than that for $\PHiggs$ production. After all, requiring a relatively energetic jet ($\pt > 20$~\GeV{}) could be expected to reduce the cross section by $\mathcal{O}(\alphas) \sim 0.1$. However, this effect is offset by the increase in cross section that comes from the opening of additional initial-state flavour channels for the $\PHiggs j$ process. While LO $\PHiggs$ production requires two initial-state gluons, the addition of an additional final-state parton allows the channels
\begin{equation*}
\begin{split}
\Pquark\Pgluon \, &\looparrow \, \PHiggs\Pquark,\\
\APquark\Pgluon \, &\looparrow \, \PHiggs\APquark,\\
\Pquark\APquark \, &\looparrow \, \PHiggs\Pgluon,
\end{split}
\end{equation*}
where $\Pquark$ represents any quark flavour (in practice excluding the top quark, which is prohibitively massive). Example Feynman diagrams for $\gluglu\,\looparrow\,\PHiggs j$ production are shown in \myfig~\ref{fig:higgs_plus_jet}. These are also contributions to the real-emission part of the NLO $\gluglu\,\looparrow\,\PHiggs$ cross section. Corresponding diagrams in the $m_{\Ptop}\to\infty$ approximation are shown in \myfig~\ref{fig:higgs_plus_jet__heft}. Hard emissions with a scale of the order of the top-quark mass or higher resolve the structure of the loop, so the $m_{\Ptop}\to\infty$ approximation is inadequate in their presence.

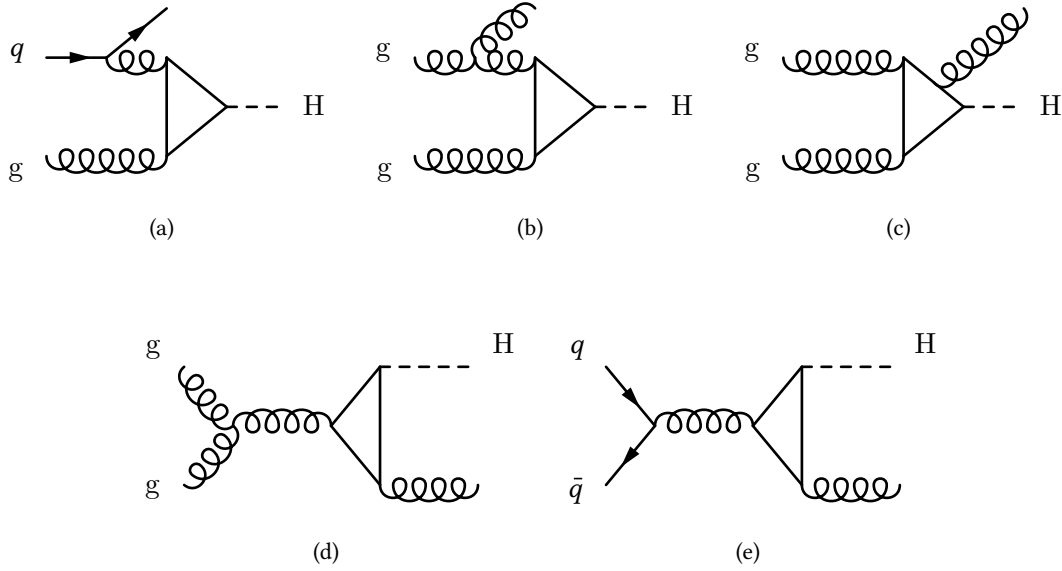
\begin{figure}[h!]
	\centering
		\subfigure[]{
		\begin{fmfgraph*}(90,65)
			\fmfset{arrow_len}{3mm}
			\fmfstraight
			\fmfleft{idummy0,i0,idummy1,i1,idummy2,i2,idummy3,i4}
			\fmfright{odummy0,o0,odummy1,o1,odummy2,o2,odummy3,o4}
			\fmflabel{\Pquark}{i2}
			\fmflabel{\Pgluon}{i0}
			\fmflabel{\PHiggs}{o1}
			\fmf{phantom}{i4,e4,o4}
			\fmf{phantom}{i2,e2,v2,dummy2,o2}
			\fmf{phantom}{i0,e0,v0,dummy0,o0}
			\fmffreeze
			\fmf{phantom}{i2,v2}
			\fmf{plain}{v2,vh}
			\fmf{gluon}{i0,v0}
			\fmf{plain}{v0,vh}
			\fmf{dashes,tension=2}{vh,o1}
			\fmffreeze
			\fmf{plain}{v0,v2}
			\fmffreeze
			\fmf{fermion}{i2,e2,e4}
			\fmf{gluon}{e2,v2}
		\end{fmfgraph*}
		}
		\hspace{1.3cm}
		\subfigure[]{
		\begin{fmfgraph*}(90,65)
			\fmfset{arrow_len}{3mm}
			\fmfstraight
			\fmfleft{idummy0,i0,idummy1,i1,idummy2,i2,idummy3,i4}
			\fmfright{odummy0,o0,odummy1,o1,odummy2,o2,odummy3,o4}
			\fmflabel{\Pgluon}{i2}
			\fmflabel{\Pgluon}{i0}
			\fmflabel{\PHiggs}{o1}
			\fmf{phantom}{i4,e4,o4}
			\fmf{phantom}{i2,e2,v2,dummy2,o2}
			\fmf{phantom}{i0,e0,v0,dummy0,o0}
			\fmffreeze
			\fmf{phantom}{i2,v2}
			\fmf{plain}{v2,vh}
			\fmf{gluon}{i0,v0}
			\fmf{plain}{v0,vh}
			\fmf{dashes,tension=2}{vh,o1}
			\fmffreeze
			\fmf{plain}{v0,v2}
			\fmffreeze
			\fmf{gluon}{i2,e2}
			\fmf{gluon}{e4,e2}
			\fmf{gluon}{e2,v2}
		\end{fmfgraph*}
		}
		\hspace{1.3cm}
		\subfigure[]{
		\begin{fmfgraph*}(90,65)
			\fmfset{arrow_len}{3mm}
			\fmfstraight
			\fmfleft{idummy0,i0,idummy1,i1,idummy2,i2,idummy3,i4}
			\fmfright{odummy0,o0,odummy1,o1,odummy2,o2,odummy3,o4}
			\fmflabel{\Pgluon}{i2}
			\fmflabel{\Pgluon}{i0}
			\fmflabel{\PHiggs}{o1}
			\fmf{phantom}{i4,e4,o4}
			\fmf{phantom}{i2,e2,v2,dummy2,o2}
			\fmf{phantom}{i0,e0,v0,dummy0,o0}
			\fmffreeze
			\fmf{phantom}{i2,v2}
			\fmf{plain}{v2,vh}
			\fmf{gluon}{i0,v0}
			\fmf{plain}{v0,vh}
			\fmf{dashes,tension=2}{vh,o1}
			\fmffreeze
			\fmf{plain}{v0,v2}
			\fmffreeze
			\fmf{gluon}{i2,v2}
			\fmf{phantom}{v2,b,vh}
			\fmffreeze
			\fmf{gluon}{b,o4}
		\end{fmfgraph*}
		\label{fig:emission_from_loop}
		}
		\vspace{1.3cm}
		\phantom{.}
		\subfigure[]{
		\begin{fmfgraph*}(110,52)
			\fmfset{arrow_len}{3mm}
			\fmfstraight
			\fmfleft{idummy0,i0,idummy1,i1,idummy2,i2,idummy3,i4}
			\fmfright{odummy0,o0,odummy1,o1,odummy2,o2,odummy3,o4}
			\fmflabel{\Pgluon}{i4}
			\fmflabel{\Pgluon}{i0}
			\fmflabel{\PHiggs}{o4}
			\fmf{phantom}{i4,e4,f4,o4}
			\fmf{phantom}{i0,e0,f0,o0}
			\fmffreeze
			\fmf{gluon}{i0,v0}
			\fmf{gluon}{i4,v0}
			\fmf{gluon}{v1,v0}
			\fmf{plain}{v1,f4,f0,v1}
			\fmf{gluon}{f0,o0}
			\fmf{dashes}{f4,o4}
		\end{fmfgraph*}
		}
		\hspace{1.3cm}
		\subfigure[]{
		\begin{fmfgraph*}(110,52)
			\fmfset{arrow_len}{3mm}
			\fmfstraight
			\fmfleft{idummy0,i0,idummy1,i1,idummy2,i2,idummy3,i4}
			\fmfright{odummy0,o0,odummy1,o1,odummy2,o2,odummy3,o4}
			\fmflabel{\Pquark}{i4}
			\fmflabel{\APquark}{i0}
			\fmflabel{\PHiggs}{o4}
			\fmf{phantom}{i4,e4,f4,o4}
			\fmf{phantom}{i0,e0,f0,o0}
			\fmffreeze
			\fmf{fermion}{v0,i0}
			\fmf{fermion}{i4,v0}
			\fmf{gluon}{v1,v0}
			\fmf{plain}{v1,f4,f0,v1}
			\fmf{gluon}{f0,o0}
			\fmf{dashes}{f4,o4}
		\end{fmfgraph*}
		}
\caption{Example LO Feynman diagrams for Higgs boson plus jet production via gluon fusion (or an $s$-channel gluon coupling to a fermion loop).}
\label{fig:higgs_plus_jet}
\end{figure}

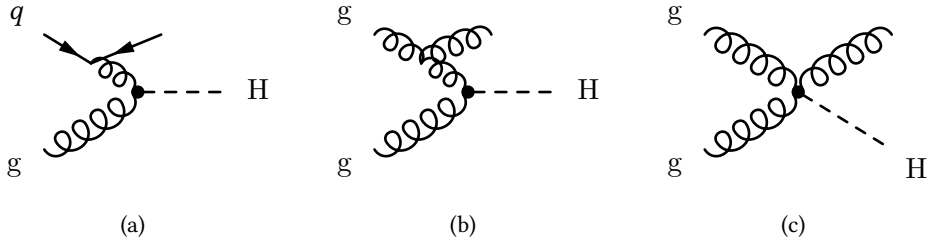
\begin{figure}[h!]
	\centering
		\subfigure[]{
		\begin{fmfgraph*}(70,65)
			\fmfset{arrow_len}{3mm}
			\fmfstraight
			\fmfleft{idummy0,i0,idummy1,i1,idummy2,i2,i4}
			\fmfright{odummy0,o0,odummy1,o1,odummy2,o2,o4}
			\fmflabel{\Pquark}{i2}
			\fmflabel{\Pgluon}{i0}
			\fmflabel{\PHiggs}{o1}
			\fmf{phantom}{i2,h1}
			\fmf{gluon}{i0,h1}
			\fmf{dashes,tension=2}{h1,o1}
			\fmffreeze
			\fmf{phantom}{h1,e,i2}
			\fmffreeze
			\fmf{phantom}{e,b,o4}
			\fmffreeze
			\fmf{gluon}{h1,e}
			\fmf{quark}{i2,e}
			\fmf{quark}{b,e}
			\fmfv{decor.shape=circle,decor.filled=full,decor.size=2thick}{h1}
		\end{fmfgraph*}
		}
		\hspace{15mm}
		\subfigure[]{
		\begin{fmfgraph*}(70,65)
			\fmfset{arrow_len}{3mm}
			\fmfstraight
			\fmfleft{idummy0,i0,idummy1,i1,idummy2,i2,i4}
			\fmfright{odummy0,o0,odummy1,o1,odummy2,o2,o4}
			\fmflabel{\Pgluon}{i2}
			\fmflabel{\Pgluon}{i0}
			\fmflabel{\PHiggs}{o1}
			\fmf{phantom}{i2,h1}
			\fmf{gluon}{i0,h1}
			\fmf{dashes,tension=2}{h1,o1}
			\fmffreeze
			\fmf{phantom}{h1,e,i2}
			\fmffreeze
			\fmf{phantom}{e,b,o4}
			\fmffreeze
			\fmf{gluon}{h1,e}
			\fmf{gluon}{e,i2}
			\fmf{gluon}{b,e}
			\fmfv{decor.shape=circle,decor.filled=full,decor.size=2thick}{h1}
		\end{fmfgraph*}
		}
		\hspace{15mm}
		\subfigure[]{
		\begin{fmfgraph*}(70,65)
			\fmfset{arrow_len}{3mm}
			\fmfstraight
			\fmfleft{idummy0,i0,idummy1,i1,idummy2,i2,idummy3}
			\fmfright{odummy0,o0,odummy1,o1,odummy2,o2,odummy3}
			\fmflabel{\Pgluon}{i2}
			\fmflabel{\Pgluon}{i0}
			\fmflabel{\PHiggs}{o0}
			\fmf{gluon}{h1,i2}
			\fmf{gluon}{i0,h1}
			\fmf{gluon}{o2,h1}
			\fmf{dashes}{h1,o0}
			\fmfv{decor.shape=circle,decor.filled=full,decor.size=2thick}{h1}
		\end{fmfgraph*}
		}
\caption{Example LO Feynman diagram for Higgs boson plus jet production via gluon fusion in the $m_{\Ptop} \to \infty$ approximation.}
\label{fig:higgs_plus_jet__heft}
\end{figure}

\subsection{Higgs boson pair production}

The inadequacy of the $m_{\Ptop} \to \infty$ approximation for single-Higgs-boson production in the presence of hard jets was discussed above. In $\PHiggs\PHiggs$ production, the approximation is even more inadequate, because $\Ptop\APtop$ production threshold effects appear when the $\PHiggs\PHiggs$ system has a mass of around $2m_{\Ptop} \approx 350$~\GeV{} \cite{Baur:2002rb,Dawson:2015oha}. Again, the breakdown of the approximation is worsened further by the presence of associated jets \cite{Dolan:2012rv,Maierhofer:2013sha}. Both of these effects are visible in \myfigs~\ref{fig:lilo_hh_hhj_mass} and \ref{fig:lilo_hh_hhj_pt}, showing differential LO $\Pproton\Pproton \to \PHiggs\PHiggs$ cross sections calculated with \HERWIG{}. All matrix elements are provided by \openloops{}. Again, the Higgs bosons are treated as stable. For this process, matrix elements including the Higgs-boson decay were not yet available in \openloops{} as of the writing of this thesis.
The LO cross section in the $m_{\Ptop} \to \infty$ approximation is around 20\% smaller than that from the full LI calculation. To study the effect of mainly soft and/or collinear radiation on the $\PHiggs\PHiggs$ kinematics, the LI prediction without parton shower is also shown. \myfig~\ref{fig:lilo_hh_hhj_mass} shows the mass of the $\PHiggs\PHiggs$ system. This observable is almost unaffected by the parton shower. The $m_{\Ptop} \to \infty$ approximation, on the other hand, predicts a very different distribution than the LI description, essentially missing the peak structure around $400$~GeV{}. Only above approximately 600~\GeV{}, it predicts a higher cross section. \myfig~\ref{fig:hh_pT} shows the transverse momentum of the $\PHiggs\PHiggs$ system. This observable is zero at LO without parton shower. Its description by the parton shower alone (i.e.~without matrix-element-level jets) behaves very differently from the $\PHiggs$ \pt{} in single-Higgs-boson production: it falls much less steeply. The $m_{\Ptop} \to \infty$ approximation yields a `tail' that is too hard, just as was observed for single-\PHiggs-boson production. The transverse momentum of the individual Higgs bosons, shown in \myfig~\ref{fig:hh_h_pT}, is shifted to slightly higher values by the parton shower. Relative to the full LI prediction, the $m_{\Ptop} \to \infty$ approximation performs worse for higher values.

\begin{figure}[h!]
\centering
\includegraphics[width=0.6\textwidth]{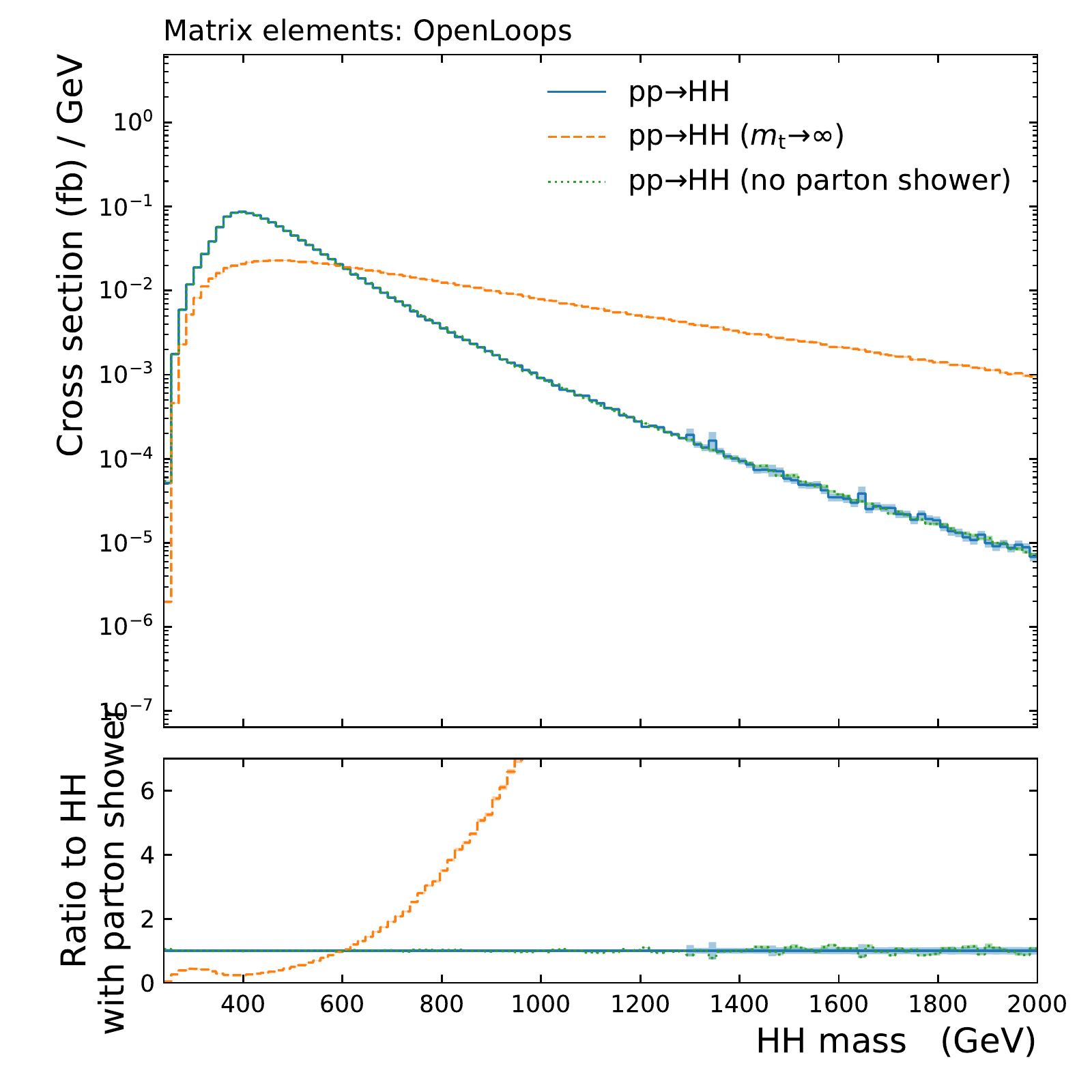}
\caption{Higgs boson pair production cross sections as a function of the mass of the $\PHiggs\PHiggs$ system. Both the LI production with full quark mass dependence (with and without parton shower) and production in the $m_{\Ptop} \to \infty$ limit are shown. The shaded bands indicate the statistical uncertainty of the MC integration.} 
\label{fig:lilo_hh_hhj_mass}
\end{figure}

\begin{figure}[h!]
\centering
\subfigure[]{\includegraphics[width=0.6\textwidth]{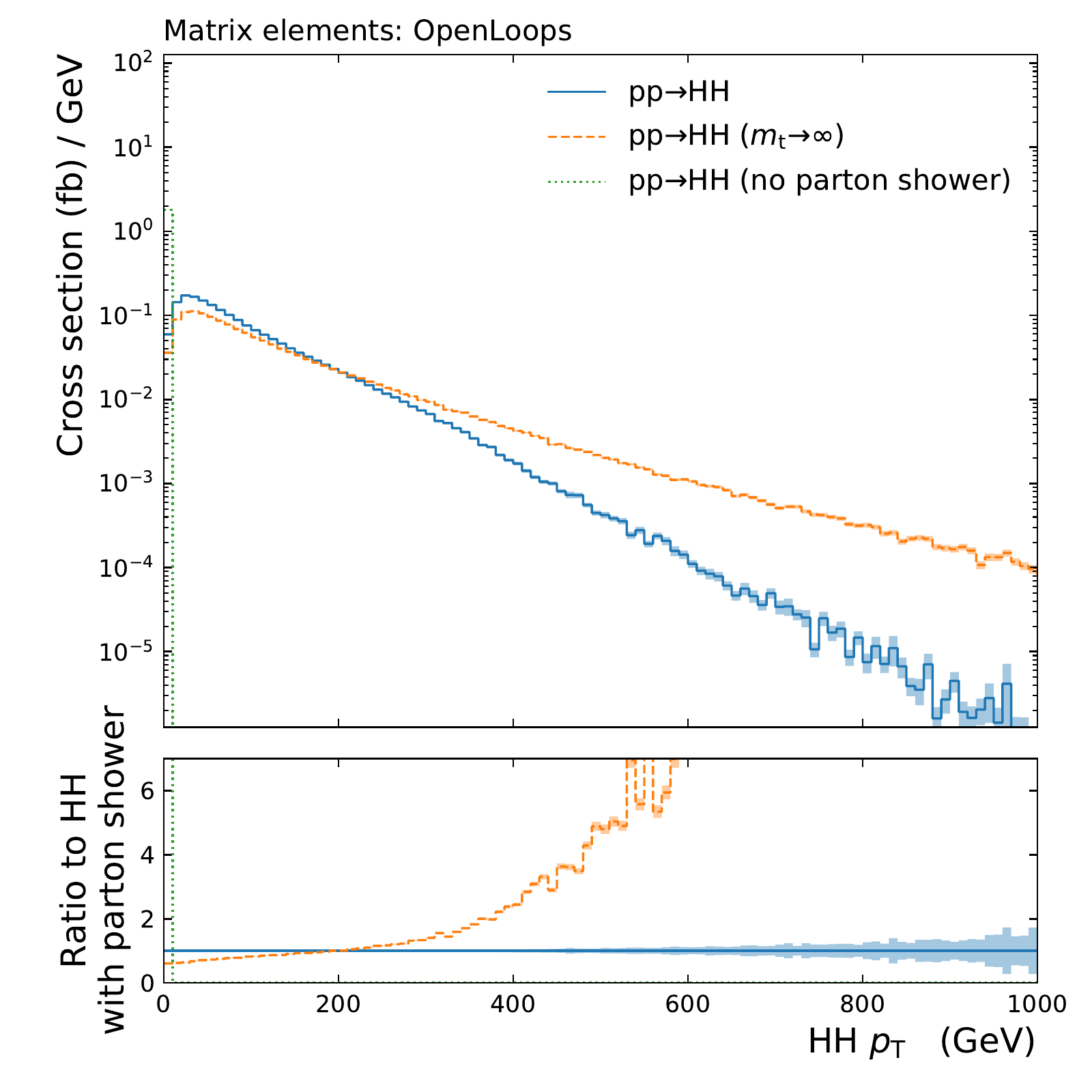}\label{fig:hh_pT}}
\subfigure[]{\includegraphics[width=0.6\textwidth]{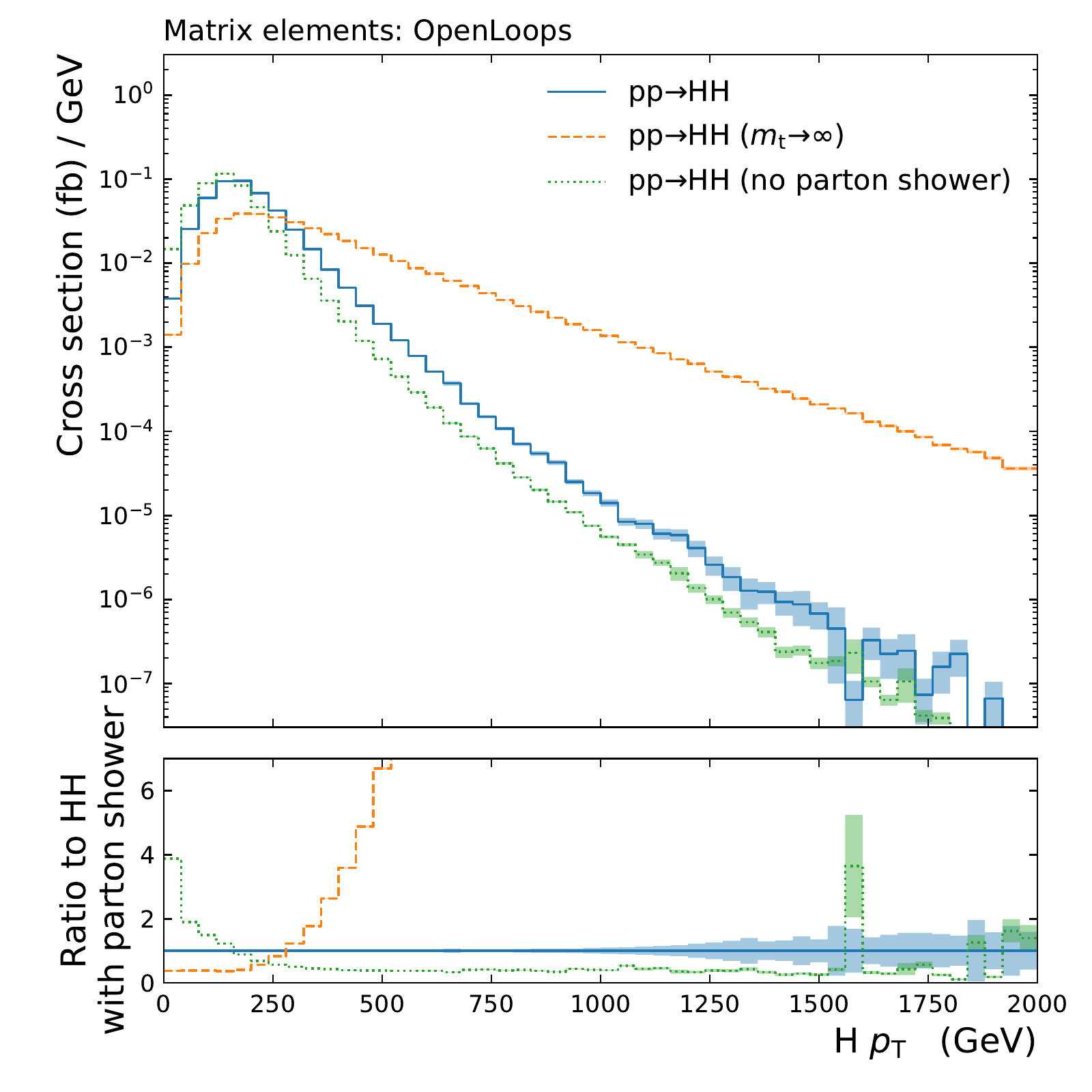}\label{fig:hh_h_pT}}
\vspace{-1mm}
\caption{Higgs boson pair production cross section as a function of the transverse momentum of (a) the $\PHiggs\PHiggs$ system and (b) the individual Higgs bosons. Both the LI production with full quark mass dependence (with and without parton shower) and production in the $m_{\Ptop} \to \infty$ limit are shown. The shaded bands indicate the statistical uncertainty of the MC integration.} 
\label{fig:lilo_hh_hhj_pt}
\end{figure}

\clearpage

\subsection{Four-lepton production}

The final example of LI generation presented in this thesis is four-lepton production, whose experimental measurement was the topic of \mypart~\ref{sec:analysis}. Compared to the production of stable Higgs bosons generated above, this process has a higher particle multiplicity at the matrix-element level. In order to benefit from existing \rivet{} analyses (with minor modifications by the author), only events in the $\Ppositron\Pelectron\APmuon\Pmuon$ channel are generated. To remove the $\Pgamma \to \ell^+\ell^-$ pole, each same-flavour lepton pair ($\Ppositron\Pelectron$, $\APmuon\Pmuon$) is required to have a mass $m_{\ell^+\ell^-} > 5$~\GeV{}. Each lepton has $\pt > 5$~\GeV{} and $|\eta| < 2.7$, reflecting the fiducial requirements of current ATLAS four-lepton analyses. 
\myfig~\ref{fig:eemm_mass} shows the mass of the four-lepton system in $\Pproton\Pproton \, \looparrow\, \Ppositron\Pelectron\APmuon\Pmuon$ events generated at LO. To speed up the event generation, no parton shower is interfaced. The inclusive production, where only the lepton selection requirements listed above are applied, shows the $\PHiggs \to \Ppositron\Pelectron\APmuon\Pmuon$ peak around a mass of 125~\GeV{}. An example Feynman diagram for this production mode was shown in \myfig~\ref{fig:feynman_h4l}. The cross section surges near the $\PZ\PZ$ diboson production threshold ($\sim$180~\GeV{}), peaking at around 220~\GeV{}, before falling off towards higher masses as the available phase space shrinks. Also shown in \myfig~\ref{fig:eemm_mass} is the four-lepton mass distribution obtained after requiring the same-flavour opposite-charge dileptons to be candidates for near-on-shell $\PZ$ bosons, by requiring $66~\GeV{}< m_{\ell^+\ell^-} < 116~\GeV{}$. This corresponds to a lowest possible four-lepton mass of 132~\GeV{}. Above this mass, the near-on-shell mode accounts for approximately half of the four-lepton production. This is a considerably lower fraction than in the tree-level process, which is dominated by $\PZ\PZ$ production. 

\begin{figure}[h!]
\centering
\includegraphics[width=0.8\textwidth]{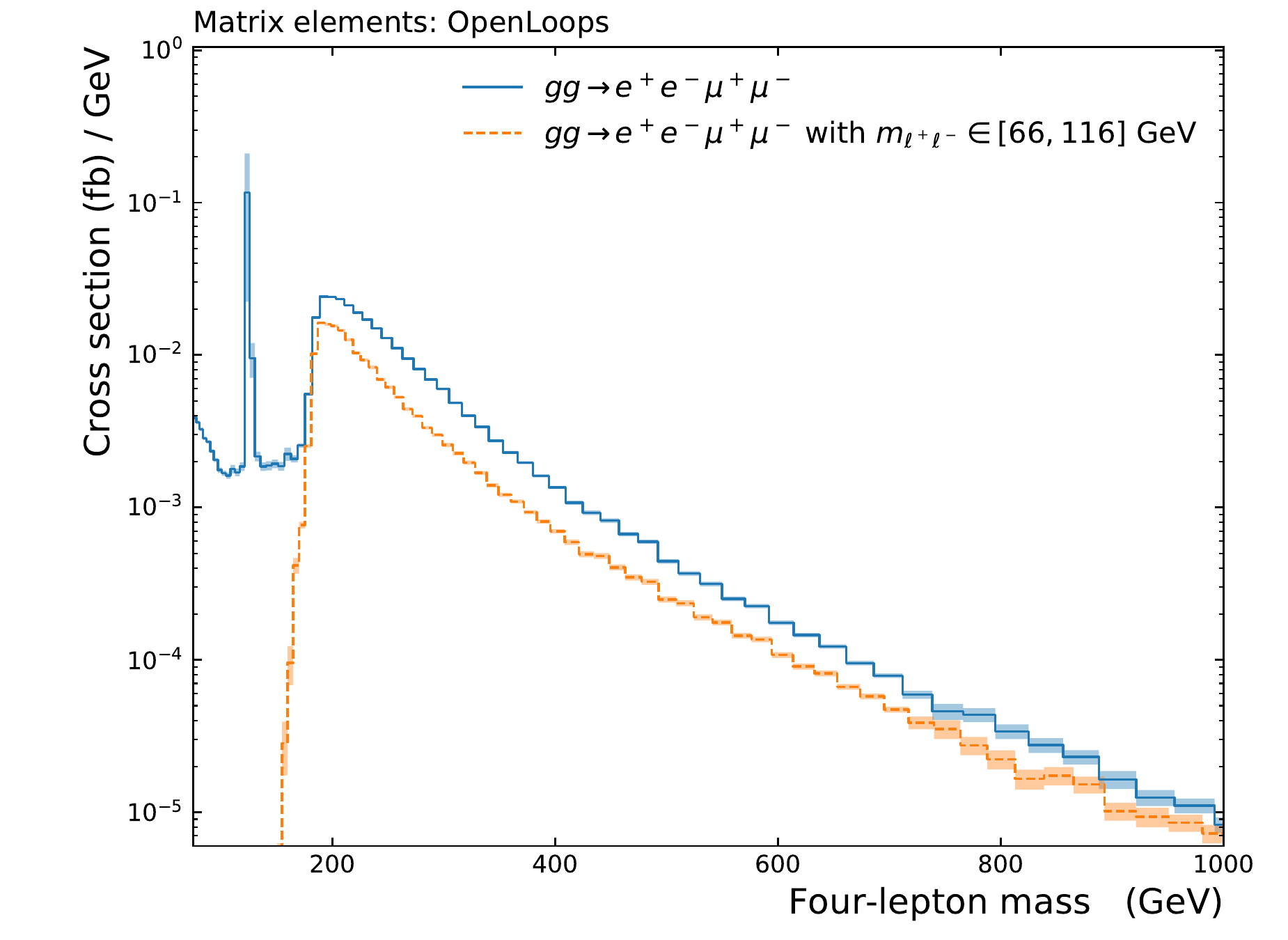}
\caption{LO production cross section of $\gluglu\, \to\, \Ppositron\Pelectron\APmuon\Pmuon$ as a function of the four-lepton mass. Both an inclusive selection (with only loose requirements) and a \ZZ{} candidate selection requiring dilepton masses $66~\GeV{}< m_{\ell^+\ell^-} < 116~\GeV{}$ is shown. No parton shower is included. The shaded bands indicate the statistical uncertainty of the MC integration.} 
\label{fig:eemm_mass}
\end{figure}

\clearpage

\section{Towards next-to-leading order}\label{sec:linlo}
Having established the importance of LI processes for the LHC physics programme above, it is not surprising that there is interest to generate these processes at higher-order accuracies. This is all the more important as some LI processes receive very large NLO corrections. In the case of $\ZZllll$ production, the NLO corrections increase the cross section by as much as $\mathcal{O}(70\%)$ \cite{Caola:2015psa}. Incorporating NLO results into a full MC event generation is therefore of immense interest to LHC physics analyses. This section describes work towards achieving (almost fully) automated NLO generation of LI processes matched with a parton shower.
The only part that cannot currently be automated in a strict sense is the computation of two-loop matrix elements. However, these may be interfaced to \HERWIG{} on a process by process basis as calculations implemented as public software become available. As long as they are available, the event generation is still fully automated from a user perspective. The work presented here is ongoing, so only the current status and a discussion of steps necessary to complete it are included.

An NLO cross section has the following anatomy:

\begin{equation*}
\sigma^{\text{NLO}} \equiv \int \dee\sigma^{\text{NLO}} = \int_{n} \dee\mathcal{B} + \int_n \alphas \dee \mathcal{V} + \int_{n+1} \alphas \dee\mathcal{R},
\end{equation*}
where $\mathcal{B}$ denotes the LO (or \emph{Born}, after the Born approximation) cross section, $\mathcal{V}$ the virtual-correction part of the NLO corrections given by the interference of the Born amplitude and the amplitude with one additional loop, and $\mathcal{R}$ is the real-emission part of the NLO corrections. They include the PDFs, and summation over initial-state flavours is implicit. The integrals are over the $n$- or $(n + 1)$-particle phase space, as indicated by the subscript. The momentum fractions $x_1$, $x_2$ of incoming partons are also integrated over. The two latter integrals are separately divergent in $d = 4$ space-time dimensions, but their sum is finite. To make the integrals well-defined, dimensional regularisation is used: the integrals are evaluated in $d = 4 - 2\varepsilon$ dimensions. In this regularisation, infrared divergences in both the virtual and real-emission correction appear as $1/\varepsilon^2$ poles (soft \emph{and} collinear divergence) and $1/\varepsilon$ poles (soft \emph{or} collinear divergence). Ultraviolet divergences may appear as $1/\varepsilon$ poles in the virtual correction $\mathcal{V}$ and are removed by renormalisation; they are not considered further in this discussion. According to the Bloch-Nordsieck and the Kinoshita-Lee-Nauenberg theorems, the infrared divergences in $\mathcal{V}$ and $\mathcal{R}$ must cancel mutually for infrared-safe observables \cite{Bloch:1937pw,Kinoshita:1962ur,Lee:1964is}.
In practice, achieving the cancellation is difficult, because the divergences appear in phase spaces of different dimensionality: $n$ particles for $\mathcal{V}$, $n + 1$ for $\mathcal{R}$. To overcome this difficulty, phase-space slicing methods \cite{Giele:1991vf,Giele:1993dj} and infrared-subtraction algorithms \cite{Catani:1996vz,Catani:2002hc,Kosower:1997zr,Kosower:2003bh,GehrmannDeRidder:2005cm,Daleo:2006xa,Frixione:1995ms,Frixione:1997np} have been developed.
In \HERWIG{}~7, the integrated \matchbox{} framework automates Catani-Seymour (CS) subtraction \cite{Catani:1996vz}. The idea of CS subtraction is to add a zero in the form of counterterms $\mathcal{A}$ to the cross section,
\begin{equation}\label{eq:nlo_after_subtraction}
\sigma^{\text{NLO}} \equiv \int \dee\sigma^{\text{NLO}} = \int_{n} \dee\mathcal{B} + \int_n \alphas \left[\dee \mathcal{V} + \int_{1} \dee\mathcal{A} \right] + \int_{n+1} \alphas \left[\dee\mathcal{R} - \dee\mathcal{A}\right],
\end{equation}
where $\dee \mathcal{A}$ possesses the same pointwise singular behaviour as $\dee \mathcal{R}$.\footnote{Cancellation of singularities is only ensured for infrared safe observables, but their exclusive use is implied throughout.} The radiation phase space always factorises from the $n$-particle phase space and $\int_1 \dee \mathcal{A}$ can be evaluated analytically in the CS construction, so that it can be used to remove the divergences in $\dee\mathcal{V}$. All remaining integrals in \myeq~\ref{eq:nlo_after_subtraction} may thereafter be efficiently evaluated using MC integration in four space-time dimensions, thus arriving at a practical NLO event generator. The power of CS subtraction lies partly in the fact that the counterterms $\dee\mathcal{A}$ are universal and do not depend on the studied process, apart from which coloured legs it contains. They are constructed by considering \emph{dipoles}, which are pairs of partons in the Born-level matrix elements that can emit a third parton. Each possible dipole has an associated factor $V_{\text{dipole}}$ incorporating the soft and collinear divergent structure of this $2 \to 3$ branching. The counterterms are then given by
\begin{equation}\label{eq:counterterm}
\dee\mathcal{A} = \sum_{\text{dipoles}} \dee\mathcal{B} \otimes \dee V_{\text{dipole}},
\end{equation}
where $\otimes$ denotes an appropriate colour and helicity projection of the LO cross section. To remove the infrared divergences of the virtual correction, one makes use of the fact that the dipole radiation phase space factorises from the $n$-particle phase space, so that
\begin{equation}\label{eq:csi_operator}
\int_1 \dee\mathcal{A} = \sum_{\text{dipoles}} \dee\mathcal{B} \otimes \int_1 \dee V_{\text{dipole}} = \dee\mathcal{B} \otimes \csi,
\end{equation}
can be inserted in \myeq~\ref{eq:nlo_after_subtraction}, where
\begin{equation*}
\csi = \sum_{\text{dipoles}} \int_1 \dee V_{\text{dipole}}
\end{equation*}
is an insertion operator containing the $\varepsilon$ poles required to cancel those in $\mathcal{V}$.
Full details of how to construct $\dee\mathcal{A}$ and \csi{} can be found in \myref~\cite{Catani:1996vz}. \matchbox{} constructs these terms automatically for a given process. The necessary colour and helicity projections of the LO cross section appearing implicitly in \myeqs~\ref{eq:counterterm} and \ref{eq:csi_operator} are given by the external matrix element provider and returned as a data structure defined in the BLHA2 conventions. Thanks to the new \HERWIG{} interface to \openloops{}, these can now be evaluated for loop-induced processes. Since the \gluglu{}-initiated LI processes considered in this work all have less than four coloured legs at LO, the colour projections are trivial. In the case of two initial-state gluons with colour indices $i$, $j$ and a colourless final state they are given by
\begin{equation}\label{eq:cc_me}
	C_{ij} = \langle \mathcal{M_B} | T_i \cdot T_j | \mathcal{M_B} \rangle = \delta_{ij} T_{\Pgluon}^2 \langle \mathcal{M_B} | \mathcal{M_B} \rangle = \delta_{ij} C_A |\mathcal{M_B}|^2,
\end{equation}
i.e.~by the squared Born matrix element $\mathcal{M_B}$ times a colour factor $C_A = 3$ if the colour indices are identical, and zero otherwise. However, the \HERWIG{} implementation does not rely on this simplification and is fully general including for processes with non-trivial colour structure.
%
Automated NLO calculations for tree processes can be performed by several event generators, for instance using the above scheme based on CS subtraction. The present work aims to extend this functionality to LI processes. The only difference for these processes at NLO is that each of $\mathcal{B}$, $\mathcal{V}$ and $\mathcal{R}$ contains one more loop than would be the case for tree processes. However, at least for the production of Higgs or electroweak bosons (possibly with matrix-element level decay), this loop engenders no infrared or ultraviolet divergences.
Therefore, no subtraction of divergences is necessary, and the NLO subtraction presented above is sufficient at the two-loop level. In this sense the process differs from other NNLO contributions, to which the CS subtraction does not extend trivially: for instance, the dipole factorisation of \myeq~\ref{eq:counterterm} does not hold in the presence of two \emph{soft} singular regions,
\begin{equation*}
\dee\mathcal{A} \neq \sum^{\text{dipoles}}_i\, \sum_j^{\text{dipoles}} \dee\mathcal{B} \otimes \dee V_i \otimes \dee V_j.
\end{equation*}
This is because the factorisation in the soft limit holds at the matrix-element level rather than the cross-section level, so multiple soft emissions do not factorise in the sense of \myeq~\ref{eq:counterterm}, but modify the radiation patterns. For an NLO description of LI processes, the limiting factor today is the availability of two-loop matrix elements needed for the virtual corrections. These have only been calculated for a handful of processes, such as Higgs-boson pair production (with full top-quark mass dependence) \cite{Heinrich:2017kxx,Borowka:2016ypz,Borowka:2016ehy} and four-lepton production \cite{vonManteuffel:2015msa,Caola:2015ila}. The author is confident that more processes will be calculated in the coming months and years.

\matchbox{} also automates the matching of the NLO matrix elements to a parton shower \cite{Platzer:2011bc}. The user can choose between a subtractive (MC@NLO-style \cite{Frixione:2002ik}) or multiplicative (\POWHEG{}-style \cite{Nason:2004rx}) matching to the dipole or angular-ordered shower. The matching schemes yield equivalent results in the leading-logarithmic approximation, but differ in subleading terms.

\subsection{Intermediate results for Higgs boson pair production}

This section contains some intermediate validation results for LI $\PHiggs\PHiggs$ production at NLO. This process, matched with a parton shower, has so far been calculated in the \POWHEG{} and \MGMCatNLO{} frameworks \cite{Heinrich:2017kxx} and very recently with \SHERPA{} \cite{Jones:2017giv}. In the \HERWIG{} description being developed here, the virtual corrections are not yet fully implemented. They are taken from the software \texttt{hhgrid} \cite{Borowka:2016ehy,Borowka:2016ypz,Heinrich:2017kxx}. It provides precomputed $\mathcal{V}$ values as a grid parametrised in the Mandelstam variables $\hat{s}$ and $\hat{t}$, as well as an interpolation algorithm to obtain the values lying between the grid points. As of the writing of this thesis, \texttt{hhgrid} has been interfaced to \HERWIG{}, but work is ongoing to match the conventions of the two softwares in order to correctly construct the full NLO cross section.

Even without the virtual corrections, the correct functioning of the Catani-Seymour subtraction for real-emission corrections can be verified. Here, this is done by studying the ratio of the dipole subtraction term $D$ over the squared-matrix element contribution $M$ as a function of the scale $Q^2$ of the parton emission. The ratio is supposed to approach unity as the singularity $Q^2 \to 0$ is approached, so that the subtraction removes the pole. For the check to be efficient, the ratio is determined using a sampling that is heavily biased towards low scales. The sampling point density is proportional to $1 / Q^2$ down until a value of $1 / Q_{\text{flat}}^2$, below which it remains constant at $1 / Q_{\text{flat}}^2$. Below another, smaller value $1 / Q_{\text{cut}}^2$, no sampling is done at all. The situation is illustrated in \myfig~\ref{fig:pole_sampling}. For the preliminary checks performed here, the sampling thresholds were set to the relatively high values of $Q_{\text{flat}} = 1$~\GeV{} and $Q_{\text{cut}} = 0.5$~\GeV{}. The reason is that the evaluation of the matrix elements for very small $Q$ takes a long time due to numerical instability. In the development phase, and because there is no reason to expect failure of the NLO subtraction, the faster preliminary check gives a first rough idea of the functioning.
\myfigs~\ref{fig:subtraction_check_collinear_0}--\ref{fig:subtraction_check_soft_gluon} show the envelopes of the ratio $D/M$ encountered during the sampling (i.e.~the ratios encountered that differed most from one) as a function of $Q$. Each figure corresponds to a different singular region. Using the numbering of external gluons shown in the example Feynman diagram in \myfig~\ref{fig:gg2hhg_notation}, the singular regions are as follows. \myfig~\ref{fig:subtraction_check_collinear_0} (\myfig~\ref{fig:subtraction_check_collinear_1}) shows the region where the emitted gluon becomes collinear with incoming gluon 0 (1), while \myfig~\ref{fig:subtraction_check_soft_gluon} shows the region where the energy of the emitted gluon in the rest frame of the emitting parton becomes very small, $E_2 \to 0$. The scale of the emission is taken to be $Q = \sqrt{s_{02}}$, $\sqrt{s_{12}}$, and $E_2$, respectively, where $\sqrt{s_{ij}}$ is the invariant mass of the system formed by legs $i$ and $j$. It can be seen that the ratios indeed converge to one as $Q$ approaches $\mathcal{O}(1~\GeV)$, as desired. The preliminary conclusion is that the NLO subtraction works properly. This is to be confirmed with more generated events and a sampling down to lower scales, $Q_{\text{cut}} \sim \mathcal{O}(1~\MeV{})$.


\begin{figure}[h!]
\vspace{5mm}
	\centering
		\begin{fmfgraph*}(90,65)
			\fmfset{arrow_len}{3mm}
			\fmfstraight
			\fmfleft{idummy0,i0,idummy1,i1,idummy2,i2,idummy3,i4}
			\fmfright{odummy0,o0,odummy1,o1,odummy2,o2,odummy3,o4}
			\fmflabel{0}{i2}
			\fmflabel{1}{i0}
			\fmflabel{2}{e4}
			\fmflabel{\PHiggs}{o2}
			\fmflabel{\PHiggs}{o0}
			\fmf{phantom}{i4,e4,o4}
			\fmf{phantom}{i2,e2,v2,h2,o2}
			\fmf{phantom}{i0,e0,v0,h0,o0}
			\fmffreeze
			\fmf{phantom}{i2,v2}
			\fmf{plain}{v2,h2}
			\fmf{gluon}{i0,v0}
			\fmf{plain}{v0,h0}
			\fmf{dashes,tension=2}{h2,o2}
			\fmf{dashes,tension=2}{h0,o0}
			\fmffreeze
			\fmf{plain}{v0,v2}
			\fmf{plain}{h0,h2}
			\fmffreeze
			\fmf{gluon}{i2,e2}
			\fmf{gluon}{e4,e2}
			\fmf{gluon}{e2,v2}
		\end{fmfgraph*}
\caption{Example Feynman diagram contributing to the NLO real-emission correction to $\gluglu\to\PHiggs\PHiggs$, showing how the gluons are numbered.}
\label{fig:gg2hhg_notation}
\end{figure}
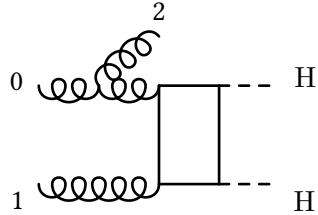

%
%




%
%
%
%
%

\begin{figure}[h!]
\centering
\begin{tikzpicture}
\node[anchor=south west,inner sep=0] (image) at (0,0) {\includegraphics[width=0.5\textwidth]{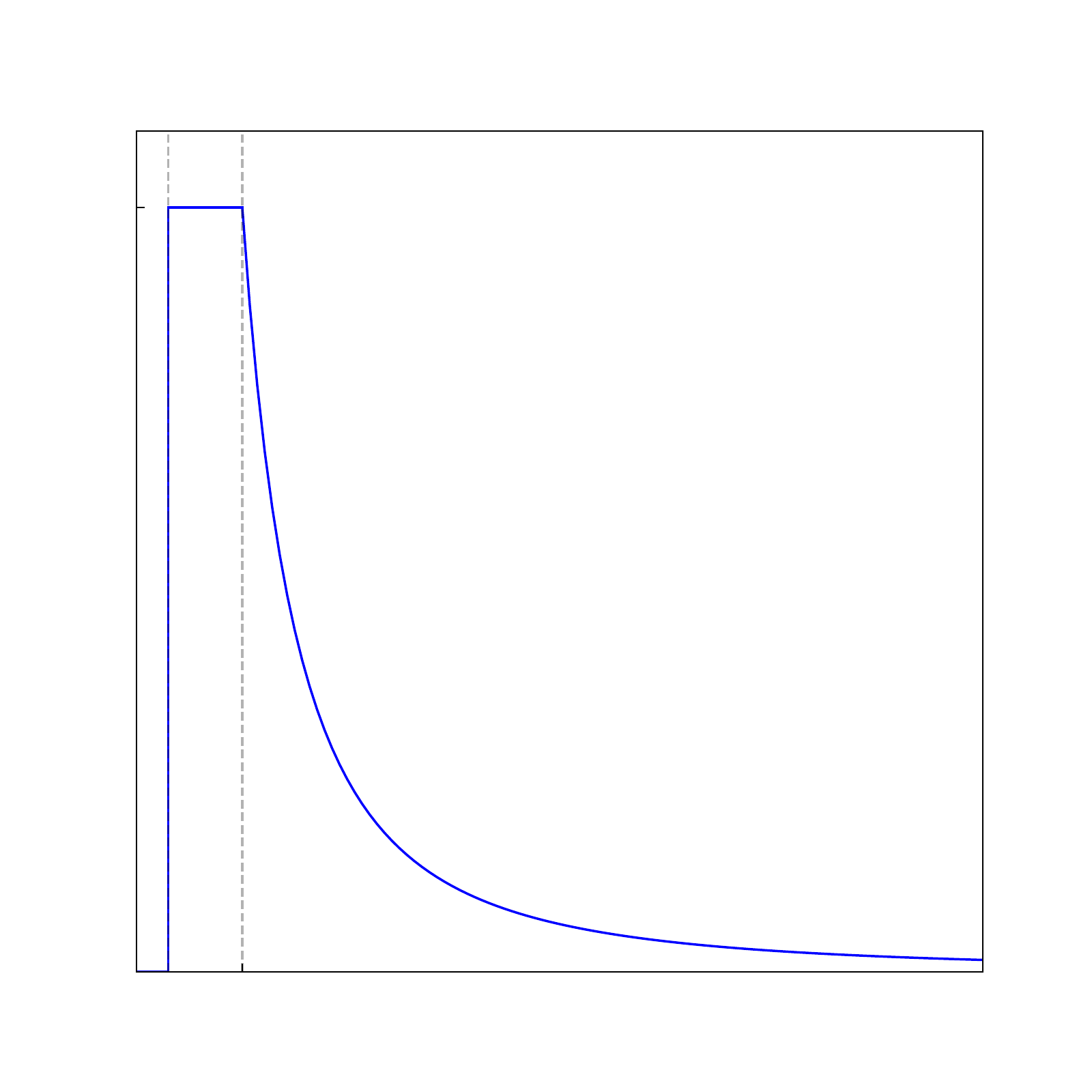}};
\begin{scope}[x={(image.south east)},y={(image.north west)}]
\node[anchor=center, rotate=90] at (0.0, 0.5) {Sampling density};
\node[anchor=west] at (0.35, 0.3) {$\propto \frac{1}{Q^2}$};
\node[anchor=east] at (0.9, 0.03) {$Q$};
\node[anchor=east] at (0.13, 0.81) {$\frac{1}{Q_{\text{flat}}^2}$};
\node[anchor=east] at (0.32, 0.07) {\footnotesize $Q_{\text{flat}}$};
\node[anchor=east] at (0.23, 0.07) {\footnotesize $Q_{\text{cut}}$};
\end{scope}
\end{tikzpicture}
\caption{Sampling density near a singular region, where $Q$ is the scale associated with the branching.}
\label{fig:pole_sampling}
\end{figure}
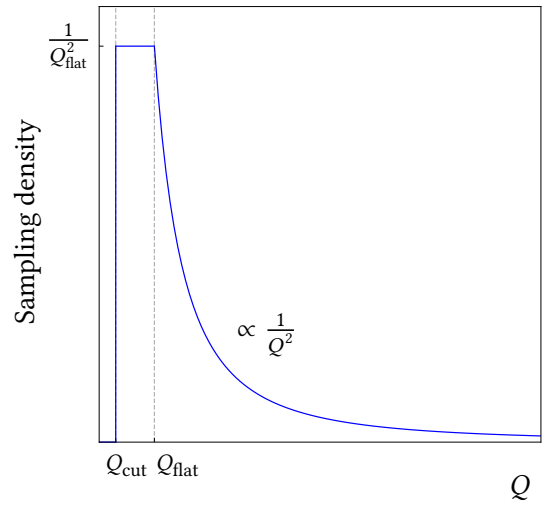

\begin{figure}[h!]
\centering
\begingroup
  \makeatletter
  \providecommand\color[2][]{%
    \GenericError{(gnuplot) \space\space\space\@spaces}{%
      Package color not loaded in conjunction with
      terminal option `colourtext'%
    }{See the gnuplot documentation for explanation.%
    }{Either use 'blacktext' in gnuplot or load the package
      color.sty in LaTeX.}%
    \renewcommand\color[2][]{}%
  }%
  \providecommand\includegraphics[2][]{%
    \GenericError{(gnuplot) \space\space\space\@spaces}{%
      Package graphicx or graphics not loaded%
    }{See the gnuplot documentation for explanation.%
    }{The gnuplot epslatex terminal needs graphicx.sty or graphics.sty.}%
    \renewcommand\includegraphics[2][]{}%
  }%
  \providecommand\rotatebox[2]{#2}%
  \@ifundefined{ifGPcolor}{%
    \newif\ifGPcolor
    \GPcolortrue
  }{}%
  \@ifundefined{ifGPblacktext}{%
    \newif\ifGPblacktext
    \GPblacktexttrue
  }{}%
  \let\gplgaddtomacro\g@addto@macro
  \gdef\gplbacktext{}%
  \gdef\gplfronttext{}%
  \makeatother
  \ifGPblacktext
    \def\colorrgb#1{}%
    \def\colorgray#1{}%
  \else
    \ifGPcolor
      \def\colorrgb#1{\color[rgb]{#1}}%
      \def\colorgray#1{\color[gray]{#1}}%
      \expandafter\def\csname LTw\endcsname{\color{white}}%
      \expandafter\def\csname LTb\endcsname{\color{black}}%
      \expandafter\def\csname LTa\endcsname{\color{black}}%
      \expandafter\def\csname LT0\endcsname{\color[rgb]{1,0,0}}%
      \expandafter\def\csname LT1\endcsname{\color[rgb]{0,1,0}}%
      \expandafter\def\csname LT2\endcsname{\color[rgb]{0,0,1}}%
      \expandafter\def\csname LT3\endcsname{\color[rgb]{1,0,1}}%
      \expandafter\def\csname LT4\endcsname{\color[rgb]{0,1,1}}%
      \expandafter\def\csname LT5\endcsname{\color[rgb]{1,1,0}}%
      \expandafter\def\csname LT6\endcsname{\color[rgb]{0,0,0}}%
      \expandafter\def\csname LT7\endcsname{\color[rgb]{1,0.3,0}}%
      \expandafter\def\csname LT8\endcsname{\color[rgb]{0.5,0.5,0.5}}%
    \else
      \def\colorrgb#1{\color{black}}%
      \def\colorgray#1{\color[gray]{#1}}%
      \expandafter\def\csname LTw\endcsname{\color{white}}%
      \expandafter\def\csname LTb\endcsname{\color{black}}%
      \expandafter\def\csname LTa\endcsname{\color{black}}%
      \expandafter\def\csname LT0\endcsname{\color{black}}%
      \expandafter\def\csname LT1\endcsname{\color{black}}%
      \expandafter\def\csname LT2\endcsname{\color{black}}%
      \expandafter\def\csname LT3\endcsname{\color{black}}%
      \expandafter\def\csname LT4\endcsname{\color{black}}%
      \expandafter\def\csname LT5\endcsname{\color{black}}%
      \expandafter\def\csname LT6\endcsname{\color{black}}%
      \expandafter\def\csname LT7\endcsname{\color{black}}%
      \expandafter\def\csname LT8\endcsname{\color{black}}%
    \fi
  \fi
    \setlength{\unitlength}{0.0750bp}%
    \ifx\gptboxheight\undefined%
      \newlength{\gptboxheight}%
      \newlength{\gptboxwidth}%
      \newsavebox{\gptboxtext}%
    \fi%
    \setlength{\fboxrule}{0.5pt}%
    \setlength{\fboxsep}{1pt}%
\begin{picture}(3600.00,3024.00)%
    \gplgaddtomacro\gplbacktext{%
      \csname LTb\endcsname
      \put(350,2600){\makebox(0,0)[r]{\strut{}$D/M$}}%
      \put(700,704){\makebox(0,0)[r]{\strut{}$0$}}%
      \put(700,1229){\makebox(0,0)[r]{\strut{}$0.5$}}%
      \put(700,1754){\makebox(0,0)[r]{\strut{}$1$}}%
      \put(700,2278){\makebox(0,0)[r]{\strut{}$1.5$}}%
      \put(700,2803){\makebox(0,0)[r]{\strut{}$2$}}%
      \put(726,600){\makebox(0,0){\strut{}$10^{-1}$}}%
      \put(1552,600){\makebox(0,0){\strut{}$10^{0}$}}%
      \put(2377,600){\makebox(0,0){\strut{}$10^{1}$}}%
      \put(3203,600){\makebox(0,0){\strut{}$10^{2}$}}%
    }%
    \gplgaddtomacro\gplfronttext{%
      \csname LTb\endcsname
      \put(2900,400){\makebox(0,0){\strut{}$\sqrt{s_{02}}$\quad(\GeV{})}}%
      \csname LTb\endcsname
      \put(2216,2630){\makebox(0,0)[r]{\strut{}$\Pgluon \Pgluon \to \PH \PH \Pgluon $}}%
    }%
    \gplbacktext
    \put(0,0){\includegraphics[scale=1.5]{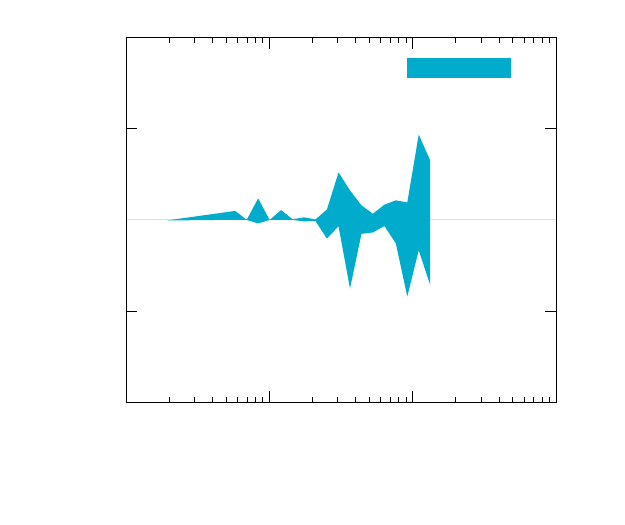}}%
    \gplfronttext
  \end{picture}%
\endgroup
\caption{Envelope of subtraction ratio $D/M$ as a function of the invariant mass of the system formed by incoming gluon 0 and the emitted gluon 2.}
\label{fig:subtraction_check_collinear_0}
\end{figure}

\begin{figure}[h!]
\centering
\begingroup
  \makeatletter
  \providecommand\color[2][]{%
    \GenericError{(gnuplot) \space\space\space\@spaces}{%
      Package color not loaded in conjunction with
      terminal option `colourtext'%
    }{See the gnuplot documentation for explanation.%
    }{Either use 'blacktext' in gnuplot or load the package
      color.sty in LaTeX.}%
    \renewcommand\color[2][]{}%
  }%
  \providecommand\includegraphics[2][]{%
    \GenericError{(gnuplot) \space\space\space\@spaces}{%
      Package graphicx or graphics not loaded%
    }{See the gnuplot documentation for explanation.%
    }{The gnuplot epslatex terminal needs graphicx.sty or graphics.sty.}%
    \renewcommand\includegraphics[2][]{}%
  }%
  \providecommand\rotatebox[2]{#2}%
  \@ifundefined{ifGPcolor}{%
    \newif\ifGPcolor
    \GPcolortrue
  }{}%
  \@ifundefined{ifGPblacktext}{%
    \newif\ifGPblacktext
    \GPblacktexttrue
  }{}%
  \let\gplgaddtomacro\g@addto@macro
  \gdef\gplbacktext{}%
  \gdef\gplfronttext{}%
  \makeatother
  \ifGPblacktext
    \def\colorrgb#1{}%
    \def\colorgray#1{}%
  \else
    \ifGPcolor
      \def\colorrgb#1{\color[rgb]{#1}}%
      \def\colorgray#1{\color[gray]{#1}}%
      \expandafter\def\csname LTw\endcsname{\color{white}}%
      \expandafter\def\csname LTb\endcsname{\color{black}}%
      \expandafter\def\csname LTa\endcsname{\color{black}}%
      \expandafter\def\csname LT0\endcsname{\color[rgb]{1,0,0}}%
      \expandafter\def\csname LT1\endcsname{\color[rgb]{0,1,0}}%
      \expandafter\def\csname LT2\endcsname{\color[rgb]{0,0,1}}%
      \expandafter\def\csname LT3\endcsname{\color[rgb]{1,0,1}}%
      \expandafter\def\csname LT4\endcsname{\color[rgb]{0,1,1}}%
      \expandafter\def\csname LT5\endcsname{\color[rgb]{1,1,0}}%
      \expandafter\def\csname LT6\endcsname{\color[rgb]{0,0,0}}%
      \expandafter\def\csname LT7\endcsname{\color[rgb]{1,0.3,0}}%
      \expandafter\def\csname LT8\endcsname{\color[rgb]{0.5,0.5,0.5}}%
    \else
      \def\colorrgb#1{\color{black}}%
      \def\colorgray#1{\color[gray]{#1}}%
      \expandafter\def\csname LTw\endcsname{\color{white}}%
      \expandafter\def\csname LTb\endcsname{\color{black}}%
      \expandafter\def\csname LTa\endcsname{\color{black}}%
      \expandafter\def\csname LT0\endcsname{\color{black}}%
      \expandafter\def\csname LT1\endcsname{\color{black}}%
      \expandafter\def\csname LT2\endcsname{\color{black}}%
      \expandafter\def\csname LT3\endcsname{\color{black}}%
      \expandafter\def\csname LT4\endcsname{\color{black}}%
      \expandafter\def\csname LT5\endcsname{\color{black}}%
      \expandafter\def\csname LT6\endcsname{\color{black}}%
      \expandafter\def\csname LT7\endcsname{\color{black}}%
      \expandafter\def\csname LT8\endcsname{\color{black}}%
    \fi
  \fi
    \setlength{\unitlength}{0.0750bp}%
    \ifx\gptboxheight\undefined%
      \newlength{\gptboxheight}%
      \newlength{\gptboxwidth}%
      \newsavebox{\gptboxtext}%
    \fi%
    \setlength{\fboxrule}{0.5pt}%
    \setlength{\fboxsep}{1pt}%
\begin{picture}(3600.00,3024.00)%
    \gplgaddtomacro\gplbacktext{%
      \csname LTb\endcsname
       \put(350,2600){\makebox(0,0)[r]{\strut{}$D/M$}}%
      \put(700,704){\makebox(0,0)[r]{\strut{}$0$}}%
      \put(700,1229){\makebox(0,0)[r]{\strut{}$0.5$}}%
      \put(700,1754){\makebox(0,0)[r]{\strut{}$1$}}%
      \put(700,2278){\makebox(0,0)[r]{\strut{}$1.5$}}%
      \put(700,2803){\makebox(0,0)[r]{\strut{}$2$}}%
      \put(726,600){\makebox(0,0){\strut{}$10^{-1}$}}%
      \put(1552,600){\makebox(0,0){\strut{}$10^{0}$}}%
      \put(2377,600){\makebox(0,0){\strut{}$10^{1}$}}%
      \put(3203,600){\makebox(0,0){\strut{}$10^{2}$}}%
    }%
    \gplgaddtomacro\gplfronttext{%
      \csname LTb\endcsname
      \put(2900,400){\makebox(0,0){\strut{}$\sqrt{s_{12}}$\quad(\GeV{})}}%
      \csname LTb\endcsname
      \put(2216,2630){\makebox(0,0)[r]{\strut{}$\Pgluon \Pgluon \to \PH \PH \Pgluon $}}%
    }%
    \gplbacktext
    \put(0,0){\includegraphics[scale=1.5]{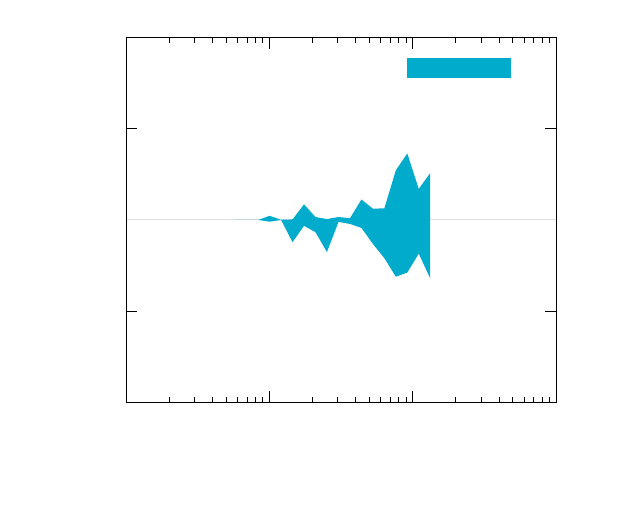}}%
    \gplfronttext
  \end{picture}%
\endgroup
\vspace{-10mm}
\caption{Envelope of subtraction ratio $D/M$ as a function of the invariant mass of the system formed by incoming gluon 1 and the emitted gluon 2.}
\label{fig:subtraction_check_collinear_1}
\end{figure}

\begin{figure}[h!]
\centering
\small
\begingroup
  \makeatletter
  \providecommand\color[2][]{%
    \GenericError{(gnuplot) \space\space\space\@spaces}{%
      Package color not loaded in conjunction with
      terminal option `colourtext'%
    }{See the gnuplot documentation for explanation.%
    }{Either use 'blacktext' in gnuplot or load the package
      color.sty in LaTeX.}%
    \renewcommand\color[2][]{}%
  }%
  \providecommand\includegraphics[2][]{%
    \GenericError{(gnuplot) \space\space\space\@spaces}{%
      Package graphicx or graphics not loaded%
    }{See the gnuplot documentation for explanation.%
    }{The gnuplot epslatex terminal needs graphicx.sty or graphics.sty.}%
    \renewcommand\includegraphics[2][]{}%
  }%
  \providecommand\rotatebox[2]{#2}%
  \@ifundefined{ifGPcolor}{%
    \newif\ifGPcolor
    \GPcolortrue
  }{}%
  \@ifundefined{ifGPblacktext}{%
    \newif\ifGPblacktext
    \GPblacktexttrue
  }{}%
  \let\gplgaddtomacro\g@addto@macro
  \gdef\gplbacktext{}%
  \gdef\gplfronttext{}%
  \makeatother
  \ifGPblacktext
    \def\colorrgb#1{}%
    \def\colorgray#1{}%
  \else
    \ifGPcolor
      \def\colorrgb#1{\color[rgb]{#1}}%
      \def\colorgray#1{\color[gray]{#1}}%
      \expandafter\def\csname LTw\endcsname{\color{white}}%
      \expandafter\def\csname LTb\endcsname{\color{black}}%
      \expandafter\def\csname LTa\endcsname{\color{black}}%
      \expandafter\def\csname LT0\endcsname{\color[rgb]{1,0,0}}%
      \expandafter\def\csname LT1\endcsname{\color[rgb]{0,1,0}}%
      \expandafter\def\csname LT2\endcsname{\color[rgb]{0,0,1}}%
      \expandafter\def\csname LT3\endcsname{\color[rgb]{1,0,1}}%
      \expandafter\def\csname LT4\endcsname{\color[rgb]{0,1,1}}%
      \expandafter\def\csname LT5\endcsname{\color[rgb]{1,1,0}}%
      \expandafter\def\csname LT6\endcsname{\color[rgb]{0,0,0}}%
      \expandafter\def\csname LT7\endcsname{\color[rgb]{1,0.3,0}}%
      \expandafter\def\csname LT8\endcsname{\color[rgb]{0.5,0.5,0.5}}%
    \else
      \def\colorrgb#1{\color{black}}%
      \def\colorgray#1{\color[gray]{#1}}%
      \expandafter\def\csname LTw\endcsname{\color{white}}%
      \expandafter\def\csname LTb\endcsname{\color{black}}%
      \expandafter\def\csname LTa\endcsname{\color{black}}%
      \expandafter\def\csname LT0\endcsname{\color{black}}%
      \expandafter\def\csname LT1\endcsname{\color{black}}%
      \expandafter\def\csname LT2\endcsname{\color{black}}%
      \expandafter\def\csname LT3\endcsname{\color{black}}%
      \expandafter\def\csname LT4\endcsname{\color{black}}%
      \expandafter\def\csname LT5\endcsname{\color{black}}%
      \expandafter\def\csname LT6\endcsname{\color{black}}%
      \expandafter\def\csname LT7\endcsname{\color{black}}%
      \expandafter\def\csname LT8\endcsname{\color{black}}%
    \fi
  \fi
    \setlength{\unitlength}{0.0750bp}%
    \ifx\gptboxheight\undefined%
      \newlength{\gptboxheight}%
      \newlength{\gptboxwidth}%
      \newsavebox{\gptboxtext}%
    \fi%
    \setlength{\fboxrule}{0.5pt}%
    \setlength{\fboxsep}{1pt}%
\begin{picture}(3600.00,3024.00)%
    \gplgaddtomacro\gplbacktext{%
      \csname LTb\endcsname
      \put(350,2600){\makebox(0,0)[r]{\strut{}$D/M$}}%
      \put(700,1229){\makebox(0,0)[r]{\strut{}$0.5$}}%
      \put(700,1754){\makebox(0,0)[r]{\strut{}$1$}}%
      \put(700,2278){\makebox(0,0)[r]{\strut{}$1.5$}}%
      \put(700,2803){\makebox(0,0)[r]{\strut{}$2$}}%
      \put(726,600){\makebox(0,0){\strut{}$10^{-5}$}}%
      \put(1139,600){\makebox(0,0){\strut{}$10^{-4}$}}%
      \put(1552,600){\makebox(0,0){\strut{}$10^{-3}$}}%
      \put(1965,600){\makebox(0,0){\strut{}$10^{-2}$}}%
      \put(2377,600){\makebox(0,0){\strut{}$10^{-1}$}}%
      \put(2790,600){\makebox(0,0){\strut{}$10^{0}$}}%
      \put(3203,600){\makebox(0,0){\strut{}$10^{1}$}}%
    }%
    \gplgaddtomacro\gplfronttext{%
      \csname LTb\endcsname
      \put(2900,400){\makebox(0,0){\strut{}$E_{2}$\quad(\GeV{})}}%
      \csname LTb\endcsname
      \put(2216,2630){\makebox(0,0)[r]{\strut{}$\Pgluon \Pgluon \to \PH \PH \Pgluon $}}%
    }%
    \gplbacktext
    \put(0,0){\includegraphics[scale=1.5]{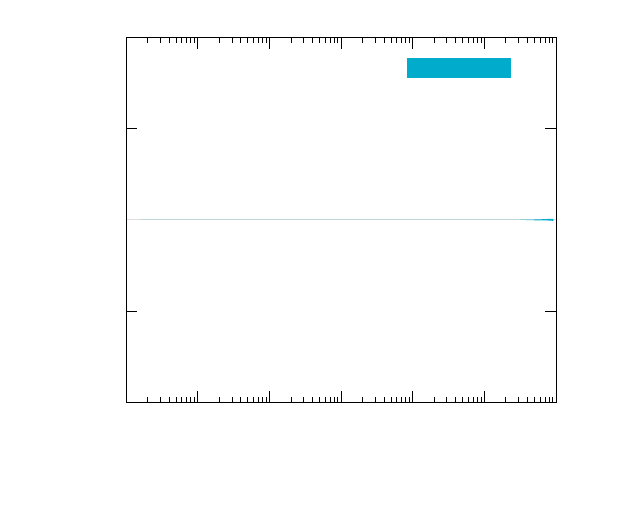}}%
    \gplfronttext
  \end{picture}%
\endgroup
\vspace{-10mm}
\caption{Envelope of subtraction ratio $D/M$ as a function of the energy of the emitted gluon. Essentially nothing can be seen, since the ratio is very close to one.}
\label{fig:subtraction_check_soft_gluon}
\end{figure}

\clearpage

\section{Conclusion and outlook}
Preliminary \HERWIG{} results for LI production at LO, possibly matched to a parton shower, were shown for Higgs-boson ($\PHiggs$, $\PHiggs\PHiggs$) and inclusive four-lepton production, in some cases with an additional jet generated at the matrix-element level. In particular, the well-known inadequacy of the common heavy-top-quark approximation in Higgs-boson production was shown here in the phase space dominated by hard associated QCD radiation. This highlights the need for calculations and event generation with the full loop dependence. With this work, the \HERWIG{} generator is now able to generate many different LI processes in a fully automated way, using matrix elements from \openloops{}.
However, the preliminary results shown here must still be validated by comparing them to other multi-purpose event generators (\SHERPA{}, \MGMCatNLO{}), different matrix-element providers (\gosam{}, \MADGRAPH{}), and, where possible, to experimental data made available in the \rivet{} framework. When this is done, the new functionality will be released in a new version of \HERWIG{}.

Going beyond LO, steps taken towards generating LI processes at NLO, matched with a parton shower if desired, were laid out. Some validation results of intermediate steps towards that goal were shown using the example of Higgs boson pair production. Once the NLO implementation has been completed for a few benchmark processes, a new \HERWIG{} software release and journal article are planned to be published. 

LI NLO generation is limited by the availability of two-loop amplitudes needed for the virtual corrections. These have been implemented in software at least for $\PHiggs\PHiggs$ production \cite{Borowka:2016ehy,Borowka:2016ypz,Heinrich:2017kxx} and four-lepton production \cite{vonManteuffel:2015msa,Caola:2015ila} ($\Pproton\Pproton\,\looparrow\,\llll$ generation at NLO matched with a parton shower has been performed using \POWHEGpy{}~8 \cite{Alioli:2016xab}, though ignoring contributions mediated by a Higgs boson). Given the current level of research activity in this direction and the urgent need by experimental collaborations at the LHC, it is very likely that many two-loop amplitudes will be calculated and made publicly available in the near future.

Once the LI NLO generation with \HERWIG{} works in principle, there are many incremental improvements that can be made: the phase space sampling for LI processes can be optimised for greater speed and precision, \HERWIG{} can select the two-loop matrix elements to use automatically where there is only one choice available, etc. Future work could also extend the functionality to arbitrary BSM models. 

\clearpage\pagebreak
\appendix
\part{Appendix}
\label{sec:appendix}

\section{Fixes applied to \SHERPA{} samples}
\label{sec:sherpa_fixes}

The \SHERPA{} NLO samples for \ZZllll{} production suffer from a bug (discovered and co-reported by the author, fixed in \SHERPA{} 2.2.2) that leads to incorrect relative contributions from the different decay channels channels (\eeee{}, \eemm{}, \mmmm{}). The observed incorrect contributions in the fiducial phase space compared to the correct ones from \POWHEG{} are shown in \mytab~\ref{tab:sherpa_channel_reweighting}, along with the reweighting factors that are applied to the \SHERPA{} events to correct for the bug. It turns out that the total relative rate of events in the channels of interest (all excluding $\tau$ leptons, which are also present in the sample) is correct to within $<1\%$, therefore it is sufficient to reweight the three signal channels. The reweighting factors are found to depend on the kinematics of the events, but the impact of not taking this into account on the differential cross sections is $\ll 1\%$ in almost all bins. The effect is at most $\mathcal{O}(1\%)$ in a few bins, but these bins have a statistical uncertainty of around 30\% or more, so the effect is considered negligible.

\begin{table}[h!]
\centering
\begin{tabular}{llll}
\toprule
Channel & \POWHEG{} (correct) & \SHERPA{} (incorrect) & Reweighting factor\\
\midrule
\eeee       &  25.48\%    &  26.18\%   &  0.9732\\
\eemm    &  49.04\%    &  49.25\%   &  0.9957\\
\mmmm      &  25.48\%    &  24.57\%   &  1.0370\\
\bottomrule
\end{tabular}
\caption{\SHERPA{} NLO sample per-channel reweighting factors. The fractions reported for \POWHEG{} and \SHERPA{} are relative to the combination of the three signal channels. The correction factor is the ratio of the aforementioned fractions.}
\label{tab:sherpa_channel_reweighting}
\end{table}

The \SHERPA{} samples also contain individual events with very large weights, which is a well-known effect. The ATLAS recommendation, sanctioned by \SHERPA{} authors \cite{frank_siegert}, of setting all event weights with absolute value above 100 to one is followed. As this still doesn't remove a few outliers in alternative weights from to scale or PDF variations, alternative weights that differ from the nominal weight by more than a factor of five (i.e.~$< 0.2$ or $> 5$) are set to the nominal weight.

\section{Photon-induced four-lepton production}\label{sec:photon_induced}

Due to the fact that photons are not massive compared to the momentum they have when participating in hard scattering, $|\vec{p_{\gamma}}| \gg m_{\gamma} = 0$, and the fact that they interact with (anti)quarks, photons can be considered to appear as partons of the colliding protons. This means that their appearance in hard interactions can be factored into photon PDFs of the proton. Available photon PDFs include MRST2004qed \cite{Martin:2004dh}, NNPDF23qed \cite{Ball:2013hta}, CT14qed \cite{Schmidt:2015zda}, as well as the recent LUXqed \cite{Manohar:2016nzj,luxqed_website}. The photon PDF itself is suppressed by $\mathcal{O}(\alpha) \sim 0.01$ with respect to the (anti)quark and gluon PDFs, so while the photon-initiated production contributes at the same order as $\Pquark\APquark$-initiated production at the \textit{matrix-element} level, it is considered an NLO EW effect overall. Despite this, it is not included in the NLO EW corrections used in the analysis that is the main topic of \mypart~\ref{sec:analysis}. This is because only weak and no photonic corrections are included there by choice, as discussed in \mysec~\ref{sec:zz_ew_corrections}. In fact, the photon-induced contributions are known to be phenomenologically unimportant in the NLO EW calculation \cite{Biedermann:2016lvg}. On the other hand, they might be of interest in the future, as more precise calculations and measurements become available.
Example Feynman diagrams for $\gamma\gamma$- and $\Pquark\gamma$-initiated production are shown in \myfig{}~\ref{fig:theory_photon_induced_zz}. The contributions from the LO diagram in \myfig~\ref{fig:theory_photon_induced_zz_LO} and the NLO diagram in \myfig~\ref{fig:theory_photon_induced_zz_NLO} are expected to be of the same effective order, because the NLO features one less initial-state photon with a corresponding PDF suppression by $\mathcal{O}(\alpha)$.

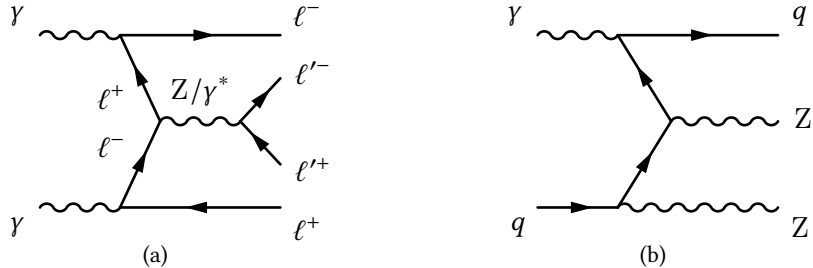
\begin{figure}[h!]
	\centering
	\subfigure[]{
	\centering
		\begin{fmfgraph*}(90,65)
		\fmfset{arrow_len}{3mm}
		\fmfstraight
		\fmfleft{i0,i1,i2,i3,i4}
		\fmfright{o0,o1,o2,o3,o4}
		\fmflabel{$\ell^{\prime +}$}{o1}
		\fmflabel{$\ell^{\prime -}$}{o3}
		\fmflabel{$\Pphoton$}{i0}
		\fmflabel{$\Pphoton$}{i4}
		\fmflabel{$\ell^+$}{o0}
		\fmflabel{$\ell^-$}{o4}
		\fmf{phantom}{i4,v4,o4}
		\fmf{phantom}{i4,v4} 
		\fmf{phantom}{i0,v0,o0}
		\fmf{phantom}{i0,v0} 
		\fmffreeze
		\fmf{photon}{i4,v4}
		\fmf{photon}{i0,v0}
		\fmffreeze
		\fmf{fermion}{v4,o4}
		\fmf{fermion,label=$\ell^+$,label.side=left}{wk2,v4}
		\fmf{fermion}{o0,v0}
		\fmf{fermion,label=$\ell^-$,label.side=left}{v0,wk2}
		\fmf{photon,label=$\PZ$$/$$\gamma^{*}$,label.side=left}{wk2,d2} 
		\fmf{fermion}{o1,d2,o3}
	\end{fmfgraph*}
	\label{fig:theory_photon_induced_zz_LO}
	}
	\hspace{3cm}
		\subfigure[]{
	\centering
		\begin{fmfgraph*}(90,65)
		\fmfset{arrow_len}{3mm}
		\fmfstraight
		\fmfleft{i0,i1,i2,i3,i4}
		\fmfright{o0,o1,o2,o3,o4}
		\fmflabel{$\PZ$}{o2}
		\fmflabel{$\Pquark$}{i0}
		\fmflabel{$\Pphoton$}{i4}
		\fmflabel{$\PZ$}{o0}
		\fmflabel{$\Pquark$}{o4}
		\fmf{phantom}{i4,v4,o4}
		\fmf{phantom}{i4,v4} 
		\fmf{phantom}{i0,v0,o0}
		\fmf{phantom}{i0,v0} 
		\fmffreeze
		\fmf{photon}{i4,v4}
		\fmf{fermion}{i0,v0}
		\fmffreeze
		\fmf{fermion}{v4,o4}
		\fmf{fermion}{wk2,v4}
		\fmf{photon}{o0,v0}
		\fmf{fermion}{v0,wk2}
		\fmf{photon}{wk2,o2} 
	\end{fmfgraph*}
	\label{fig:theory_photon_induced_zz_NLO}
	}
	\caption{Example Feynman diagrams for photon-induced four-lepton production. (a) A LO diagram, corresponding to non-resonant production. (b) An NLO diagram with an emitted quark. The leptonic decays of the \PZ{} bosons are not shown.}
	  \label{fig:theory_photon_induced_zz}
\end{figure}

If the photon-induced component is considered in combining QCD and electroweak corrections, its contribution is sometimes added linearly (e.g.~in \myref{}~\cite{Yong:2016njr}), as it corresponds to distinct initial states,
\begin{equation*}
	\sigma^{\text{NNLO $\otimes$ NLO EW}} = \sigma^{LO} \big[ (1 + \delta_{\text{QCD}})(1 + \delta_{\text{EW excl. $\gamma$-ind.}}) + \delta_{\text{$\gamma$-ind.}} \big].
\end{equation*}
This differs from the approach where QCD and EW corrections are combined purely multiplicatively, which is used in the analysis in \mypart~\ref{sec:analysis}.

\section{Additional and intermediate $\PZ\PZ$ results}
\label{sec:zz_aux_materials}

\subsection{Comparisons of data and predictions}
\label{sec:zz_aux_datamc}

The following sections show additional intermediate steps and results of the unfolding. Figures that were already shown in \mypart~\ref{sec:analysis} are not repeated here.

\myfigs~\ref{fig:reco_plots_first}--\ref{fig:reco_plots_last} show the measured distributions before correcting for detector effects.

\begin{figure}[h!]
\centering
\subfigure{\includegraphics[width=0.45\textwidth]{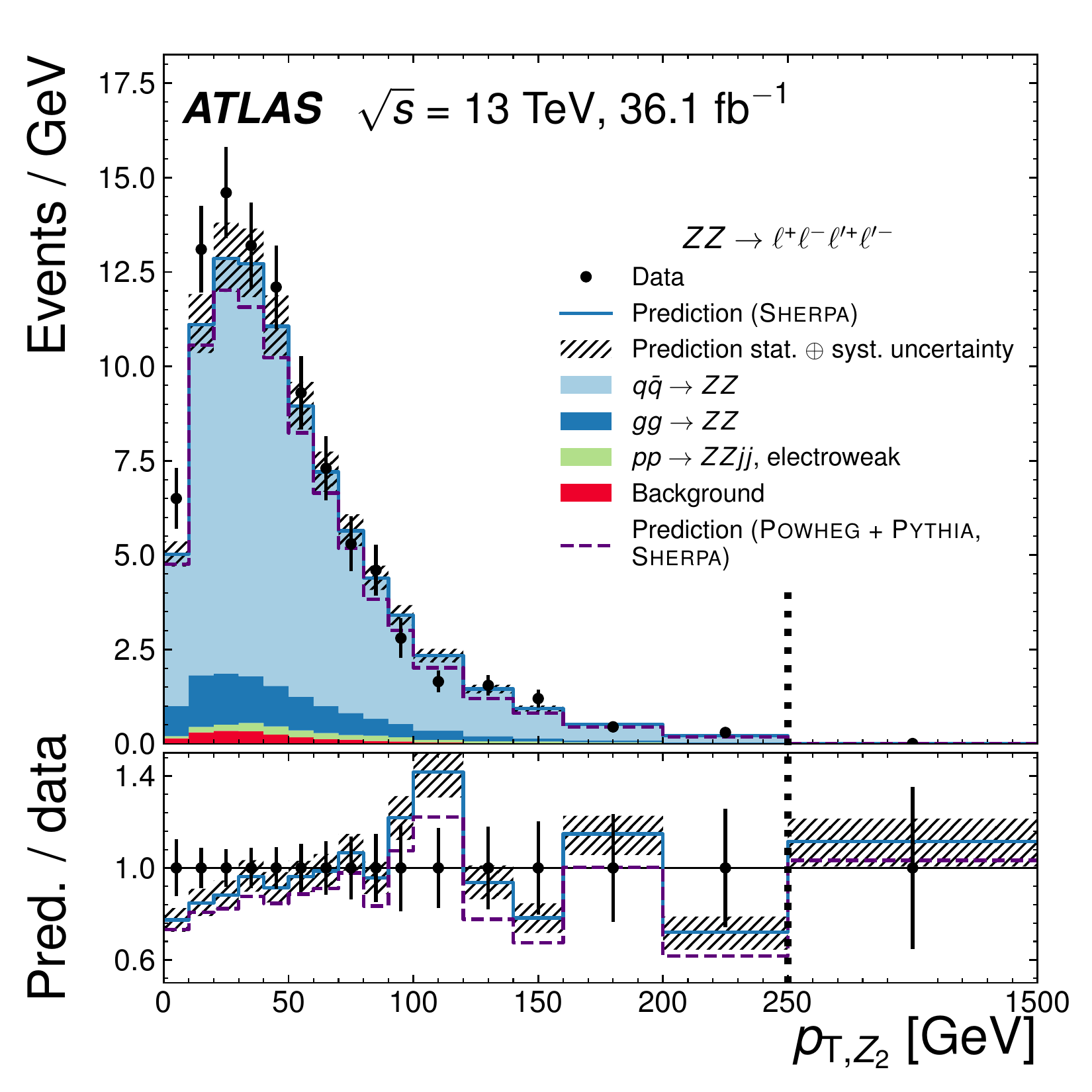}}
\subfigure{\includegraphics[width=0.45\textwidth]{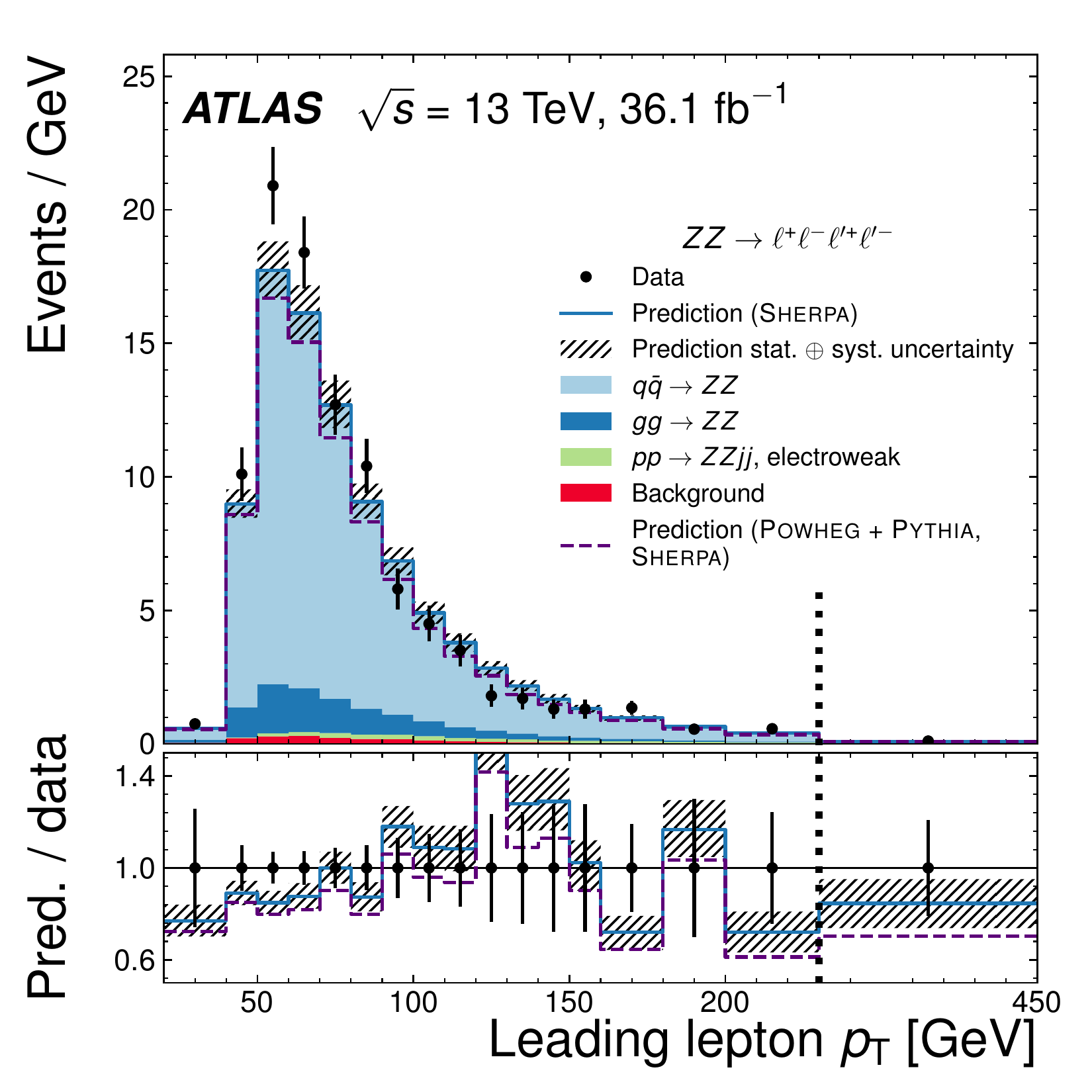}}
\subfigure{\includegraphics[width=0.45\textwidth]{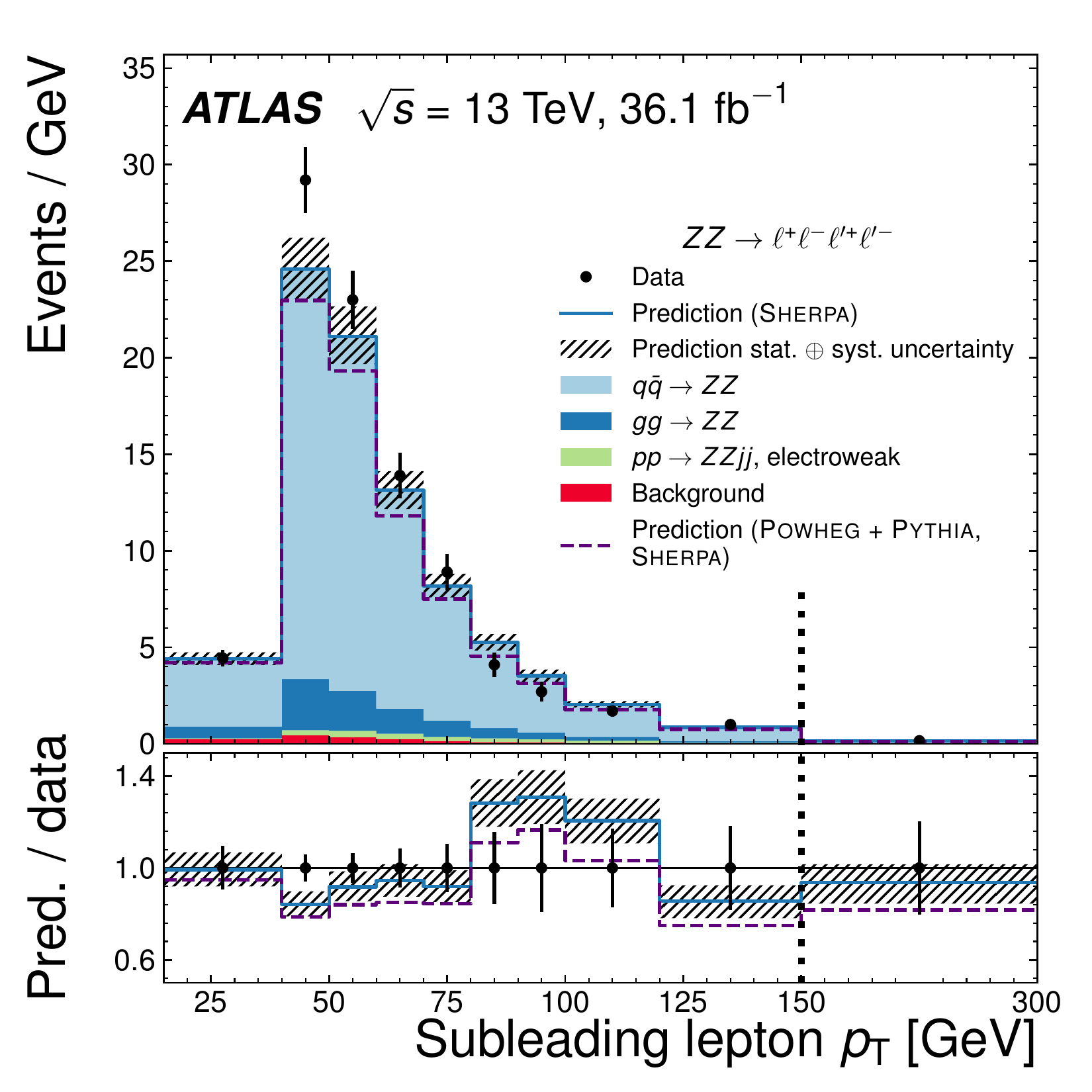}}
\subfigure{\includegraphics[width=0.45\textwidth]{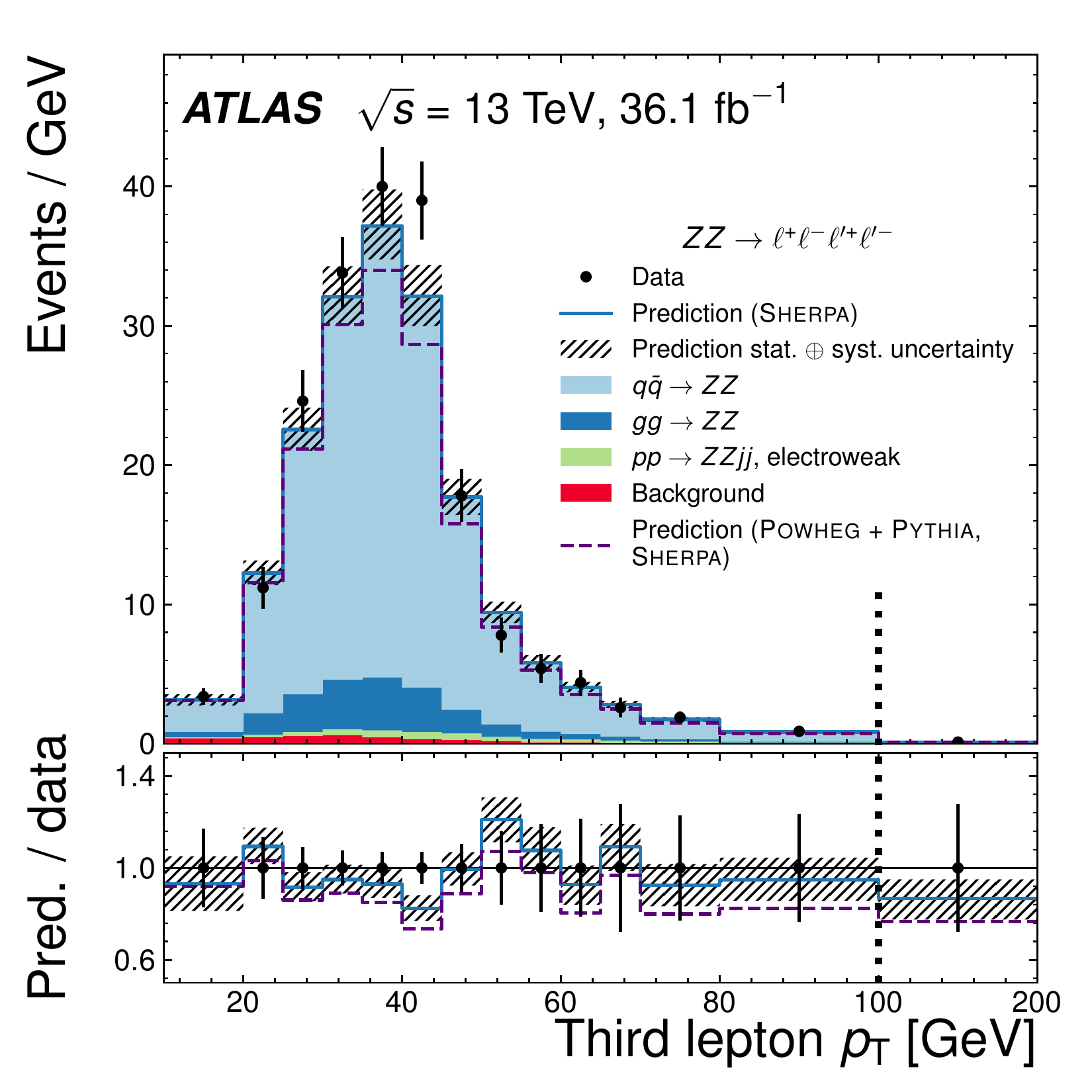}}
\subfigure{\includegraphics[width=0.45\textwidth]{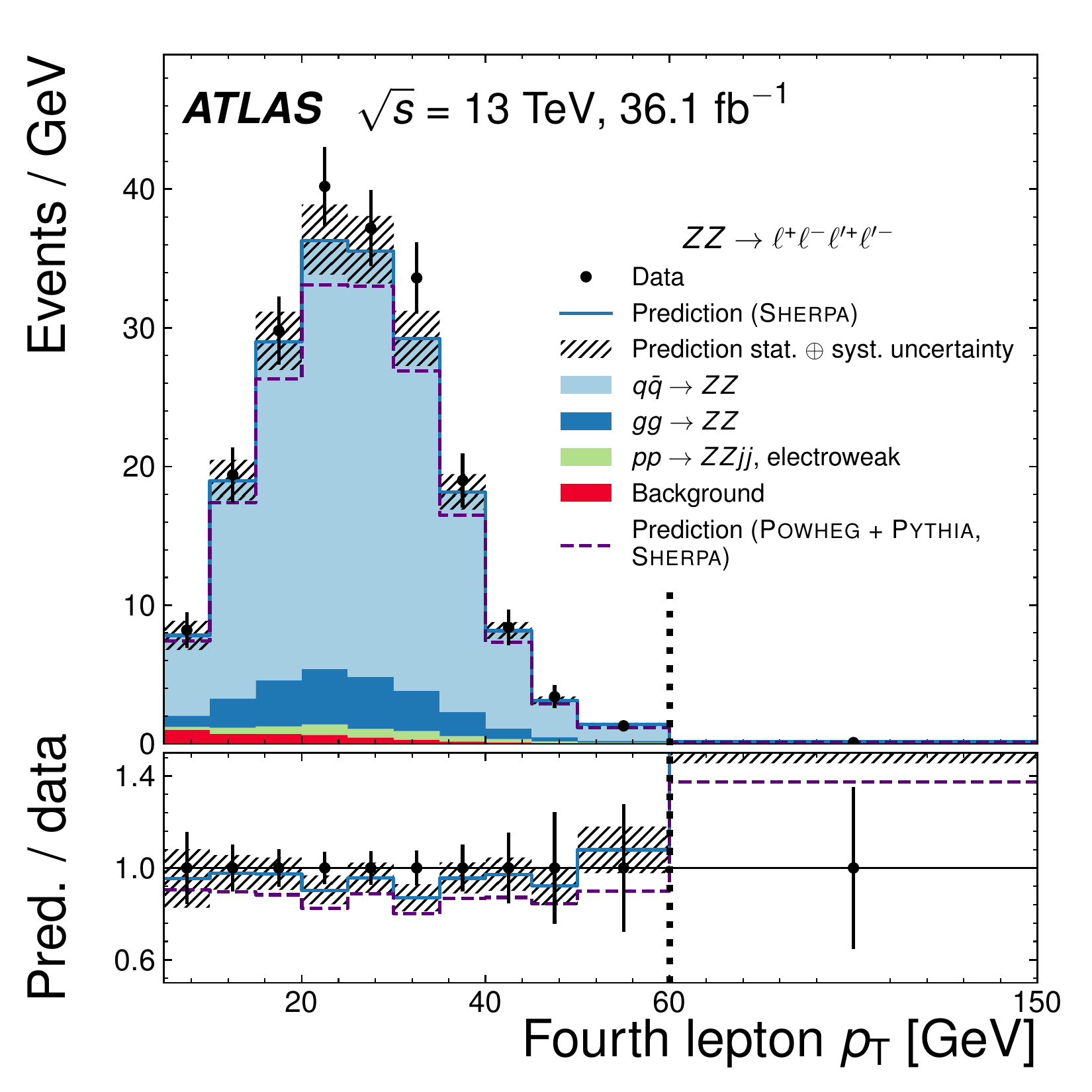}}
\subfigure{\includegraphics[width=0.45\textwidth]{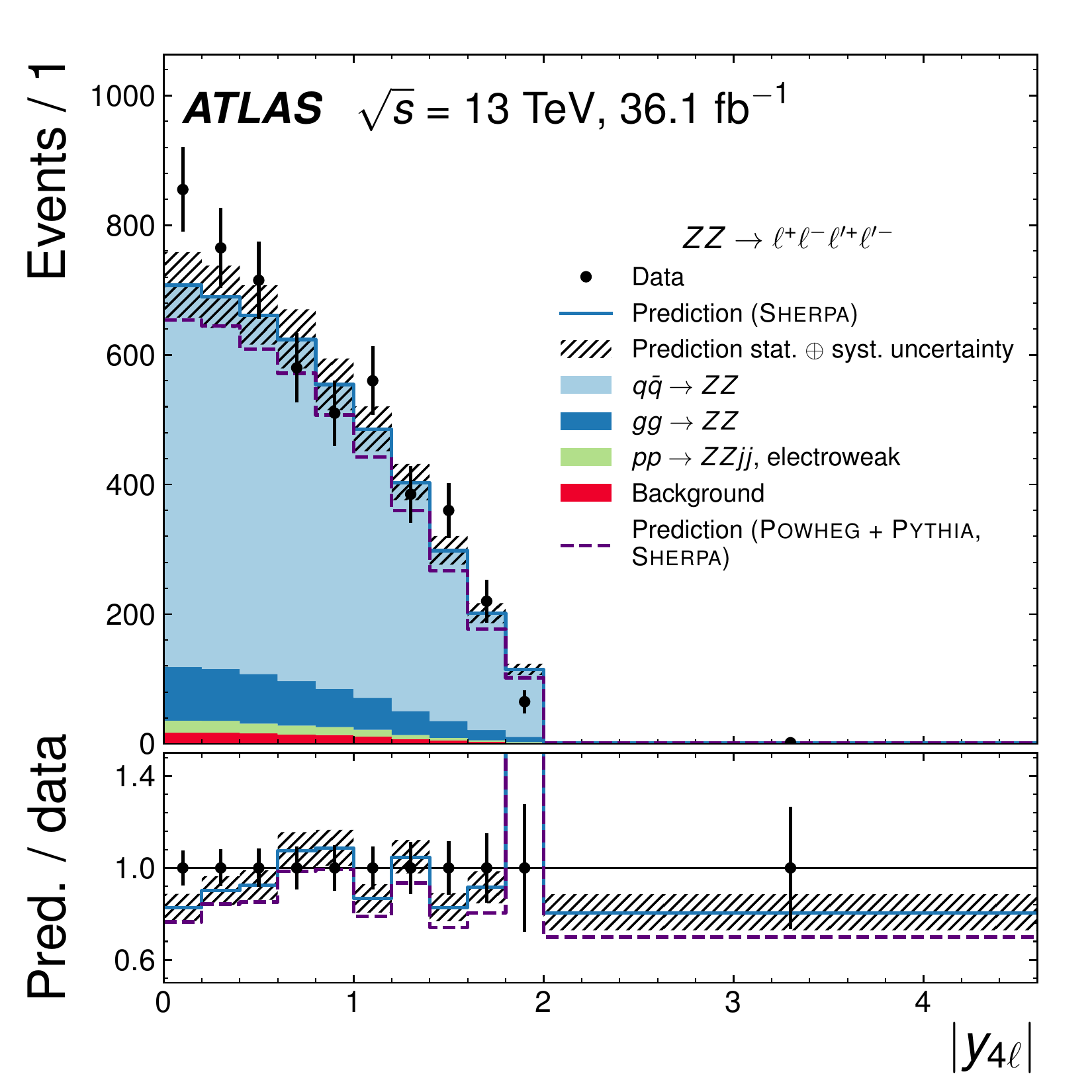}}
\caption{Measured distributions of the selected data events along with predictions in bins of various observables. The nominal prediction uses the nominal \SHERPA{} setup. Different signal contributions and the background are shown, as is an alternative prediction that uses \POWHEGpy{} to generate the $\Pquark\APquark$-initiated subprocess. For better visualisation, the last bin is shown using a different x-axis scale where indicated by a dashed vertical line. Published in the auxiliary materials of \myref~\cite{STDM-2016-15}.}
\label{fig:reco_plots_first}
\end{figure}

\begin{figure}[h!]
\centering
\subfigure{\includegraphics[width=0.45\textwidth]{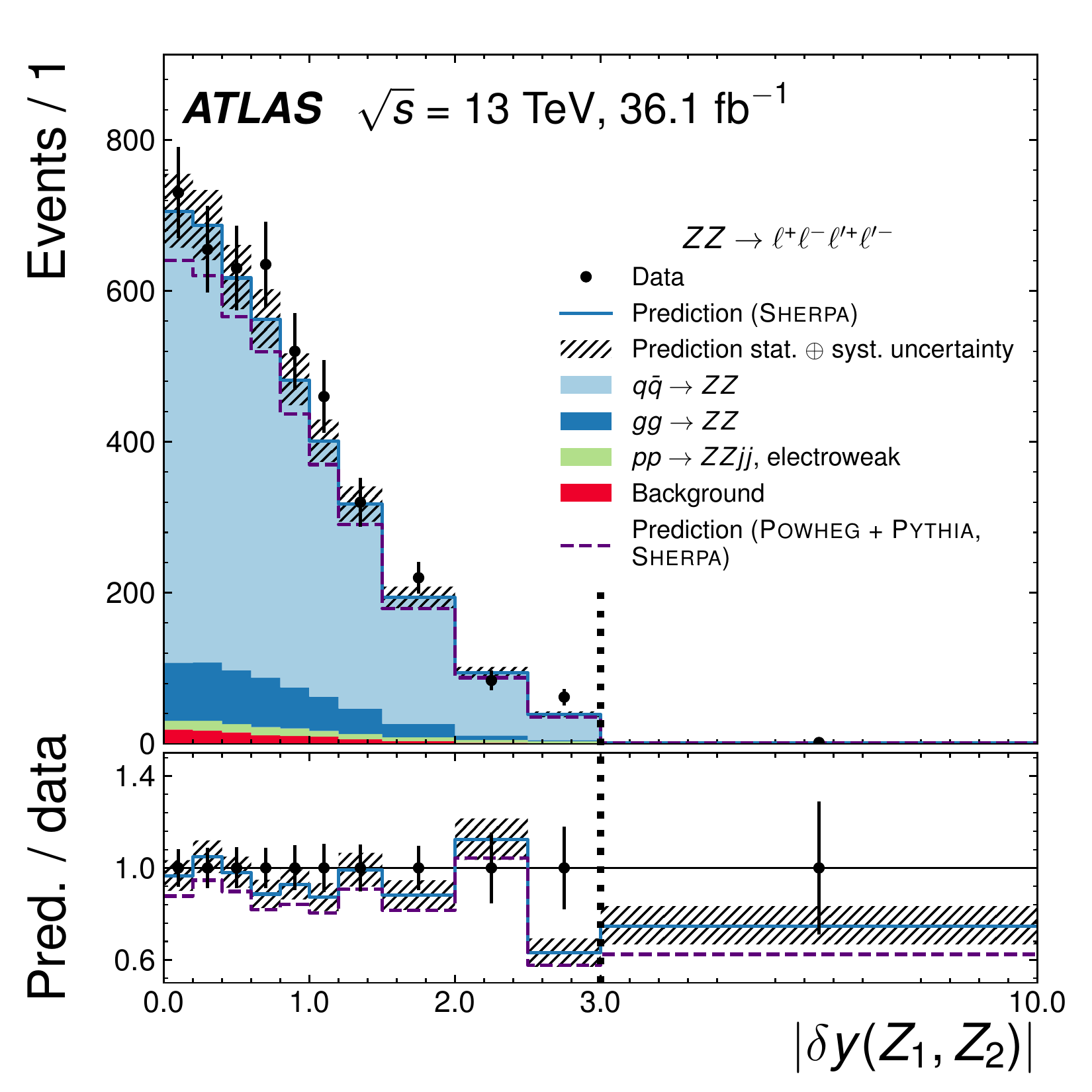}}
\subfigure{\includegraphics[width=0.45\textwidth]{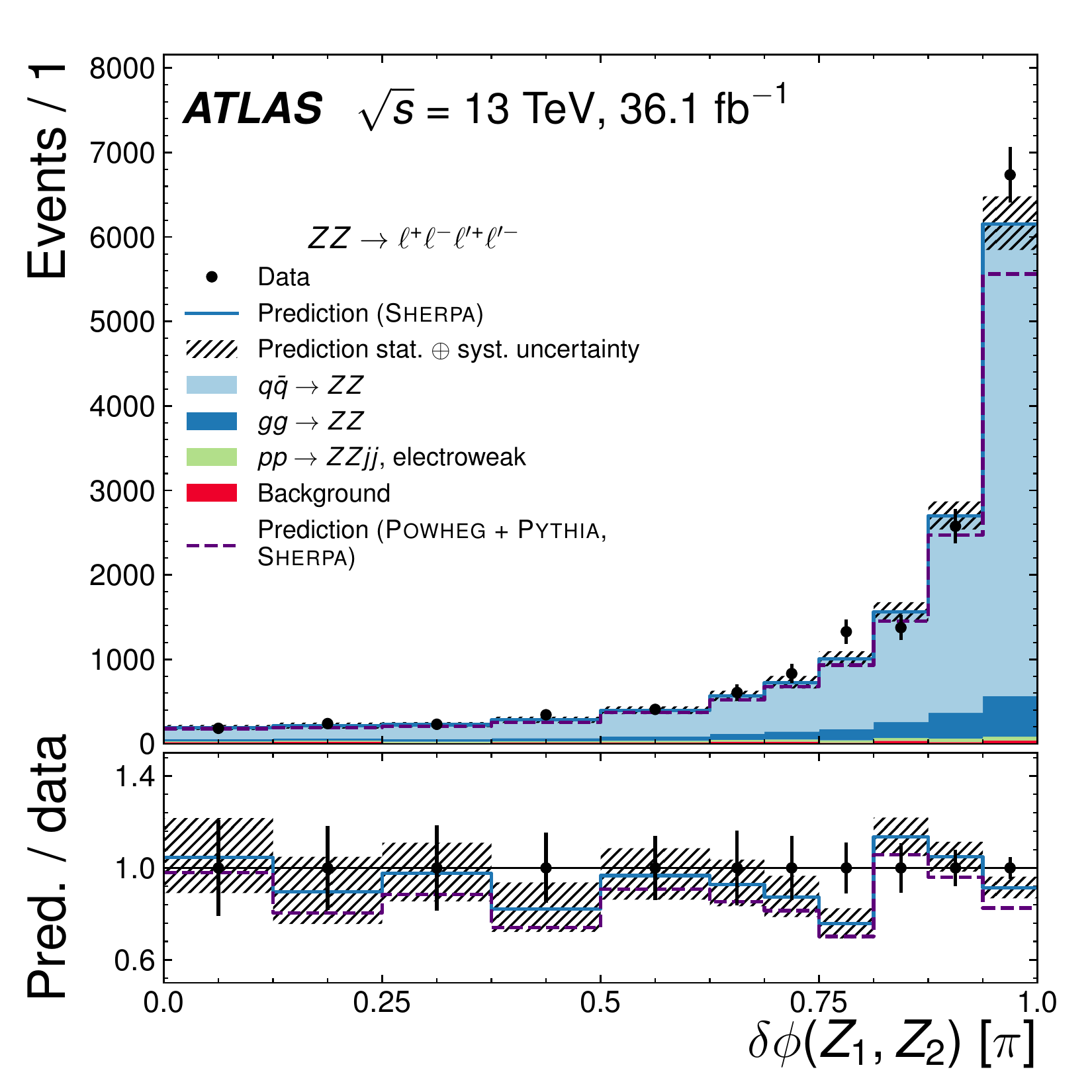}}
\subfigure{\includegraphics[width=0.45\textwidth]{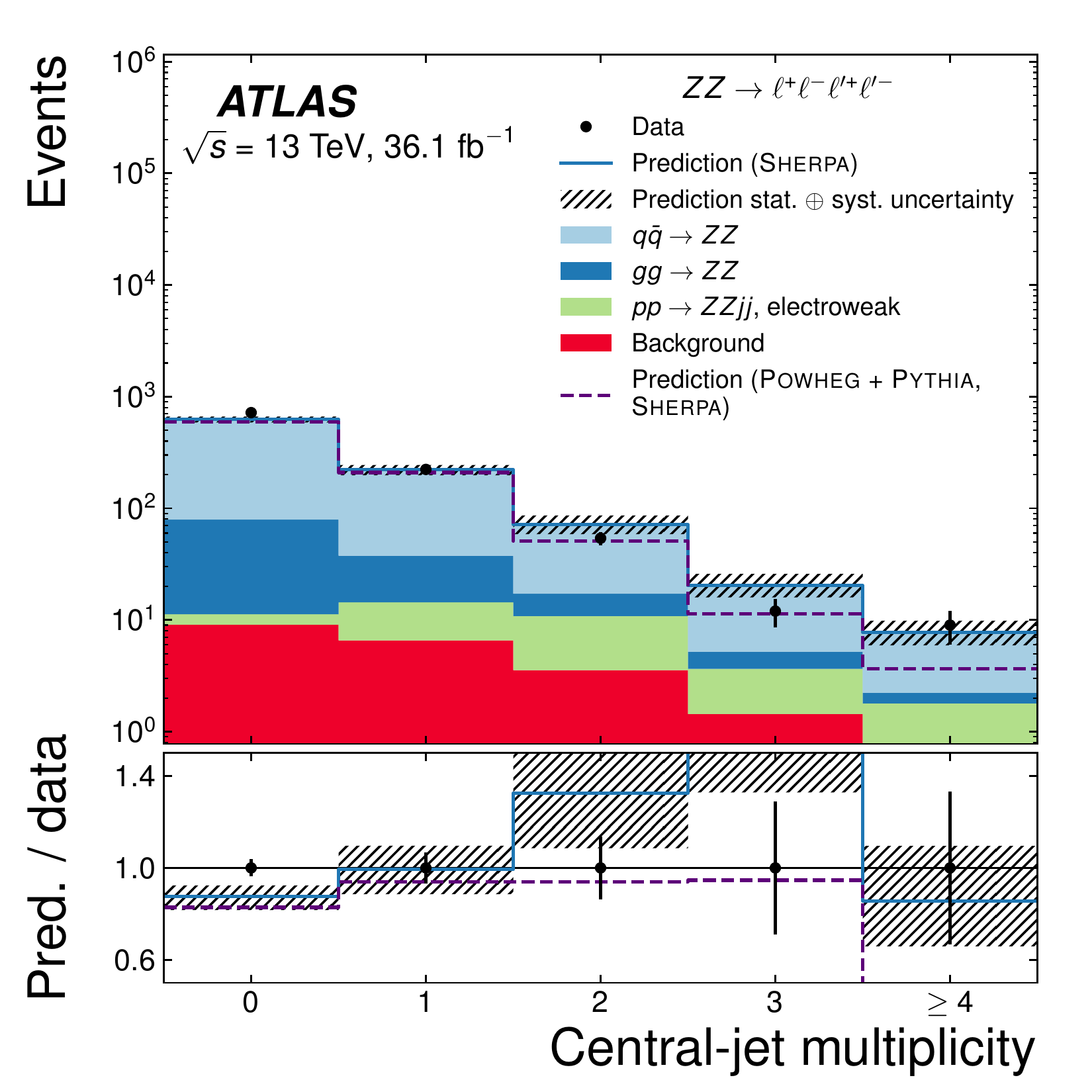}}
\subfigure{\includegraphics[width=0.45\textwidth]{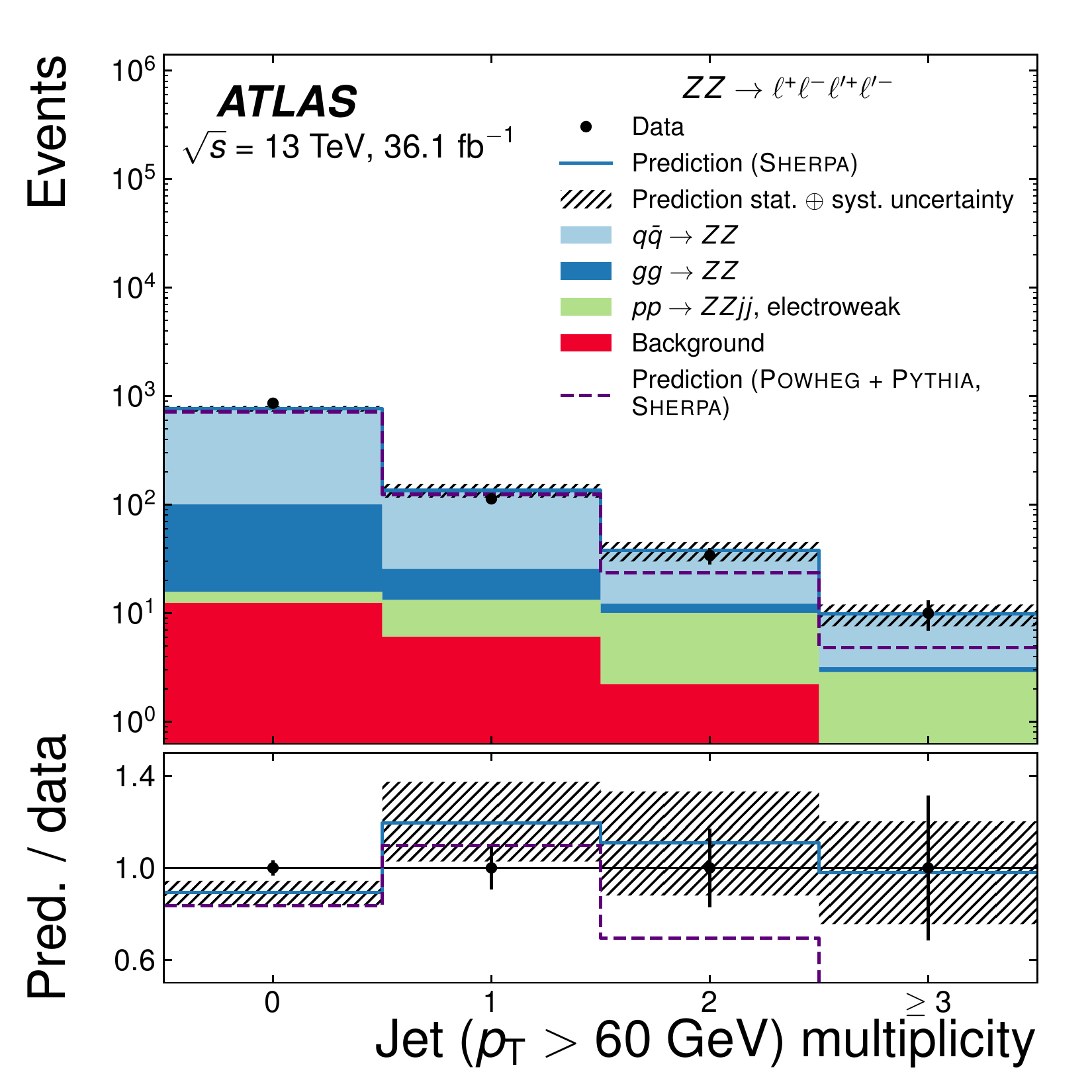}}
\subfigure{\includegraphics[width=0.45\textwidth]{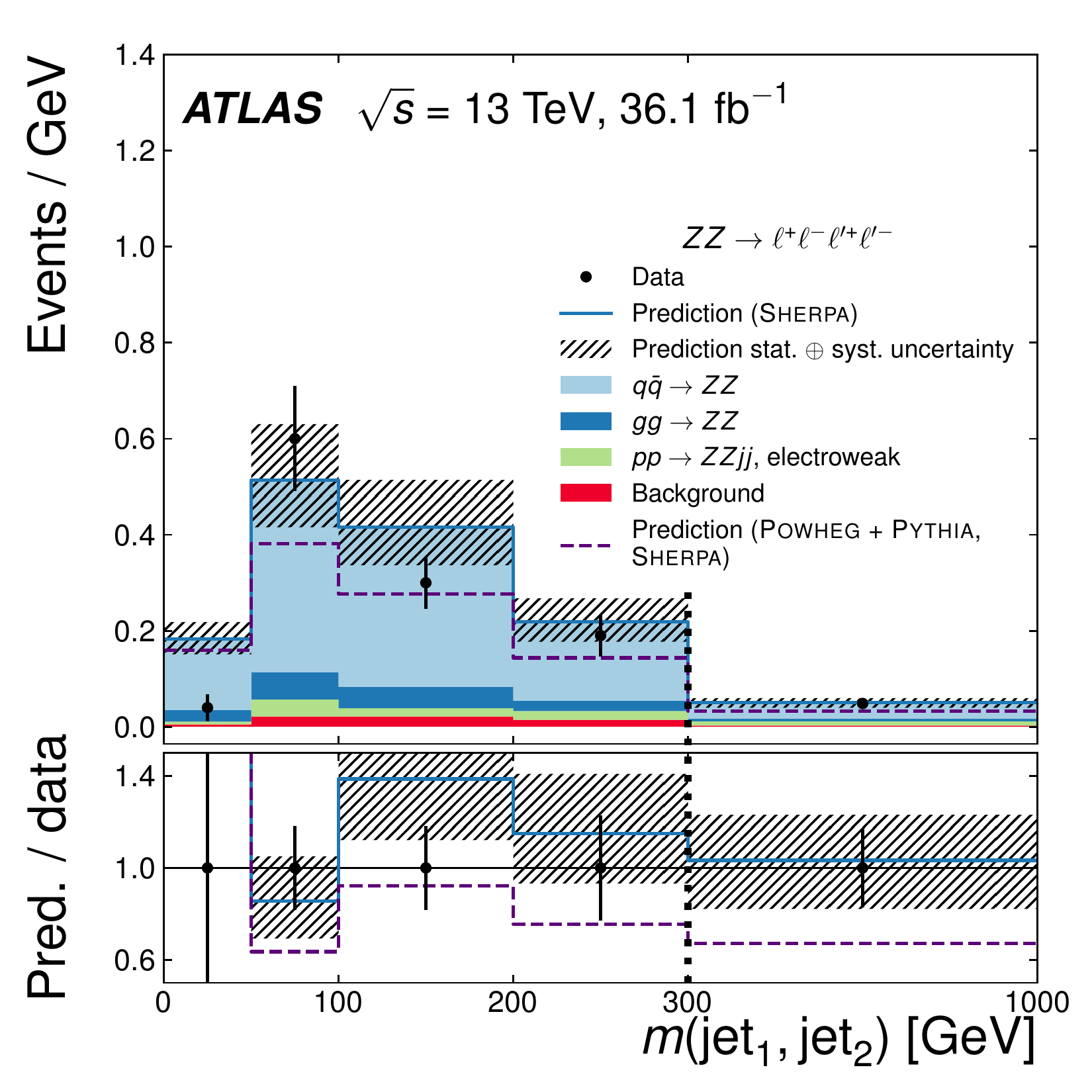}}
\subfigure{\includegraphics[width=0.45\textwidth]{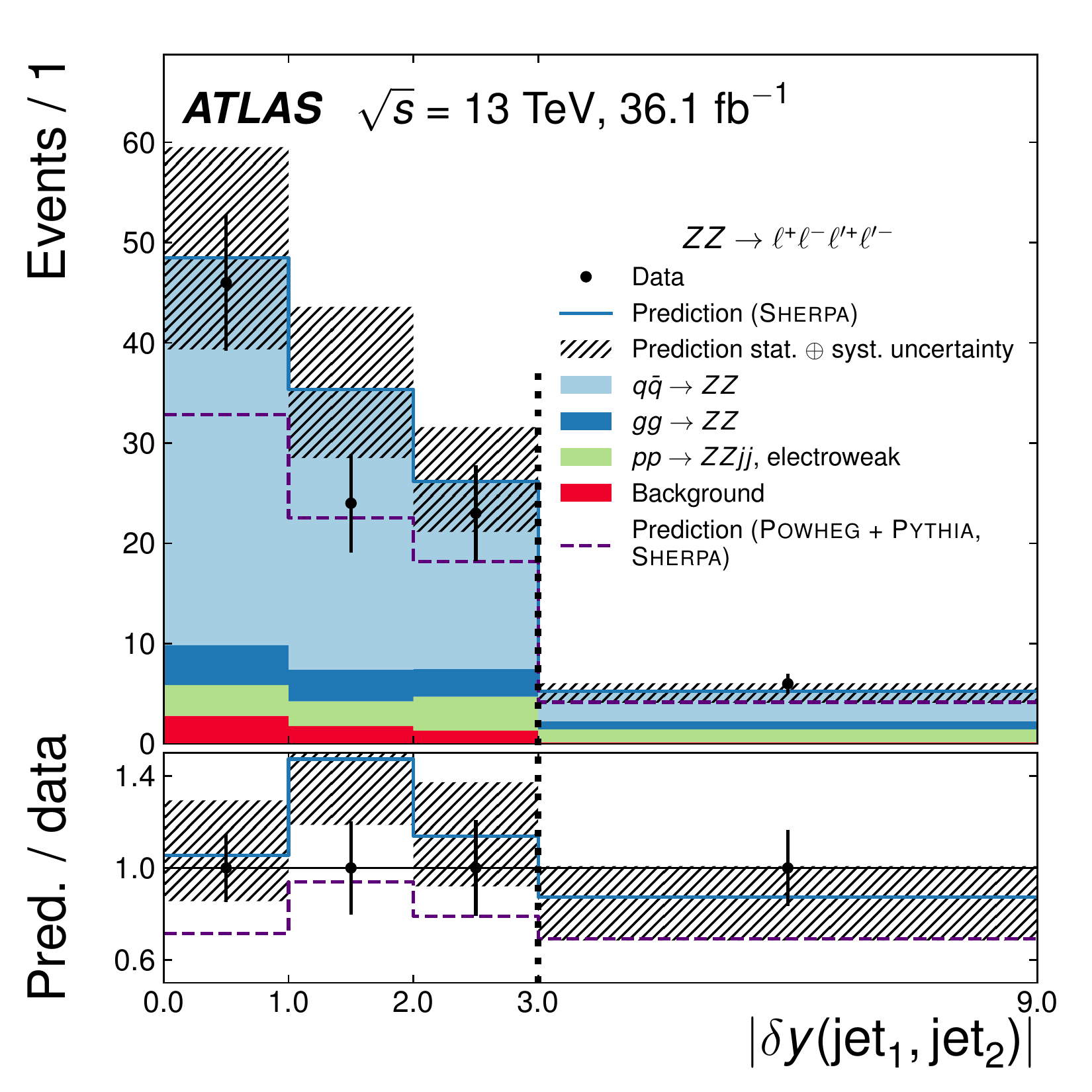}}
\caption{Measured distributions of the selected data events along with predictions in bins of various observables. The nominal prediction uses the nominal \SHERPA{} setup. Different signal contributions and the background are shown, as is an alternative prediction that uses \POWHEGpy{} to generate the $\Pquark\APquark$-initiated subprocess. For better visualisation, the last bin is shown using a different x-axis scale where indicated by a dashed vertical line. Published in the auxiliary materials of \myref~\cite{STDM-2016-15}.}
\end{figure}

\begin{figure}[h!]
\centering
\subfigure{\includegraphics[width=0.45\textwidth]{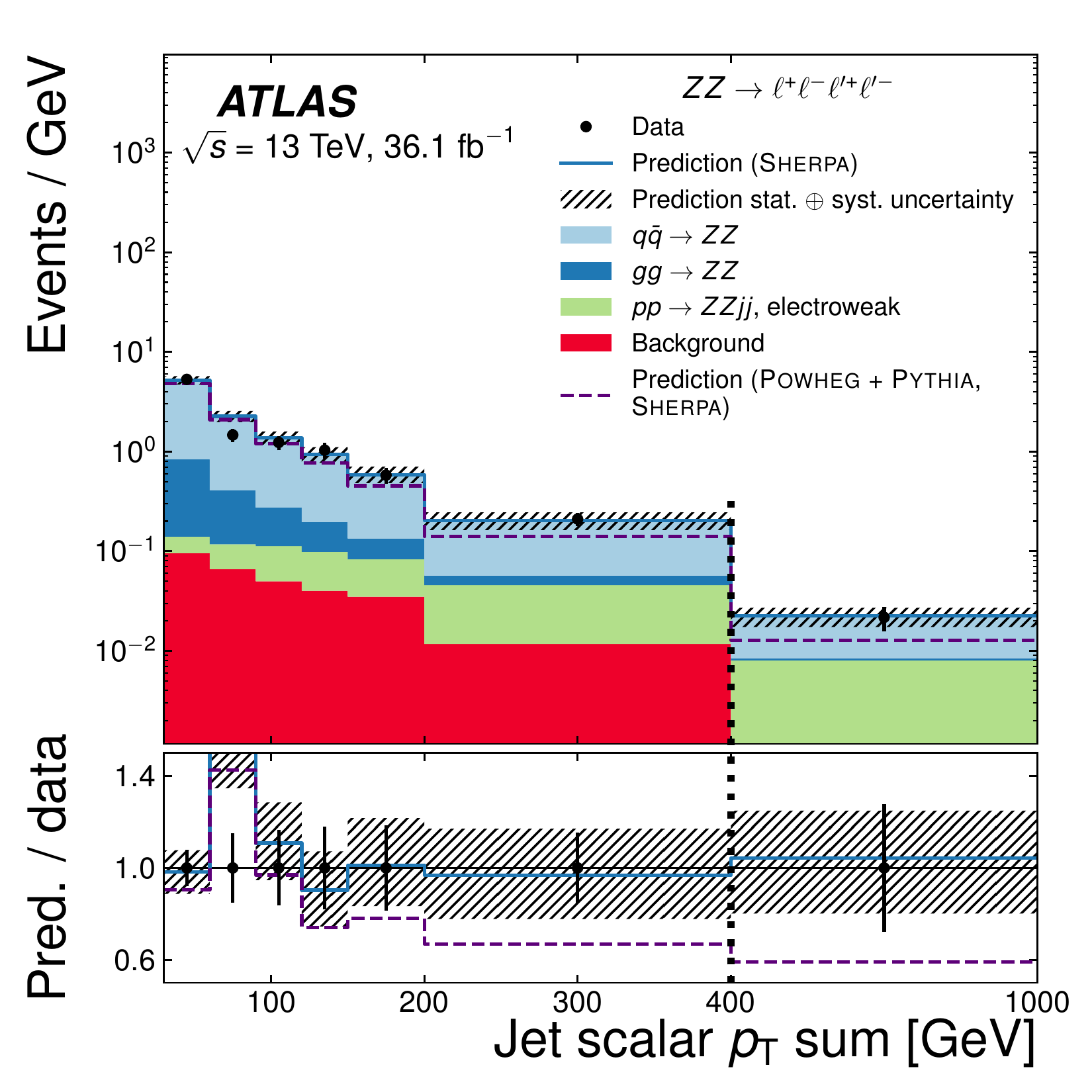}}
\subfigure{\includegraphics[width=0.45\textwidth]{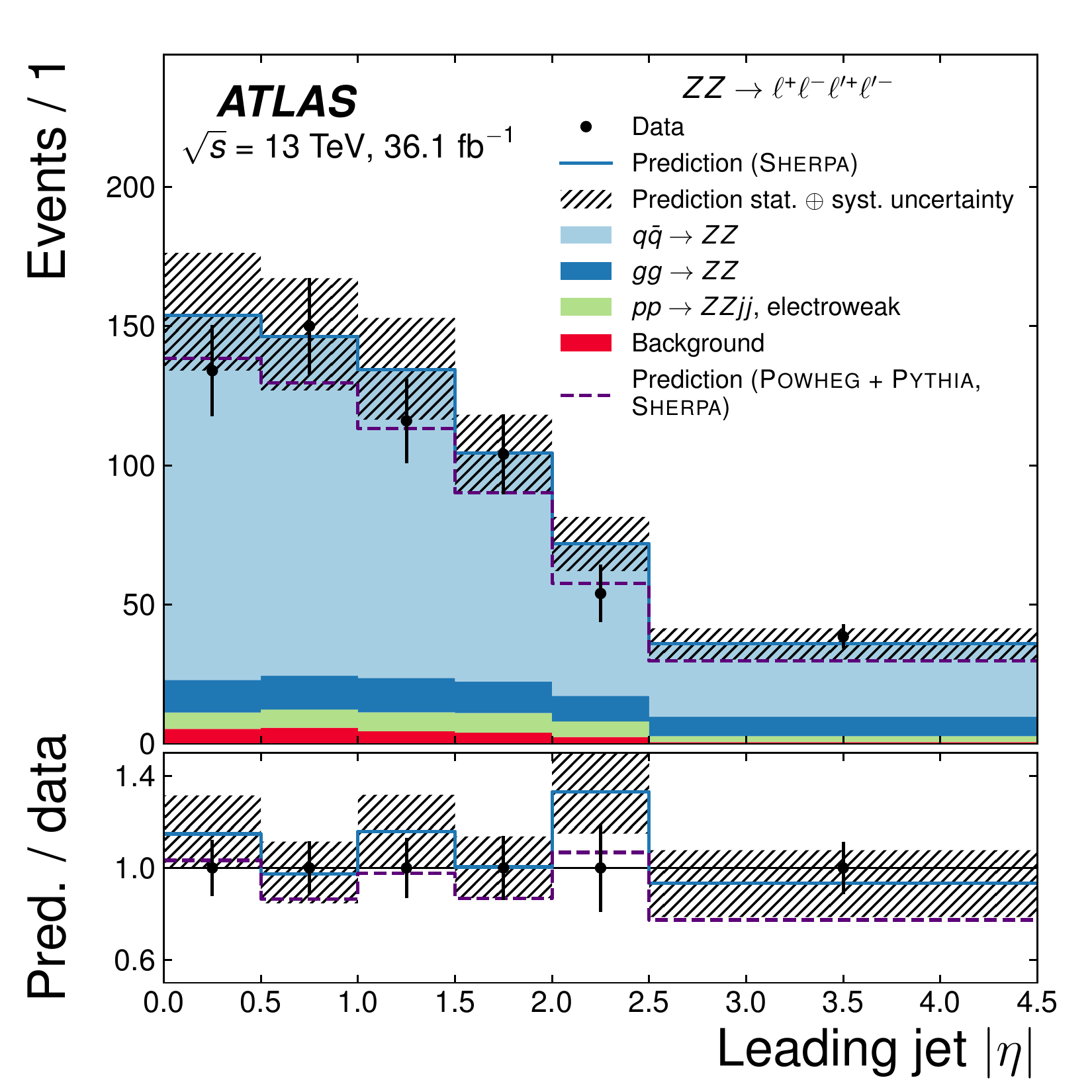}}
\subfigure{\includegraphics[width=0.45\textwidth]{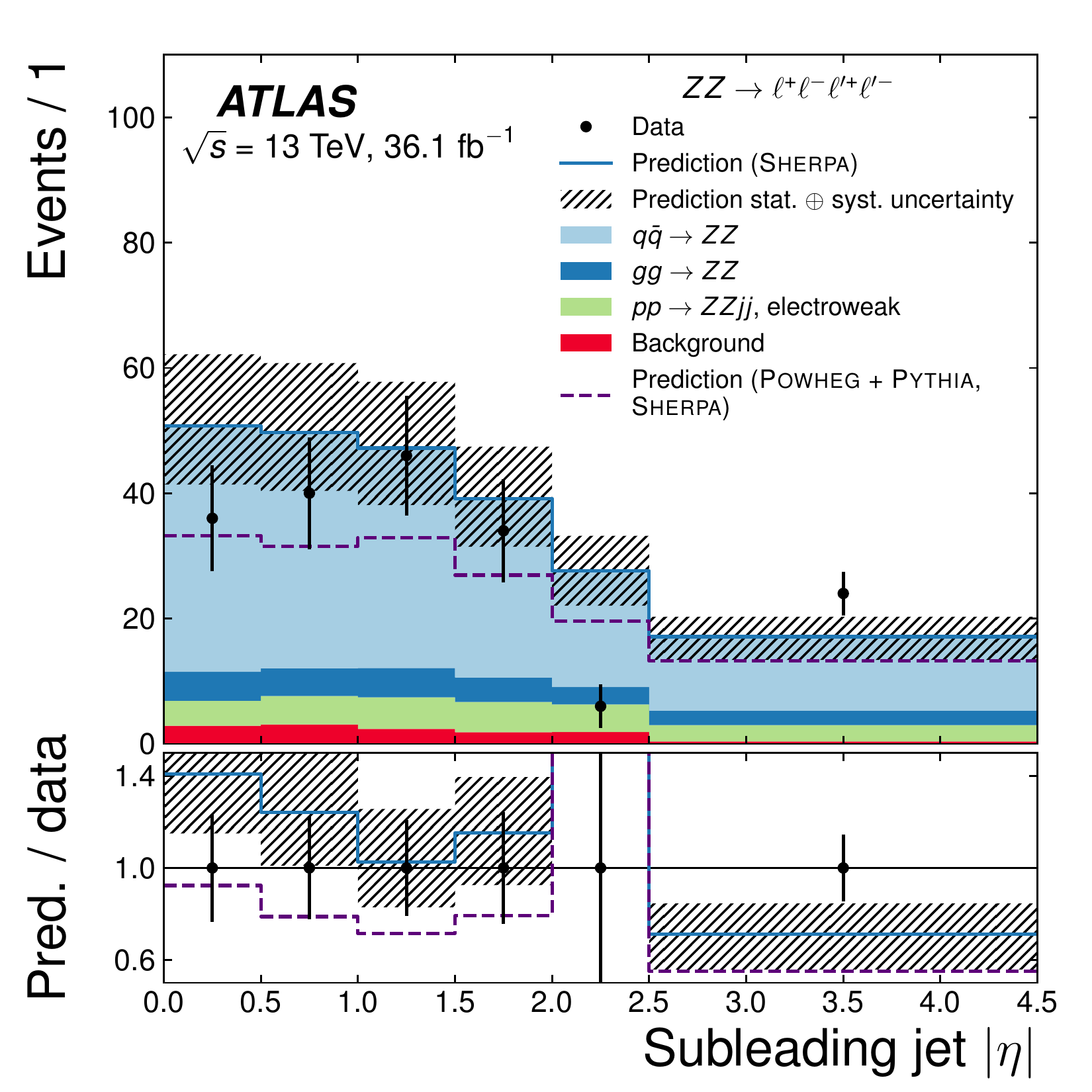}}
\subfigure{\includegraphics[width=0.45\textwidth]{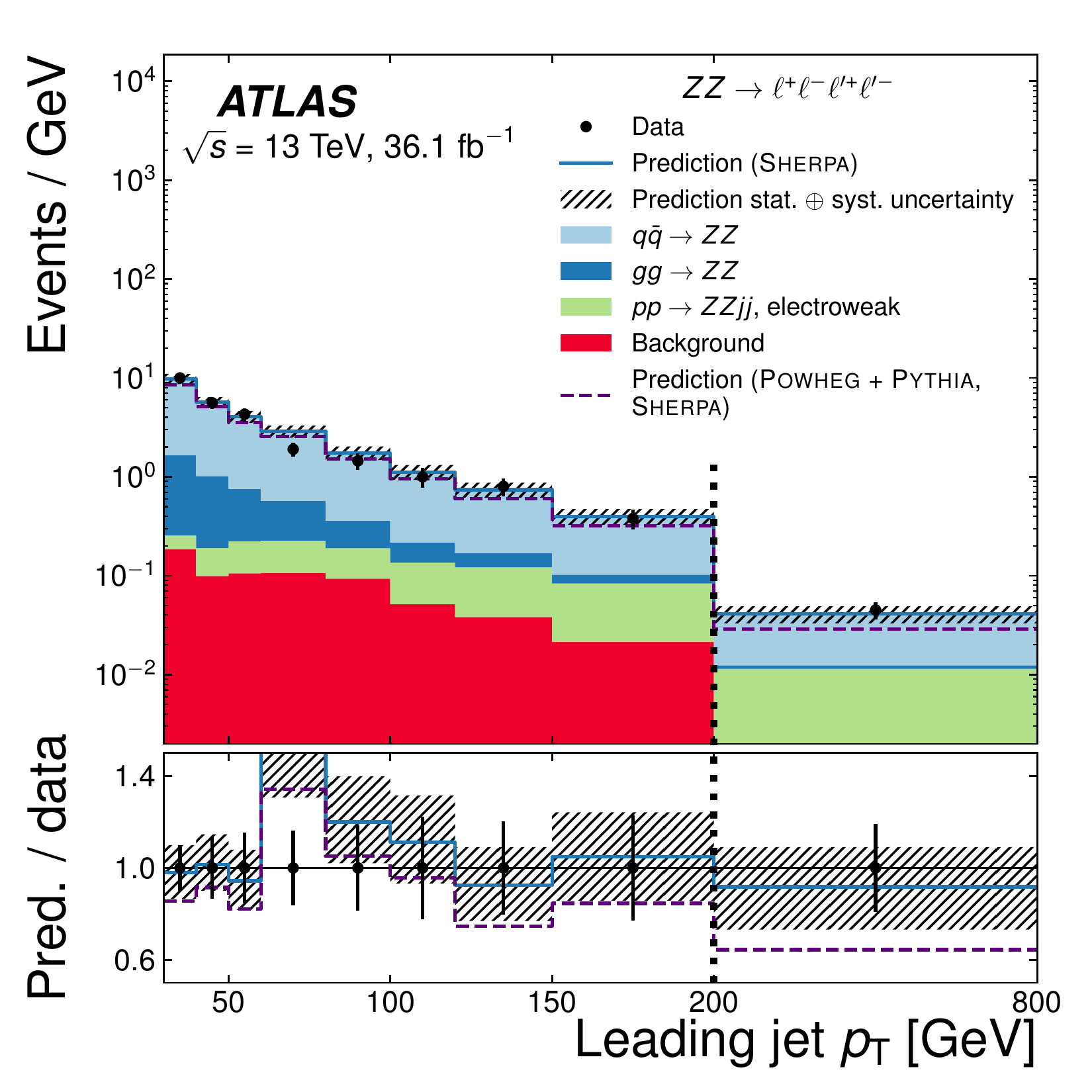}}
\subfigure{\includegraphics[width=0.45\textwidth]{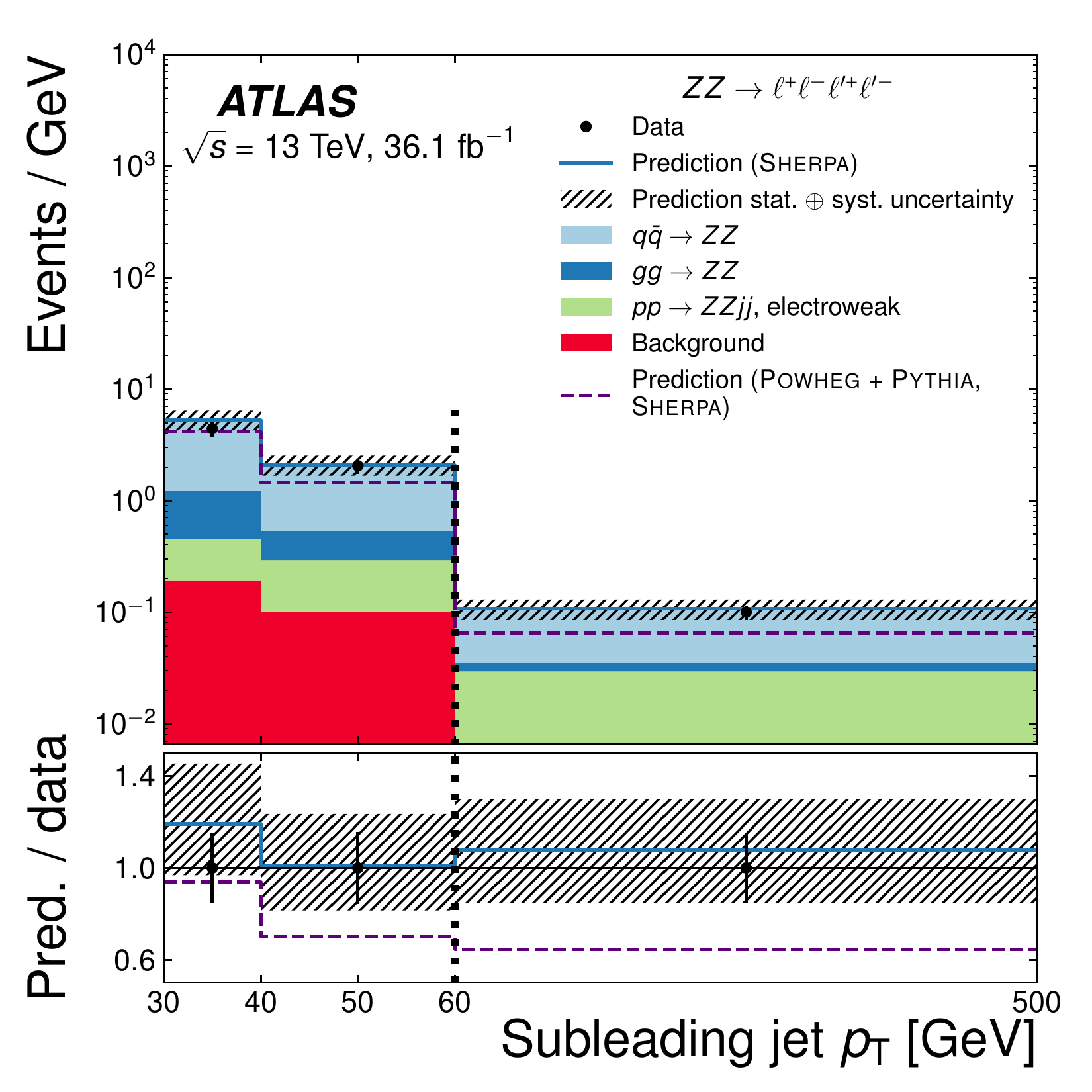}}
\caption{Measured distributions of the selected data events along with predictions in bins of various observables. The nominal prediction uses the nominal \SHERPA{} setup. Different signal contributions and the background are shown, as is an alternative prediction that uses \POWHEGpy{} to generate the $\Pquark\APquark$-initiated subprocess. For better visualisation, the last bin is shown using a different x-axis scale where indicated by a dashed vertical line. Published in the auxiliary materials of \myref~\cite{STDM-2016-15}.}
\label{fig:reco_plots_last}
\end{figure}

\clearpage

\subsection{Reweighting for the unfolding bias assessment}
\label{sec:zz_aux_reweighting}
\myfigs~\ref{fig:polyfit_plots_first}--\ref{fig:polyfit_plots_last} show the polynomials fitted to the ratio of background-subtracted data over reconstruction-level prediction used to reweight the predictions for the unfolding bias estimation, as described in \mysec~\ref{sec:iteropt}.

\begin{figure}[h!]
\centering
\subfigure{\includegraphics[width=0.49\textwidth]{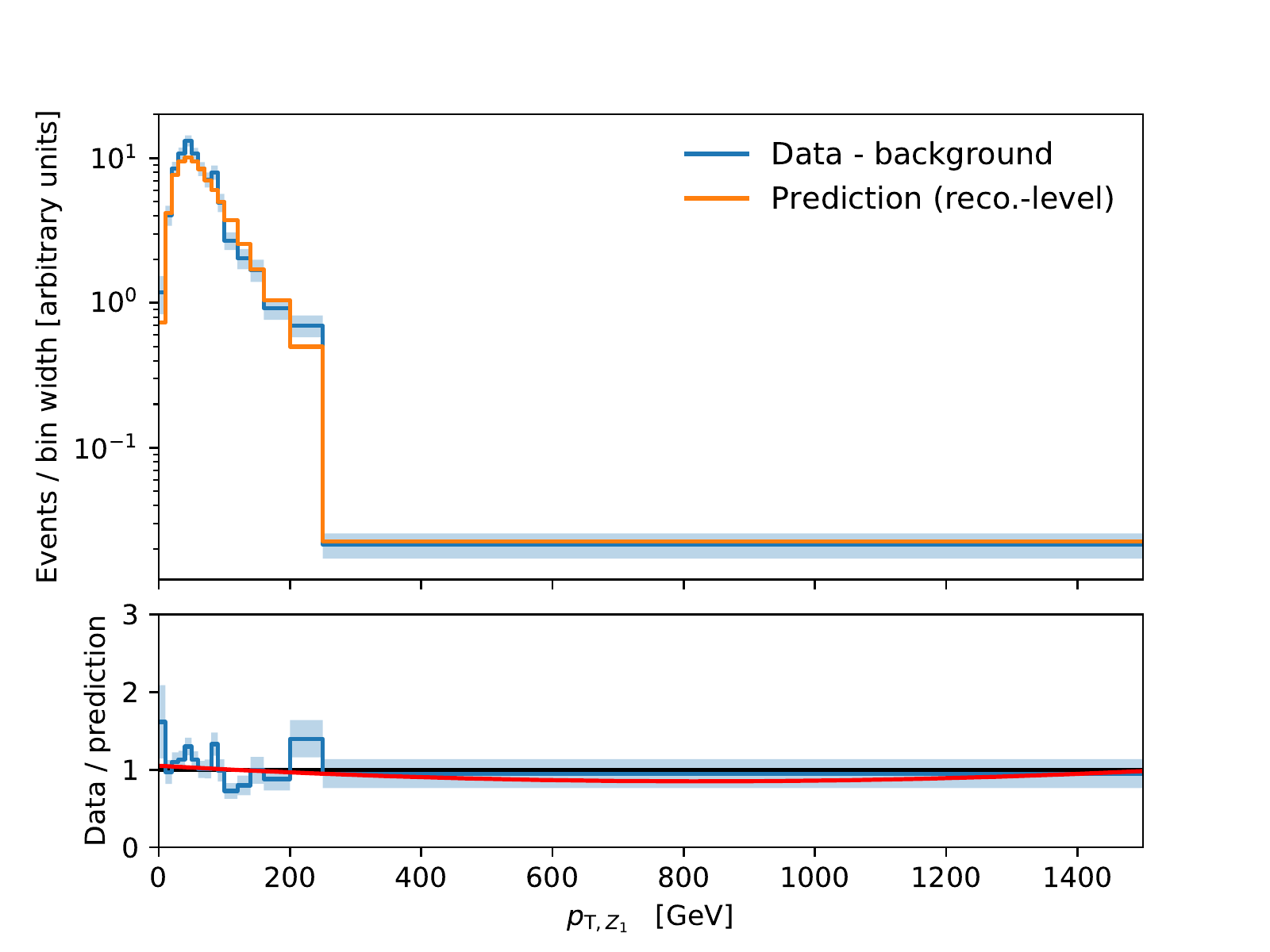}}
\subfigure{\includegraphics[width=0.49\textwidth]{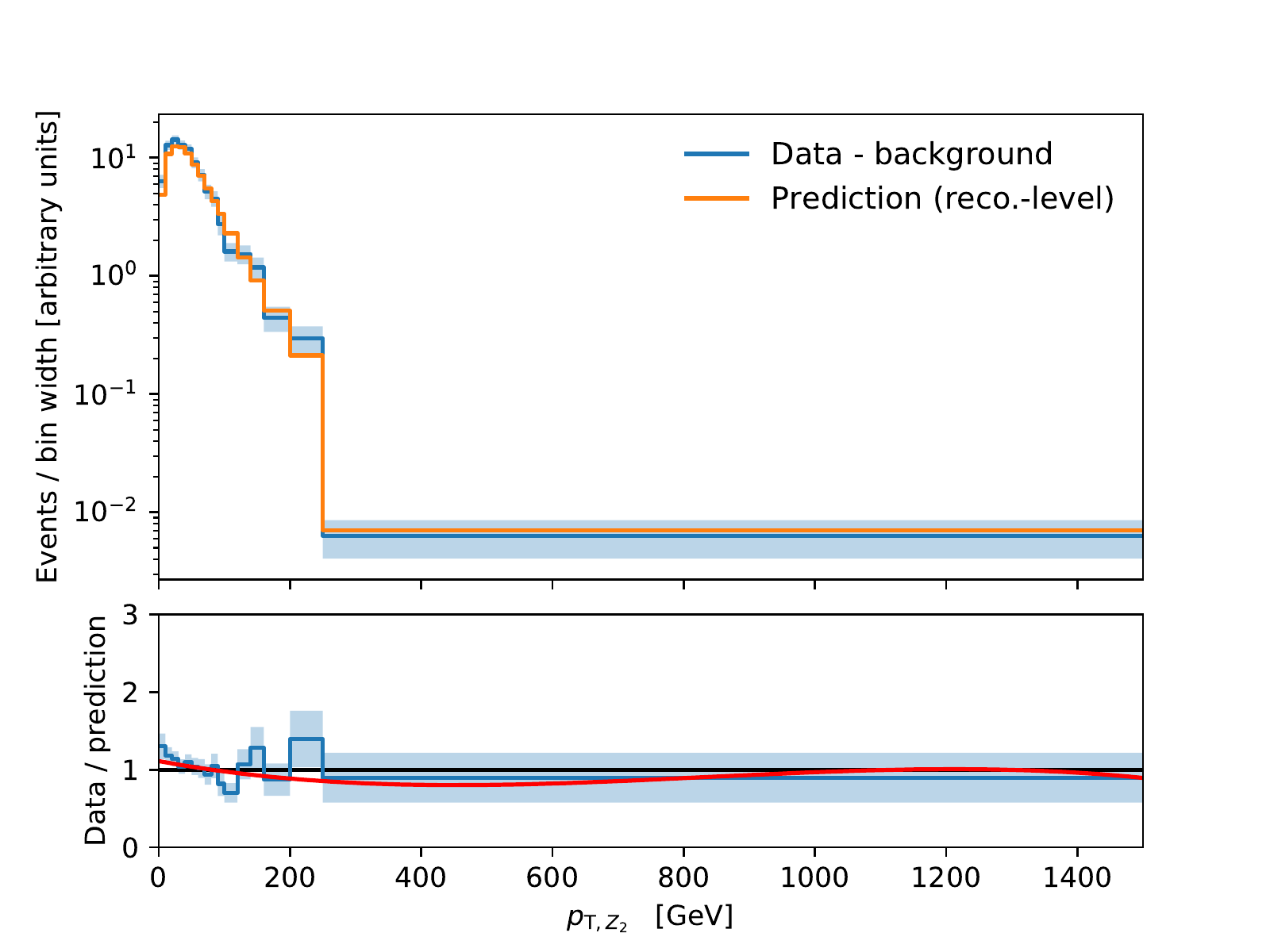}}
\subfigure{\includegraphics[width=0.49\textwidth]{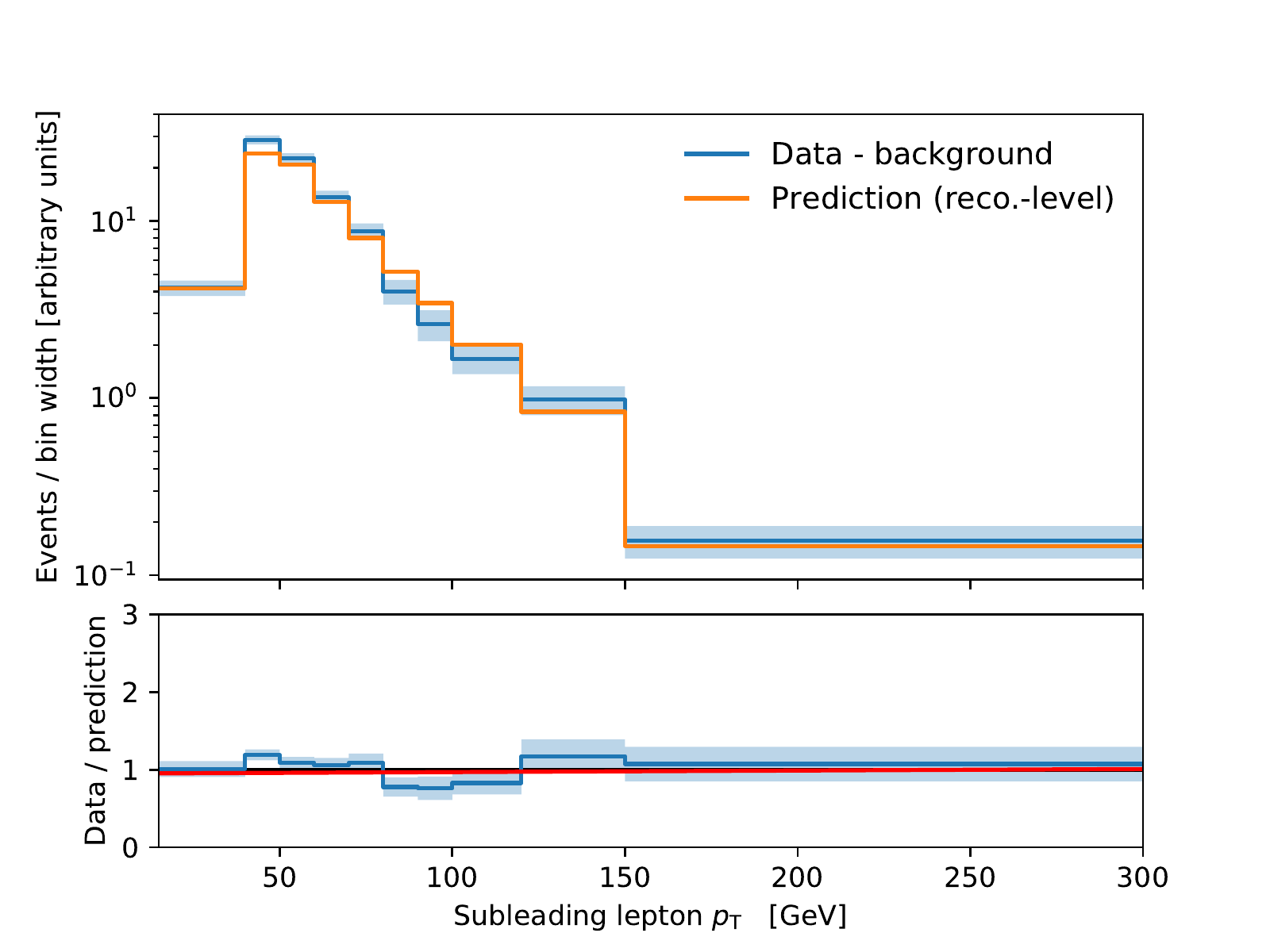}}
\subfigure{\includegraphics[width=0.49\textwidth]{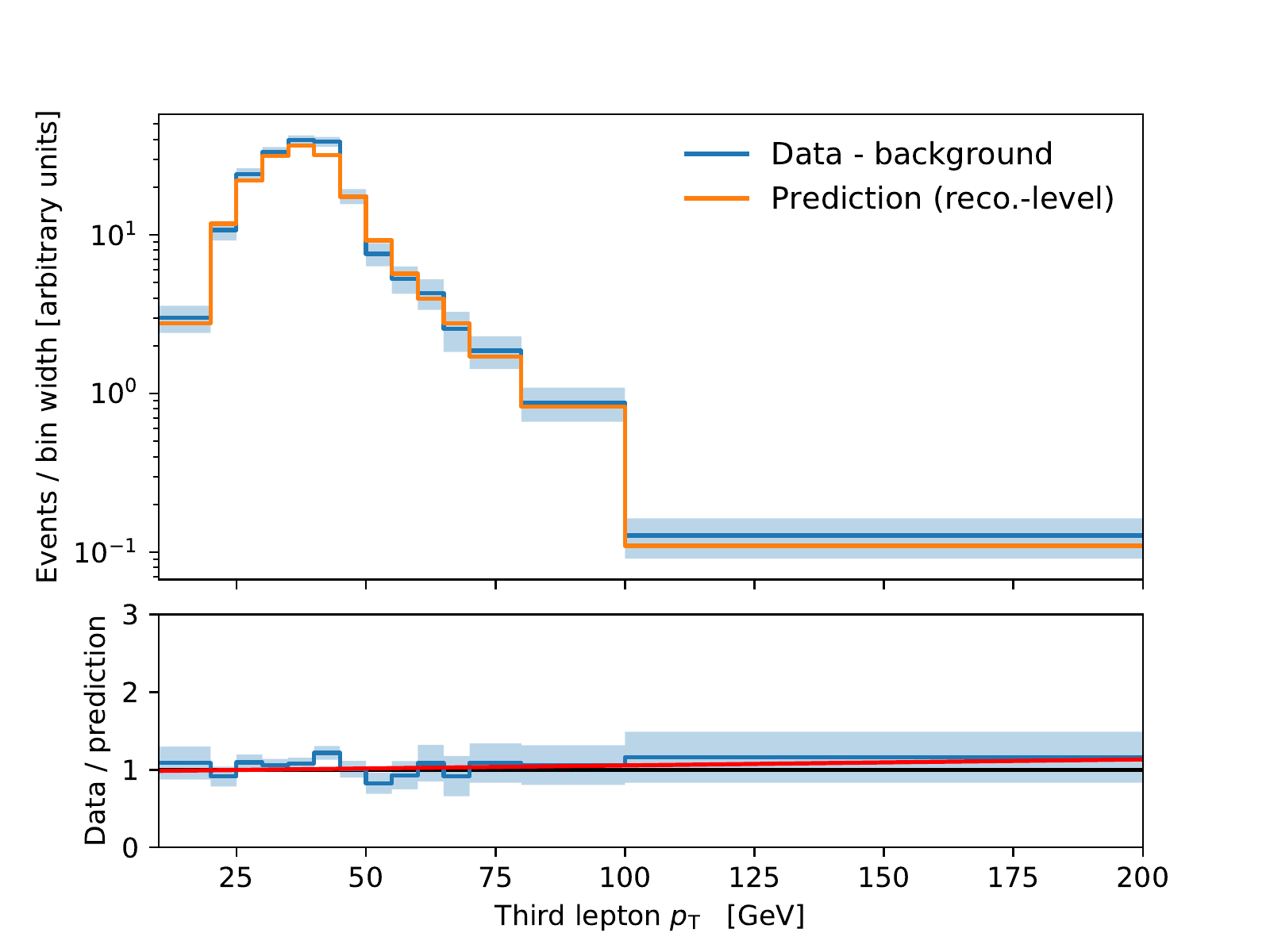}}
\subfigure{\includegraphics[width=0.49\textwidth]{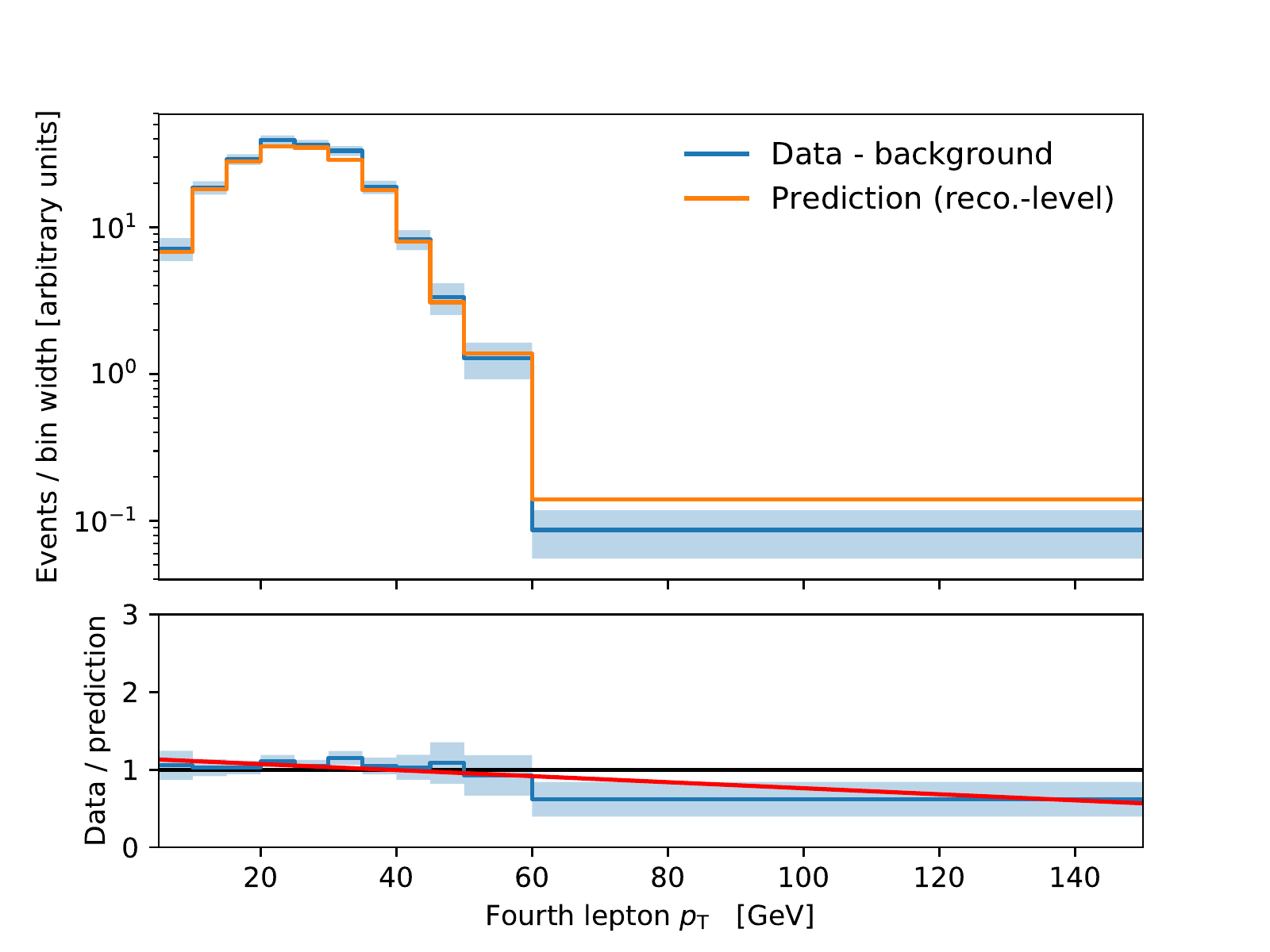}}
\subfigure{\includegraphics[width=0.49\textwidth]{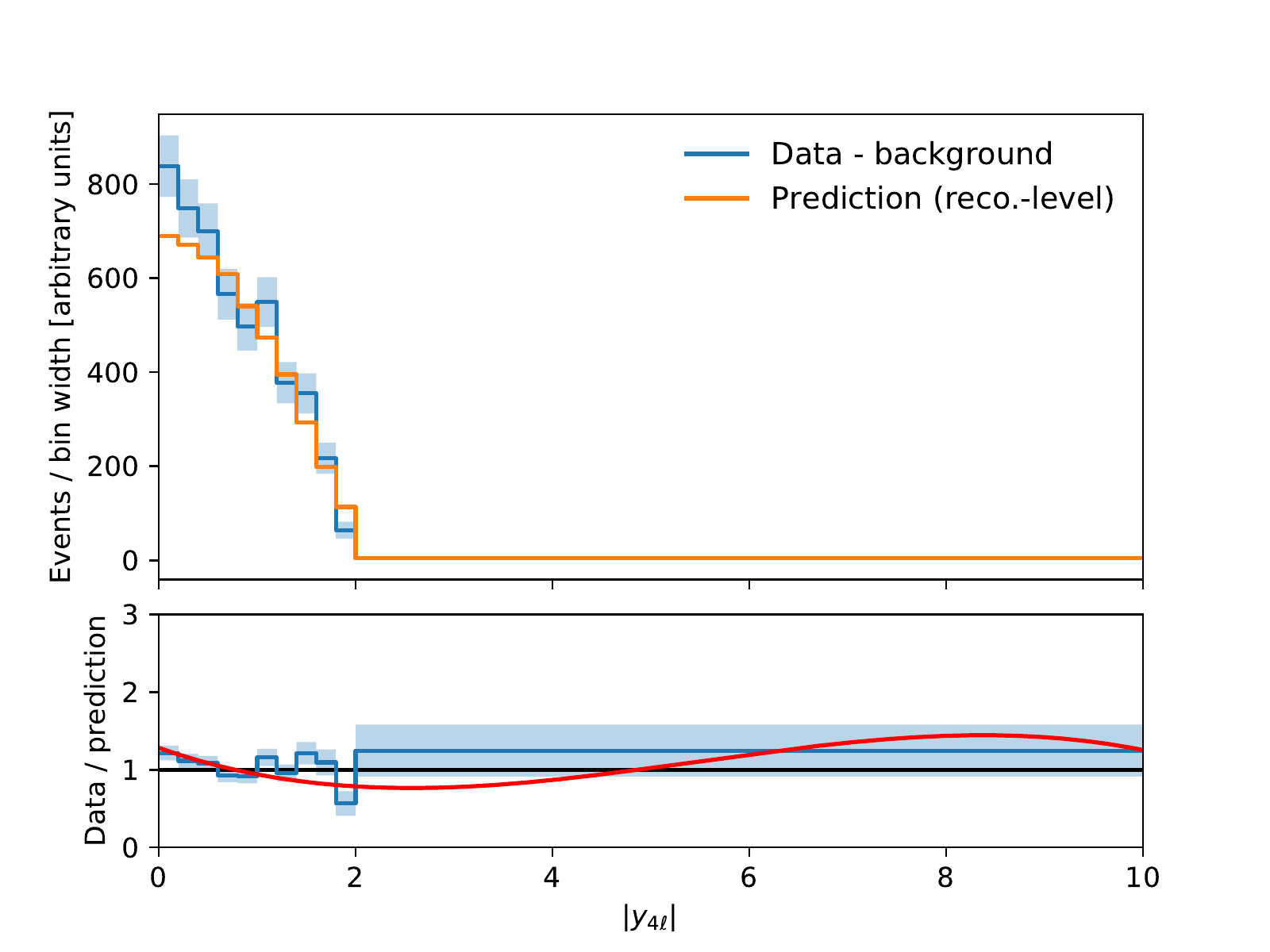}}
\caption{Background-subtracted data compared to the reconstruction-level prediction for various observables. The ratio is fitted with a polynomial, visualised as a red curve.}
\label{fig:polyfit_plots_first}
\end{figure}

\begin{figure}[h!]
\centering
\subfigure{\includegraphics[width=0.49\textwidth]{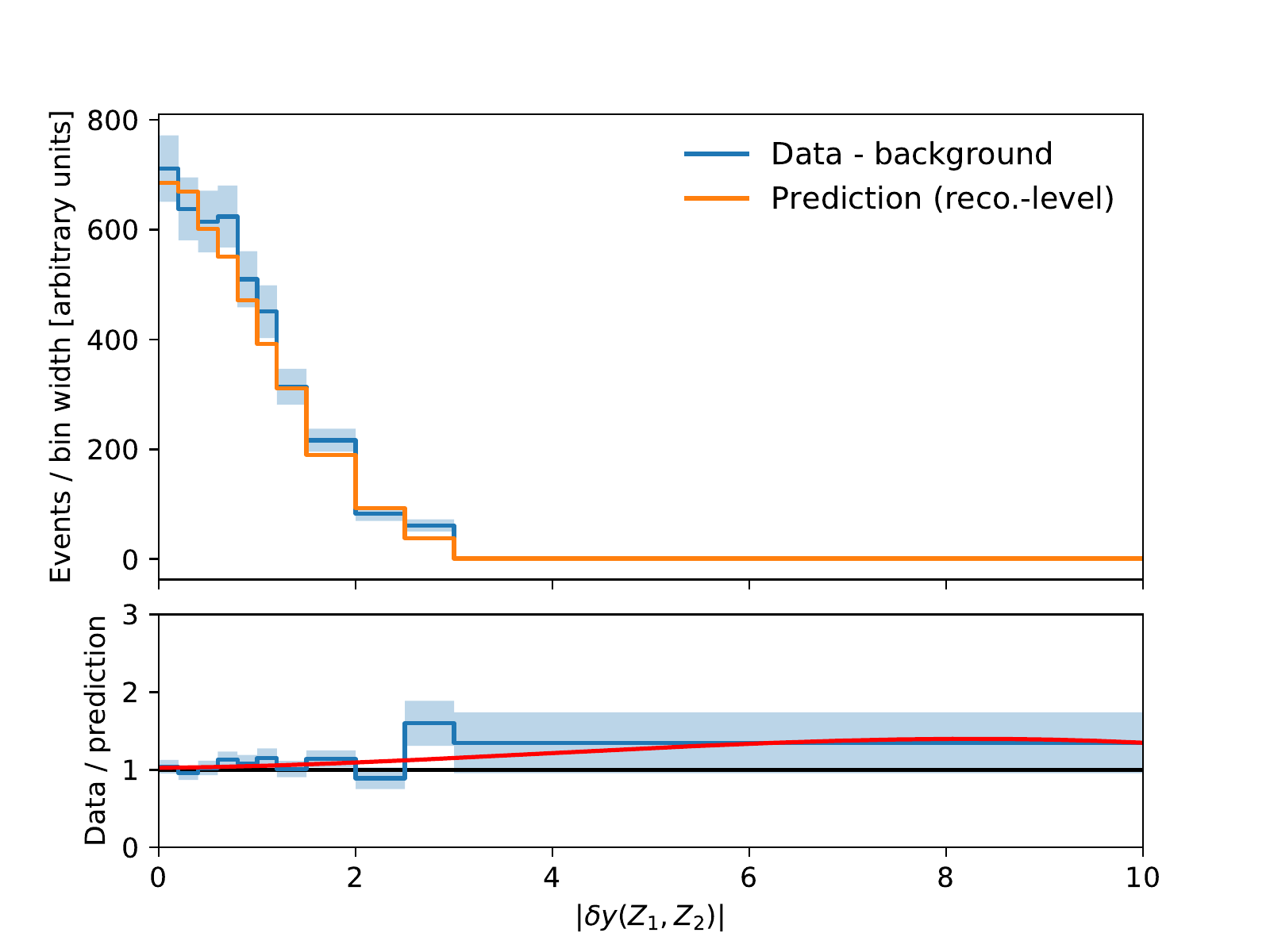}}
\subfigure{\includegraphics[width=0.49\textwidth]{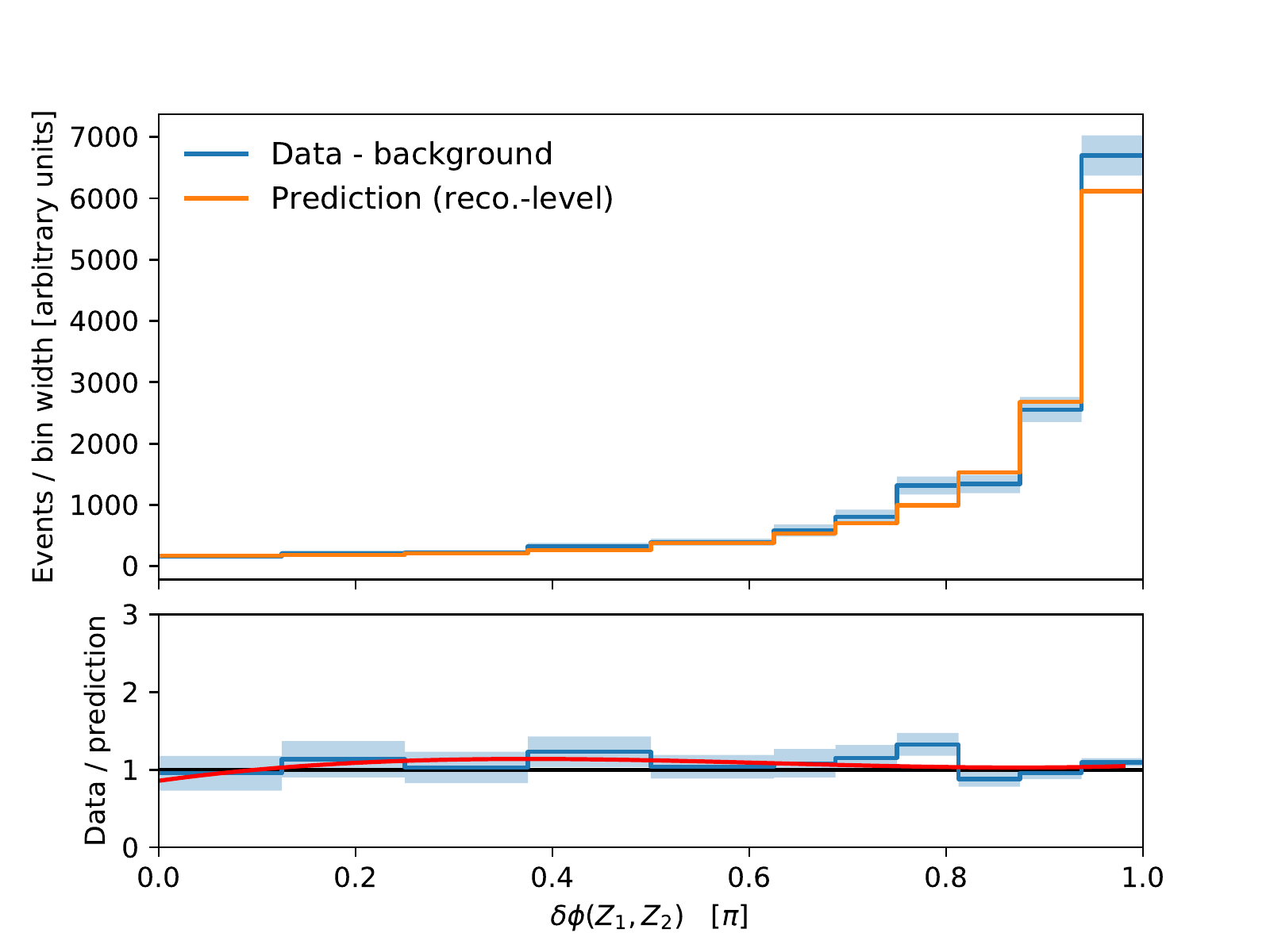}}
\subfigure{\includegraphics[width=0.49\textwidth]{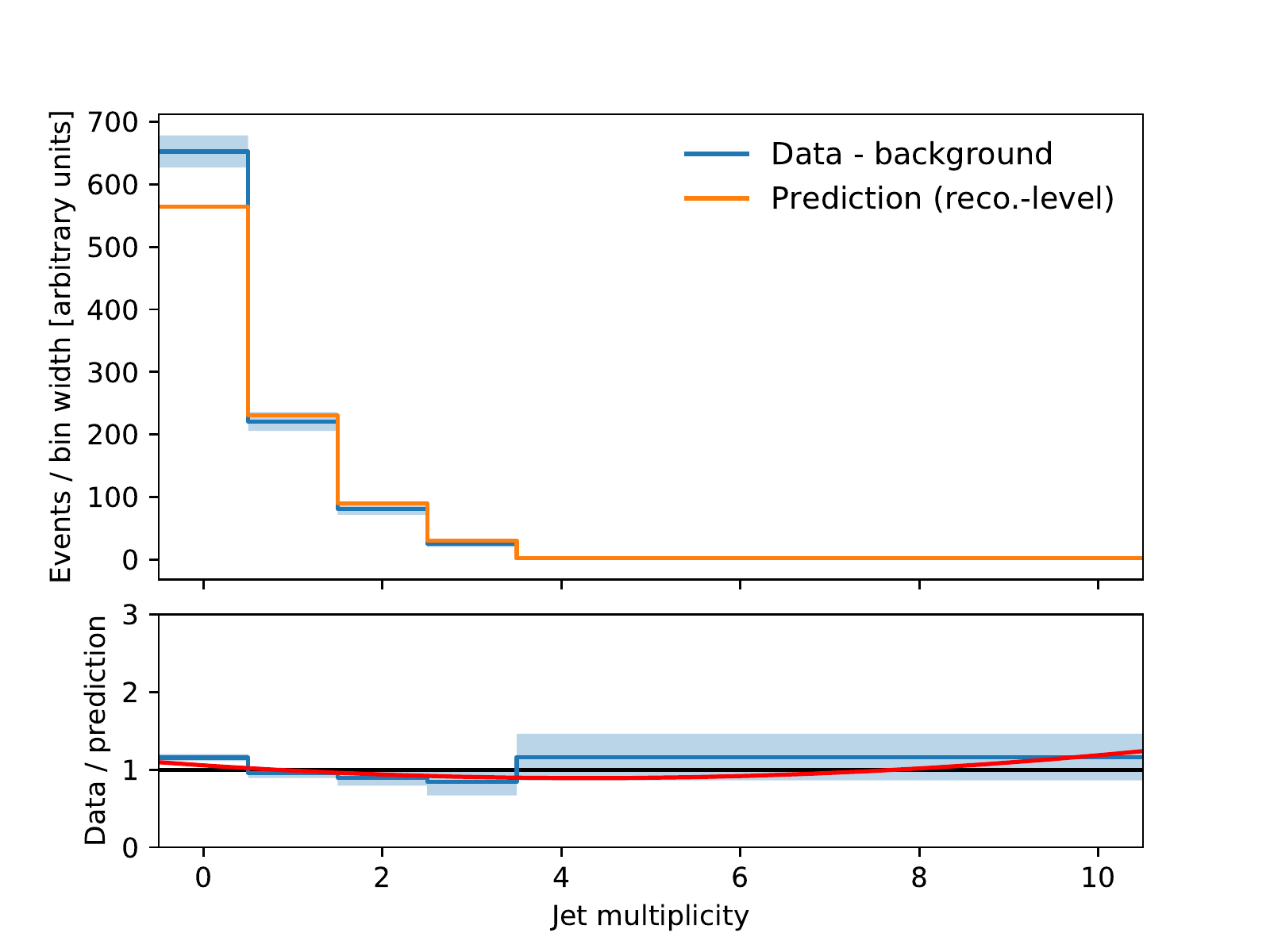}}
\subfigure{\includegraphics[width=0.49\textwidth]{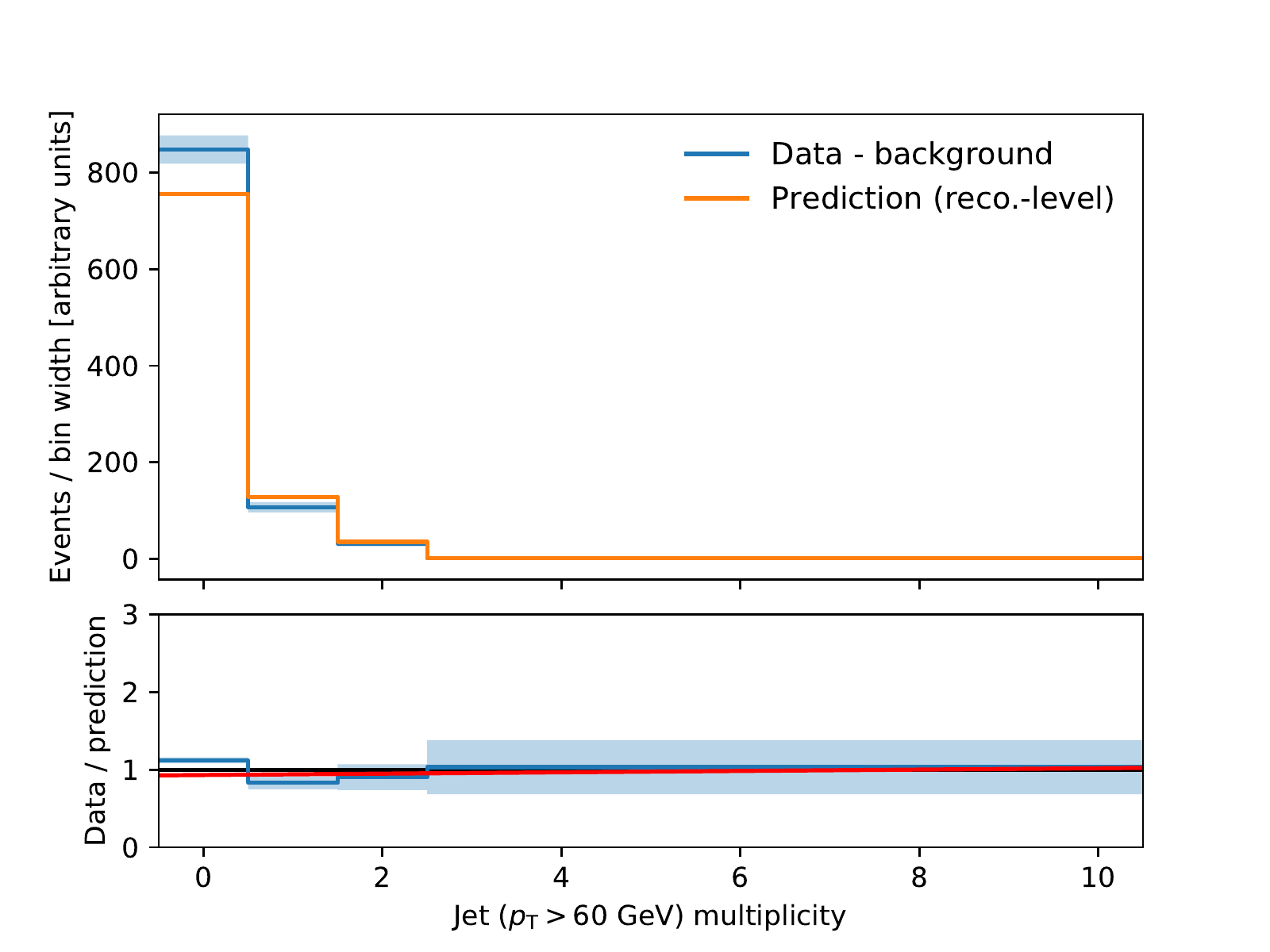}}
\subfigure{\includegraphics[width=0.49\textwidth]{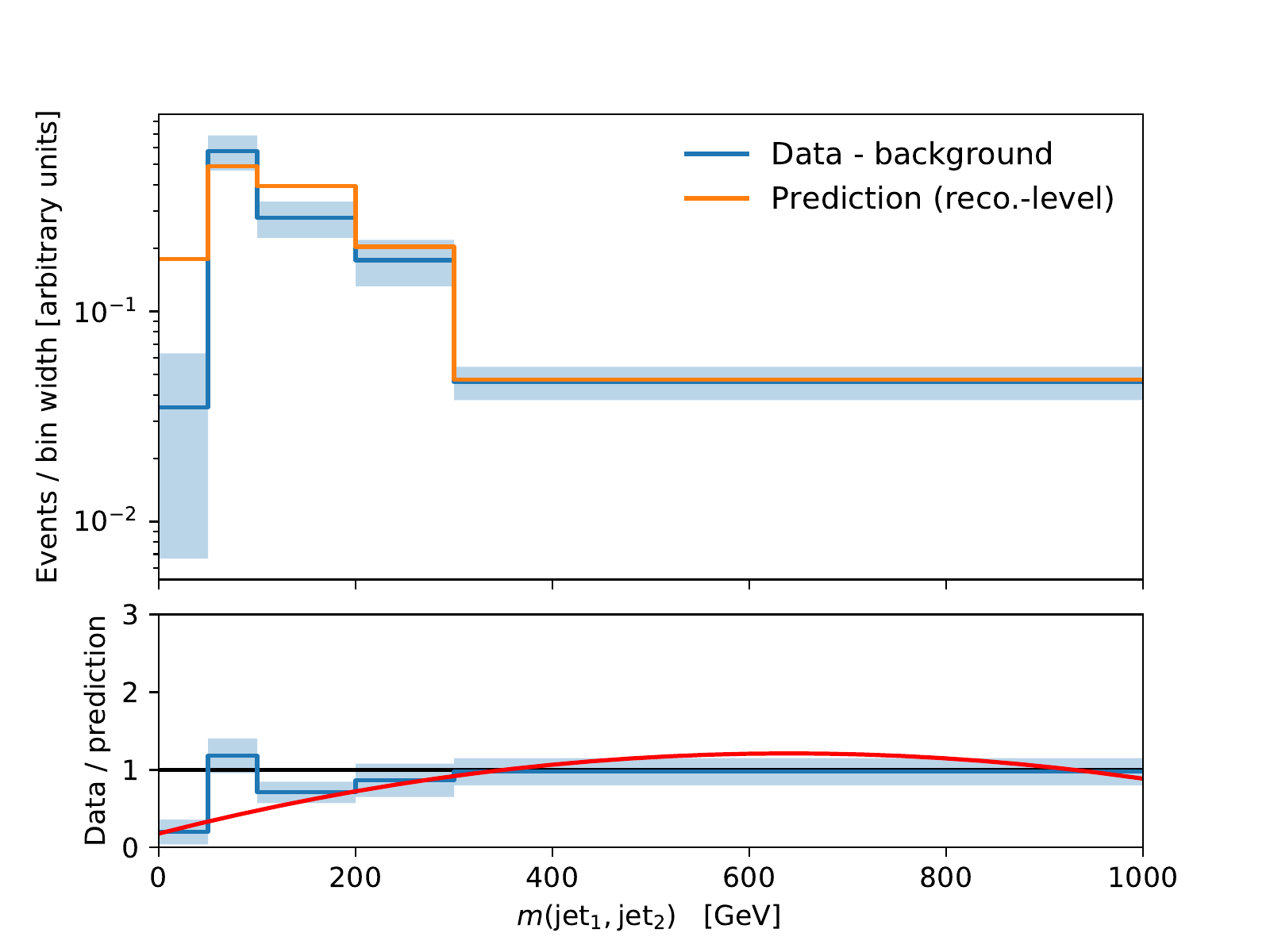}}
\subfigure{\includegraphics[width=0.49\textwidth]{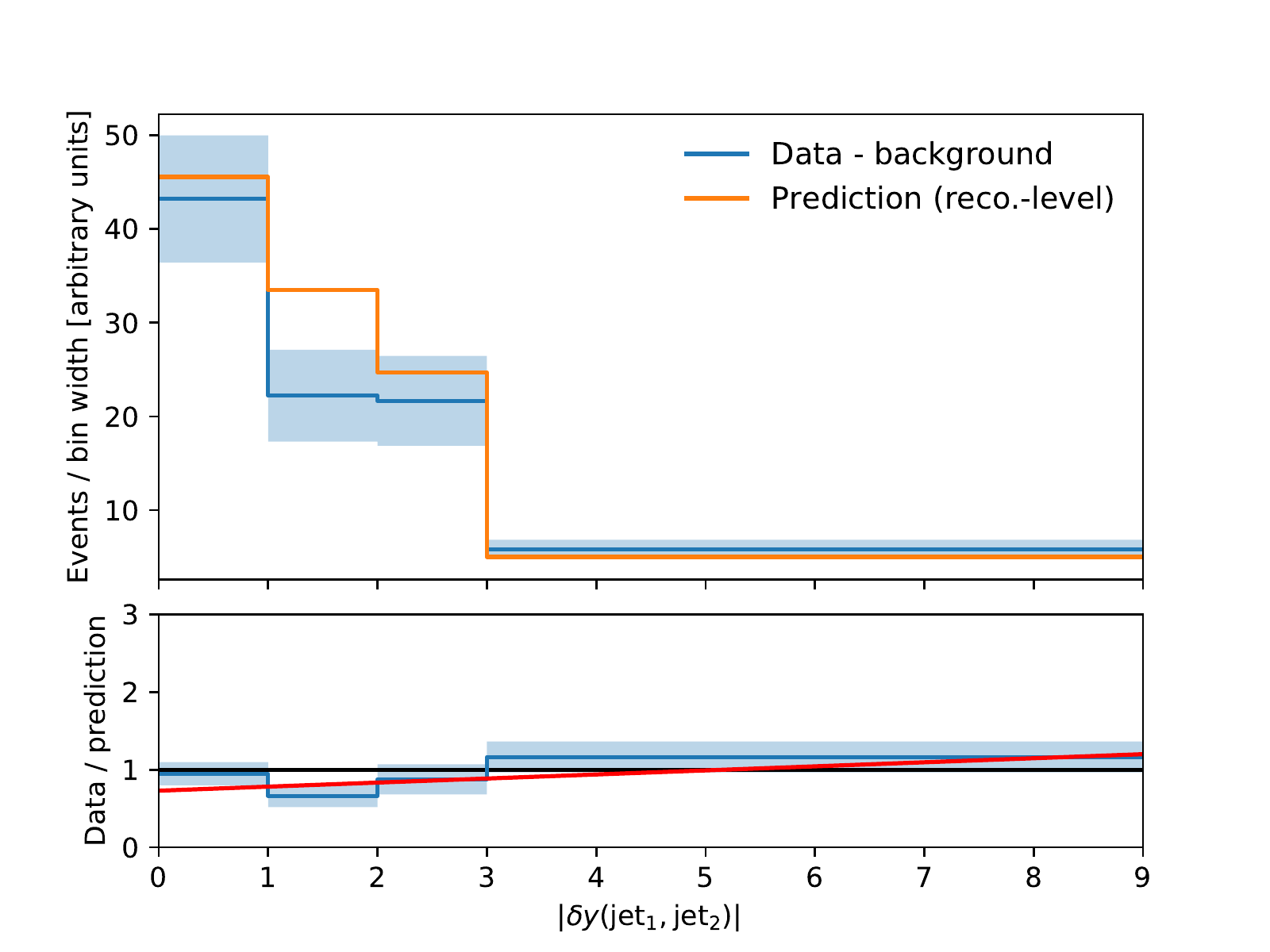}}
\caption{Background-subtracted data compared to the reconstruction-level prediction for various observables. The ratio is fitted with a polynomial, visualised as a red curve.}
\end{figure}

\begin{figure}[h!]
\centering
\subfigure{\includegraphics[width=0.49\textwidth]{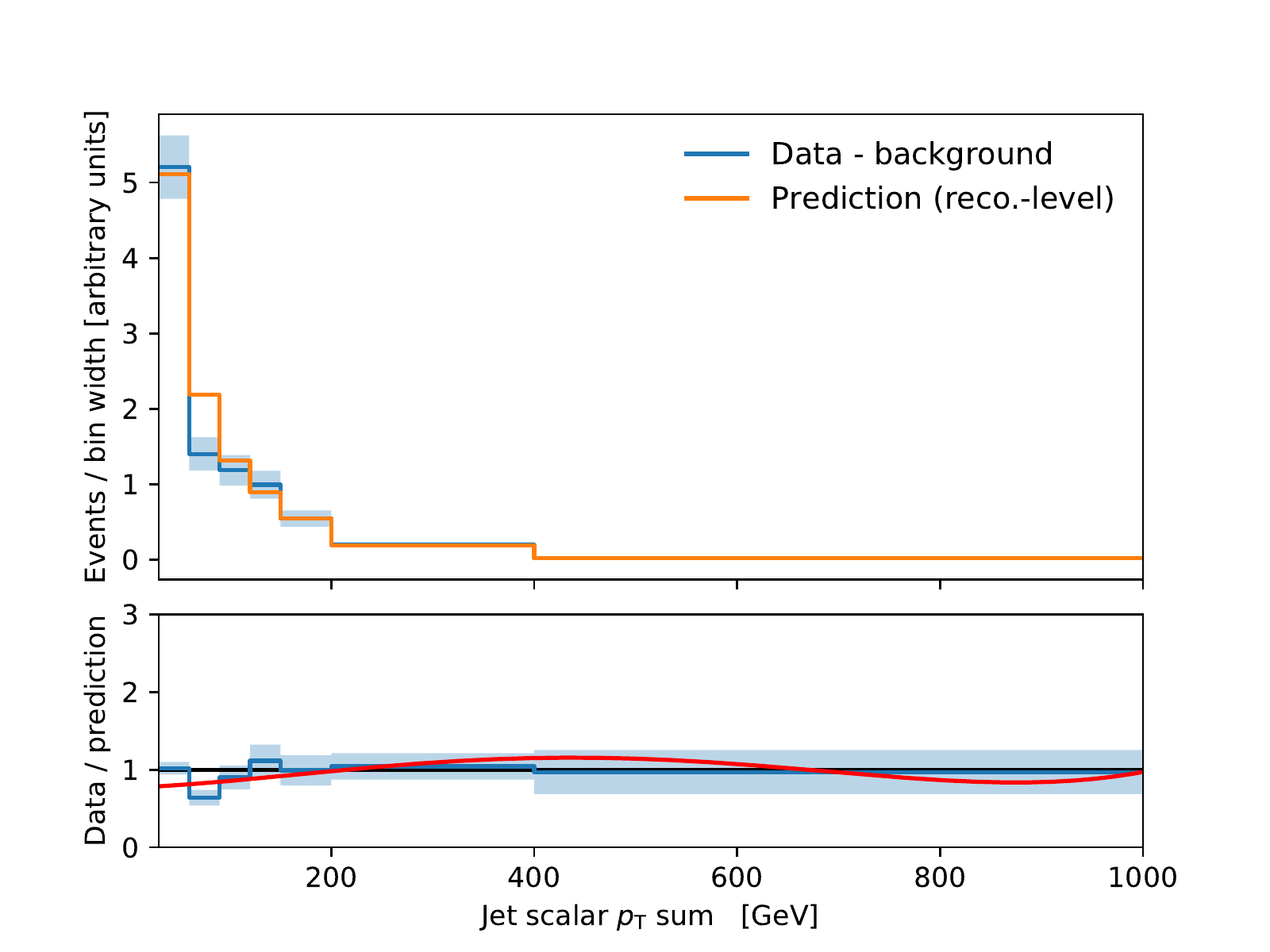}}
\subfigure{\includegraphics[width=0.49\textwidth]{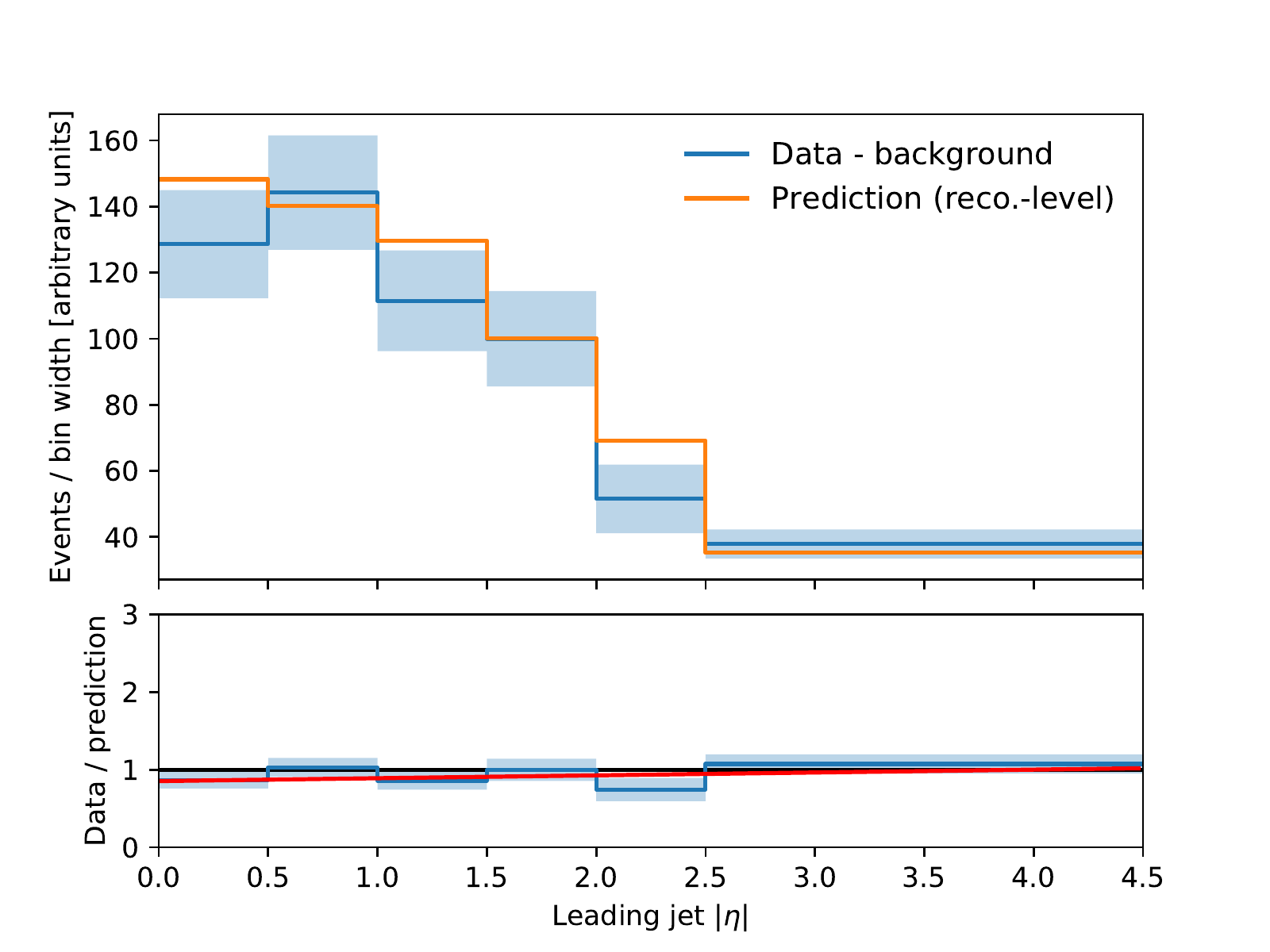}}
\subfigure{\includegraphics[width=0.49\textwidth]{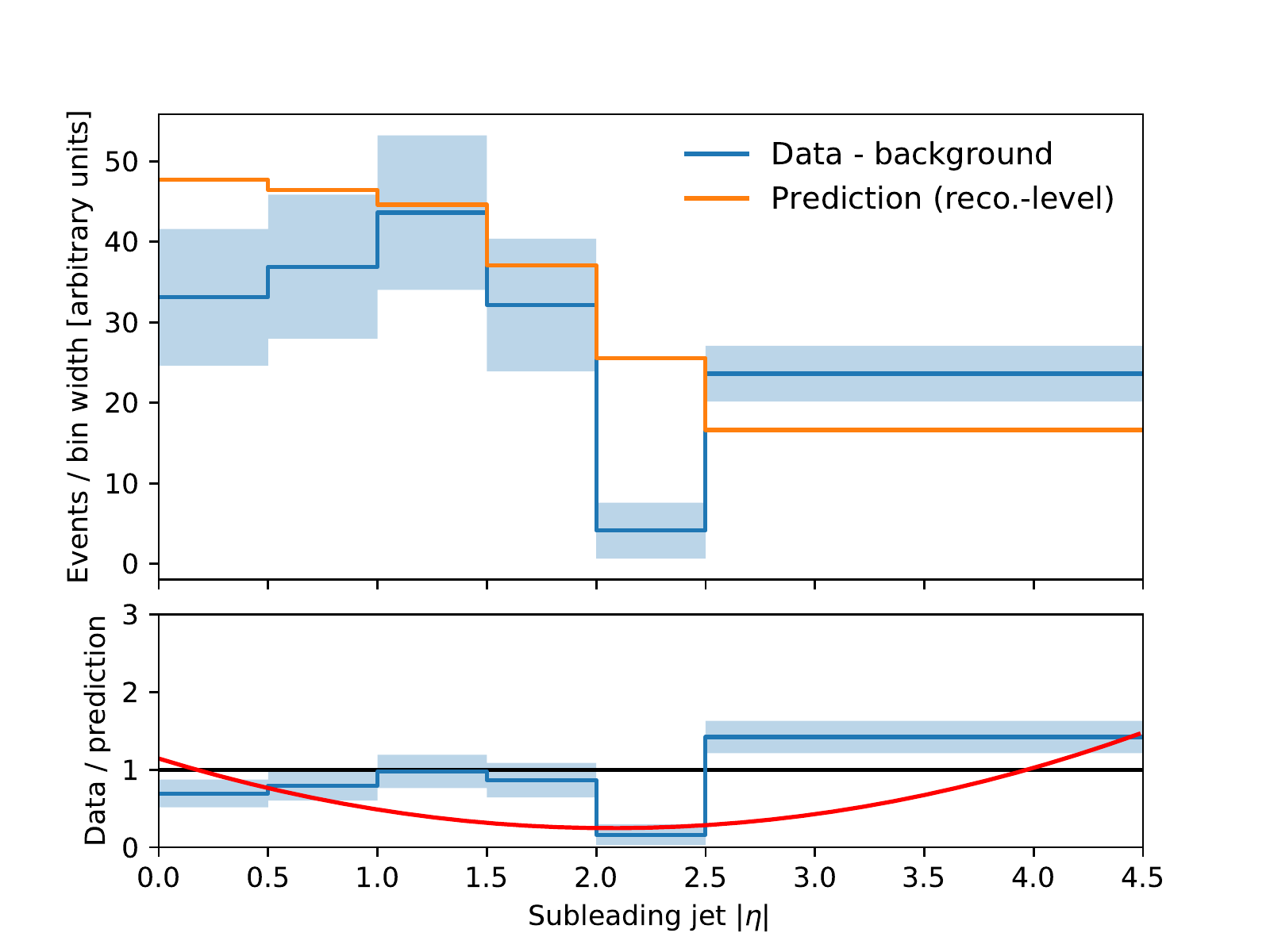}}
\subfigure{\includegraphics[width=0.49\textwidth]{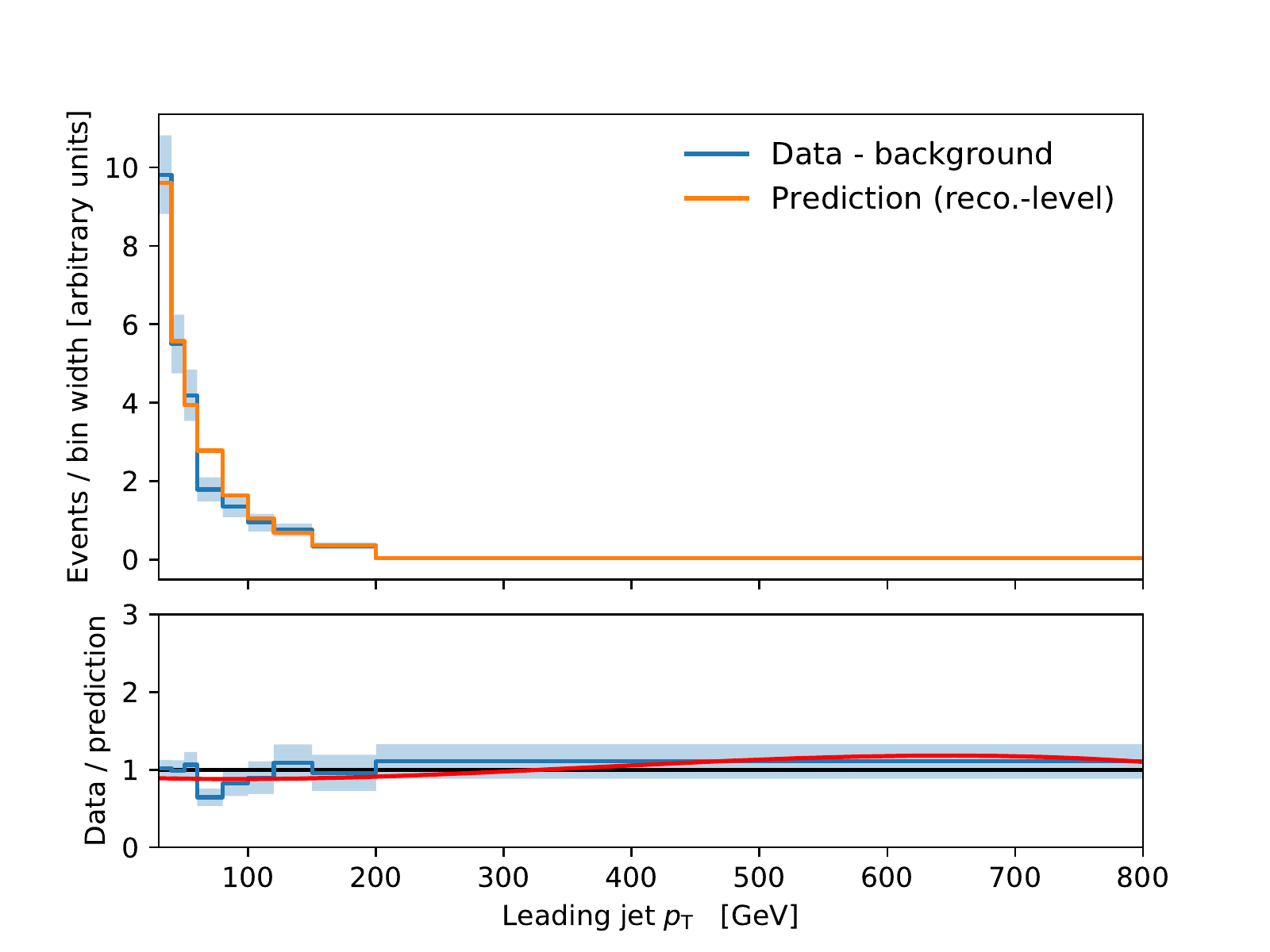}}
\subfigure{\includegraphics[width=0.49\textwidth]{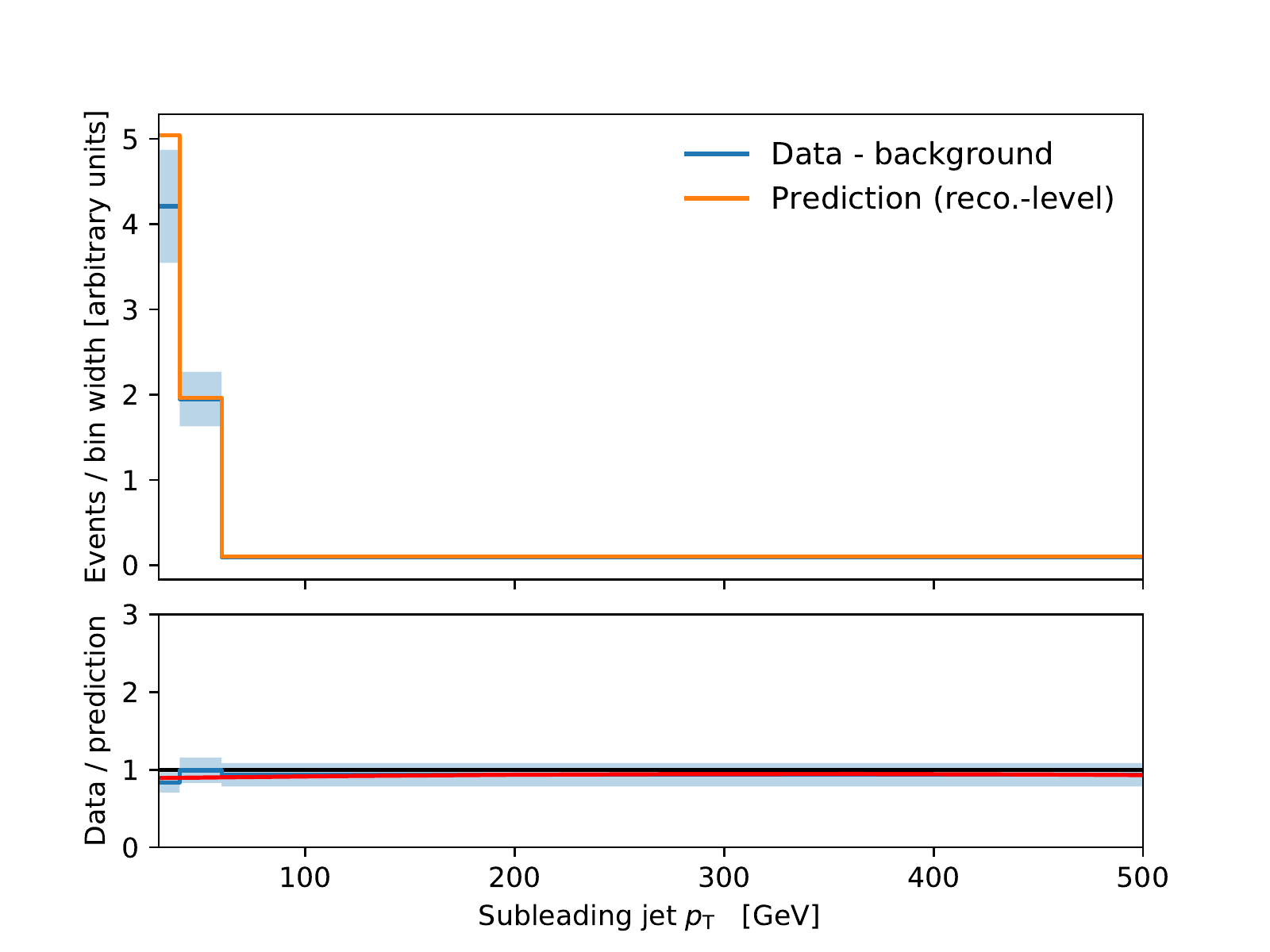}}
\caption{Background-subtracted data compared to the reconstruction-level prediction for various observables. The ratio is fitted with a polynomial, visualised as a red curve.}
\label{fig:polyfit_plots_last}
\end{figure}

\clearpage

\subsection{Statistical uncertainty versus unfolding bias}
\label{sec:zz_aux_statbias}

\myfigs~\ref{fig:statbias_plots_first}--\ref{fig:statbias_plots_last} show the statistical uncertainty and the unfolding method uncertainty for various numbers of iterations, as a function of different observables. The estimation of the unfolding method uncertainty is described in \mysec~\ref{sec:iteropt}.

\begin{figure}[h!]
\centering
\subfigure{\includegraphics[width=0.49\textwidth]{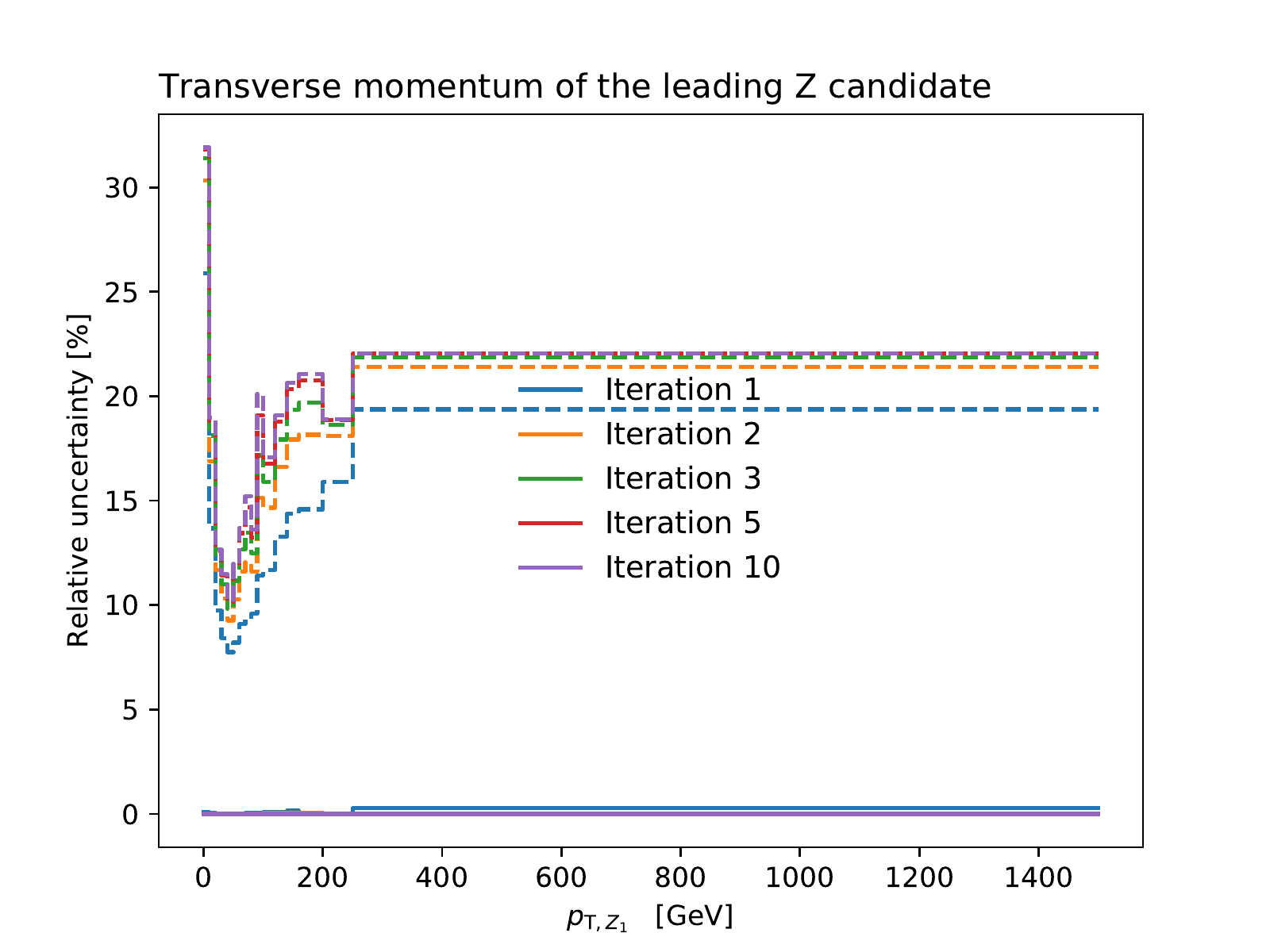}}
\subfigure{\includegraphics[width=0.49\textwidth]{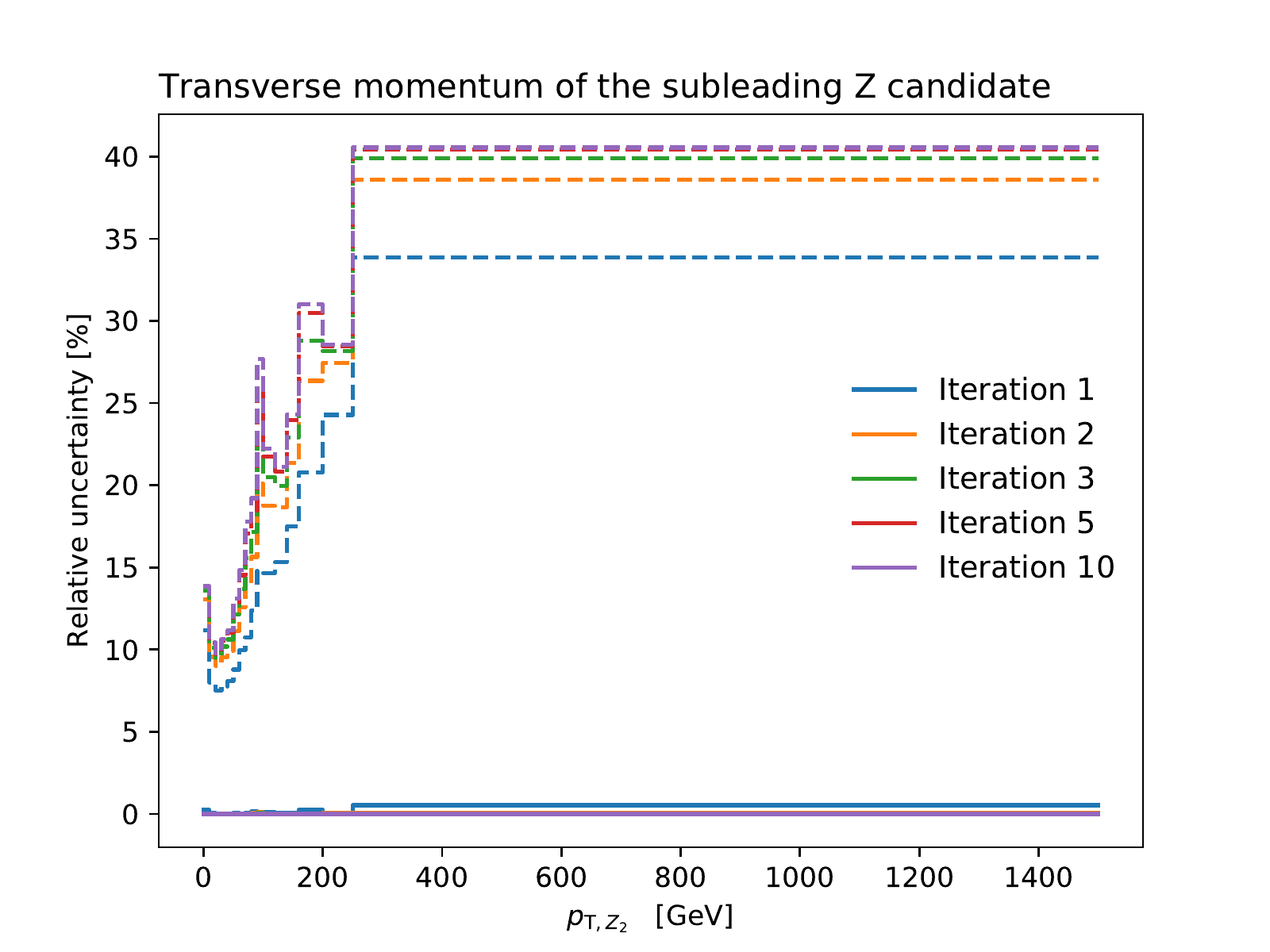}}
\subfigure{\includegraphics[width=0.49\textwidth]{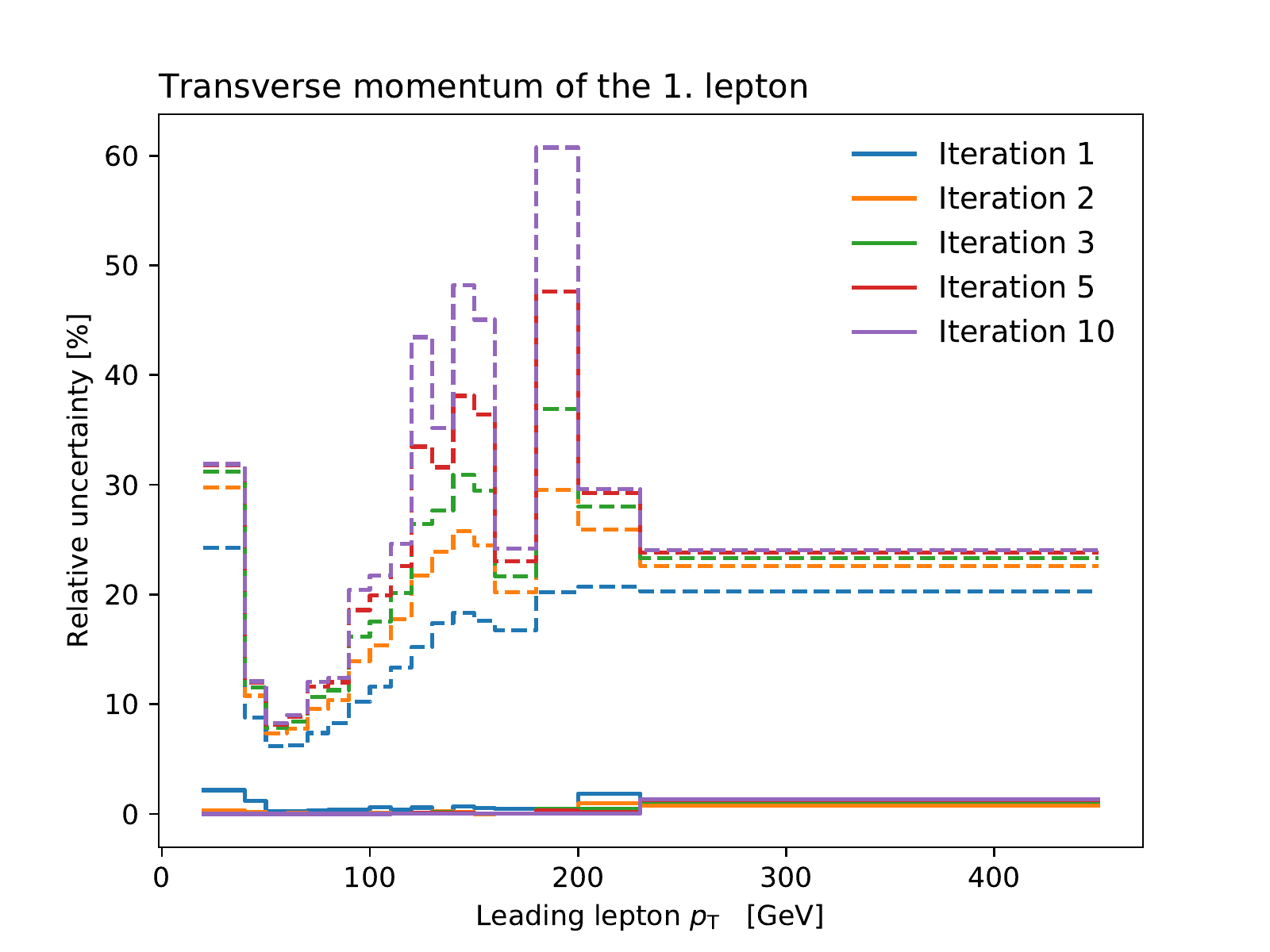}}
\subfigure{\includegraphics[width=0.49\textwidth]{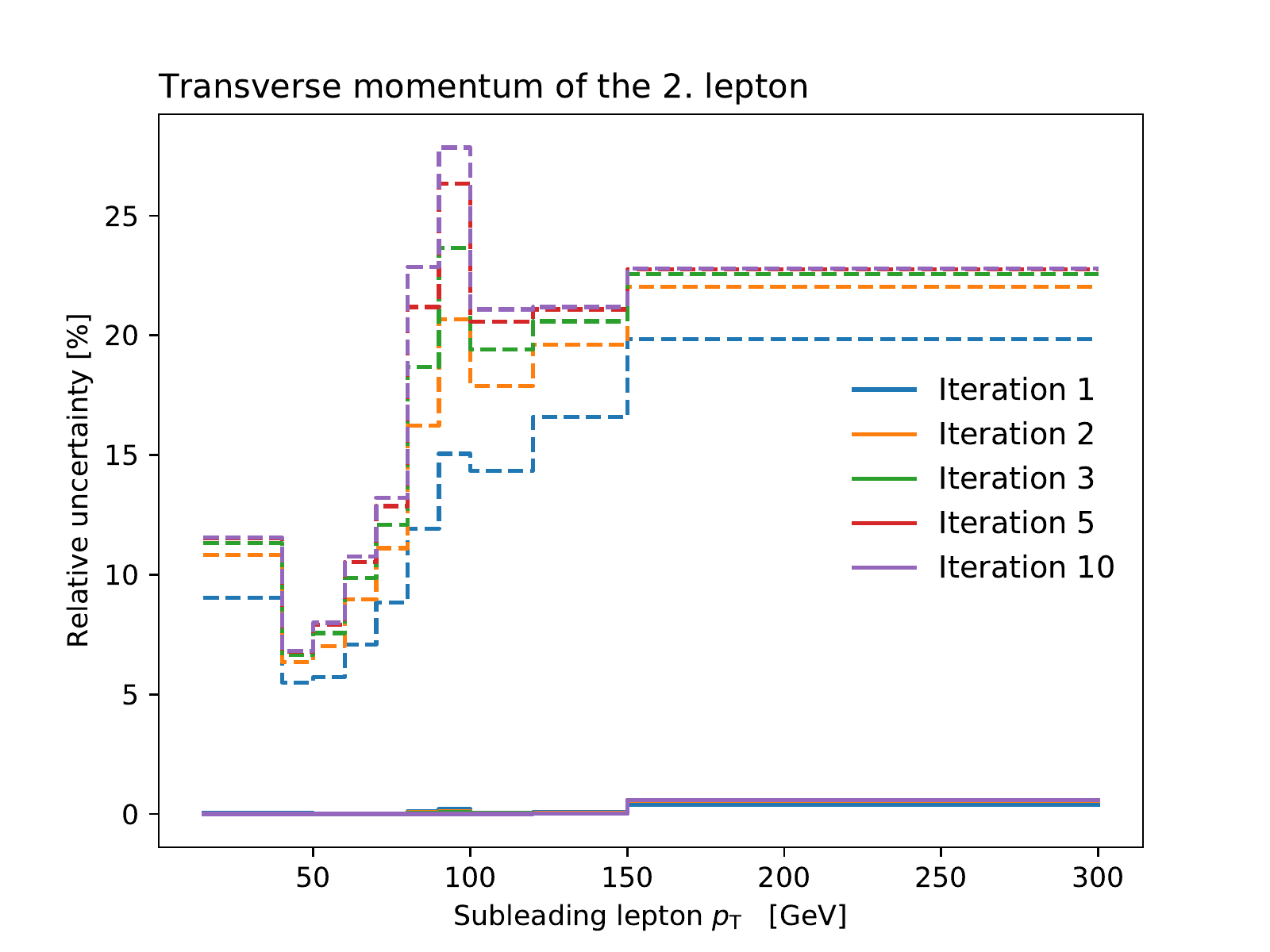}}
\subfigure{\includegraphics[width=0.49\textwidth]{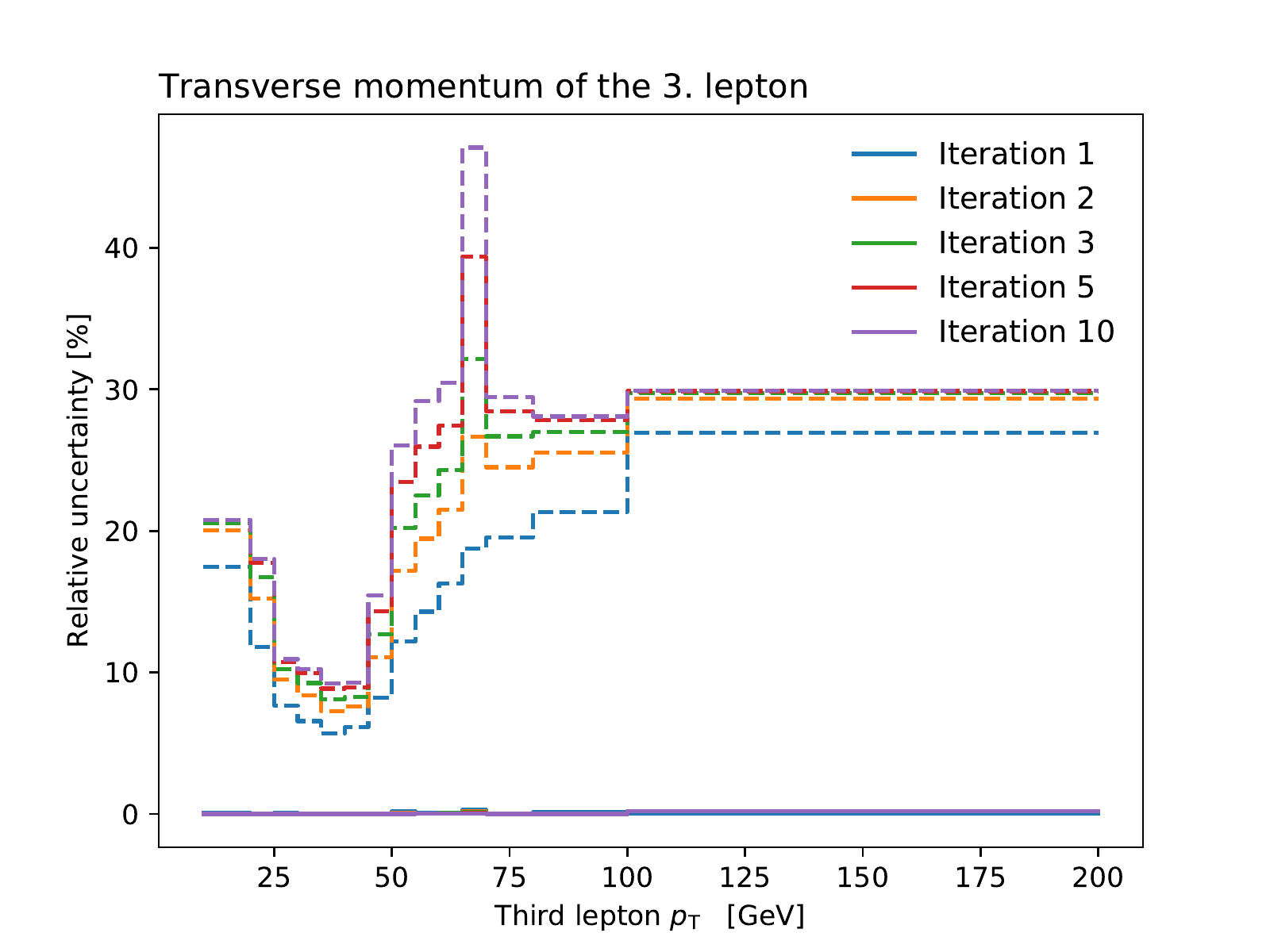}}
\subfigure{\includegraphics[width=0.49\textwidth]{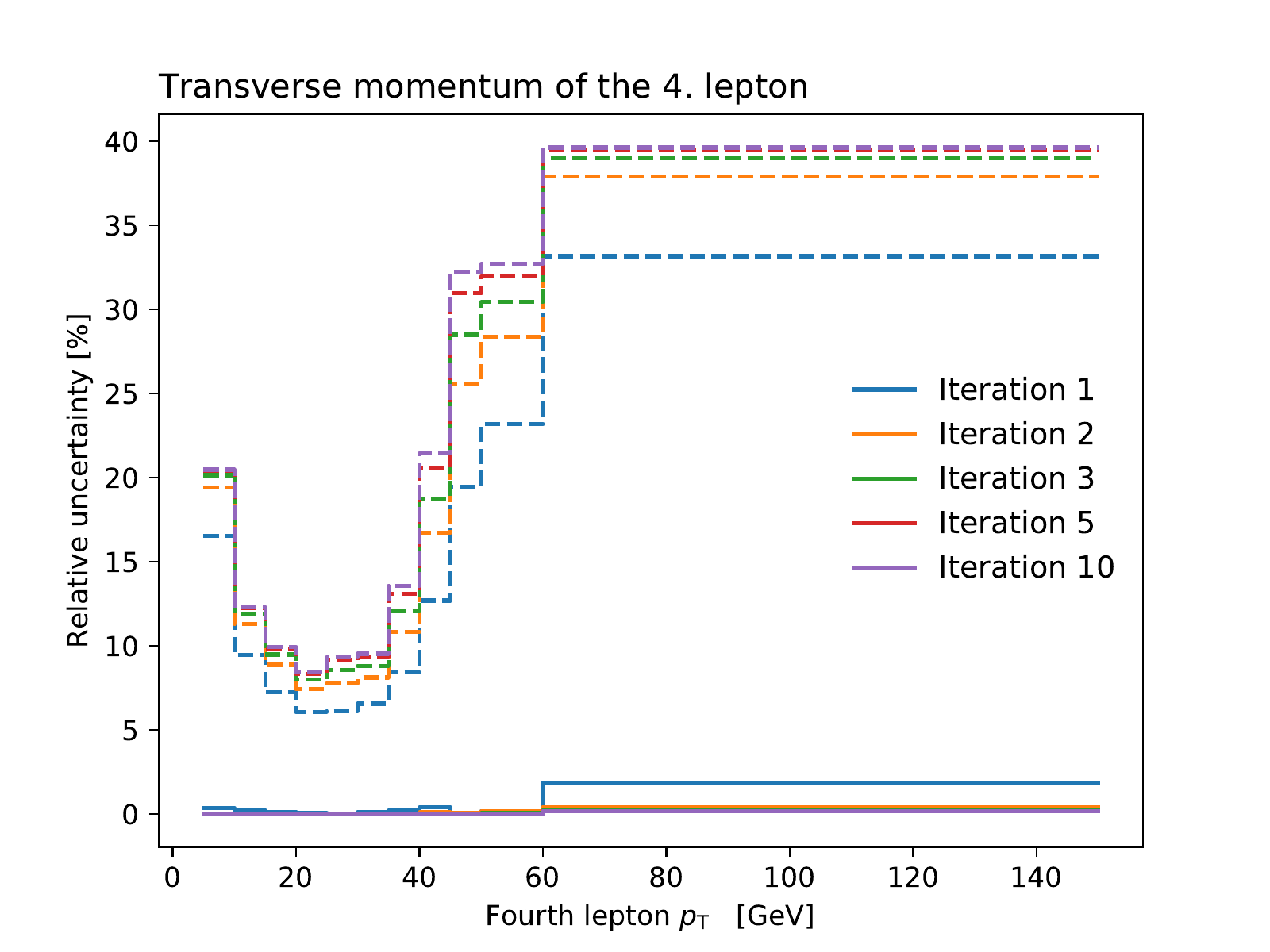}}
\caption{Statistical uncertainty (dashed) and unfolding method uncertainty (solid) for various numbers of iterations for various observables.}
\label{fig:statbias_plots_first}
\end{figure}

\begin{figure}[h!]
\centering
\subfigure{\includegraphics[width=0.49\textwidth]{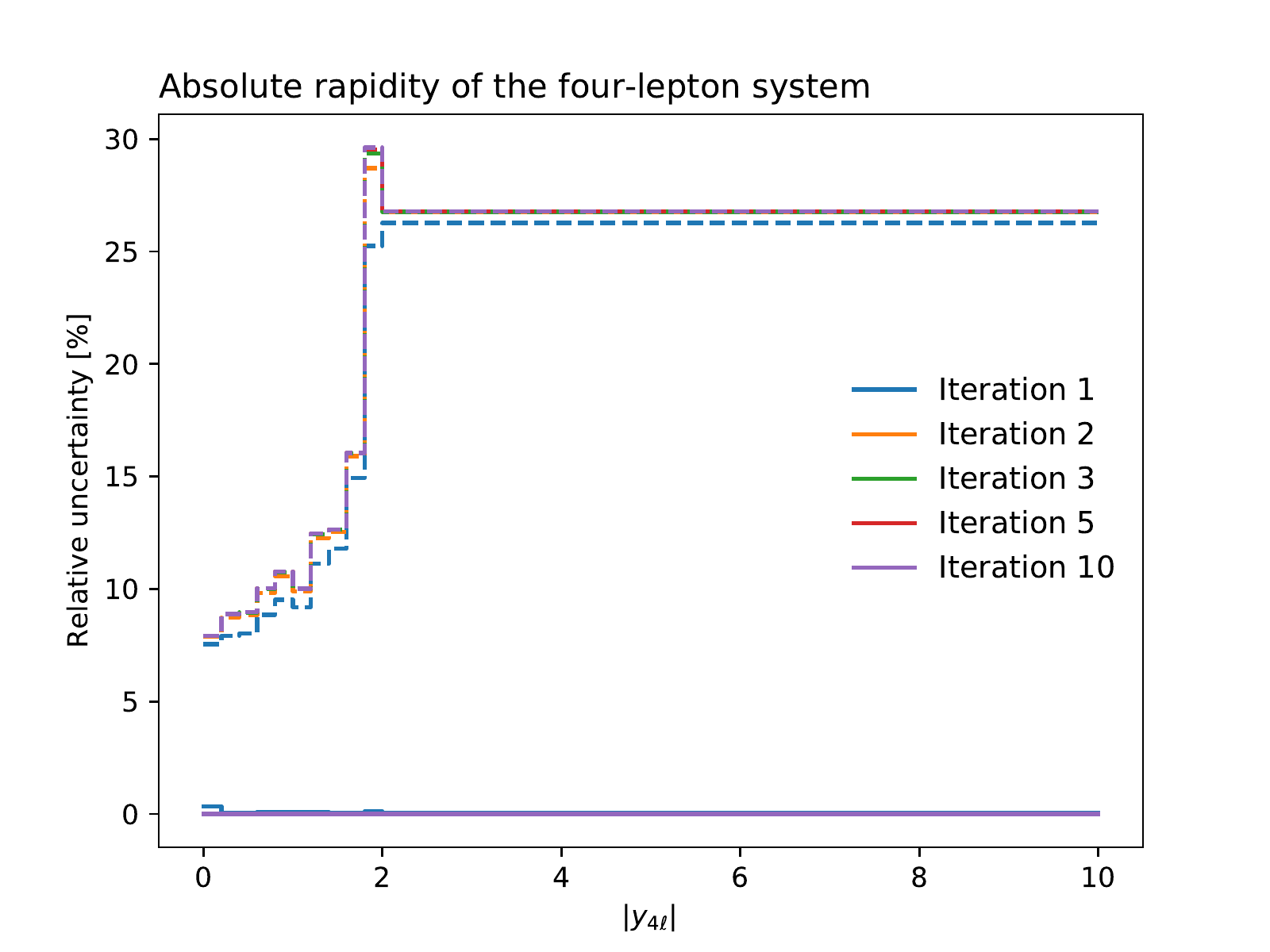}}
\subfigure{\includegraphics[width=0.49\textwidth]{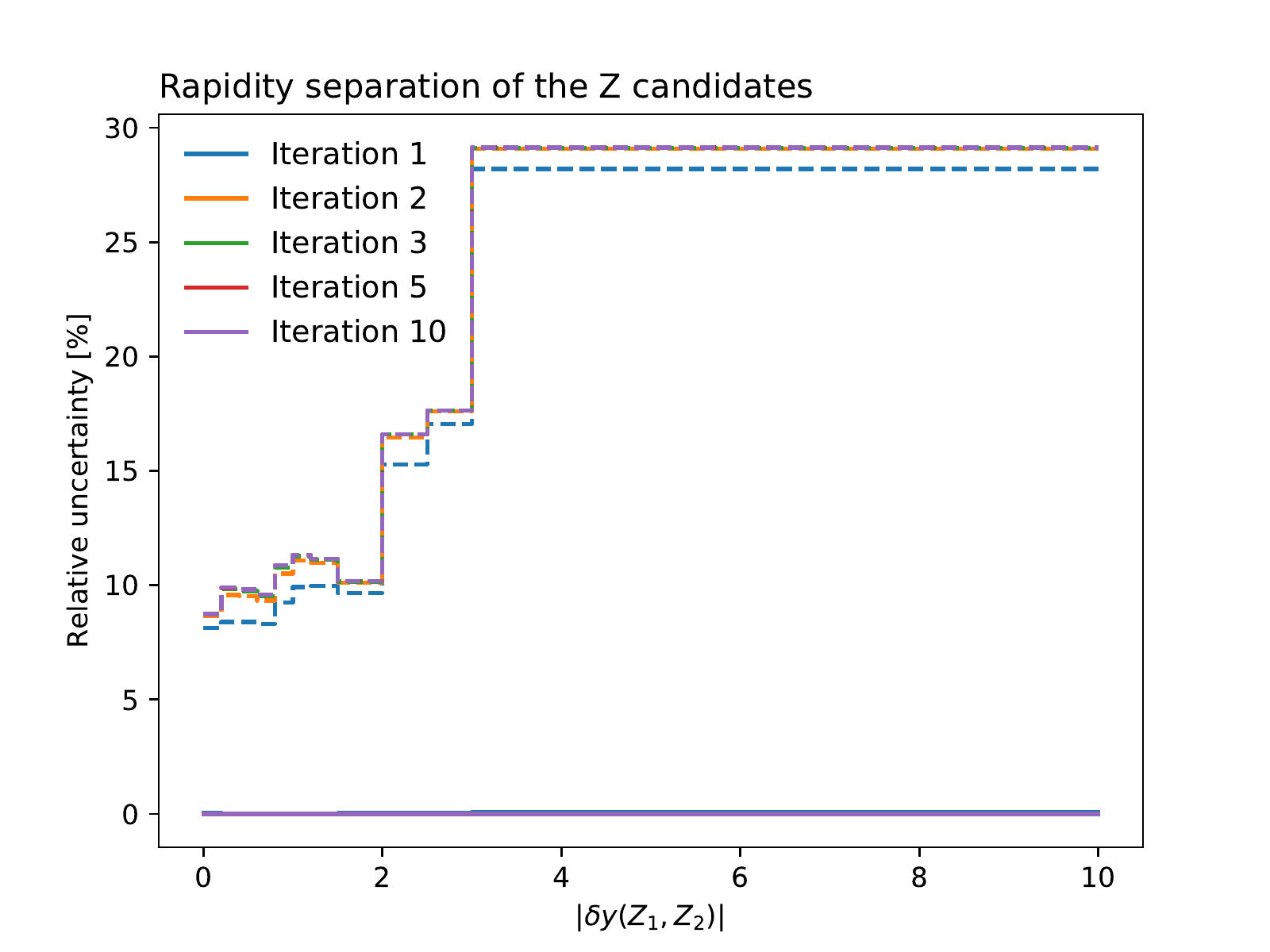}}
\subfigure{\includegraphics[width=0.49\textwidth]{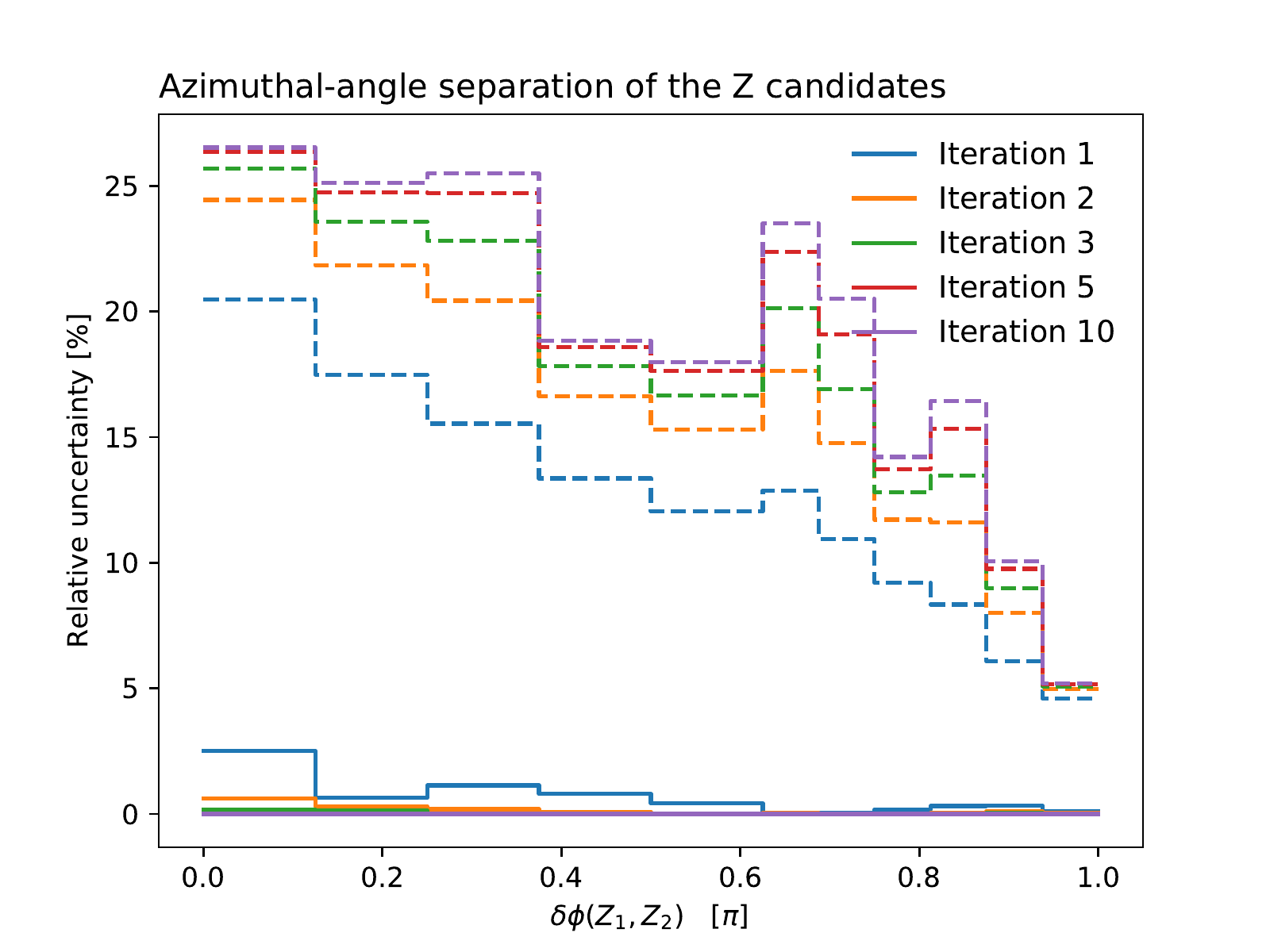}}
\subfigure{\includegraphics[width=0.49\textwidth]{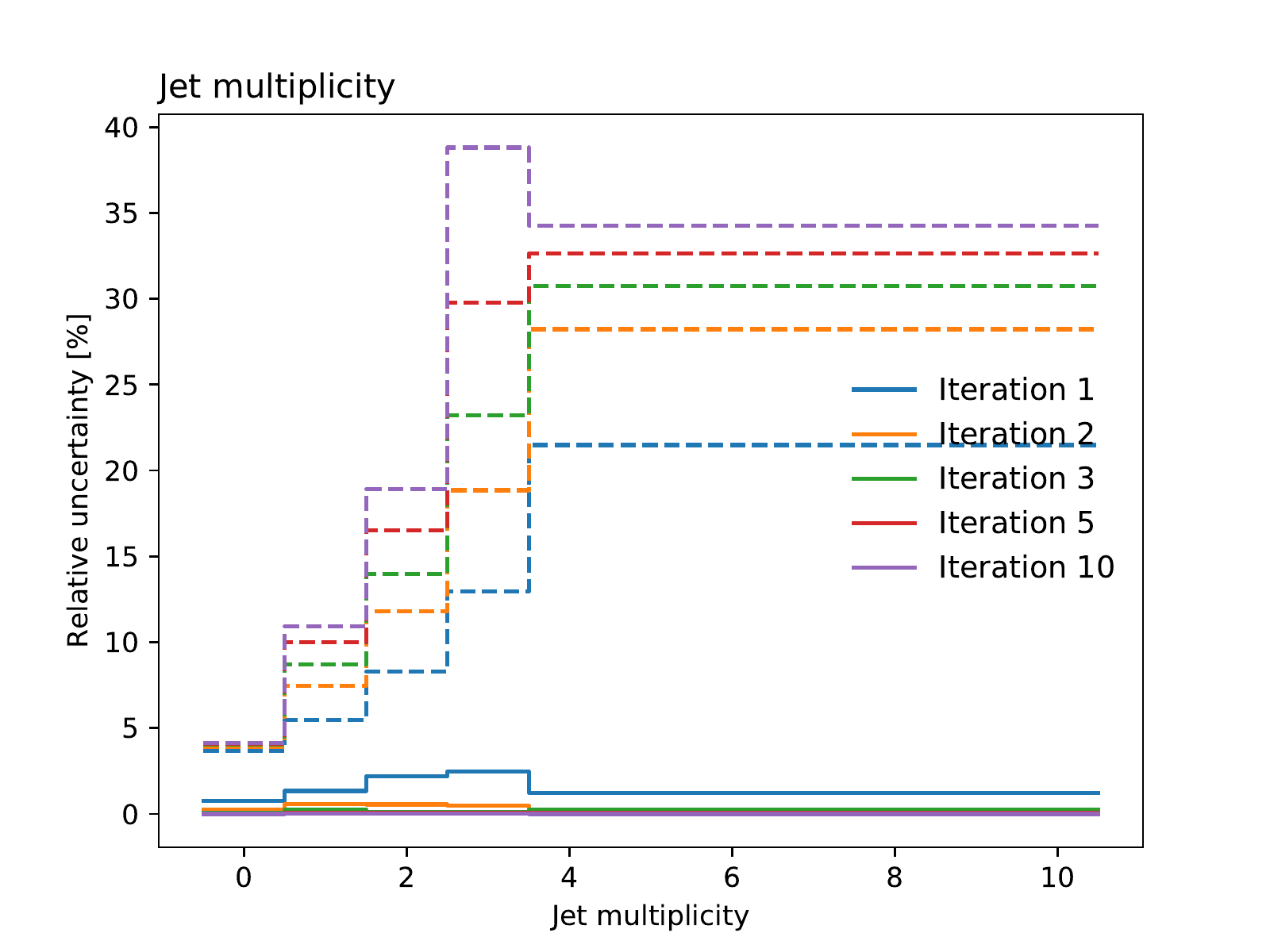}}
\subfigure{\includegraphics[width=0.49\textwidth]{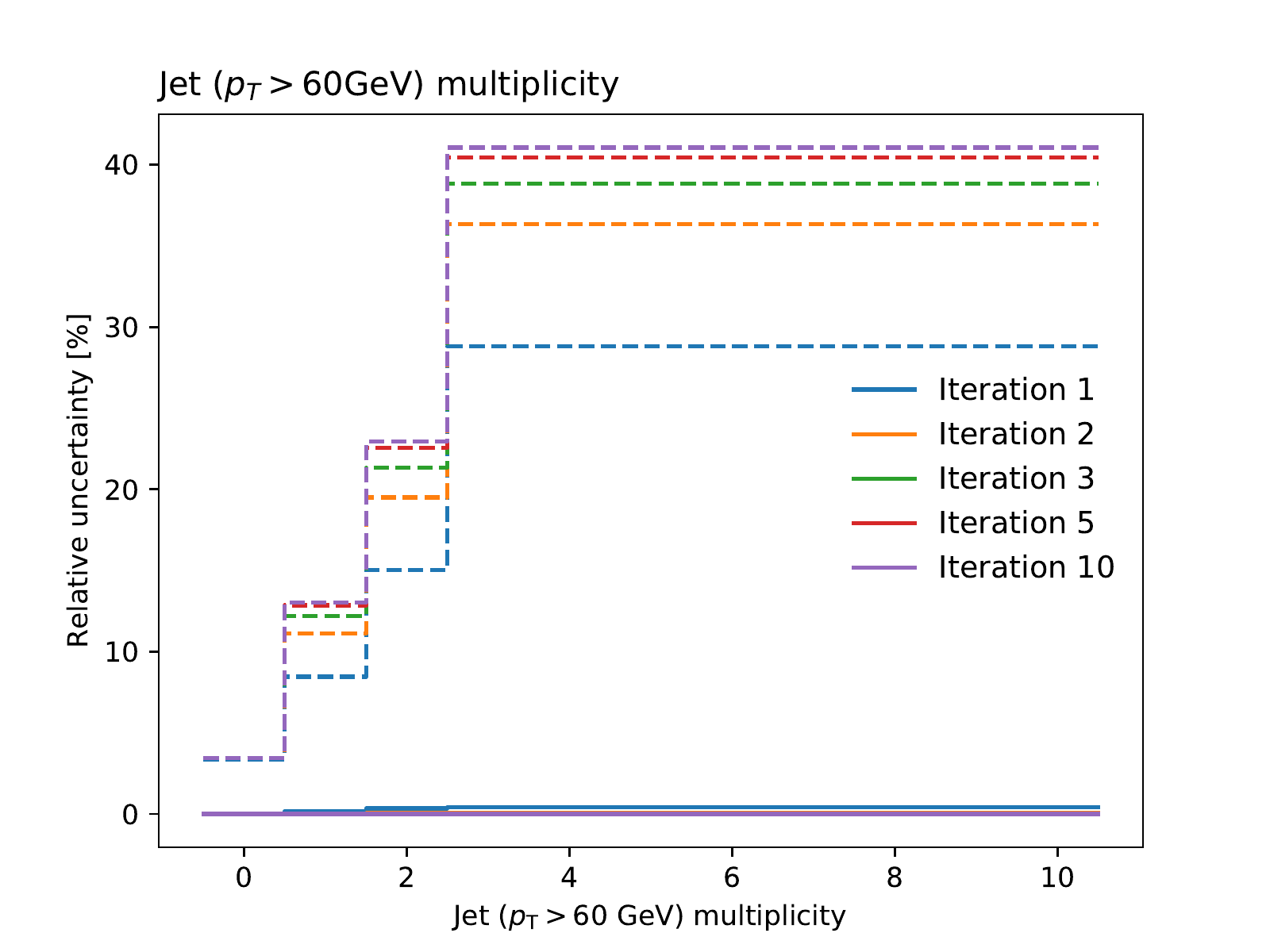}}
\subfigure{\includegraphics[width=0.49\textwidth]{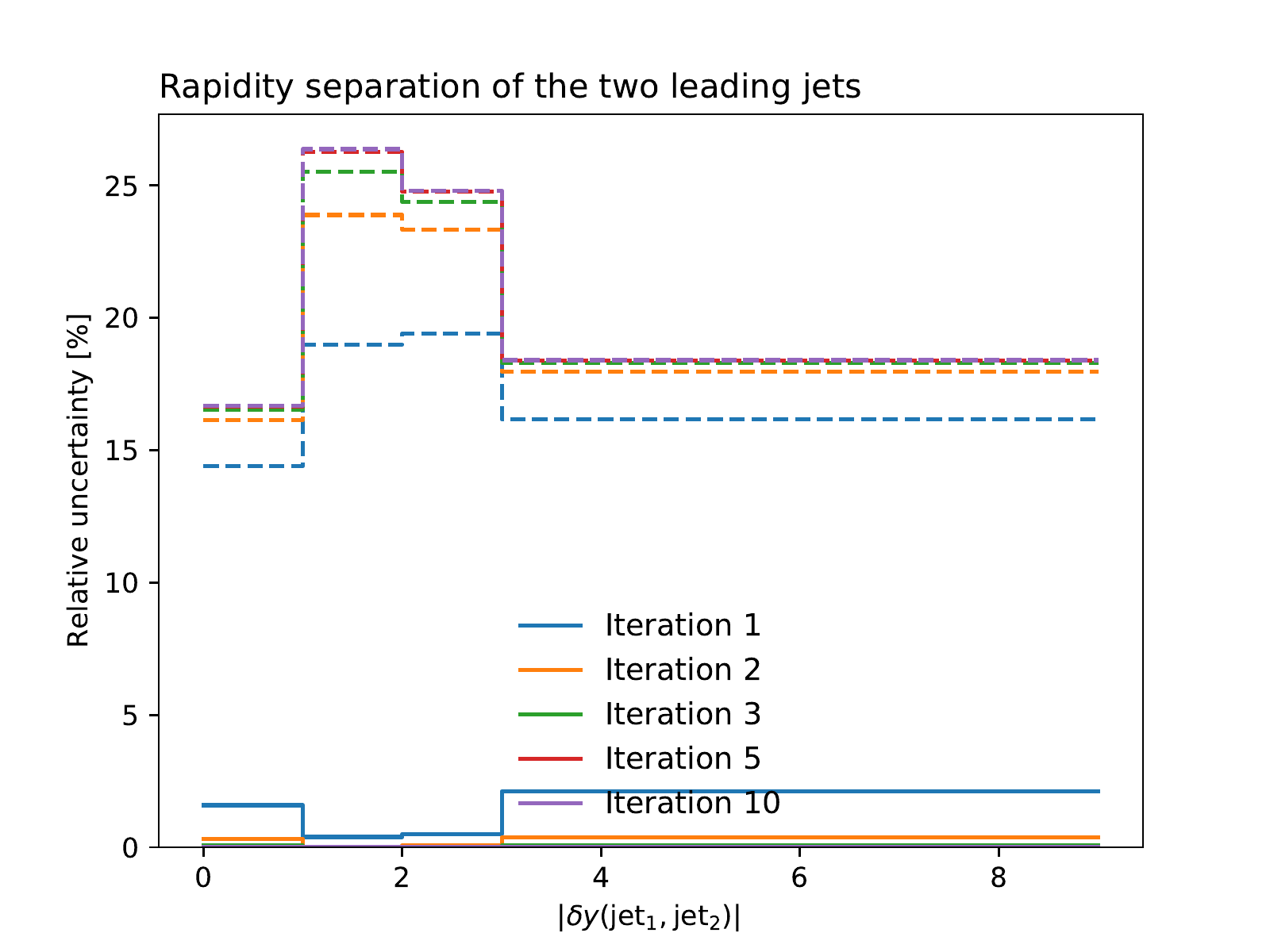}}
\caption{Statistical uncertainty (dashed) and unfolding method uncertainty (solid) for various numbers of iterations for various observables.}
\end{figure}

\begin{figure}[h!]
\centering
\subfigure{\includegraphics[width=0.49\textwidth]{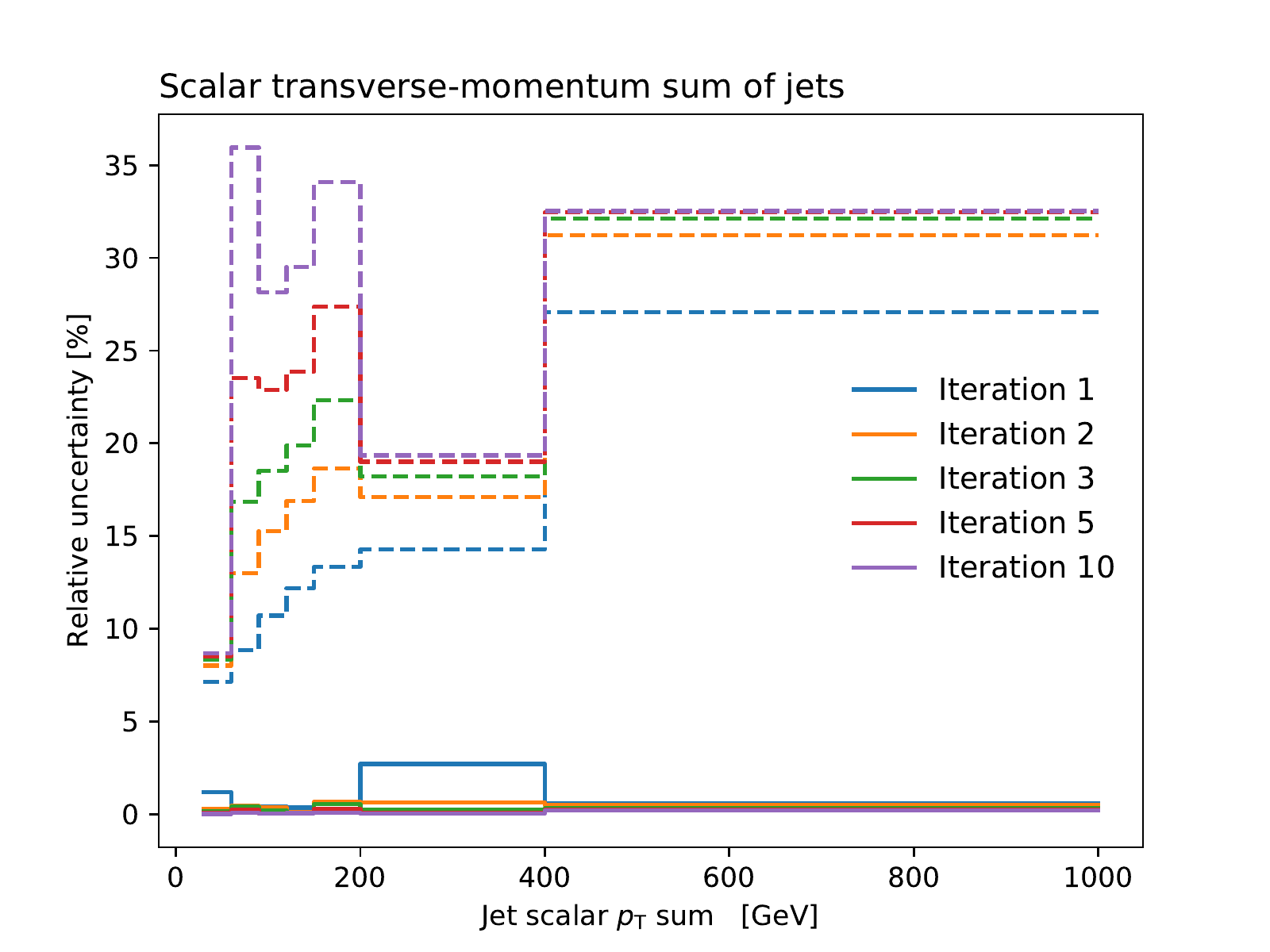}}
\subfigure{\includegraphics[width=0.49\textwidth]{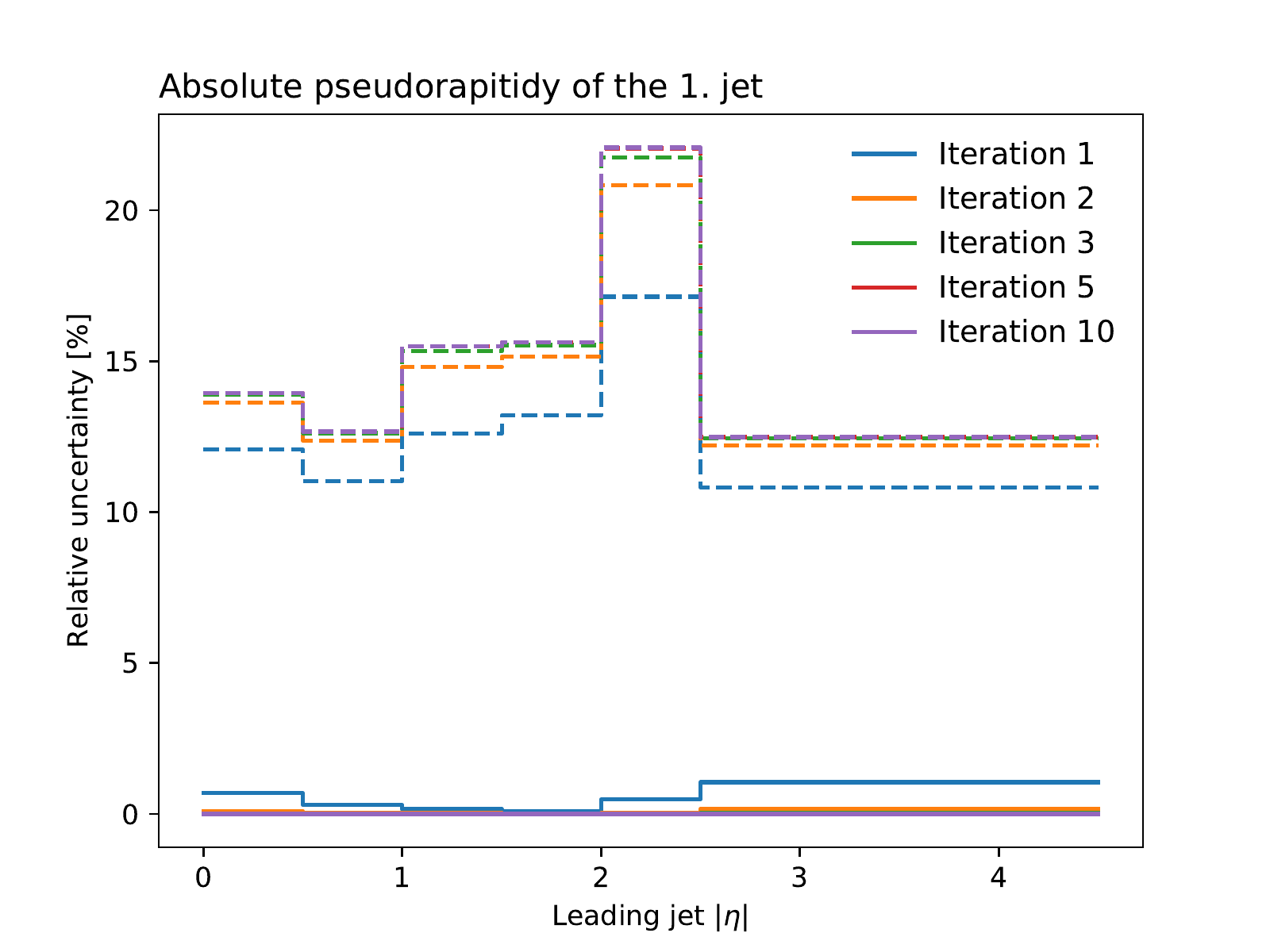}}
\subfigure{\includegraphics[width=0.49\textwidth]{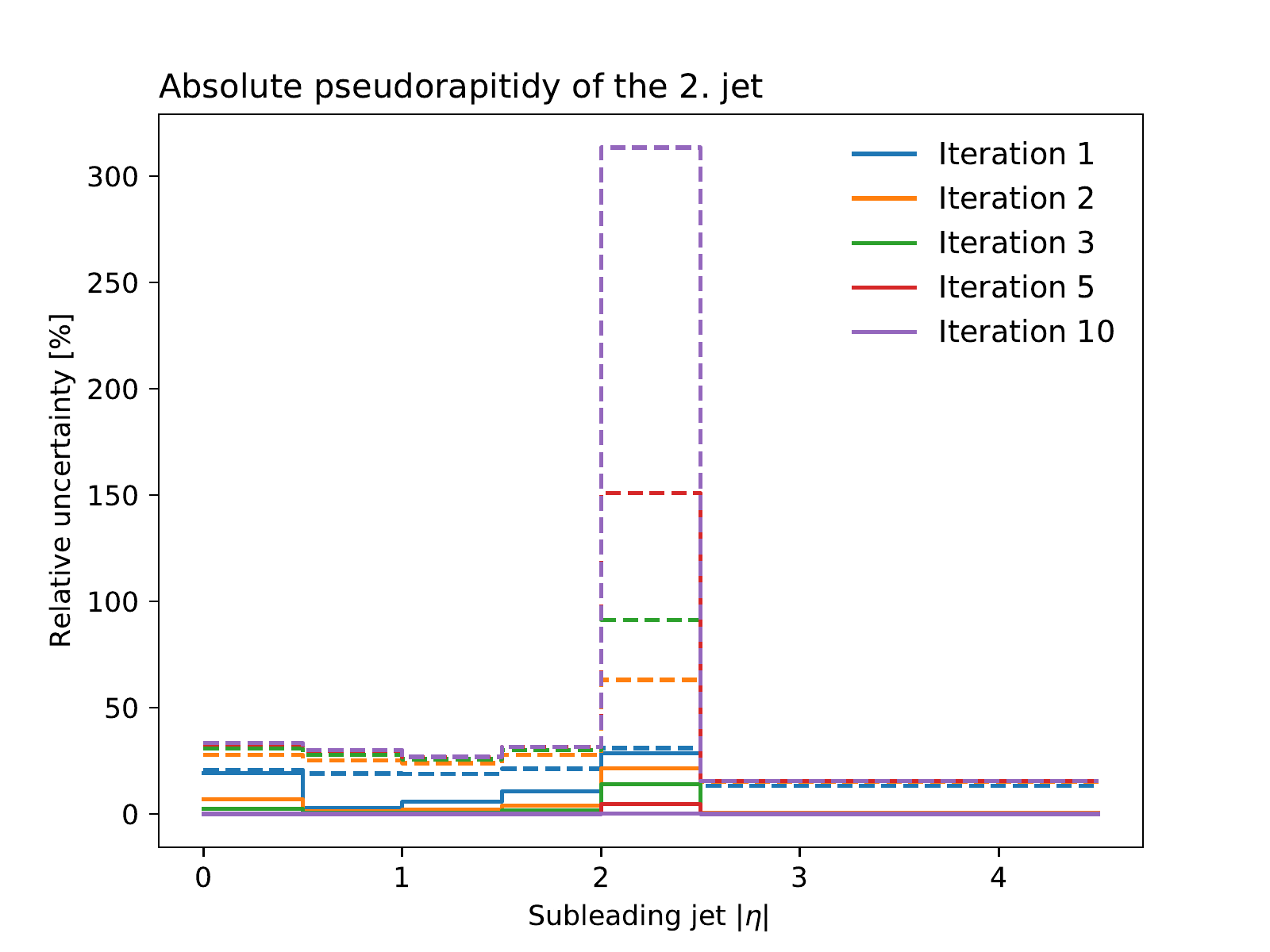}}
\subfigure{\includegraphics[width=0.49\textwidth]{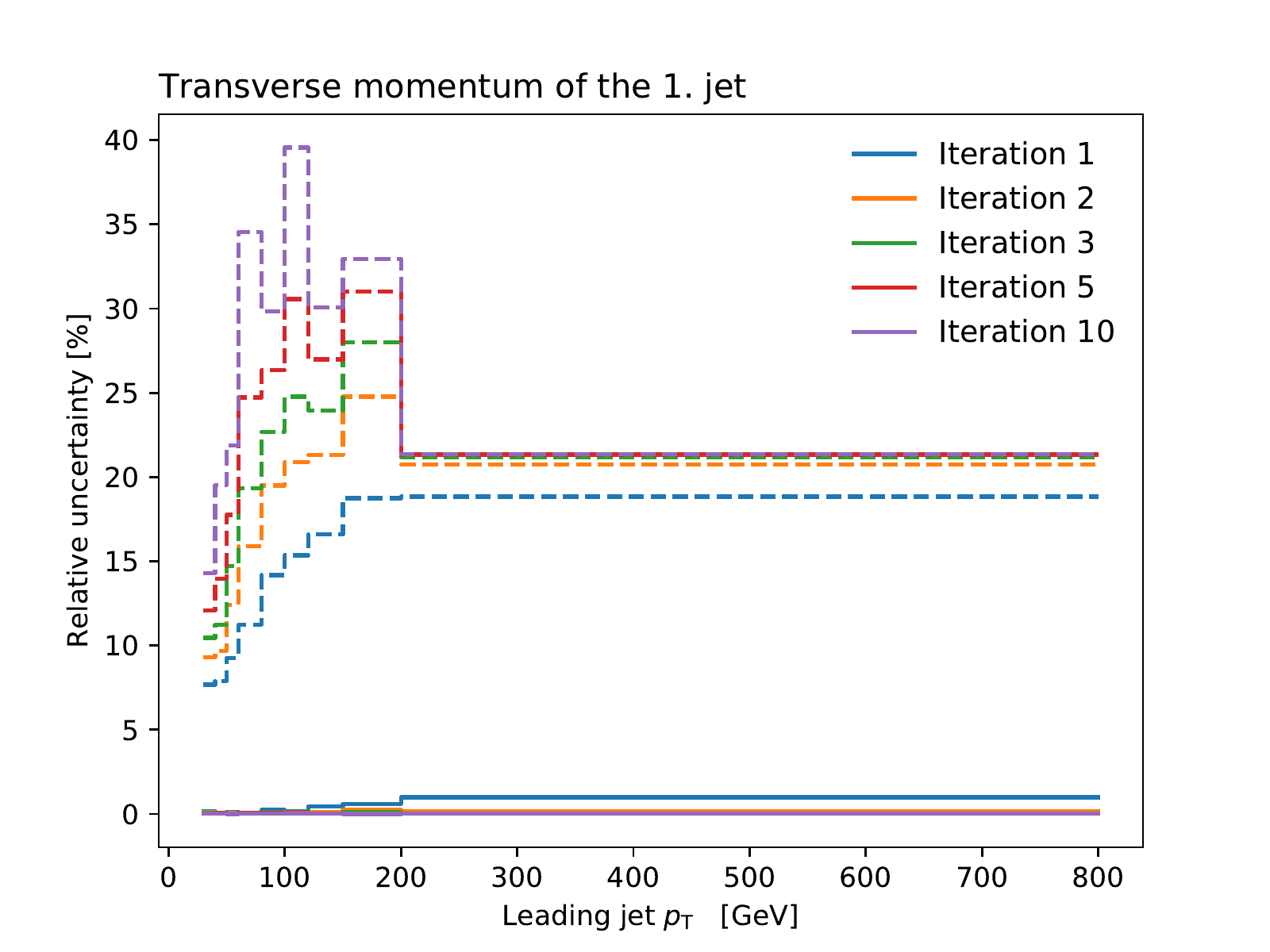}}
\subfigure{\includegraphics[width=0.49\textwidth]{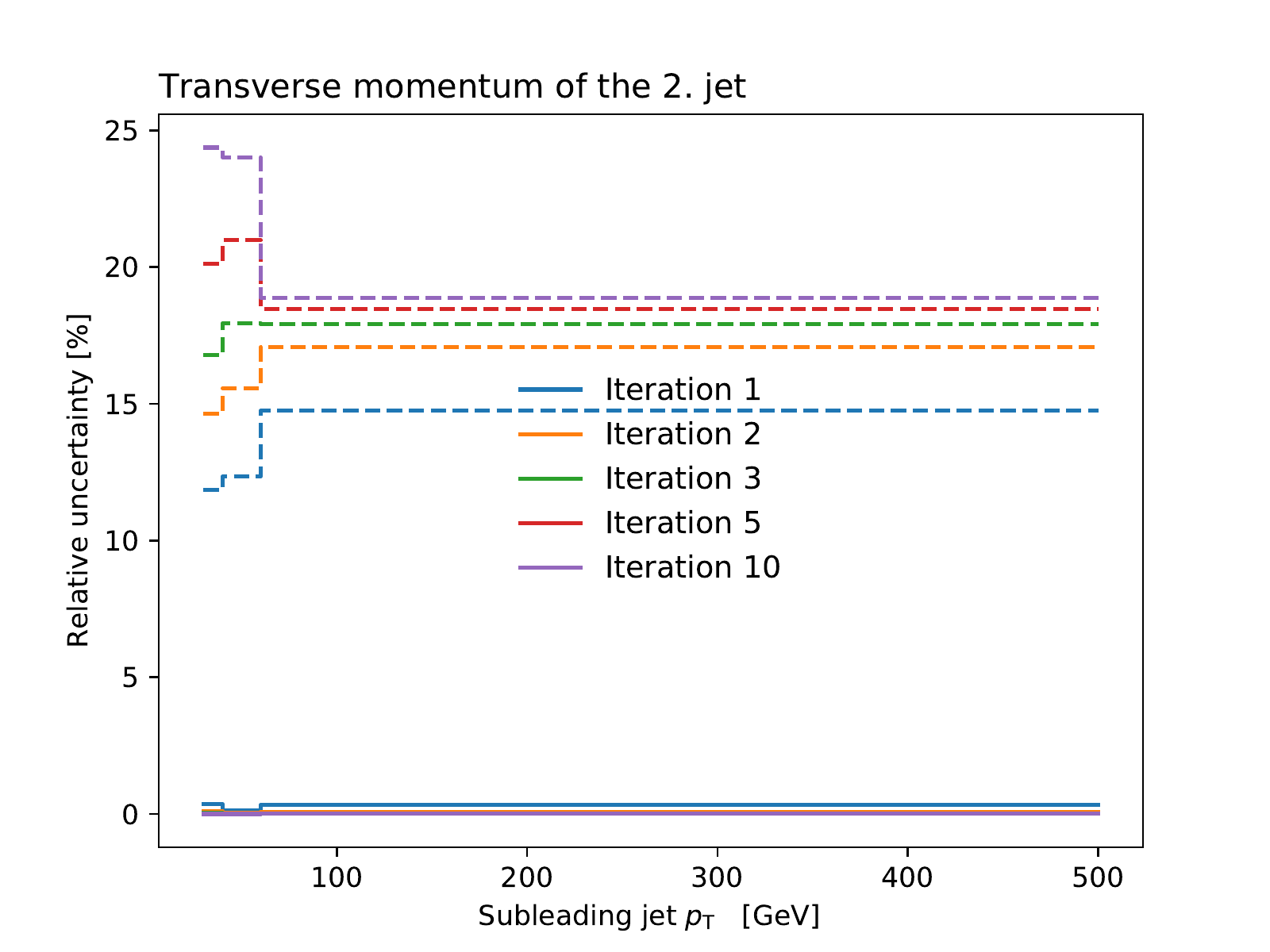}}
\caption{Statistical uncertainty (dashed) and unfolding method uncertainty (solid) for various numbers of iterations for various observables.}
\label{fig:statbias_plots_last}
\end{figure}

\clearpage

\subsection{Response matrices}
\label{sec:zz_aux_response}
\myfigs~\ref{fig:response_matrix_first}--\ref{fig:response_matrix_last} show the response matrices used in the unfolding.

\begin{figure}[h!]
\centering
\subfigure{\includegraphics[width=0.49\textwidth]{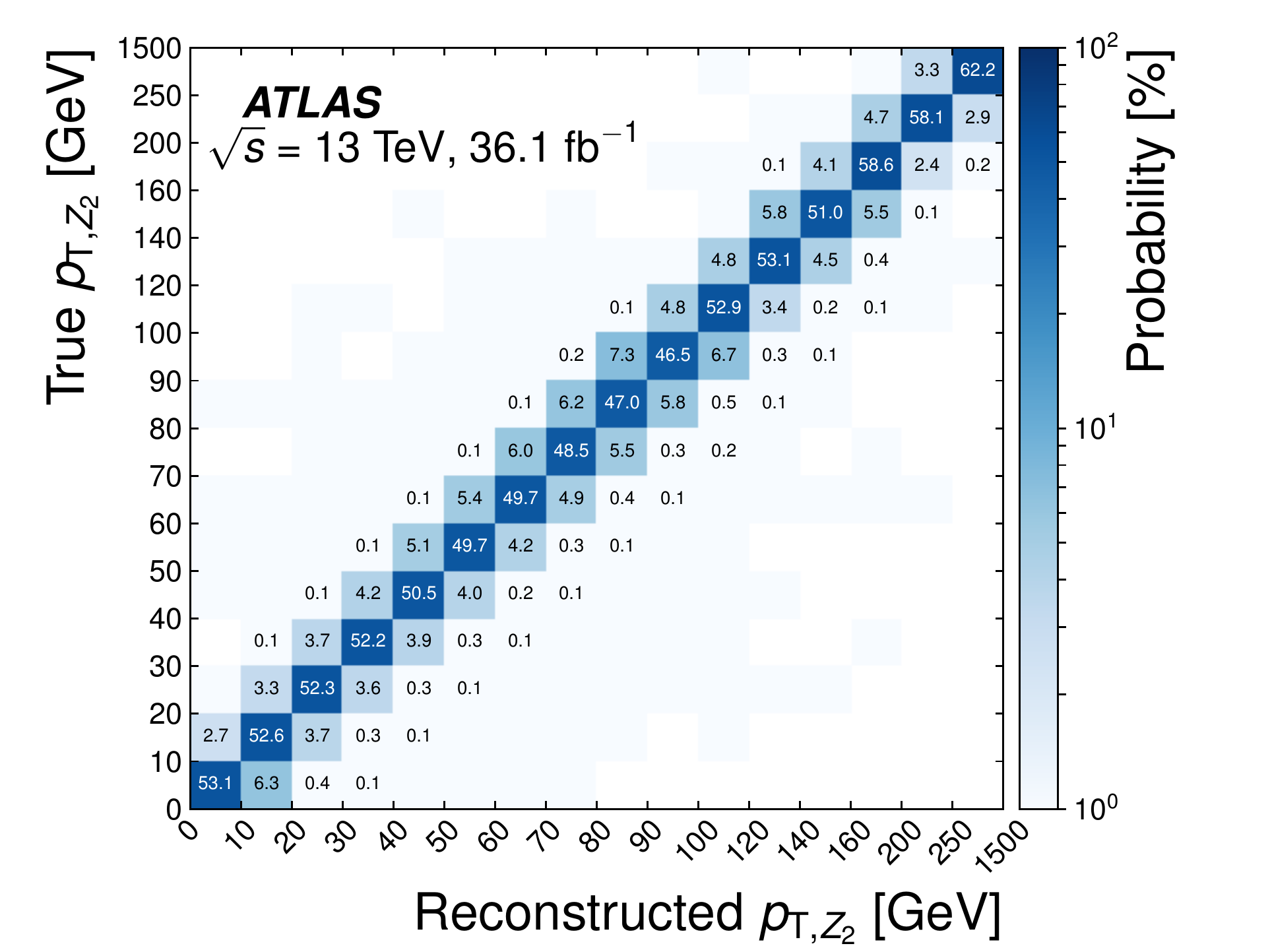}}
\subfigure{\includegraphics[width=0.49\textwidth]{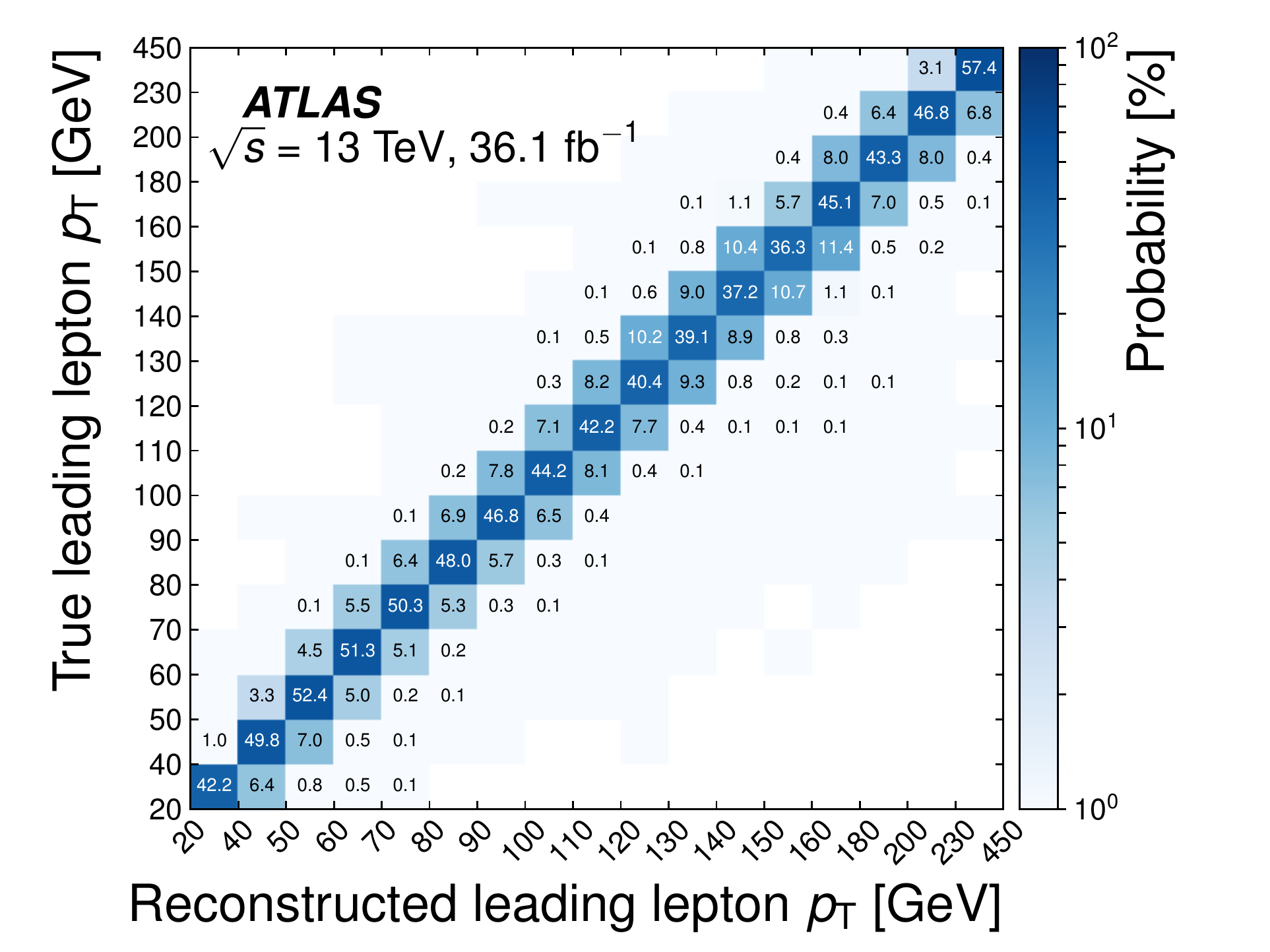}}
\subfigure{\includegraphics[width=0.49\textwidth]{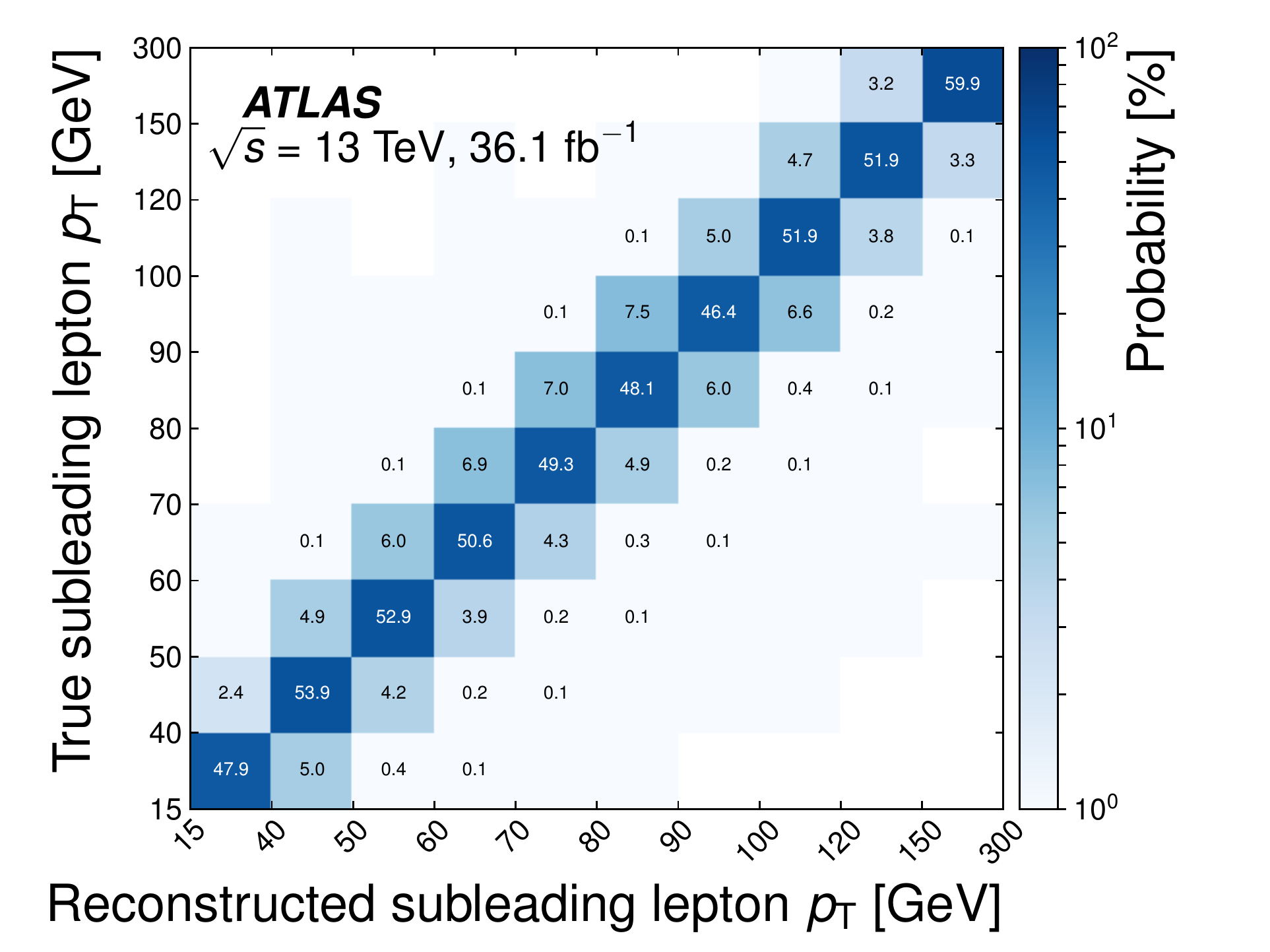}}
\subfigure{\includegraphics[width=0.49\textwidth]{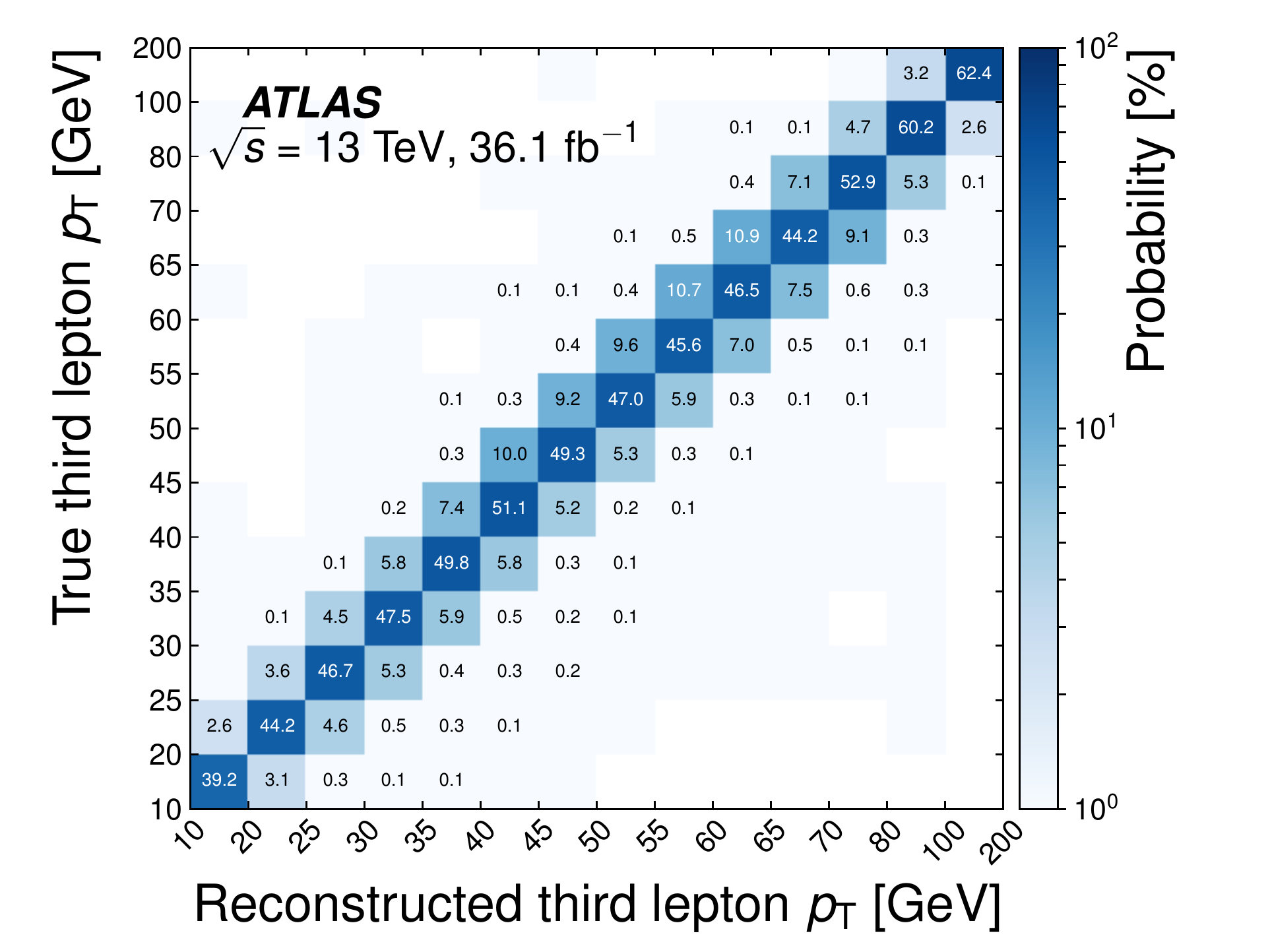}}
\subfigure{\includegraphics[width=0.49\textwidth]{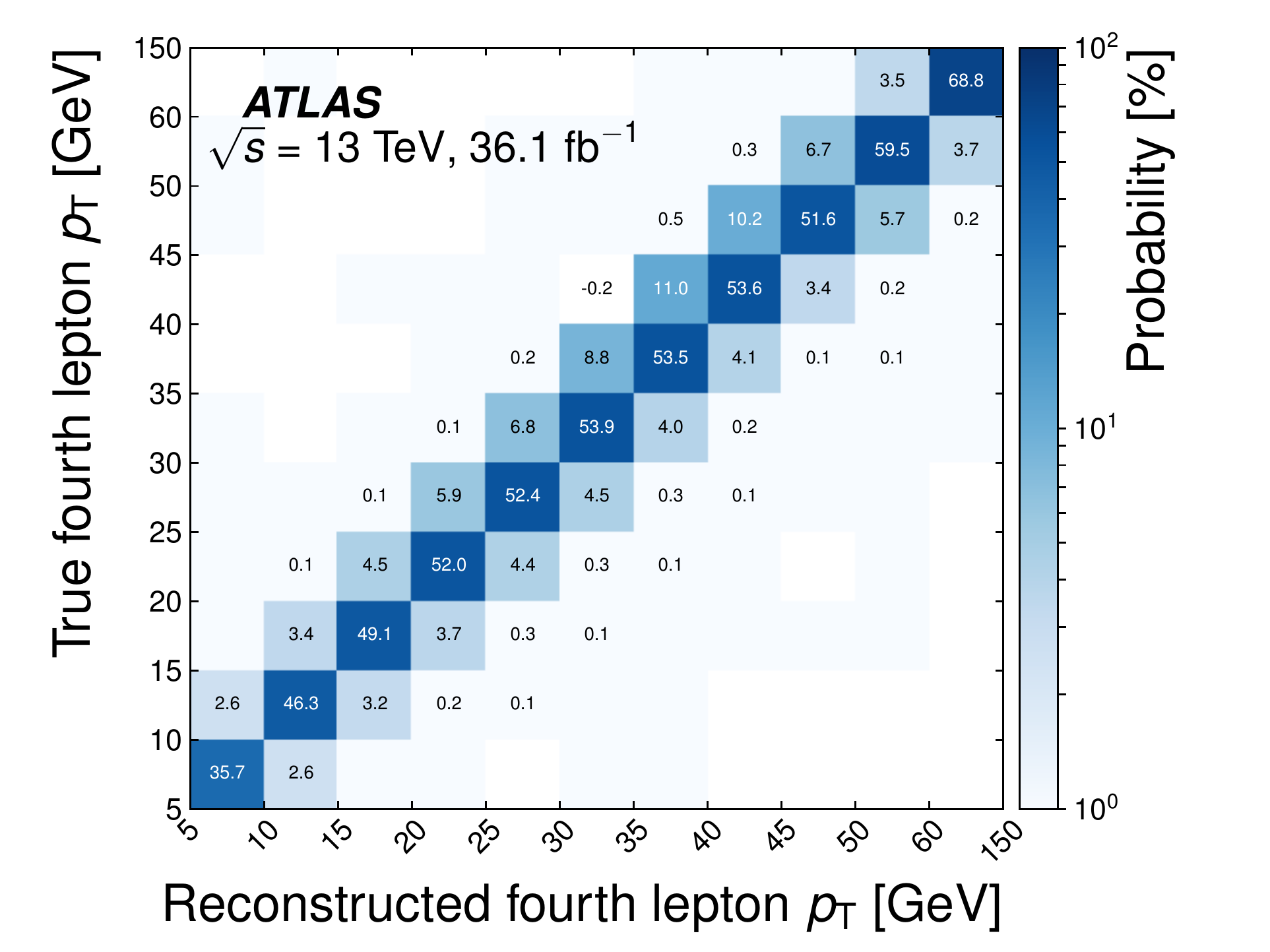}}
\subfigure{\includegraphics[width=0.49\textwidth]{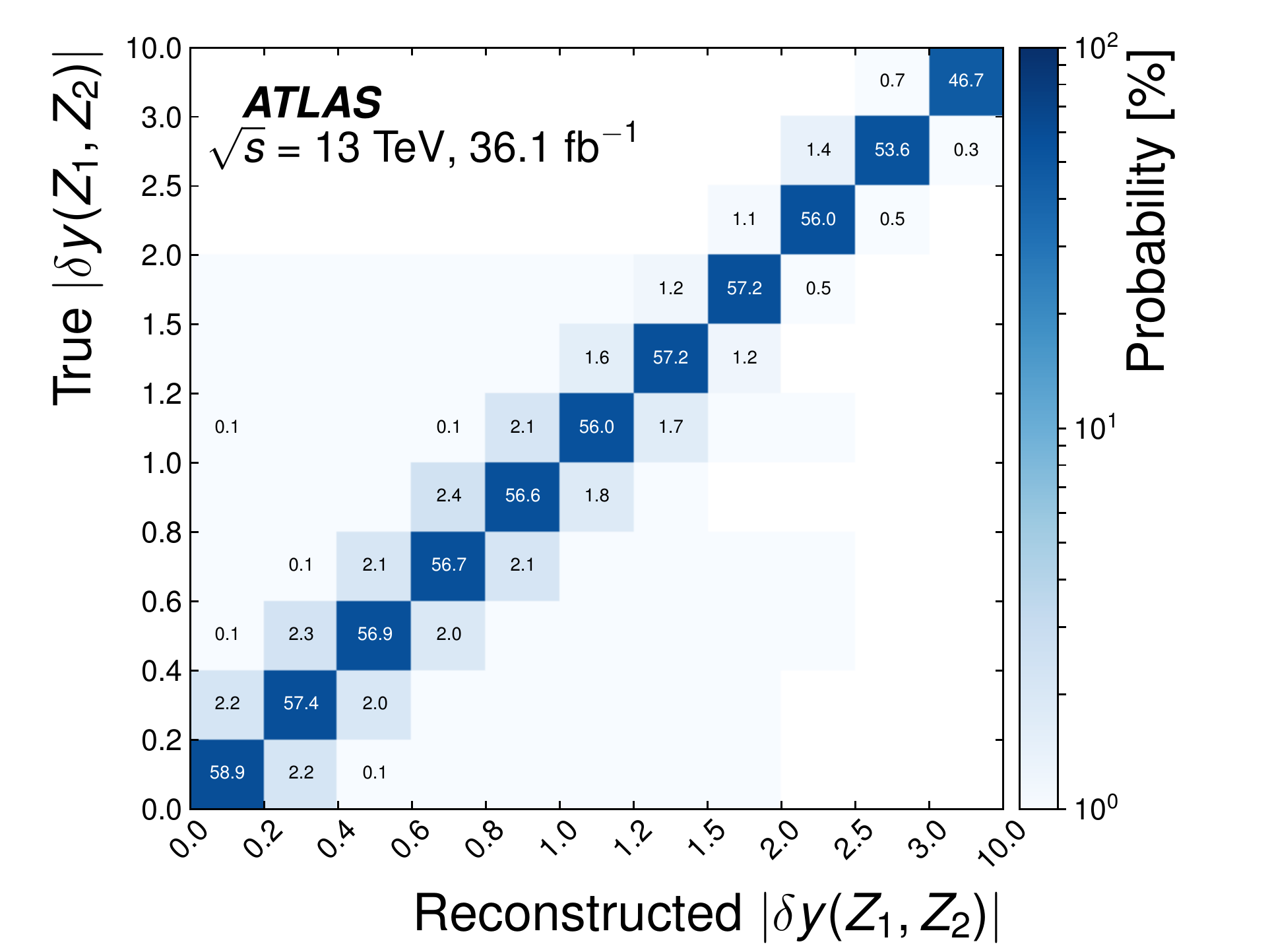}}
\caption{Response matrices of various observables. Published in the auxiliary materials of \myref~\cite{STDM-2016-15}.}
\label{fig:response_matrix_first}
\end{figure}

\begin{figure}[h!]
\centering
\subfigure{\includegraphics[width=0.49\textwidth]{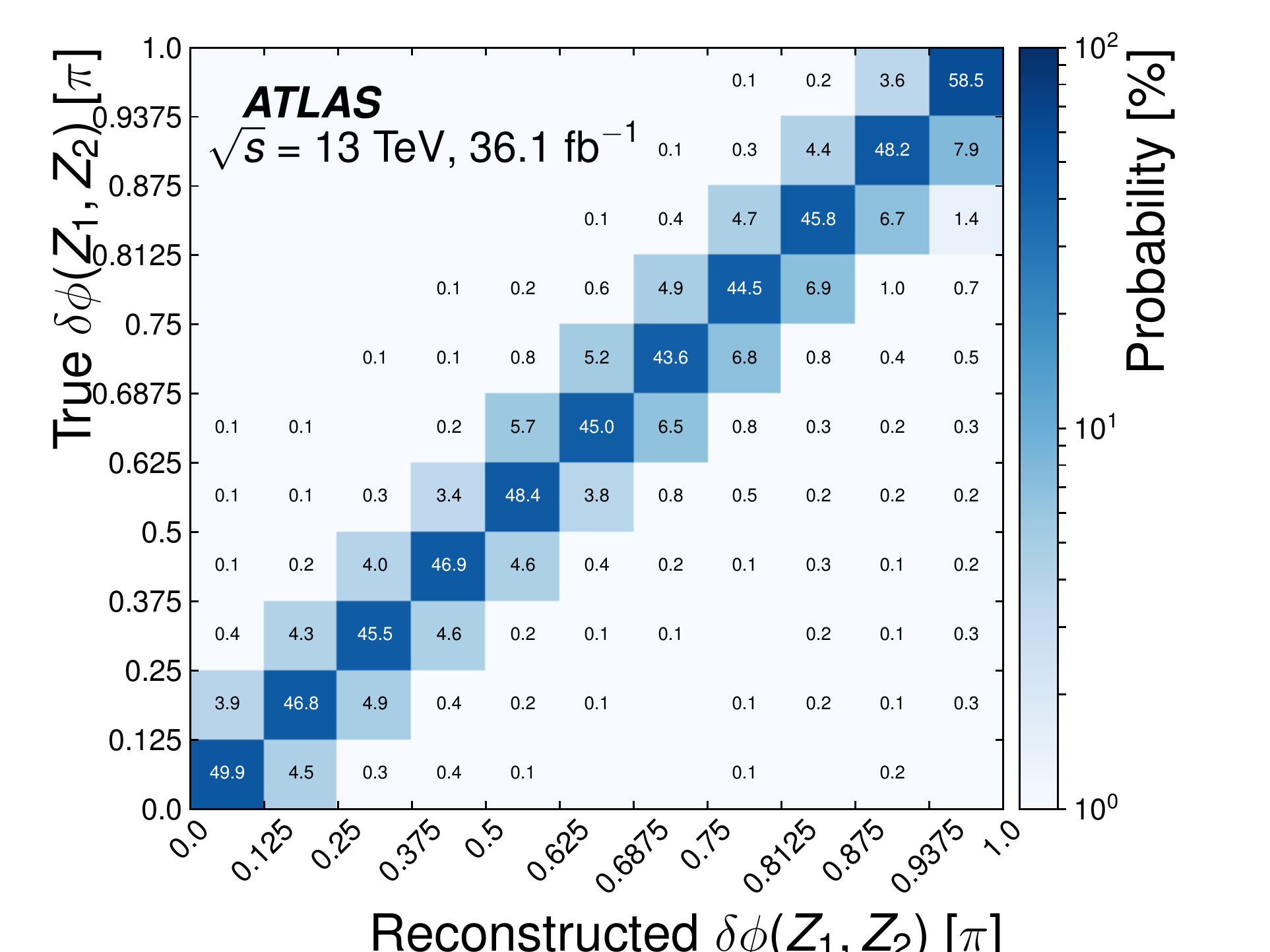}}
\subfigure{\includegraphics[width=0.49\textwidth]{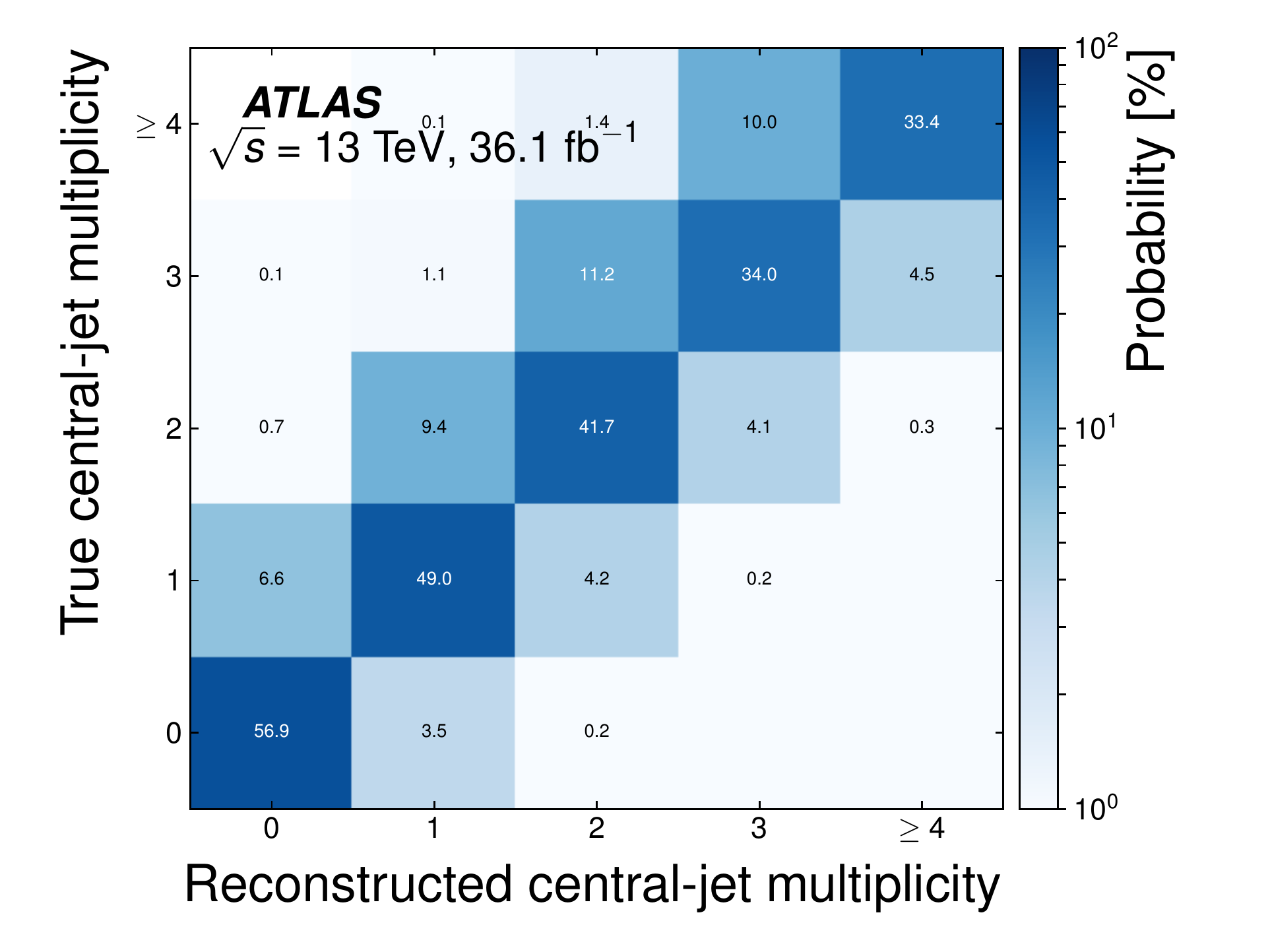}}
\subfigure{\includegraphics[width=0.49\textwidth]{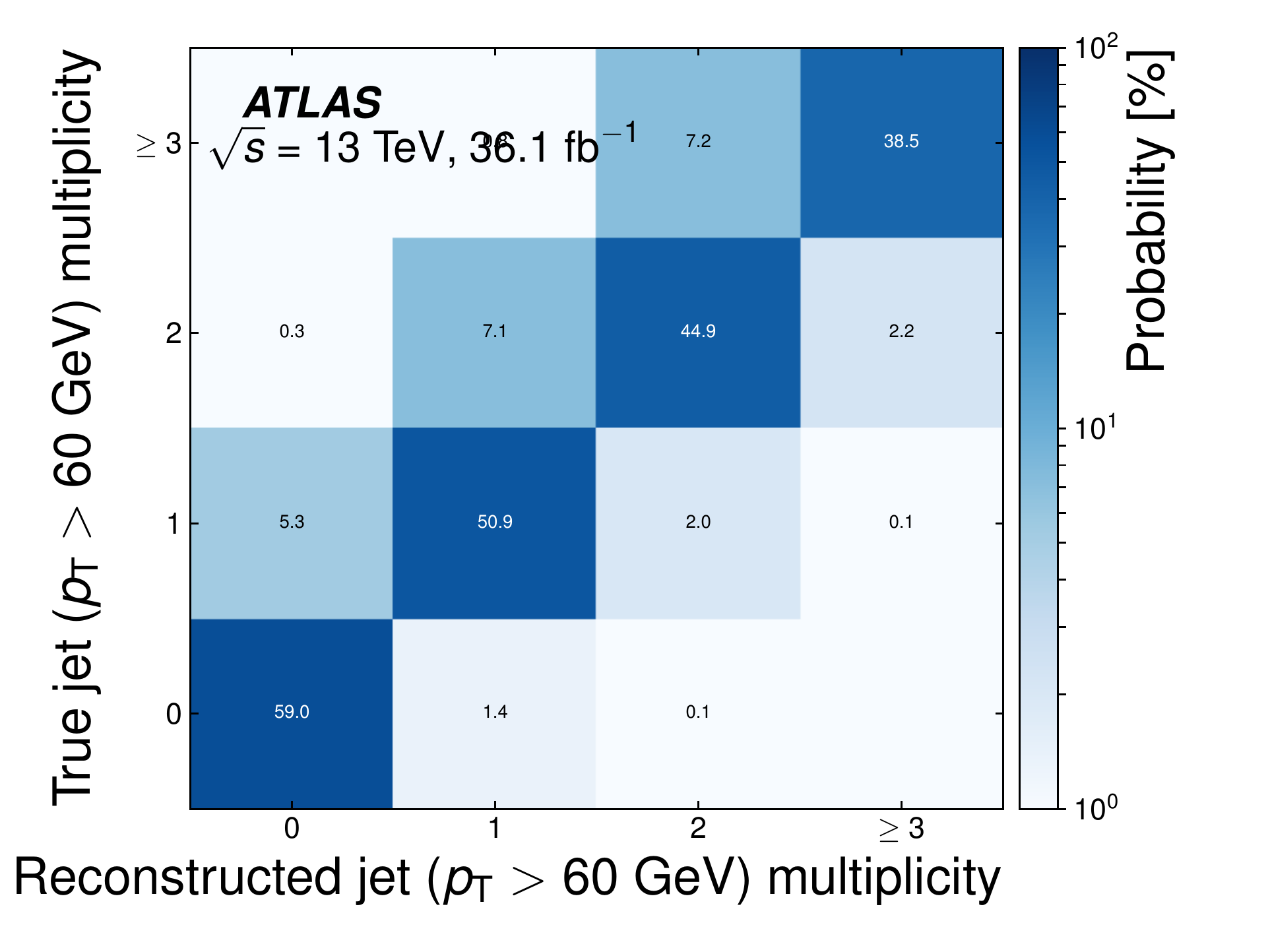}}
\subfigure{\includegraphics[width=0.49\textwidth]{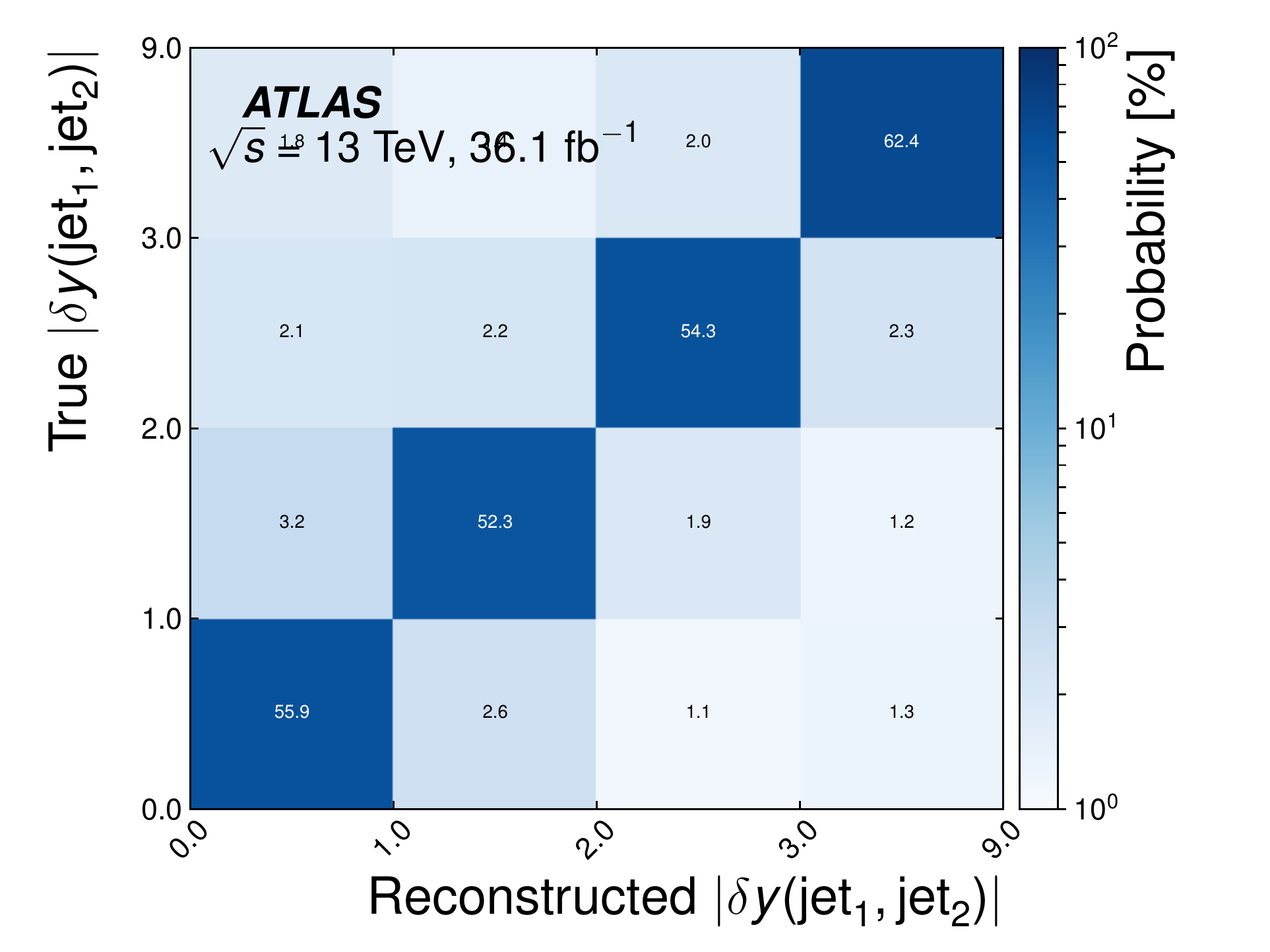}}
\subfigure{\includegraphics[width=0.49\textwidth]{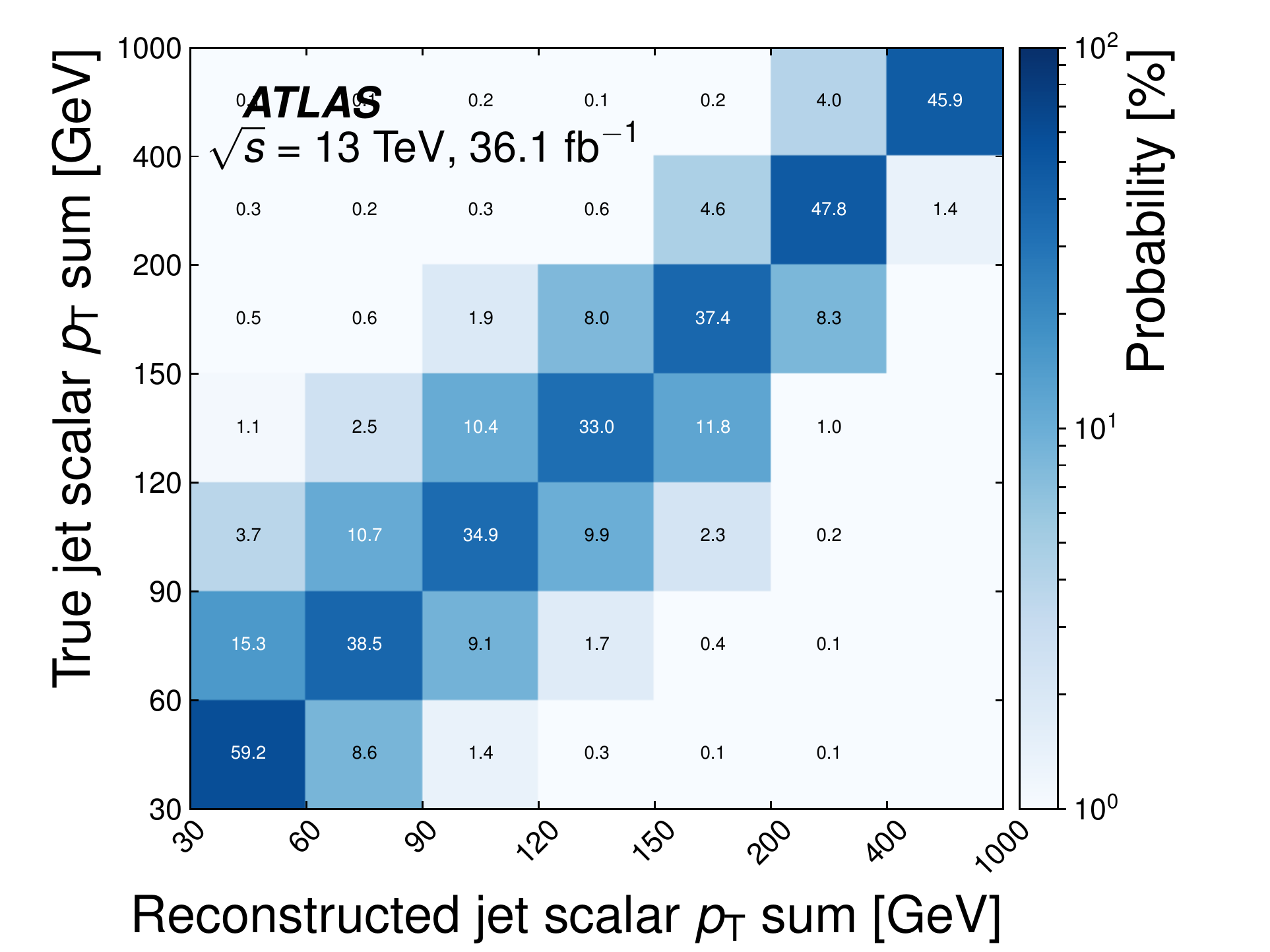}}
\subfigure{\includegraphics[width=0.49\textwidth]{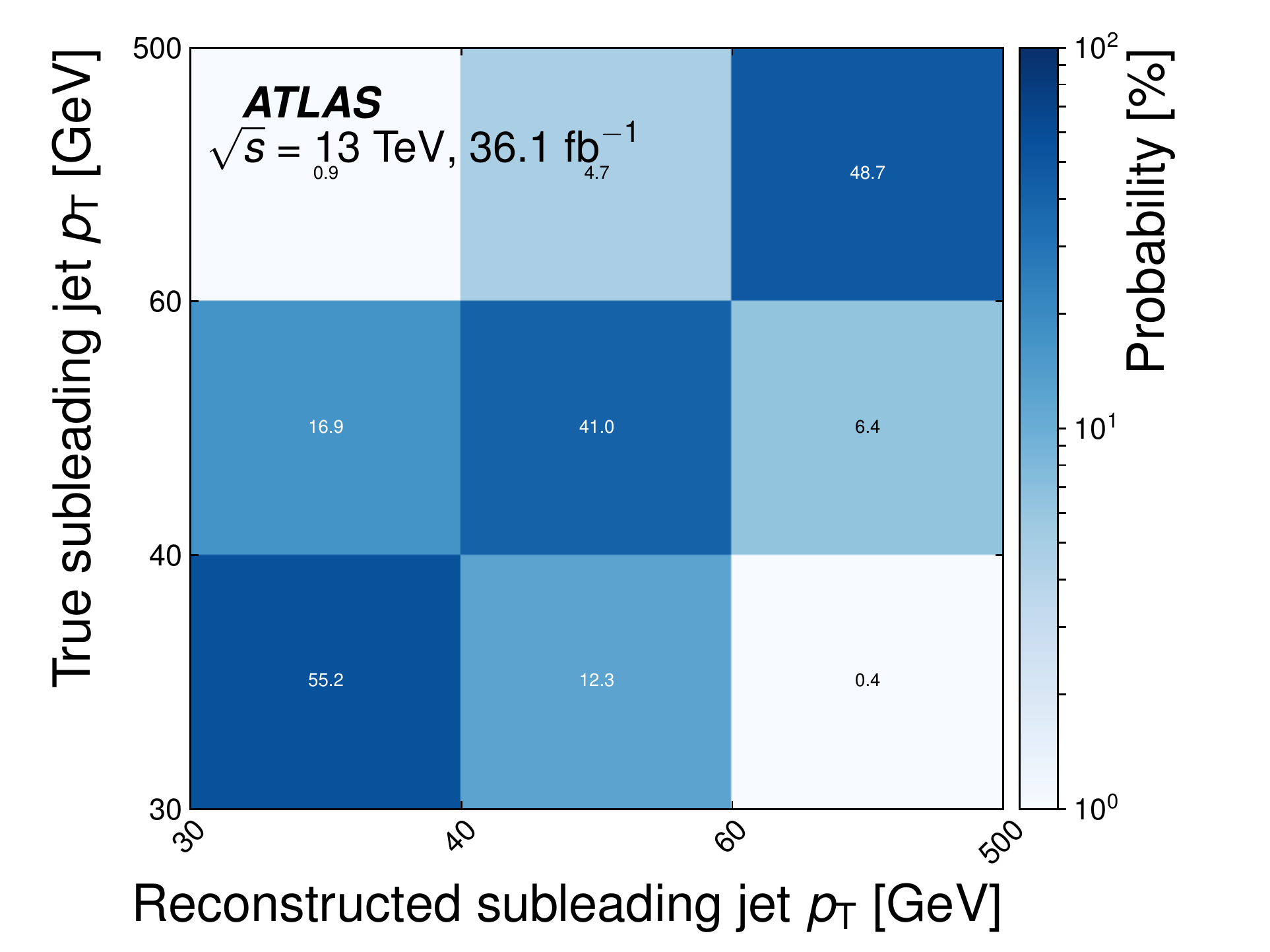}}
\caption{Response matrices of various observables. Published in the auxiliary materials of \myref~\cite{STDM-2016-15}.}
\end{figure}

\begin{figure}[h!]
\centering
\subfigure{\includegraphics[width=0.49\textwidth]{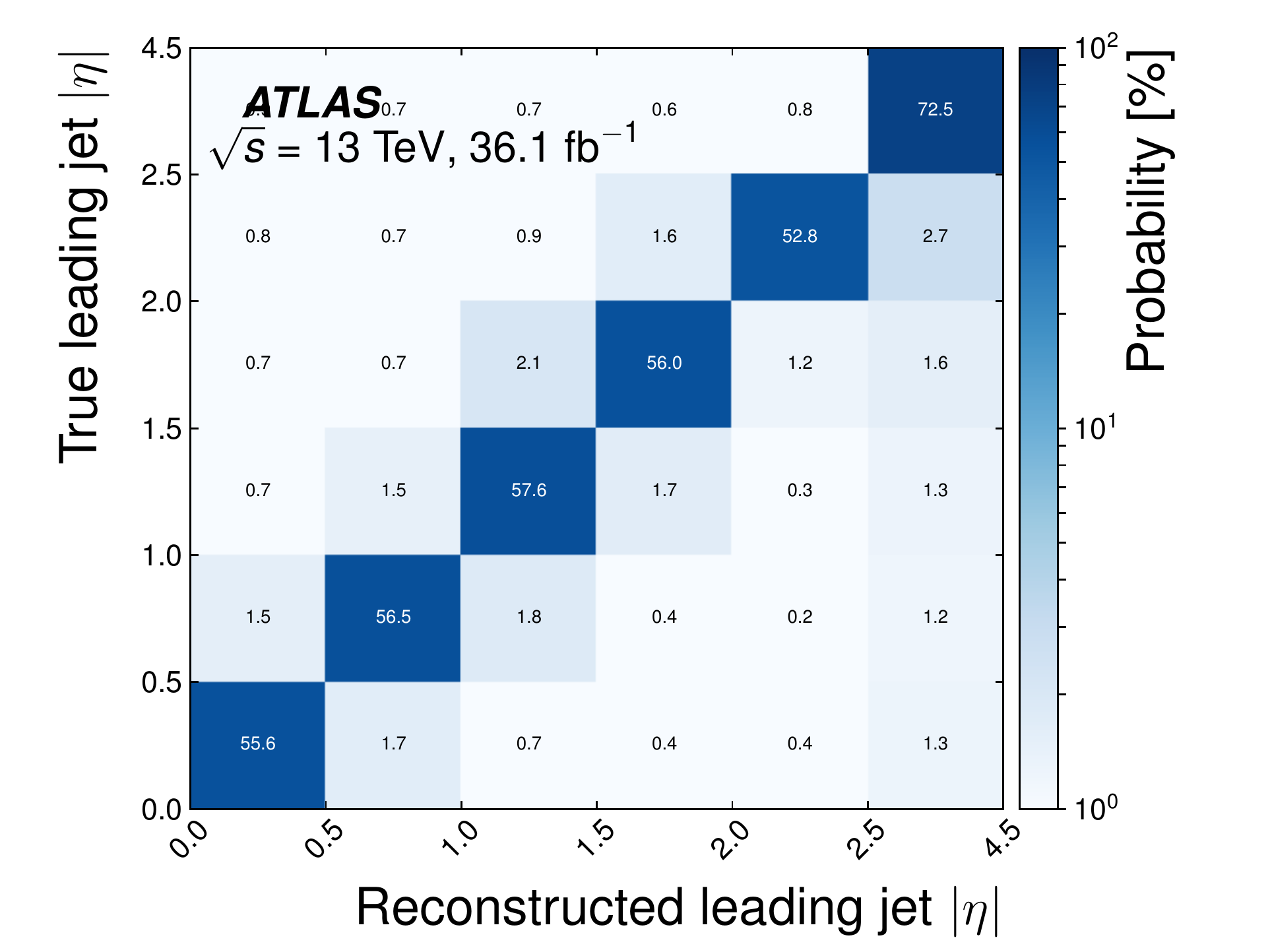}}
\subfigure{\includegraphics[width=0.49\textwidth]{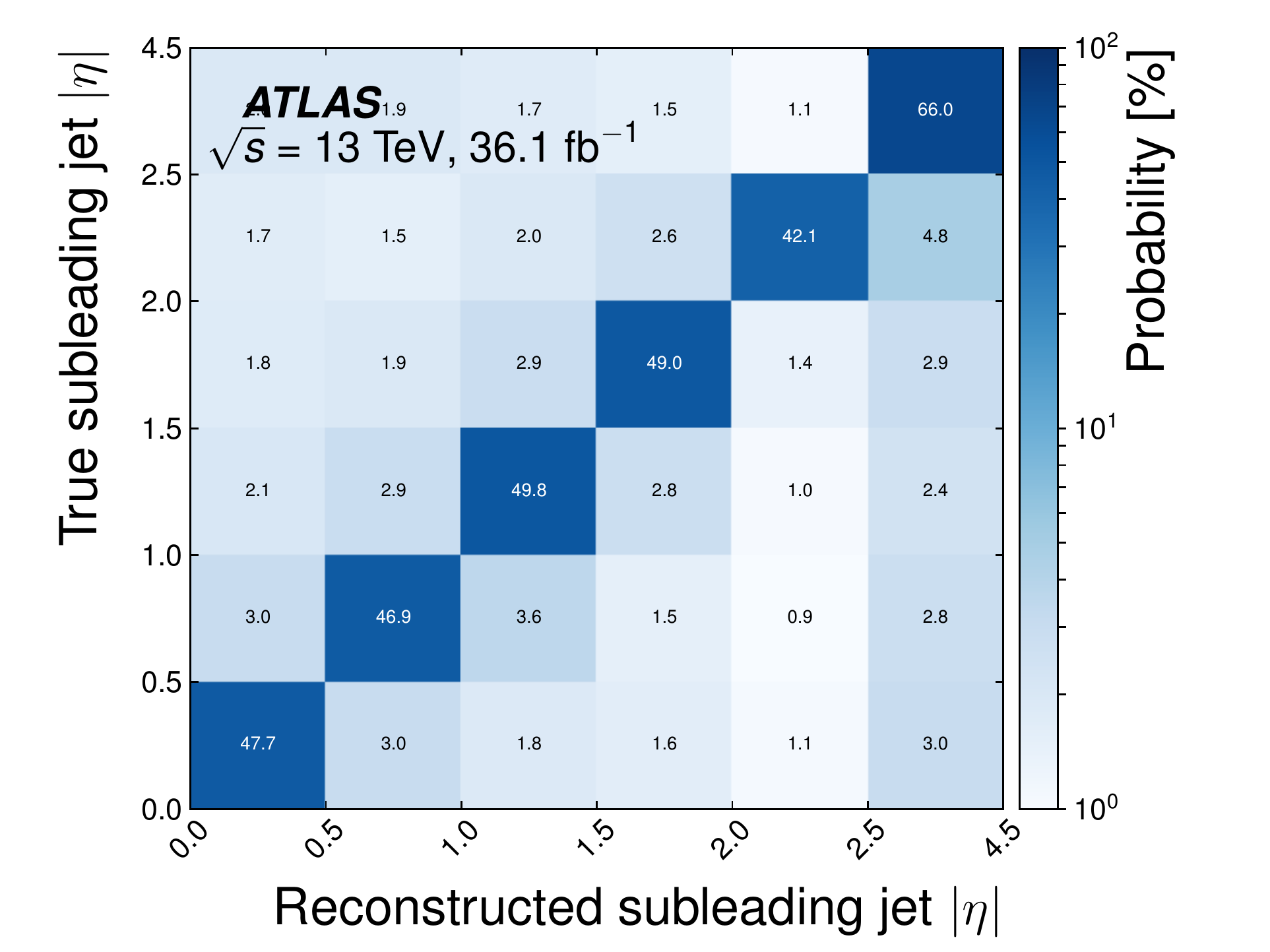}}
\caption{Response matrices of various observables. Published in the auxiliary materials of \myref~\cite{STDM-2016-15}.}
\label{fig:response_matrix_last}
\end{figure}

\clearpage

\subsection{Uncertainty breakdowns}
\label{sec:zz_aux_uncerts}
\myfigs~\ref{fig:diffuncert_plots_first}--\ref{fig:diffuncert_plots_last} show the bin-by-bin uncertainty contributions after unfolding of the various observables.

\begin{figure}[h!]
\centering
\subfigure{\includegraphics[width=0.48\textwidth]{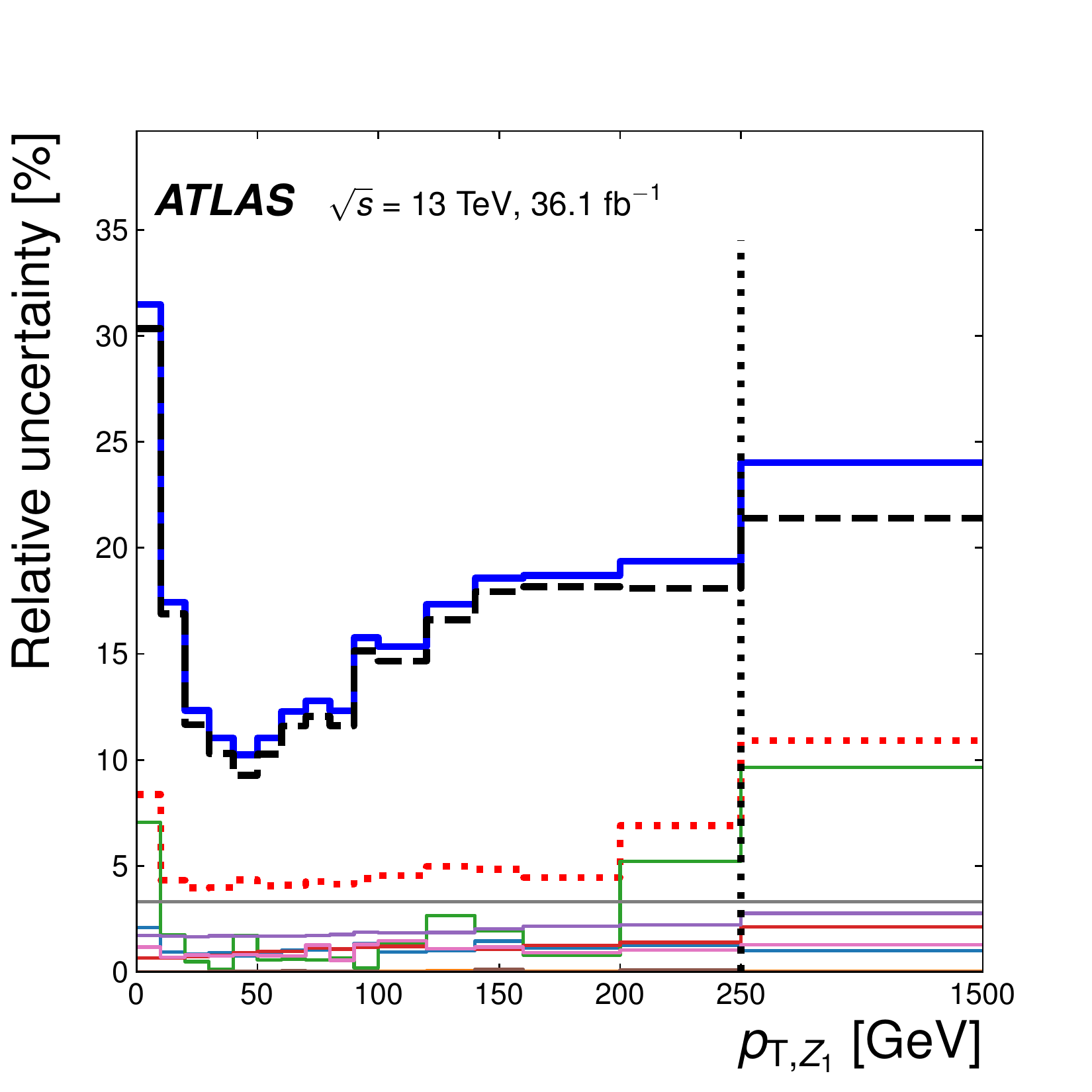}}
\subfigure{\includegraphics[width=0.48\textwidth]{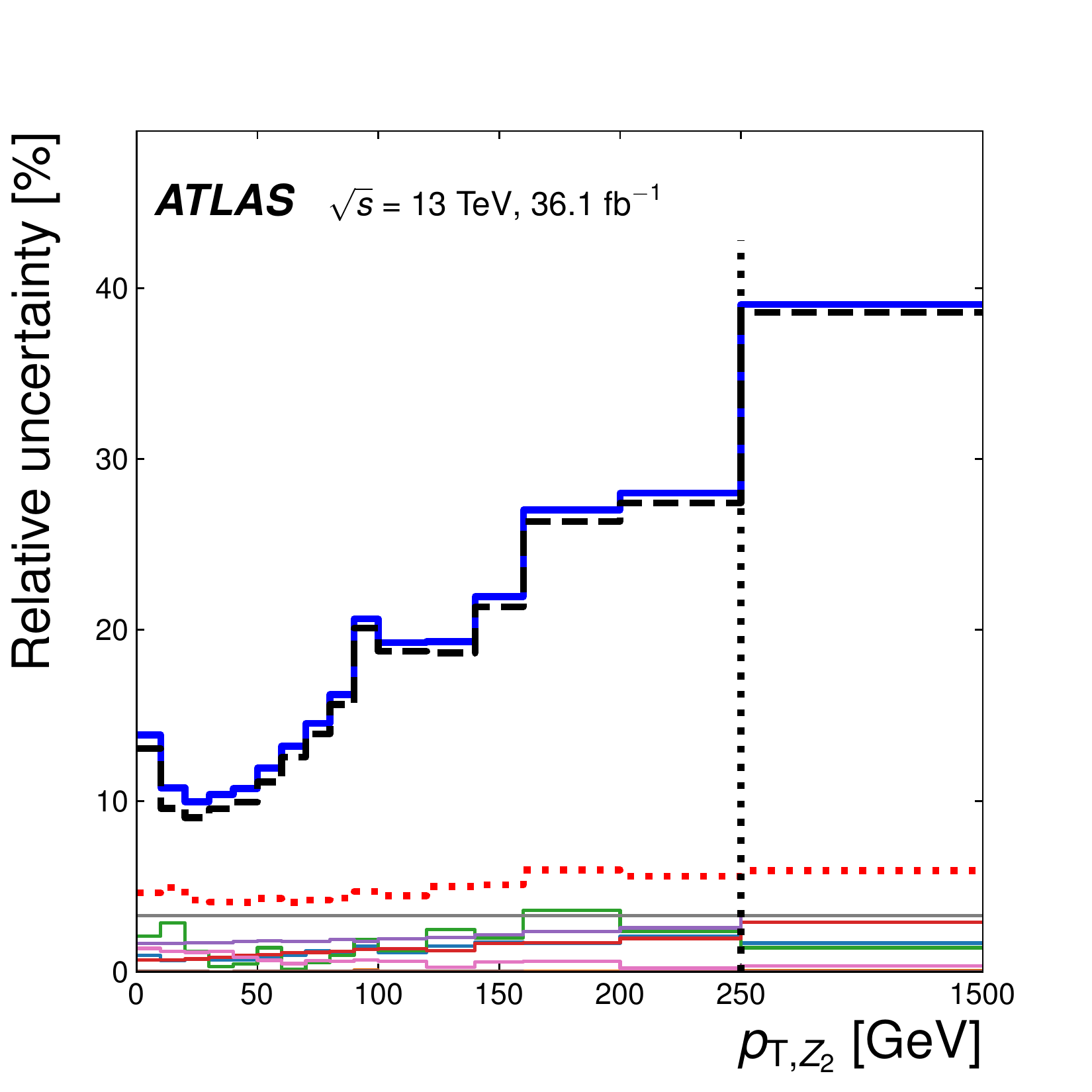}}
\subfigure{\includegraphics[width=0.48\textwidth]{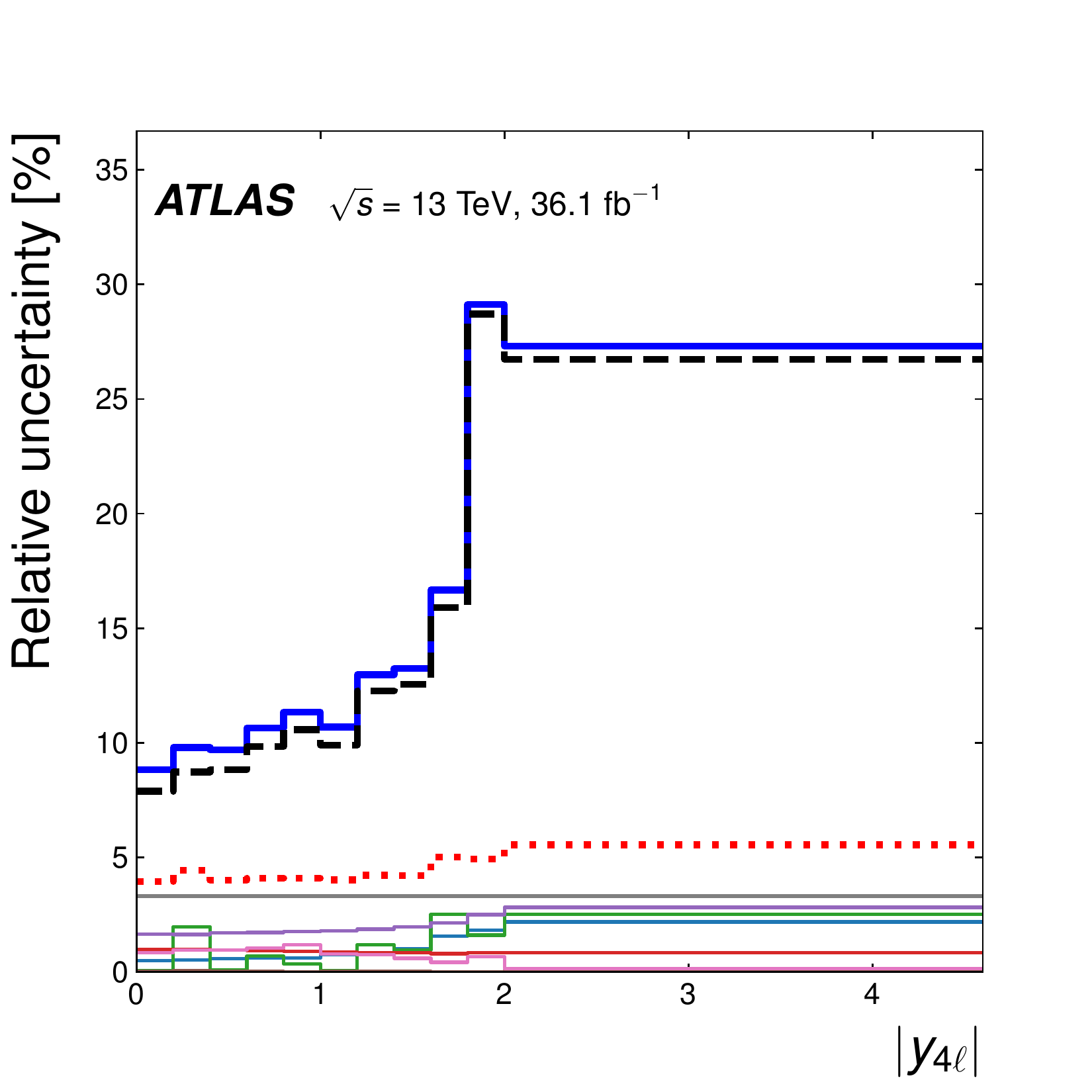}}
\subfigure{\includegraphics[width=0.48\textwidth]{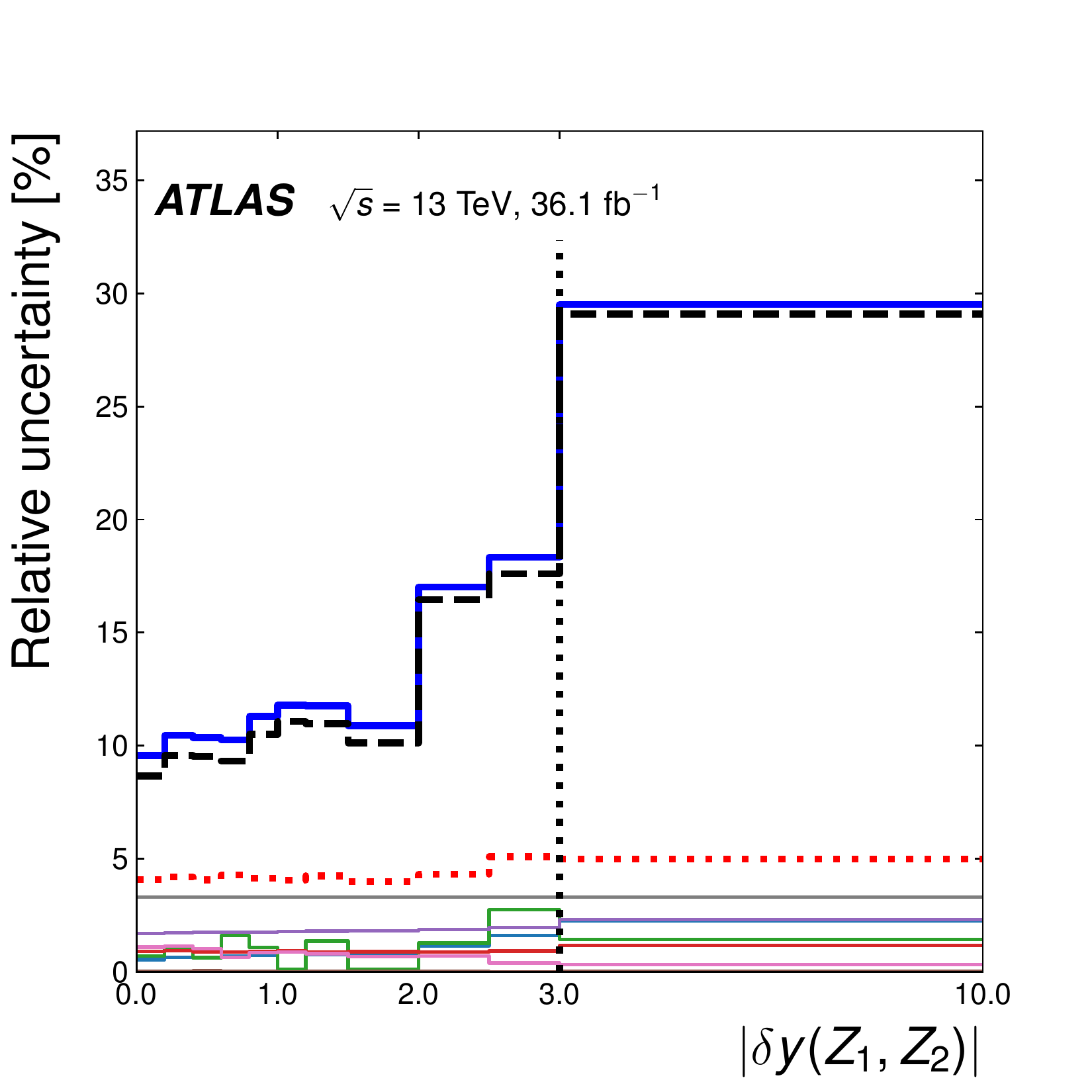}}
\subfigure{\includegraphics[width=0.48\textwidth]{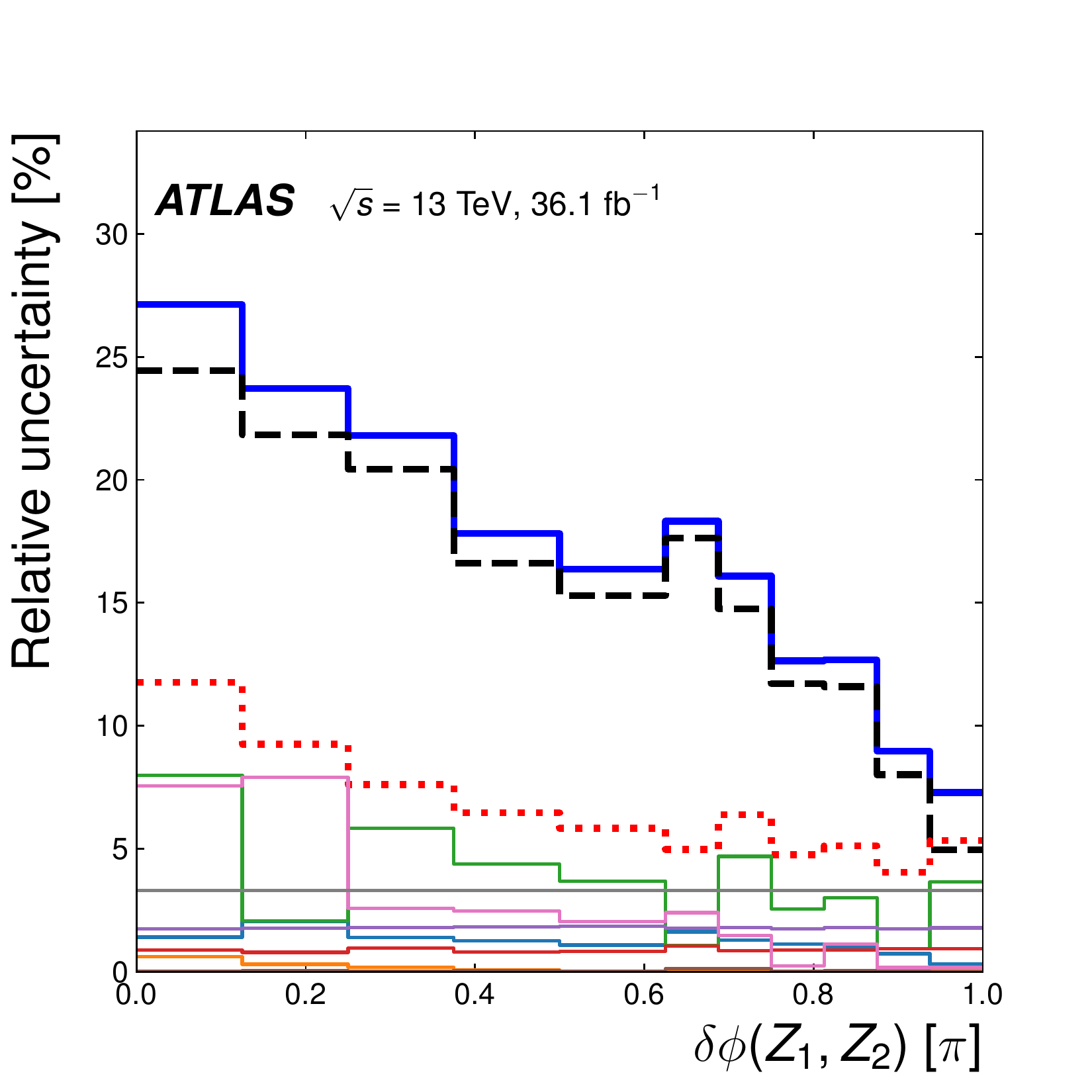}}
\subfigure{\includegraphics[width=0.48\textwidth]{diffuncerts_legend.pdf}}
\caption{Bin-by-bin uncertainty breakdown for various observables. For better visualisation, the last bin is shown using a different $x$-axis scale where indicated by a dashed vertical line.}
\label{fig:diffuncert_plots_first}
\end{figure}

\begin{figure}[h!]
\centering
\subfigure{\includegraphics[width=0.48\textwidth]{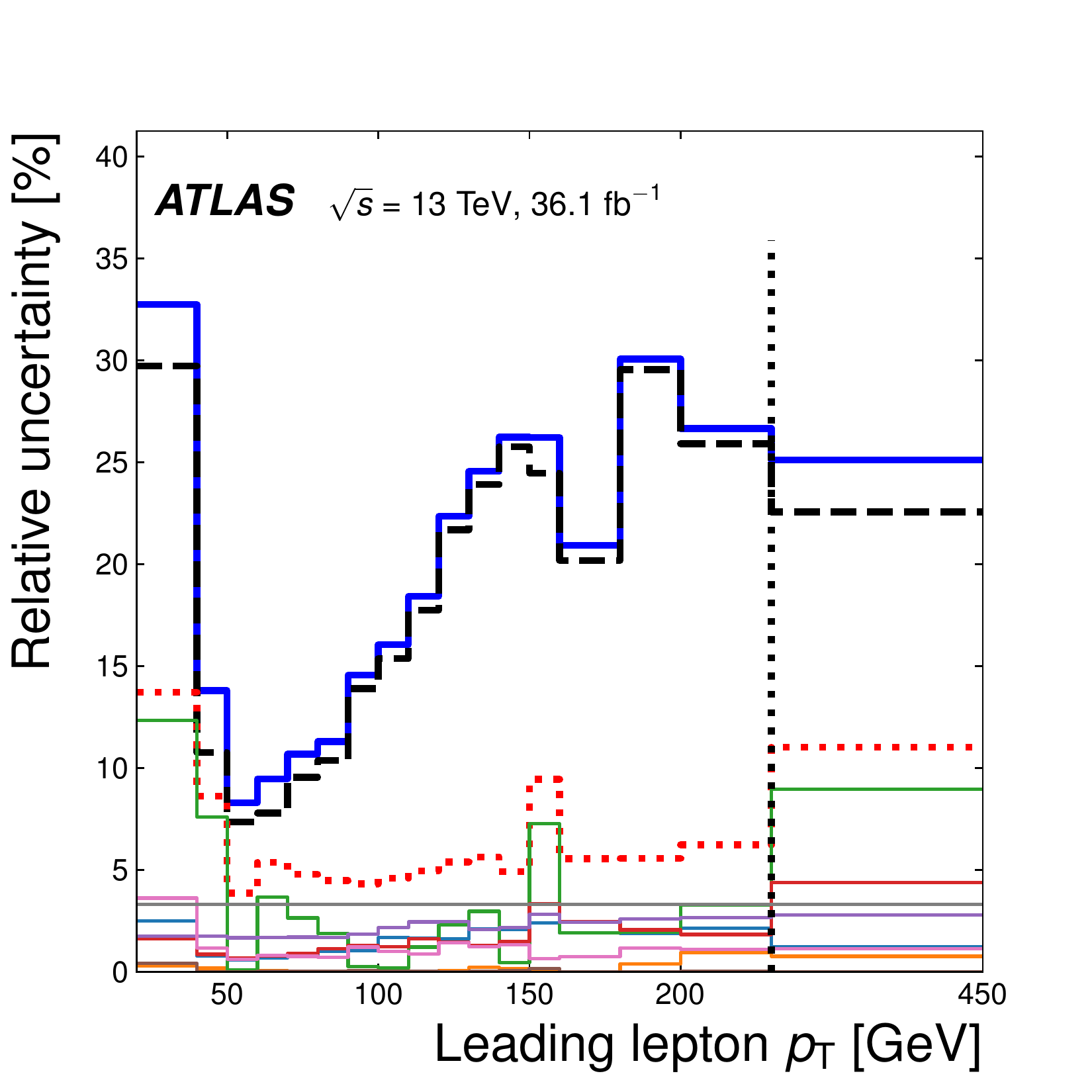}}
\subfigure{\includegraphics[width=0.48\textwidth]{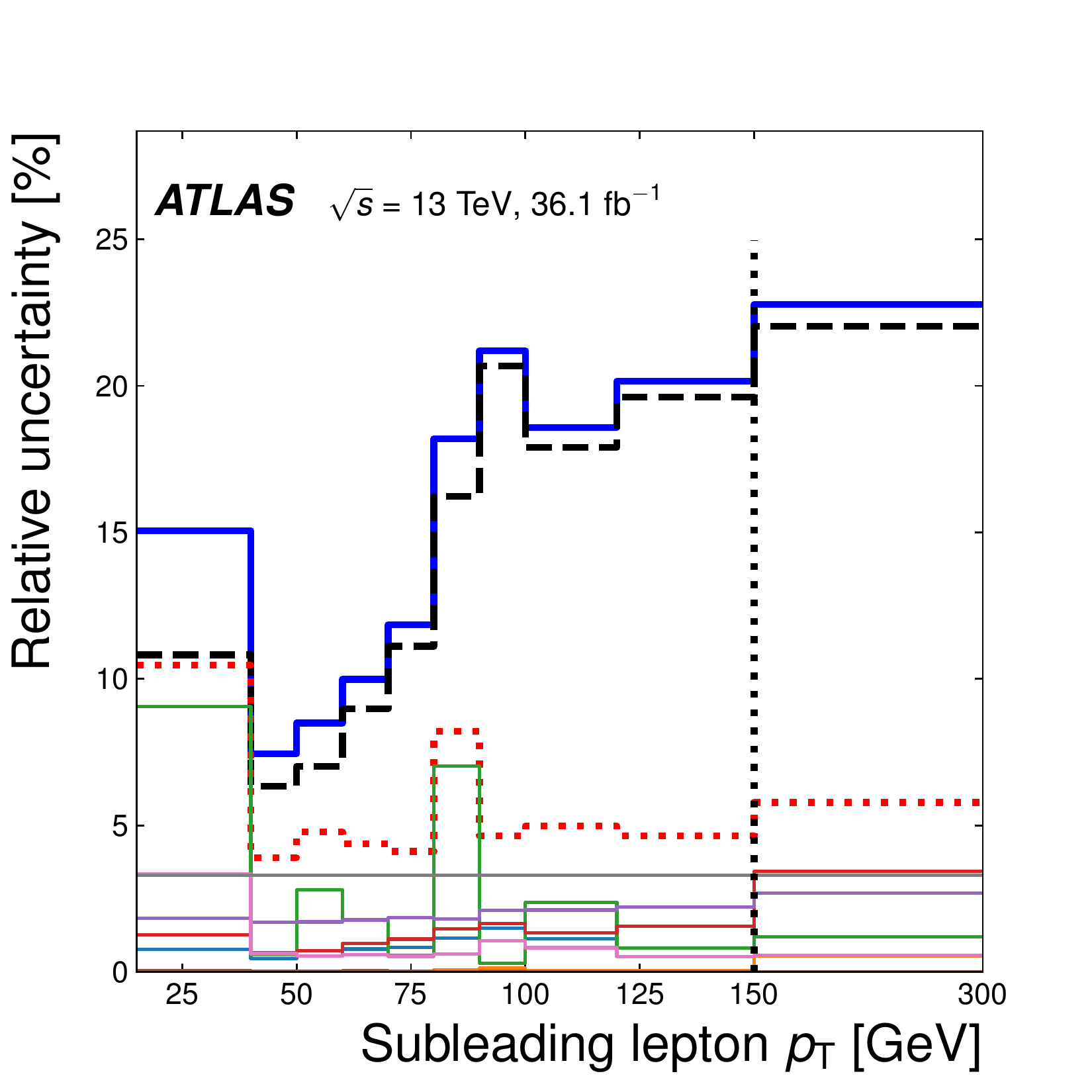}}
\subfigure{\includegraphics[width=0.48\textwidth]{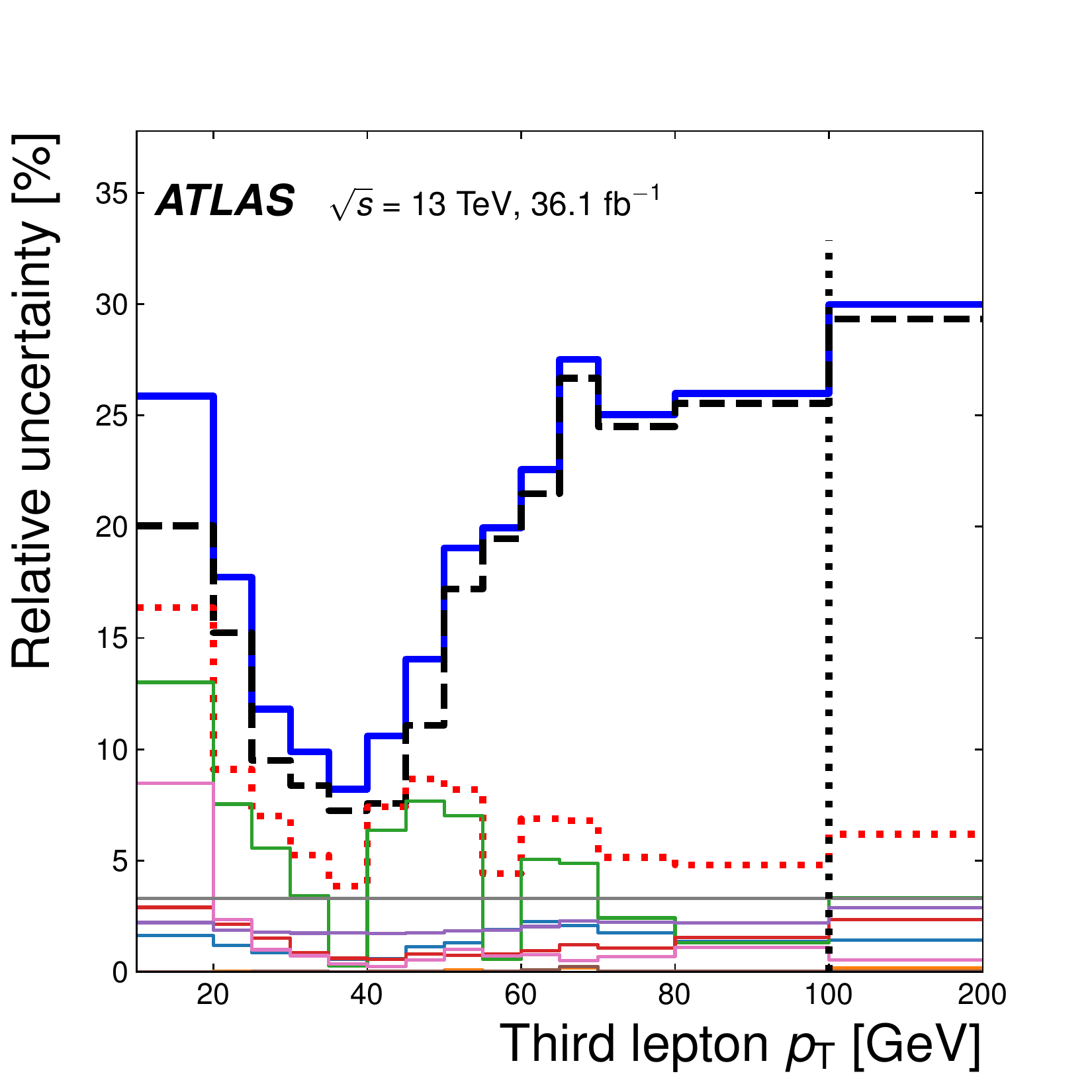}}
\subfigure{\includegraphics[width=0.48\textwidth]{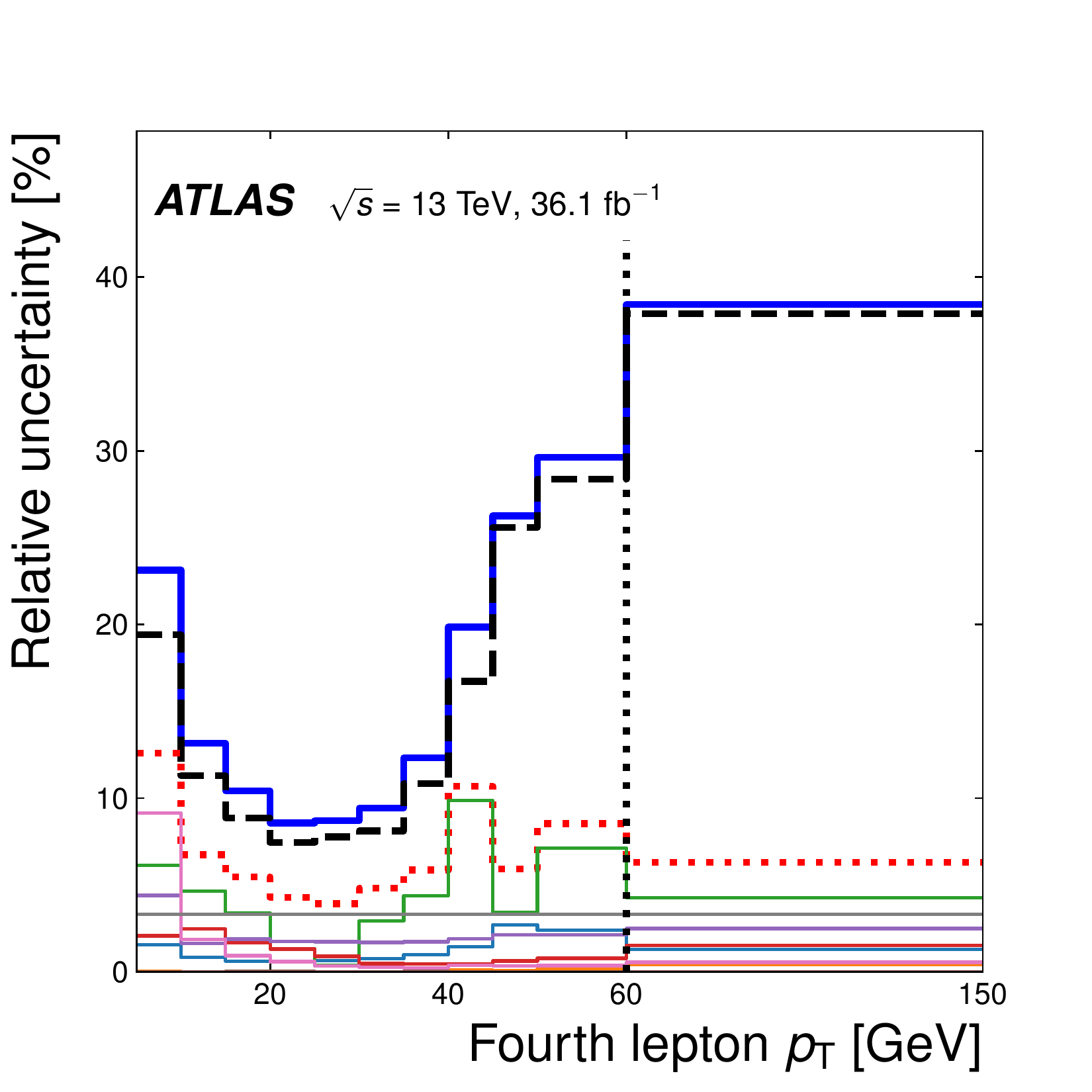}}
\subfigure{\includegraphics[width=0.48\textwidth]{diffuncerts_legend.pdf}}
\caption{Bin-by-bin uncertainty breakdown for various observables. For better visualisation, the last bin is shown using a different $x$-axis scale where indicated by a dashed vertical line.}
\end{figure}

\begin{figure}[h!]
\centering
\subfigure{\includegraphics[width=0.48\textwidth]{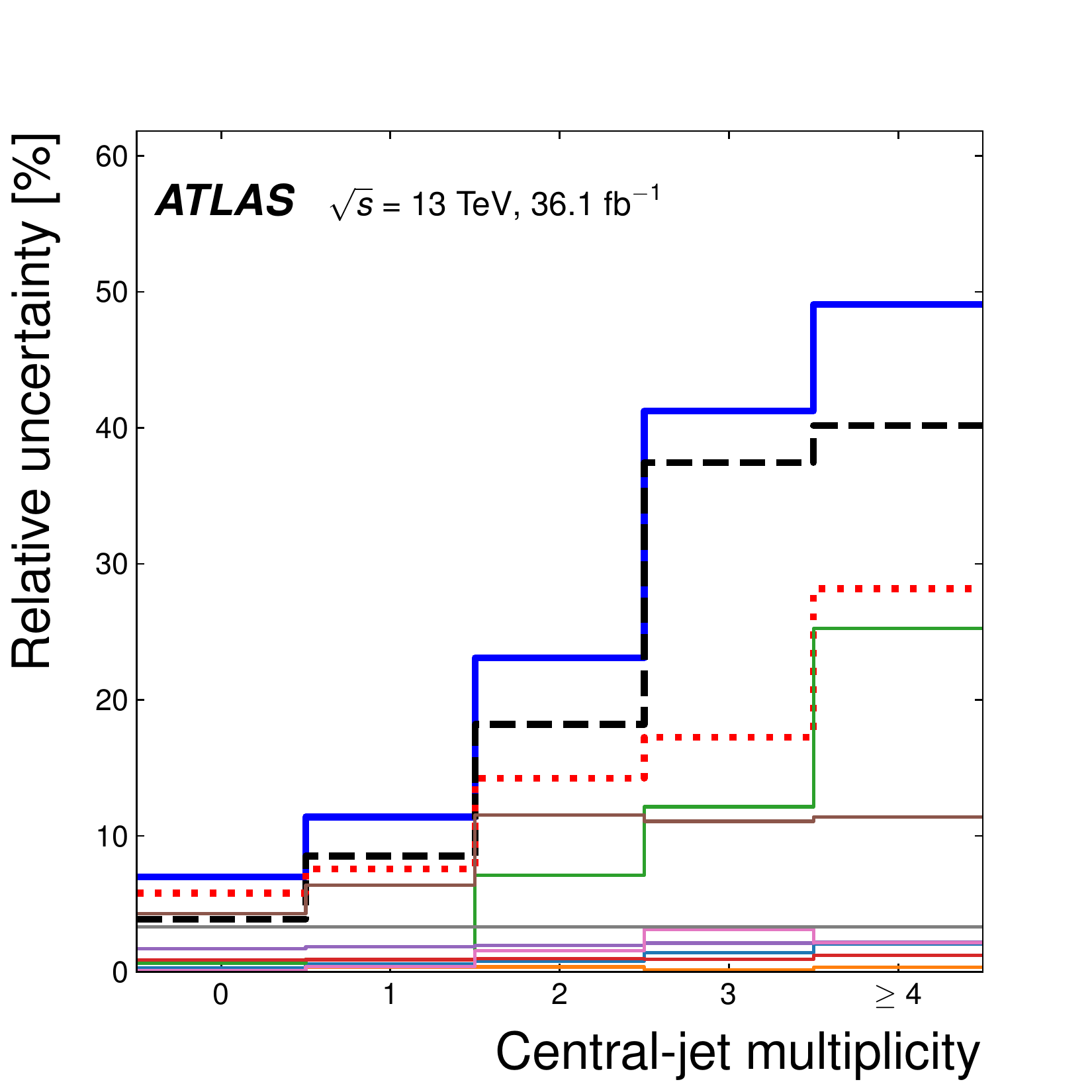}}
\subfigure{\includegraphics[width=0.48\textwidth]{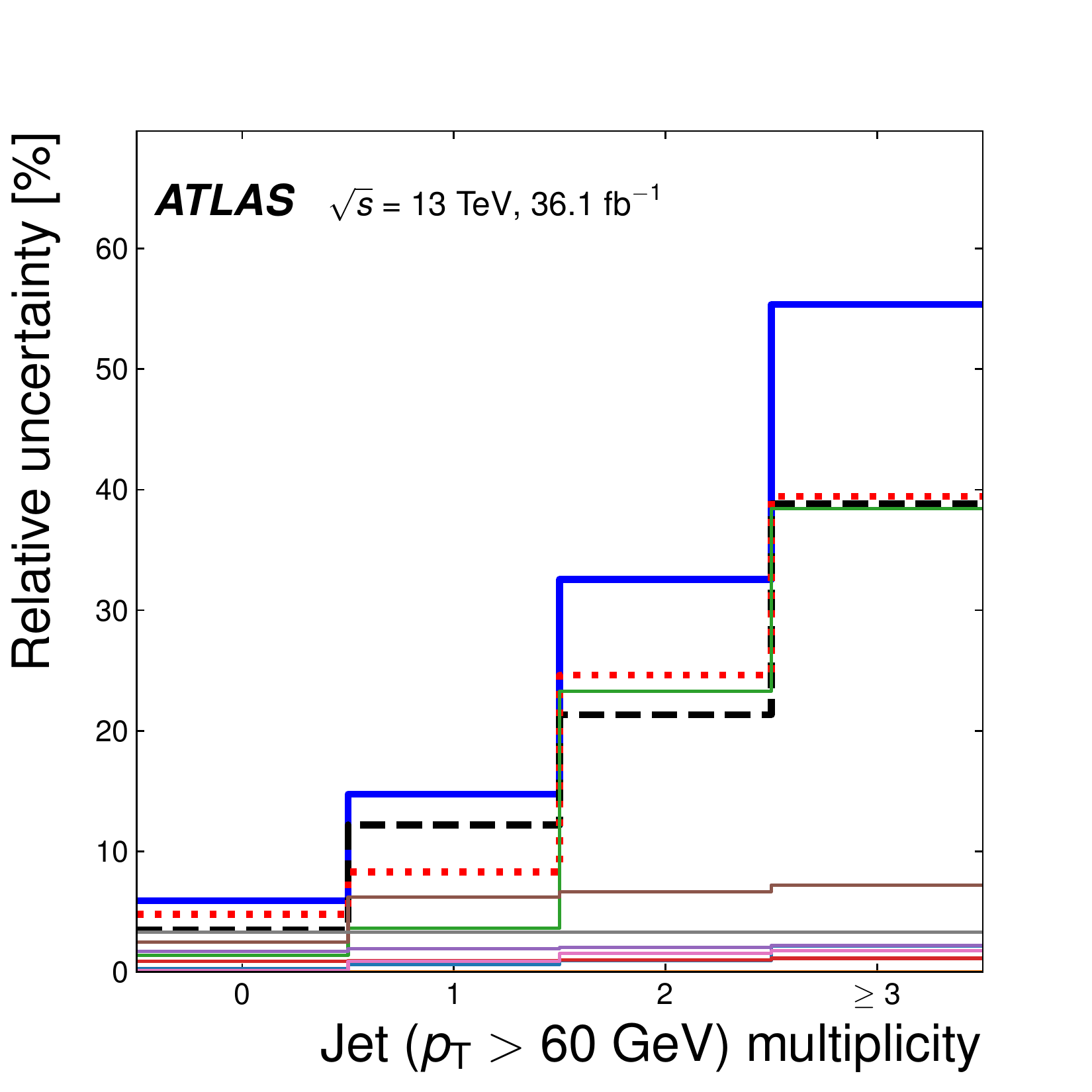}}
\subfigure{\includegraphics[width=0.48\textwidth]{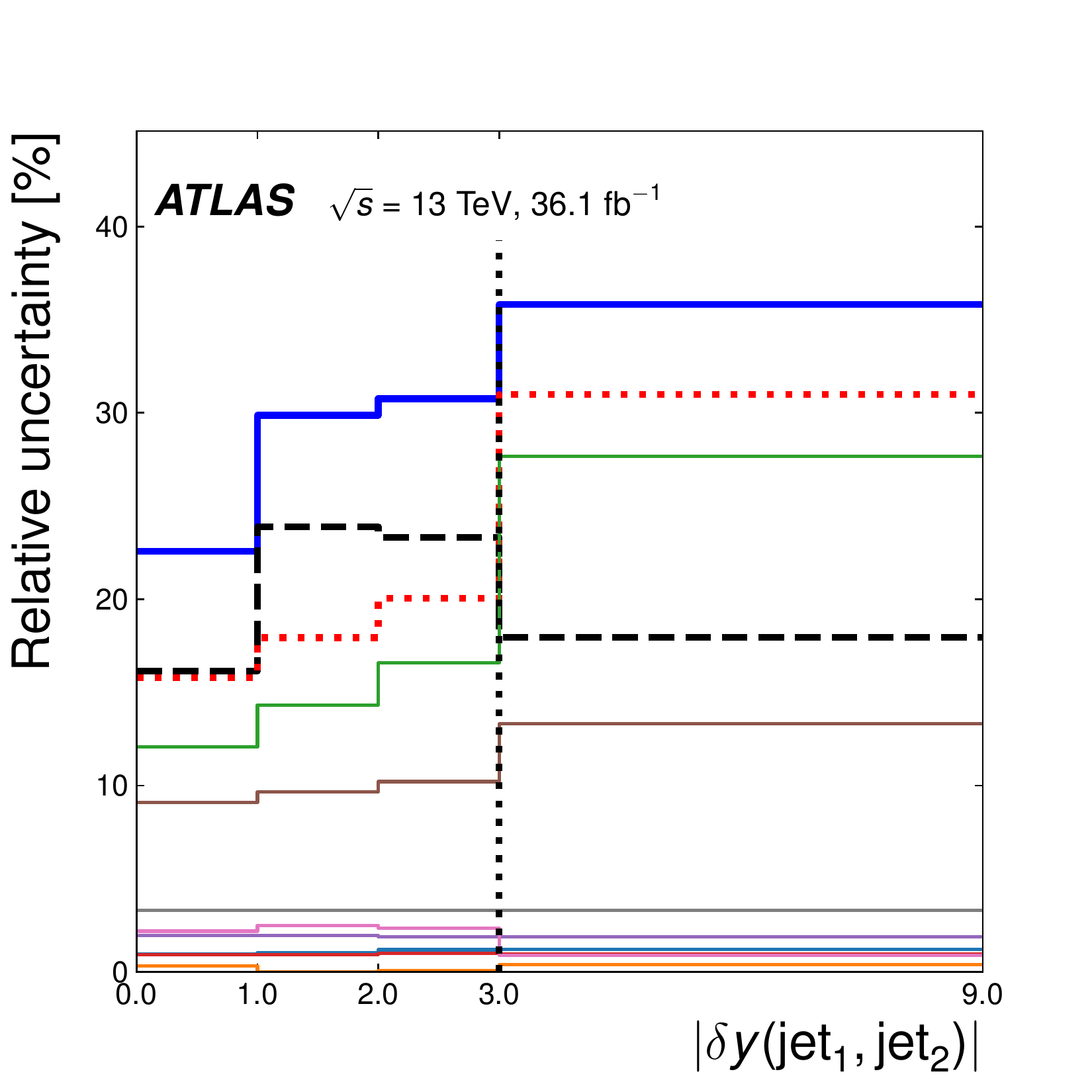}}
\subfigure{\includegraphics[width=0.48\textwidth]{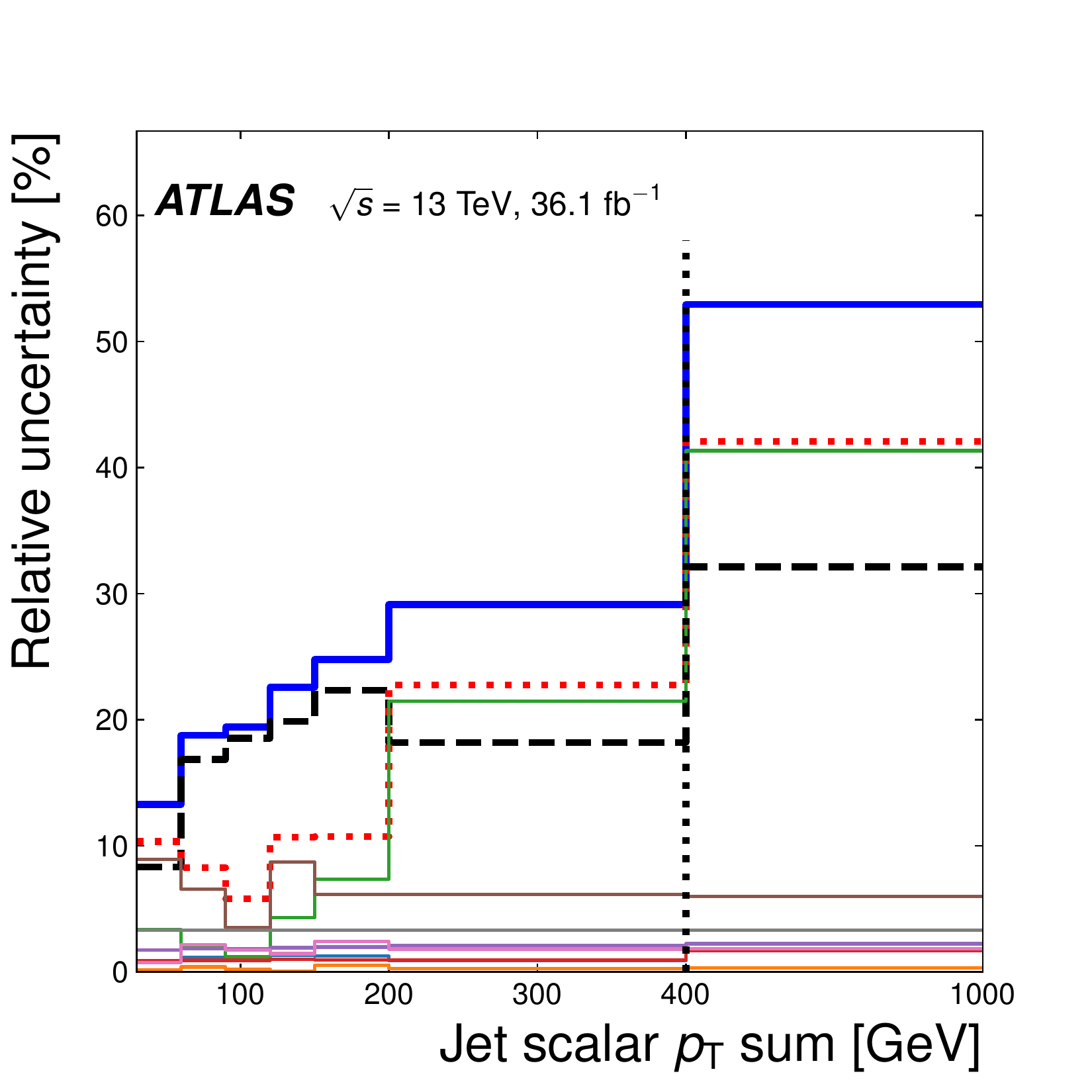}}
\subfigure{\includegraphics[width=0.48\textwidth]{diffuncerts_legend.pdf}}
\caption{Bin-by-bin uncertainty breakdown for various observables. For better visualisation, the last bin is shown using a different $x$-axis scale where indicated by a dashed vertical line.}
\end{figure}

\begin{figure}[h!]
\centering
\subfigure{\includegraphics[width=0.48\textwidth]{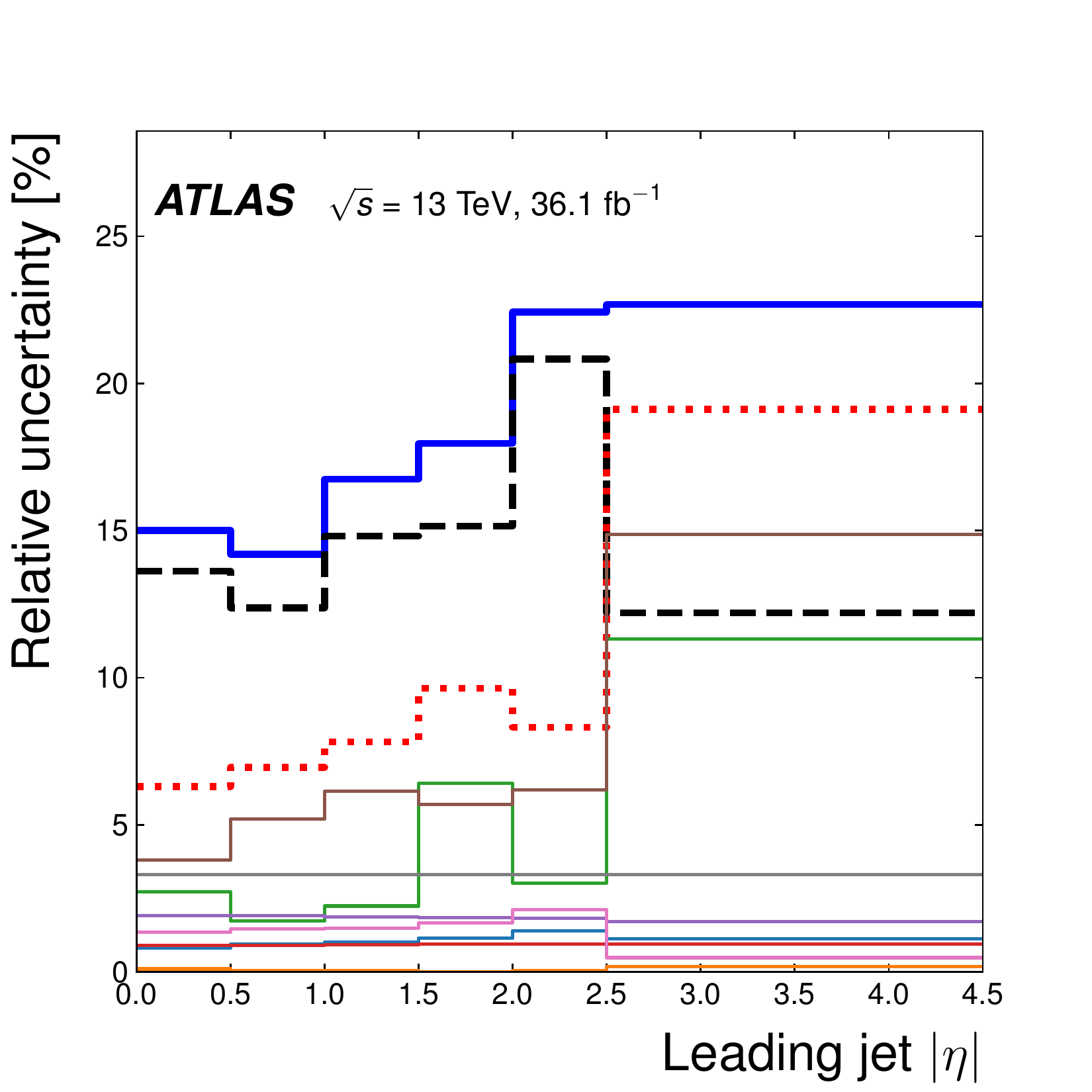}}
\subfigure{\includegraphics[width=0.48\textwidth]{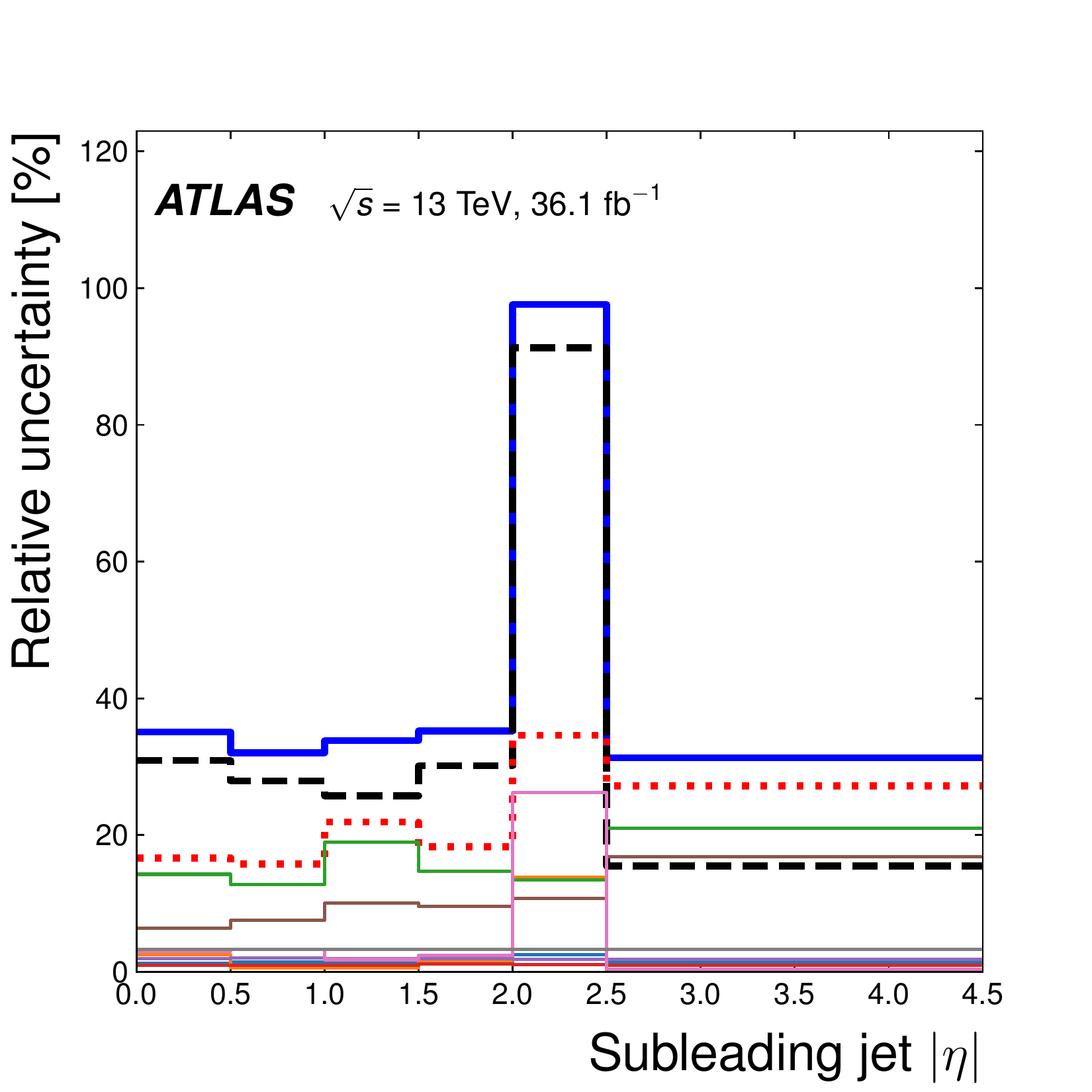}}
\subfigure{\includegraphics[width=0.48\textwidth]{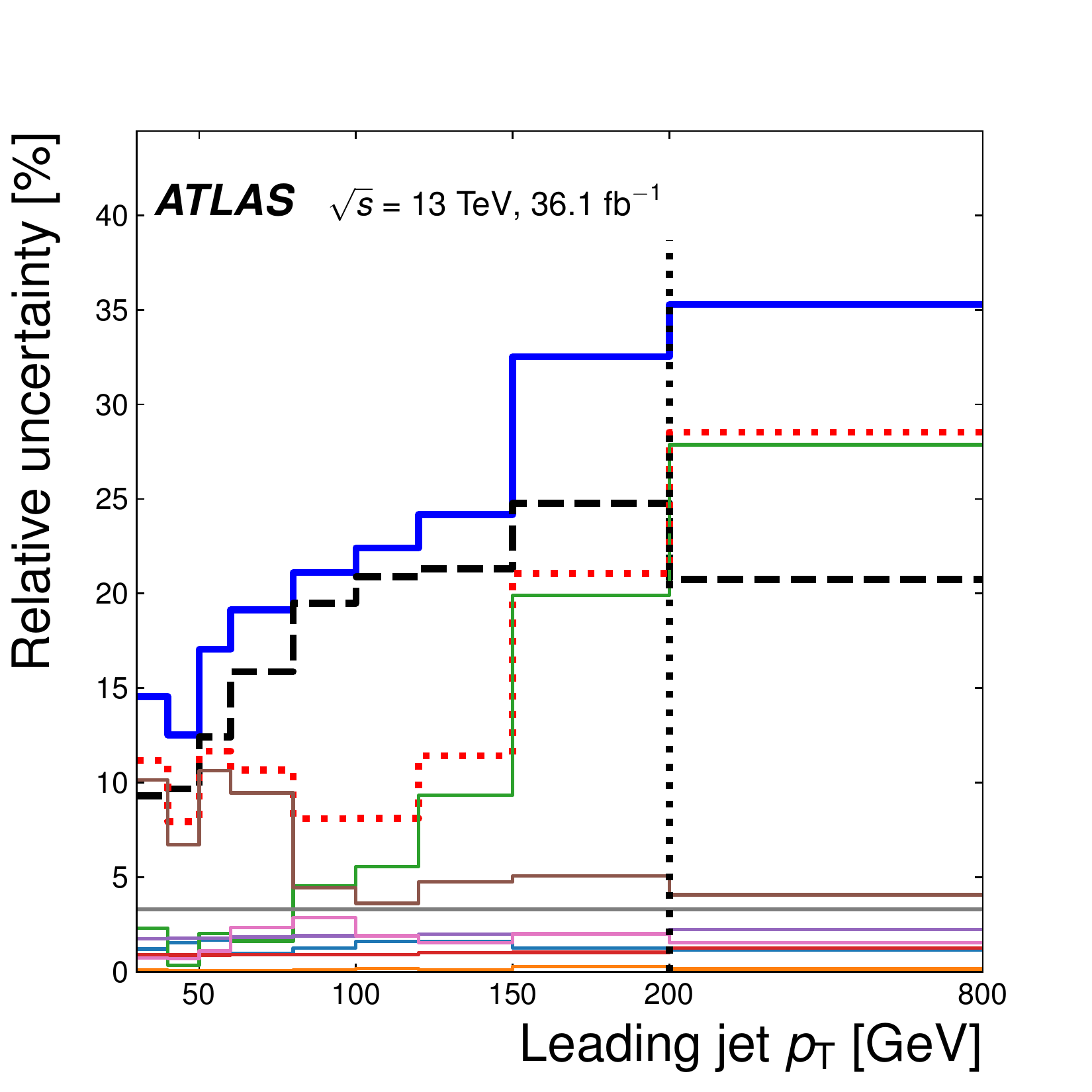}}
\subfigure{\includegraphics[width=0.48\textwidth]{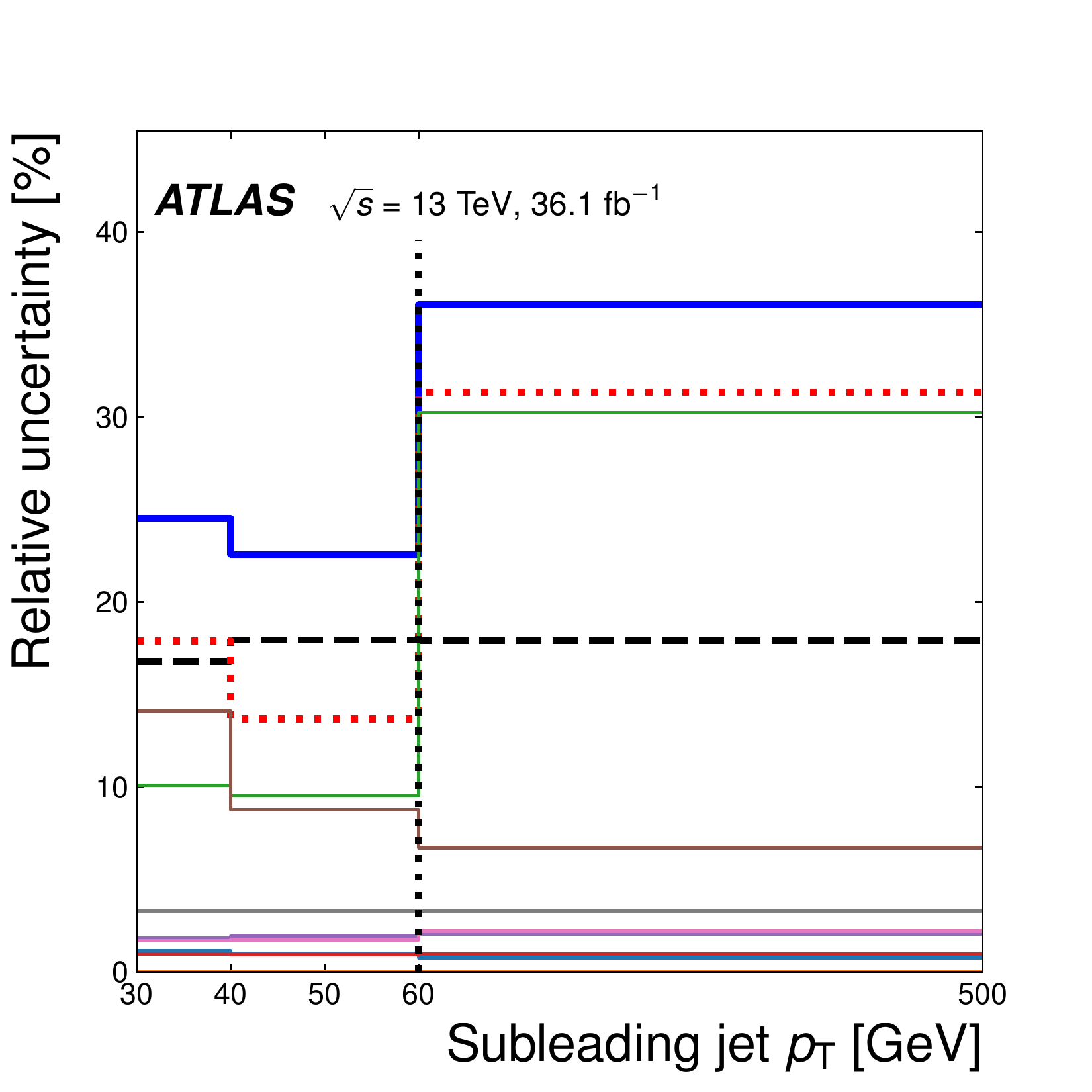}}
\subfigure{\includegraphics[width=0.48\textwidth]{diffuncerts_legend.pdf}}
\caption{Bin-by-bin uncertainty breakdown for various observables. For better visualisation, the last bin is shown using a different $x$-axis scale where indicated by a dashed vertical line.}
\label{fig:diffuncert_plots_last}
\end{figure}

\clearpage

\subsection{Bin-by-bin correlations of differential cross sections}
\label{sec:zz_aux_correlations}
\mytabs~\ref{tab:results_unfolded_corr_first}--\ref{tab:results_unfolded_corr_last} show the total uncertainty correlations between bins of the measured differential cross sections.
\mytabs~\ref{tab:results_unfolded_statcorr_first}--\ref{tab:results_unfolded_statcorr_last} show the statistical uncertainty correlations between bins of the measured differential cross sections.

\begin{landscape}

\begin{table}[h!]
\centering
\tiny
\input{Final_CombinedCorrelationMatrix_fourLepton_pT_binning0.tex}
\caption{Bin-to-bin uncertainty correlations in the differential cross section as a function of the four-lepton transverse momentum, including the statistical uncertainty of the data as well as the systematic uncertainties of the signal and background predictions.}
\label{tab:results_unfolded_corr_first}
\end{table}\clearpage

\begin{table}[h!]
\centering
\tiny
\input{Final_CombinedCorrelationMatrix_fourLepton_absy_binning0.tex}
\caption{Bin-to-bin uncertainty correlations in the differential cross section as a function of the absolute four-lepton rapidity, including the statistical uncertainty of the data as well as the systematic uncertainties of the signal and background predictions.}
\label{}
\end{table}

\begin{table}[h!]
\centering
\tiny
\input{Final_CombinedCorrelationMatrix_leadingDilepton_pT_binning0.tex}
\caption{Bin-to-bin uncertainty correlations in the differential cross section as a function of the transverse momentum of the leading-\pt{} \PZ{} candidate, including the statistical uncertainty of the data as well as the systematic uncertainties of the signal and background predictions.}
\label{}
\end{table}\clearpage

\begin{table}[h!]
\centering
\tiny
\input{Final_CombinedCorrelationMatrix_subleadingDilepton_pT_binning0.tex}
\caption{Bin-to-bin uncertainty correlations in the differential cross section as a function of the transverse momentum of the subleading-\pt{} \PZ{} candidate, including the statistical uncertainty of the data as well as the systematic uncertainties of the signal and background predictions.}
\label{}
\end{table}\clearpage

\begin{table}[h!]
\centering
\tiny
\input{Final_CombinedCorrelationMatrix_lepton1_pT_binning0.tex}
\caption{Bin-to-bin uncertainty correlations in the differential cross section as a function of the transverse momentum of the leading-\pt{} lepton in the final selected quadruplet, including the statistical uncertainty of the data as well as the systematic uncertainties of the signal and background predictions.}
\label{}
\end{table}

\begin{table}[h!]
\centering
\tiny
\input{Final_CombinedCorrelationMatrix_lepton2_pT_binning0.tex}
\caption{Bin-to-bin uncertainty correlations in the differential cross section as a function of the transverse momentum of the subleading-\pt{} lepton in the final selected quadruplet, including the statistical uncertainty of the data as well as the systematic uncertainties of the signal and background predictions.}
\label{}
\end{table}\clearpage

\begin{table}[h!]
\centering
\tiny
\input{Final_CombinedCorrelationMatrix_lepton3_pT_binning0.tex}
\caption{Bin-to-bin uncertainty correlations in the differential cross section as a function of the transverse momentum of the third lepton in the final selected quadruplet, including the statistical uncertainty of the data as well as the systematic uncertainties of the signal and background predictions.}
\label{}
\end{table}

\begin{table}[h!]
\centering
\tiny
\input{Final_CombinedCorrelationMatrix_lepton4_pT_binning0.tex}
\caption{Bin-to-bin uncertainty correlations in the differential cross section as a function of the transverse momentum of the fourth lepton in the final selected quadruplet, including the statistical uncertainty of the data as well as the systematic uncertainties of the signal and background predictions.}
\label{}
\end{table}\clearpage

\begin{table}[h!]
\centering
\tiny
\input{Final_CombinedCorrelationMatrix_dileptons_dy_binning0.tex}
\caption{Bin-to-bin uncertainty correlations in the differential cross section as a function of the rapidity difference between the two selected $\PZ$ candidates, including the statistical uncertainty of the data as well as the systematic uncertainties of the signal and background predictions.}
\label{}
\end{table}

\begin{table}[h!]
\centering
\tiny
\input{Final_CombinedCorrelationMatrix_dileptons_dphi_binning0.tex}
\caption{Bin-to-bin uncertainty correlations in the differential cross section as a function of the azimuthal-angle difference between the two selected $\PZ$ candidates, including the statistical uncertainty of the data as well as the systematic uncertainties of the signal and background predictions. The bin edges are given in units of $\pi$.}
\label{}
\end{table}\clearpage

\begin{table}[h!]
\centering
\tiny
\input{Final_CombinedCorrelationMatrix_jets_N_binning0.tex}
\caption{Bin-to-bin uncertainty correlations in the differential cross section as a function of the number of jets, including the statistical uncertainty of the data as well as the systematic uncertainties of the signal and background predictions.}
\label{}
\end{table}

\begin{table}[h!]
\centering
\tiny
\input{Final_CombinedCorrelationMatrix_centralJets_N_binning0.tex}
\caption{Bin-to-bin uncertainty correlations in the differential cross section as a function of the number of central jets, including the statistical uncertainty of the data as well as the systematic uncertainties of the signal and background predictions.}
\label{}
\end{table}

\begin{table}[h!]
\centering
\tiny
\input{Final_CombinedCorrelationMatrix_jets60_N_binning0.tex}
\caption{Bin-to-bin uncertainty correlations in the differential cross section as a function of the number of jets with $\pt > 60$~\GeV{}, including the statistical uncertainty of the data as well as the systematic uncertainties of the signal and background predictions.}
\label{}
\end{table}

\begin{table}[h!]
\centering
\tiny
\input{Final_CombinedCorrelationMatrix_hardestJetsDijet_mass_binning0.tex}
\caption{Bin-to-bin uncertainty correlations in the differential cross section as a function of the mass of the dijet formed of the two leading jets, including the statistical uncertainty of the data as well as the systematic uncertainties of the signal and background.}
\label{}
\end{table}

\begin{table}[h!]
\centering
\tiny
\input{Final_CombinedCorrelationMatrix_hardestJetsDijet_dy_binning0.tex}
\caption{Bin-to-bin uncertainty correlations in the differential cross section as a function of the rapidity difference of the two leading jets, including the statistical uncertainty of the data as well as the systematic uncertainties of the signal and background predictions.}
\label{}
\end{table}
\clearpage

\begin{table}[h!]
\centering
\tiny
\input{Final_CombinedCorrelationMatrix_jets_scalarPtSum_binning0.tex}
\caption{Bin-to-bin uncertainty correlations in the differential cross section as a function of the scalar transverse momentum sum of all jets, including the statistical uncertainty of the data as well as the systematic uncertainties of the signal and background.}
\label{}
\end{table}

\begin{table}[h!]
\centering
\tiny
\input{Final_CombinedCorrelationMatrix_jet1_pt_binning0.tex}
\caption{Bin-to-bin uncertainty correlations in the differential cross section as a function of the leading-\pt{} jet transverse momentum, including the statistical uncertainty of the data as well as the systematic uncertainties of the signal and background.}
\label{}
\end{table}

\begin{table}[h!]
\centering
\tiny
\input{Final_CombinedCorrelationMatrix_jet2_pt_binning0.tex}
\caption{Bin-to-bin uncertainty correlations in the differential cross section as a function of the subleading-\pt{} jet transverse momentum, including the statistical uncertainty of the data as well as the systematic uncertainties of the signal and background.}
\label{}
\end{table}\clearpage

\begin{table}[h!]
\centering
\tiny
\input{Final_CombinedCorrelationMatrix_jet1_eta_binning0.tex}
\caption{Bin-to-bin uncertainty correlations in the differential cross section as a function of the leading-\pt{} jet absolute pseudorapidity, including the statistical uncertainty of the data as well as the systematic uncertainties of the signal and background predictions.}
\label{}
\end{table}

\begin{table}[h!]
\centering
\tiny
\input{Final_CombinedCorrelationMatrix_jet2_eta_binning0.tex}
\caption{Bin-to-bin uncertainty correlations in the differential cross section as a function of the subleading-\pt{} jet absolute pseudorapidity, including the statistical uncertainty of the data as well as the systematic uncertainties of the signal and background predictions.}
\label{tab:results_unfolded_corr_last}
\end{table}\clearpage

\begin{table}[h!]
\centering
\tiny
\input{Final_StatCorrelationMatrix_fourLepton_pT_binning0.tex}
\caption{Bin-to-bin data statistical uncertainty correlations in the differential cross section as a function of the four-lepton transverse momentum.}
\label{tab:results_unfolded_statcorr_first}
\end{table}\clearpage

\begin{table}[h!]
\centering
\tiny
\input{Final_StatCorrelationMatrix_fourLepton_absy_binning0.tex}
\caption{Bin-to-bin data statistical uncertainty correlations in the differential cross section as a function of the absolute four-lepton rapidity.}
\label{}
\end{table}

\begin{table}[h!]
\centering
\tiny
\input{Final_StatCorrelationMatrix_leadingDilepton_pT_binning0.tex}
\caption{Bin-to-bin data statistical uncertainty correlations in the differential cross section as a function of the transverse momentum of the leading-\pt{} \PZ{} candidate.}
\label{}
\end{table}\clearpage

\begin{table}[h!]
\centering
\tiny
\input{Final_StatCorrelationMatrix_subleadingDilepton_pT_binning0.tex}
\caption{Bin-to-bin data statistical uncertainty correlations in the differential cross section as a function of the transverse momentum of the subleading-\pt{} \PZ{} candidate.}
\label{}
\end{table}\clearpage

\begin{table}[h!]
\centering
\tiny
\input{Final_StatCorrelationMatrix_lepton1_pT_binning0.tex}
\caption{Bin-to-bin data statistical uncertainty correlations in the differential cross section as a function of the transverse momentum of the leading-\pt{} lepton in the final selected quadruplet.}
\label{}
\end{table}

\begin{table}[h!]
\centering
\tiny
\input{Final_StatCorrelationMatrix_lepton2_pT_binning0.tex}
\caption{Bin-to-bin data statistical uncertainty correlations in the differential cross section as a function of the transverse momentum of the subleading-\pt{} lepton in the final selected quadruplet.}
\label{}
\end{table}\clearpage

\begin{table}[h!]
\centering
\tiny
\input{Final_StatCorrelationMatrix_lepton3_pT_binning0.tex}
\caption{Bin-to-bin data statistical uncertainty correlations in the differential cross section as a function of the transverse momentum of the third lepton in the final selected quadruplet.}
\label{}
\end{table}

\begin{table}[h!]
\centering
\tiny
\input{Final_StatCorrelationMatrix_lepton4_pT_binning0.tex}
\caption{Bin-to-bin data statistical uncertainty correlations in the differential cross section as a function of the transverse momentum of the fourth lepton in the final selected quadruplet.}
\label{}
\end{table}\clearpage

\begin{table}[h!]
\centering
\tiny
\input{Final_StatCorrelationMatrix_dileptons_dy_binning0.tex}
\caption{Bin-to-bin data statistical uncertainty correlations in the differential cross section as a function of the rapidity difference between the two selected $\PZ$ candidates.}
\label{}
\end{table}

\begin{table}[h!]
\centering
\tiny
\input{Final_StatCorrelationMatrix_dileptons_dphi_binning0.tex}
\caption{Bin-to-bin data statistical uncertainty correlations in the differential cross section as a function of the azimuthal-angle difference between the two selected $\PZ$ candidates. The bin edges are given in units of $\pi$.}
\label{}
\end{table}\clearpage

\begin{table}[h!]
\centering
\tiny
\input{Final_StatCorrelationMatrix_jets_N_binning0.tex}
\caption{Bin-to-bin data statistical uncertainty correlations in the differential cross section as a function of the number of jets.}
\label{}
\end{table}

\begin{table}[h!]
\centering
\tiny
\input{Final_StatCorrelationMatrix_centralJets_N_binning0.tex}
\caption{Bin-to-bin data statistical uncertainty correlations in the differential cross section as a function of the number of central jets.}
\label{}
\end{table}

\begin{table}[h!]
\centering
\tiny
\input{Final_StatCorrelationMatrix_jets60_N_binning0.tex}
\caption{Bin-to-bin data statistical uncertainty correlations in the differential cross section as a function of the number of jets with $\pt > 60$~\GeV{}.}
\label{}
\end{table}

\begin{table}[h!]
\centering
\tiny
\input{Final_StatCorrelationMatrix_hardestJetsDijet_mass_binning0.tex}
\caption{Bin-to-bin data statistical uncertainty correlations in the differential cross section as a function of the mass of the dijet formed of the two leading jets.}
\label{}
\end{table}

\begin{table}[h!]
\centering
\tiny
\input{Final_StatCorrelationMatrix_hardestJetsDijet_dy_binning0.tex}
\caption{Bin-to-bin data statistical uncertainty correlations in the differential cross section as a function of the rapidity difference of the two leading jets.}
\label{}
\end{table}
\clearpage

\begin{table}[h!]
\centering
\tiny
\input{Final_StatCorrelationMatrix_jets_scalarPtSum_binning0.tex}
\caption{Bin-to-bin data statistical uncertainty correlations in the differential cross section as a function of the scalar transverse momentum sum of all jets.}
\label{}
\end{table}

\begin{table}[h!]
\centering
\tiny
\input{Final_StatCorrelationMatrix_jet1_pt_binning0.tex}
\caption{Bin-to-bin data statistical uncertainty correlations in the differential cross section as a function of the leading-\pt{} jet transverse momentum.}
\label{}
\end{table}

\begin{table}[h!]
\centering
\tiny
\input{Final_StatCorrelationMatrix_jet2_pt_binning0.tex}
\caption{Bin-to-bin data statistical uncertainty correlations in the differential cross section as a function of the subleading-\pt{} jet transverse momentum.}
\label{}
\end{table}\clearpage

\begin{table}[h!]
\centering
\tiny
\input{Final_StatCorrelationMatrix_jet1_eta_binning0.tex}
\caption{Bin-to-bin data statistical uncertainty correlations in the differential cross section as a function of the leading-\pt{} jet absolute pseudorapidity.}
\label{}
\end{table}

\begin{table}[h!]
\centering
\tiny
\input{Final_StatCorrelationMatrix_jet2_eta_binning0.tex}
\caption{Bin-to-bin data statistical uncertainty correlations in the differential cross section as a function of the subleading-\pt{} jet absolute pseudorapidity.}
\label{tab:results_unfolded_statcorr_last}
\end{table}\clearpage

\end{landscape}


%
%
%
%

\clearpage\pagebreak
\addcontentsline{toc}{section}{References}
\printbibliography
\end{fmffile}
\end{document}

%% file: data.tex
249 & 465 & 303 & 1017\\

%% file: sherpaprediction.tex
$198\phantom{.0\pm}^{+16}_{-14}$ & $469\phantom{.0\pm}^{+35}_{-31}$ & $290\phantom{.0\pm}^{+22}_{-21}$ & \phantom{0}$958\phantom{.0\pm}^{+70}_{-63}$\\

%% file: sherpaqqbar.tex
$168\phantom{.0\pm}^{+14}_{-13}$ & $400\phantom{.0\pm}^{+31}_{-28}$ & $246\phantom{.0\pm}^{+19}_{-18}$ & \phantom{0}$814\phantom{.0\pm}^{+63}_{-57}$\\

%% file: gg.tex
$\phantom{0}21.3 \phantom{0}\pm 3.5$ & $\phantom{0}50.2\phantom{0}\pm 8.2$ & $\phantom{0}29.7\phantom{0}\pm 4.9$ & \phantom{0}$101\phantom{.00}\pm 17$\\

%% file: zzjj.tex
$\phantom{00}4.36\pm 0.42$ & $\phantom{0}10.23\pm 0.72$ & $\phantom{00}6.43\pm 0.55$ & \phantom{0}$\phantom{0}21.0\phantom{0}\pm1.2$\\

%% file: taubkg.tex
$\phantom{00}0.59 \pm 0.09$ & $\phantom{00}0.55 \pm 0.08$ & $\phantom{00}0.55 \pm 0.09$ & \phantom{0}$\phantom{00}1.69 \pm 0.16$\\ 

%% file: tribosonbkg.tex
$\phantom{00}0.68 \pm 0.21$ & $\phantom{00}1.50 \pm 0.46$ & $\phantom{00}0.96 \pm 0.30$ & \phantom{0}$\phantom{00}3.14 \pm 0.30$\\

%% file: ttzbkg.tex
$\phantom{00}0.81 \pm 0.25$ & $\phantom{00}1.86 \pm 0.56$ & $\phantom{00}1.42 \pm 0.43$ & \phantom{0}$\phantom{00}4.1\phantom{0} \pm 1.2$\\

%% file: fakebkg.tex
$\phantom{00}2.1\phantom{0}\pm 2.1$ & $\phantom{00}4.9\phantom{0}\pm 3.9$ & $\phantom{00}5.3\phantom{0}\pm 5.2$ & \phantom{0}$\phantom{0}12.3\phantom{0}\pm 8.3$\\

%% file: powhegprediction_withKfac.tex
$193 \phantom{.00}\pm 11$ & $456 \phantom{.00}\pm 24$ & $286 \phantom{.00}\pm 17$ & \phantom{0}$934 \phantom{.00}\pm 50$\\

%% file: CZZ_powhegqqbar.tex
$0.4821 \pm 0.0033$ & $0.5925 \pm 0.0024$ & $0.7103 \pm 0.0032$ & $0.5946 \pm 0.0017$\\

%% file: CZZ_sherpaqqbar.tex
$0.4928 \pm 0.0018$ & $0.6051 \pm 0.0010$ & $0.7114 \pm 0.0010$ & $0.6040 \pm 0.0007$\\

%% file: CZZ_sherpagg.tex
$0.5069 \pm 0.0037$ & $0.6025 \pm 0.0025$ & $0.7086 \pm 0.0031$ & $0.6050 \pm 0.0018$\\

%% file: CZZ_sherpazzjj.tex
$0.4801 \pm 0.0229$ & $0.5790 \pm 0.0123$ & $0.6480 \pm 0.0121$ & $0.5731 \pm 0.0087$\\

%% file: CZZ_line.tex
$0.4940 \pm 0.0017$ & $0.6042 \pm 0.0010$ & $0.7095 \pm 0.0010$ & $0.6033 \pm 0.0007$\\

%% file: Final_CombinedCorrelationMatrix_fourLepton_pT_binning0.tex
\begin{tabular}{lrrrrrrrrrrrrrrr}
\toprule
$p_{\mathrm{T}, 4\ell}$ [GeV] &  0--5 &  5--15 &  15--25 &  25--35 &  35--45 &  45--55 &  55--65 &  65--75 &  75--85 &  85--100 &  100--125 &  125--150 &  150--200 &  200--250 &  250--1500 \\
\midrule
0--5      &  1.00 &   0.28 &   -0.15 &   -0.19 &   -0.11 &   -0.05 &   -0.16 &    0.01 &   -0.01 &    -0.15 &     -0.12 &     -0.20 &     -0.27 &     -0.27 &      -0.39 \\
5--15     &  0.28 &   1.00 &    0.02 &   -0.01 &    0.03 &    0.09 &   -0.03 &    0.03 &    0.07 &     0.01 &      0.02 &     -0.02 &     -0.02 &     -0.04 &      -0.05 \\
15--25    & -0.15 &   0.02 &    1.00 &    0.12 &    0.02 &    0.10 &    0.07 &    0.03 &    0.04 &     0.10 &      0.04 &      0.09 &      0.09 &      0.10 &       0.07 \\
25--35    & -0.19 &  -0.01 &    0.12 &    1.00 &    0.13 &    0.02 &    0.10 &    0.01 &    0.06 &     0.08 &      0.11 &      0.10 &      0.14 &      0.12 &       0.15 \\
35--45    & -0.11 &   0.03 &    0.02 &    0.13 &    1.00 &    0.10 &    0.04 &    0.05 &    0.06 &     0.08 &      0.06 &      0.05 &      0.06 &      0.08 &       0.10 \\
45--55    & -0.05 &   0.09 &    0.10 &    0.02 &    0.10 &    1.00 &    0.13 &    0.04 &    0.03 &     0.08 &      0.04 &      0.01 &      0.06 &      0.13 &       0.08 \\
55--65    & -0.16 &  -0.03 &    0.07 &    0.10 &    0.04 &    0.13 &    1.00 &    0.13 &   -0.01 &     0.07 &      0.07 &      0.10 &      0.07 &      0.12 &       0.12 \\
\bottomrule
\end{tabular}

%% file: Final_CombinedCorrelationMatrix_fourLepton_absy_binning0.tex
\begin{tabular}{lrrrrrrrrrrr}
\toprule
$|y_{4\ell}|$ &  0--0.2 &  0.2--0.4 &  0.4--0.6 &  0.6--0.8 &  0.8--1 &  1--1.2 &  1.2--1.4 &  1.4--1.6 &  1.6--1.8 &  1.8--2 &  2--4.6 \\
\midrule
0--0.2   &    1.00 &      0.04 &      0.05 &      0.03 &    0.09 &    0.09 &      0.08 &      0.08 &      0.02 &    0.02 &    0.03 \\
0.2--0.4 &    0.04 &      1.00 &      0.03 &      0.08 &    0.03 &    0.09 &      0.01 &      0.02 &      0.02 &    0.07 &    0.05 \\
0.4--0.6 &    0.05 &      0.03 &      1.00 &      0.01 &    0.06 &    0.09 &      0.07 &      0.03 &      0.02 &    0.04 &    0.00 \\
0.6--0.8 &    0.03 &      0.08 &      0.01 &      1.00 &    0.02 &    0.05 &      0.05 &      0.08 &      0.02 &    0.00 &    0.08 \\
0.8--1   &    0.09 &      0.03 &      0.06 &      0.02 &    1.00 &    0.06 &      0.05 &      0.07 &      0.05 &    0.06 &    0.07 \\
1--1.2   &    0.09 &      0.09 &      0.09 &      0.05 &    0.06 &    1.00 &     -0.01 &      0.09 &      0.05 &    0.04 &    0.07 \\
1.2--1.4 &    0.08 &      0.01 &      0.07 &      0.05 &    0.05 &   -0.01 &      1.00 &      0.03 &      0.04 &    0.05 &    0.07 \\
1.4--1.6 &    0.08 &      0.02 &      0.03 &      0.08 &    0.07 &    0.09 &      0.03 &      1.00 &      0.03 &    0.02 &    0.04 \\
1.6--1.8 &    0.02 &      0.02 &      0.02 &      0.02 &    0.05 &    0.05 &      0.04 &      0.03 &      1.00 &    0.02 &    0.09 \\
1.8--2   &    0.02 &      0.07 &      0.04 &      0.00 &    0.06 &    0.04 &      0.05 &      0.02 &      0.02 &    1.00 &    0.03 \\
2--4.6   &    0.03 &      0.05 &      0.00 &      0.08 &    0.07 &    0.07 &      0.07 &      0.04 &      0.09 &    0.03 &    1.00 \\
\bottomrule
\end{tabular}

%% file: Final_CombinedCorrelationMatrix_leadingDilepton_pT_binning0.tex
\begin{tabular}{lrrrrrrrrrrrrrrrr}
\toprule
$p_{\mathrm{T}, Z_{1}}$ [GeV] &  0--10 &  10--20 &  20--30 &  30--40 &  40--50 &  50--60 &  60--70 &  70--80 &  80--90 &  90--100 &  100--120 &  120--140 &  140--160 &  160--200 &  200--250 &  250--1500 \\
\midrule
0--10     &   1.00 &   -0.02 &   -0.03 &    0.03 &    0.07 &    0.03 &   -0.01 &    0.03 &    0.05 &     0.01 &     -0.04 &      0.01 &     -0.00 &      0.06 &     -0.02 &      -0.07 \\
10--20    &  -0.02 &    1.00 &   -0.04 &   -0.00 &    0.01 &    0.04 &    0.00 &    0.05 &    0.04 &     0.02 &      0.01 &      0.05 &      0.05 &      0.02 &     -0.00 &      -0.02 \\
20--30    &  -0.03 &   -0.04 &    1.00 &    0.01 &   -0.04 &    0.04 &    0.07 &    0.02 &    0.03 &     0.08 &      0.03 &      0.04 &      0.05 &      0.05 &      0.03 &       0.03 \\
30--40    &   0.03 &   -0.00 &    0.01 &    1.00 &    0.01 &    0.06 &    0.06 &    0.05 &    0.05 &     0.07 &      0.04 &     -0.01 &      0.05 &      0.04 &      0.05 &       0.03 \\
40--50    &   0.07 &    0.01 &   -0.04 &    0.01 &    1.00 &    0.02 &    0.03 &    0.08 &    0.10 &     0.05 &      0.00 &      0.10 &      0.03 &      0.04 &     -0.00 &      -0.02 \\
50--60    &   0.03 &    0.04 &    0.04 &    0.06 &    0.02 &    1.00 &    0.03 &    0.09 &    0.06 &     0.05 &      0.08 &      0.01 &      0.05 &      0.02 &      0.08 &       0.09 \\
60--70    &  -0.01 &    0.00 &    0.07 &    0.06 &    0.03 &    0.03 &    1.00 &    0.11 &   -0.01 &     0.03 &      0.10 &      0.04 &      0.01 &      0.05 &      0.03 &       0.09 \\
\bottomrule
\end{tabular}

%% file: Final_CombinedCorrelationMatrix_subleadingDilepton_pT_binning0.tex
\begin{tabular}{lrrrrrrrrrrrrrrrr}
\toprule
$p_{\mathrm{T}, Z_{2}}$ [GeV] &  0--10 &  10--20 &  20--30 &  30--40 &  40--50 &  50--60 &  60--70 &  70--80 &  80--90 &  90--100 &  100--120 &  120--140 &  140--160 &  160--200 &  200--250 &  250--1500 \\
\midrule
0--10     &   1.00 &    0.03 &    0.01 &    0.05 &    0.05 &    0.07 &    0.01 &    0.02 &    0.03 &     0.01 &     -0.01 &     -0.03 &      0.05 &      0.03 &      0.03 &       0.04 \\
10--20    &   0.03 &    1.00 &   -0.02 &    0.03 &    0.05 &    0.10 &    0.02 &    0.07 &    0.05 &     0.01 &      0.00 &      0.01 &      0.11 &     -0.02 &      0.02 &       0.04 \\
20--30    &   0.01 &   -0.02 &    1.00 &    0.06 &    0.01 &    0.04 &    0.06 &    0.01 &    0.07 &     0.10 &      0.05 &      0.07 &      0.06 &      0.06 &      0.05 &       0.01 \\
30--40    &   0.05 &    0.03 &    0.06 &    1.00 &    0.04 &    0.05 &    0.05 &    0.04 &    0.05 &     0.06 &      0.05 &     -0.00 &      0.05 &      0.05 &      0.05 &       0.02 \\
40--50    &   0.05 &    0.05 &    0.01 &    0.04 &    1.00 &    0.04 &    0.03 &    0.05 &    0.09 &     0.04 &      0.02 &      0.07 &      0.06 &      0.02 &      0.04 &       0.05 \\
50--60    &   0.07 &    0.10 &    0.04 &    0.05 &    0.04 &    1.00 &   -0.01 &    0.05 &    0.05 &     0.01 &      0.04 &     -0.01 &      0.06 &     -0.02 &      0.04 &       0.06 \\
60--70    &   0.01 &    0.02 &    0.06 &    0.05 &    0.03 &   -0.01 &    1.00 &    0.08 &   -0.01 &    -0.01 &      0.05 &      0.03 &     -0.01 &      0.03 &      0.00 &       0.05 \\
\bottomrule
\end{tabular}

%% file: Final_CombinedCorrelationMatrix_lepton1_pT_binning0.tex
\begin{tabular}{lrrrrrrrrrrrrrrrrr}
\toprule
Leading lepton $p_{\mathrm{T}}$ [GeV] &  20--40 &  40--50 &  50--60 &  60--70 &  70--80 &  80--90 &  90--100 &  100--110 &  110--120 &  120--130 &  130--140 &  140--150 &  150--160 &  160--180 &  180--200 &  200--230 &  230--450 \\
\midrule
20--40   &    1.00 &    0.22 &   -0.01 &   -0.12 &   -0.04 &   -0.04 &     0.02 &      0.05 &     -0.04 &      0.06 &     -0.04 &      0.04 &      0.13 &      0.05 &      0.06 &      0.07 &      0.16 \\
40--50   &    0.22 &    1.00 &    0.06 &   -0.22 &   -0.14 &   -0.03 &     0.02 &      0.01 &     -0.02 &      0.09 &     -0.02 &      0.03 &      0.17 &      0.08 &      0.07 &      0.09 &      0.21 \\
50--60   &   -0.01 &    0.06 &    1.00 &    0.04 &    0.04 &    0.09 &     0.10 &      0.08 &      0.05 &      0.03 &     -0.01 &      0.03 &      0.03 &      0.07 &      0.03 &      0.06 &      0.08 \\
60--70   &   -0.12 &   -0.22 &    0.04 &    1.00 &    0.19 &    0.11 &     0.06 &      0.05 &      0.03 &      0.04 &      0.10 &      0.06 &     -0.06 &      0.00 &      0.01 &     -0.03 &     -0.11 \\
70--80   &   -0.04 &   -0.14 &    0.04 &    0.19 &    1.00 &    0.12 &    -0.01 &      0.07 &      0.09 &      0.02 &      0.06 &      0.03 &     -0.05 &      0.03 &      0.02 &      0.01 &     -0.06 \\
80--90   &   -0.04 &   -0.03 &    0.09 &    0.11 &    0.12 &    1.00 &     0.08 &      0.01 &      0.06 &      0.01 &      0.07 &      0.05 &     -0.01 &      0.05 &      0.04 &      0.02 &      0.00 \\
90--100  &    0.02 &    0.02 &    0.10 &    0.06 &   -0.01 &    0.08 &     1.00 &      0.11 &     -0.04 &     -0.01 &      0.02 &      0.04 &      0.06 &      0.07 &      0.02 &      0.01 &      0.02 \\
\bottomrule
\end{tabular}

%% file: Final_CombinedCorrelationMatrix_lepton2_pT_binning0.tex
\begin{tabular}{lrrrrrrrrrr}
\toprule
Subleading lepton $p_{\mathrm{T}}$ [GeV] &  15--40 &  40--50 &  50--60 &  60--70 &  70--80 &  80--90 &  90--100 &  100--120 &  120--150 &  150--300 \\
\midrule
15--40   &    1.00 &    0.10 &   -0.15 &   -0.05 &   -0.01 &   -0.21 &     0.03 &      0.13 &      0.04 &      0.08 \\
40--50   &    0.10 &    1.00 &    0.04 &    0.05 &    0.10 &    0.01 &     0.01 &      0.02 &      0.06 &      0.04 \\
50--60   &   -0.15 &    0.04 &    1.00 &    0.17 &    0.06 &    0.19 &     0.09 &     -0.00 &      0.05 &      0.04 \\
60--70   &   -0.05 &    0.05 &    0.17 &    1.00 &    0.05 &    0.10 &     0.04 &      0.06 &      0.08 &      0.03 \\
70--80   &   -0.01 &    0.10 &    0.06 &    0.05 &    1.00 &    0.08 &    -0.01 &      0.04 &      0.06 &      0.04 \\
80--90   &   -0.21 &    0.01 &    0.19 &    0.10 &    0.08 &    1.00 &     0.07 &     -0.02 &     -0.01 &      0.05 \\
90--100  &    0.03 &    0.01 &    0.09 &    0.04 &   -0.01 &    0.07 &     1.00 &      0.05 &     -0.05 &      0.01 \\
100--120 &    0.13 &    0.02 &   -0.00 &    0.06 &    0.04 &   -0.02 &     0.05 &      1.00 &      0.02 &      0.02 \\
120--150 &    0.04 &    0.06 &    0.05 &    0.08 &    0.06 &   -0.01 &    -0.05 &      0.02 &      1.00 &     -0.00 \\
150--300 &    0.08 &    0.04 &    0.04 &    0.03 &    0.04 &    0.05 &     0.01 &      0.02 &     -0.00 &      1.00 \\
\bottomrule
\end{tabular}

%% file: Final_CombinedCorrelationMatrix_lepton3_pT_binning0.tex
\begin{tabular}{lrrrrrrrrrrrrrr}
\toprule
Third lepton $p_{\mathrm{T}}$ [GeV] &  10--20 &  20--25 &  25--30 &  30--35 &  35--40 &  40--45 &  45--50 &  50--55 &  55--60 &  60--65 &  65--70 &  70--80 &  80--100 &  100--200 \\
\midrule
10--20   &    1.00 &    0.29 &    0.27 &    0.27 &    0.11 &   -0.28 &   -0.25 &   -0.14 &    0.05 &   -0.06 &    0.14 &    0.08 &     0.06 &      0.08 \\
20--25   &    0.29 &    1.00 &    0.25 &    0.20 &    0.08 &   -0.23 &   -0.24 &   -0.12 &    0.03 &   -0.07 &    0.10 &    0.07 &     0.04 &      0.07 \\
25--30   &    0.27 &    0.25 &    1.00 &    0.25 &    0.07 &   -0.27 &   -0.21 &   -0.13 &    0.03 &   -0.08 &    0.12 &    0.09 &     0.07 &      0.08 \\
30--35   &    0.27 &    0.20 &    0.25 &    1.00 &    0.11 &   -0.20 &   -0.15 &   -0.07 &    0.04 &   -0.06 &    0.11 &    0.06 &     0.05 &      0.04 \\
35--40   &    0.11 &    0.08 &    0.07 &    0.11 &    1.00 &    0.05 &    0.01 &   -0.02 &    0.07 &    0.04 &    0.04 &    0.05 &     0.05 &      0.02 \\
40--45   &   -0.28 &   -0.23 &   -0.27 &   -0.20 &    0.05 &    1.00 &    0.41 &    0.22 &    0.02 &    0.12 &   -0.08 &   -0.01 &     0.00 &     -0.05 \\
45--50   &   -0.25 &   -0.24 &   -0.21 &   -0.15 &    0.01 &    0.41 &    1.00 &    0.29 &   -0.06 &    0.13 &   -0.07 &   -0.05 &    -0.00 &     -0.03 \\
\bottomrule
\end{tabular}

%% file: Final_CombinedCorrelationMatrix_lepton4_pT_binning0.tex
\begin{tabular}{lrrrrrrrrrrr}
\toprule
Fourth lepton $p_{\mathrm{T}}$ [GeV] &  5--10 &  10--15 &  15--20 &  20--25 &  25--30 &  30--35 &  35--40 &  40--45 &  45--50 &  50--60 &  60--150 \\
\midrule
5--10   &   1.00 &    0.18 &    0.16 &    0.10 &    0.08 &   -0.03 &   -0.03 &   -0.10 &    0.03 &   -0.06 &    -0.04 \\
10--15  &   0.18 &    1.00 &    0.19 &    0.14 &    0.03 &   -0.05 &   -0.11 &   -0.18 &    0.09 &   -0.04 &     0.00 \\
15--20  &   0.16 &    0.19 &    1.00 &    0.12 &    0.06 &   -0.03 &   -0.05 &   -0.17 &    0.05 &   -0.04 &    -0.05 \\
20--25  &   0.10 &    0.14 &    0.12 &    1.00 &    0.06 &   -0.02 &    0.00 &   -0.02 &    0.05 &   -0.04 &     0.05 \\
25--30  &   0.08 &    0.03 &    0.06 &    0.06 &    1.00 &    0.09 &    0.04 &    0.05 &    0.03 &    0.07 &     0.04 \\
30--35  &  -0.03 &   -0.05 &   -0.03 &   -0.02 &    0.09 &    1.00 &    0.14 &    0.18 &   -0.02 &    0.12 &     0.06 \\
35--40  &  -0.03 &   -0.11 &   -0.05 &    0.00 &    0.04 &    0.14 &    1.00 &    0.23 &   -0.06 &    0.12 &     0.08 \\
40--45  &  -0.10 &   -0.18 &   -0.17 &   -0.02 &    0.05 &    0.18 &    0.23 &    1.00 &   -0.03 &    0.11 &     0.04 \\
45--50  &   0.03 &    0.09 &    0.05 &    0.05 &    0.03 &   -0.02 &   -0.06 &   -0.03 &    1.00 &    0.02 &     0.01 \\
50--60  &  -0.06 &   -0.04 &   -0.04 &   -0.04 &    0.07 &    0.12 &    0.12 &    0.11 &    0.02 &    1.00 &     0.05 \\
60--150 &  -0.04 &    0.00 &   -0.05 &    0.05 &    0.04 &    0.06 &    0.08 &    0.04 &    0.01 &    0.05 &     1.00 \\
\bottomrule
\end{tabular}

%% file: Final_CombinedCorrelationMatrix_dileptons_dy_binning0.tex
\begin{tabular}{lrrrrrrrrrrr}
\toprule
$|\delta y(Z_{1}, Z_{2})|$ &  0.0--0.2 &  0.2--0.4 &  0.4--0.6 &  0.6--0.8 &  0.8--1.0 &  1.0--1.2 &  1.2--1.5 &  1.5--2.0 &  2.0--2.5 &  2.5--3.0 &  3.0--10.0 \\
\midrule
0.0--0.2  &      1.00 &      0.03 &      0.03 &      0.06 &      0.08 &      0.09 &      0.10 &      0.08 &      0.02 &      0.06 &       0.01 \\
0.2--0.4  &      0.03 &      1.00 &      0.03 &      0.06 &      0.04 &      0.08 &      0.02 &      0.04 &      0.05 &      0.06 &       0.07 \\
0.4--0.6  &      0.03 &      0.03 &      1.00 &      0.00 &      0.05 &      0.07 &      0.06 &      0.02 &      0.00 &      0.04 &      -0.01 \\
0.6--0.8  &      0.06 &      0.06 &      0.00 &      1.00 &      0.00 &      0.06 &      0.09 &      0.10 &      0.04 &      0.05 &       0.08 \\
0.8--1.0  &      0.08 &      0.04 &      0.05 &      0.00 &      1.00 &      0.04 &      0.03 &      0.06 &      0.03 &      0.07 &       0.06 \\
1.0--1.2  &      0.09 &      0.08 &      0.07 &      0.06 &      0.04 &      1.00 &      0.00 &      0.08 &      0.03 &      0.08 &       0.05 \\
1.2--1.5  &      0.10 &      0.02 &      0.06 &      0.09 &      0.03 &      0.00 &      1.00 &      0.03 &      0.04 &      0.08 &       0.05 \\
1.5--2.0  &      0.08 &      0.04 &      0.02 &      0.10 &      0.06 &      0.08 &      0.03 &      1.00 &      0.01 &      0.05 &      -0.00 \\
2.0--2.5  &      0.02 &      0.05 &      0.00 &      0.04 &      0.03 &      0.03 &      0.04 &      0.01 &      1.00 &      0.05 &       0.06 \\
2.5--3.0  &      0.06 &      0.06 &      0.04 &      0.05 &      0.07 &      0.08 &      0.08 &      0.05 &      0.05 &      1.00 &       0.05 \\
3.0--10.0 &      0.01 &      0.07 &     -0.01 &      0.08 &      0.06 &      0.05 &      0.05 &     -0.00 &      0.06 &      0.05 &       1.00 \\
\bottomrule
\end{tabular}

%% file: Final_CombinedCorrelationMatrix_dileptons_dphi_binning0.tex
\begin{tabular}{lrrrrrrrrrrr}
\toprule
$\delta \phi(Z_{1}, Z_{2})$ [$\pi$] &  0.0--0.125 &  0.125--0.25 &  0.25--0.375 &  0.375--0.5 &  0.5--0.625 &  0.625--0.6875 &  0.6875--0.75 &  0.75--0.8125 &  0.8125--0.875 &  0.875--0.9375 &  0.9375--1.0 \\
\midrule
0.0--0.125    &        1.00 &         0.11 &         0.09 &        0.09 &        0.13 &           0.07 &          0.16 &          0.09 &           0.09 &           0.06 &        -0.13 \\
0.125--0.25   &        0.11 &         1.00 &         0.07 &        0.07 &        0.05 &           0.08 &          0.06 &          0.02 &           0.08 &           0.07 &         0.02 \\
0.25--0.375   &        0.09 &         0.07 &         1.00 &        0.09 &        0.08 &           0.06 &          0.14 &          0.05 &           0.08 &           0.09 &        -0.11 \\
0.375--0.5    &        0.09 &         0.07 &         0.09 &        1.00 &        0.06 &           0.01 &          0.12 &          0.11 &           0.08 &           0.03 &        -0.05 \\
0.5--0.625    &        0.13 &         0.05 &         0.08 &        0.06 &        1.00 &           0.07 &          0.07 &          0.06 &           0.09 &           0.10 &        -0.06 \\
0.625--0.6875 &        0.07 &         0.08 &         0.06 &        0.01 &        0.07 &           1.00 &          0.07 &          0.01 &           0.04 &           0.07 &         0.01 \\
0.6875--0.75  &        0.16 &         0.06 &         0.14 &        0.12 &        0.07 &           0.07 &          1.00 &          0.13 &           0.07 &           0.08 &        -0.07 \\
\bottomrule
\end{tabular}

%% file: Final_CombinedCorrelationMatrix_jets_N_binning0.tex
\begin{tabular}{lrrrrr}
\toprule
Jet multiplicity &     0 &     1 &     2 &     3 &  $\geq 4$ \\
\midrule
0        &  1.00 & -0.21 & -0.41 & -0.33 &     -0.31 \\
1        & -0.21 &  1.00 &  0.30 &  0.08 &      0.02 \\
2        & -0.41 &  0.30 &  1.00 &  0.65 &      0.53 \\
3        & -0.33 &  0.08 &  0.65 &  1.00 &      0.66 \\
$\geq 4$ & -0.31 &  0.02 &  0.53 &  0.66 &      1.00 \\
\bottomrule
\end{tabular}

%% file: Final_CombinedCorrelationMatrix_centralJets_N_binning0.tex
\begin{tabular}{lrrrrr}
\toprule
Central-jet multiplicity &     0 &     1 &     2 &     3 &  $\geq 4$ \\
\midrule
0        &  1.00 & -0.19 & -0.32 & -0.19 &     -0.14 \\
1        & -0.19 &  1.00 &  0.27 &  0.14 &      0.05 \\
2        & -0.32 &  0.27 &  1.00 &  0.37 &      0.18 \\
3        & -0.19 &  0.14 &  0.37 &  1.00 &      0.18 \\
$\geq 4$ & -0.14 &  0.05 &  0.18 &  0.18 &      1.00 \\
\bottomrule
\end{tabular}

%% file: Final_CombinedCorrelationMatrix_jets60_N_binning0.tex
\begin{tabular}{lrrrr}
\toprule
Jet ($p_{\mathrm{T}} > 60$ GeV) multiplicity &     0 &     1 &     2 &  $\geq 3$ \\
\midrule
0        &  1.00 & -0.03 & -0.19 &     -0.13 \\
1        & -0.03 &  1.00 &  0.23 &      0.17 \\
2        & -0.19 &  0.23 &  1.00 &      0.48 \\
$\geq 3$ & -0.13 &  0.17 &  0.48 &      1.00 \\
\bottomrule
\end{tabular}

%% file: Final_CombinedCorrelationMatrix_hardestJetsDijet_mass_binning0.tex
\begin{tabular}{lrrrrr}
\toprule
$m(\mathrm{jet}_{1}, \mathrm{jet}_{2})$ [GeV] &  0--50 &  50--100 &  100--200 &  200--300 &  300--1000 \\
\midrule
0--50     &   1.00 &     0.08 &     -0.03 &      0.01 &       0.03 \\
50--100   &   0.08 &     1.00 &      0.35 &      0.27 &       0.38 \\
100--200  &  -0.03 &     0.35 &      1.00 &      0.32 &       0.42 \\
200--300  &   0.01 &     0.27 &      0.32 &      1.00 &       0.44 \\
300--1000 &   0.03 &     0.38 &      0.42 &      0.44 &       1.00 \\
\bottomrule
\end{tabular}

%% file: Final_CombinedCorrelationMatrix_hardestJetsDijet_dy_binning0.tex
\begin{tabular}{lrrrr}
\toprule
$|\delta y(\mathrm{jet}_{1}, \mathrm{jet}_{2})|$ &  0--1 &  1--2 &  2--3 &  3--9 \\
\midrule
0--1 &  1.00 &  0.37 &  0.41 &  0.58 \\
1--2 &  0.37 &  1.00 &  0.37 &  0.48 \\
2--3 &  0.41 &  0.37 &  1.00 &  0.54 \\
3--9 &  0.58 &  0.48 &  0.54 &  1.00 \\
\bottomrule
\end{tabular}

%% file: Final_CombinedCorrelationMatrix_jets_scalarPtSum_binning0.tex
\begin{tabular}{lrrrrrrr}
\toprule
Jet scalar $p_{\mathrm{T}}$ sum [GeV] &  30--60 &  60--90 &  90--120 &  120--150 &  150--200 &  200--400 &  400--1000 \\
\midrule
30--60    &    1.00 &    0.20 &    -0.06 &      0.12 &      0.06 &     -0.04 &      -0.13 \\
60--90    &    0.20 &    1.00 &     0.18 &      0.06 &      0.07 &      0.18 &       0.10 \\
90--120   &   -0.06 &    0.18 &     1.00 &      0.12 &     -0.10 &     -0.04 &      -0.04 \\
120--150  &    0.12 &    0.06 &     0.12 &      1.00 &      0.29 &      0.17 &       0.21 \\
150--200  &    0.06 &    0.07 &    -0.10 &      0.29 &      1.00 &      0.24 &       0.27 \\
200--400  &   -0.04 &    0.18 &    -0.04 &      0.17 &      0.24 &      1.00 &       0.59 \\
400--1000 &   -0.13 &    0.10 &    -0.04 &      0.21 &      0.27 &      0.59 &       1.00 \\
\bottomrule
\end{tabular}

%% file: Final_CombinedCorrelationMatrix_jet1_pt_binning0.tex
\begin{tabular}{lrrrrrrrrr}
\toprule
Leading jet $p_{\mathrm{T}}$ [GeV] &  30--40 &  40--50 &  50--60 &  60--80 &  80--100 &  100--120 &  120--150 &  150--200 &  200--800 \\
\midrule
30--40   &    1.00 &    0.59 &    0.30 &    0.17 &     0.03 &     -0.00 &      0.00 &     -0.10 &     -0.09 \\
40--50   &    0.59 &    1.00 &    0.37 &    0.23 &    -0.01 &      0.09 &      0.08 &      0.04 &      0.07 \\
50--60   &    0.30 &    0.37 &    1.00 &    0.32 &     0.14 &     -0.08 &      0.03 &     -0.03 &     -0.03 \\
60--80   &    0.17 &    0.23 &    0.32 &    1.00 &     0.23 &      0.04 &      0.11 &      0.14 &      0.13 \\
80--100  &    0.03 &   -0.01 &    0.14 &    0.23 &     1.00 &      0.31 &      0.08 &      0.15 &      0.20 \\
100--120 &   -0.00 &    0.09 &   -0.08 &    0.04 &     0.31 &      1.00 &      0.32 &      0.12 &      0.25 \\
120--150 &    0.00 &    0.08 &    0.03 &    0.11 &     0.08 &      0.32 &      1.00 &      0.34 &      0.30 \\
150--200 &   -0.10 &    0.04 &   -0.03 &    0.14 &     0.15 &      0.12 &      0.34 &      1.00 &      0.49 \\
200--800 &   -0.09 &    0.07 &   -0.03 &    0.13 &     0.20 &      0.25 &      0.30 &      0.49 &      1.00 \\
\bottomrule
\end{tabular}

%% file: Final_CombinedCorrelationMatrix_jet2_pt_binning0.tex
\begin{tabular}{lrrr}
\toprule
Subleading jet $p_{\mathrm{T}}$ [GeV] &  30--40 &  40--60 &  60--500 \\
\midrule
30--40  &    1.00 &    0.39 &     0.39 \\
40--60  &    0.39 &    1.00 &     0.37 \\
60--500 &    0.39 &    0.37 &     1.00 \\
\bottomrule
\end{tabular}

%% file: Final_CombinedCorrelationMatrix_jet1_eta_binning0.tex
\begin{tabular}{lrrrrrr}
\toprule
Leading jet $|\eta|$ &  0--0.5 &  0.5--1.0 &  1.0--1.5 &  1.5--2.0 &  2.0--2.5 &  2.5--4.5 \\
\midrule
0--0.5   &    1.00 &      0.15 &      0.16 &      0.18 &      0.14 &      0.20 \\
0.5--1.0 &    0.15 &      1.00 &      0.18 &      0.19 &      0.12 &      0.24 \\
1.0--1.5 &    0.16 &      0.18 &      1.00 &      0.21 &      0.15 &      0.26 \\
1.5--2.0 &    0.18 &      0.19 &      0.21 &      1.00 &      0.12 &      0.37 \\
2.0--2.5 &    0.14 &      0.12 &      0.15 &      0.12 &      1.00 &      0.24 \\
2.5--4.5 &    0.20 &      0.24 &      0.26 &      0.37 &      0.24 &      1.00 \\
\bottomrule
\end{tabular}

%% file: Final_CombinedCorrelationMatrix_jet2_eta_binning0.tex
\begin{tabular}{lrrrrrr}
\toprule
Subleading jet $|\eta|$ &  0--0.5 &  0.5--1.0 &  1.0--1.5 &  1.5--2.0 &  2.0--2.5 &  2.5--4.5 \\
\midrule
0--0.5   &    1.00 &      0.19 &      0.26 &      0.21 &      0.18 &      0.28 \\
0.5--1.0 &    0.19 &      1.00 &      0.26 &      0.21 &      0.13 &      0.32 \\
1.0--1.5 &    0.26 &      0.26 &      1.00 &      0.30 &      0.09 &      0.50 \\
1.5--2.0 &    0.21 &      0.21 &      0.30 &      1.00 &      0.09 &      0.39 \\
2.0--2.5 &    0.18 &      0.13 &      0.09 &      0.09 &      1.00 &      0.04 \\
2.5--4.5 &    0.28 &      0.32 &      0.50 &      0.39 &      0.04 &      1.00 \\
\bottomrule
\end{tabular}

%% file: Final_StatCorrelationMatrix_fourLepton_pT_binning0.tex
\begin{tabular}{lrrrrrrrrrrrrrrr}
\toprule
$p_{\mathrm{T}, 4\ell}$ [GeV] &  0--5 &  5--15 &  15--25 &  25--35 &  35--45 &  45--55 &  55--65 &  65--75 &  75--85 &  85--100 &  100--125 &  125--150 &  150--200 &  200--250 &  250--1500 \\
\midrule
0--5      &  1.00 &   0.15 &   -0.07 &   -0.05 &   -0.00 &    0.03 &    0.02 &    0.00 &    0.01 &    -0.01 &     -0.02 &      0.03 &     -0.02 &      0.01 &       0.03 \\
5--15     &  0.15 &   1.00 &   -0.03 &   -0.04 &   -0.02 &    0.04 &   -0.04 &   -0.03 &    0.03 &    -0.00 &     -0.00 &      0.00 &     -0.00 &     -0.02 &       0.01 \\
15--25    & -0.07 &  -0.03 &    1.00 &    0.02 &   -0.07 &    0.02 &   -0.02 &   -0.02 &   -0.00 &     0.03 &     -0.04 &      0.02 &      0.00 &      0.00 &      -0.05 \\
25--35    & -0.05 &  -0.04 &    0.02 &    1.00 &    0.04 &   -0.07 &    0.00 &   -0.04 &    0.01 &    -0.01 &      0.03 &      0.02 &      0.04 &      0.01 &       0.01 \\
35--45    & -0.00 &  -0.02 &   -0.07 &    0.04 &    1.00 &    0.04 &   -0.05 &    0.00 &    0.00 &     0.00 &     -0.01 &     -0.03 &     -0.02 &     -0.02 &      -0.01 \\
45--55    &  0.03 &   0.04 &    0.02 &   -0.07 &    0.04 &    1.00 &    0.07 &   -0.02 &   -0.03 &     0.01 &     -0.02 &     -0.05 &     -0.00 &      0.05 &      -0.00 \\
55--65    &  0.02 &  -0.04 &   -0.02 &    0.00 &   -0.05 &    0.07 &    1.00 &    0.09 &   -0.07 &    -0.01 &      0.00 &      0.00 &     -0.03 &      0.02 &      -0.03 \\
\bottomrule
\end{tabular}

%% file: Final_StatCorrelationMatrix_fourLepton_absy_binning0.tex
\begin{tabular}{lrrrrrrrrrrr}
\toprule
$|y_{4\ell}|$ &  0--0.2 &  0.2--0.4 &  0.4--0.6 &  0.6--0.8 &  0.8--1 &  1--1.2 &  1.2--1.4 &  1.4--1.6 &  1.6--1.8 &  1.8--2 &  2--4.6 \\
\midrule
0--0.2   &    1.00 &     -0.04 &     -0.03 &     -0.05 &    0.01 &    0.01 &      0.03 &      0.01 &     -0.04 &   -0.01 &   -0.03 \\
0.2--0.4 &   -0.04 &      1.00 &     -0.04 &      0.00 &   -0.04 &    0.02 &     -0.03 &     -0.03 &      0.01 &    0.03 &    0.03 \\
0.4--0.6 &   -0.03 &     -0.04 &      1.00 &     -0.06 &   -0.01 &    0.02 &      0.02 &     -0.04 &     -0.03 &    0.01 &   -0.05 \\
0.6--0.8 &   -0.05 &      0.00 &     -0.06 &      1.00 &   -0.05 &   -0.02 &      0.01 &      0.03 &     -0.02 &   -0.04 &    0.04 \\
0.8--1   &    0.01 &     -0.04 &     -0.01 &     -0.05 &    1.00 &   -0.02 &     -0.01 &     -0.00 &     -0.00 &    0.03 &    0.01 \\
1--1.2   &    0.01 &      0.02 &      0.02 &     -0.02 &   -0.02 &    1.00 &     -0.06 &      0.02 &     -0.01 &    0.02 &    0.01 \\
1.2--1.4 &    0.03 &     -0.03 &      0.02 &      0.01 &   -0.01 &   -0.06 &      1.00 &     -0.02 &     -0.01 &    0.02 &    0.03 \\
1.4--1.6 &    0.01 &     -0.03 &     -0.04 &      0.03 &   -0.00 &    0.02 &     -0.02 &      1.00 &     -0.04 &   -0.00 &   -0.04 \\
1.6--1.8 &   -0.04 &      0.01 &     -0.03 &     -0.02 &   -0.00 &   -0.01 &     -0.01 &     -0.04 &      1.00 &    0.01 &    0.04 \\
1.8--2   &   -0.01 &      0.03 &      0.01 &     -0.04 &    0.03 &    0.02 &      0.02 &     -0.00 &      0.01 &    1.00 &    0.02 \\
2--4.6   &   -0.03 &      0.03 &     -0.05 &      0.04 &    0.01 &    0.01 &      0.03 &     -0.04 &      0.04 &    0.02 &    1.00 \\
\bottomrule
\end{tabular}

%% file: Final_StatCorrelationMatrix_leadingDilepton_pT_binning0.tex
\begin{tabular}{lrrrrrrrrrrrrrrrr}
\toprule
$p_{\mathrm{T}, Z_{1}}$ [GeV] &  0--10 &  10--20 &  20--30 &  30--40 &  40--50 &  50--60 &  60--70 &  70--80 &  80--90 &  90--100 &  100--120 &  120--140 &  140--160 &  160--200 &  200--250 &  250--1500 \\
\midrule
0--10     &   1.00 &   -0.05 &   -0.04 &    0.01 &   -0.01 &    0.01 &   -0.02 &   -0.02 &    0.00 &    -0.01 &     -0.03 &     -0.04 &     -0.00 &      0.04 &      0.01 &       0.01 \\
10--20    &  -0.05 &    1.00 &   -0.06 &   -0.03 &   -0.04 &    0.02 &   -0.03 &    0.02 &    0.02 &    -0.00 &     -0.02 &      0.01 &      0.05 &     -0.01 &      0.01 &       0.00 \\
20--30    &  -0.04 &   -0.06 &    1.00 &   -0.03 &   -0.09 &   -0.01 &    0.02 &   -0.03 &   -0.00 &     0.04 &     -0.02 &      0.01 &      0.01 &      0.02 &     -0.01 &      -0.02 \\
30--40    &   0.01 &   -0.03 &   -0.03 &    1.00 &   -0.04 &    0.00 &    0.01 &   -0.00 &    0.00 &     0.02 &      0.00 &     -0.04 &      0.01 &      0.01 &      0.01 &      -0.01 \\
40--50    &  -0.01 &   -0.04 &   -0.09 &   -0.04 &    1.00 &   -0.05 &   -0.02 &    0.00 &    0.04 &    -0.00 &     -0.03 &      0.04 &     -0.00 &      0.00 &     -0.01 &       0.02 \\
50--60    &   0.01 &    0.02 &   -0.01 &    0.00 &   -0.05 &    1.00 &   -0.05 &    0.01 &    0.01 &    -0.02 &      0.01 &     -0.04 &     -0.00 &     -0.03 &      0.02 &       0.02 \\
60--70    &  -0.02 &   -0.03 &    0.02 &    0.01 &   -0.02 &   -0.05 &    1.00 &    0.05 &   -0.04 &    -0.04 &      0.02 &      0.00 &     -0.04 &      0.01 &     -0.02 &       0.02 \\
\bottomrule
\end{tabular}

%% file: Final_StatCorrelationMatrix_subleadingDilepton_pT_binning0.tex
\begin{tabular}{lrrrrrrrrrrrrrrrr}
\toprule
$p_{\mathrm{T}, Z_{2}}$ [GeV] &  0--10 &  10--20 &  20--30 &  30--40 &  40--50 &  50--60 &  60--70 &  70--80 &  80--90 &  90--100 &  100--120 &  120--140 &  140--160 &  160--200 &  200--250 &  250--1500 \\
\midrule
0--10     &   1.00 &   -0.07 &   -0.04 &    0.01 &   -0.01 &    0.01 &   -0.02 &   -0.02 &    0.00 &    -0.01 &     -0.03 &     -0.04 &     -0.00 &      0.04 &      0.01 &       0.01 \\
10--20    &  -0.07 &    1.00 &   -0.07 &   -0.03 &   -0.04 &    0.02 &   -0.03 &    0.02 &    0.02 &    -0.00 &     -0.02 &      0.01 &      0.05 &     -0.01 &      0.01 &       0.01 \\
20--30    &  -0.04 &   -0.07 &    1.00 &   -0.03 &   -0.09 &   -0.02 &    0.02 &   -0.03 &   -0.00 &     0.04 &     -0.02 &      0.01 &      0.01 &      0.02 &     -0.01 &      -0.02 \\
30--40    &   0.01 &   -0.03 &   -0.03 &    1.00 &   -0.03 &   -0.00 &    0.01 &   -0.00 &    0.00 &     0.02 &      0.00 &     -0.04 &      0.01 &      0.01 &      0.01 &      -0.01 \\
40--50    &  -0.01 &   -0.04 &   -0.09 &   -0.03 &    1.00 &   -0.04 &   -0.02 &    0.00 &    0.03 &    -0.00 &     -0.03 &      0.04 &     -0.00 &      0.00 &     -0.01 &       0.02 \\
50--60    &   0.01 &    0.02 &   -0.02 &   -0.00 &   -0.04 &    1.00 &   -0.05 &    0.01 &    0.01 &    -0.02 &      0.01 &     -0.04 &     -0.00 &     -0.03 &      0.02 &       0.03 \\
60--70    &  -0.02 &   -0.03 &    0.02 &    0.01 &   -0.02 &   -0.05 &    1.00 &    0.05 &   -0.04 &    -0.04 &      0.02 &      0.00 &     -0.04 &      0.01 &     -0.02 &       0.02 \\
\bottomrule
\end{tabular}

%% file: Final_StatCorrelationMatrix_lepton1_pT_binning0.tex
\begin{tabular}{lrrrrrrrrrrrrrrrrr}
\toprule
Leading lepton $p_{\mathrm{T}}$ [GeV] &  20--40 &  40--50 &  50--60 &  60--70 &  70--80 &  80--90 &  90--100 &  100--110 &  110--120 &  120--130 &  130--140 &  140--150 &  150--160 &  160--180 &  180--200 &  200--230 &  230--450 \\
\midrule
20--40   &    1.00 &   -0.02 &   -0.06 &    0.01 &    0.03 &   -0.00 &     0.00 &      0.02 &     -0.04 &      0.00 &     -0.01 &      0.01 &      0.01 &     -0.01 &      0.02 &      0.00 &      0.01 \\
40--50   &   -0.02 &    1.00 &   -0.01 &   -0.06 &   -0.05 &    0.03 &    -0.01 &     -0.04 &     -0.01 &      0.01 &      0.03 &     -0.01 &     -0.01 &     -0.00 &      0.02 &     -0.01 &     -0.02 \\
50--60   &   -0.06 &   -0.01 &    1.00 &   -0.05 &   -0.05 &    0.02 &     0.04 &      0.01 &      0.01 &     -0.03 &     -0.04 &     -0.02 &     -0.03 &      0.02 &     -0.02 &      0.01 &      0.03 \\
60--70   &    0.01 &   -0.06 &   -0.05 &    1.00 &    0.02 &   -0.03 &    -0.01 &     -0.01 &     -0.05 &      0.04 &      0.02 &      0.03 &      0.02 &     -0.01 &      0.00 &     -0.03 &     -0.02 \\
70--80   &    0.03 &   -0.05 &   -0.05 &    0.02 &    1.00 &    0.01 &    -0.07 &      0.00 &      0.03 &     -0.00 &     -0.00 &     -0.01 &     -0.02 &      0.01 &      0.00 &     -0.01 &     -0.02 \\
80--90   &   -0.00 &    0.03 &    0.02 &   -0.03 &    0.01 &    1.00 &     0.03 &     -0.06 &      0.00 &     -0.02 &      0.02 &      0.02 &      0.00 &      0.02 &      0.01 &      0.00 &      0.03 \\
90--100  &    0.00 &   -0.01 &    0.04 &   -0.01 &   -0.07 &    0.03 &     1.00 &      0.06 &     -0.09 &     -0.04 &     -0.01 &      0.01 &      0.04 &      0.03 &     -0.00 &     -0.02 &     -0.00 \\
\bottomrule
\end{tabular}

%% file: Final_StatCorrelationMatrix_lepton2_pT_binning0.tex
\begin{tabular}{lrrrrrrrrrr}
\toprule
Subleading lepton $p_{\mathrm{T}}$ [GeV] &  15--40 &  40--50 &  50--60 &  60--70 &  70--80 &  80--90 &  90--100 &  100--120 &  120--150 &  150--300 \\
\midrule
15--40   &    1.00 &   -0.04 &   -0.02 &   -0.01 &   -0.03 &   -0.02 &    -0.02 &      0.00 &     -0.03 &      0.03 \\
40--50   &   -0.04 &    1.00 &   -0.04 &   -0.04 &    0.03 &    0.01 &    -0.03 &     -0.05 &     -0.01 &     -0.02 \\
50--60   &   -0.02 &   -0.04 &    1.00 &    0.01 &   -0.03 &    0.01 &     0.05 &     -0.02 &     -0.00 &      0.01 \\
60--70   &   -0.01 &   -0.04 &    0.01 &    1.00 &   -0.03 &   -0.02 &    -0.01 &      0.03 &      0.03 &     -0.00 \\
70--80   &   -0.03 &    0.03 &   -0.03 &   -0.03 &    1.00 &    0.03 &    -0.04 &      0.01 &      0.02 &      0.00 \\
80--90   &   -0.02 &    0.01 &    0.01 &   -0.02 &    0.03 &    1.00 &     0.05 &     -0.01 &     -0.02 &      0.04 \\
90--100  &   -0.02 &   -0.03 &    0.05 &   -0.01 &   -0.04 &    0.05 &     1.00 &      0.02 &     -0.07 &     -0.01 \\
100--120 &    0.00 &   -0.05 &   -0.02 &    0.03 &    0.01 &   -0.01 &     0.02 &      1.00 &     -0.02 &     -0.01 \\
120--150 &   -0.03 &   -0.01 &   -0.00 &    0.03 &    0.02 &   -0.02 &    -0.07 &     -0.02 &      1.00 &     -0.04 \\
150--300 &    0.03 &   -0.02 &    0.01 &   -0.00 &    0.00 &    0.04 &    -0.01 &     -0.01 &     -0.04 &      1.00 \\
\bottomrule
\end{tabular}

%% file: Final_StatCorrelationMatrix_lepton3_pT_binning0.tex
\begin{tabular}{lrrrrrrrrrrrrrr}
\toprule
Third lepton $p_{\mathrm{T}}$ [GeV] &  10--20 &  20--25 &  25--30 &  30--35 &  35--40 &  40--45 &  45--50 &  50--55 &  55--60 &  60--65 &  65--70 &  70--80 &  80--100 &  100--200 \\
\midrule
10--20   &    1.00 &   -0.01 &   -0.08 &    0.02 &    0.03 &   -0.02 &   -0.04 &    0.01 &    0.01 &    0.02 &    0.02 &    0.00 &     0.01 &     -0.00 \\
20--25   &   -0.01 &    1.00 &   -0.02 &   -0.02 &    0.01 &    0.00 &   -0.06 &    0.00 &   -0.01 &   -0.01 &    0.01 &    0.00 &    -0.01 &      0.00 \\
25--30   &   -0.08 &   -0.02 &    1.00 &    0.00 &   -0.05 &   -0.04 &    0.01 &    0.01 &   -0.02 &   -0.03 &    0.00 &    0.01 &     0.01 &      0.00 \\
30--35   &    0.02 &   -0.02 &    0.00 &    1.00 &    0.01 &   -0.05 &   -0.01 &    0.03 &   -0.00 &   -0.03 &    0.02 &   -0.02 &     0.01 &     -0.04 \\
35--40   &    0.03 &    0.01 &   -0.05 &    0.01 &    1.00 &    0.01 &   -0.05 &   -0.06 &    0.03 &    0.00 &   -0.01 &    0.01 &     0.02 &     -0.02 \\
40--45   &   -0.02 &    0.00 &   -0.04 &   -0.05 &    0.01 &    1.00 &    0.04 &   -0.05 &    0.02 &   -0.07 &   -0.00 &    0.03 &     0.02 &     -0.01 \\
45--50   &   -0.04 &   -0.06 &    0.01 &   -0.01 &   -0.05 &    0.04 &    1.00 &    0.06 &   -0.09 &   -0.05 &   -0.00 &   -0.04 &     0.00 &      0.01 \\
\bottomrule
\end{tabular}

%% file: Final_StatCorrelationMatrix_lepton4_pT_binning0.tex
\begin{tabular}{lrrrrrrrrrrr}
\toprule
Fourth lepton $p_{\mathrm{T}}$ [GeV] &  5--10 &  10--15 &  15--20 &  20--25 &  25--30 &  30--35 &  35--40 &  40--45 &  45--50 &  50--60 &  60--150 \\
\midrule
5--10   &   1.00 &   -0.04 &   -0.05 &   -0.05 &    0.01 &    0.01 &    0.03 &    0.01 &   -0.04 &   -0.02 &    -0.03 \\
10--15  &  -0.04 &    1.00 &   -0.04 &   -0.02 &   -0.04 &    0.02 &   -0.03 &   -0.03 &    0.01 &    0.03 &     0.02 \\
15--20  &  -0.05 &   -0.04 &    1.00 &   -0.05 &   -0.04 &    0.01 &    0.02 &   -0.04 &   -0.03 &    0.01 &    -0.05 \\
20--25  &  -0.05 &   -0.02 &   -0.05 &    1.00 &   -0.04 &   -0.05 &    0.01 &    0.03 &   -0.02 &   -0.04 &     0.04 \\
25--30  &   0.01 &   -0.04 &   -0.04 &   -0.04 &    1.00 &   -0.00 &   -0.04 &   -0.01 &   -0.00 &    0.03 &     0.01 \\
30--35  &   0.01 &    0.02 &    0.01 &   -0.05 &   -0.00 &    1.00 &   -0.04 &   -0.03 &   -0.01 &    0.02 &     0.01 \\
35--40  &   0.03 &   -0.03 &    0.02 &    0.01 &   -0.04 &   -0.04 &    1.00 &    0.02 &   -0.05 &    0.02 &     0.03 \\
40--45  &   0.01 &   -0.03 &   -0.04 &    0.03 &   -0.01 &   -0.03 &    0.02 &    1.00 &    0.02 &   -0.04 &    -0.04 \\
45--50  &  -0.04 &    0.01 &   -0.03 &   -0.02 &   -0.00 &   -0.01 &   -0.05 &    0.02 &    1.00 &    0.04 &     0.02 \\
50--60  &  -0.02 &    0.03 &    0.01 &   -0.04 &    0.03 &    0.02 &    0.02 &   -0.04 &    0.04 &    1.00 &     0.01 \\
60--150 &  -0.03 &    0.02 &   -0.05 &    0.04 &    0.01 &    0.01 &    0.03 &   -0.04 &    0.02 &    0.01 &     1.00 \\
\bottomrule
\end{tabular}

%% file: Final_StatCorrelationMatrix_dileptons_dy_binning0.tex
\begin{tabular}{lrrrrrrrrrrr}
\toprule
$|\delta y(Z_{1}, Z_{2})|$ &  0.0--0.2 &  0.2--0.4 &  0.4--0.6 &  0.6--0.8 &  0.8--1.0 &  1.0--1.2 &  1.2--1.5 &  1.5--2.0 &  2.0--2.5 &  2.5--3.0 &  3.0--10.0 \\
\midrule
0.0--0.2  &      1.00 &     -0.05 &     -0.04 &     -0.05 &      0.01 &      0.01 &      0.03 &      0.01 &     -0.04 &     -0.01 &      -0.03 \\
0.2--0.4  &     -0.05 &      1.00 &     -0.05 &      0.00 &     -0.04 &      0.02 &     -0.03 &     -0.03 &      0.01 &      0.03 &       0.03 \\
0.4--0.6  &     -0.04 &     -0.05 &      1.00 &     -0.06 &     -0.01 &      0.02 &      0.02 &     -0.04 &     -0.03 &      0.01 &      -0.05 \\
0.6--0.8  &     -0.05 &      0.00 &     -0.06 &      1.00 &     -0.06 &     -0.02 &      0.01 &      0.03 &     -0.02 &     -0.04 &       0.04 \\
0.8--1.0  &      0.01 &     -0.04 &     -0.01 &     -0.06 &      1.00 &     -0.02 &     -0.01 &     -0.00 &     -0.00 &      0.03 &       0.01 \\
1.0--1.2  &      0.01 &      0.02 &      0.02 &     -0.02 &     -0.02 &      1.00 &     -0.06 &      0.02 &     -0.01 &      0.02 &       0.01 \\
1.2--1.5  &      0.03 &     -0.03 &      0.02 &      0.01 &     -0.01 &     -0.06 &      1.00 &     -0.02 &     -0.01 &      0.02 &       0.03 \\
1.5--2.0  &      0.01 &     -0.03 &     -0.04 &      0.03 &     -0.00 &      0.02 &     -0.02 &      1.00 &     -0.03 &     -0.00 &      -0.04 \\
2.0--2.5  &     -0.04 &      0.01 &     -0.03 &     -0.02 &     -0.00 &     -0.01 &     -0.01 &     -0.03 &      1.00 &      0.01 &       0.04 \\
2.5--3.0  &     -0.01 &      0.03 &      0.01 &     -0.04 &      0.03 &      0.02 &      0.02 &     -0.00 &      0.01 &      1.00 &       0.02 \\
3.0--10.0 &     -0.03 &      0.03 &     -0.05 &      0.04 &      0.01 &      0.01 &      0.03 &     -0.04 &      0.04 &      0.02 &       1.00 \\
\bottomrule
\end{tabular}

%% file: Final_StatCorrelationMatrix_dileptons_dphi_binning0.tex
\begin{tabular}{lrrrrrrrrrrr}
\toprule
$\delta \phi(Z_{1}, Z_{2})$ [$\pi$] &  0.0--0.125 &  0.125--0.25 &  0.25--0.375 &  0.375--0.5 &  0.5--0.625 &  0.625--0.6875 &  0.6875--0.75 &  0.75--0.8125 &  0.8125--0.875 &  0.875--0.9375 &  0.9375--1.0 \\
\midrule
0.0--0.125    &        1.00 &        -0.03 &        -0.06 &       -0.06 &        0.01 &           0.01 &          0.03 &          0.01 &          -0.04 &          -0.02 &        -0.03 \\
0.125--0.25   &       -0.03 &         1.00 &        -0.01 &       -0.03 &       -0.04 &           0.01 &         -0.03 &         -0.03 &           0.01 &           0.03 &         0.02 \\
0.25--0.375   &       -0.06 &        -0.01 &         1.00 &       -0.03 &       -0.03 &           0.01 &          0.02 &         -0.04 &          -0.04 &           0.01 &        -0.05 \\
0.375--0.5    &       -0.06 &        -0.03 &        -0.03 &        1.00 &       -0.04 &          -0.04 &          0.00 &          0.03 &          -0.02 &          -0.04 &         0.04 \\
0.5--0.625    &        0.01 &        -0.04 &        -0.03 &       -0.04 &        1.00 &           0.02 &         -0.04 &         -0.02 &          -0.01 &           0.03 &         0.01 \\
0.625--0.6875 &        0.01 &         0.01 &         0.01 &       -0.04 &        0.02 &           1.00 &          0.02 &         -0.03 &          -0.02 &           0.02 &         0.01 \\
0.6875--0.75  &        0.03 &        -0.03 &         0.02 &        0.00 &       -0.04 &           0.02 &          1.00 &          0.03 &          -0.05 &           0.01 &         0.02 \\
\bottomrule
\end{tabular}

%% file: Final_StatCorrelationMatrix_jets_N_binning0.tex
\begin{tabular}{lrrrrr}
\toprule
Jet multiplicity &     0 &     1 &     2 &     3 &  $\geq 4$ \\
\midrule
0        &  1.00 & -0.02 & -0.10 & -0.04 &      0.01 \\
1        & -0.02 &  1.00 &  0.11 & -0.03 &     -0.05 \\
2        & -0.10 &  0.11 &  1.00 &  0.23 &     -0.01 \\
3        & -0.04 & -0.03 &  0.23 &  1.00 &      0.21 \\
$\geq 4$ &  0.01 & -0.05 & -0.01 &  0.21 &      1.00 \\
\bottomrule
\end{tabular}

%% file: Final_StatCorrelationMatrix_centralJets_N_binning0.tex
\begin{tabular}{lrrrrr}
\toprule
Central-jet multiplicity &     0 &     1 &     2 &     3 &  $\geq 4$ \\
\midrule
0        &  1.00 & -0.11 & -0.07 & -0.02 &      0.02 \\
1        & -0.11 &  1.00 & -0.06 & -0.04 &     -0.04 \\
2        & -0.07 & -0.06 &  1.00 &  0.08 &     -0.04 \\
3        & -0.02 & -0.04 &  0.08 &  1.00 &      0.02 \\
$\geq 4$ &  0.02 & -0.04 & -0.04 &  0.02 &      1.00 \\
\bottomrule
\end{tabular}

%% file: Final_StatCorrelationMatrix_jets60_N_binning0.tex
\begin{tabular}{lrrrr}
\toprule
Jet ($p_{\mathrm{T}} > 60$ GeV) multiplicity &     0 &     1 &     2 &  $\geq 3$ \\
\midrule
0        &  1.00 & -0.09 & -0.03 &      0.05 \\
1        & -0.09 &  1.00 & -0.07 &     -0.04 \\
2        & -0.03 & -0.07 &  1.00 &     -0.09 \\
$\geq 3$ &  0.05 & -0.04 & -0.09 &      1.00 \\
\bottomrule
\end{tabular}

%% file: Final_StatCorrelationMatrix_hardestJetsDijet_mass_binning0.tex
\begin{tabular}{lrrrrr}
\toprule
$m(\mathrm{jet}_{1}, \mathrm{jet}_{2})$ [GeV] &  0--50 &  50--100 &  100--200 &  200--300 &  300--1000 \\
\midrule
0--50     &   1.00 &     0.11 &     -0.04 &     -0.01 &       0.01 \\
50--100   &   0.11 &     1.00 &     -0.03 &     -0.04 &      -0.06 \\
100--200  &  -0.04 &    -0.03 &      1.00 &      0.03 &      -0.00 \\
200--300  &  -0.01 &    -0.04 &      0.03 &      1.00 &      -0.02 \\
300--1000 &   0.01 &    -0.06 &     -0.00 &     -0.02 &       1.00 \\
\bottomrule
\end{tabular}

%% file: Final_StatCorrelationMatrix_hardestJetsDijet_dy_binning0.tex
\begin{tabular}{lrrrr}
\toprule
$|\delta y(\mathrm{jet}_{1}, \mathrm{jet}_{2})|$ &  0--1 &  1--2 &  2--3 &  3--9 \\
\midrule
0--1 &  1.00 & -0.06 & -0.06 &  0.01 \\
1--2 & -0.06 &  1.00 & -0.02 & -0.04 \\
2--3 & -0.06 & -0.02 &  1.00 & -0.04 \\
3--9 &  0.01 & -0.04 & -0.04 &  1.00 \\
\bottomrule
\end{tabular}

%% file: Final_StatCorrelationMatrix_jets_scalarPtSum_binning0.tex
\begin{tabular}{lrrrrrrr}
\toprule
Jet scalar $p_{\mathrm{T}}$ sum [GeV] &  30--60 &  60--90 &  90--120 &  120--150 &  150--200 &  200--400 &  400--1000 \\
\midrule
30--60    &    1.00 &   -0.01 &    -0.15 &     -0.11 &     -0.03 &      0.04 &       0.05 \\
60--90    &   -0.01 &    1.00 &     0.18 &     -0.09 &     -0.07 &      0.04 &      -0.04 \\
90--120   &   -0.15 &    0.18 &     1.00 &      0.12 &     -0.13 &     -0.02 &      -0.02 \\
120--150  &   -0.11 &   -0.09 &     0.12 &      1.00 &      0.13 &     -0.12 &       0.04 \\
150--200  &   -0.03 &   -0.07 &    -0.13 &      0.13 &      1.00 &     -0.09 &       0.00 \\
200--400  &    0.04 &    0.04 &    -0.02 &     -0.12 &     -0.09 &      1.00 &      -0.05 \\
400--1000 &    0.05 &   -0.04 &    -0.02 &      0.04 &      0.00 &     -0.05 &       1.00 \\
\bottomrule
\end{tabular}

%% file: Final_StatCorrelationMatrix_jet1_pt_binning0.tex
\begin{tabular}{lrrrrrrrrr}
\toprule
Leading jet $p_{\mathrm{T}}$ [GeV] &  30--40 &  40--50 &  50--60 &  60--80 &  80--100 &  100--120 &  120--150 &  150--200 &  200--800 \\
\midrule
30--40   &    1.00 &    0.35 &   -0.16 &   -0.11 &    -0.02 &     -0.01 &     -0.03 &     -0.04 &     -0.03 \\
40--50   &    0.35 &    1.00 &    0.58 &   -0.04 &    -0.05 &      0.01 &     -0.01 &     -0.02 &     -0.00 \\
50--60   &   -0.16 &    0.58 &    1.00 &    0.40 &    -0.07 &     -0.04 &     -0.01 &      0.02 &      0.01 \\
60--80   &   -0.11 &   -0.04 &    0.40 &    1.00 &     0.24 &     -0.07 &     -0.04 &      0.01 &      0.00 \\
80--100  &   -0.02 &   -0.05 &   -0.07 &    0.24 &     1.00 &      0.26 &     -0.06 &     -0.02 &     -0.00 \\
100--120 &   -0.01 &    0.01 &   -0.04 &   -0.07 &     0.26 &      1.00 &      0.22 &     -0.10 &      0.04 \\
120--150 &   -0.03 &   -0.01 &   -0.01 &   -0.04 &    -0.06 &      0.22 &      1.00 &      0.08 &     -0.07 \\
150--200 &   -0.04 &   -0.02 &    0.02 &    0.01 &    -0.02 &     -0.10 &      0.08 &      1.00 &     -0.05 \\
200--800 &   -0.03 &   -0.00 &    0.01 &    0.00 &    -0.00 &      0.04 &     -0.07 &     -0.05 &      1.00 \\
\bottomrule
\end{tabular}

%% file: Final_StatCorrelationMatrix_jet2_pt_binning0.tex
\begin{tabular}{lrrr}
\toprule
Subleading jet $p_{\mathrm{T}}$ [GeV] &  30--40 &  40--60 &  60--500 \\
\midrule
30--40  &    1.00 &   -0.02 &    -0.07 \\
40--60  &   -0.02 &    1.00 &    -0.13 \\
60--500 &   -0.07 &   -0.13 &     1.00 \\
\bottomrule
\end{tabular}

%% file: Final_StatCorrelationMatrix_jet1_eta_binning0.tex
\begin{tabular}{lrrrrrr}
\toprule
Leading jet $|\eta|$ &  0--0.5 &  0.5--1.0 &  1.0--1.5 &  1.5--2.0 &  2.0--2.5 &  2.5--4.5 \\
\midrule
0--0.5   &    1.00 &     -0.03 &      0.01 &     -0.00 &      0.03 &     -0.03 \\
0.5--1.0 &   -0.03 &      1.00 &     -0.02 &     -0.01 &     -0.02 &     -0.03 \\
1.0--1.5 &    0.01 &     -0.02 &      1.00 &     -0.00 &     -0.00 &     -0.04 \\
1.5--2.0 &   -0.00 &     -0.01 &     -0.00 &      1.00 &     -0.05 &     -0.01 \\
2.0--2.5 &    0.03 &     -0.02 &     -0.00 &     -0.05 &      1.00 &     -0.01 \\
2.5--4.5 &   -0.03 &     -0.03 &     -0.04 &     -0.01 &     -0.01 &      1.00 \\
\bottomrule
\end{tabular}

%% file: Final_StatCorrelationMatrix_jet2_eta_binning0.tex
\begin{tabular}{lrrrrrr}
\toprule
Subleading jet $|\eta|$ &  0--0.5 &  0.5--1.0 &  1.0--1.5 &  1.5--2.0 &  2.0--2.5 &  2.5--4.5 \\
\midrule
0--0.5   &    1.00 &     -0.06 &     -0.02 &     -0.03 &      0.04 &     -0.07 \\
0.5--1.0 &   -0.06 &      1.00 &     -0.06 &     -0.04 &     -0.00 &     -0.07 \\
1.0--1.5 &   -0.02 &     -0.06 &      1.00 &     -0.04 &      0.01 &     -0.08 \\
1.5--2.0 &   -0.03 &     -0.04 &     -0.04 &      1.00 &     -0.02 &     -0.05 \\
2.0--2.5 &    0.04 &     -0.00 &      0.01 &     -0.02 &      1.00 &      0.03 \\
2.5--4.5 &   -0.07 &     -0.07 &     -0.08 &     -0.05 &      0.03 &      1.00 \\
\bottomrule
\end{tabular}